\documentclass[onecolumn]{aastex62}

\usepackage[counterclockwise, figuresleft]{rotating}
\usepackage{multirow}
\usepackage{graphicx}
\usepackage{threeparttable} 
\usepackage{hyperref}
\usepackage[maxfloats=256]{morefloats}
\usepackage{wrapfig}

\makeatletter
\newcommand{\manlabel}[2]{#2\def\@currentlabel{#2}\label{#1}}
\makeatother

\newcommand\hst{{\it HST \/}}
\newcommand\HST{{\it Hubble Space Telescope \/}}

\def\simgt{\lower.5ex\hbox{$\; \buildrel > \over \sim \;$}}
\def\simlt{\lower.5ex\hbox{$\; \buildrel < \over \sim \;$}}

\graphicspath{{./}{figures/}}

\journalinfo{}

\shorttitle{M87 Novae}
\shortauthors{Shara, Lessing, Hounsell et al.}

\begin{document}

\title[M87 NUV novae]{A 9-Month {\it Hubble Space Telescope} Near-UV Survey of M87. I. Light and Color Curves of 94 Novae, and a Re-determination of the Nova Rate}

\correspondingauthor{Michael Shara  mshara@amnh.org\\This paper is respectfully dedicated to Jay Gallagher and the late Art Code, who first characterized the striking ultraviolet behaviour of a nova a half-century ago.}

\author[0000-0003-0155-2539]{Michael M. Shara}
\affiliation{Department of Astrophysics, American Museum of Natural History, New York, NY 10024, USA}

\author[0000-0001-6714-2706]{Alec M. Lessing}
\affiliation{Department of Physics, Stanford CA 94305, CA, USA}

\author[0000-0002-0476-4206]{Rebekah Hounsell}
\affiliation{University of Maryland Baltimore County, 1000 Hilltop Cir, Baltimore, MD 21250, USA}
\affiliation{NASA Goddard Space Flight Center, Greenbelt, MD 20771, USA} 

\author[0000-0002-6126-7409]{Shifra Mandel}
\affil{Department of Astronomy, Columbia University, New York City, NY 10024, USA}

\author[0009-0008-4599-2935]{David Zurek}
\affiliation{Department of Astrophysics, American Museum of Natural History, New York, NY 10024, USA}

\author[ 0000-0003-0156-3377]{Matthew J. Darnley}
\affiliation{Liverpool John Moores University, Astrophysics Research Institute, Liverpool L3 5UG, UK}

\author[0000-0002-4391-6137]{Or Graur}
\affil{University of Portsmouth, Institute of Cosmology and Gravitation, Portsmouth PO1 2UP, UK}
\affiliation{Department of Astrophysics, American Museum of Natural History, New York, NY 10024, USA}

\author[0000-0002-0023-0485]{Yael Hillman}
\affil{Technion - Israel Institute of Technology, Department of Physics, Haifa 3200003, Israel} 

\author[0000-0002-7676-9962]{Eileen T. Meyer}
\affil{University of Maryland Baltimore County, Department of Physics, Baltimore, MD 21250, USA }

\author[0000-0003-3457-0020]{Joanna Mikolajewska}
\affil{Nicolaus Copernicus Astronomical Center, Warsaw, 00-716, Poland}

\author[0000-0002-0466-1119]{James D. Neill}
\affil{California Institute of Technology,Division of Physics, Math \& Astronomy, Pasadena, CA 91125}

\author[0000-0002-6317-4839]{Dina Prialnik}
\affil{Tel-Aviv University, Department of Earth and Planetary Sciences, Ramat-Aviv, Israel 6997801}

\author[0000-0002-9011-6829]{William Sparks}
\affil{SETI Institute, Mountain View, CA 94043, USA}

\begin{abstract}

M87 has been monitored with a cadence of 5 days over a 9 month-long span through the near-ultraviolet (NUV:F275W) and optical (F606W) filters of the Wide Field Camera 3 (WFC3) of the \HST. This unprecedented dataset yields the NUV and optical light and color curves of 94 M87 novae, characterizing the outburst and decline properties of the largest extragalactic nova dataset in the literature (after M31 and M81).  We test and confirm nova modelers' prediction that recurrent novae cannot erupt more frequently that once every 45 days; show that there are zero rapidly recurring novae in the central $\sim$ 1/3 of M87 with recurrence times $ < $ 130 days; demonstrate that novae closely follow the K-band light of M87 to within a few arcsec of the galaxy nucleus; show that nova NUV light curves are as heterogeneous as their optical counterparts, and usually peak 5 to 30 days after visible light maximum; {  determine our observations' {annual} detection completeness to be {71 - 77\%}; and measure the rate Rnova of nova eruptions in M87 as {$352_{-37}^{+37}$}/yr. The corresponding luminosity-specific classical nova rate for this galaxy is {$7.91_{-1.20}^{+1.20}/yr/10^{10}L_\odot,_{K}$}.} These rates confirm that ground-based observations of extragalactic novae miss most faint, fast novae and those near the centers of galaxies. An annual M87 nova rate of 300 or more seems inescapable. A luminosity-specific nova rate of $\sim$ $7 - 10/yr/10^{10}L_\odot,_{K}$ in {\it all} types of galaxies is indicated by the data available in 2023.

\end{abstract} 

\section{Introduction and Motivation} \label{sec:intro}
All cataclysmic variables (CVs) are binaries containing a white dwarf (WD) which accretes matter from a close companion. A nova eruption is a luminous (up to $10^6$ $L_{\Sun}$) transient that erupts when the envelope accreted onto the WD's surface undegoes a thermonuclear runaway.  The recurrence rate, peak luminosity, and brightness decay timescale of a nova depend on the WD mass and the binary mass transfer rate during the time (usually millenia) between nova eruptions \citep{Yaron2005,Hillman2016,Hillman2020,Hillman2021}, as well as the chemical compositions of the two stars. 

Novae are our only means of detecting and studying studying CV populations (and indeed most binary populations except for X-ray binaries) in galaxies beyond the Local Group. Differences in CV populations in different types of galaxies would indicate different binary fractions and/or stellar evolution pathways. Additionally, very rapidly accreting WDs in nova binaries can give rise to exploding WD ``standard candle'' type Ia supernovae \citep{Maoz2014,Hillman2016,Jha2019,Liu2023}. Thus the importance of CVs extends beyond the field of binary star formation and evolution to cosmology. 

 Despite CVs' importance, a lack of consensus on one of the most basic parameters that characterize them - the annual nova eruption rates in galaxies - has persisted for two decades. \citet{Shafter2000,Shafter2014,Shafter2021} claimed that the luminosity specific nova rates (LSNR, i.e. annual rate of novae per unit K-band luminosity) in different galaxy types are all similar, $\sim$ 1--3 novae/year/$10^{10}L_{\Sun,K}$ (solar luminosities in the K-band). This conclusion is based on relatively time-sparse, ground-based optical surveys of multiple galaxies, most recently summarized in \citet{Shafter2021}. 
 
 Population synthesis studies of \citet{Matteucci2003}, \citet{Claeys2014} and \citet{Chen2016} predicted a very different behaviour: spiral, and especially starburst galaxies should exhibit an order-of-magnitude higher nova rates and LSNR than elliptical galaxies. This is because newborn binaries containing high-mass WDs should be most common in spiral and starburst galaxies characterised by recent massive star formation. Novae which erupt on those high-mass WDs need only accrete relatively low-mass envelopes in order to initiate thermonuclear runaways \citep{Shara1981,Yaron2005}, hence they outburst more frequently than those associated with the mostly low mass WDs in the nova binaries of elliptical galaxies.

A daily imaging \HST (\textit{HST}) - based survey of the massive elliptical galaxy M87 \citep{Shara2016}, spanning 72 days, showed that ground-based surveys of external galaxies fail to detect fainter novae, those with short decline times and those near the bright centers of galaxies. {\it These effects cause ground-based surveys to systematically and significantly underestimate the true nova rates in galaxies.} The \hst- determined LSNR of M87 was shown to be $7.88_{-2.6}^{+2.3}$ novae/yr/$10^{10}L_\odot,_{K}$ \citep{Shara2016}. This is 2-4 times larger than previous ground-based surveys' results. \citet{Mroz2016} demonstrated that the LSNR in the Large Magellanic Cloud (LMC) is much higher than previous ground-based estimates, thereby confirming that it is comparable to the M87 LSNR. \citet{De2021} discovered a sizeable population of Galactic novae (in the infrared) that have gone undetected in over a century of optical searches, and \citet{Kawash2021} found that approximately half of all Galactic novae are hidden by extinction from current surveys. Most recently, \citet{Mandel2023} used a year-long \hst survey of M51 to determine its nova rate to be $172_{-37}^{+46}$ novae/yr,  corresponding to a luminosity specific nova rate (LSNR) of $10.4_{-2.2}^{+2.8}$ novae/yr/$10^{10}L_\odot,_{K}$. Both of these rates are $\sim\times10$ larger than the ground-based-determined nova rates for M51 \citep{Shafter2000}. 

These discoveries (of much higher than previously claimed LSNR) in a giant elliptical (M87), a barred spiral (the Galaxy), a dwarf irregular galaxy (the LMC), and a giant Sc-type spiral galaxy (M51) were carried out via surveys with much longer baselines, denser time coverage and/or deeper magnitude limits than all previous surveys. They argue strongly against the claim that the LSNR is relatively low ($\sim$ $1 - 3$ novae/yr/$10^{10}L_\odot,_{K}$) in all galaxies, as the earlier, shallower and sparser cadence coverage suggested. 

\hst is especially well-suited to detecting extragalactic novae  because of its unparalleled angular resolution and consequent sensitivity, its very small and nearly-constant point-spread function, its insensitivity to lunar phase and its immunity to atmospheric seeing. In addition, \hst operates effectively in the near-ultraviolet (NUV), a property that has only rarely been exploited in extragalactic nova searches \citep{Sohn2006, Madrid2007}. Novae erupting on WDs with masses $\gtrsim$ 1.0 $M_\odot$ are expected to be NUV-bright \citep{Hillman2014}, and NUV observations greatly suppress the light of red giants, which dominate the optical output of elliptical galaxies, so that novae even near ellipticals' bright cores should be detectable in the NUV.

Motivated by the high nova rate in M87 that only the \HST could have determined, we applied for and were awarded 53 \hst orbits (GO-14618, PI:M. Shara)  to survey that galaxy for transients with a 5 day cadence for 9 months. Among the questions we proposed to answer were:

1. Do novae continue to follow the light of M87 all the way to the galaxy nucleus? Would a definitive measurement of the M87 nova rate, using an optimal set of filters (both NUV and visible) change the remarkably high rate? 

A non-optimal (for novae) choice of filters (F814W and F606W, chosen for detecting microlensing in M87) meant that even the \citet{Shara2016} \hst survey of M87 for novae is incomplete within 20 arcsec of its bright nucleus. 

2. How do the NUV light curves differ from the optical light curves of novae? What is the distribution of time differences of maximum luminosity in NUV and visible light? Are these correlated with other nova properties?

Only nine UV nova light curves have ever been observed: one via the Orbiting Astronomical Observatory \citep{Gallagher1974} and eight via the \textit{Galaxy Evolution Explorer (GALEX)} satellite \citep{Cao2012}. In groundbreaking work a half-century ago, the nova FH Serpentis was shown to brighten in NUV light much later than in the optical \citep{Gallagher1974}, while four decades later \citet{Cao2012} detected two novae in M31 (of eight with visual and NUV light curves) that achieved peak brightness in the NUV {\it before} visible maximum. Theoretical UV (and optical) light curves have been published \citep{Hillman2014}, but no large-scale test of them has been possible due to the paucity of observed NUV nova light curves. Extrapolating from \citet{Shara2016}, of order 100 NUV nova light curves should emerge from a 9-month \hst survey.

3.  Do ultra-rapidly recurring novae exist? 

\citet{Hillman2015}'s models of the most massive, rapidly accreting WDs (1.399 $M_\odot$ accreting near the Eddington limit) predict that novae can never recur more frequently than once every 45 days, and that such rapidly recurring novae are extremely NUV-bright. The 260 day baseline of 5 day cadence observations of M87 of GO-14618 is sufficient to detect any such ultra-rapidly recurring novae multiple times, which would be a serious challenge to the theory and models of novae.

Section \ref{sec:data} describes the data collected during the M87 \hst observing campaign. In Section \ref{sec:novae} we describe our searches for and identifications of nova candidates and their properties. We derive the nova rate in M87 in Section \ref{sec:sim_and_rate}, where we compare it to previous measurements. In section \ref{sec:RRNs} we place constraints on the possible incidence of the most rapidly recurring novae. We use our large sample of novae to contrast their NUV and optical behaviors in Section \ref{sec:NUV-Behaviour}. We present our other findings in sections \ref{sec:mag_decline} and \ref{sec:radial_dist_cors}, and summarize our results in Section \ref{sec:conclusion}. In the \hyperref[sec:appendices]{Appendix} we present the tabular data that describes all 94 novae; display a montage of the field of each nova in each filter at each epoch; and the corresponding light and color curves of all of the novae of this study.

\section{\hst Imaging Data} \label{sec:data}

The \hst observing campaign of M87 (Proposal ID = 14618; PI: Shara) was conducted over the course of 260 days using the Wide Field Camera 3 (WFC3) $F275W$ and $F606$ filters (hereafter U and V respectively). The first observations were taken on 13 November 2016, with the last completed on 31 July 2017. During each of the 53 epochs \hst was scheduled to observe the center of M87 for a total 720 sec exposure in the F606W filter and 1500 sec in the F275W filter; just a few epochs were 1-3\% shorter in exposure time. Figure \ref{fig:finderchart} shows the \hst fields of view (FOV) of those 53 epochs.  Note that the FOV rotates to maintain optimal pointing of \hst's solar panels throughout the course of the year. Because of this rotation some novae were rotated into or out of the \hst FOV during their eruptions, and some novae were almost certainly entirely missed (see section 4). We refer to the area {within 81.1'' (the half-width of the WFC3 UVIS chip) of M87's nucleus as the ``inner cirlce'' (see Figure \ref{fig:finderchart}).}

The survey's magnitudes are on the \textit{STMAG} system.  All absolute magnitudes were computed using an M87 distance modulus of {31.03 \citep{deGrijs2019}} with Galactic extinctions in the direction of M87 of $A_{606} = 0.05$ and $A_{275} = 0.12$ mags \citep{Harris2009,Cardelli1989}.

\section{M87 Nova Search and identification} \label{sec:novae}

We independently conducted two separate searches for novae in the 53 epochs' images of M87: a search based on visual inspection of difference images and a search based on statistical classification of photometric results.

\subsection{Difference Images}

All WFC3/UVIS data were reduced using the STScI \texttt{calwf3} pipeline \citep{Dressel2019}. Individual FLC exposures were aligned and combined using Dizzlepacs's \texttt{astrodrizzle} package to create individual DRC epoch images in each filter. All images were aligned to the same WCS and pixel scale to easily enable comparison and subtraction.

Next, template epochs were established using combined DRC data at both an early and late times during the observation cycle. These images were then subtracted from each epoch DRZ image and the resultant subtracted images used to search for nova events. 

The search for novae from the subtracted data set was conducted both manually via visual inspection and via the use of SExtractor \citep{Bertin96}. This resulted in the detection of 125 nova candidates.

\newpage

\subsection{The Statistical Method}
The statistical search proceeded in multiple steps. First, we used astrodrizzle to combine all {FLC} images of a given filter in a given epoch (so-called “1epoch” images) or set of three subsequent epochs (so-called “3epoch” images). Additionally, we created median images of all the data from all epochs in each filter. For each use of astrodrizzle, we used its cosmic ray cleaning functionality to eliminate cosmic rays. Other preliminary targets for photometry were identified running DAOFIND with liberal rejection criteria on all drizzled images. Due to the strong gradient in the background because of the presence of M87 in the images, 10 annuli centered on M87’s center were created. For each image, and for each annulus, an average sky standard deviation was calculated and used for the DAOFIND statistical significance criteria for detection within that annulus. This process yielded 27,061 detected sources, most of which were globular clusters, giant stars within M87, resolved features of M87, background galaxies, random statistical fluctuations of the background, and residual noise or cosmic rays. 

On each of these targets, we used PyRAF’s PHOT function to measure magnitudes in 1epoch and 3epoch images in both filters. We applied a differential correction to photometric measurements of 3epochs, finding the average change in magnitude of the hundred nearest sources of similar magnitude to every source and subtracting out overall fluctuations from each source’s light-curve. We then calculated an average statistical variation $\sigma$ in the magnitude of each light-curve. To find transients, light-curves with {  either F606W or F275W } peak magnitudes greater than 3$\sigma$ above the median in one passband or $2\sigma$ in both passbands were selected as candidates. This eliminated the vast majority of candidates. Each remaining candidate was examined by eye to eliminate remaining noise or cosmic ray events. These were evident through highly irregular or resolved PSFs upon visual examination, or large differences in the magnitudes measured in {FLC} images from the same epoch and passband. This left a list of 122 candidate novae.

\subsection{The combined list of novae}\label{ssec:comb-list}
The lists from the two different search methods were then combined, yielding a total of 151 candidates. We then closely inspected each of these, eliminating candidates that only appeared in a single image {(in one band and epoch)}; that had irregular PSFs; were too dim to be confident of photometric statistical significance given local background characteristics, particularly those candidates that only appeared in one epoch; or had light-curve or color characteristics inconsistent with being a nova. This left the final list of 94 novae presented in this paper.

There was excellent overlap between the lists generated by visual inspection of difference images and the statistical method, suggesting detection efficiency near the limit of what is possible with the dataset. Of the final 94 novae, 89 were found using visual inspection of difference images and 91 were found using the statistical method. Notably, the statistical method failed to detect a few novae very close to M87’s jet, whereas difference imaging missed a few novae close to M87's bright nucleus.

\begin{figure}
    \centering
    \includegraphics[width=5.2in]{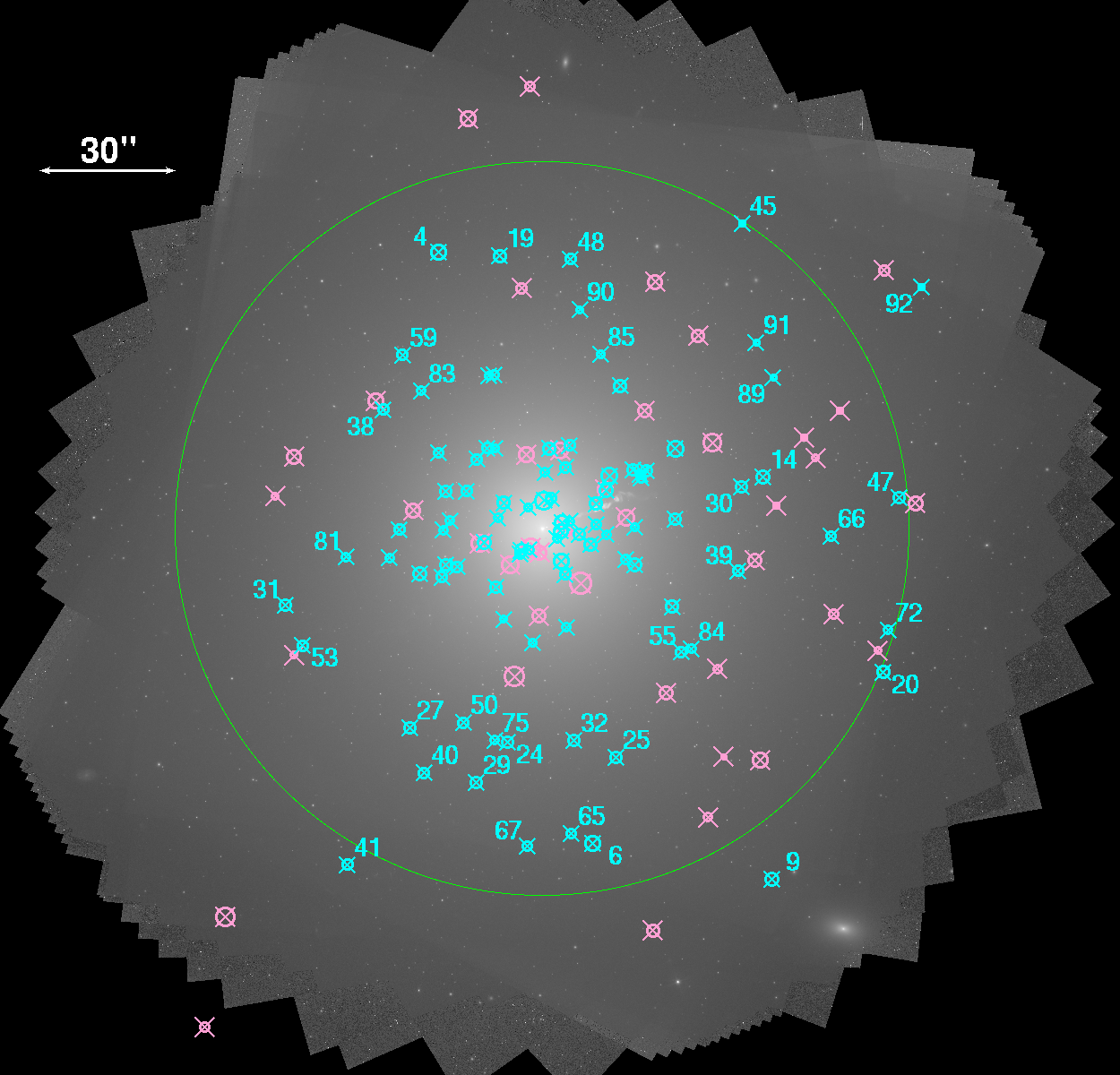}
    \vspace{0.7in}
    \includegraphics[width=5.2in]{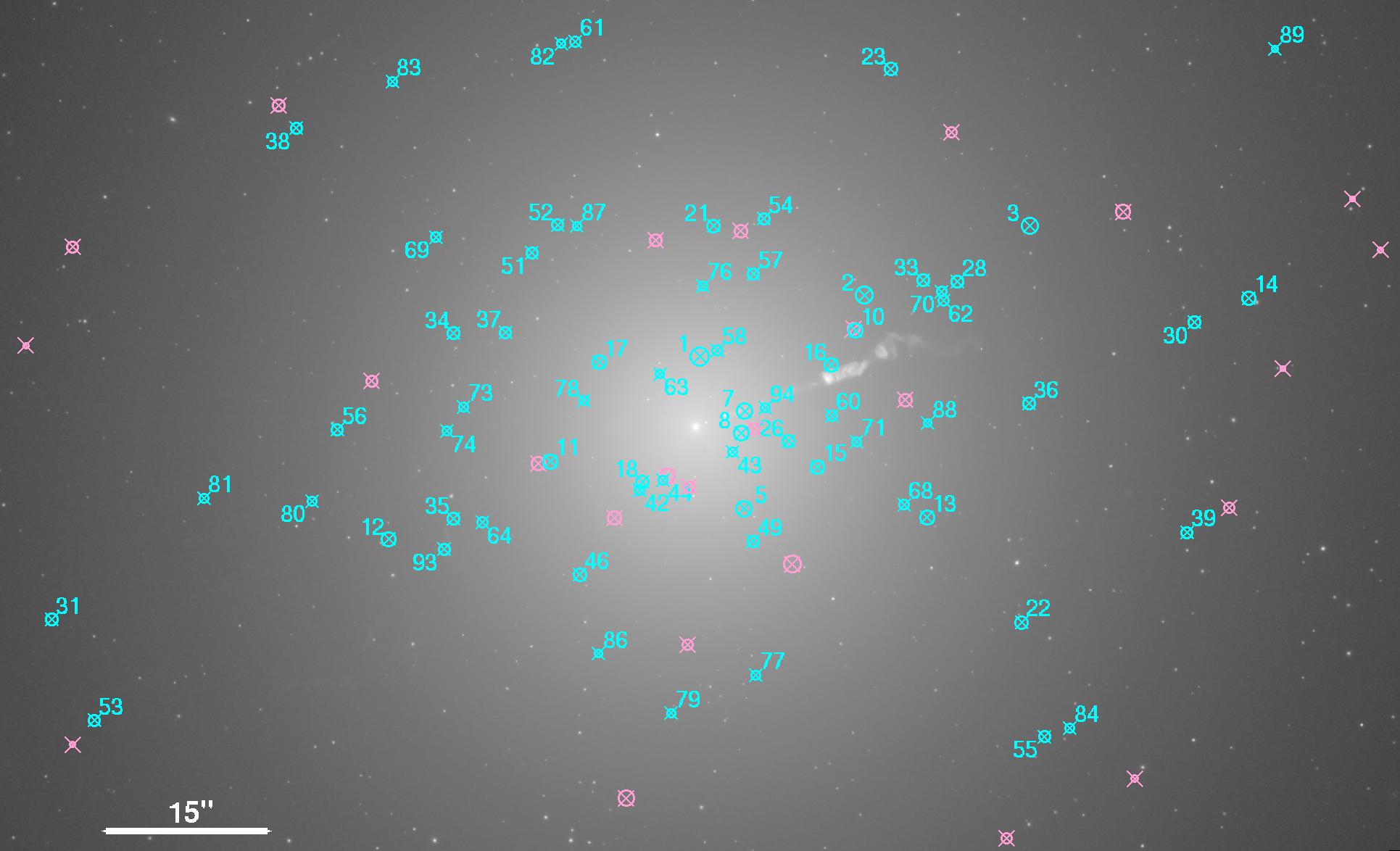}
    \vspace*{-1.5cm}
    \caption{Top: The Field of View of 53 \HST pointings, and locations (cyan crosses) of all 94 novae detected in M87. North is up and East is left. Also shown as pink crosses are the 32 {``certain"} novae of \citet{Shara2016}. The size of each nova's circle scales linearly with the brightest observed $F606W$ magnitude of that nova. Markers for novae whose peaks were not observed do not have a circle. {The region encompassed by the large green circle is the ``inner circle" defined in section \ref{sec:data} and used throughout the paper.} Bottom: A close-up of the nuclear region of M87 and its novae.}
    \label{fig:finderchart}
\end{figure}

The locations of all 94 novae (and those of 32 {``certain''} novae from the \citet{Shara2016} survey) are shown in Figure \ref{fig:finderchart}. The log of observations is given in Table \ref{tab:obs}. The positions, magnitudes, colors and rates of decline of the novae are listed in Tables \ref{tab:summary}, and photometric measurements are presented in Table \ref{tab:big_phot}. 

\newpage
\pagebreak

\section{M87 Nova rate} \label{sec:sim_and_rate}

To measure the nova rate in M87, we must first determine our survey's incompleteness: the fraction of novae that erupted within {\it HST}'s field of view in M87 during our survey but which were not detected. Peak luminosity, decline time, and the \emph{shape} of a nova light curve all play a significant role in an individual nova's detectability, as demonstrated in Figures 5 and 6 of \citet{Mandel2023}. The rates of change in luminosity as well as the shapes of light curves vary significantly among well-sampled Galactic novae \citep{Strope2010}. Thus the regularly spaced epochs of this survey must be convolved with a set of realistic light curves, representative of M87 novae, to determine our survey's incompleteness, as described below. 

\subsection{Limiting Magnitudes of Detectable M87 Novae} \label{ssec:lim_mag_detectable}

\begin{figure}[h]
    \centering
    \includegraphics[width=6in]{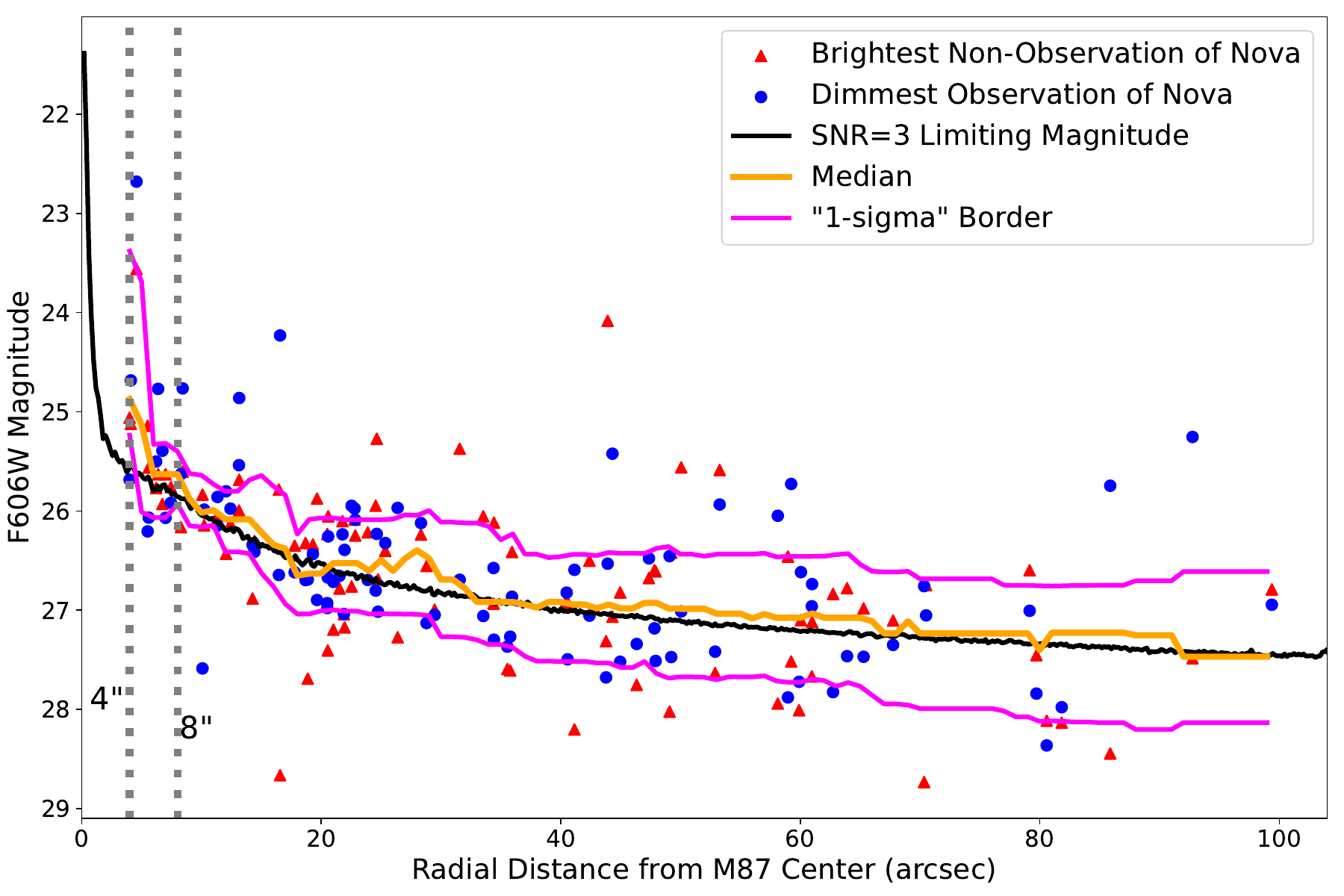}
    \caption{The dimmest F606W magnitude {(blue filled circle)} at which each of the 94 M87 novae was observable via direct inspection of the 1epoch images and brightest magnitude {(red filled triangle)} where each was not observable. {The median (orange curve) and ``1-sigma'' (15.9th and 84.1th) percentiles (magenta lines) of these points are plotted as a function of radial distance.} Also plotted (the black curve) is the magnitude of a point source that would have SNR=3 given the average local background noise measured in the F606W 1epoch images at a given radial distance from the galactic center. {The SNR=3 line is observed to match the median line well outside of the inner 8''.} See text for details.}
    \label{fig:cutoff-line}
\end{figure}

To determine a detectability criterion for our simulations we estimated a cutoff magnitude, fainter than which a nova would not reliably be considered observable (visually distinguishable from noise) to a human inspector in a given epoch, in both F606W and F275W bandpasses, as a function of distance from M87's center. The distribution of light in M87 is close to radially symmetric \citep{Harris1978} within a few arcminutes of M87’s core (which covers our entire FOV), so its galactic radius can be calibrated as a good proxy for local limiting magnitude. We plotted the dimmest and brightest magnitudes, as measured through aperture photometry, at which our human inspector marked each nova as detectable and nondetectable as a function of radial distance (see Figure \ref{fig:cutoff-line}). At a given radius, the central magnitude at which the distributions of the two sets of points overlapped was taken as an estimate for the local limiting magnitude by the confident human detectability criterion, which is the ultimate criterion we used in the real survey. Outside the inner 8{'' from the nucleus }the local SNR=3 limiting magnitude as a function of radius approximated the center of the overlap region well. This limiting magnitude was computed by measuring the image background noise as a function of distance from the galactic center and determining the magnitude of a point source that would have an SNR of 3 when placed upon that background. 

Inspection of Figure \ref{fig:cutoff-line} and examination of the 1epoch images showed that the SNR=3 curve overestimated the local human detection cutoff magnitude closer than 8 arcsec from M87's nucleus. This is likely due to the effects of the very strong gradient of the M87 galactic background on the PSFs of stars very near the galactic center. Accordingly, we defined the SNR=3 line as the simulation cutoff magnitude outside 8 arcsec, while between 4 arcsec and 8 arcsec we estimated the cutoff based on empirical detection/nondetection of novae. In the inner 3-4 arcsec of M87 novae are nearly undetectable with the current dataset.

\subsection{Placement of Simulated Novae}

We selected coordinates for simulated novae within the area of our study's footprint where the detectability of novae could be quantified -- inside {the ``inner circle"} region {(shown in Figure \ref{fig:finderchart}) with}  a radius of 2048 WFC3 pixels, or half the detector's width, from M87's center and outside 4'' from M87's center. Coordinates were sampled with probabilities proportional to M87's local 2MASS K-band surface brightness, in accordance with the observation that the distribution of novae closely follows the K-band light in M87 (\citet{Shara2016} and see below).

\subsection{Nova Template Light curves}

\begin{wrapfigure}[34]{r}{0.55\textwidth}
    \centering
    \hspace{-0.4cm}
    \includegraphics[width=3.8in]{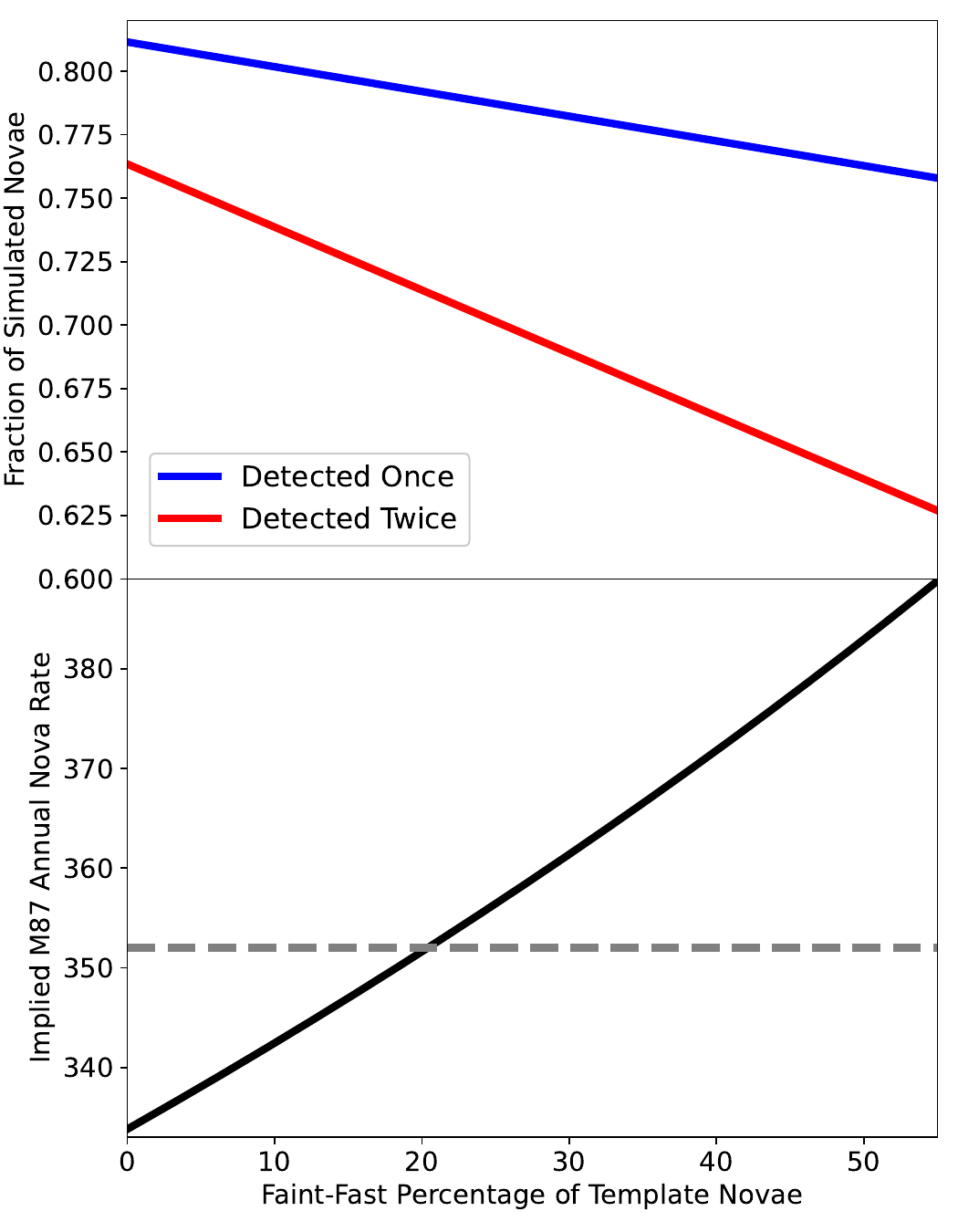}
    \caption{{Top: The fraction of simulated novae detected once (blue) and twice (red) in F606W as a function of the percentage of template novae that are faint-fast ($t_2 < 10$ days and peak $M(F606W) < -7$). Bottom: the annual nova rate for the entirety of M87 implied by the recovery fraction, following the procedure of section \ref{subsec:deteff}. Also shown (horizontal line) is the annual nova rate adopted in section \ref{subsec:deteff}.}}
    \label{fig:faint-fast-frac}
\end{wrapfigure}

While exquisitely detailed light curves exist for hundreds of Galactic novae, observational bias results in very few faint, fast novae \citep{Kasliwal2011} being included in the Galactic sample. Over 1000 novae have been detected in M31, and excellent visible-light light curves of novae there include very long-duration, faint novae, and faint-fast novae \citep{Kasliwal2011}. A remarkable $\sim$50\% of the \citet{Kasliwal2011} M31 novae are of the faint-fast variety, and these novae are also ubiquitous in M87 \citep{Shara2016}. To be conservative we drew the 59 best-sampled template light curves for our simulation from the \citet{Mandel2023} compilation of best-sampled M31 novae, which are mostly bulge novae and which includes only $\sim$ {21}\% faint-fast novae. (Our nova detection completeness fraction, used to derive the M87 nova rate is weakly dependent on the faint-fast nova fraction that we adopt; a 50\% adopted faint-fast nova fraction would have lead to a few percent higher incompleteness fraction and {an 8\%} percent higher deduced nova rate, {as shown in Figure \ref{fig:faint-fast-frac}}). We corrected the 59 nova light curves to the distance and reddening of M87 \citep{Shara2016}. Light curves were discarded if they had less than 20 days of complete data and/or did not reach as faint as an F606W magnitude of 26.5. We used $g$, $r$, and $m_{pg}$ light curves, and assumed that novae have colors close enough to 0.0 \citep{vandenbergh1987,Shara2016} that we could use these data to simulate M87 nova $F606W$ light curves.

\newpage

\subsection{Detection Fractions}

Each of {5}00,000 simulated novae was randomly assigned both a template light curve and a day of peak brightness during {a one year interval begining 80 days before the start of our survey's 260 day window.} The template light curves were used to determine the magnitude of each simulated nova in each of the 53 epochs. Simulated novae were deemed ``detectable'' in a given epoch if they were brighter than the local cutoff magnitude (see section \ref{ssec:lim_mag_detectable} and Figure \ref{fig:cutoff-line}).

{71.1\%} of simulated novae in the surveyed M87 area were detectable in at least two visible epochs. A further {8.0\%} of were detectable in precisely one visible epoch; this would have warranted exclusion as a nova candidate {under our real survey's criterion that a nova be observed at least twice (see section \ref{ssec:comb-list})}. Using our own nova F275W + F606W light-curves as templates for {this 8.0\%} of once-only detected novae, we estimate $\sim$ {75}\% to have also been detectable at least once in the near-ultraviolet (and thereby to have been confirmable using our real survey's ``seen twice'' criterion), for an overall annual detection fraction of {$71.1\text{\%} + 75\text{\%} \cdot 8.0\text{\%} = 77.1\text{\%}$. We also note that 95.8\% of novae that peaked during the 260 day survey window were detected.}

\subsection{The M87 Nova Rate} \label{subsec:deteff}

The simulations described in section \ref{sec:sim_and_rate} indicate that between {71.1 and 77.1} percent of nova eruptions {in a one year interval}, inside the inner circle and further than 4{''} from the nucleus, were detected. Of the 94 novae detected in our survey, {90 were in this region}.

{To model} the actual number of novae that peaked in this region {in the one-year interval}, given that {90} were detected with a detection rate between {71.1} and {77.1} percent, {we consider how many novae would have to have peaked in order for us to have detected 90. This quantity} is modelled by a $\Gamma(90, .711\text{ or }.771)$ random variable, {which} implies that with 68.2 percent (``1-sigma'') confidence the average {annual} nova {rate} in the {inner circle} and outside 4{''} is {between }$116.7_{-12.3}^{+12.3}$ and $126.5_{-13.3}^{+13.3}$. {We adopt a simple average of these two rates -- $121.6_{-12.8}^{+12.8}$ -- as our best estimate of the annual nova rate in the inner circle outside the inner 4''.}

 An entire WFC3 frame covers only the central portion of M87, so there is no  region in our images that we can use to empirically determine the sky background. Thus we drizzled all of the $F606W$ images in our study to create a master $F606W$ image, and then used a least squares fit to determine the sky background level and photometric zeropoint needed to fit this drizzled image to the M87 ellipsoidal surface brightness photometry profile of \citet{Kormendy2009}. We then measured the $F606W$ magnitude of M87's light within the {``inner circle'' and } outside 4'' from the nucleus to be m(F606W) = 9.45. \citet{Kormendy2009}'s published total M87 V magnitude is 8.30, so that 34.5 percent of M87's light is contained in {this region}. 

Applying this correction, we find that the annual nova rate within all of M87 is {Rnova = $352_{-37}^{+37}$/yr. }

This measurement of the overall nova rate is in very good agreement with the finding of \citet{Shara2016}, whose shorter-duration, 72-day survey detected 32 ``certain'' and 9 ``possible'' novae, yielded Rnova = $363_{-45}^{+33}$/yr. It is more than 3X larger than the M87 nova rate of the much sparser-cadence, ground-based rate published by \citet{Shafter2000} ($91\pm 34$/yr) and more than double the ground-based rate of \citet{Curtin2015} ($154_{-19}^{+23}$/yr). 

\subsection{The Luminosity-Specific Nova Rate of M87}

By adopting an M87 distance of $15.2 \pm 1.4$ Mpc, \citet{Shafter2000} derived a K-band luminosity for M87 of $39.8 \pm 8.2 $ x $10^{10}L_\odot,_{K}$. Correcting that luminosity to our adopted distance of {$16.07 \pm 1.03$ Mpc \citep{deGrijs2019}}, and combining with this study's nova rate of {$352_{-37}^{+37}$/yr}, we derive an M87 luminosity-specific nova rate (LSNR) of $7.91_{-1.20}^{+1.20}/yr/10^{10}L_\odot,_{K}$. This agrees closely with the value of \citet{Shara2016}: $7.88_{-.98}^{+.72}/yr/10^{10}L_\odot,_{K}$.

\subsection{M87 nova rates - Two recent criticisms answered}
In a response to the $363_{-45}^{+33}$/yr rate of \citet{Shara2016}, \citet{Shafter2017b} undertook an independent review of the {\it HST} dataset. They stated that...``Our results are in broad agreement with those of Shara et al., although we argue that the global nova rate in M87 remains uncertain, both due to the difficulty in identifying bona fide novae from incomplete light-curves, and in extrapolating observations near the center of M87 to the entire galaxy. {\it We conclude that nova rates as low as $\sim$200 per year remain plausible}.'' (Italics are ours.) We respond to these two suggestions as follows.

1. Almost all of the 94 novae reported in the present work are detected in both F606W {\it and} F275W images. (The few missing NUV light and color curves belong to novae which erupted late in our 9-month observing window. Our observations ended before these novae became detectable in the NUV). These transients' {F275W - F606W} colors are so blue (typically m(F275W) - m(F606w) = -2; see below) that we can preclude their being anything but classical novae, dwarf novae or AGN in eruption. Their spatial distribution follows the light of M87 so closely (see below) that they cannot be foreground dwarf novae or background AGN. They {\it can only} be erupting classical novae in M87.

\begin{figure}[h]
    \centering
    \includegraphics[width=5.3in]{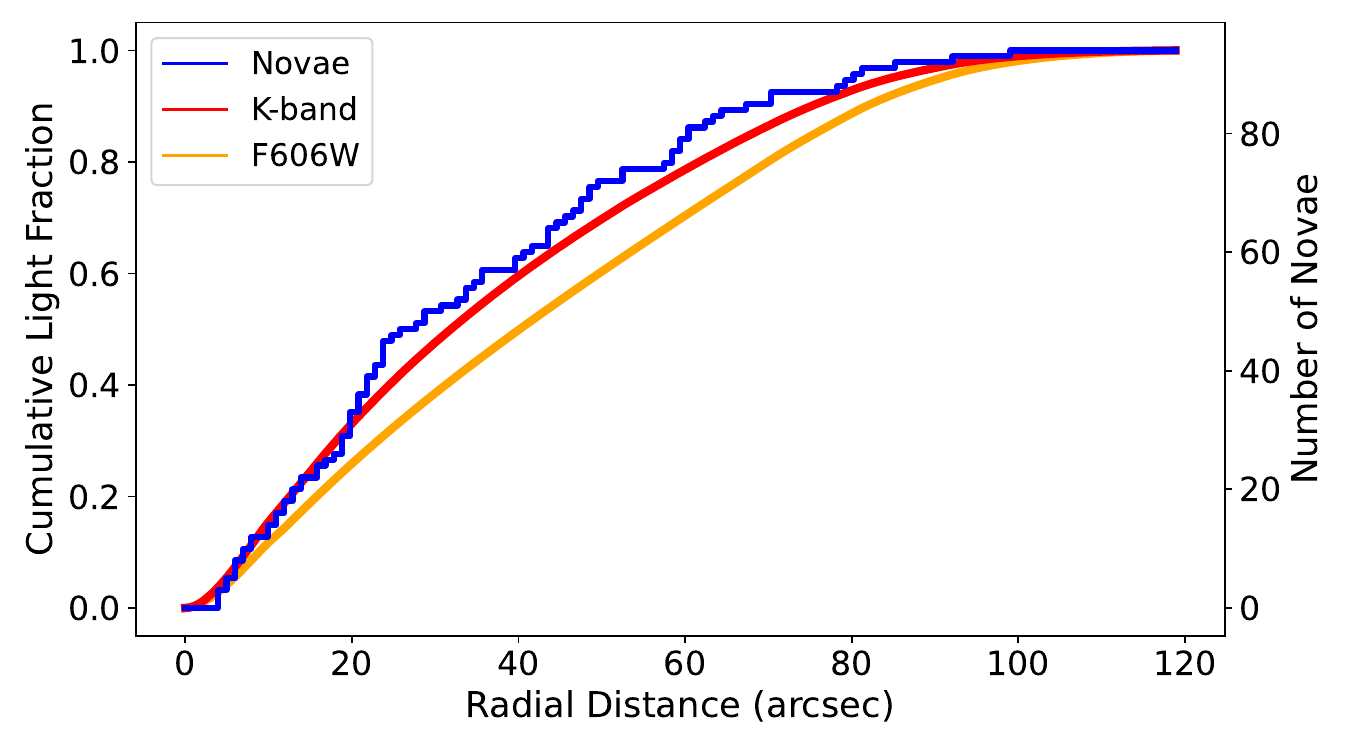}
    \caption{The normalized radial distributions of 94 M87 novae, and the K-band and F606W-band light of the galaxy. The novae are seen to closely track the galaxy K-band light. This is confirmed by the KS test which returns the statistic p = 0.61.}
    \label{fig:cumul}
\end{figure}

2. Figure 3 of \citet{Curtin2015} demonstrated that novae follow the light of M87 with high fidelity (KS test statistic = 0.81) from $\sim$ 1 arcmin out to 10 arcmin. In Figure \ref{fig:cumul} we plot the cumulative distributions of novae from this study, as well as the K-band and visible light in M87. The 94 novae we have detected follow the K-band light of M87 closely (KS test statistic = 0.61). In particular, there is no discontinuity in the cumulative number of novae in the radial distance range 1' to 2' where our data overlaps those of \citet{Curtin2015}. We conclude that extrapolating nova rates in the region encompassing the inner $\sim$ 1/3 of M87's light to the whole galaxy is entirely justified by Figure \ref{fig:cumul} of this paper and Figure 3 of \citet{Curtin2015}. 

In summary, the extraordinarily blue F275W - F606W colors {\it and} spatial concentration around M87 of the transients reported in this paper uniquely identify these objects as erupting novae. The lack of any discontinuity in the cumulative radial distributions of M87 novae, which closely follow the K-band light in ground-based and space-based nova samples stretching from 4 arcsec to 10 arcmin from M87's nucleus, argues strongly that the {\HST}- determined nova rate apply throughout M87.  

{We again emphasize that ground-based, sparsely sampled surveys hampered by the moon, clouds, variable seeing and irregular cadence, as well as simplistic simulations which model novae instead of using real-world nova light curves, and which omit faint-fast novae, have all contributed to very significant underestimates of nova rates in galaxies.} 

An annual M87 nova rate of 300 or more seems inescapable. {A luminosity-specific nova rate of 
$\sim$ $7 - 10/yr/10^{10}L_\odot,_{K}$ in {\it all} types of galaxies is indicated by the dense time-coverage and \hst data available in 2023.}

\section{Rapidly recurring novae}\label{sec:RRNs}

As the mass of a white dwarf approaches the Chandrasekhar mass, the accreted envelope mass required to trigger a thermonuclear runaway decreases monotonically (see Figure 4 of \citet{Hillman2016}). The simulations indicate that the time between nova eruptions can become as short as 45 days in the final years before a WD erupts as an SNIa. As of 2023, the most rapidly recurring nova known is located in M31 \citep{Darnley2014}. That nova, M31-2008-12a, erupts annually. One of the prime scientific drivers of the current study was to find, or place strong limits on the number of even more rapidly recurring novae in M87.

The time baseline of the present survey is 260 days, so all novae recurring more frequently than every 130/86/65/52 days should have been seen to erupt at least twice/3 times/4 times/5 times. None of the 94 novae detected in the current survey erupted more than once.
We conclude that there are zero novae (in the inner $\sim$ 1/3 of M87) recurring more frequently than once every 130 days.  We defer modelling and a detailed discussion to place stringent limits on more infrequently recurring novae in M87 to a subsequent paper.

\section{The NUV, Visible and NIR Light behaviors of Novae} \label{sec:NUV-Behaviour}

\subsection{NUV and Visible Light Curves}

\begin{figure}[h]
    \centering
    \includegraphics[width=5.7in]{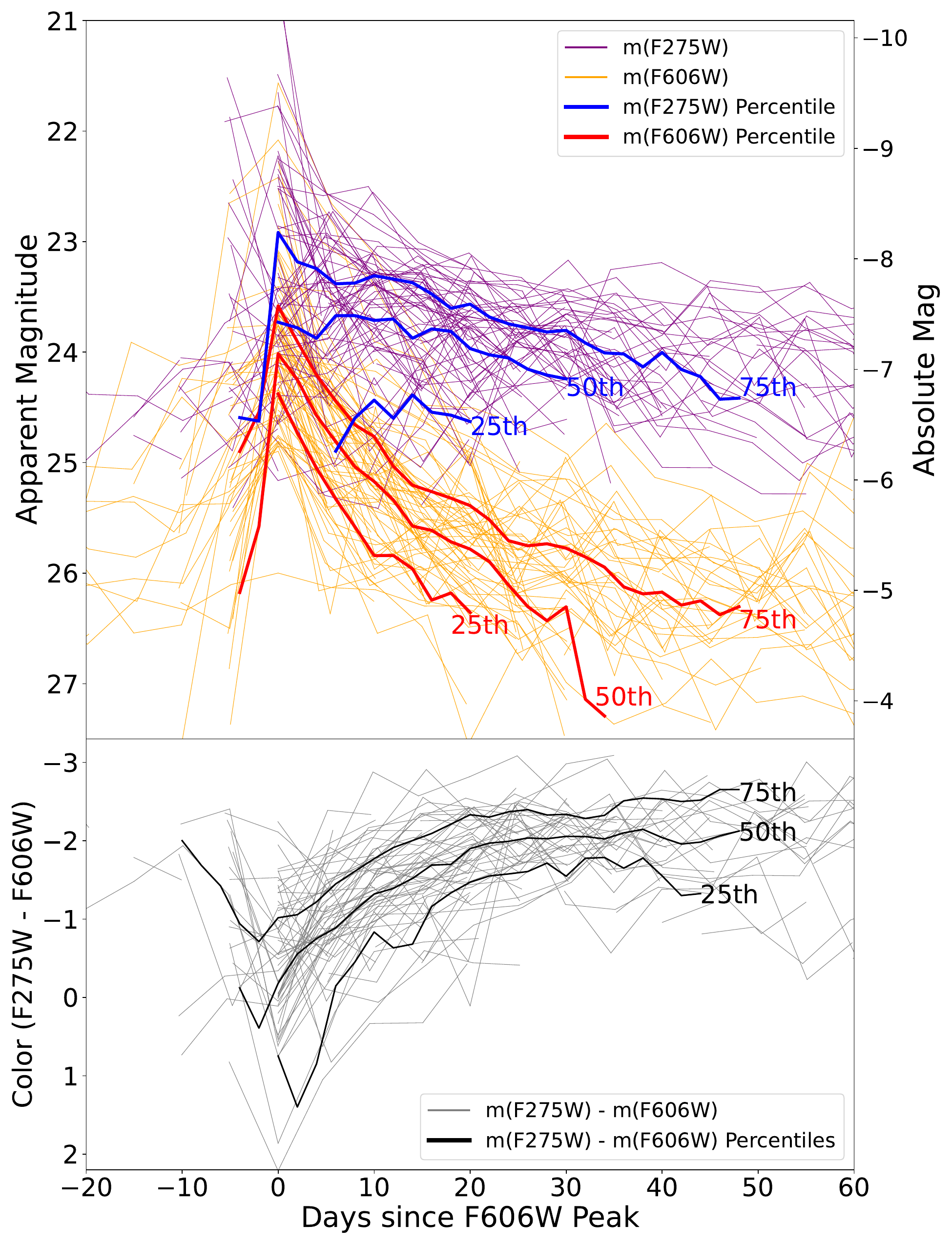}
    \caption{Top: F275W and F606W light-curves (purple and orange, respectively) of 77 novae with observed brightness peaks in M87. Over-plotted are the 25th, 50th, and 75th percentiles (in order of brightness) of all of these F275W (in blue) and F606W (in red) light curves. Note that the computation of a percentile at a given time takes into account upper limit magnitude datapoints in individual nova light curves. To avoid clutter in this plot, those individual limit data points are not shown, but they can be seen in Table \ref{tab:big_phot} and as arrows in Figures \ref{fig:nova-1}-\ref{fig:nova-94}. The higher luminosities and slower rates of decline of novae in the NUV are apparent. Bottom: the F275W - F606W color curves of the M87 novae, as well as the 25th, 50th, and 75th percentiles of the color-curves. Novae near maximum light exhibit m(F275W) - m(F606W) $\sim$ 0 $\pm 1$, then become increasingly blue during the ensuing $\sim$30 days. After $\sim$ 30 days they remain at m(F275W) - m(F606W) $\sim$ -2 $\pm$ 0.5 .}
    \label{fig:overlay_lc_cc}
\end{figure}

In the top section of Figure \ref{fig:overlay_lc_cc} we plot the F275W and F606W light-curves of 94 M87 novae. Collectively, novae are seen to be $\sim$ 0.9 mag more luminous in F275W than in F606W near maximum light. But over the course of the ensuing 2-3 weeks, the gap widens to $\sim$ 2 magnitudes. This is reflected in the bottom section of Figure \ref{fig:overlay_lc_cc}, where novae are seen to be reddest at or shortly after maximum light, then become increasingly blue, reaching m(F275W) - m(F606W) $\sim$ -2 $\pm 0.5$ about 3 weeks later. Similar striking color behavior is seen in Figure 8 of \citet{Shara2016}, where the (F606W - F814W) colors of 32 M87 novae are reddest at and shortly after maximum light, then become $\sim$ a magnitude bluer in the ensuing month.  

\newpage

\subsection{NUV, Visible and NIR light-curves}

\begin{figure}[h]
    \centering
    \includegraphics[width=6in]{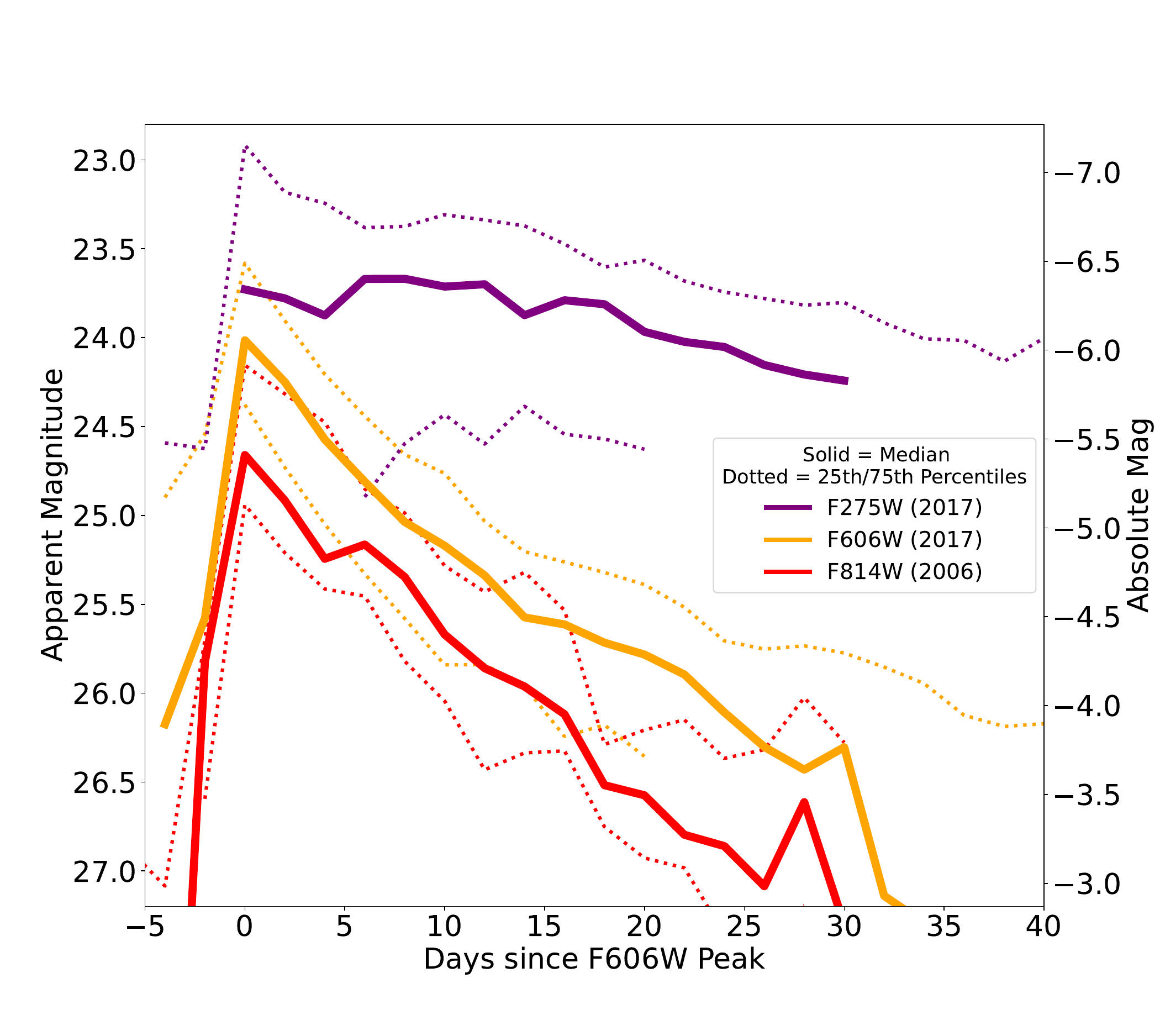}
    \caption{Median {\it HST} light curve of 77 M87 novae with observed peaks in F275W bandpass (solid purple) and F606W bandpass (solid orange), both based on 5-day cadence observations. Also shown are the median daily-cadence light curves of 32 M87 novae in the F814W bandpass {(red)} of {\it HST} from \cite{Shara2016}. {For each band, the 25th and 75th percentiles of all lightcurves in that band are shown as dotted lines.} The much slower declines of novae in the NUV are apparent, {as is the later rise to peak of many novae in the NUV.}}
    \label{fig:trish_overlay_lc}
\end{figure}

In Figure \ref{fig:trish_overlay_lc} we plot the median F275W and F606W light curves of 77 M87 novae with observed peak brightnesses (from the present survey, with a 5-day cadence) and the F606W and F814W (near-infrared) light curves of 32 more novae (from the \citet{Shara2016} survey with 1-day cadence). This plot re-enforces the facts that {1)} novae are $\sim$ 1 magnitude more luminous in NUV than visible light, {2) they are $\sim$ 1 magnitude brighter in visible than near-infrared light,} and 3) that they fade much more slowly in NUV {than in visible or NIR light. }

\newpage

\subsection{The nova peak absolute magnitude distributions} \label{ssec:peak mags}

\begin{figure}[h]
    
    \centering
    \includegraphics[width=6in]{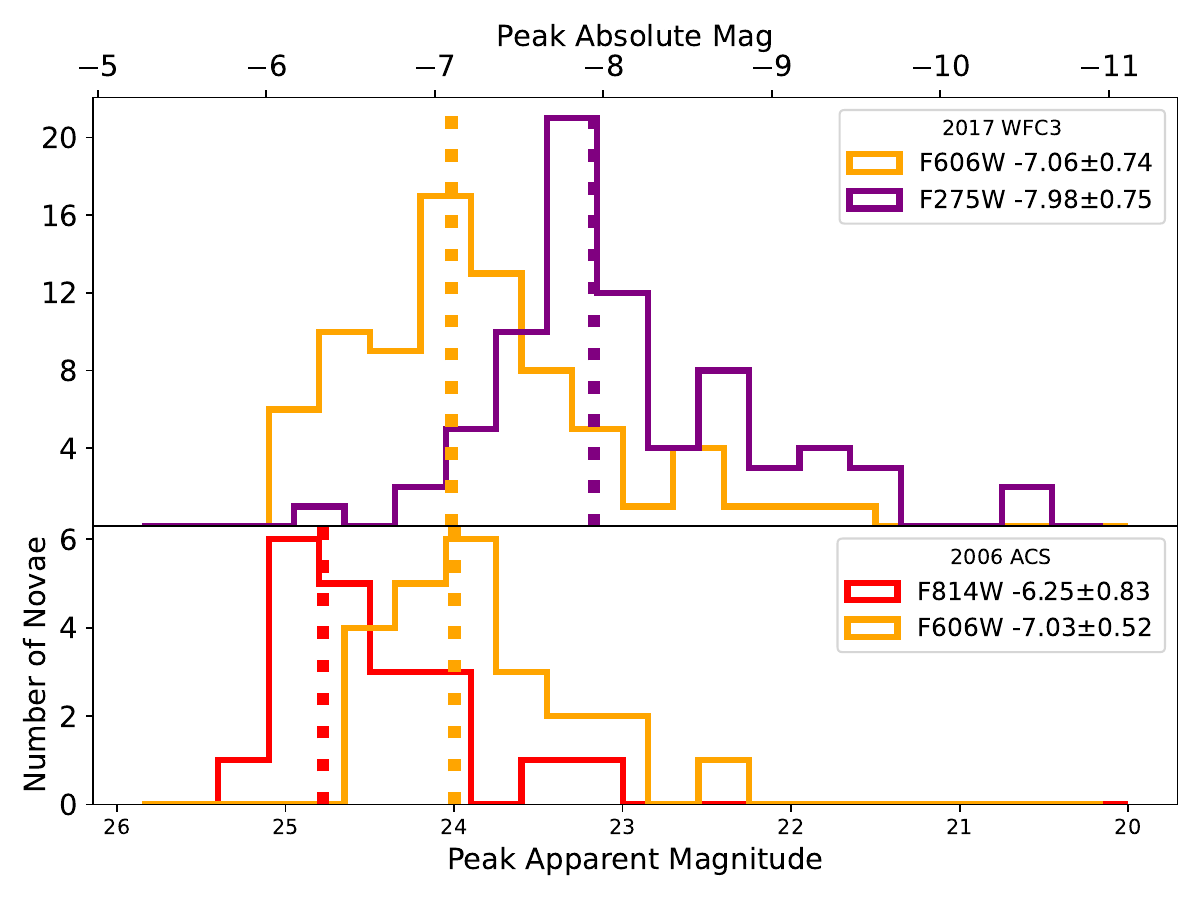}
    \caption{Top: The peak absolute magnitude distributions of 77 novae with observed peaks in 5-day cadence imagery of M87 in the F275W and F606W filters of {\it HST}, along with the median and standard deviations of the distribution. Novae are 0.9 magnitude more luminous at peak brightness in the NUV than in the visible. Bottom: A histogram of the peak F606W and F814W magnitudes from 1-day cadence imagery of M87 novae \citet{Shara2016}. Note that the median peak F606W magnitude from this 1-day cadence sample is {almost identical to that} of the 5-day cadence sample.}
    \label{fig:peak_mag_hist}
\end{figure}

In Figure \ref{fig:peak_mag_hist} we plot the histograms of the observed peak absolute magnitudes of M87 novae. Novae peak at M(F275W) = -8.0 $\pm$ 0.8 and at M(F606W) = 7.1 $\pm$ 0.7 in the current, 5-day cadence data set. They peak at M(606W) = {-7.0} $\pm$ 0.5 and and at M(814W) = {-6.3} $\pm$ 0.8 in the \citet{Shara2016} 1-day cadence data set. {One-day versus 5-day temporal sampling barely changes the detected absolute magnitudes of novae despite their being seen closer in time (on average) to the epoch of maximum light in the 1-day cadence data.}

\newpage

\subsection{Correlations between peak magnitudes} \label{sec:peak_mag_correlations}

\begin{figure}[h]
    \centering
    \includegraphics[width=7.5in]{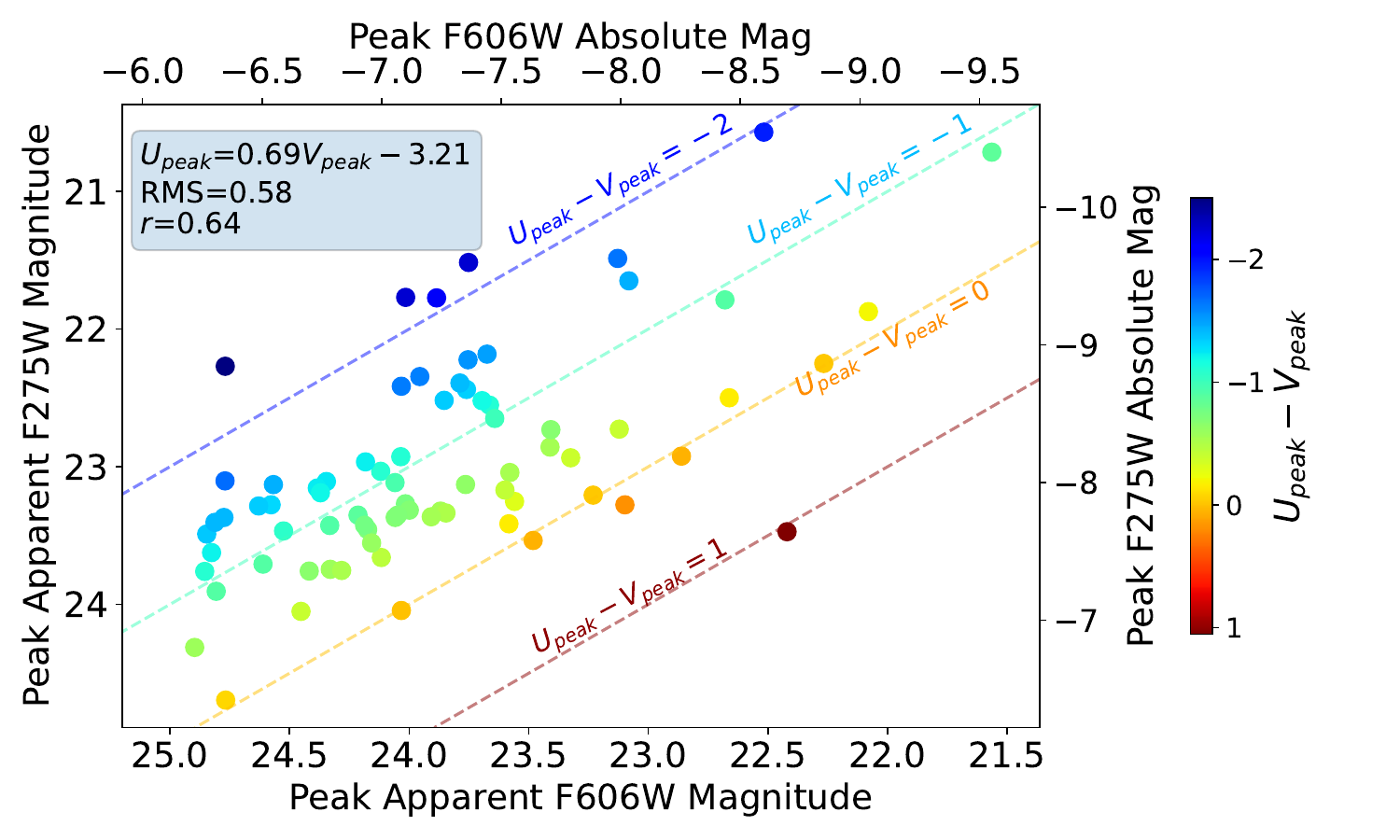}
    \caption{Peak F275W vs peak F606W absolute magnitudes of the 77 novae with observed peaks in M87. The correlations are given in the figure legend.}
    \label{fig:peak_mag_scatter}
\end{figure}

In Figure \ref{fig:peak_mag_scatter} we plot the peak F275W magnitudes versus the peak F606W magnitudes for 77 M87 novae with both quantities observed. The strong correlation between the peak magnitudes (with r = 0.64) is apparent, and it would likely be even stronger if we had daily cadence data available. {The data of Figures \ref{fig:overlay_lc_cc}, \ref{fig:trish_overlay_lc}, \ref{fig:peak_mag_hist} and \ref{fig:peak_mag_scatter} will allow tests of the predicted multi-wavelength light curves of novae of \citep{Hillman2014}, but are beyond the scope of this paper.}

\newpage

\subsection{Lag between NUV and Visible peak magnitudes}\label{ssec:uv-lag}

\begin{figure}[h]
    \centering
    \includegraphics[width=6.8in]{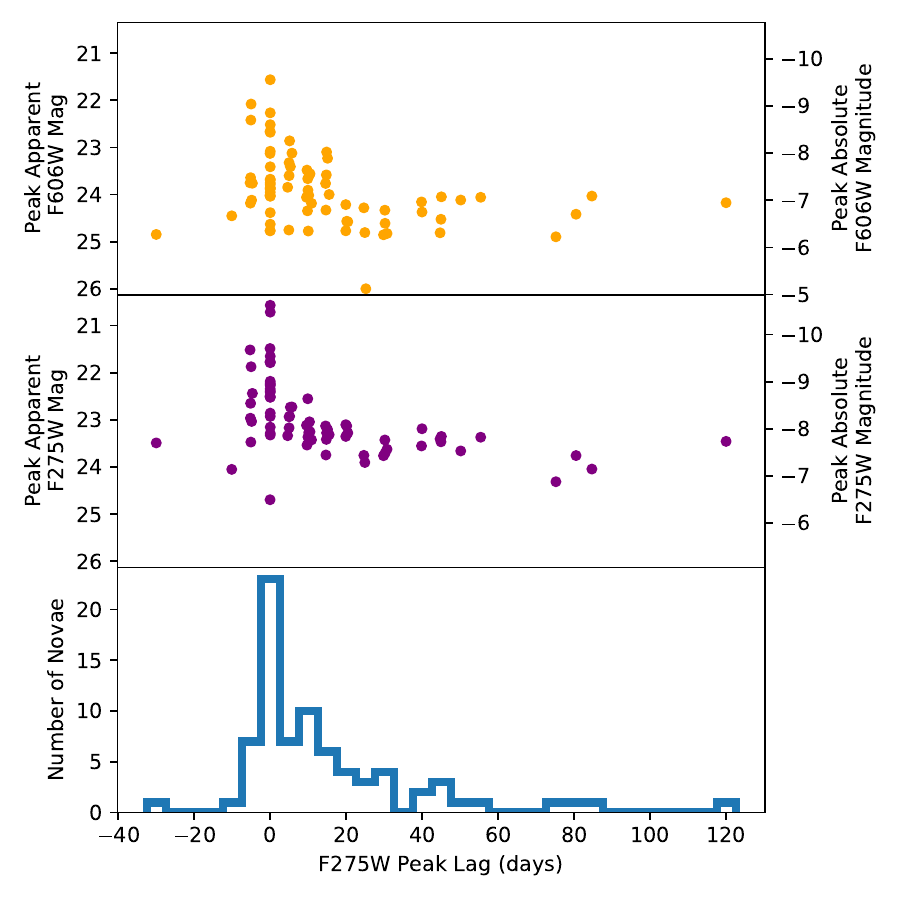}
    \caption{The number of days after the observed F606W peak that the F275W peak was observed plotted versus peak F606W magnitude (top), versus peak F275W magnitude (middle), and as a histogram (bottom). Data from the 77 novae whose peaks were observed are included in the plots.}
    \label{fig:uv-lag-stack}
\end{figure}

In Figure \ref{fig:uv-lag-stack} we plot the number of days after the observed F606W peak that the F275W peak was observed versus peak F606W magnitude (top), versus the peak F275W magnitude (middle), and as a histogram (bottom). Just 9 of 77 novae peak in F275W {\it before} peaking in F606W, and just by 5-10 days. In contrast, while F275W maximum is typically observed 5-30 days after F606W maximum, a few lags of 40-80 days are observed, as is one extreme event {(Nova \# 54)} with a 120 day lag. {As noted above, these lags are a direct test of models of nova light-curves \citep{Hillman2014}, beyond the scope of this paper. }

\newpage

\subsection{Color-magnitude correlations}

\begin{figure}[h]
    \centering
    \includegraphics[width=6in]{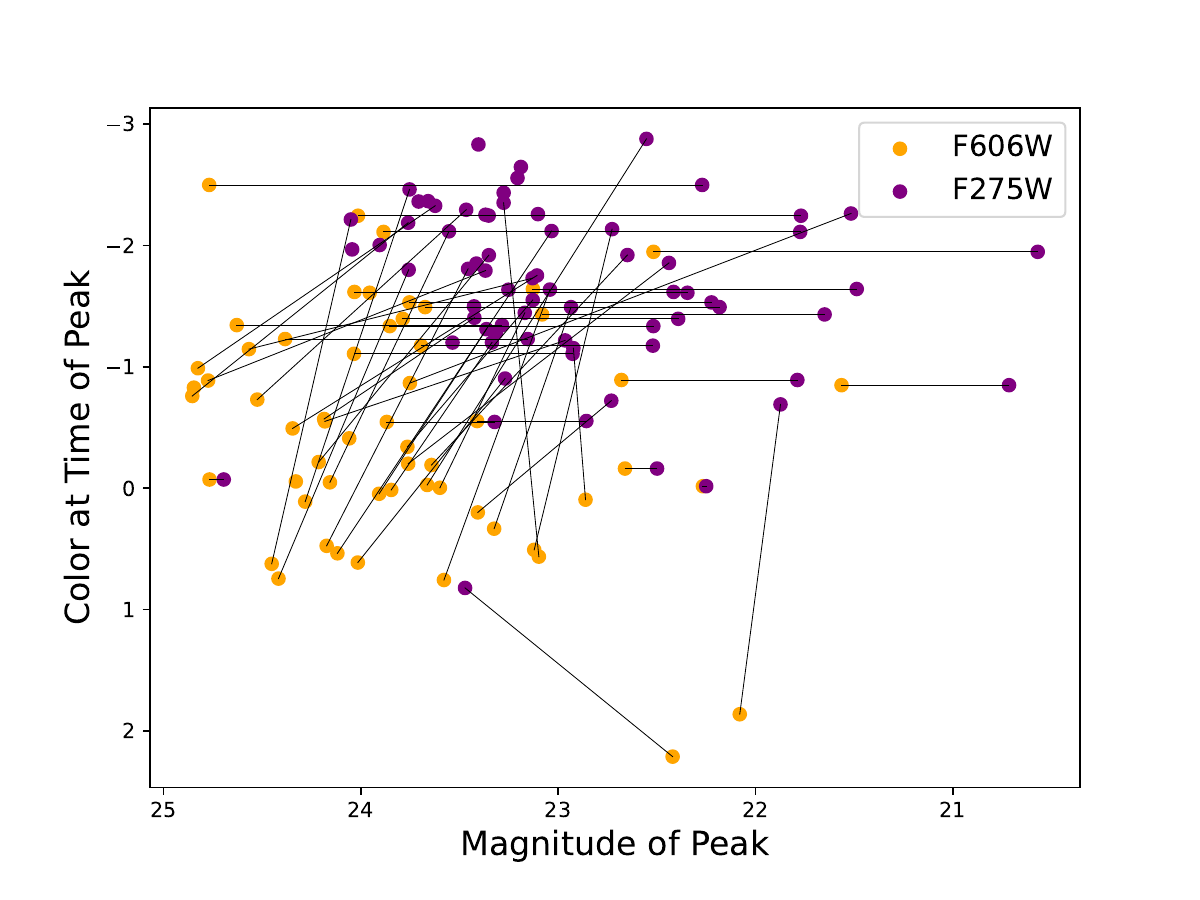}
    \caption{The m(F275W) - m(F606W) color at the time of the observed visible peak vs the visible peak magnitude (orange) and the color at the time of the observed NUV peak vs the NUV peak magnitude (purple). The two data points for each of the 55 novae that are observed in both bands at the time of both peaks are connected by a thin line. The horizontal lines represent novae whose observed visible and NUV peaks occurred at the same epoch. The most luminous novae (those with peak magnitudes $<$ 23) achieve peak F275W and F606W brightnesses very close in time. Most (less luminous) novae exhibit fainter, redder F606W brightness peaks followed by more luminous, bluer F275W peaks.}
    \label{fig:color_vs_peak}
\end{figure}

In Figure \ref{fig:color_vs_peak} we plot the F275W - F606W color at the time of the observed F606W peak vs the F606W peak magnitude (orange) and the same color at the time of the observed F275W peak vs the F275W peak magnitude (purple). The two datapoints for each of the 55 novae that are observed in both bands in the time of both peaks are connected by a thin line. The horizontal lines represent novae whose observed visible and NUV peaks occurred at the same epoch.

\begin{figure}[h]
    \centering
    \includegraphics[width=5.5in]{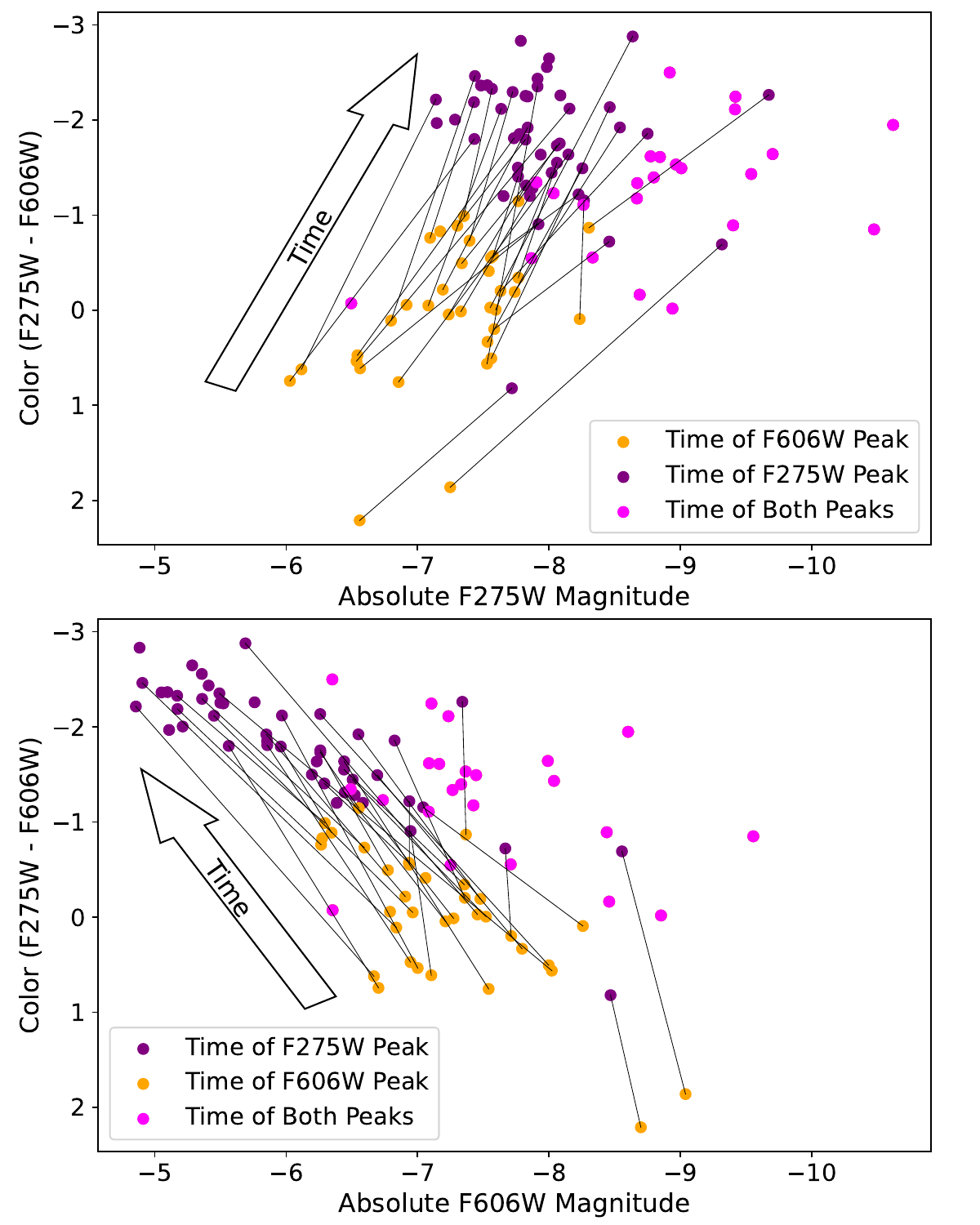}
    \caption{F275W - F606W colors of 55 novae vs the absolute peak F275W and F606W magnitudes. The arrows labeled ``Time'' indicate directions of color and magnitude change between the F606W and F275W peak brightnesses. As in Figure \ref{fig:color_vs_peak}, we see that the most luminous novae (those with peak absolute magnitudes $<-8$) achieve peak F275W and F606W brightnesses very close in time.}
    \label{fig:stack_color_and_peaks}
\end{figure}

\newpage
In Figure \ref{fig:stack_color_and_peaks} we plot the F275W - F606W colors of 55 novae vs the absolute peak F275W and F606W magnitudes. As in Figure \ref{fig:color_vs_peak}, we see that the most luminous novae (those with peak absolute magnitudes $<$ -8) achieve peak F275W and F606W brightnesses very close in time. Most (less luminous) novae exhibit fainter, redder F606W brightness peaks followed by more luminous, bluer F275W peaks.

\newpage

\section{Absolute magnitude versus decline time} \label{sec:mag_decline}

Novae have been investigated as possible ``standard candles'' for over a century \citep{Lundmark1919,McLaughlin1945,Arp1956,Shara1981,Darnley2006,Della_Valle2020,Schaefer2022}. If the eruptions of novae were controlled by just one parameter - the underlying WD mass - then the absolute magnitudes of novae at peak brightness would all be strongly correlated with their rates of decline \citep{Shara1981}, displaying an RMS scatter of just $\sim$ 0.5 mag. In fact the critical masses of the thermonuclear-powered envelopes on white dwarfs in nova binaries depend strongly on white dwarf mass {\it and} mass accretion history, and to a lesser extent on the underlying white dwarf luminosity \citep{Yaron2005} and possibly the metalicity of accreted matter. It is thus no surprise that nova light curves are highly inhomogeneous. The definitive ``snuffing out'' of novae as standard candles came with \citet{Yaron2005}'s predictions of, and \citet{Kasliwal2011}'s observational discovery (in M31) of ``faint-fast'' novae. These objects are as common in the giant elliptical galaxy M87 \citep{Shara2016} as in the giant spiral galaxy M31, strongly increasing the RMS scatter in the peak visual absolute magnitude vs. decline time relationship of Galactic novae (\citet{Schaefer2022}, Figure 4). In the current era of precision cosmology, with Cepheid and tip-of-the-red giant branch distance indicators yielding $\sim$ 1\% accurate distances, the nova visual absolute magnitude - t2 relationship is of little value as a distance indicator. Is the same true in the NUV?

\begin{figure}
    \centering
    \gridline{\fig{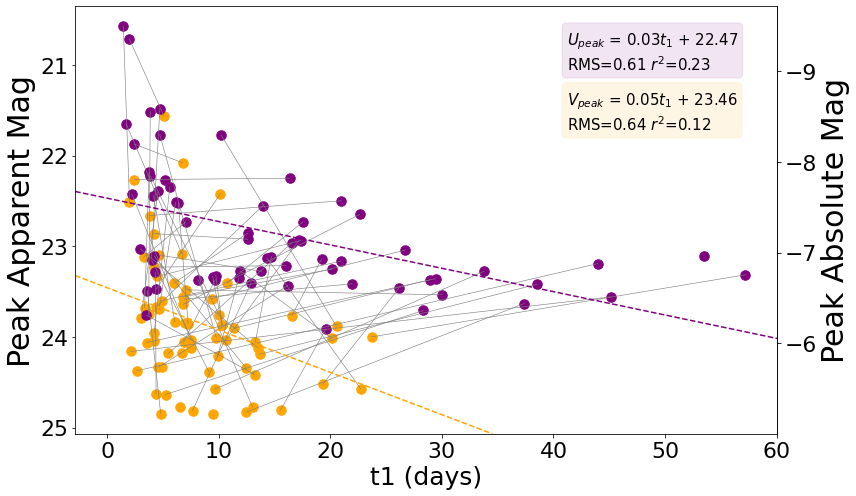}{0.98\textwidth}{(a)}}
    \gridline{\fig{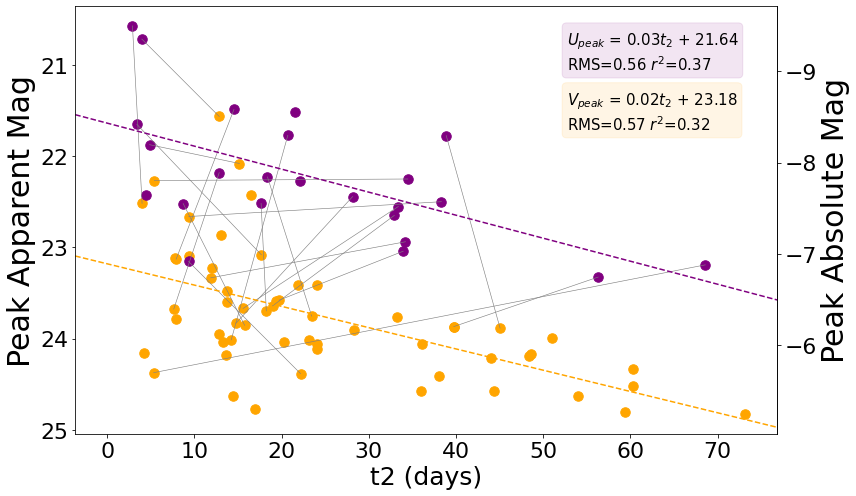}{0.98\textwidth}{(b)}}
    \caption{Top: F275W ({violet}) and F606W ({yellow}) peak magnitudes versus t1 decline time for novae with observed peak brightnesses. Each nova is plotted in both filters, and the pairs of points for each nova are connected by a thin line. Faint, fast novae are as prevalent in the F275W bandpass as in the F606W bandpass. Bottom: Same as above except F275W and F606W magnitudes versus t2 decline time.}
    \label{fig:mag-decline-time}
\end{figure}

In Figure \ref{fig:mag-decline-time} we plot the peak absolute magnitude vs. t1 and t2 decline time relationships for 77 M87 novae with observed peak brightnesses. The RMS scatter of each plot is close to 0.6 magnitudes. Faint, fast novae are the strongest sources of scatter in both F275W and F606W bandpasses. This diagram demonstrates that, just as for F606W-detected novae, novae are not useful distance indicators in the F275W bandpass.

\newpage

\subsection{Decline time histograms}

\begin{figure}[h]
    \centering
    \gridline{
        \fig{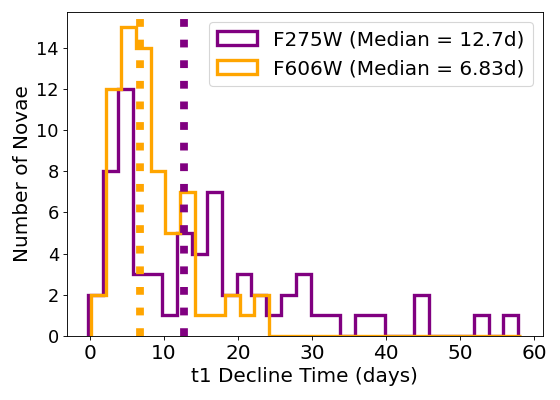}{0.5\textwidth}{(a)}
        \fig{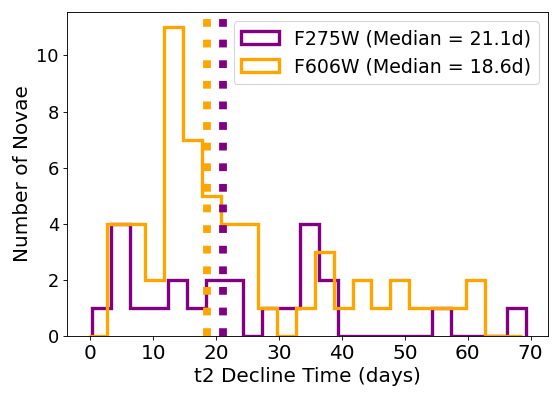}{0.5\textwidth}{(b)}
    }
    \caption{Left: Histograms of the t1 decline times in F275W and F606W of 77 novae in M87. The much slower t1 declines in the NUV are apparent. Right: Histograms of the t2 decline times in F275W and F606W of 77 novae in M87. The declines in the NUV are slightly slower than in the visible, but much less so than for the t1 decline times.}
    \label{fig:t-hist}
\end{figure}

In Figure \ref{fig:t-hist} we plot the histograms of one and two-magnitude decline times (t1 and t2, respectively) of the 77 novae in the present sample with well-defined decline times. The mean t1 of the F275W and F606W light curves of these novae are 12.7 and 6.83 days, respectively, while the corresponding t2 are 21.1 and 18.6 days, respectively. 

\newpage

\section{Correlations with galactocentric distance} \label{sec:radial_dist_cors}

\begin{figure}[h]
    \centering
    \includegraphics[height=6.8in]{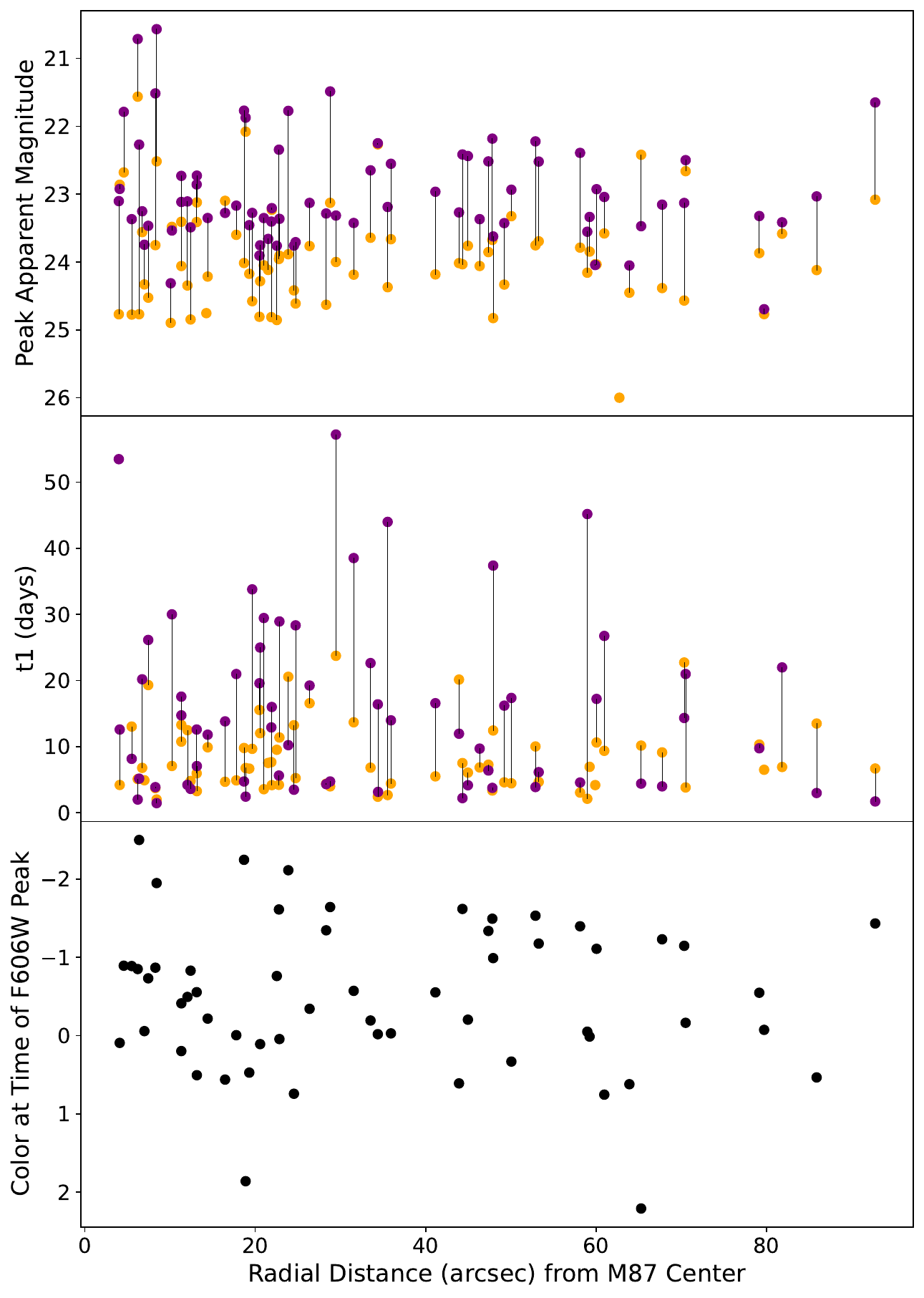}
    \caption{Peak apparent magnitude, t1 decline time and (F275W - F606W) color at peak brightness of the novae in M87 that had observed peaks. The lack of correlations means that novae of all luminosities, speed classes are origins are thoroughly mixed in the galaxy.}
    \label{fig:radial_dist_stack}
\end{figure}

In Figure \ref{fig:radial_dist_stack} we plot the peak apparent magnitudes, t1 decline times and (F275W - F606W) color at peak brightness for 77 M87 novae vs radial distance from the nucleus of M87. The three plots are ``scatter diagrams'', indicating that novae of all types, whether descendants of primordial binaries born in M87, or binaries captured during galaxy cannibalistic episodes, are thoroughly mixed in the galaxy.

\newpage

\section{Conclusions} \label{sec:conclusion}

We conducted a 9-month-long, 5-day cadence near-ultraviolet and visible-light \hst survey for erupting M87 novae. The survey covered the inner $\sim$ 35\% of M87's light. Simulations using ``real-world" nova light curves showed the nova detection efficiency of the survey to be $\sim$ {75}\%. {Taking a conservative 21\% as the fraction of ``faint-fast'' novae in M87, we find the nova rate in M87 to be $352_{-37}^{+37}$/yr, which is the value we adopt. (That rate would have increased to $\sim$ 380/yr if we had adopted a faint-fast nova fraction of $\sim$50\%, suggested by the M31 survey data of \citet{Kasliwal2011})}. The {M87} luminosity-specific nova rate is $7.91_{-1.20}^{+1.20}/yr/10^{10}L_\odot,_{K}$.  Both these rates are within {0.5} standard deviations of the rates previously derived in \citet{Shara2016}, confirming their claim that previous ground-based surveys of M87 - and by implication other ground-based surveys of all galaxies - are significantly incomplete.  The radial distribution of novae closely followed M87's light to within $\sim$ 4'' of the galaxy's nucleus. While theory predicts that novae can recur as often as every 45 days, we detect zero novae in the surveyed area erupting more frequently than once every 130 days. Novae are $\sim$ 1 magnitude brighter in the NUV than in the visible at maximum light, {and $\sim$ 2 magnitudes brighter in NUV than in near-IR light at maximum light. Novae are $\sim$ 2 magnitudes brighter in NUV light than in visible light $\sim$ 3 weeks after peak brightness.} The peak visible and NUV luminosities are strongly and positively correlated. Just 9 of 77 novae achieved peak brightness in NUV light light before visible light peak brightness was reached, with observed time lags between peak visible light and peak NUV light as long as 120 days. {The most luminous novae (those with peak absolute magnitudes $<$ -8) achieve peak F275W and F606W brightnesses very close in time. Most (less luminous) novae exhibit fainter, redder F606W brightness peaks followed by more luminous, bluer F275W peaks.}

\acknowledgments
This research is based on observations made with the NASA/ESA Hubble Space Telescope obtained from the Space Telescope Science Institute, which is operated by the Association of Universities for Research in Astronomy, Inc., under NASA contract NAS 5–26555.  These observations are associated with programs 10543 (PI:Baltz) and 14618 (PI:Shara). The specific observations analyzed can be accessed via \dataset[https://doi.org/10.17909/natk-ks60]{https://doi.org/10.17909/natk-ks60}.  Support to MAST for these data is provided by the NASA Office of Space Science via grant NAG5–7584 and by other grants and contracts. MMS and RH were funded by NASA/STScI grant GO-14651. The paper also is based upon work of RH supported by NASA under award number 80GSFC21M0002. JM acknowledges the National Science Centre, Poland, grant OPUS 2017/27/B/ST9/01940. We thank the schedulers of {\it HST} for successfully obtaining the extremely regularly spaced 53 epochs of observations which made the analyses presented here possible. 
\vspace{5mm}
\facilities{HST(WFC3 and ACS)} \\

Software: \textsc{calwf3} \citep{Dressel2019}, \textsc{astrodrizzle} \citep{Avila2015}, \textsc{DAOFIND} \citep{Stetson2011}, \textsc{PyRAF} \citep{Greenfield2000}.

\section*{Data Availability}
All data described here are available on the STScI/MAST web site:

\url{https://mast.stsci.edu/search/ui/#/hst} 
and setting Proposal ID = 10543,14618

\newpage    

\clearpage

\bibliography{M87}

\begin{thebibliography}{}
\expandafter\ifx\csname natexlab\endcsname\relax\def\natexlab#1{#1}\fi
\providecommand{\url}[1]{\href{#1}{#1}}

\bibitem[{{Arp}(1956)}]{Arp1956}
{Arp}, H.~C. 1956, \aj, 61, 15

\bibitem[{{Avila} {et~al.}(2015){Avila}, {Hack}, {Cara}, {Borncamp}, {Mack}, {Smith}, \& {Ubeda}}]{Avila2015}
{Avila}, R.~J., {Hack}, W., {Cara}, M., {et~al.} 2015, in Astronomical Society of the Pacific Conference Series, Vol. 495, Astronomical Data Analysis Software an Systems XXIV (ADASS XXIV), ed. A.~R. {Taylor} \& E.~{Rosolowsky}, 281

\bibitem[{{Bertin} \& {Arnouts}(1996)}]{Bertin96}
{Bertin}, E., \& {Arnouts}, S. 1996, \aaps, 117, 393

\bibitem[{{Cao} {et~al.}(2012){Cao}, {Kasliwal}, {Neill}, {Kulkarni}, {Lou}, {Ben-Ami}, {Bloom}, {Cenko}, {Law}, {Nugent}, {Ofek}, {Poznanski}, \& {Quimby}}]{Cao2012}
{Cao}, Y., {Kasliwal}, M.~M., {Neill}, J.~D., {et~al.} 2012, \apj, 752, 133

\bibitem[{{Cardelli} {et~al.}(1989){Cardelli}, {Clayton}, \& {Mathis}}]{Cardelli1989}
{Cardelli}, J.~A., {Clayton}, G.~C., \& {Mathis}, J.~S. 1989, \apj, 345, 245

\bibitem[{{Chen} {et~al.}(2016){Chen}, {Woods}, {Yungelson}, {Gilfanov}, \& {Han}}]{Chen2016}
{Chen}, H.-L., {Woods}, T.~E., {Yungelson}, L.~R., {Gilfanov}, M., \& {Han}, Z. 2016, \mnras, 458, 2916

\bibitem[{{Claeys} {et~al.}(2014){Claeys}, {Pols}, {Izzard}, {Vink}, \& {Verbunt}}]{Claeys2014}
{Claeys}, J.~S.~W., {Pols}, O.~R., {Izzard}, R.~G., {Vink}, J., \& {Verbunt}, F.~W.~M. 2014, \aap, 563, A83

\bibitem[{{Curtin} {et~al.}(2015){Curtin}, {Shafter}, {Pritchet}, {Neill}, {Kundu}, \& {Maccarone}}]{Curtin2015}
{Curtin}, C., {Shafter}, A.~W., {Pritchet}, C.~J., {et~al.} 2015, \apj, 811, 34

\bibitem[{{Darnley} {et~al.}(2014){Darnley}, {Williams}, {Bode}, {Henze}, {Ness}, {Shafter}, {Hornoch}, \& {Votruba}}]{Darnley2014}
{Darnley}, M.~J., {Williams}, S.~C., {Bode}, M.~F., {et~al.} 2014, \aap, 563, L9

\bibitem[{{Darnley} {et~al.}(2006){Darnley}, {Bode}, {Kerins}, {Newsam}, {An}, {Baillon}, {Belokurov}, {Calchi Novati}, {Carr}, {Cr{\'e}z{\'e}}, {Evans}, {Giraud-H{\'e}raud}, {Gould}, {Hewett}, {Jetzer}, {Kaplan}, {Paulin-Henriksson}, {Smartt}, {Tsapras}, \& {Weston}}]{Darnley2006}
{Darnley}, M.~J., {Bode}, M.~F., {Kerins}, E., {et~al.} 2006, \mnras, 369, 257

\bibitem[{{De} {et~al.}(2021){De}, {Kasliwal}, {Hankins}, {Sokoloski}, {Adams}, {Ashley}, {Babul}, {Bagdasaryan}, {Delacroix}, {Dekany}, {Greffe}, {Hale}, {Jencson}, {Karambelkar}, {Lau}, {Mahabal}, {McKenna}, {Moore}, {Ofek}, {Sharma}, {Smith}, {Soon}, {Soria}, {Srinivasaragavan}, {Tinyanont}, {Travouillon}, {Tzanidakis}, \& {Yao}}]{De2021}
{De}, K., {Kasliwal}, M.~M., {Hankins}, M.~J., {et~al.} 2021, \apj, 912, 19

\bibitem[{de~Grijs \& Bono(2019)}]{deGrijs2019}
de~Grijs, R., \& Bono, G. 2019, The Astrophysical Journal Supplement Series, 246, 3.
\newblock \url{https://dx.doi.org/10.3847/1538-4365/ab5711}

\bibitem[{{Della Valle} \& {Izzo}(2020)}]{Della_Valle2020}
{Della Valle}, M., \& {Izzo}, L. 2020, \aapr, 28, 3

\bibitem[{{Dressel}(2019)}]{Dressel2019}
{Dressel}, L. 2019, in WFC3 Instrument Handbook for Cycle 27 v. 11, Vol.~11, 11

\bibitem[{{Gallagher} \& {Code}(1974)}]{Gallagher1974}
{Gallagher}, J.~S., I., \& {Code}, A.~D. 1974, \apj, 189, 303

\bibitem[{{Greenfield} \& {White}(2000)}]{Greenfield2000}
{Greenfield}, P., \& {White}, R.~L. 2000, in Astronomical Society of the Pacific Conference Series, Vol. 216, Astronomical Data Analysis Software and Systems IX, ed. N.~{Manset}, C.~{Veillet}, \& D.~{Crabtree}, 59

\bibitem[{{Harris}(2009)}]{Harris2009}
{Harris}, W.~E. 2009, \apj, 703, 939

\bibitem[{{Harris} \& {Petrie}(1978)}]{Harris1978}
{Harris}, W.~E., \& {Petrie}, P.~L. 1978, \apj, 223, 88

\bibitem[{{Hillman}(2021)}]{Hillman2021}
{Hillman}, Y. 2021, \mnras, 505, 3260

\bibitem[{{Hillman} {et~al.}(2015){Hillman}, {Prialnik}, {Kovetz}, \& {Shara}}]{Hillman2015}
{Hillman}, Y., {Prialnik}, D., {Kovetz}, A., \& {Shara}, M.~M. 2015, \mnras, 446, 1924

\bibitem[{{Hillman} {et~al.}(2016){Hillman}, {Prialnik}, {Kovetz}, \& {Shara}}]{Hillman2016}
---. 2016, \apj, 819, 168

\bibitem[{{Hillman} {et~al.}(2014){Hillman}, {Prialnik}, {Kovetz}, {Shara}, \& {Neill}}]{Hillman2014}
{Hillman}, Y., {Prialnik}, D., {Kovetz}, A., {Shara}, M.~M., \& {Neill}, J.~D. 2014, \mnras, 437, 1962

\bibitem[{{Hillman} {et~al.}(2020){Hillman}, {Shara}, {Prialnik}, \& {Kovetz}}]{Hillman2020}
{Hillman}, Y., {Shara}, M.~M., {Prialnik}, D., \& {Kovetz}, A. 2020, Nature Astronomy, 4, 886

\bibitem[{{Jha} {et~al.}(2019){Jha}, {Maguire}, \& {Sullivan}}]{Jha2019}
{Jha}, S.~W., {Maguire}, K., \& {Sullivan}, M. 2019, Nature Astronomy, 3, 706

\bibitem[{{Kasliwal} {et~al.}(2011){Kasliwal}, {Cenko}, {Kulkarni}, {Ofek}, {Quimby}, \& {Rau}}]{Kasliwal2011}
{Kasliwal}, M.~M., {Cenko}, S.~B., {Kulkarni}, S.~R., {et~al.} 2011, \apj, 735, 94

\bibitem[{{Kawash} {et~al.}(2021){Kawash}, {Chomiuk}, {Rodriguez}, {Strader}, {Sokolovsky}, {Aydi}, {Kochanek}, {Stanek}, {Mukai}, {De}, {Shappee}, {Holoien}, {Prieto}, \& {Thompson}}]{Kawash2021}
{Kawash}, A., {Chomiuk}, L., {Rodriguez}, J.~A., {et~al.} 2021, \apj, 922, 25

\bibitem[{{Kormendy} {et~al.}(2009){Kormendy}, {Fisher}, {Cornell}, \& {Bender}}]{Kormendy2009}
{Kormendy}, J., {Fisher}, D.~B., {Cornell}, M.~E., \& {Bender}, R. 2009, \apjs, 182, 216

\bibitem[{{Liu} {et~al.}(2023){Liu}, {R{\"o}pke}, \& {Han}}]{Liu2023}
{Liu}, Z.-W., {R{\"o}pke}, F.~K., \& {Han}, Z. 2023, Research in Astronomy and Astrophysics, 23, 082001

\bibitem[{{Lundmark}(1919)}]{Lundmark1919}
{Lundmark}, K. 1919, Astronomische Nachrichten, 209, 369

\bibitem[{{Madrid} {et~al.}(2007){Madrid}, {Sparks}, {Ferguson}, {Livio}, \& {Macchetto}}]{Madrid2007}
{Madrid}, J.~P., {Sparks}, W.~B., {Ferguson}, H.~C., {Livio}, M., \& {Macchetto}, D. 2007, \apjl, 654, L41

\bibitem[{{Mandel} {et~al.}(2023){Mandel}, {Shara}, {Zurek}, {Conroy}, \& {van Dokkum}}]{Mandel2023}
{Mandel}, S., {Shara}, M.~M., {Zurek}, D., {Conroy}, C., \& {van Dokkum}, P. 2023, \mnras, 518, 5279

\bibitem[{{Maoz} {et~al.}(2014){Maoz}, {Mannucci}, \& {Nelemans}}]{Maoz2014}
{Maoz}, D., {Mannucci}, F., \& {Nelemans}, G. 2014, \araa, 52, 107

\bibitem[{{Matteucci} {et~al.}(2003){Matteucci}, {Renda}, {Pipino}, \& {Della Valle}}]{Matteucci2003}
{Matteucci}, F., {Renda}, A., {Pipino}, A., \& {Della Valle}, M. 2003, \aap, 405, 23

\bibitem[{{Mclaughlin}(1945)}]{McLaughlin1945}
{Mclaughlin}, D.~B. 1945, \pasp, 57, 69

\bibitem[{{Mr{\'o}z} {et~al.}(2016){Mr{\'o}z}, {Udalski}, {Poleski}, {Soszy{\'n}ski}, {Szyma{\'n}ski}, {Pietrzy{\'n}ski}, {Wyrzykowski}, {Ulaczyk}, {Koz{\l}owski}, {Pietrukowicz}, \& {Skowron}}]{Mroz2016}
{Mr{\'o}z}, P., {Udalski}, A., {Poleski}, R., {et~al.} 2016, \apjs, 222, 9

\bibitem[{{Schaefer}(2022)}]{Schaefer2022}
{Schaefer}, B.~E. 2022, \mnras, 517, 6150

\bibitem[{{Shafter} {et~al.}(2000){Shafter}, {Ciardullo}, \& {Pritchet}}]{Shafter2000}
{Shafter}, A.~W., {Ciardullo}, R., \& {Pritchet}, C.~J. 2000, \apj, 530, 193

\bibitem[{{Shafter} {et~al.}(2014){Shafter}, {Curtin}, {Pritchet}, {Bode}, \& {Darnley}}]{Shafter2014}
{Shafter}, A.~W., {Curtin}, C., {Pritchet}, C.~J., {Bode}, M.~F., \& {Darnley}, M.~J. 2014, in Astronomical Society of the Pacific Conference Series, Vol. 490, Stellar Novae: Past and Future Decades, ed. P.~A. {Woudt} \& V.~A.~R.~M. {Ribeiro}, 77

\bibitem[{{Shafter} {et~al.}(2017){Shafter}, {Kundu}, \& {Henze}}]{Shafter2017b}
{Shafter}, A.~W., {Kundu}, A., \& {Henze}, M. 2017, Research Notes of the American Astronomical Society, 1, 11

\bibitem[{{Shafter} {et~al.}(2021){Shafter}, {Hornoch}, {Ben{\'a}{\v{c}}ek}, {Gal{\'a}d}, {Jan{\'\i}k}, {Jury{\v{s}}ek}, {Kotkov{\'a}}, {Kurf{\"u}rst}, {Ku{\v{c}}{\'a}kov{\'a}}, {Ku{\v{s}}nir{\'a}k}, {Li{\v{s}}ka}, {Paunzen}, {Skarka}, {{\v{S}}koda}, {Wolf}, {Zasche}, \& {Zejda}}]{Shafter2021}
{Shafter}, A.~W., {Hornoch}, K., {Ben{\'a}{\v{c}}ek}, J., {et~al.} 2021, \apj, 923, 239

\bibitem[{{Shara}(1981)}]{Shara1981}
{Shara}, M.~M. 1981, \apj, 243, 926

\bibitem[{{Shara} {et~al.}(2016){Shara}, {Doyle}, {Lauer}, {Zurek}, {Neill}, {Madrid}, {Miko{\l}ajewska}, {Welch}, \& {Baltz}}]{Shara2016}
{Shara}, M.~M., {Doyle}, T.~F., {Lauer}, T.~R., {et~al.} 2016, \apjs, 227, 1

\bibitem[{{Sohn} {et~al.}(2006){Sohn}, {O'Connell}, {Kundu}, {Landsman}, {Burstein}, {Bohlin}, {Frogel}, \& {Rose}}]{Sohn2006}
{Sohn}, S.~T., {O'Connell}, R.~W., {Kundu}, A., {et~al.} 2006, \aj, 131, 866

\bibitem[{{Stetson}(2011)}]{Stetson2011}
{Stetson}, P.~B. 2011, {DAOPHOT: Crowded-field Stellar Photometry Package}, Astrophysics Source Code Library, record ascl:1104.011, , , ascl:1104.011

\bibitem[{{Strope} {et~al.}(2010){Strope}, {Schaefer}, \& {Henden}}]{Strope2010}
{Strope}, R.~J., {Schaefer}, B.~E., \& {Henden}, A.~A. 2010, \aj, 140, 34

\bibitem[{{van den Bergh} \& {Younger}(1987)}]{vandenbergh1987}
{van den Bergh}, S., \& {Younger}, P.~F. 1987, \aaps, 70, 125

\bibitem[{{Yaron} {et~al.}(2005){Yaron}, {Prialnik}, {Shara}, \& {Kovetz}}]{Yaron2005}
{Yaron}, O., {Prialnik}, D., {Shara}, M.~M., \& {Kovetz}, A. 2005, \apj, 623, 398

\end{thebibliography}

\section{Appendix: The Data} \label{sec:appendices}

In Table \ref{tab:obs} we list the {\it HST} image root names, observation dates, passbands and exposure times collected for program GO-14618.

In Table \ref{tab:summary} we list the positions, peak magnitudes, colors and decline times of the 94 novae we detected in M87.

Table \ref{tab:big_phot} lists the light curve data for every nova: the epochs and corresponding date of detection, as well as the U (F275W) and V (F606W) magnitudes, (F275W-F606W) color, and their errors.

\begin{sidewaysfigure}
\label{fig:big_individual_novae}
    \begin{tabular}{cc}
     \multirow{2}{*}[2.6in]{\includegraphics[height=6in]{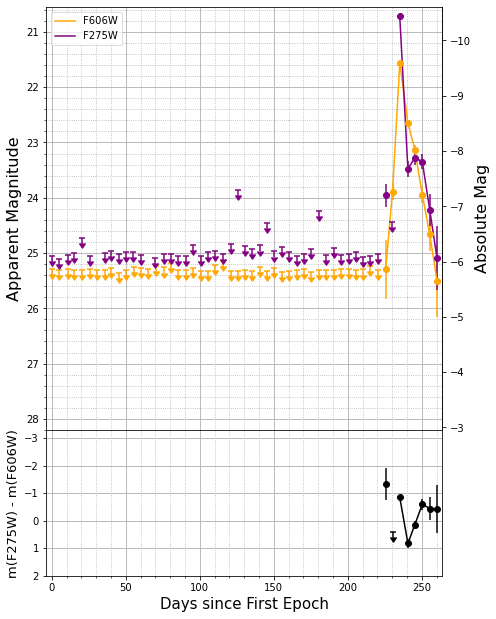}} & \includegraphics[height=3.4in]{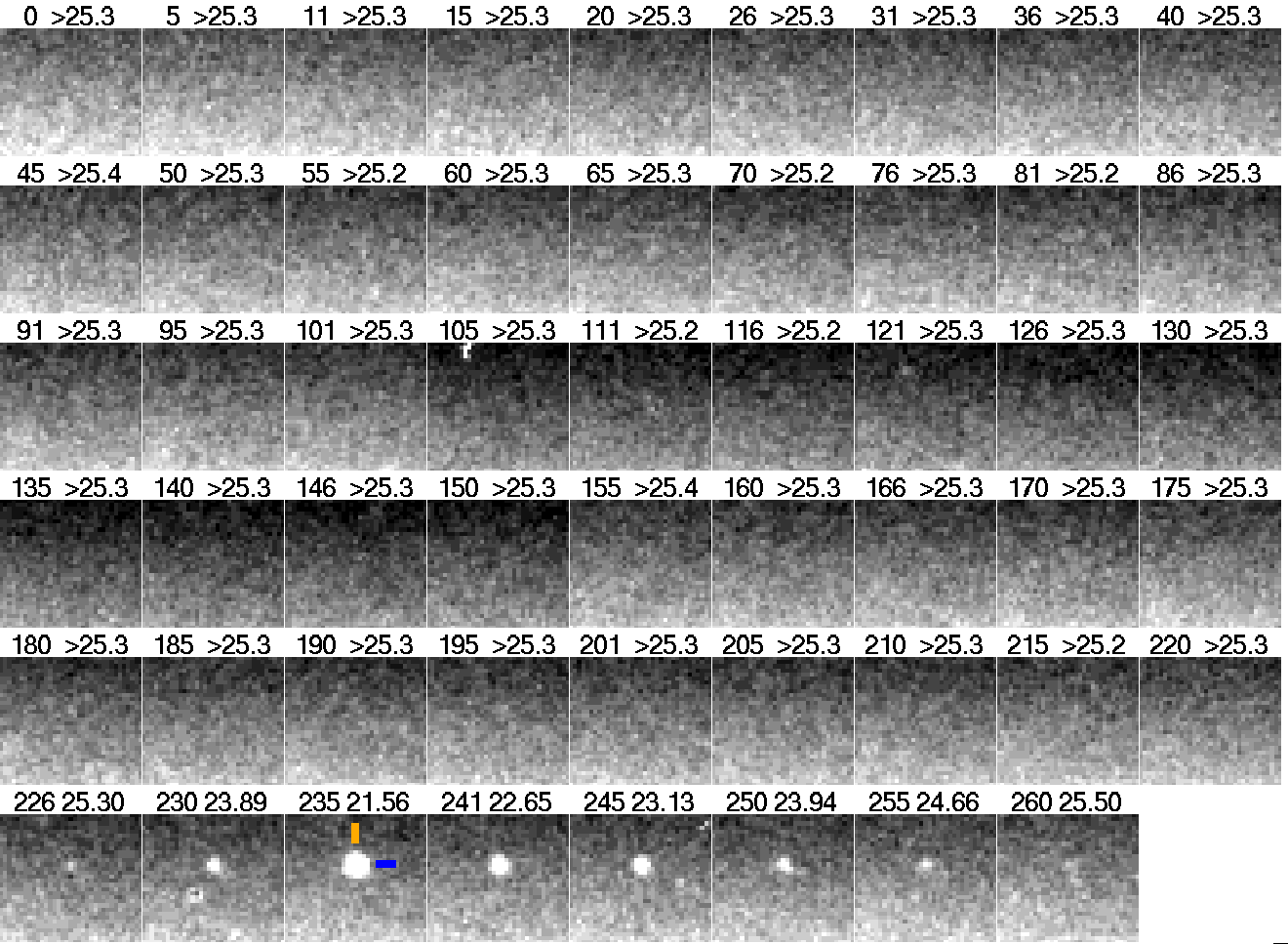} \\
     & \includegraphics[height=3.4in]{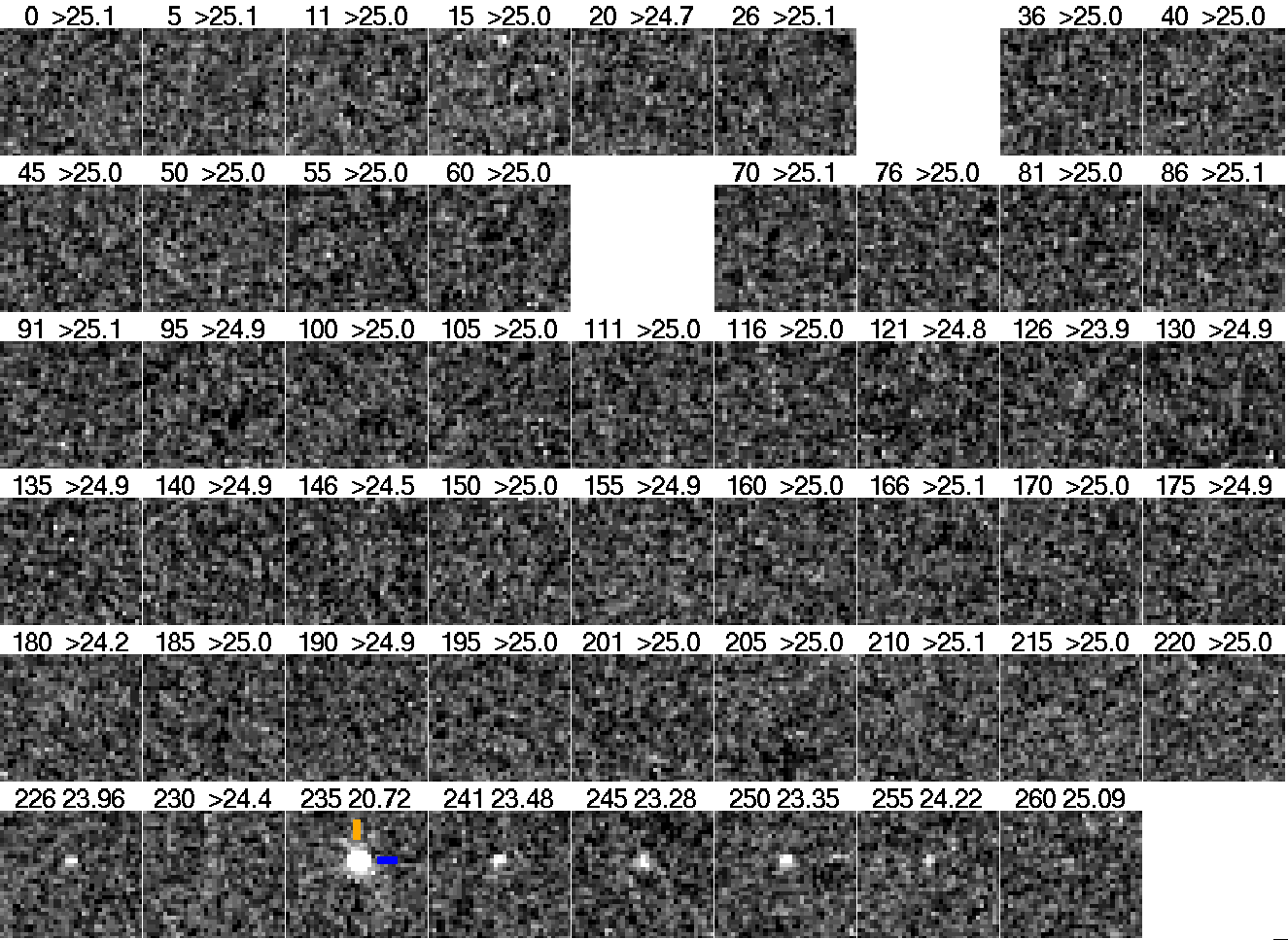}
    \end{tabular}
    \caption{Nova 1. See Section \hyperref[sec:appendices]{Appendix} for a description of this and the next 93 Figures. }
    \label{fig:nova-1}
\end{sidewaysfigure}

In Figures \ref{fig:nova-1} through \ref{fig:nova-94} we provide light and color curves and ``postage stamp'' images of each nova. The 94 novae are ordered by peak brightness in the F606W bandpass, where Nova 1 is the most luminous and nova 94 is the least luminous.

The top left section of each figure contains the F606W and F275W light curves of the nova. The bottom left section is the (F275W - F606W) color curve of the nova. The top right section is a series of 1.1 x 1.3 arcsec ``postage stamp'' F606W images of the nova in every epoch in which it was imaged by {\it HST}. N is up, E is left. The day of observation (0,5,10,15 etc) and the observed magnitude are just above each image. The bottom right section is the same as the top right section, except that it displays the F275W images of the same nova. The nova is marked with an orange (blue) tic mark on the day it reaches maximum light in F606W (F275W).

\startlongtable
\begin{deluxetable}{ccccc|ccccc}
\tablewidth{0pt}
\tabletypesize{\footnotesize}
\tablecaption{HST GO-14618 WFC3 Observations of M87}
\tablehead{ \colhead{Epoch} & \colhead{Root Name} & \colhead{Obs Date} & \colhead{Band} & \colhead{Exp Time} & \colhead{Epoch} & \colhead{Root Name} & \colhead{Obs Date} & \colhead{Band} & \colhead{Exp Time}\\ \# & & YYYY-MM-DD & & sec& \# & & YYYY-MM-DD & & sec}
\startdata
1 & id5o01rdq & 2016-11-13 & F275W & 500.0 & 2 & id5o02hmq & 2016-11-18 & F275W & 520.0\\
1 & id5o01req & 2016-11-13 & F275W & 500.0 & 2 & id5o02hoq & 2016-11-18 & F275W & 520.0\\
1 & id5o01rhq & 2016-11-13 & F275W & 500.0 & 2 & id5o02hqq & 2016-11-18 & F275W & 520.0\\
1 & id5o01rjq & 2016-11-13 & F606W & 360.0 & 2 & id5o02hsq & 2016-11-18 & F606W & 360.0\\
1 & id5o01rlq & 2016-11-13 & F606W & 360.0 & 2 & id5o02hxq & 2016-11-18 & F606W & 360.0\\
3 & id5o03diq & 2016-11-23 & F275W & 520.0 & 4 & id5o04avq & 2016-11-28 & F275W & 520.0\\
3 & id5o03djq & 2016-11-23 & F275W & 520.0 & 4 & id5o04axq & 2016-11-28 & F275W & 520.0\\
3 & id5o03dlq & 2016-11-23 & F275W & 520.0 & 4 & id5o04azq & 2016-11-28 & F275W & 520.0\\
\enddata
\end{deluxetable}\manlabel{tab:obs}{1}

\begin{longrotatetable}\begin{deluxetable}{lcccccccccccc}
\tablewidth{0pt}
\tabletypesize{\footnotesize}
\tablecaption{Positions, Peak Magnitudes, and Decline Times of M87 Novae}
\tablehead{ \colhead{Nova} & \colhead{Radial Distance}  & \colhead{RA} & \colhead{DEC} & \colhead{V$_{\mathrm{peak}}$} & \colhead{$U_{\mathrm{peak}}$} & $(U-V)_{Vpeak}$ &  $(U-V)_{Upeak}$ & $U_{peak} -V_{peak}$ & V $t_1$ & U $t_1$ & V $t_2$ & U $t_2$ \\ \# & arcsec & HH:MM:SS.ss & HH:MM:SS.ss & mag & mag & mag & mag & mag & days & days & days & days}
\startdata
1 & 6.21 & 12:30:49.40 & +12:23:34.24&$21.56$ & $20.72$ & -0.85 & -0.85 & -0.85 & 5.09 & 1.99 & 12.86 & 3.99\\
2 & 18.88 & 12:30:48.41 & +12:23:39.63&$22.08$ & $21.87$ & 1.86 & -0.69 & -0.21 & 6.76 & 2.43 & 15.17 & 4.87\\
3 & 34.39 & 12:30:47.41 & +12:23:45.75&$22.27$ & $22.25$ & -0.02 & -0.02 & -0.02 & 2.42 & 16.39 & 5.41 & 34.48\\
4 & 65.29 & 12:30:50.98 & +12:24:29.20&$22.42$ & $23.47$ & 2.21 & 0.82 & 1.05 & 10.15 & 4.40 & 16.49 & $\cdots$\\
5 & 8.43 & 12:30:49.13 & +12:23:20.79&$22.52$ & $20.57$ & -1.95 & -1.95 & -1.95 & 2.00 & 1.45 & 4.01 & 2.91\\
6 & 70.51 & 12:30:48.66 & +12:22:18.43&$22.66$ & $22.50$ & -0.16 & -0.16 & -0.16 & 3.82 & 20.98 & 9.35 & 38.33\\
7 & 4.58 & 12:30:49.13 & +12:23:29.41&$22.68$ & $21.79$ & -0.89 & -0.89 & -0.89 & $\cdots$ & $\cdots$ & $\cdots$ & $\cdots$\\
8 & 4.10 & 12:30:49.15 & +12:23:27.49&$22.86$ & $22.92$ & 0.09 & -1.15 & 0.06 & 4.20 & 12.58 & 13.04 & $\cdots$\\
\enddata
\end{deluxetable}\end{longrotatetable}\manlabel{tab:summary}{2}

\startlongtable
\begin{deluxetable}{lccccccccc}
\tablewidth{0pt}
\tabletypesize{\footnotesize}
\tablecolumns{9}
\tablecaption{M87 Nova Lightcurve Data}
\tablecomments{Table \ref{tab:big_phot} is published in its entirety, with full data for all 53 epochs for all 94 novae, in the machine-readable format. A portion is shown here for guidance regarding its form and content. The precise time of the ``First Ep" is MJD 57705.0605.}
\tablehead{ \colhead{Epoch} & \colhead{Days since} & \colhead{Days since} &\colhead{Days since} &  \colhead{$V$} & \colhead{$V_{err}$} & \colhead{$U$} & \colhead{$U_{err}$} & \colhead{$(U-V)$} & \colhead{$(U-V)_{err}$} \\ \# & First Ep & F606W Peak & F275W Peak & mag & mag & mag & mag & mag & mag}
\startdata
\multicolumn{10}{c}{M87 Nova 1} \\
\hline\\
44 & 215.26 & -19.87 & -19.87 & $>25.23$ & $\cdots$ & $>25.05$ & $\cdots$ & $\cdots$ & $\cdots$ \\
45 & 220.48 & -14.65 & -14.65 & $>25.31$ & $\cdots$ & $>25.02$ & $\cdots$ & $\cdots$ & $\cdots$ \\
46 & 225.58 & -9.55 & -9.55 & $25.30$ & $0.53$ & $23.96$ & $0.21$ & $-1.34$ & $0.57$ \\
47 & 230.23 & -4.90 & -4.90 & $23.89$ & $0.15$ & $>24.43$ & $\cdots$ & $\cdots$ & $\cdots$ \\
48 & 235.13 & 0.00 & 0.00 & $21.56$ & $0.02$ & $20.72$ & $0.02$ & $-0.85$ & $0.03$ \\
49 & 240.66 & 5.55 & 5.51 & $22.65$ & $0.05$ & $23.48$ & $0.14$ & $0.82$ & $0.15$ \\
\enddata
\end{deluxetable}\manlabel{tab:big_phot}{3}

\begin{sidewaysfigure}
    \begin{tabular}{cc}
     \multirow{2}{*}[2.6in]{\includegraphics[height=6in]{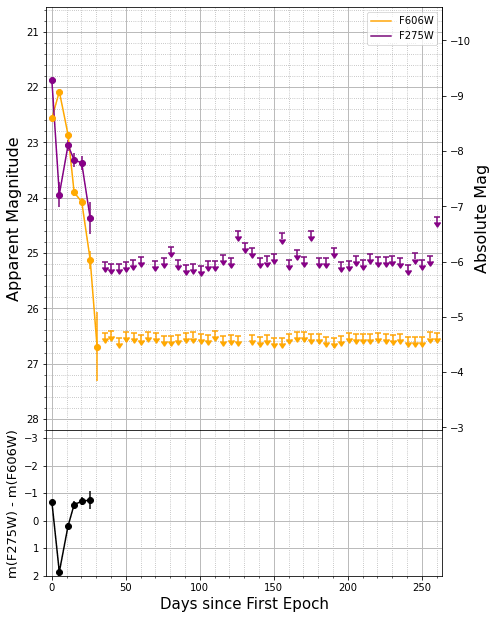}} & \includegraphics[height=3.6in]{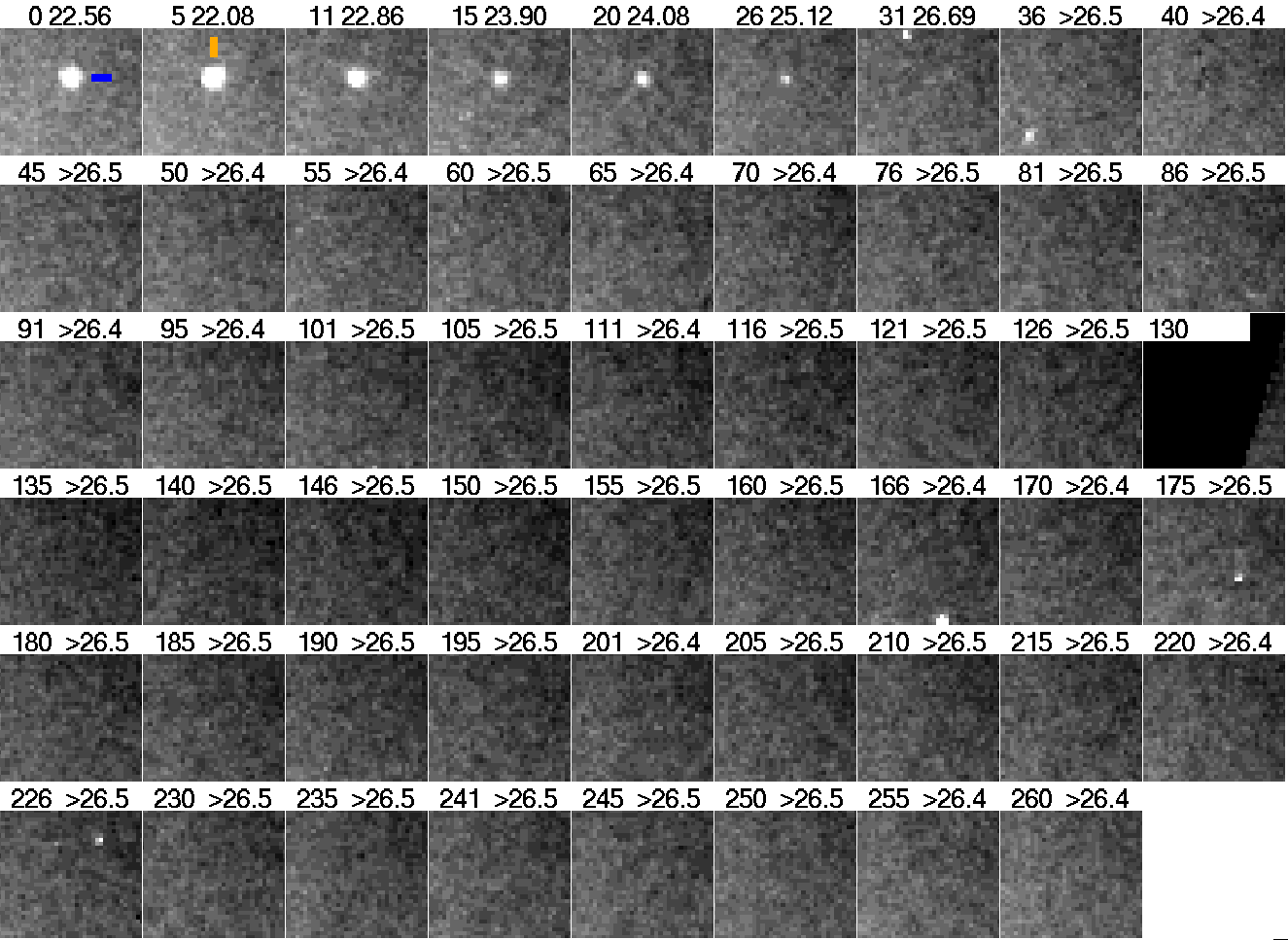} \\
     & \includegraphics[height=3.6in]{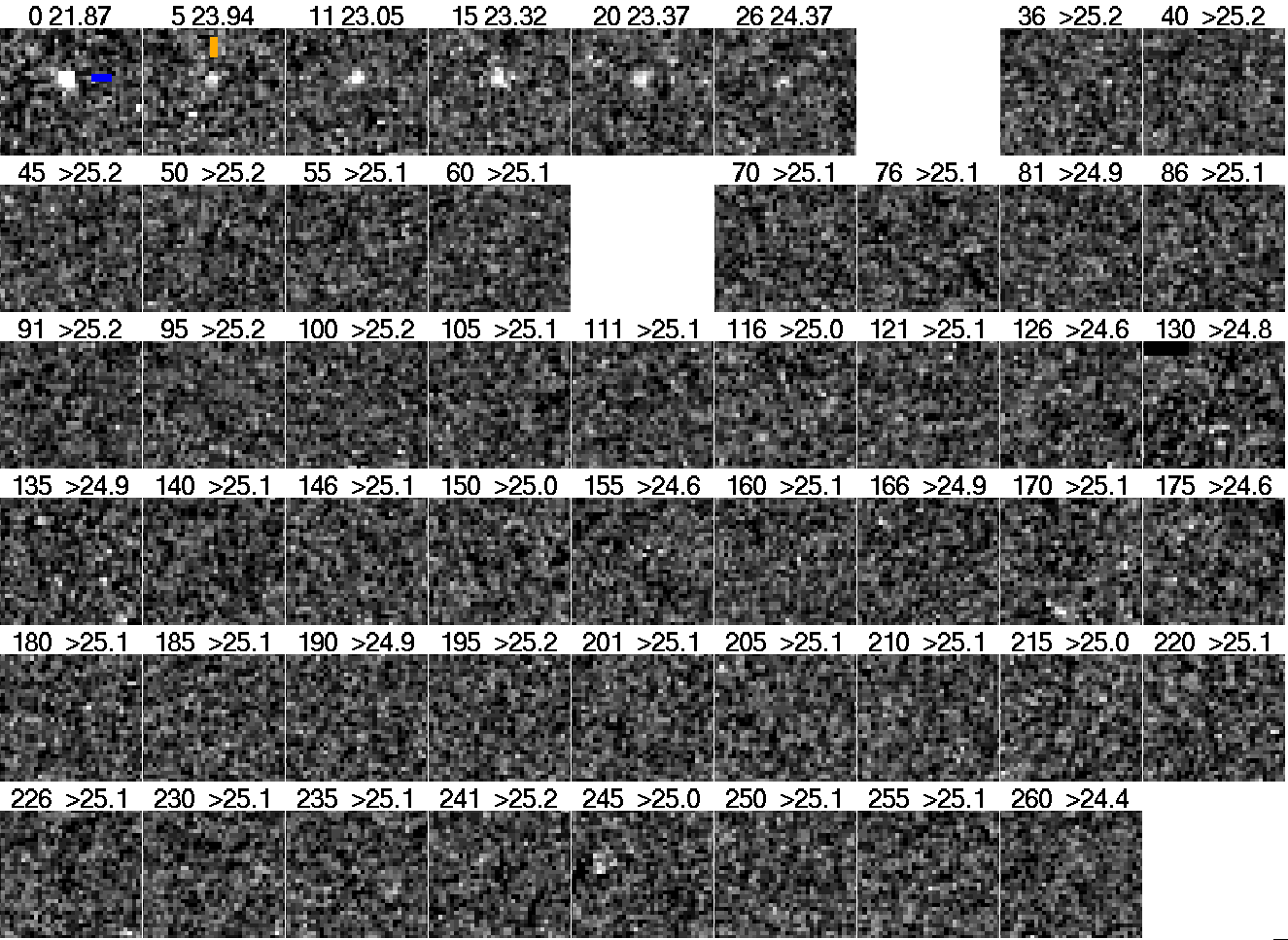}
    \end{tabular}
    \caption{Nova 2}
\end{sidewaysfigure}
\begin{sidewaysfigure}
    \begin{tabular}{cc}
     \multirow{2}{*}[2.6in]{\includegraphics[height=6in]{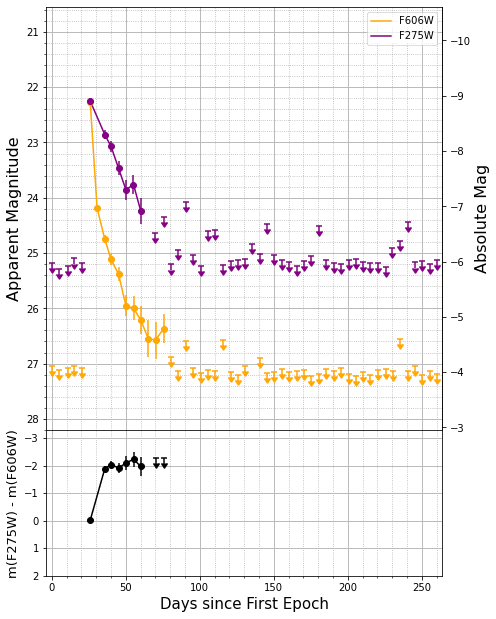}} & \includegraphics[height=3.6in]{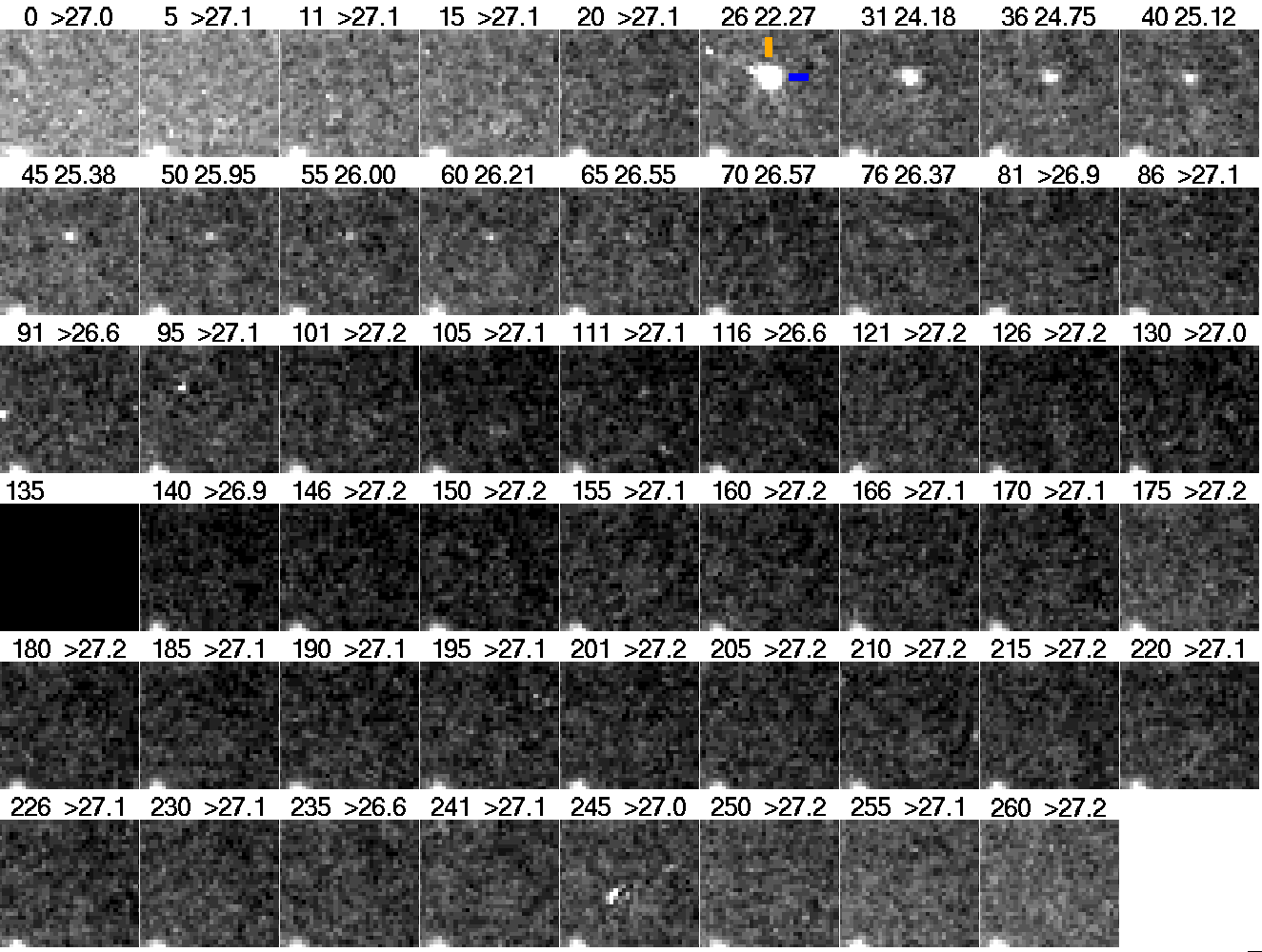} \\
     & \includegraphics[height=3.6in]{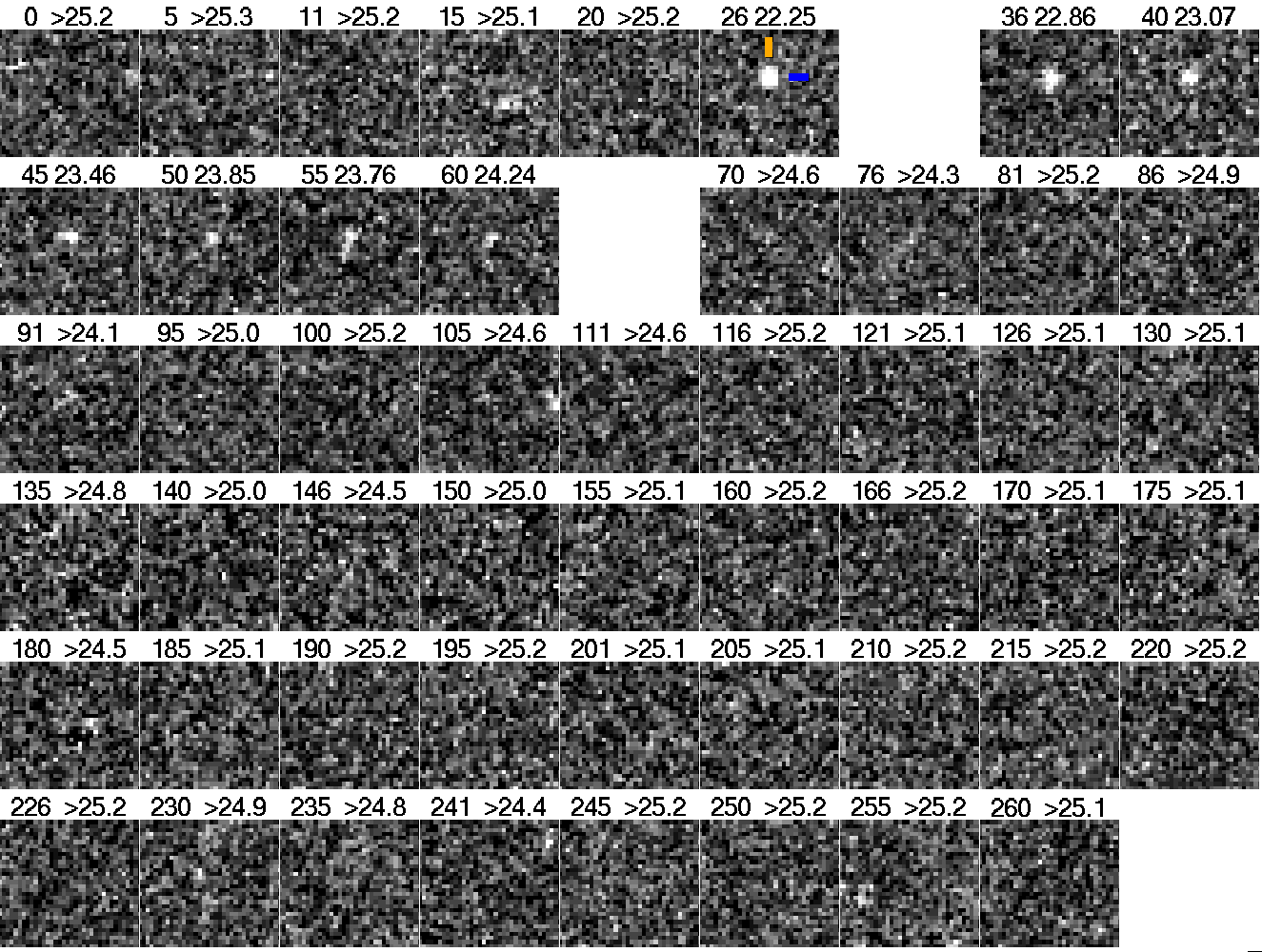}
    \end{tabular}
    \caption{Nova 3}
\end{sidewaysfigure}
\begin{sidewaysfigure}
    \begin{tabular}{cc}
     \multirow{2}{*}[2.6in]{\includegraphics[height=6in]{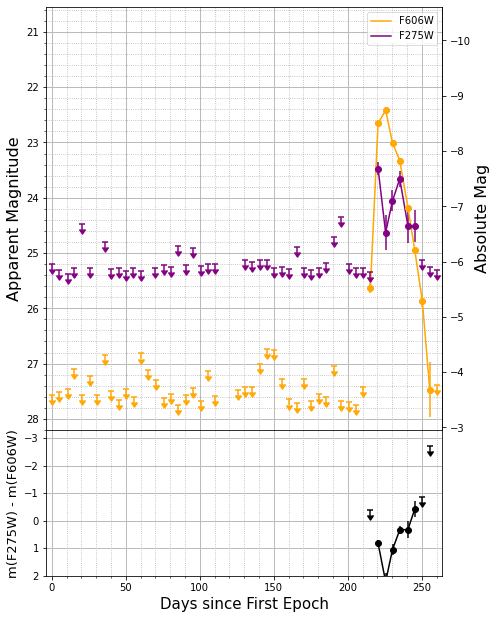}} & \includegraphics[height=3.6in]{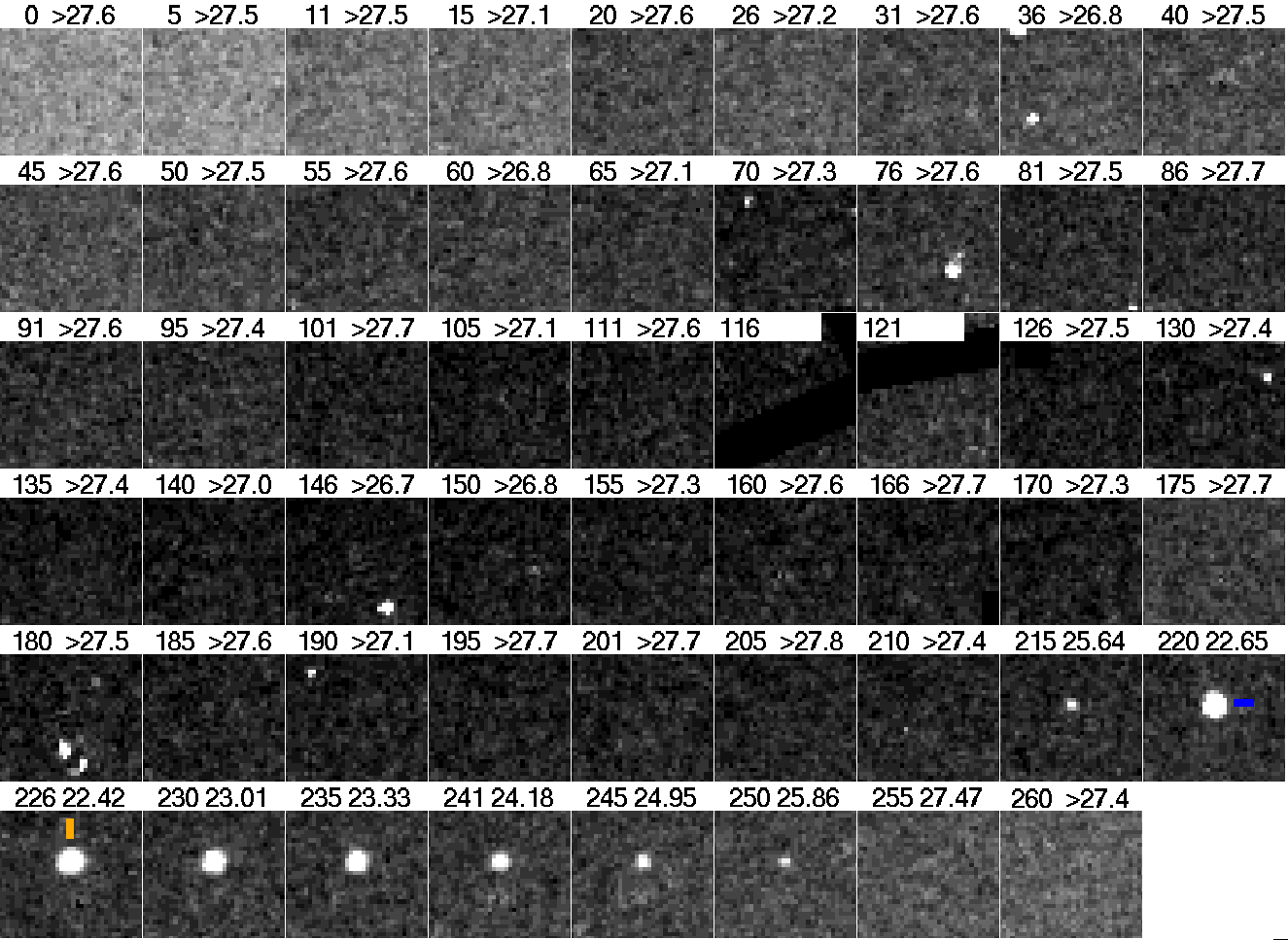} \\
     & \includegraphics[height=3.6in]{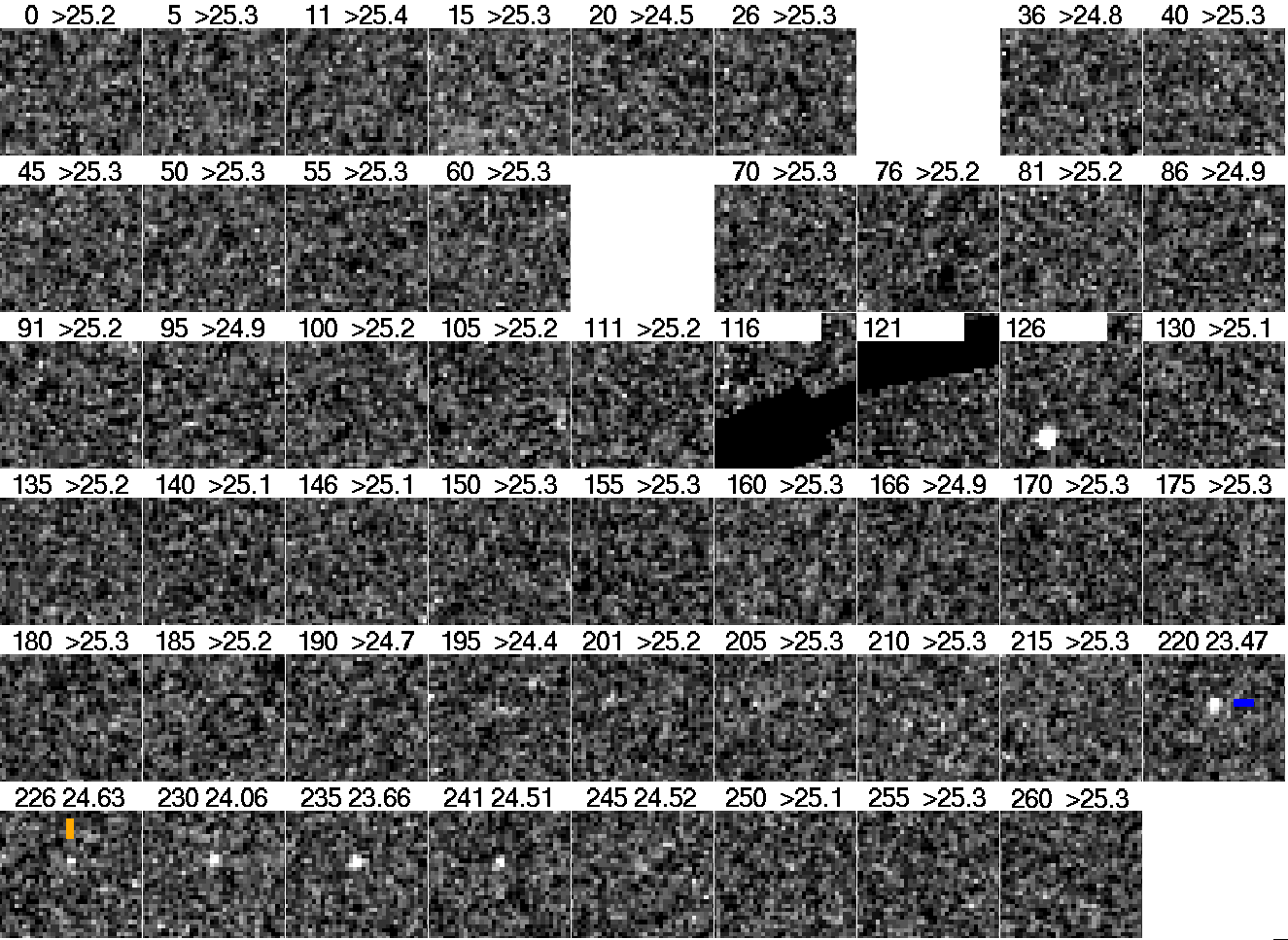}
    \end{tabular}
    \caption{Nova 4}
\end{sidewaysfigure}
\begin{sidewaysfigure}
    \begin{tabular}{cc}
     \multirow{2}{*}[2.6in]{\includegraphics[height=6in]{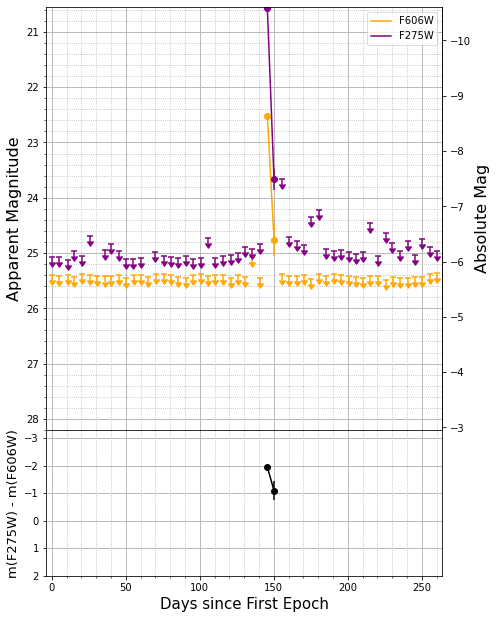}} & \includegraphics[height=3.6in]{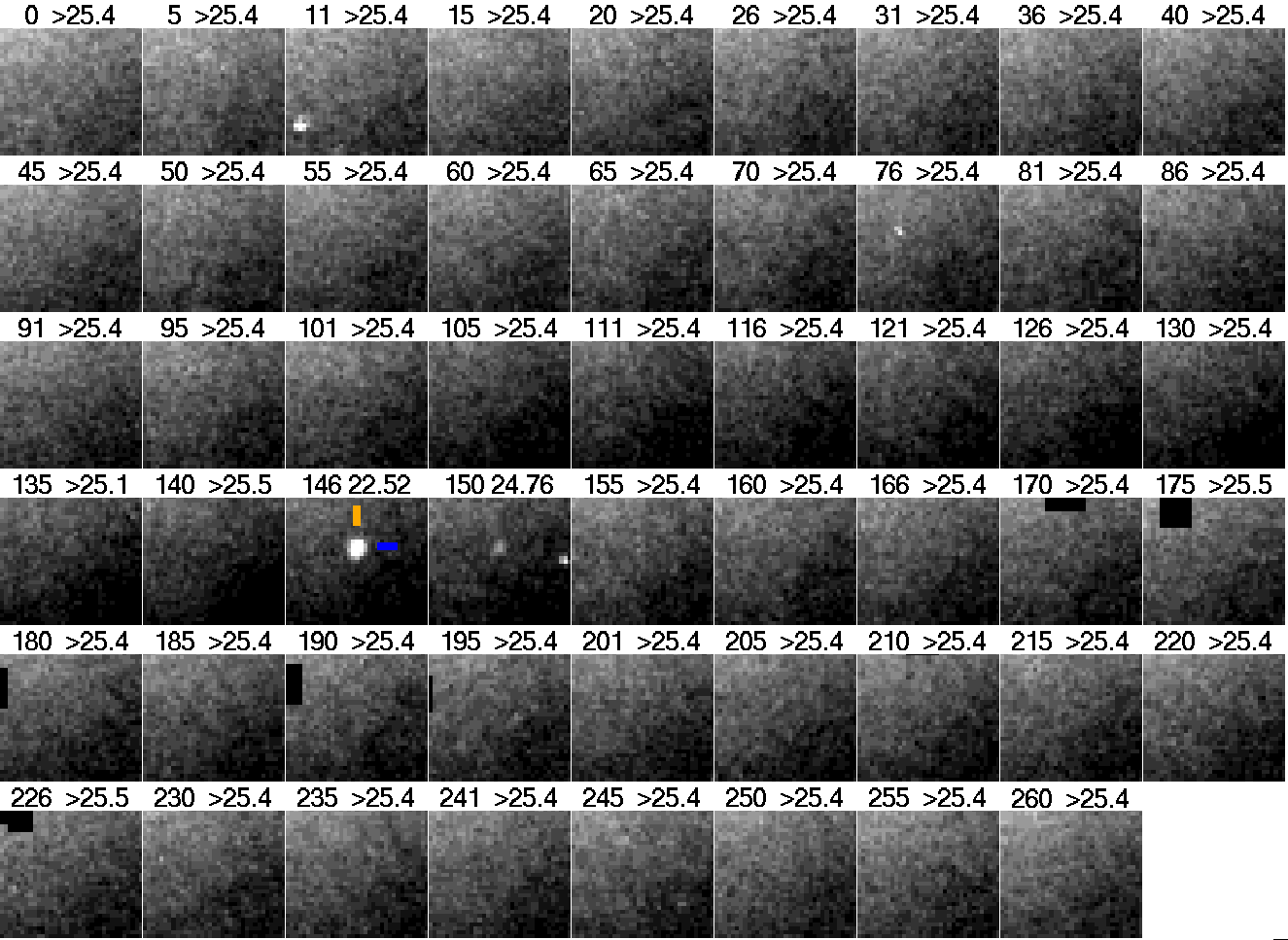} \\
     & \includegraphics[height=3.6in]{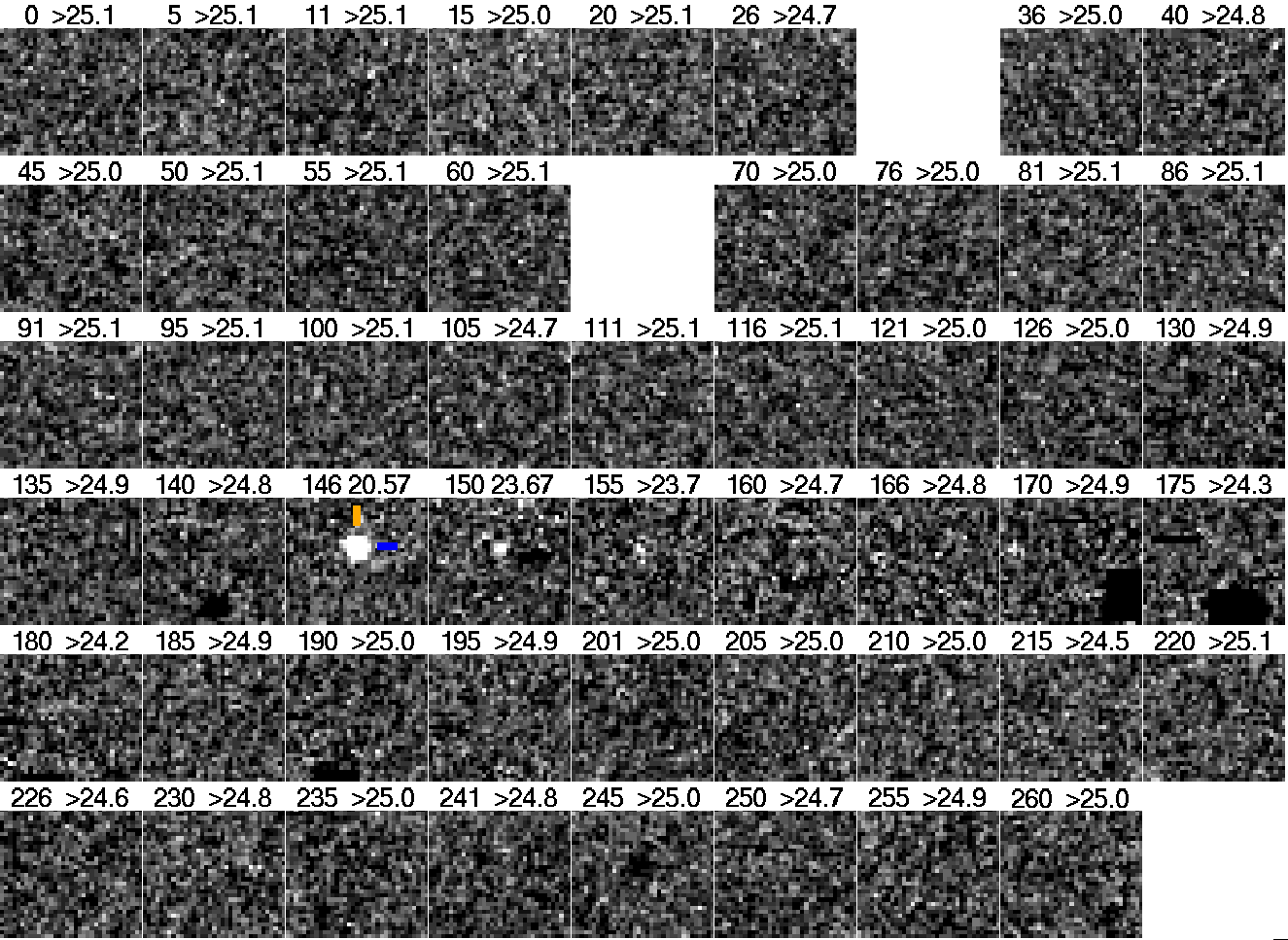}
    \end{tabular}
    \caption{Nova 5}
\end{sidewaysfigure}
\begin{sidewaysfigure}
    \begin{tabular}{cc}
     \multirow{2}{*}[2.6in]{\includegraphics[height=6in]{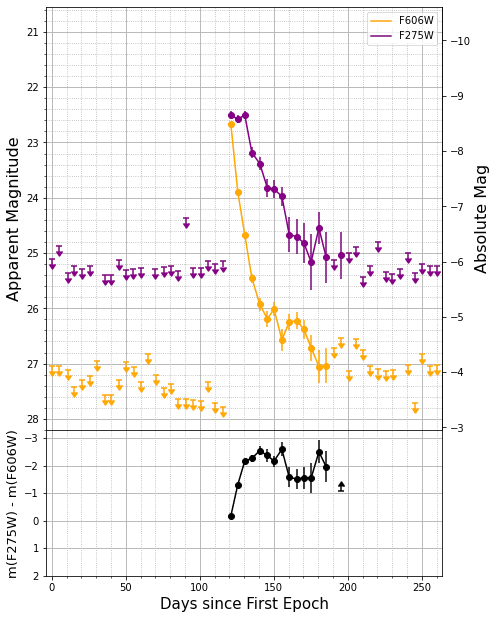}} & \includegraphics[height=3.6in]{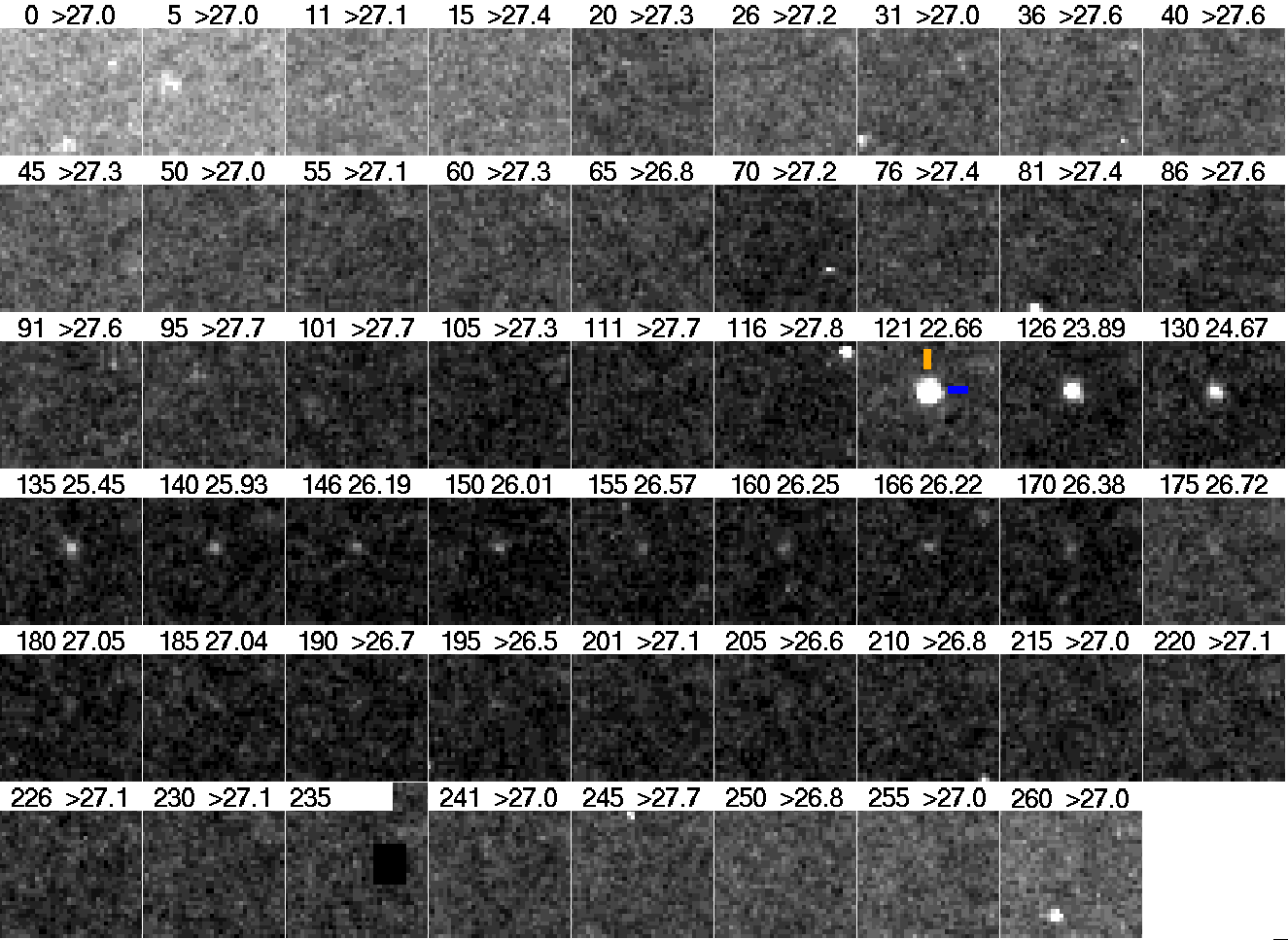} \\
     & \includegraphics[height=3.6in]{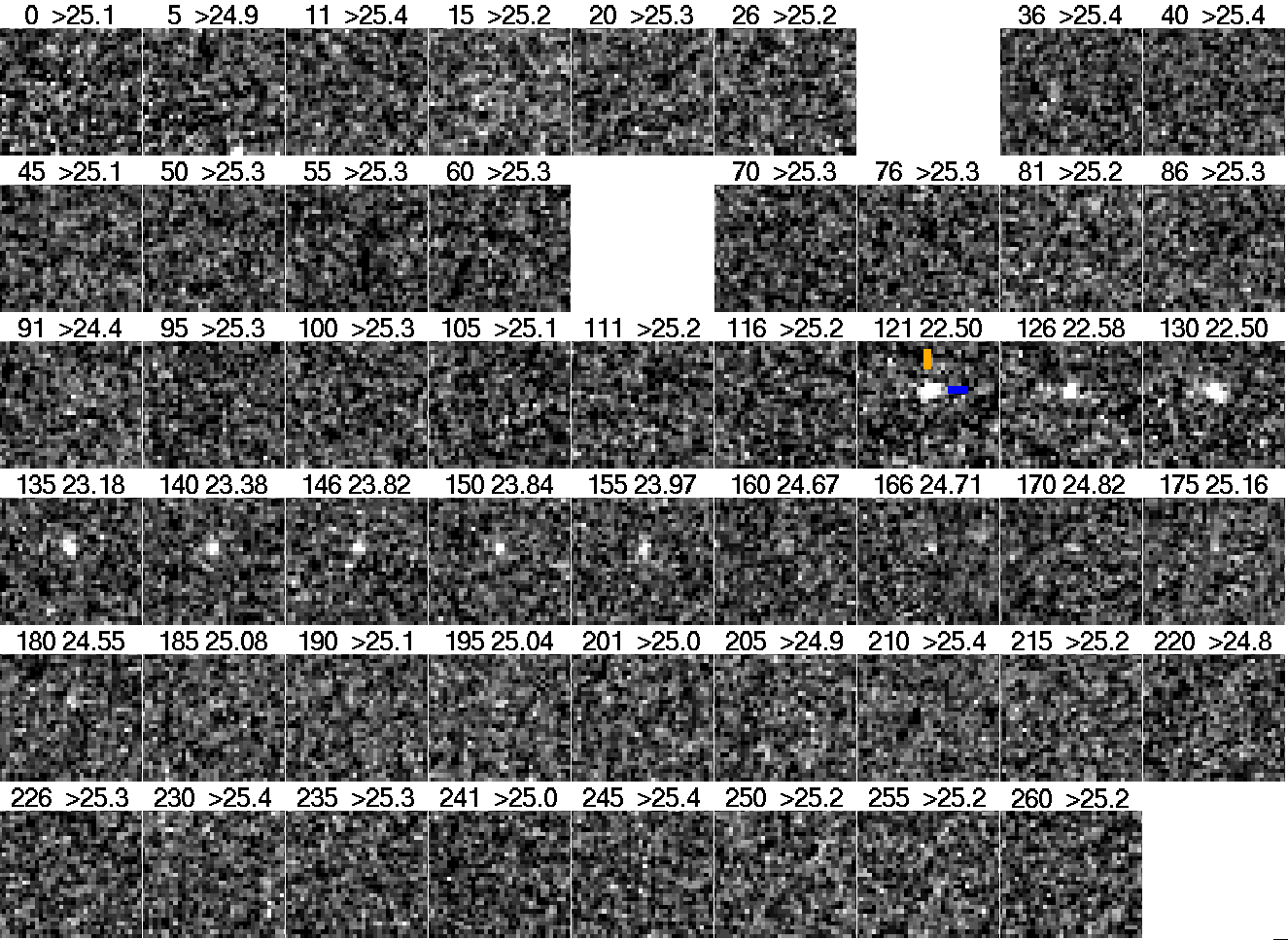}
    \end{tabular}
    \caption{Nova 6}
\end{sidewaysfigure}
\begin{sidewaysfigure}
    \begin{tabular}{cc}
     \multirow{2}{*}[2.6in]{\includegraphics[height=6in]{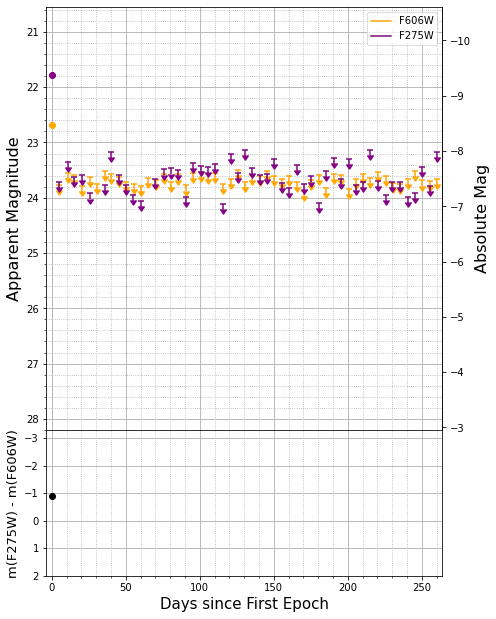}} & \includegraphics[height=3.6in]{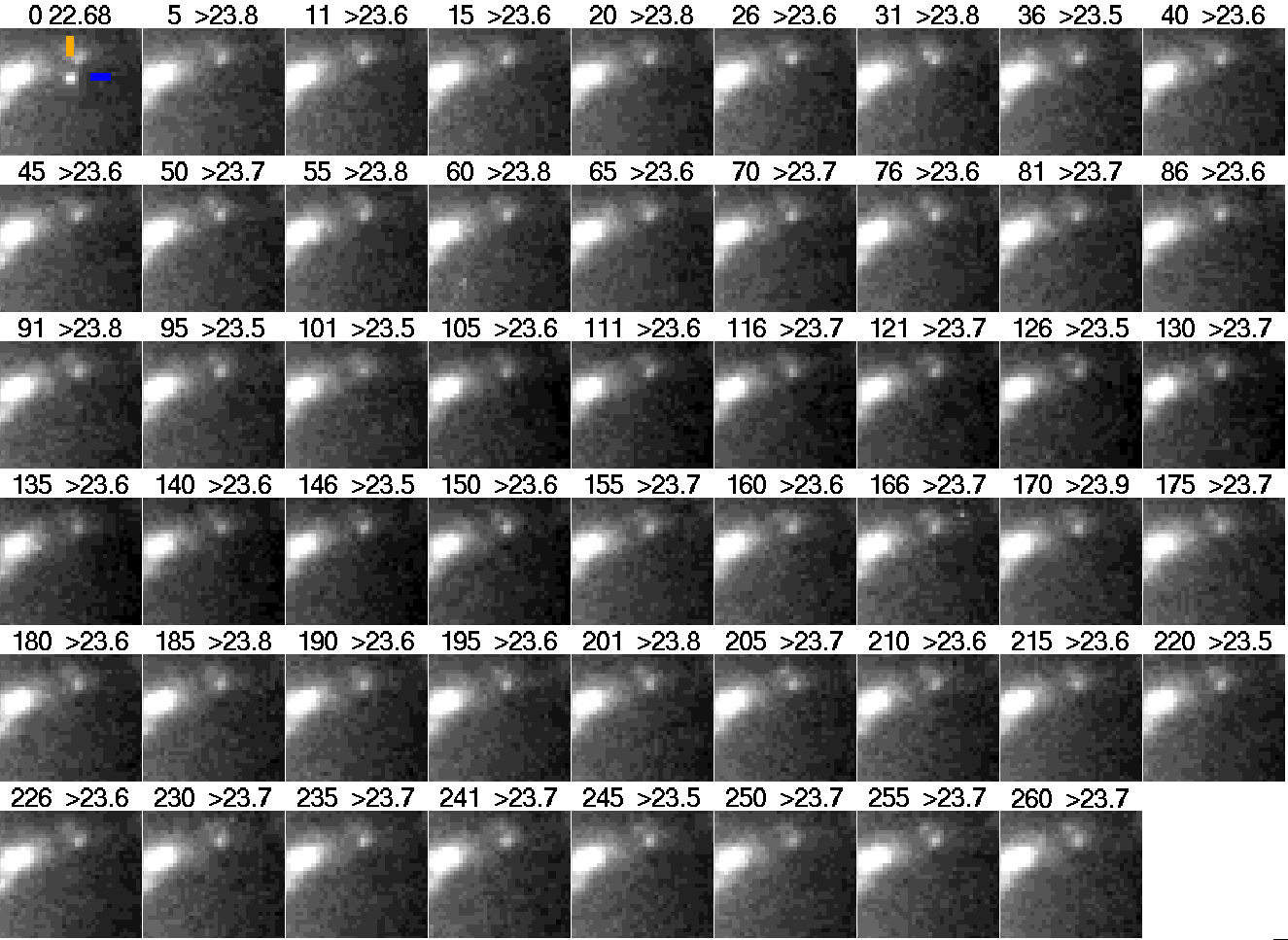} \\
     & \includegraphics[height=3.6in]{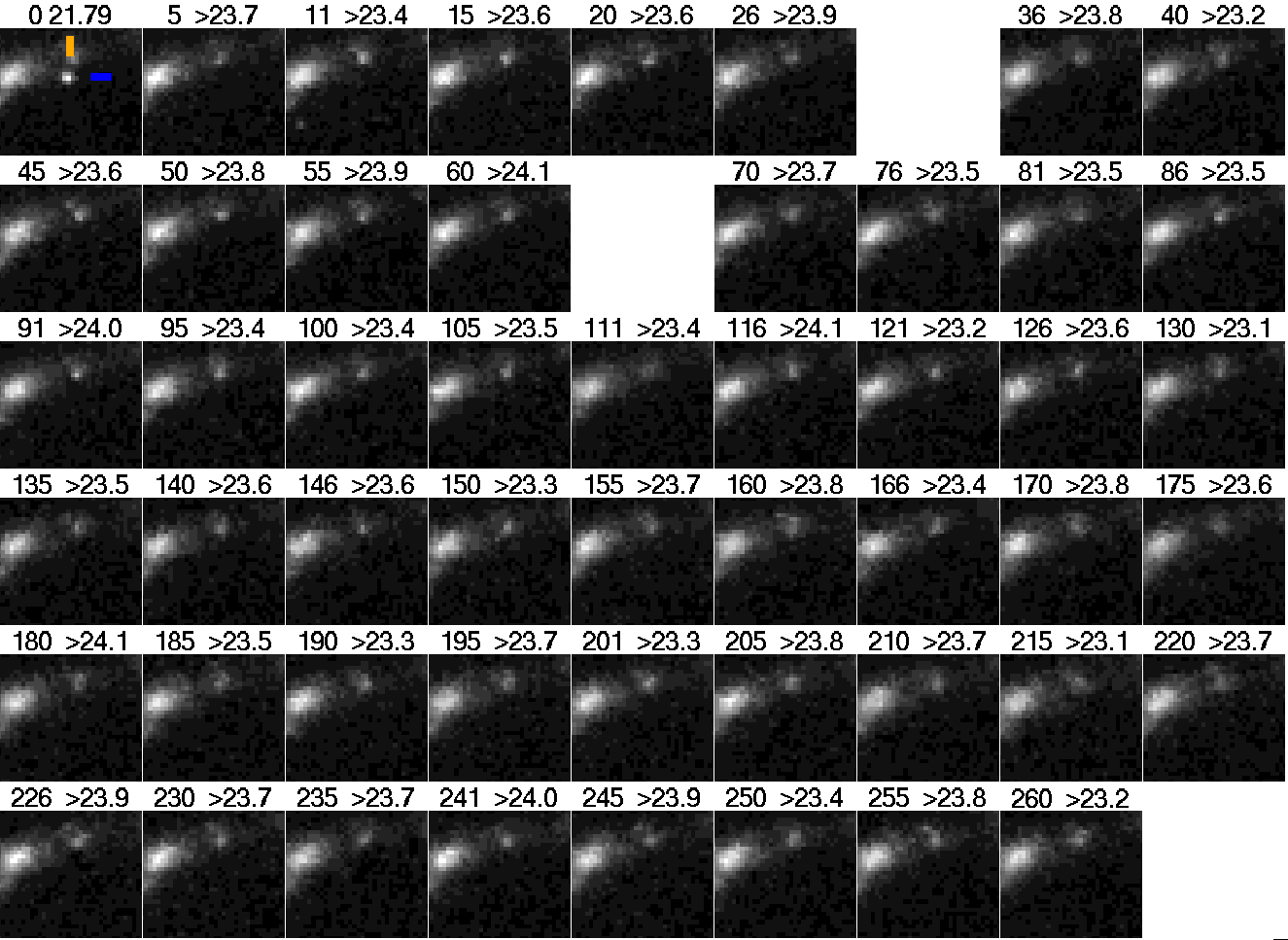}
    \end{tabular}
    \caption{Nova 7}
\end{sidewaysfigure}
\begin{sidewaysfigure}
    \begin{tabular}{cc}
     \multirow{2}{*}[2.6in]{\includegraphics[height=6in]{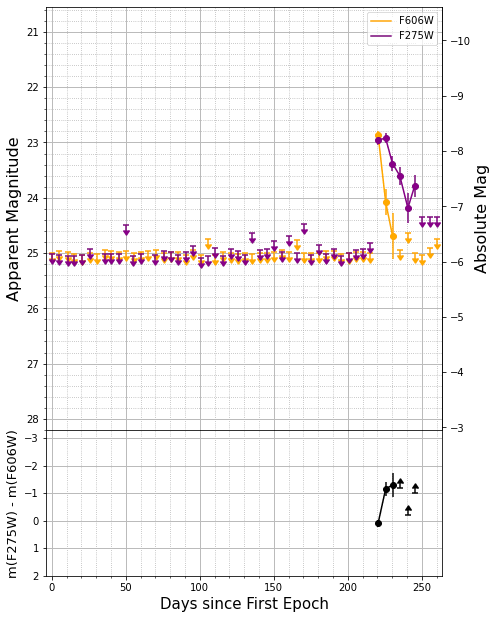}} & \includegraphics[height=3.6in]{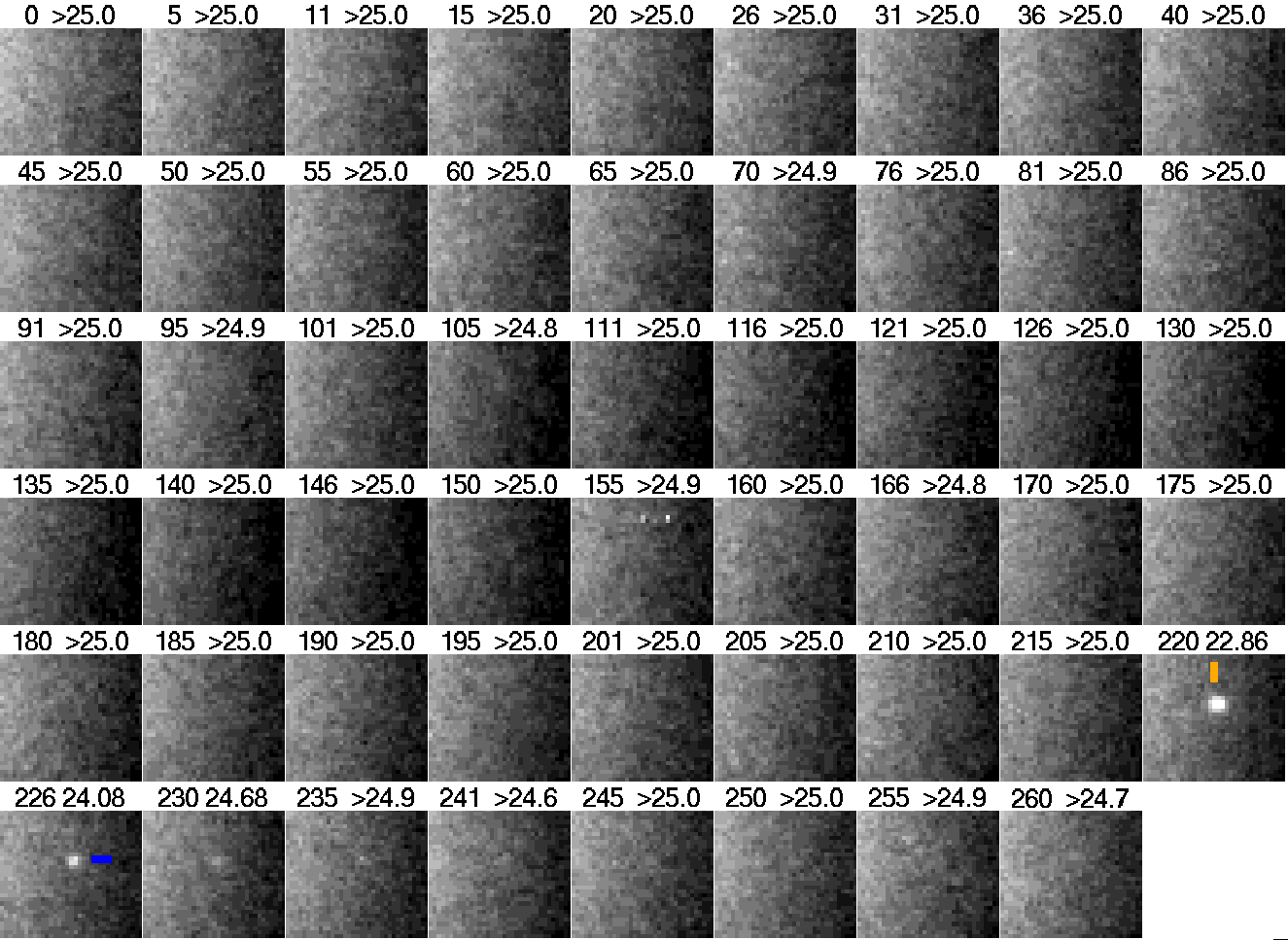} \\
     & \includegraphics[height=3.6in]{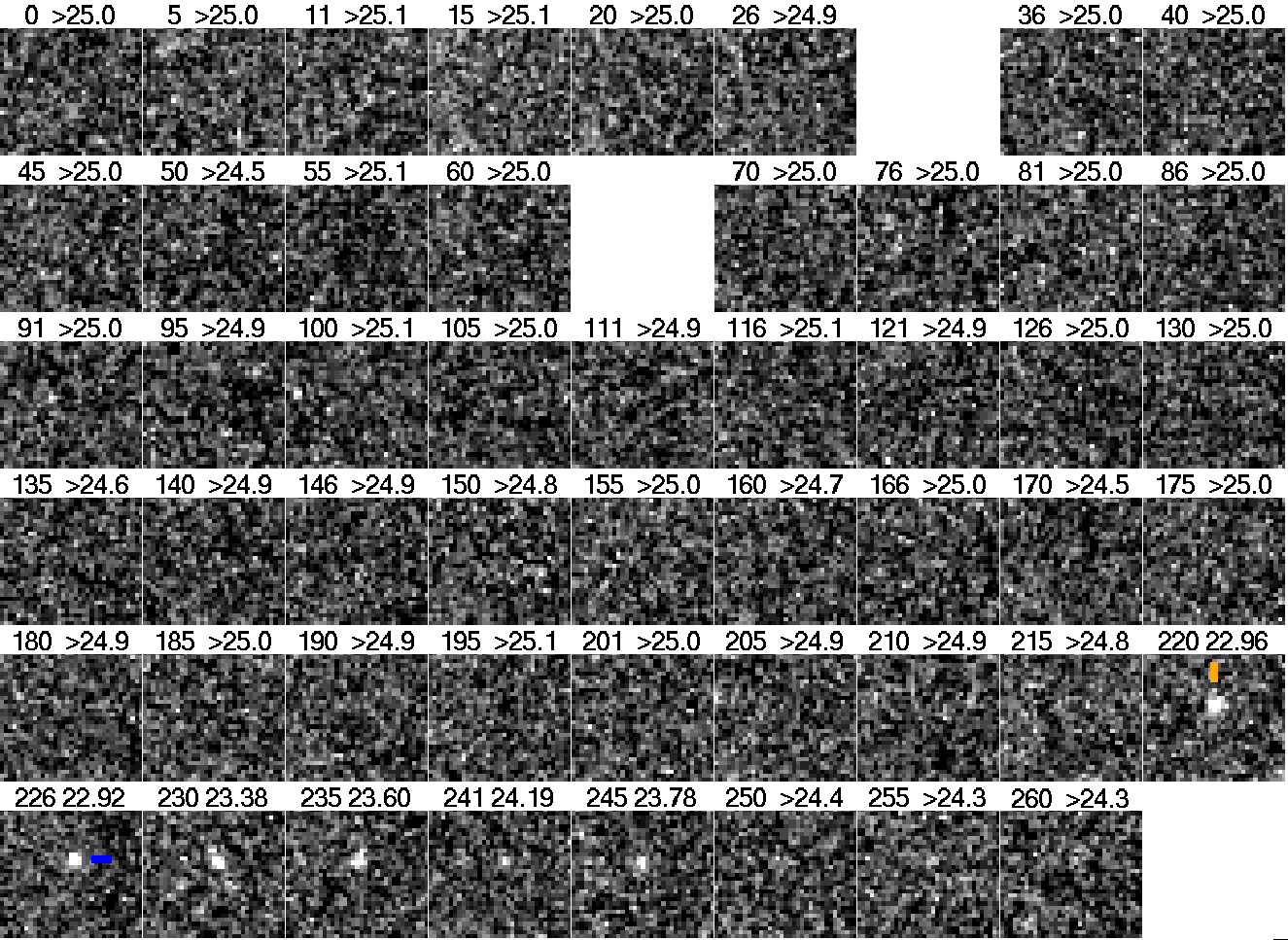}
    \end{tabular}
    \caption{Nova 8}
\end{sidewaysfigure}
\begin{sidewaysfigure}
    \begin{tabular}{cc}
     \multirow{2}{*}[2.6in]{\includegraphics[height=6in]{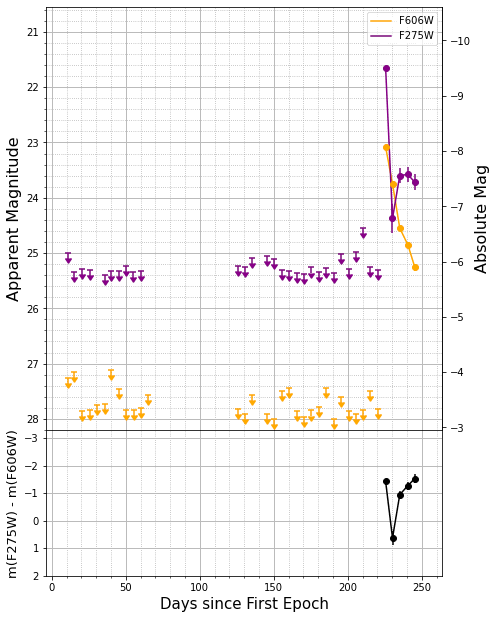}} & \includegraphics[height=3.6in]{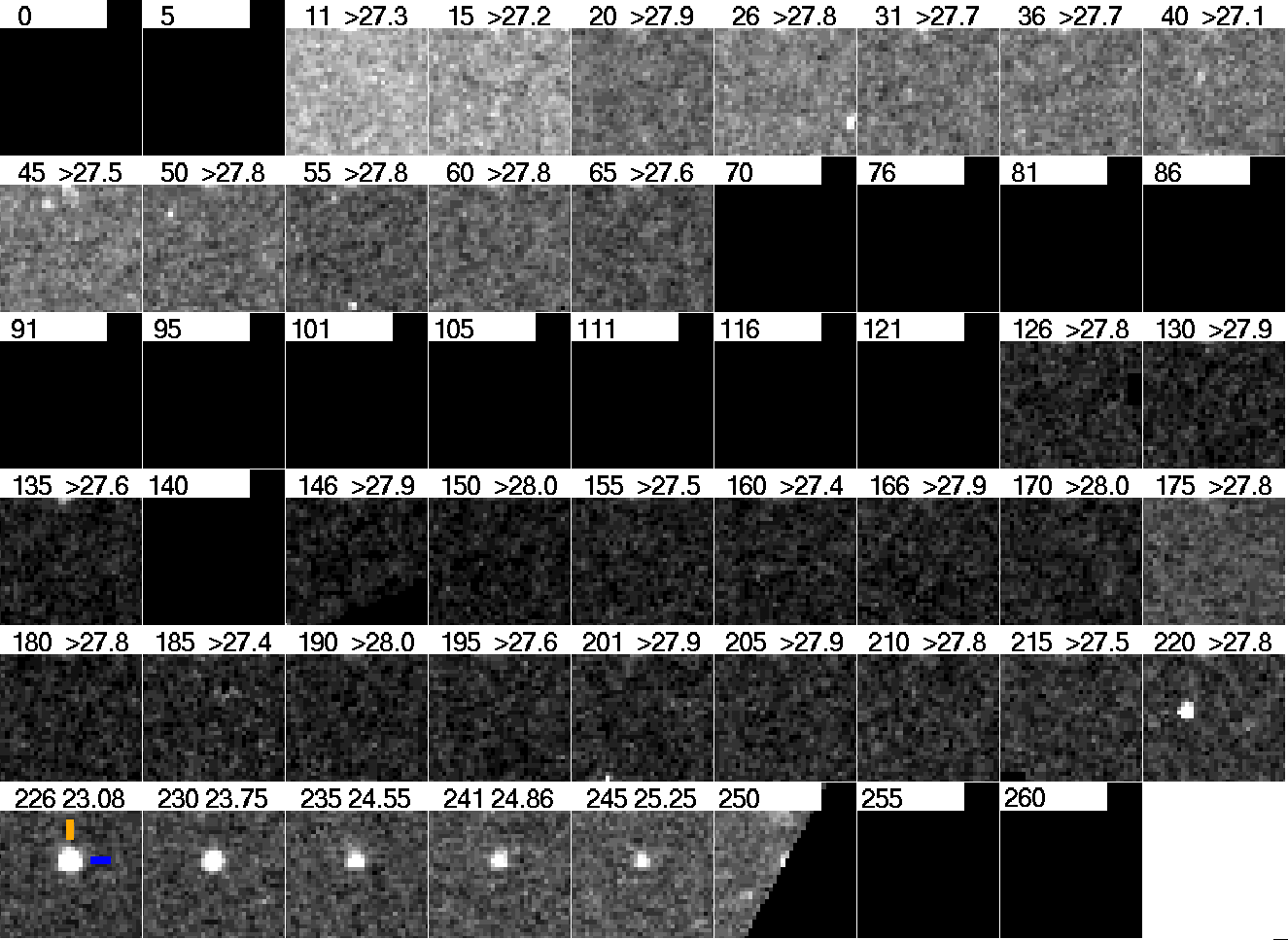} \\
     & \includegraphics[height=3.6in]{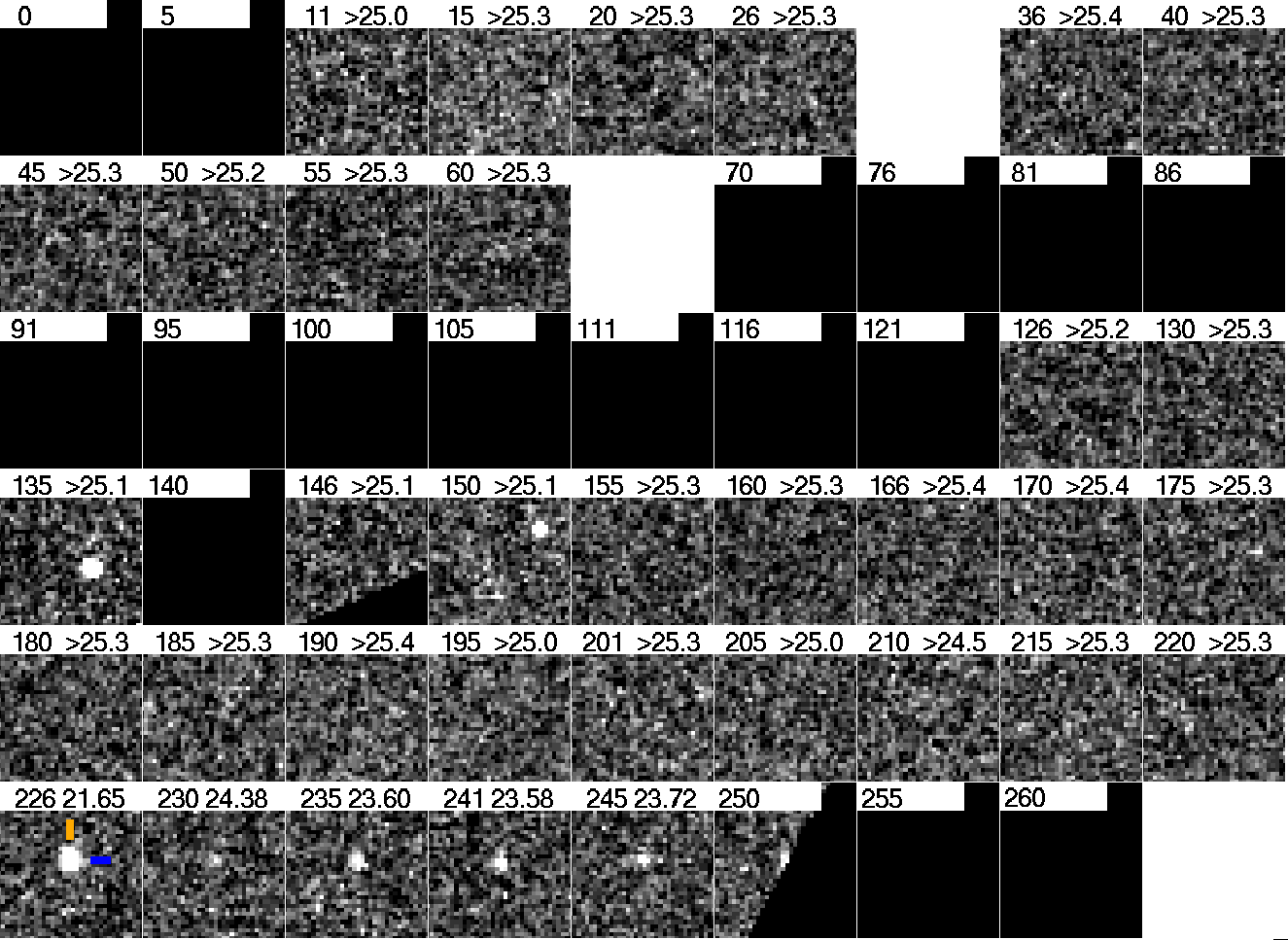}
    \end{tabular}
    \caption{Nova 9}
\end{sidewaysfigure}
\begin{sidewaysfigure}
    \begin{tabular}{cc}
     \multirow{2}{*}[2.6in]{\includegraphics[height=6in]{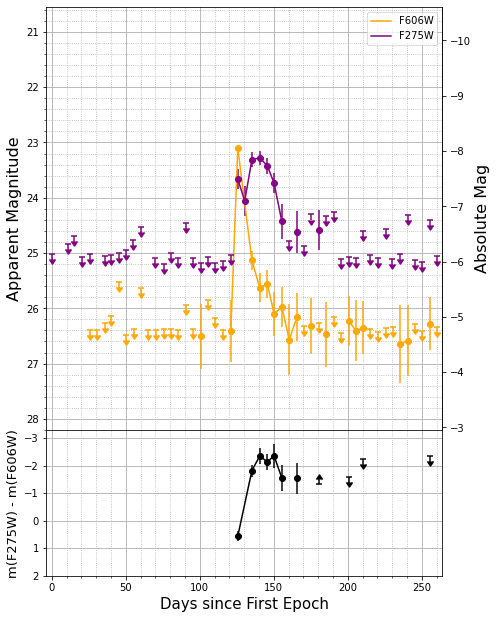}} & \includegraphics[height=3.6in]{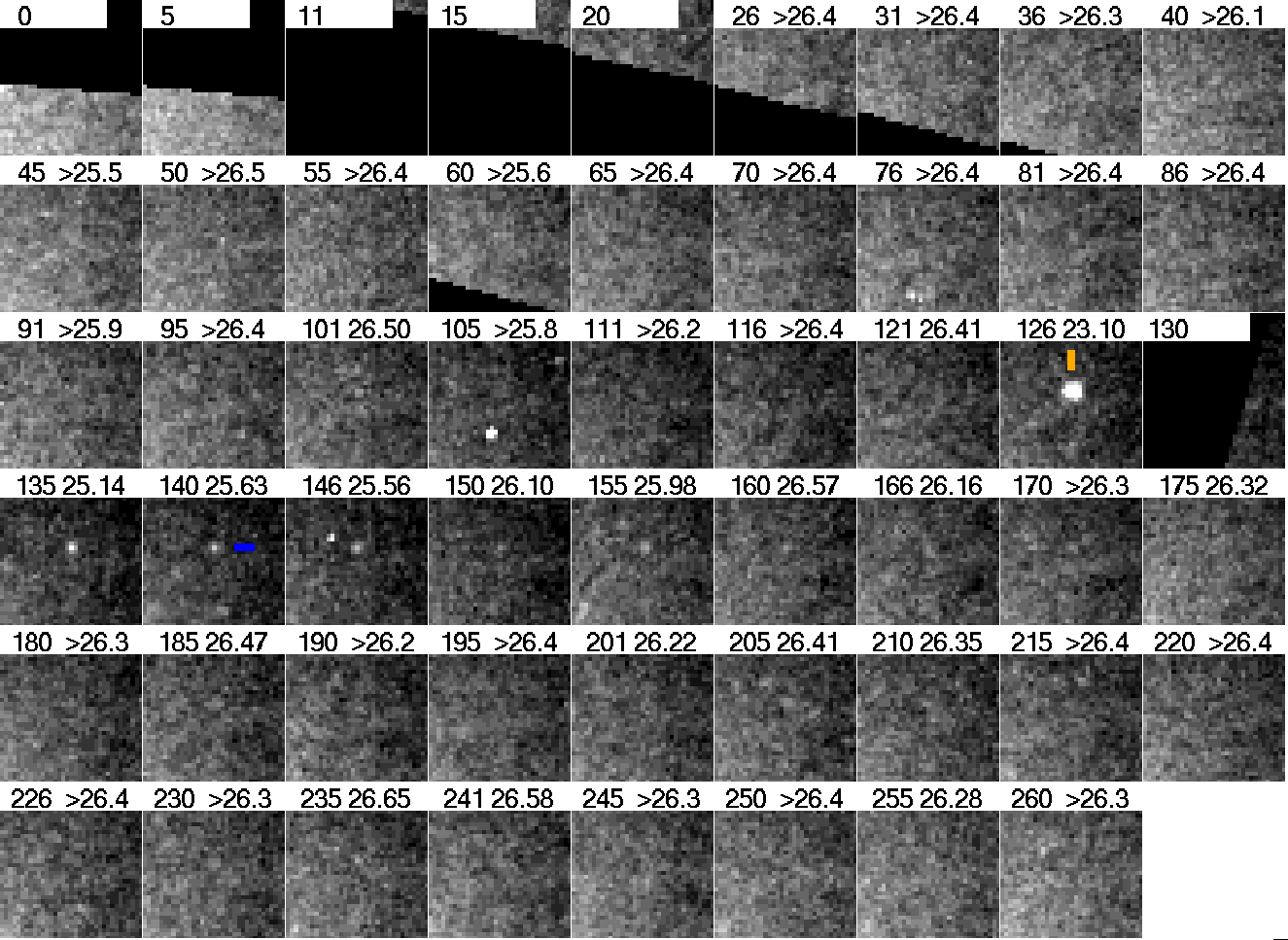} \\
     & \includegraphics[height=3.6in]{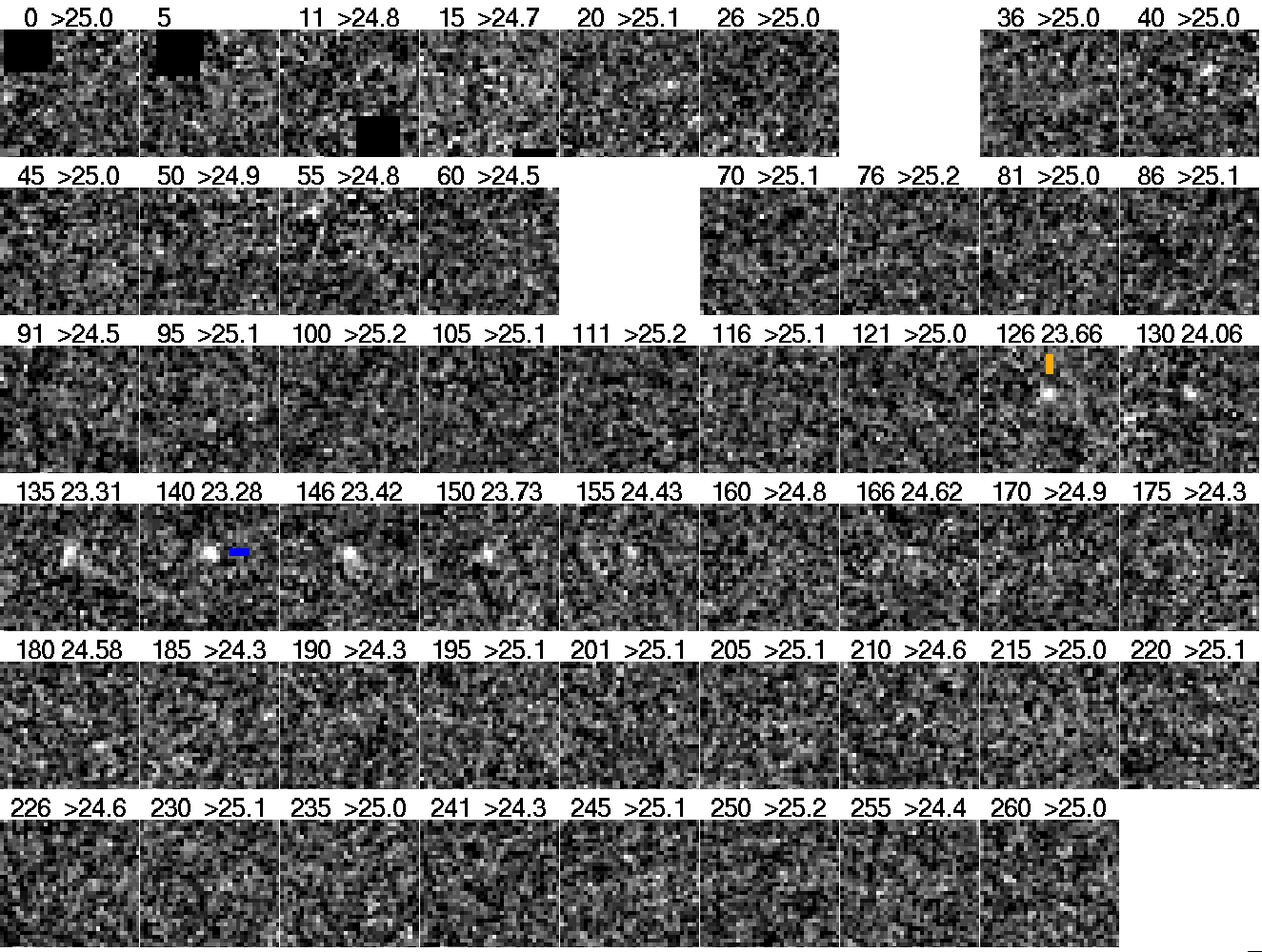}
    \end{tabular}
    \caption{Nova 10}
\end{sidewaysfigure}
\begin{sidewaysfigure}
    \begin{tabular}{cc}
     \multirow{2}{*}[2.6in]{\includegraphics[height=6in]{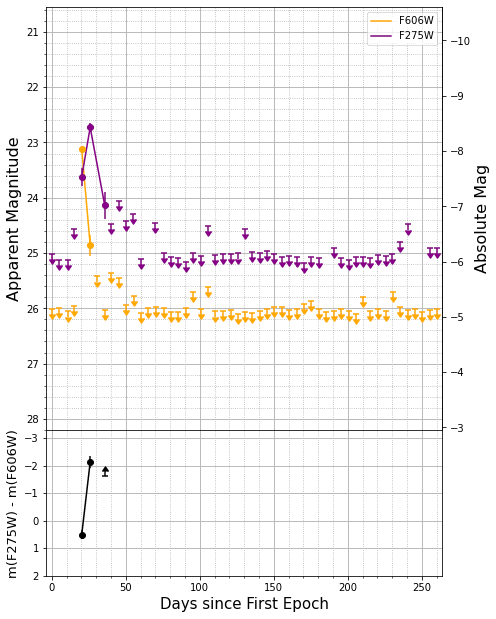}} & \includegraphics[height=3.6in]{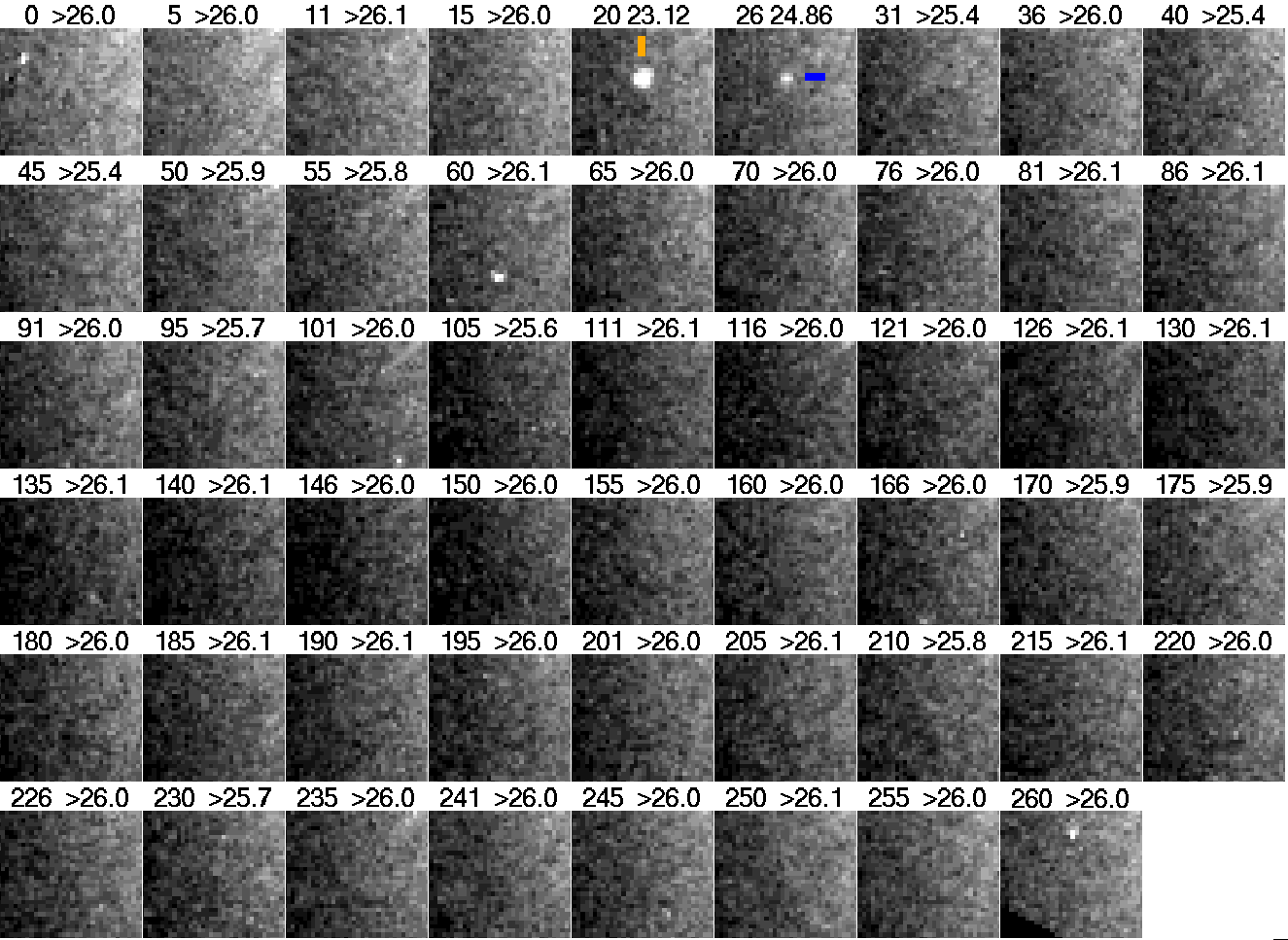} \\
     & \includegraphics[height=3.6in]{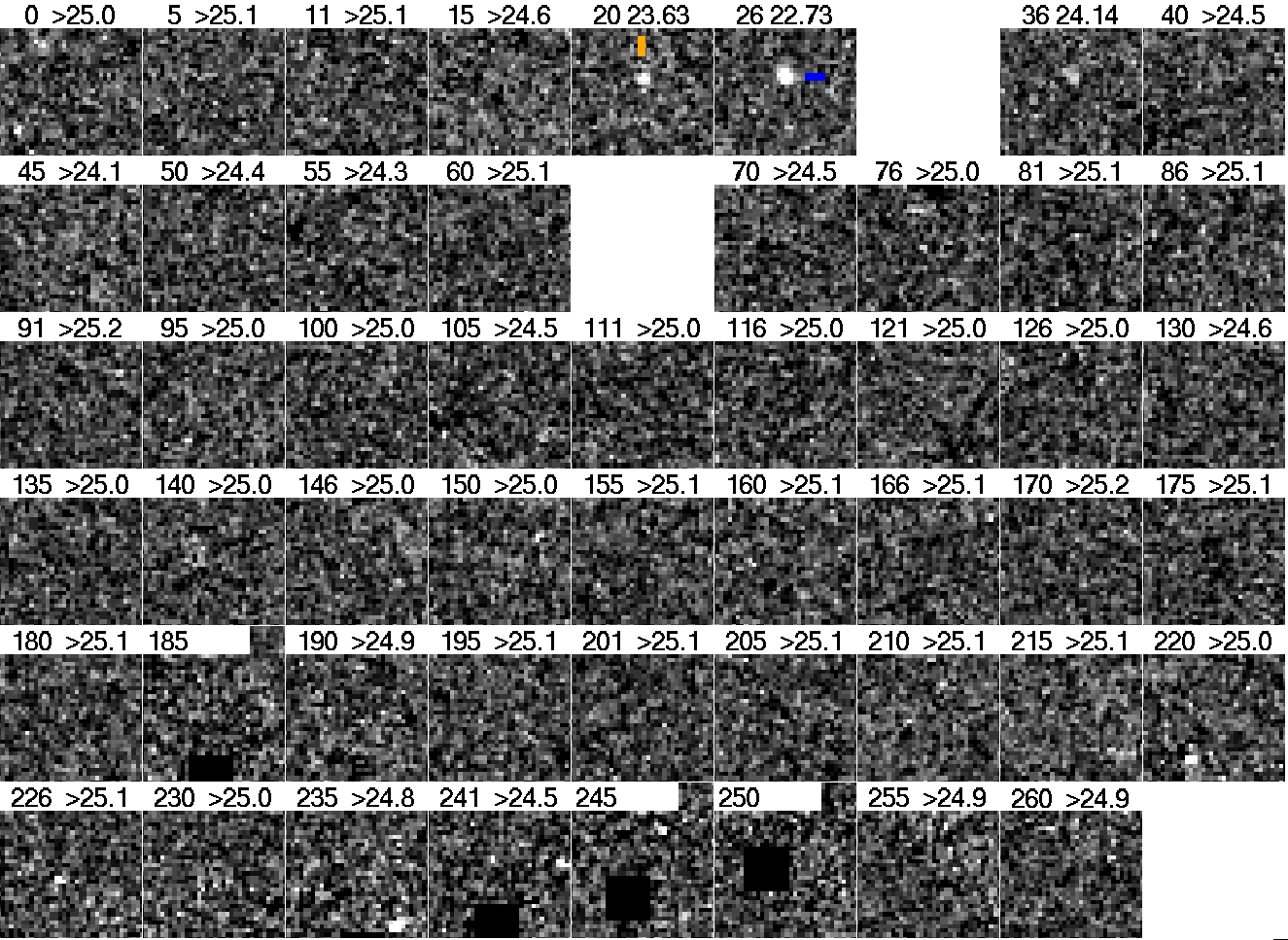}
    \end{tabular}
    \caption{Nova 11}
\end{sidewaysfigure}
\begin{sidewaysfigure}
    \begin{tabular}{cc}
     \multirow{2}{*}[2.6in]{\includegraphics[height=6in]{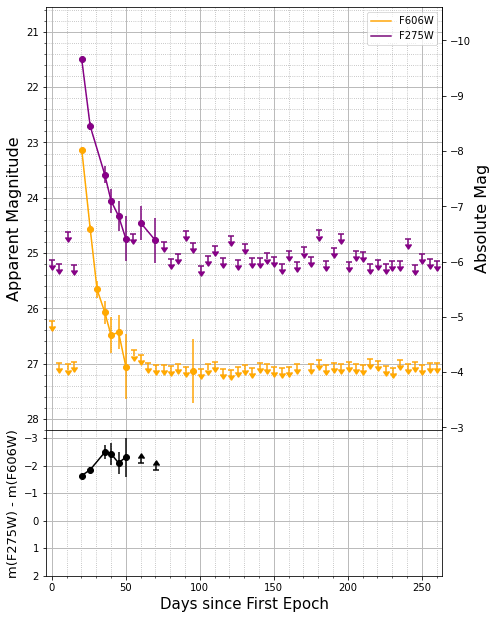}} & \includegraphics[height=3.6in]{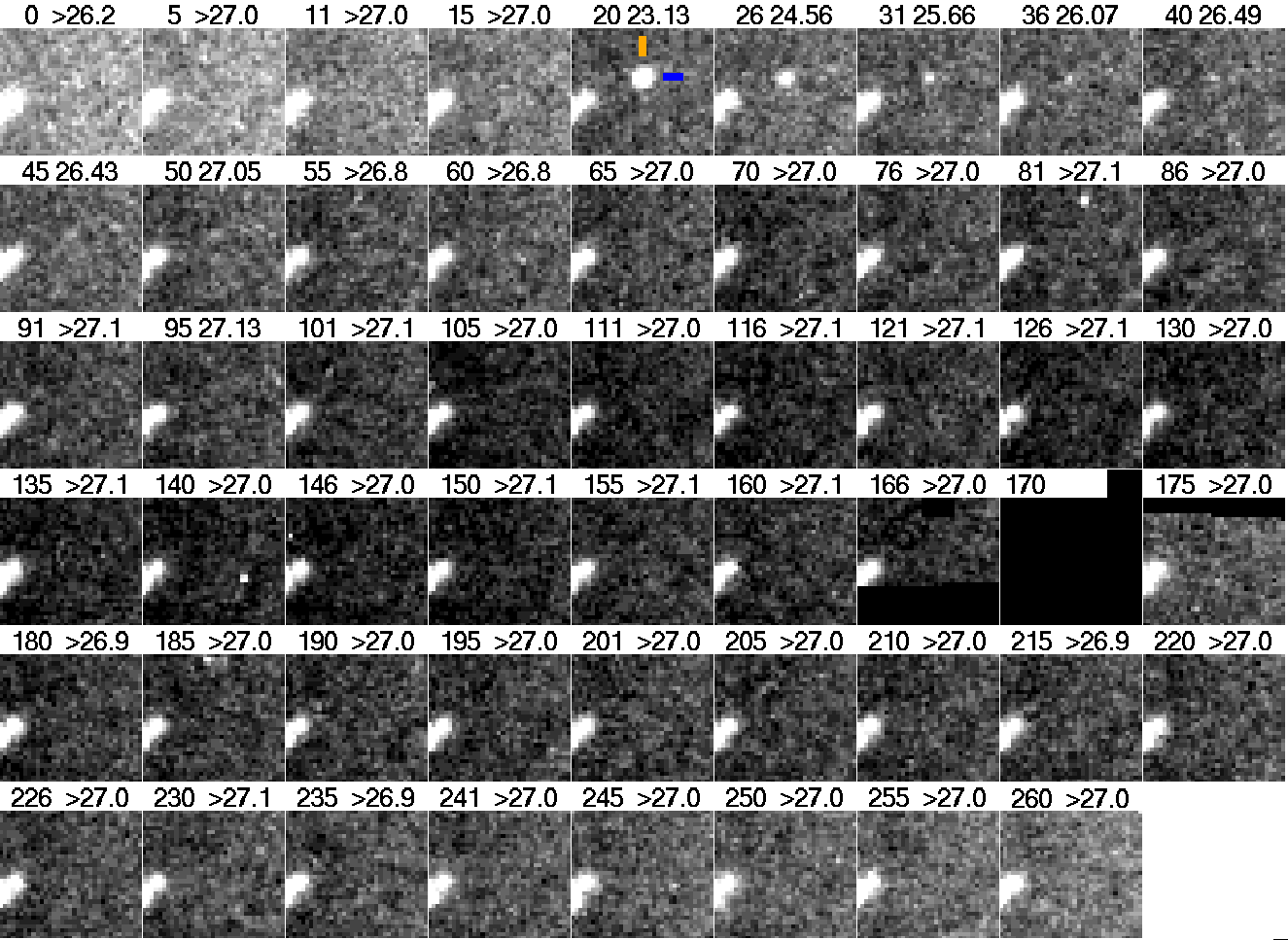} \\
     & \includegraphics[height=3.6in]{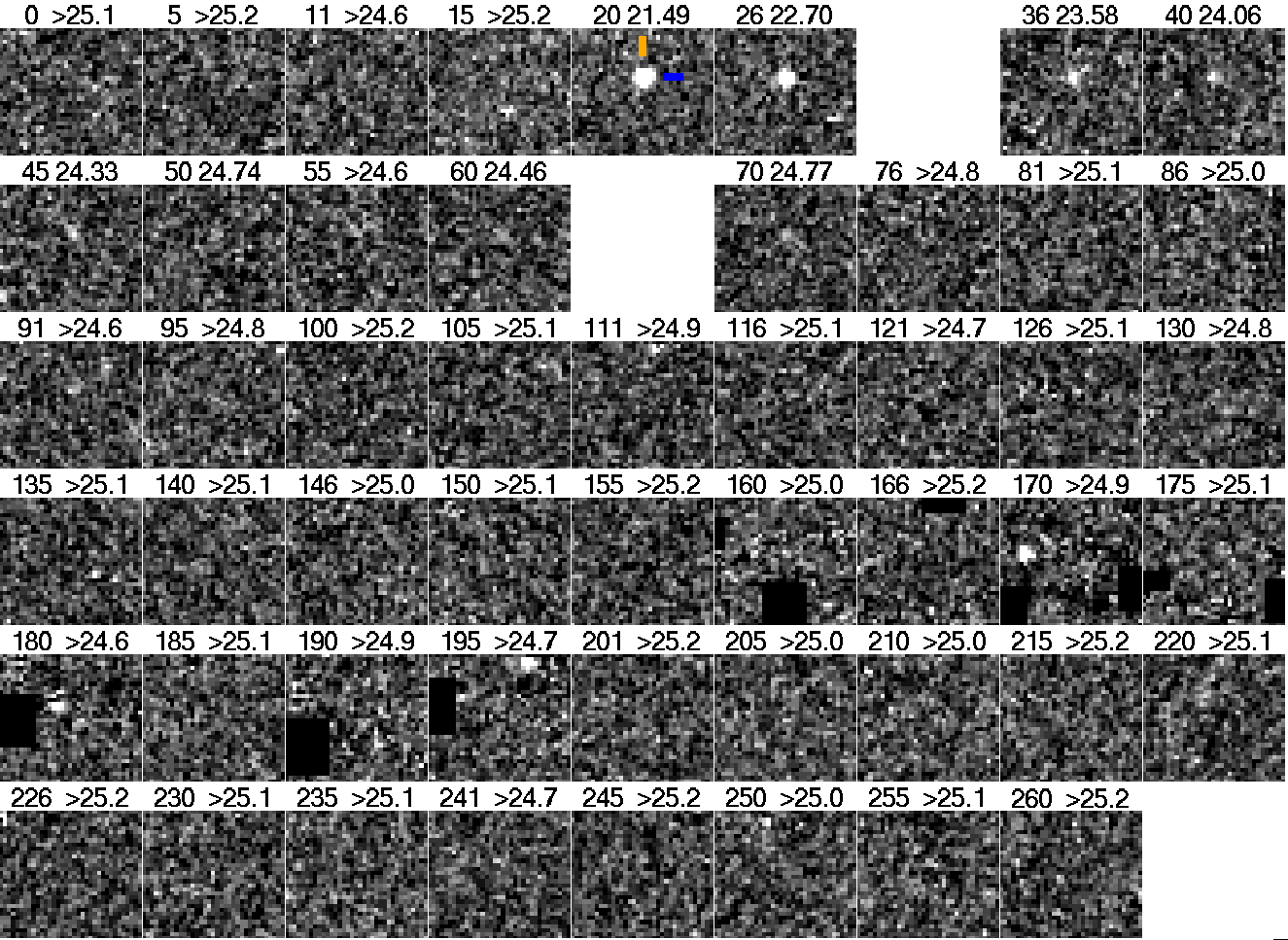}
    \end{tabular}
    \caption{Nova 12}
\end{sidewaysfigure}
\begin{sidewaysfigure}
    \begin{tabular}{cc}
     \multirow{2}{*}[2.6in]{\includegraphics[height=6in]{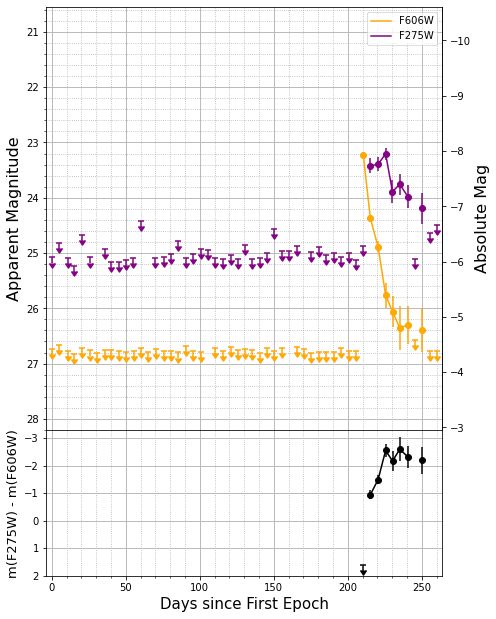}} & \includegraphics[height=3.6in]{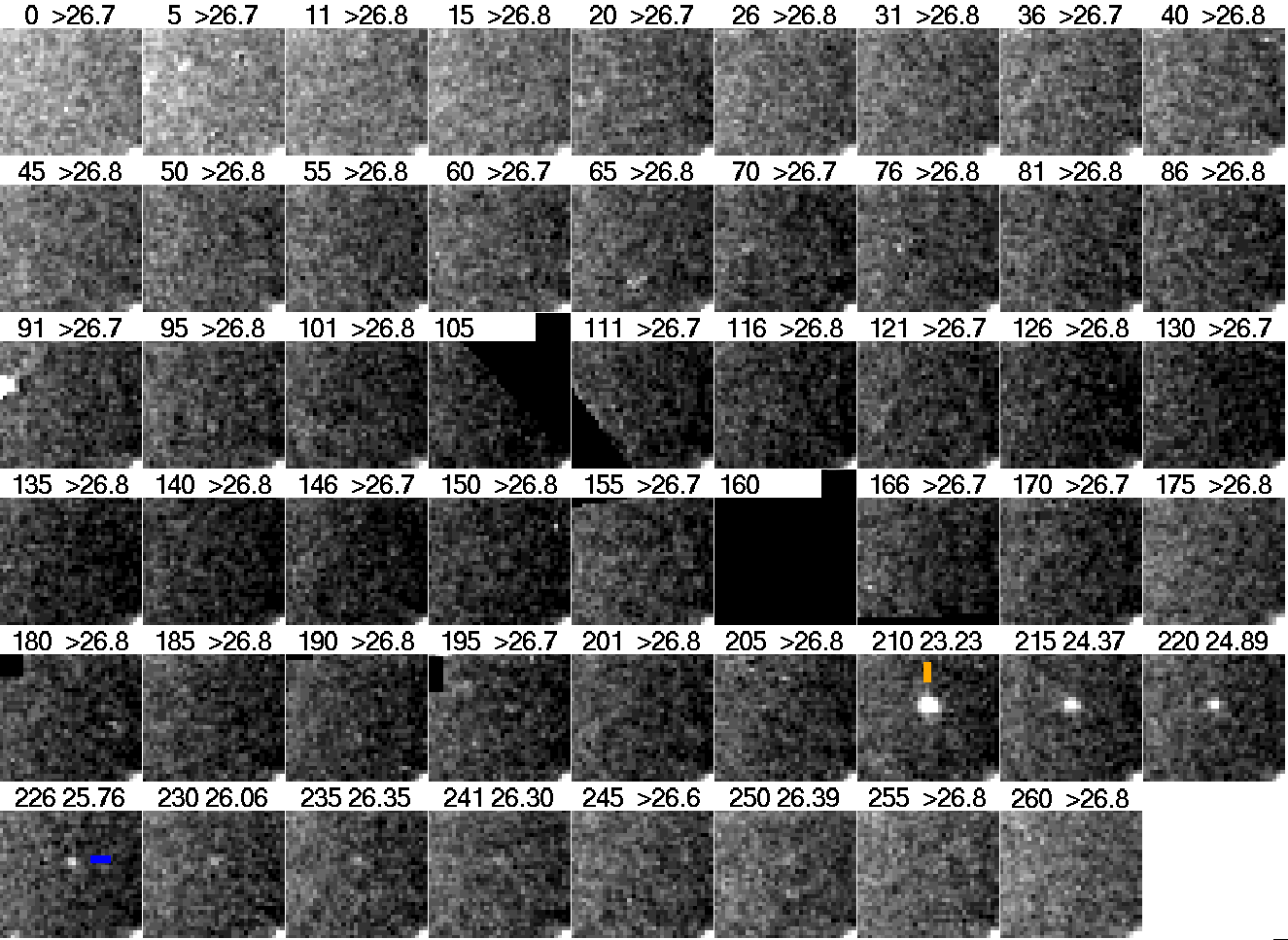} \\
     & \includegraphics[height=3.6in]{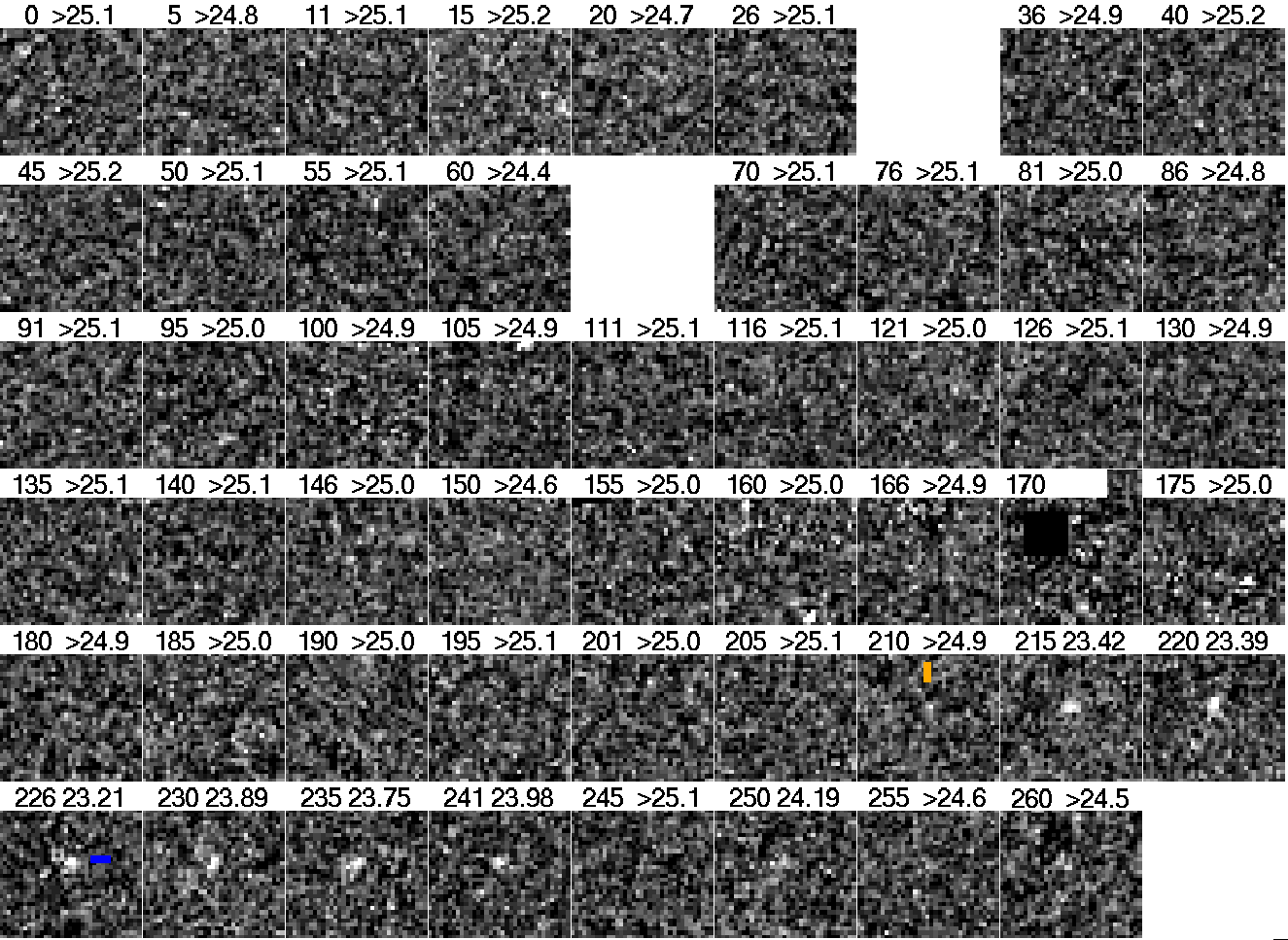}
    \end{tabular}
    \caption{Nova 13}
\end{sidewaysfigure}
\begin{sidewaysfigure}
    \begin{tabular}{cc}
     \multirow{2}{*}[2.6in]{\includegraphics[height=6in]{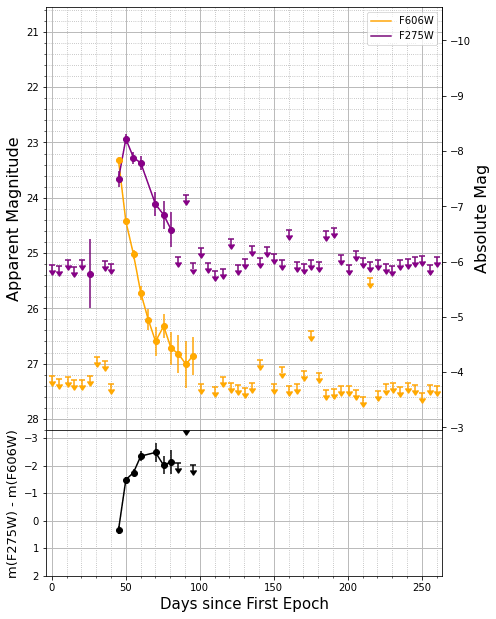}} & \includegraphics[height=3.6in]{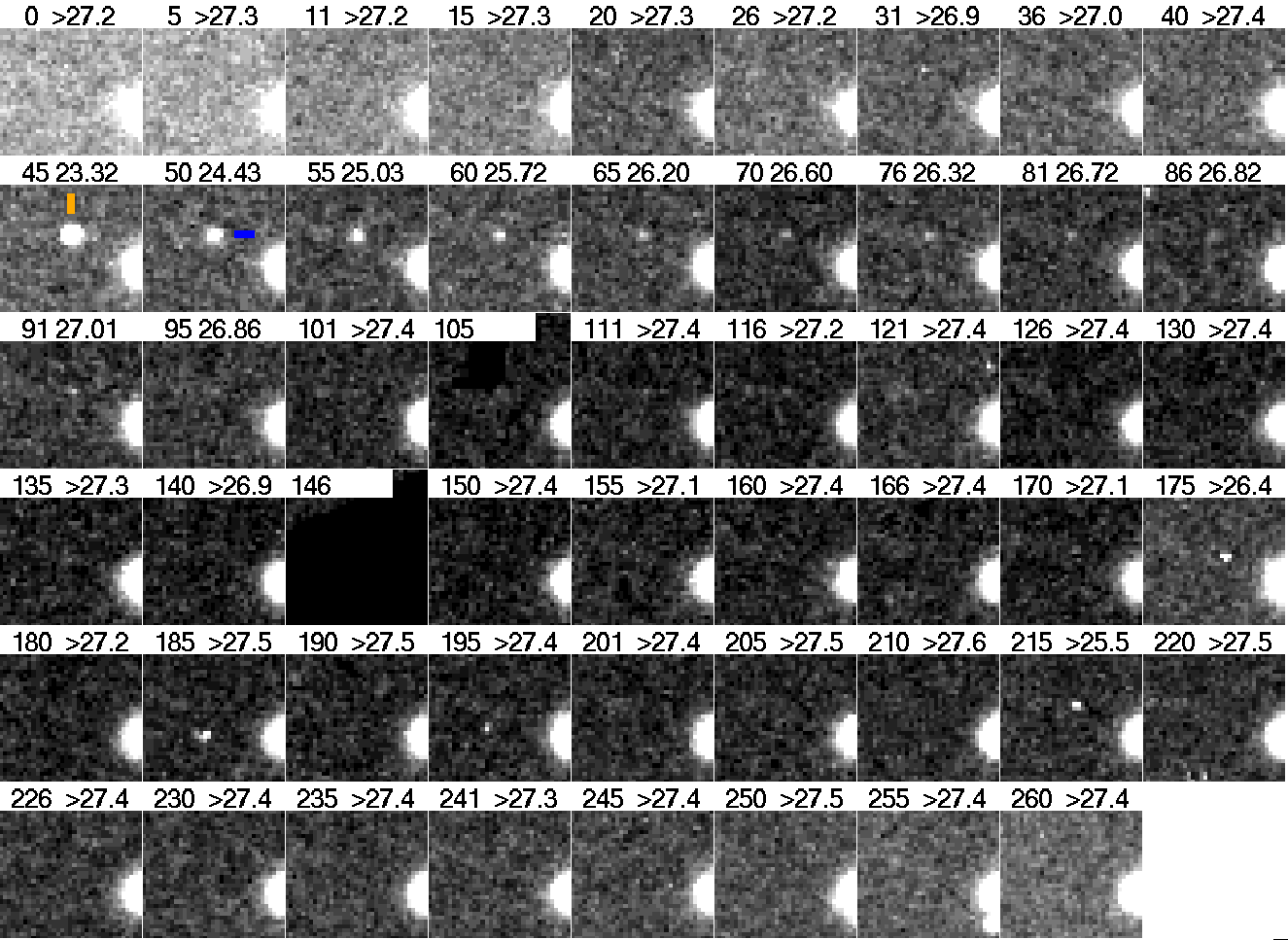} \\
     & \includegraphics[height=3.6in]{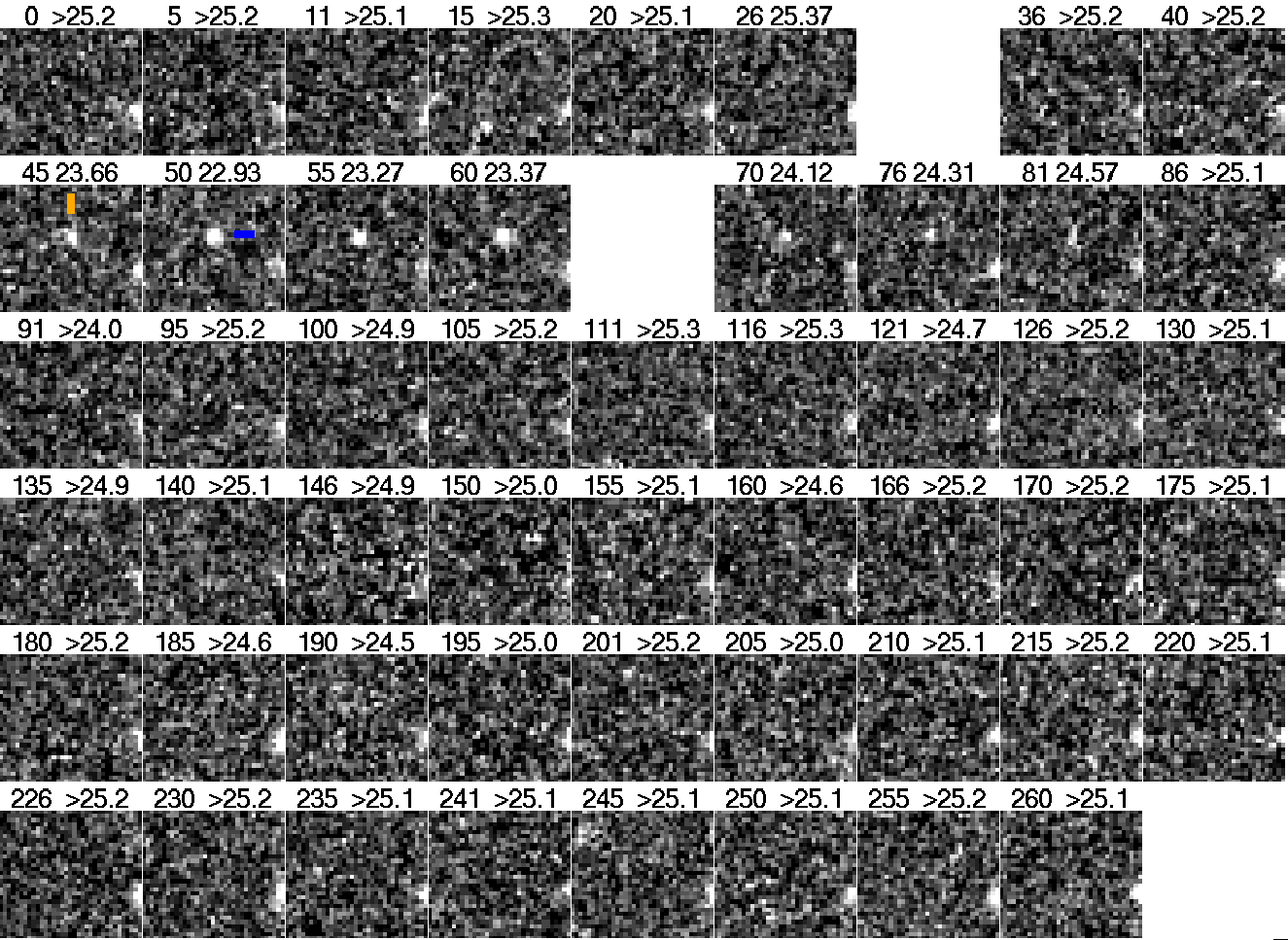}
    \end{tabular}
    \caption{Nova 14}
\end{sidewaysfigure}
\begin{sidewaysfigure}
    \begin{tabular}{cc}
     \multirow{2}{*}[2.6in]{\includegraphics[height=6in]{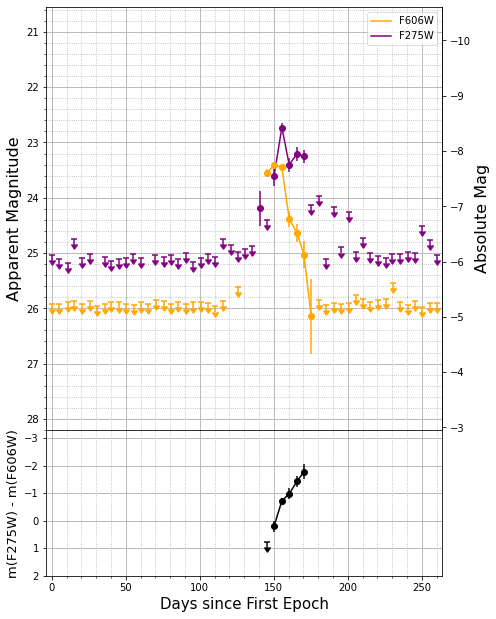}} & \includegraphics[height=3.6in]{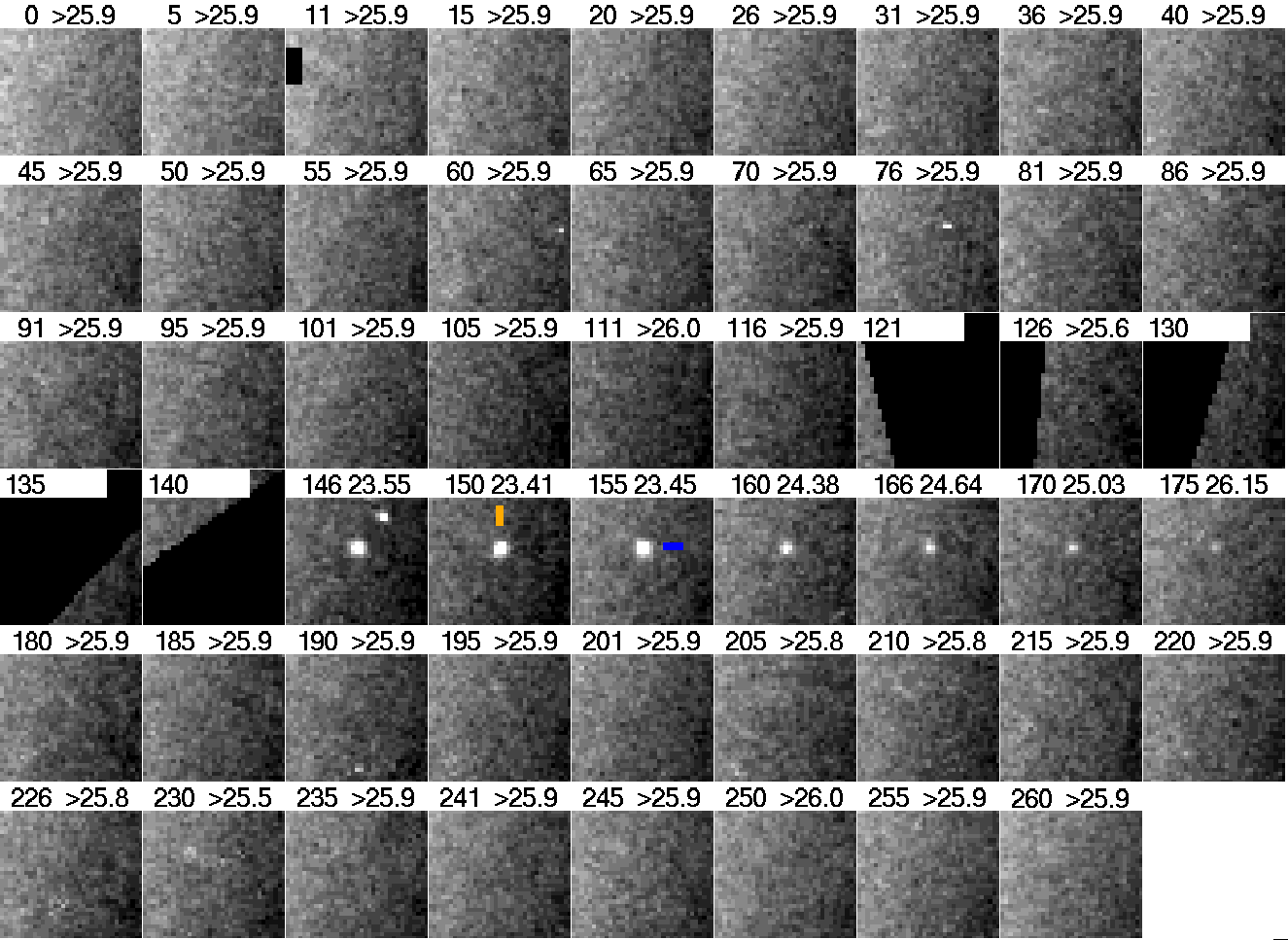} \\
     & \includegraphics[height=3.6in]{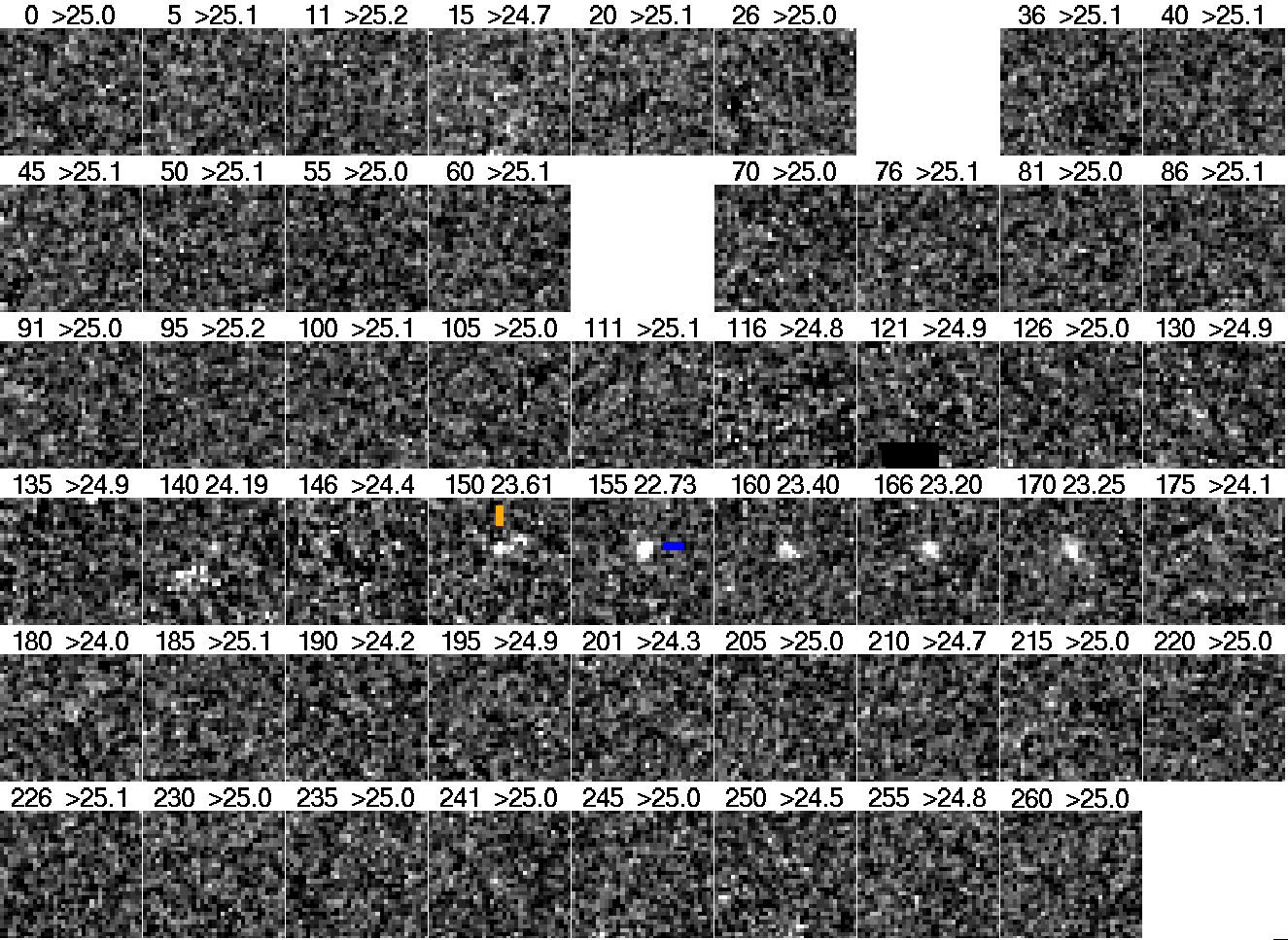}
    \end{tabular}
    \caption{Nova 15}
\end{sidewaysfigure}
\begin{sidewaysfigure}
    \begin{tabular}{cc}
     \multirow{2}{*}[2.6in]{\includegraphics[height=6in]{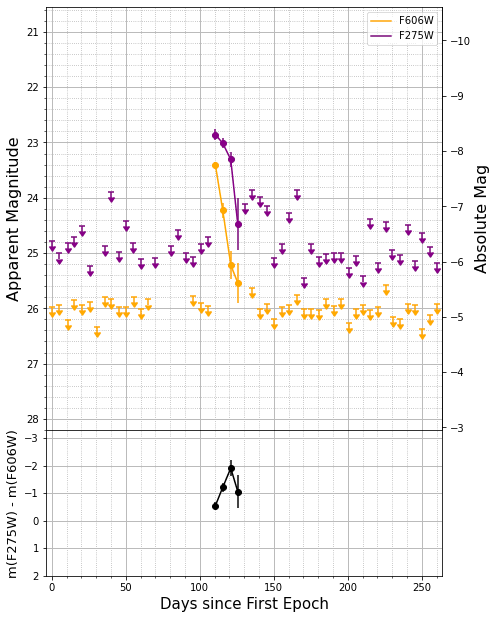}} & \includegraphics[height=3.6in]{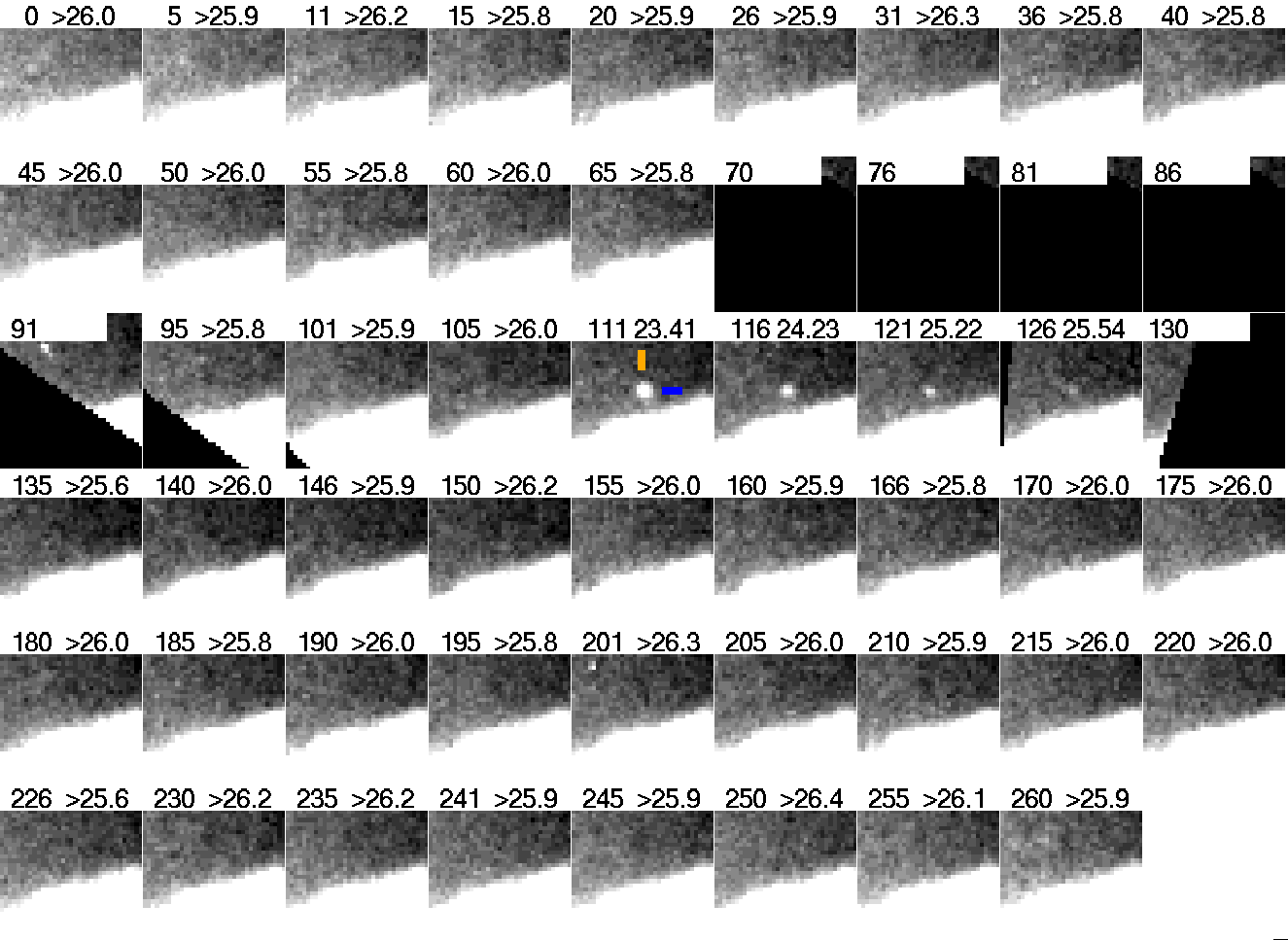} \\
     & \includegraphics[height=3.6in]{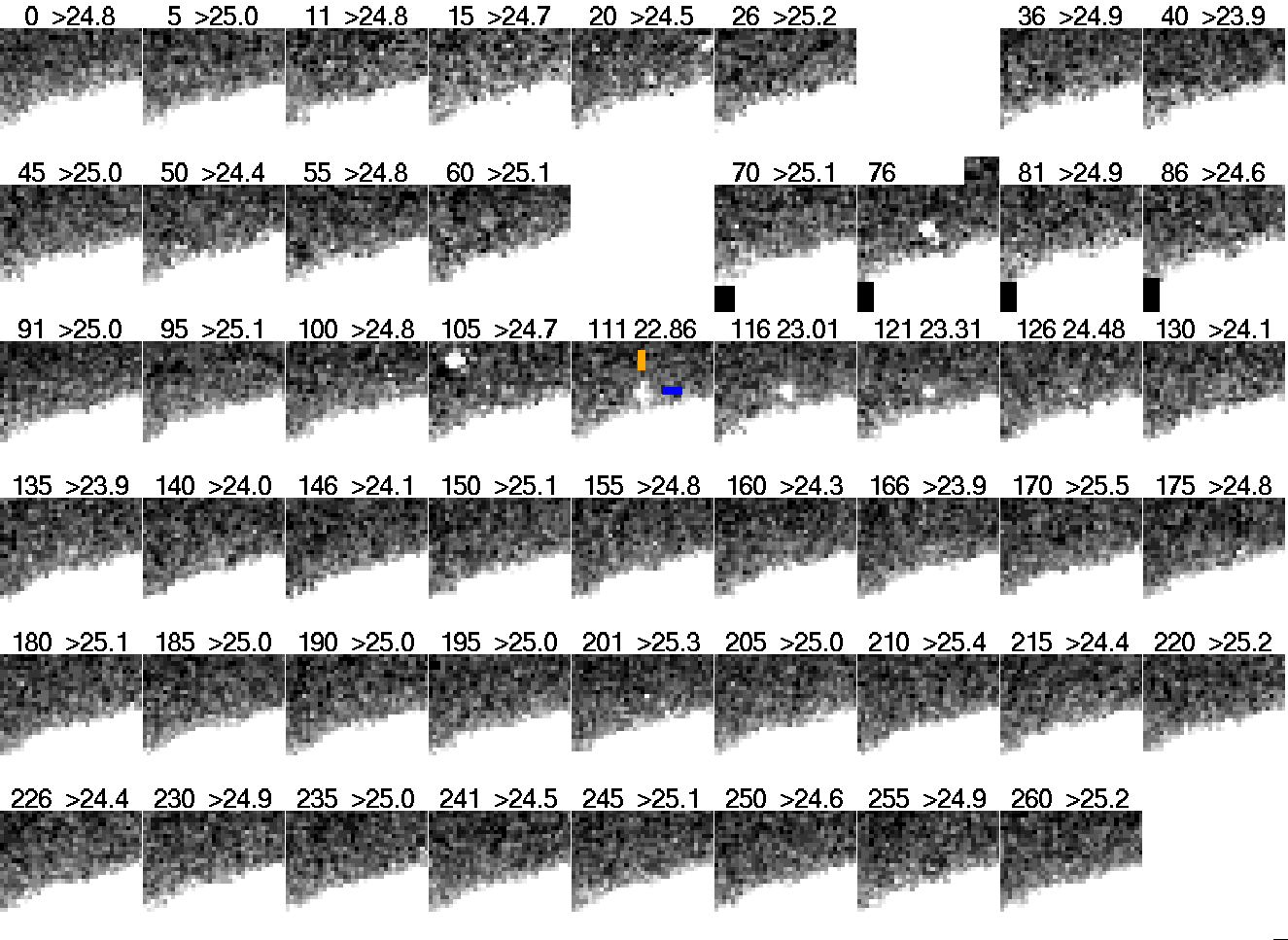}
    \end{tabular}
    \caption{Nova 16}
\end{sidewaysfigure}
\begin{sidewaysfigure}
    \begin{tabular}{cc}
     \multirow{2}{*}[2.6in]{\includegraphics[height=6in]{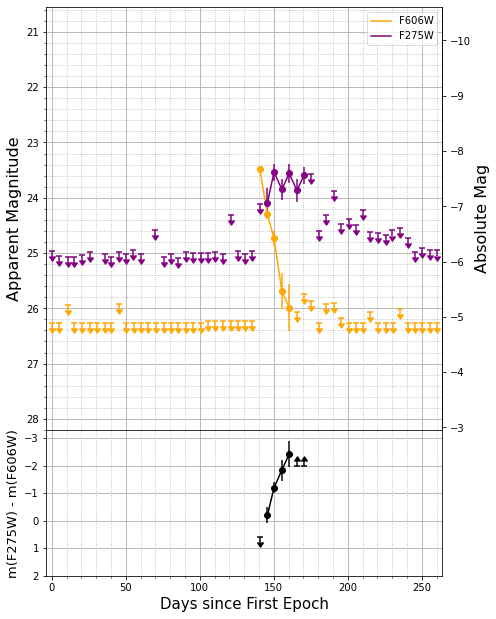}} & \includegraphics[height=3.6in]{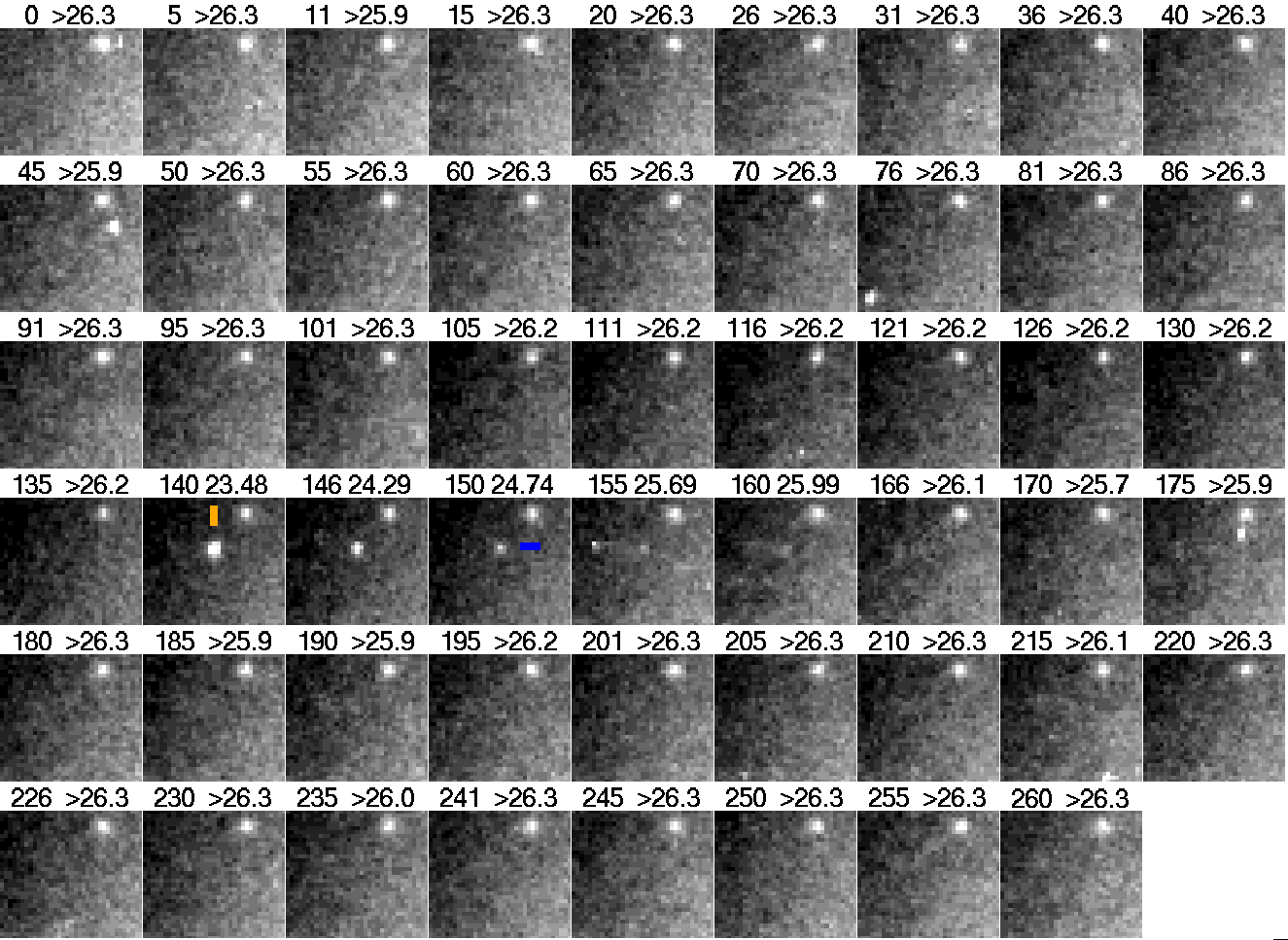} \\
     & \includegraphics[height=3.6in]{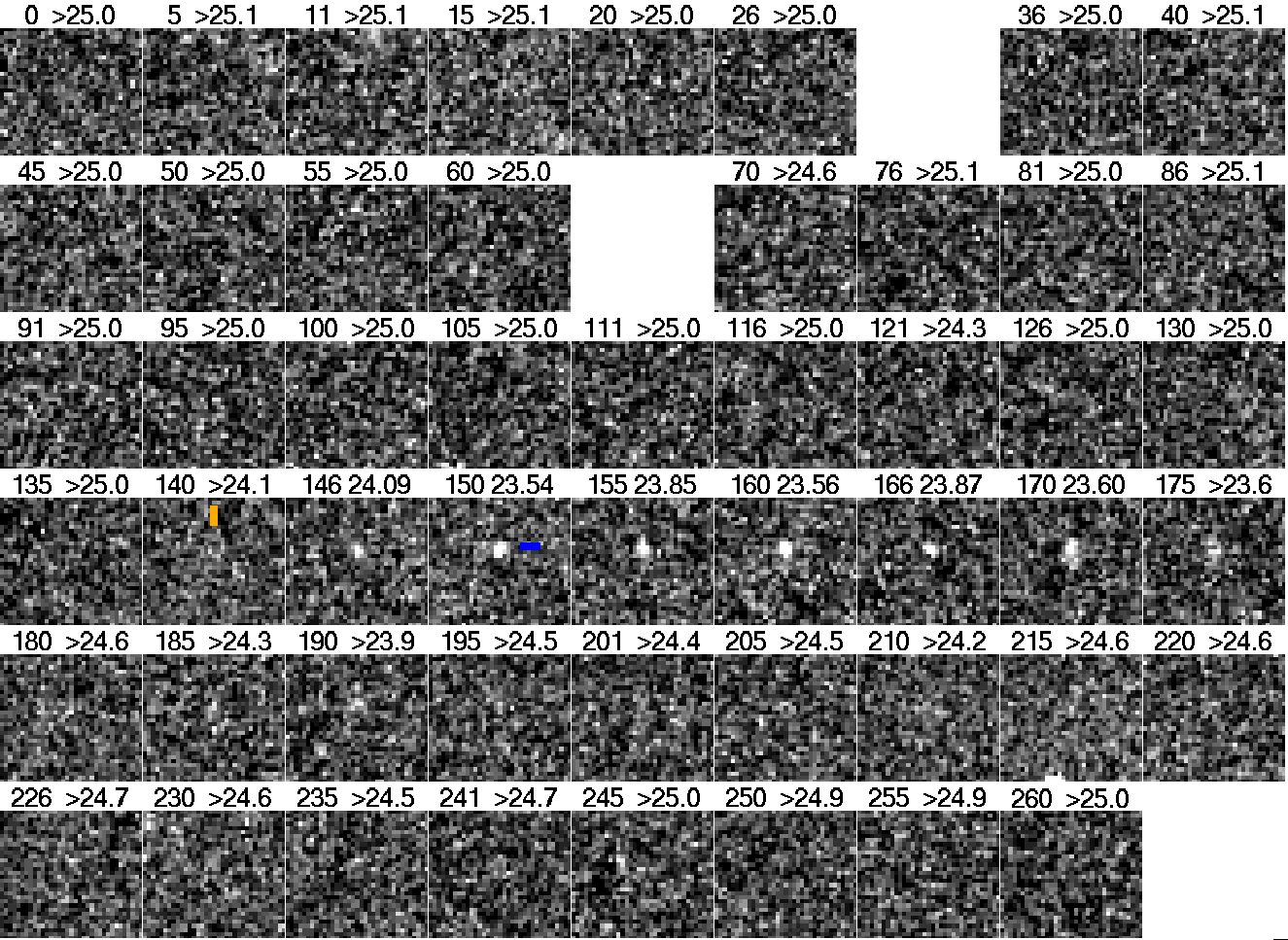}
    \end{tabular}
    \caption{Nova 17}
\end{sidewaysfigure}
\begin{sidewaysfigure}
    \begin{tabular}{cc}
     \multirow{2}{*}[2.6in]{\includegraphics[height=6in]{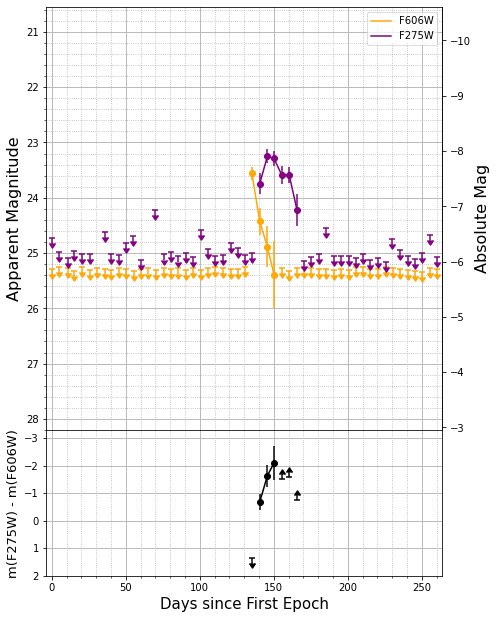}} & \includegraphics[height=3.6in]{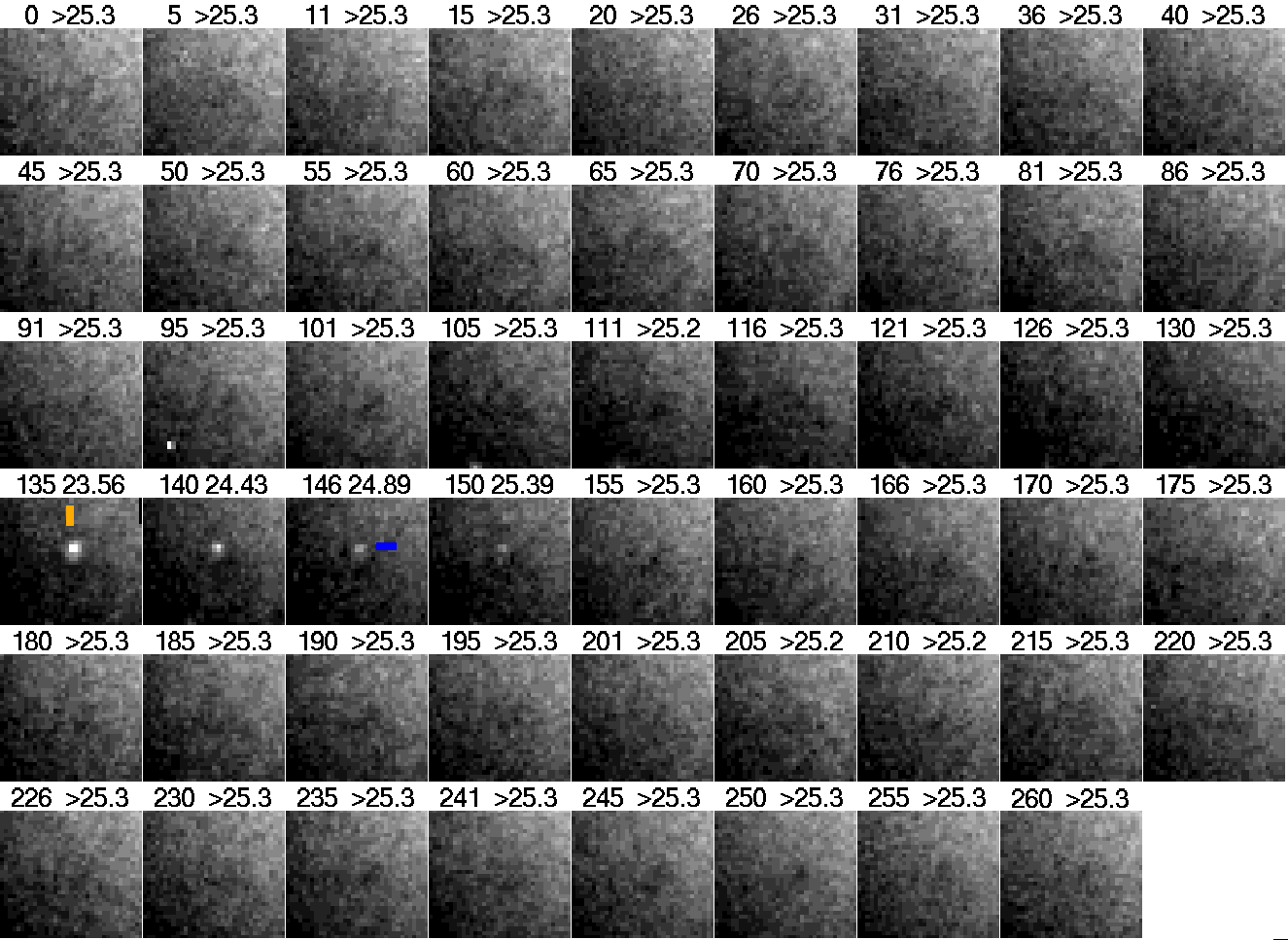} \\
     & \includegraphics[height=3.6in]{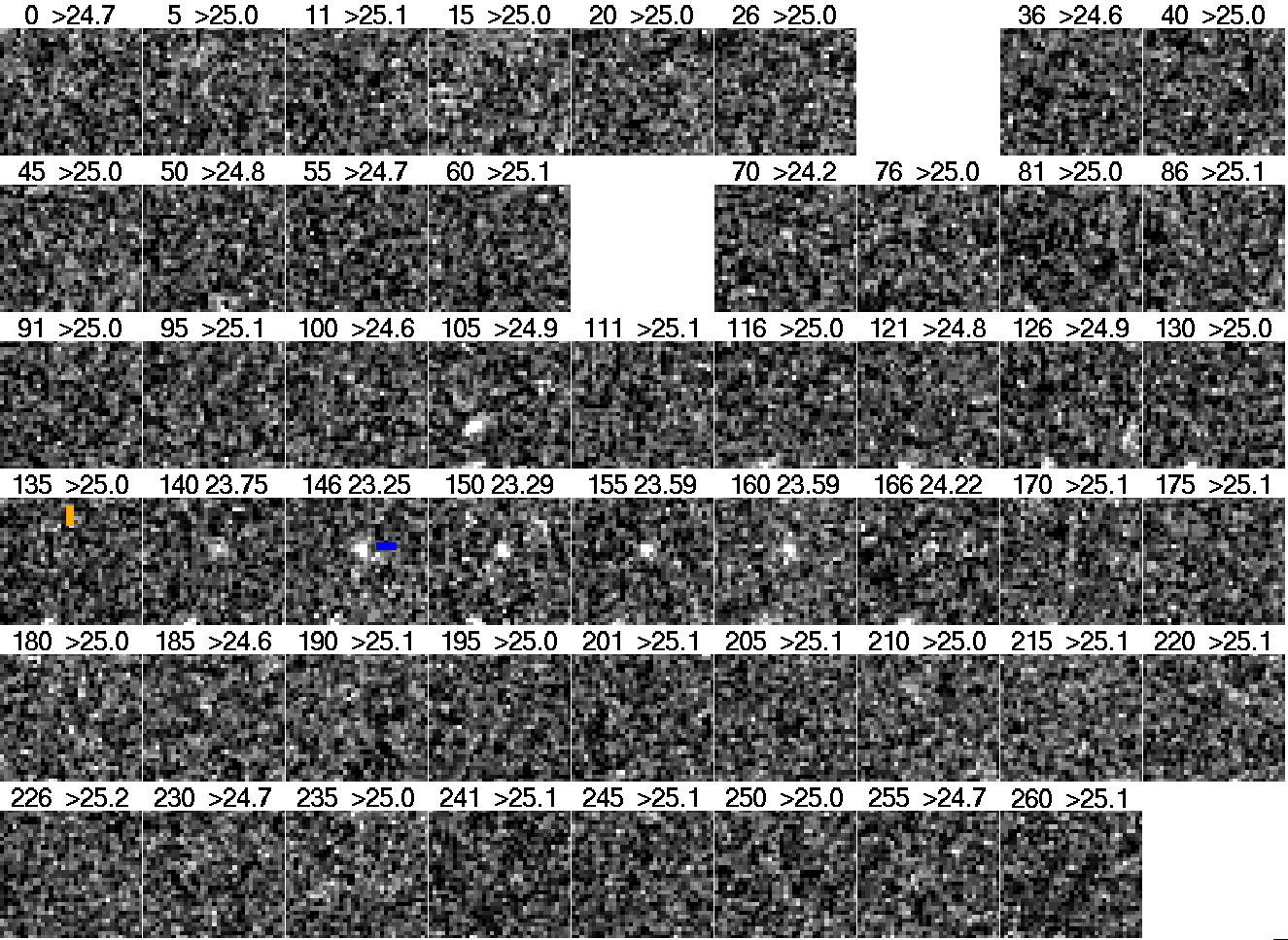}
    \end{tabular}
    \caption{Nova 18}
\end{sidewaysfigure}
\begin{sidewaysfigure}
    \begin{tabular}{cc}
     \multirow{2}{*}[2.6in]{\includegraphics[height=6in]{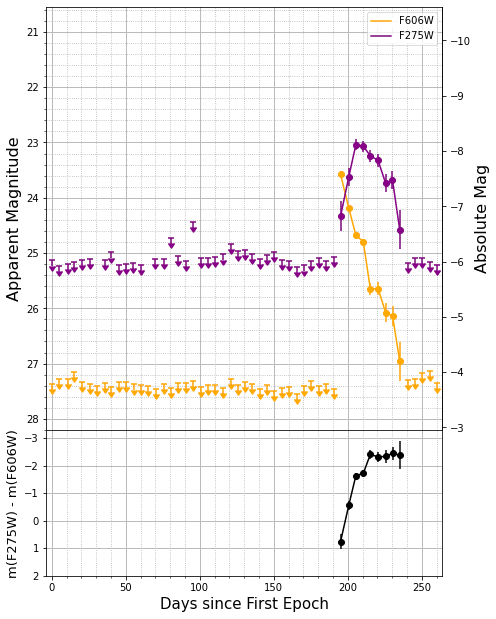}} & \includegraphics[height=3.6in]{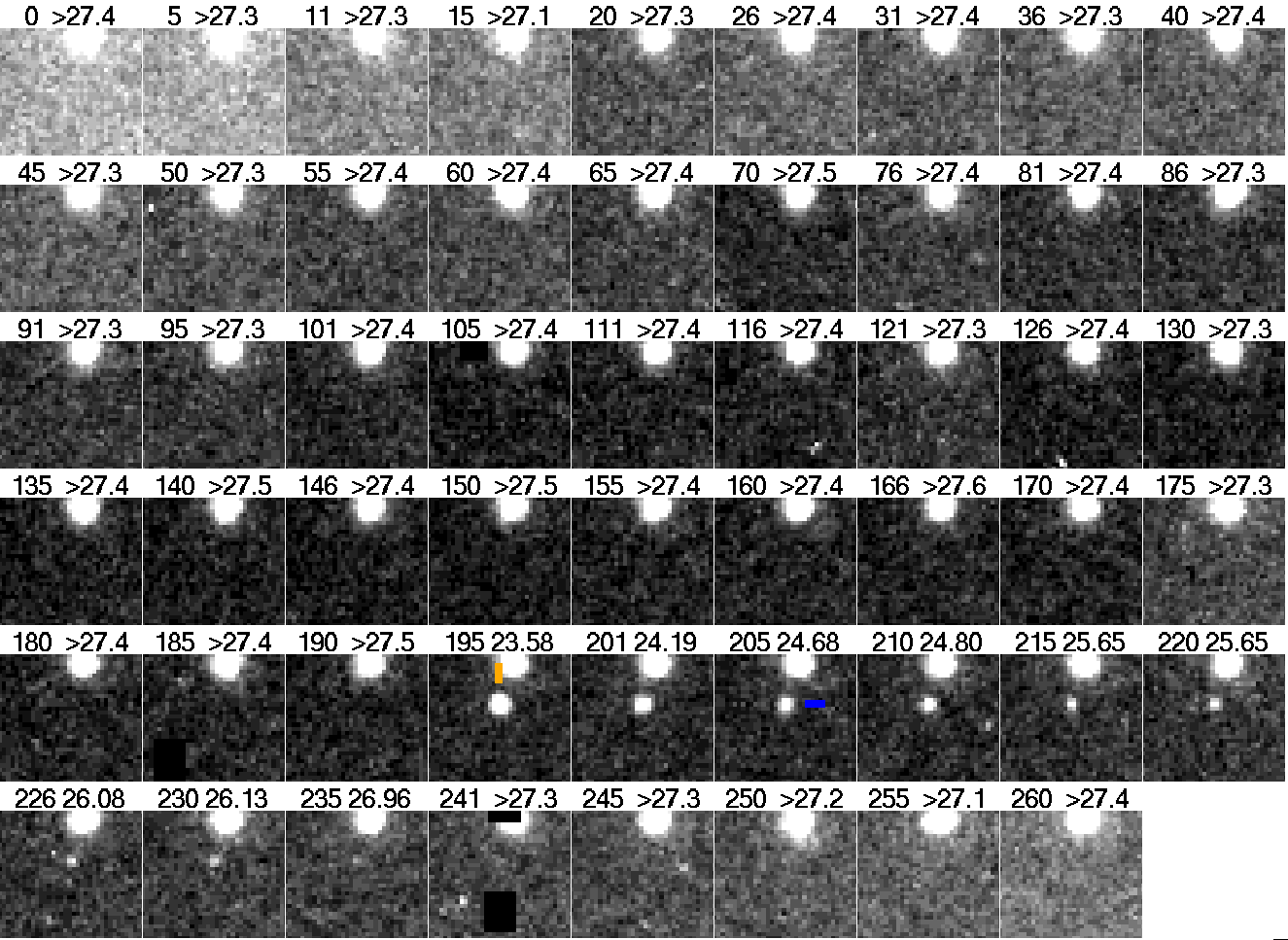} \\
     & \includegraphics[height=3.6in]{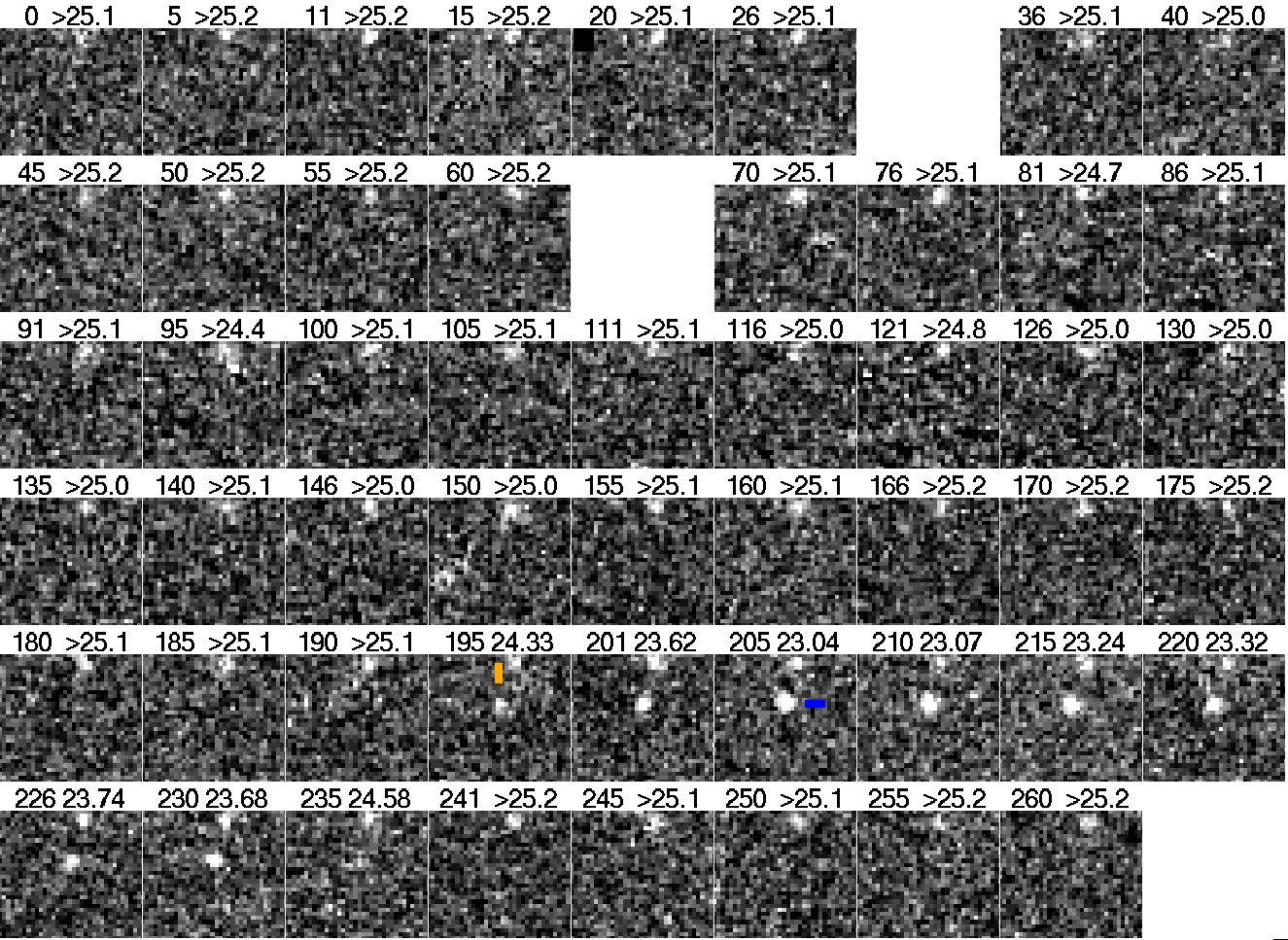}
    \end{tabular}
    \caption{Nova 19}
\end{sidewaysfigure}
\begin{sidewaysfigure}
    \begin{tabular}{cc}
     \multirow{2}{*}[2.6in]{\includegraphics[height=6in]{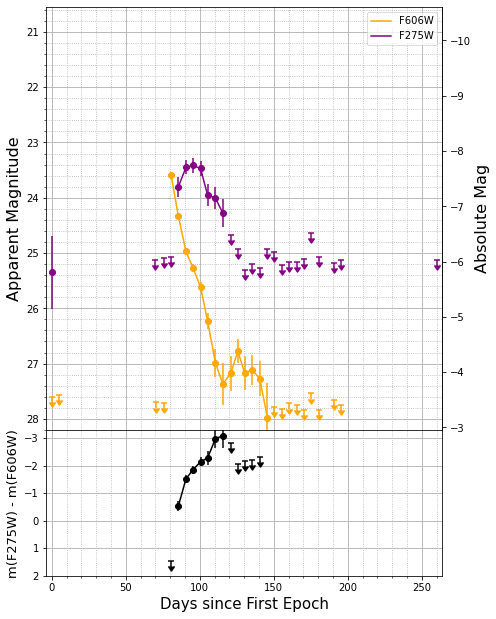}} & \includegraphics[height=3.6in]{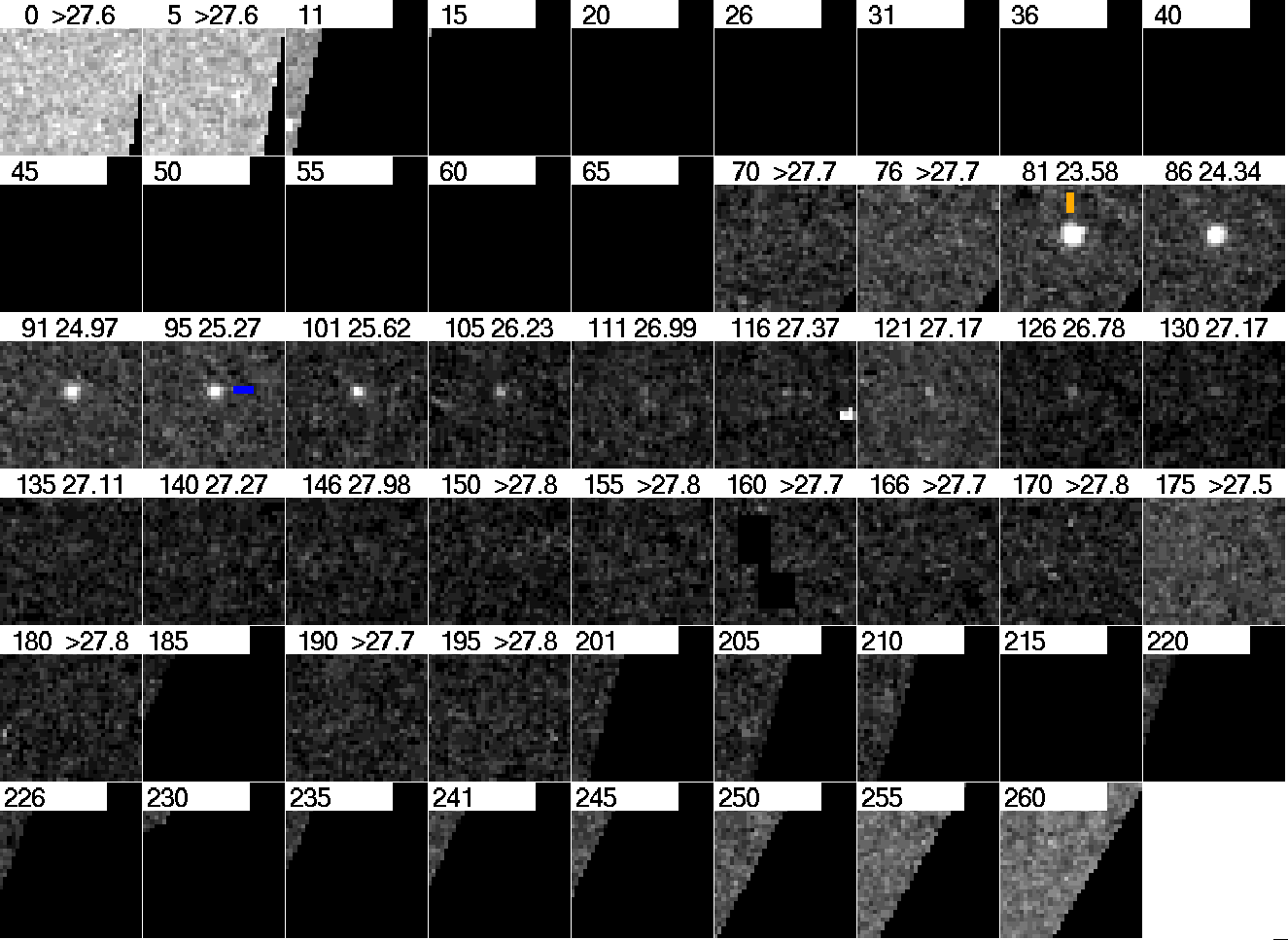} \\
     & \includegraphics[height=3.6in]{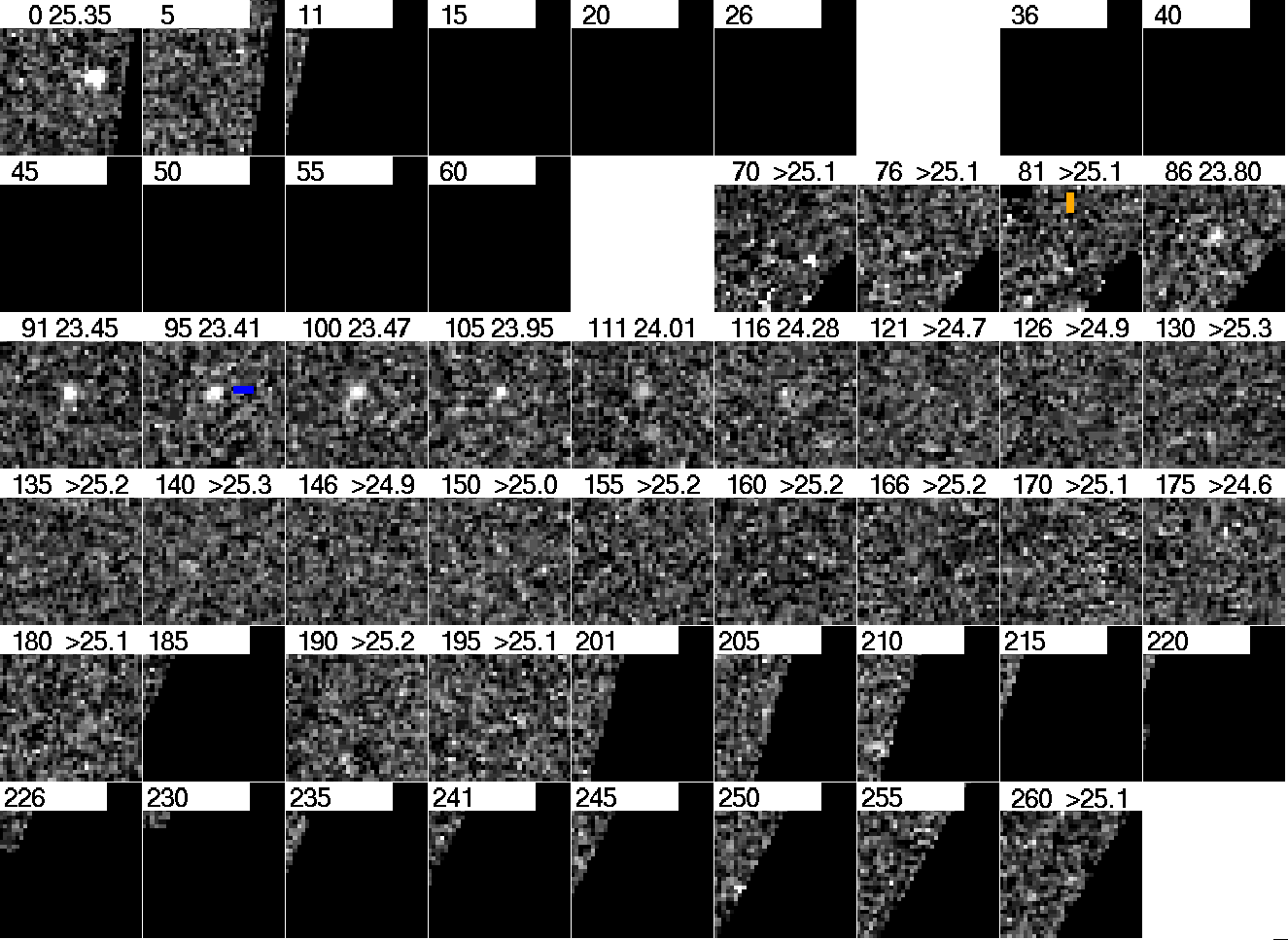}
    \end{tabular}
    \caption{Nova 20}
\end{sidewaysfigure}
\begin{sidewaysfigure}
    \begin{tabular}{cc}
     \multirow{2}{*}[2.6in]{\includegraphics[height=6in]{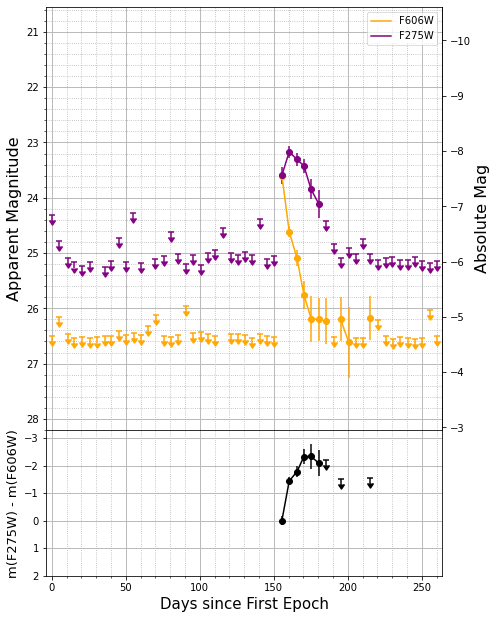}} & \includegraphics[height=3.6in]{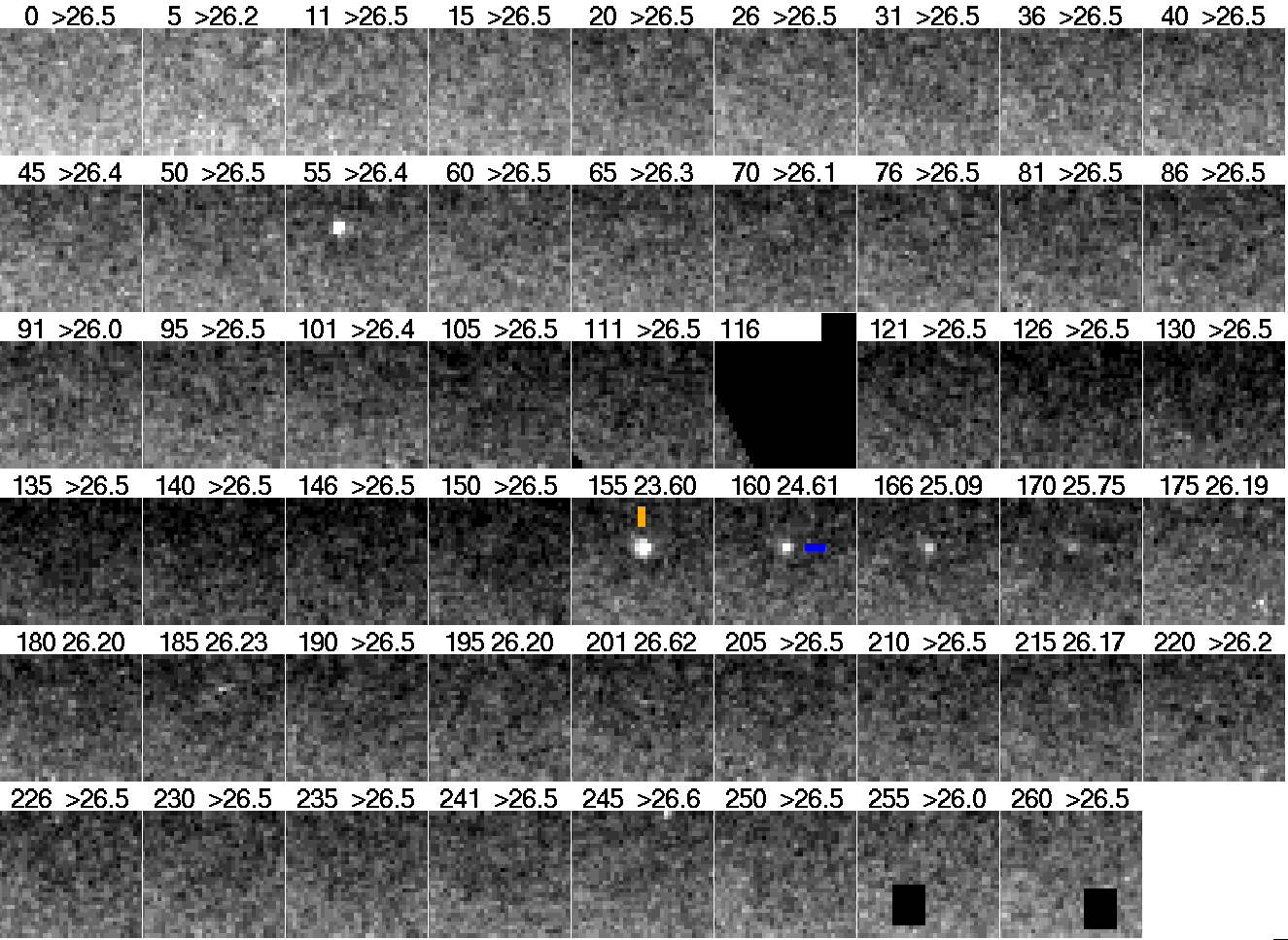} \\
     & \includegraphics[height=3.6in]{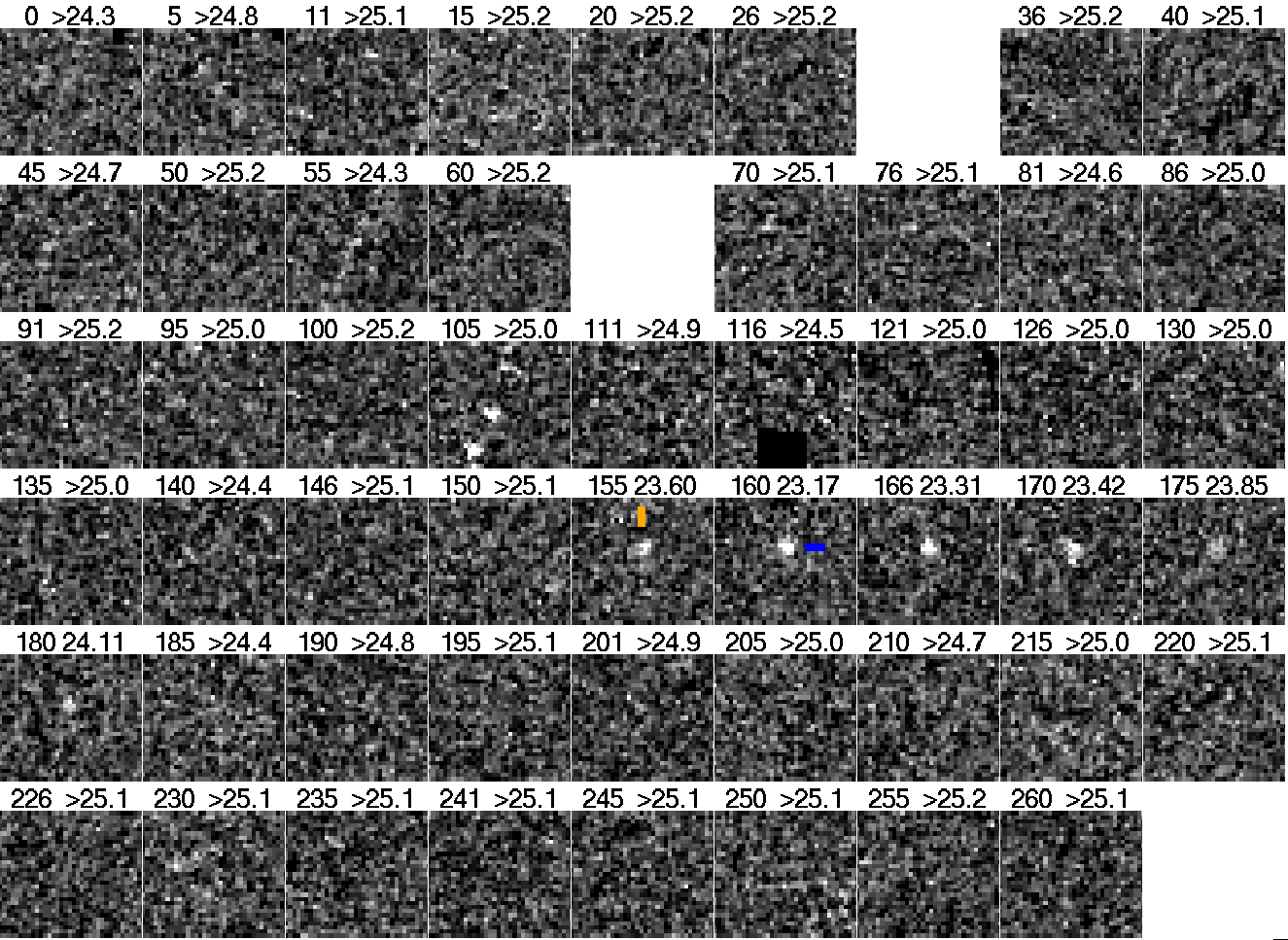}
    \end{tabular}
    \caption{Nova 21}
\end{sidewaysfigure}
\begin{sidewaysfigure}
    \begin{tabular}{cc}
     \multirow{2}{*}[2.6in]{\includegraphics[height=6in]{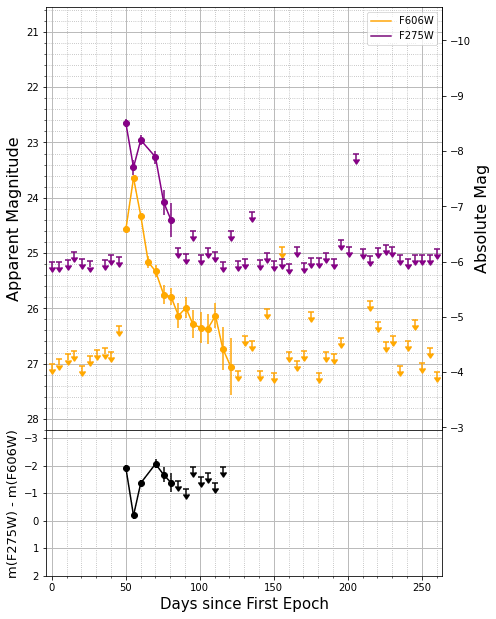}} & \includegraphics[height=3.6in]{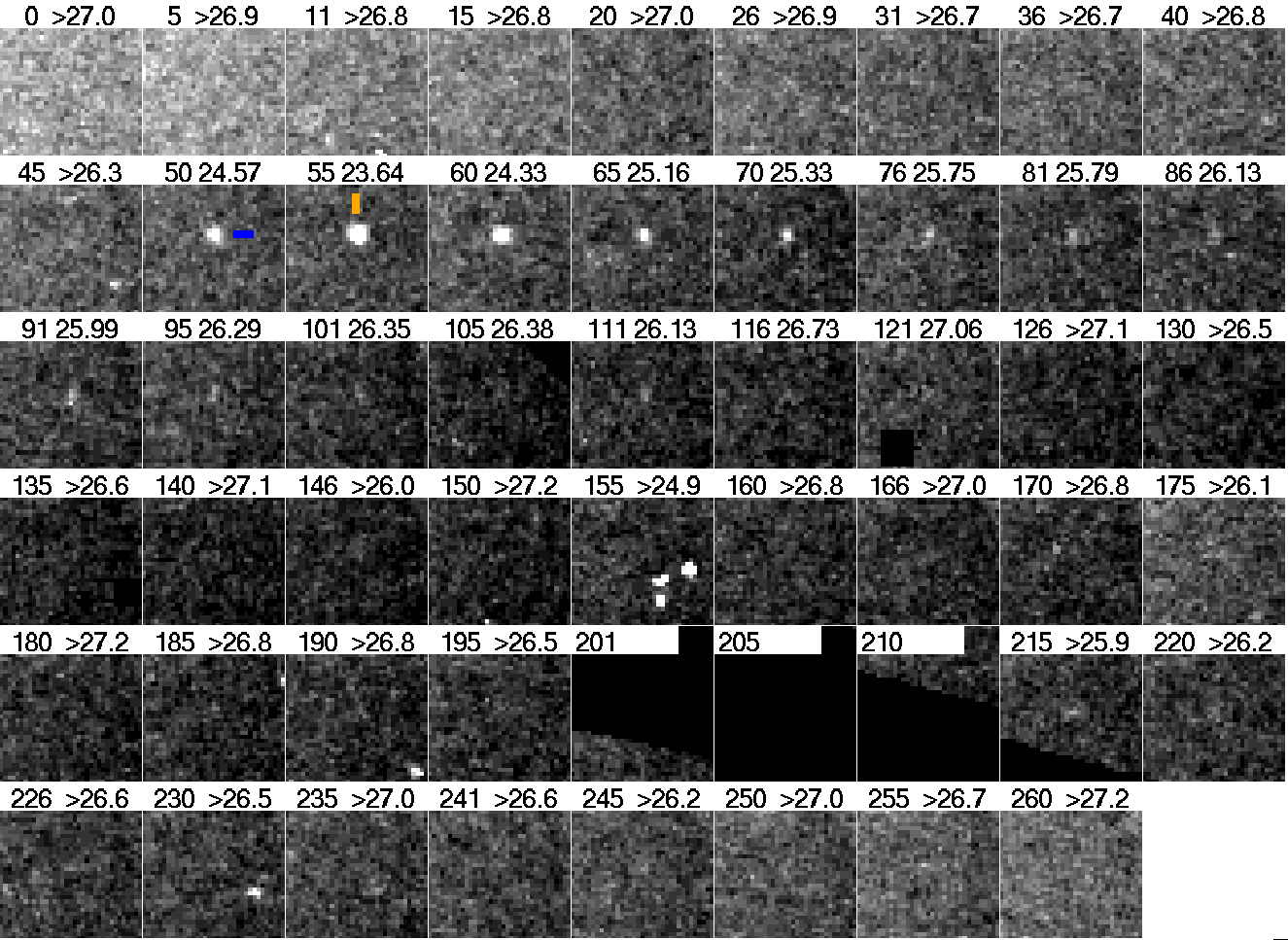} \\
     & \includegraphics[height=3.6in]{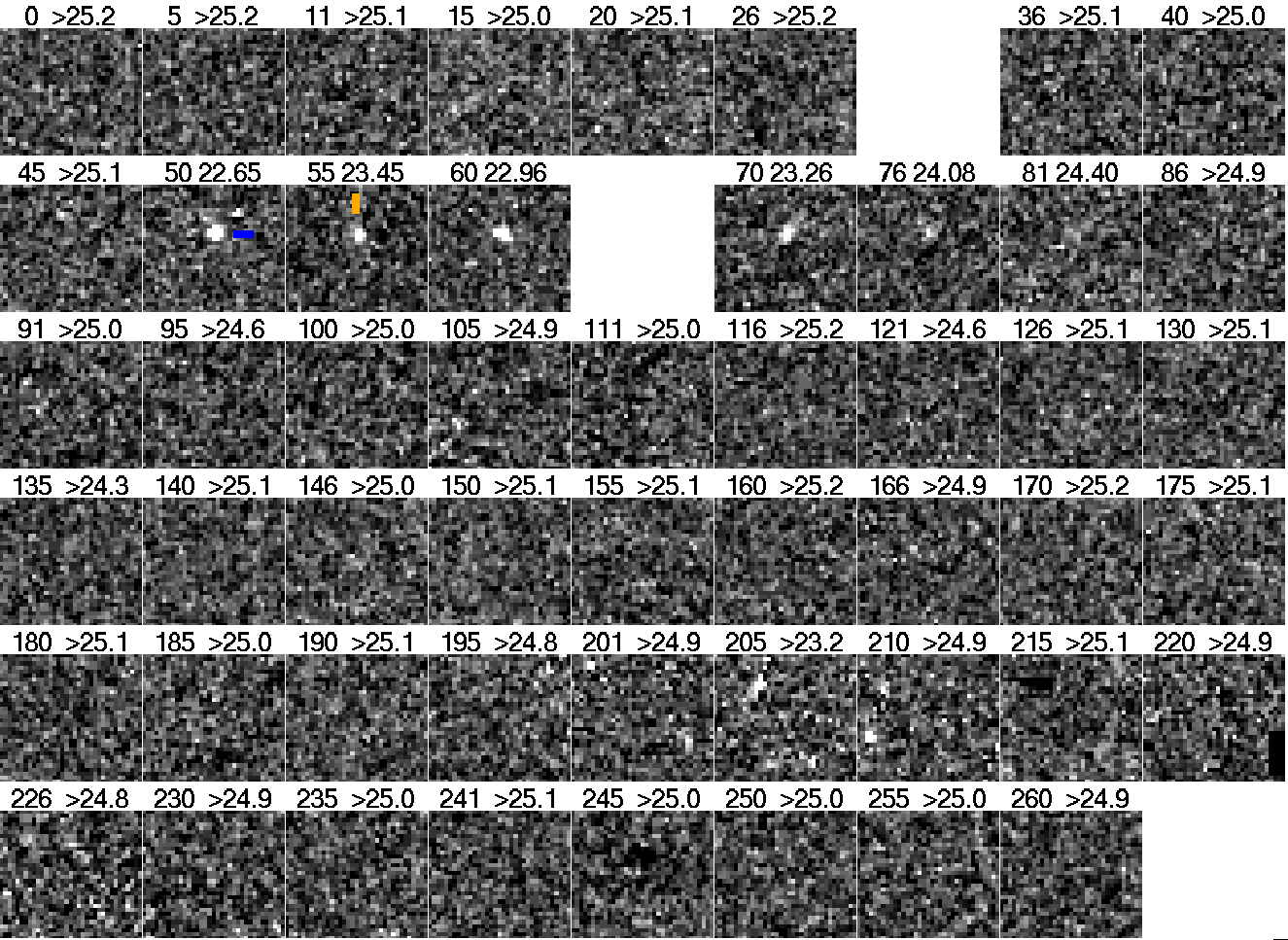}
    \end{tabular}
    \caption{Nova 22}
\end{sidewaysfigure}
\begin{sidewaysfigure}
    \begin{tabular}{cc}
     \multirow{2}{*}[2.6in]{\includegraphics[height=6in]{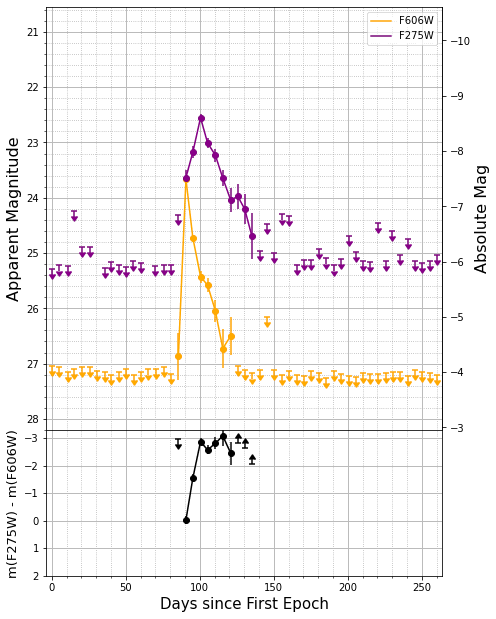}} & \includegraphics[height=3.6in]{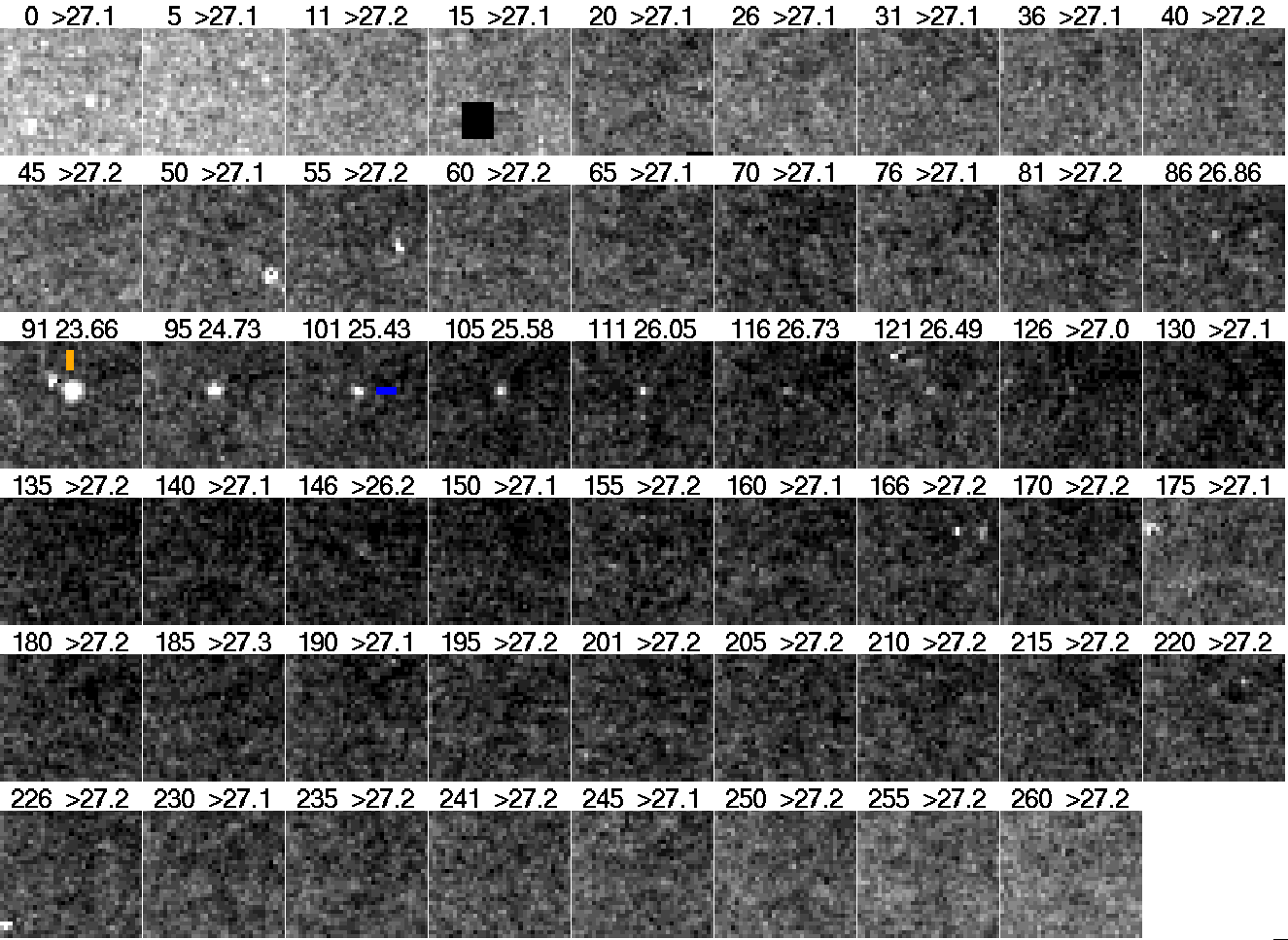} \\
     & \includegraphics[height=3.6in]{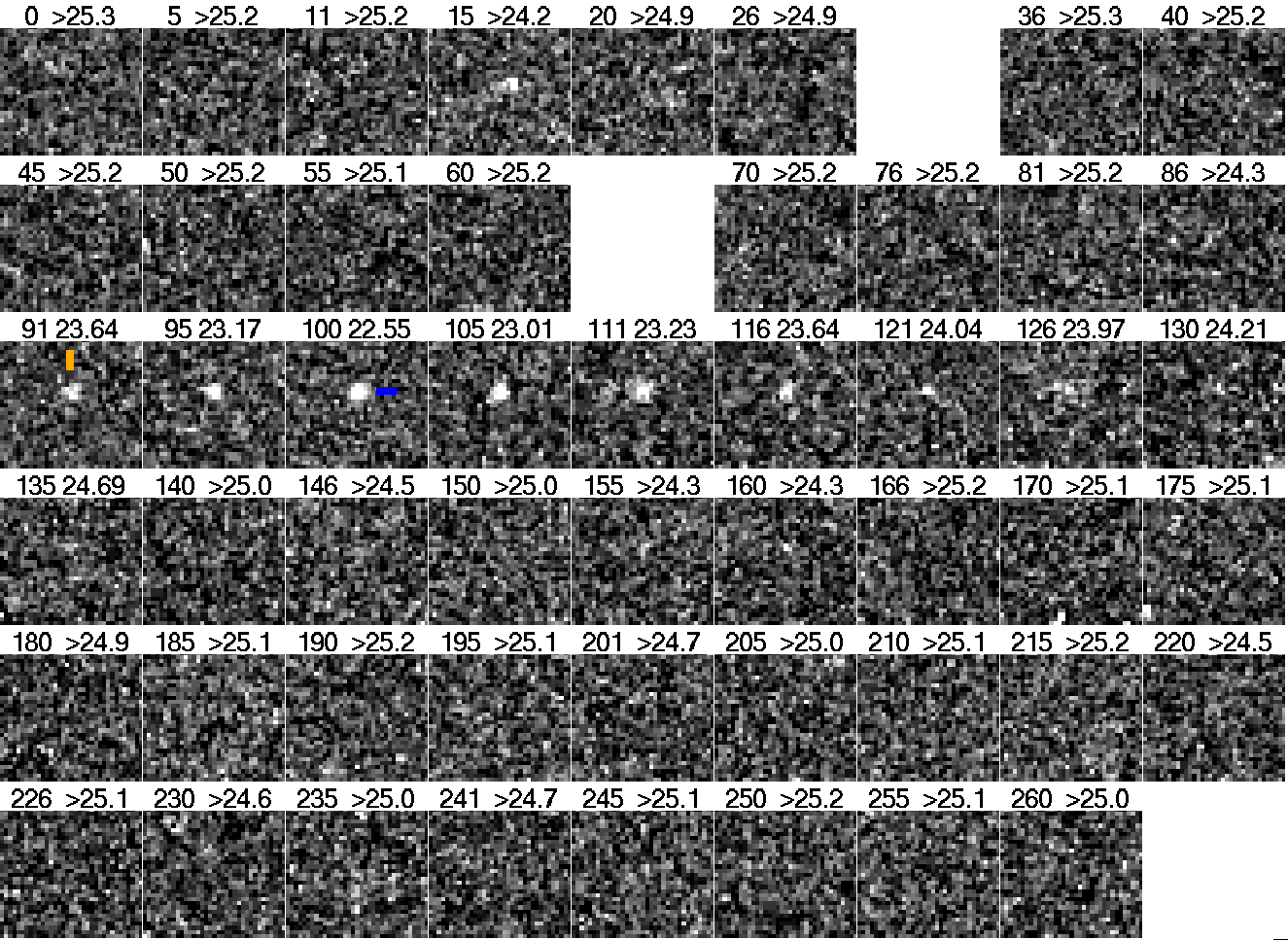}
    \end{tabular}
    \caption{Nova 23}
\end{sidewaysfigure}
\begin{sidewaysfigure}
    \begin{tabular}{cc}
     \multirow{2}{*}[2.6in]{\includegraphics[height=6in]{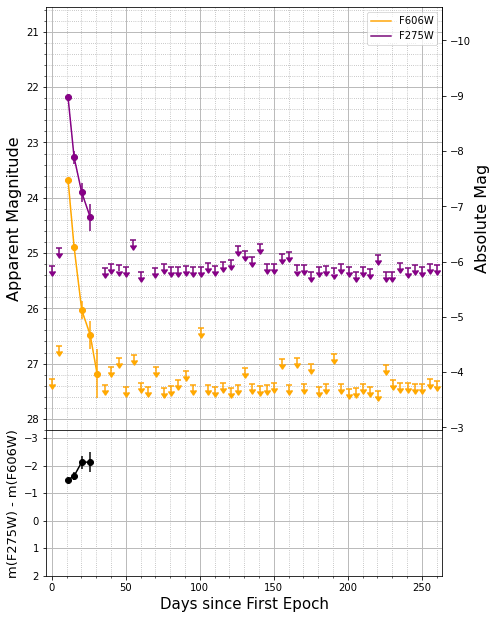}} & \includegraphics[height=3.6in]{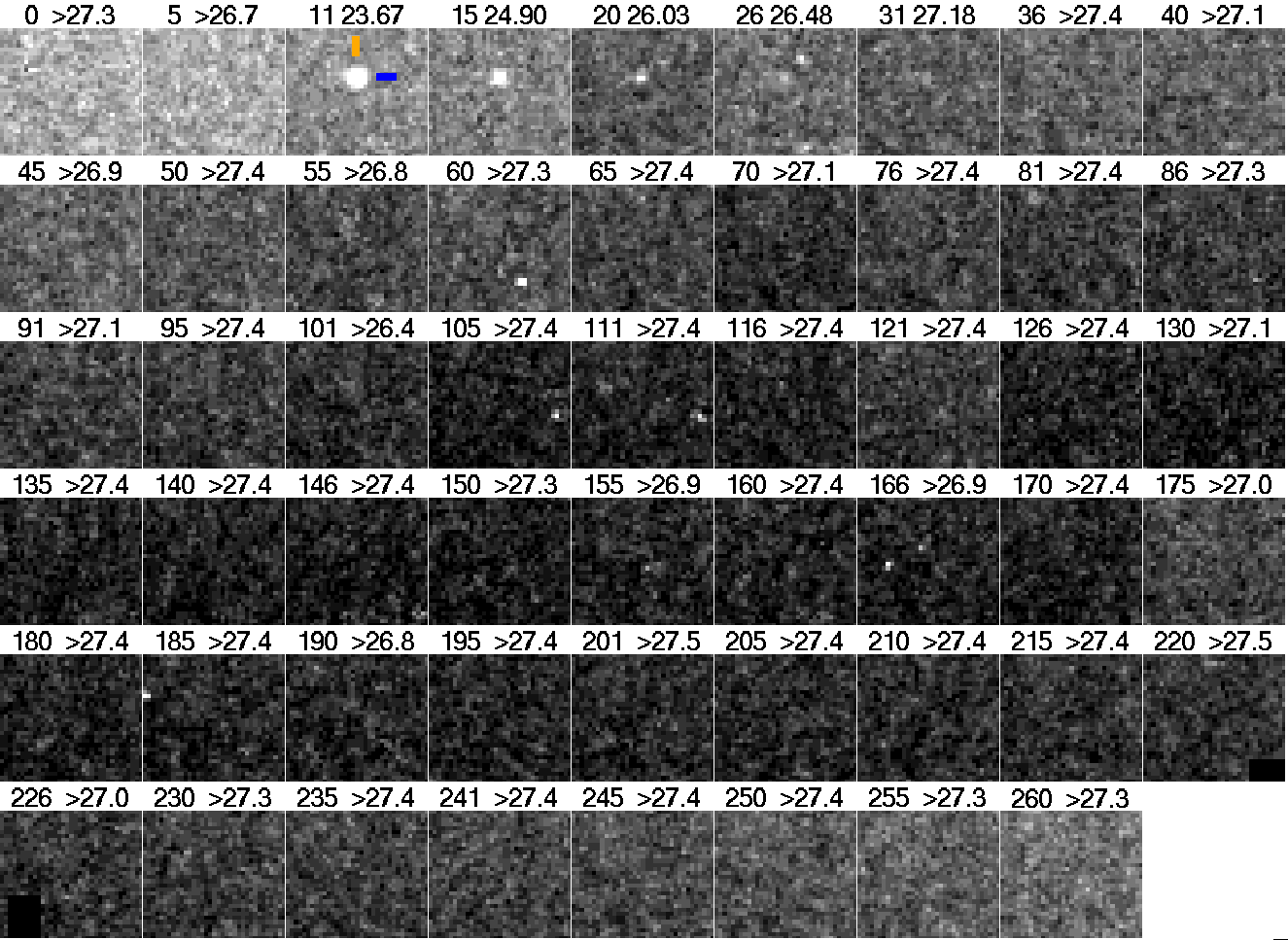} \\
     & \includegraphics[height=3.6in]{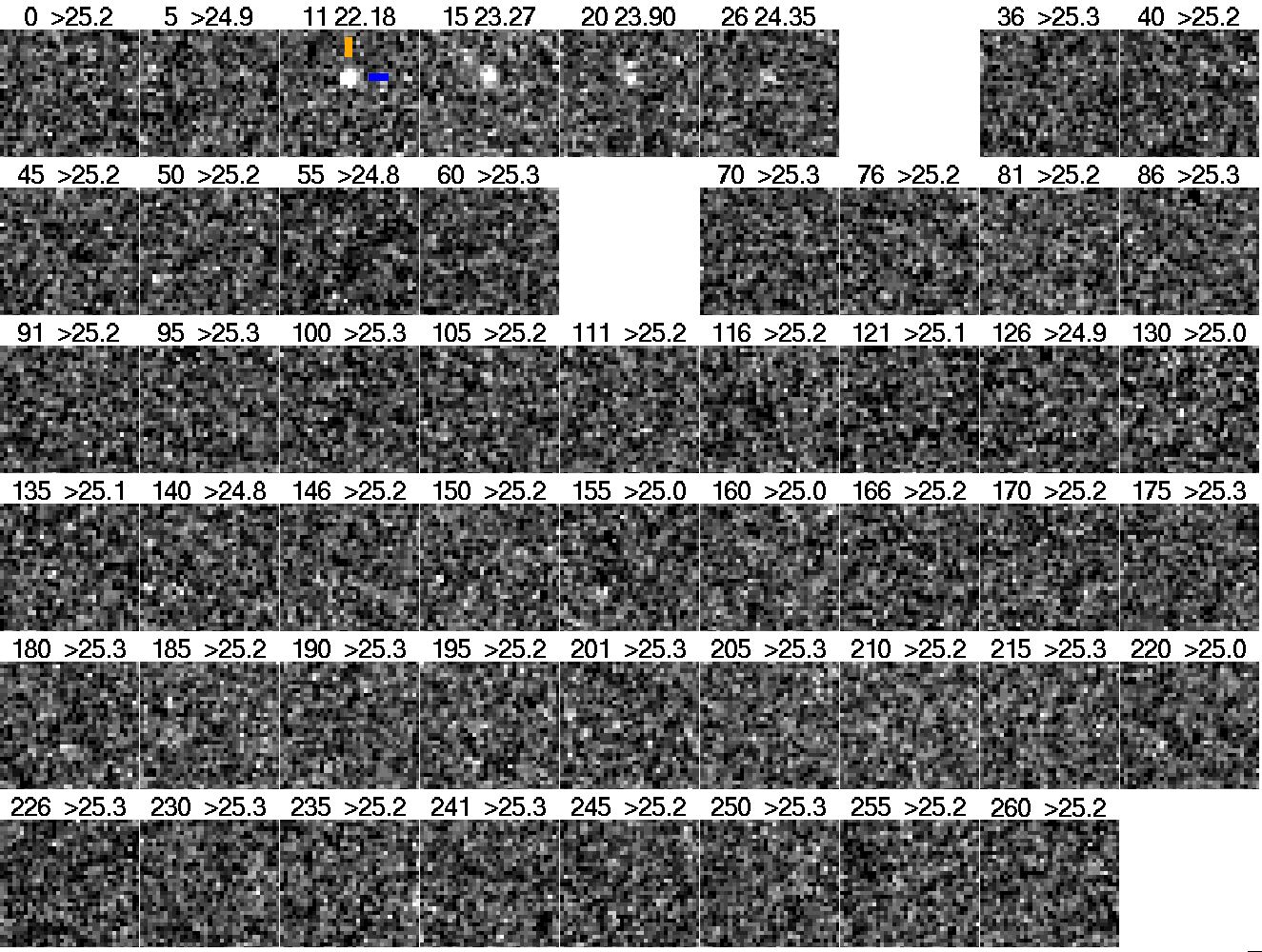}
    \end{tabular}
    \caption{Nova 24}
\end{sidewaysfigure}
\begin{sidewaysfigure}
    \begin{tabular}{cc}
     \multirow{2}{*}[2.6in]{\includegraphics[height=6in]{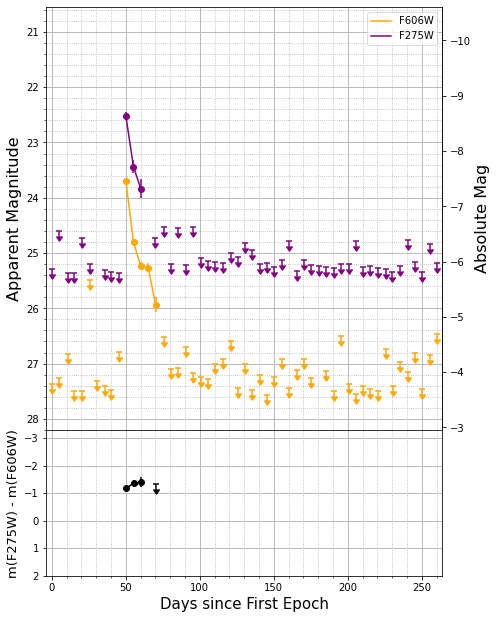}} & \includegraphics[height=3.6in]{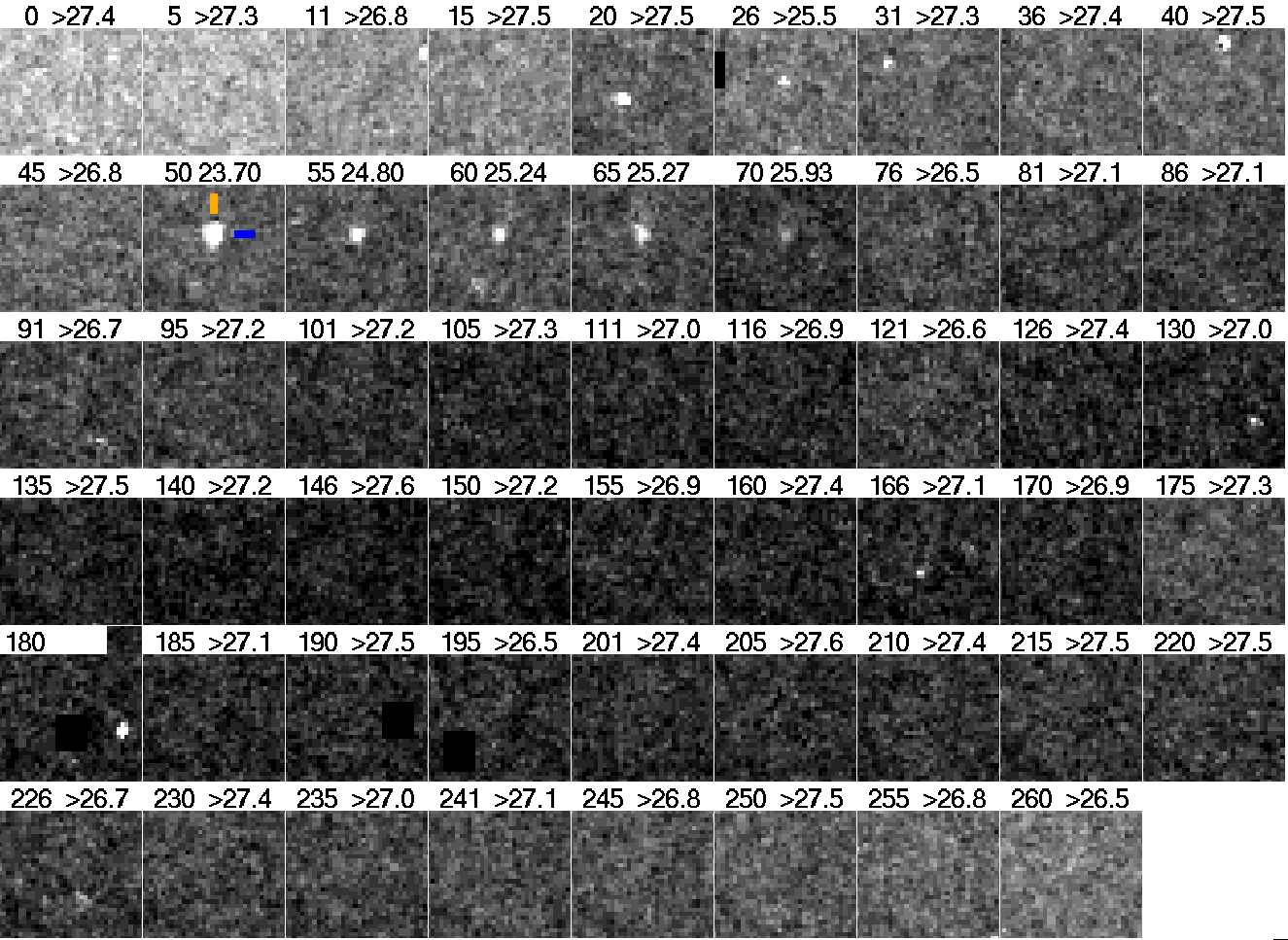} \\
     & \includegraphics[height=3.6in]{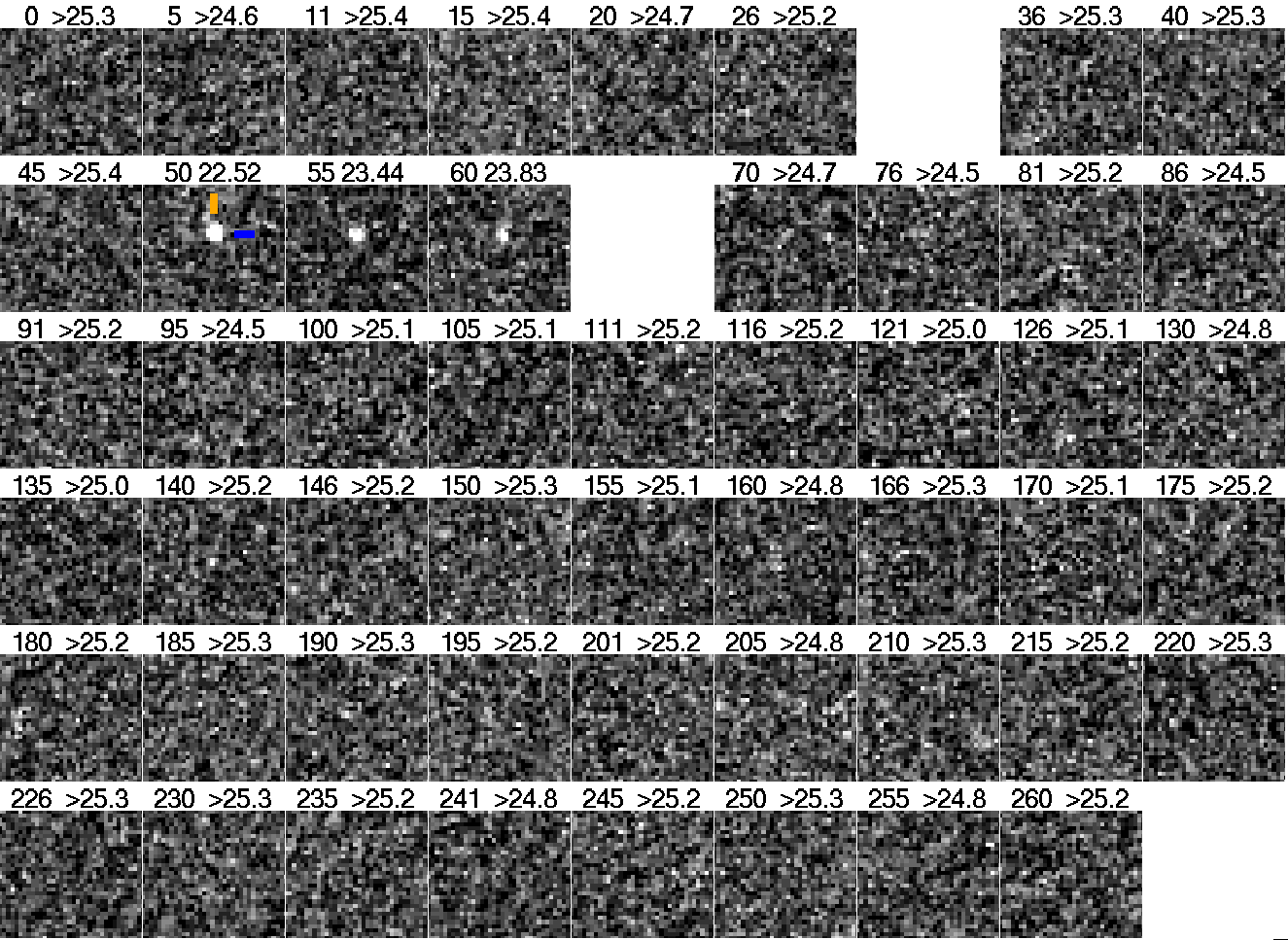}
    \end{tabular}
    \caption{Nova 25}
\end{sidewaysfigure}
\begin{sidewaysfigure}
    \begin{tabular}{cc}
     \multirow{2}{*}[2.6in]{\includegraphics[height=6in]{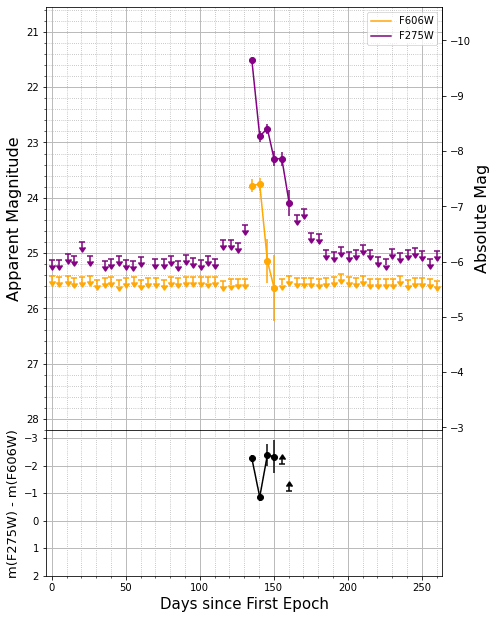}} & \includegraphics[height=3.6in]{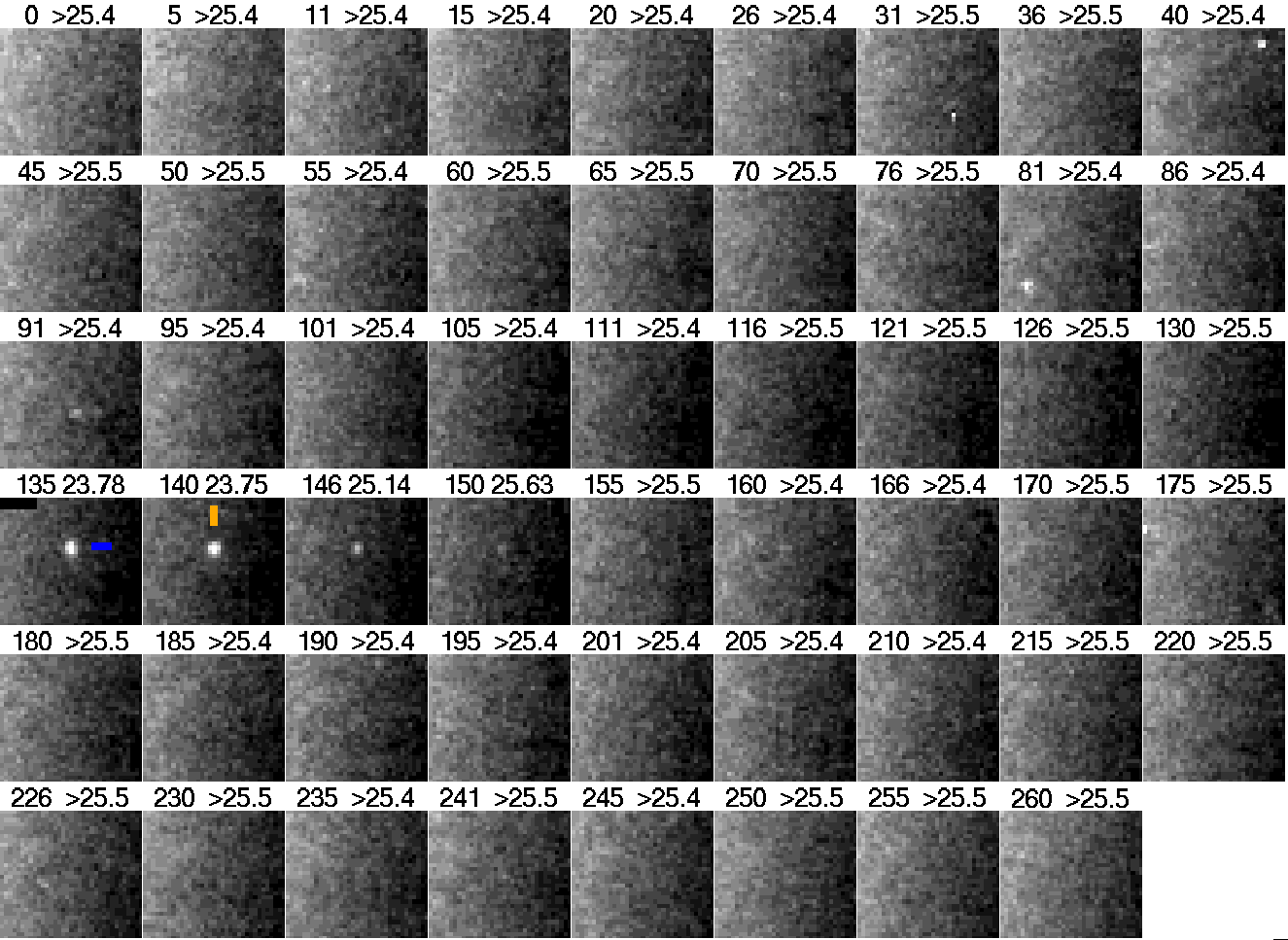} \\
     & \includegraphics[height=3.6in]{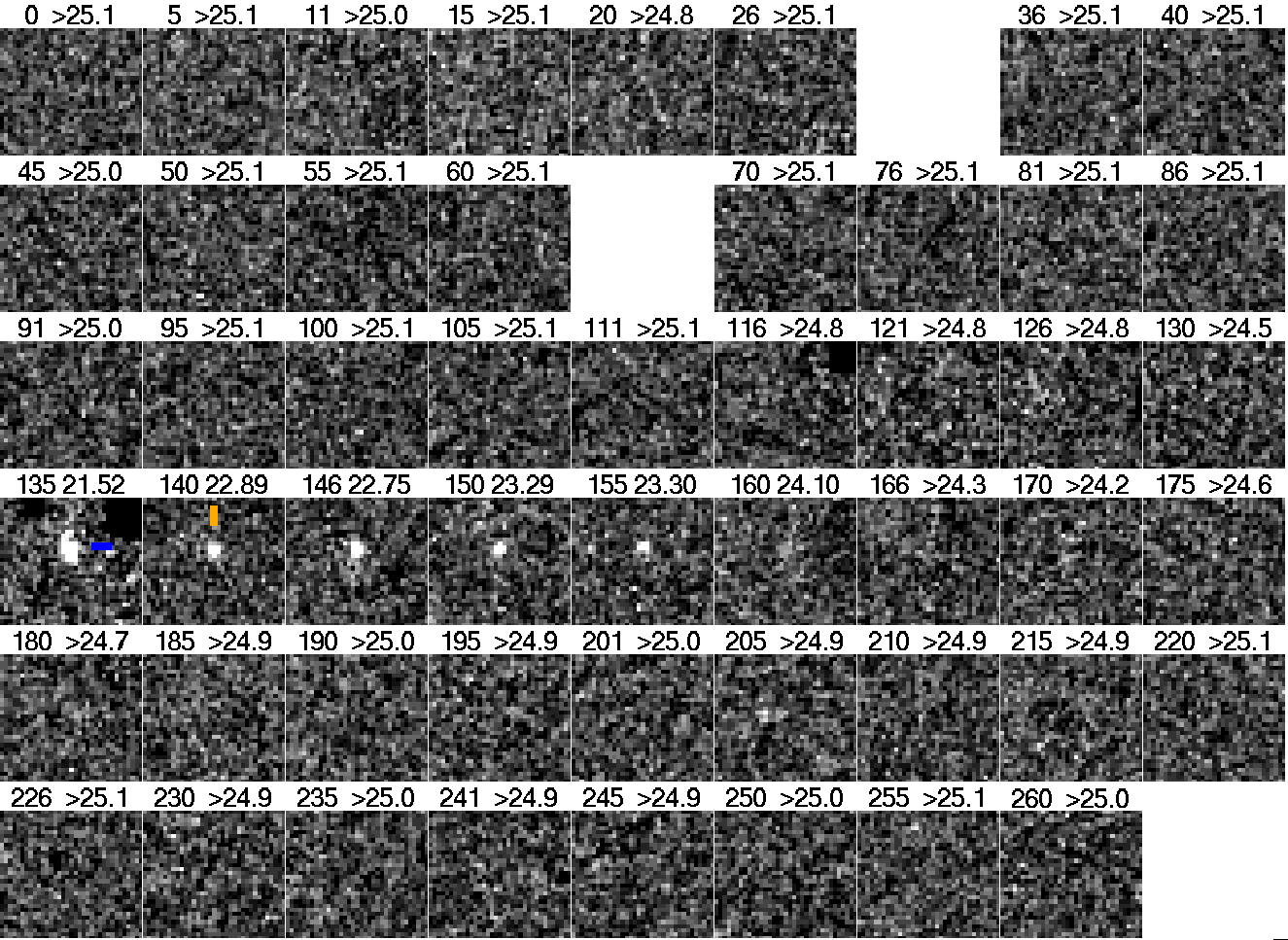}
    \end{tabular}
    \caption{Nova 26}
\end{sidewaysfigure}
\begin{sidewaysfigure}
    \begin{tabular}{cc}
     \multirow{2}{*}[2.6in]{\includegraphics[height=6in]{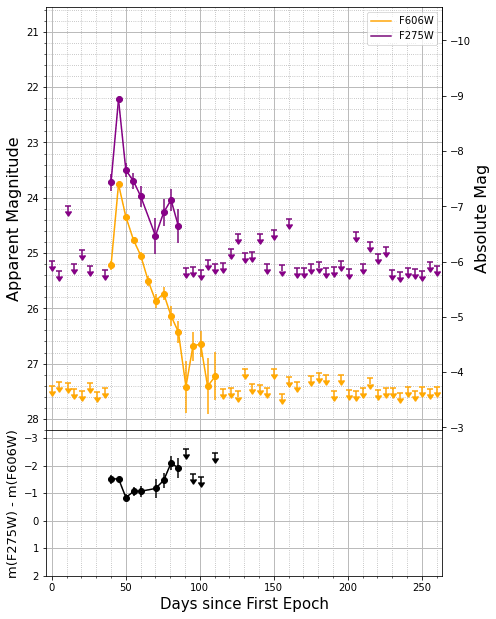}} & \includegraphics[height=3.6in]{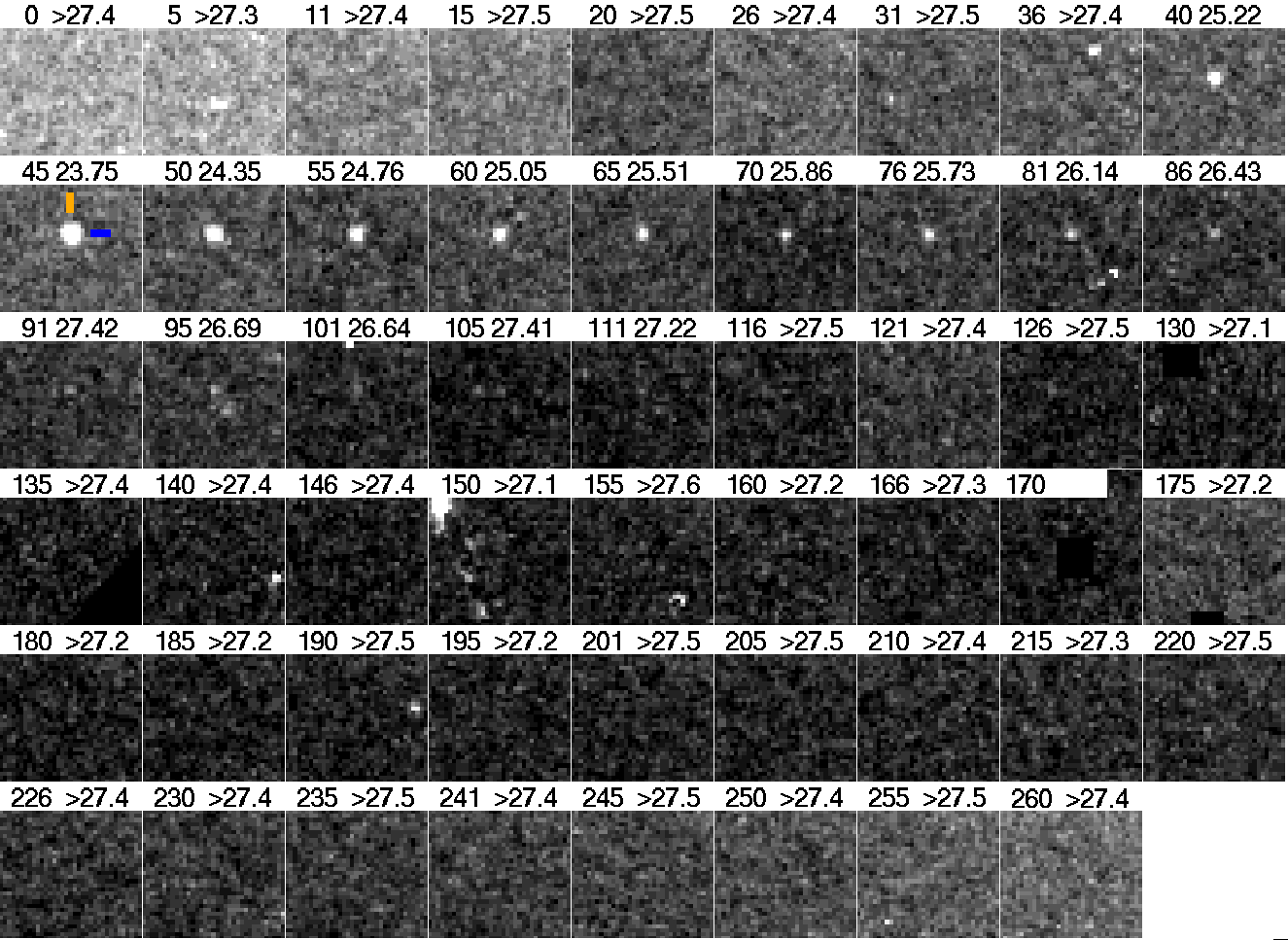} \\
     & \includegraphics[height=3.6in]{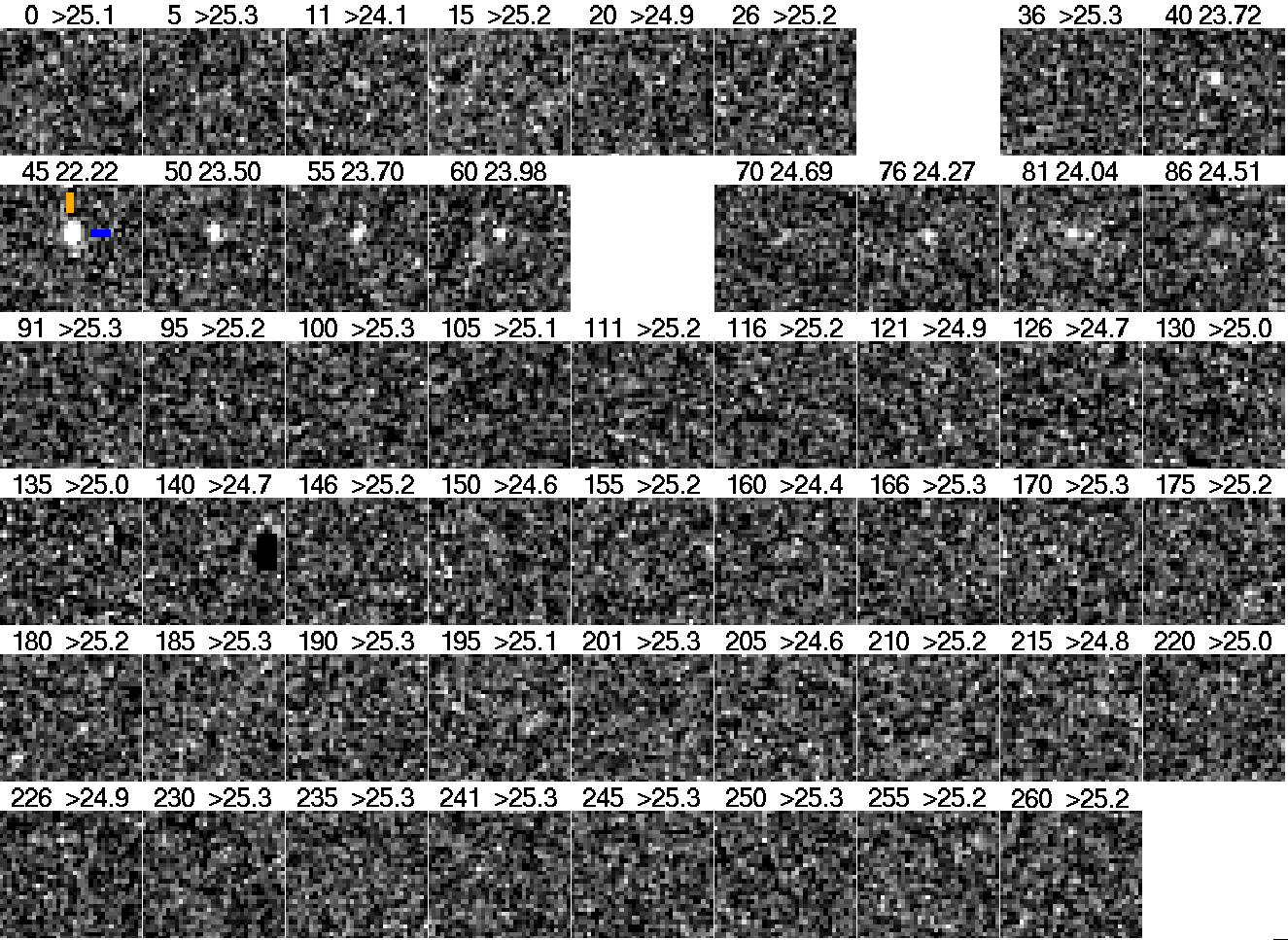}
    \end{tabular}
    \caption{Nova 27}
\end{sidewaysfigure}
\begin{sidewaysfigure}
    \begin{tabular}{cc}
     \multirow{2}{*}[2.6in]{\includegraphics[height=6in]{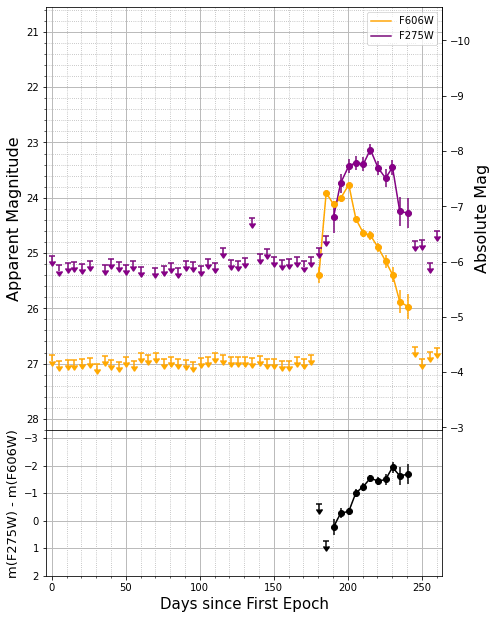}} & \includegraphics[height=3.6in]{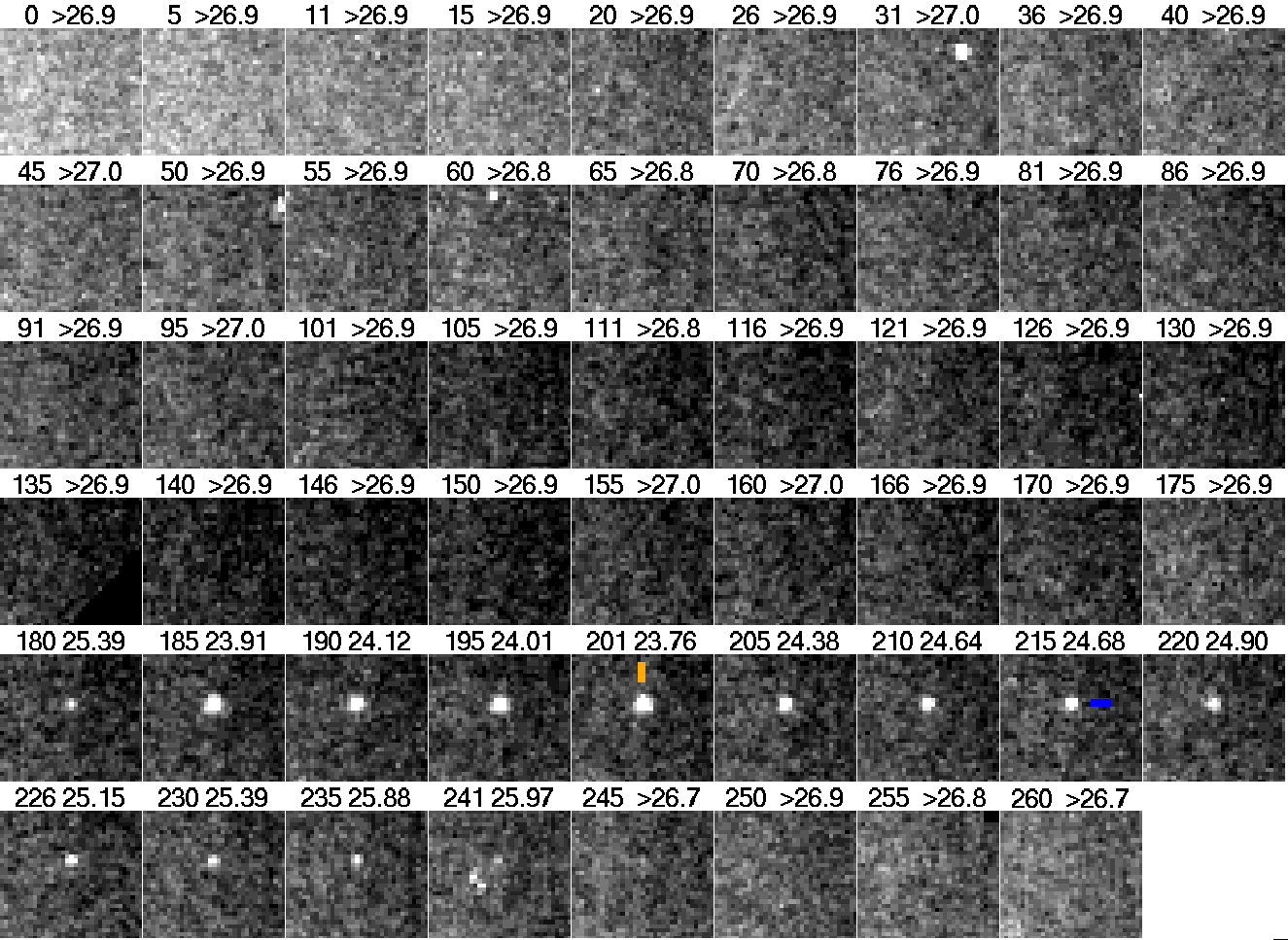} \\
     & \includegraphics[height=3.6in]{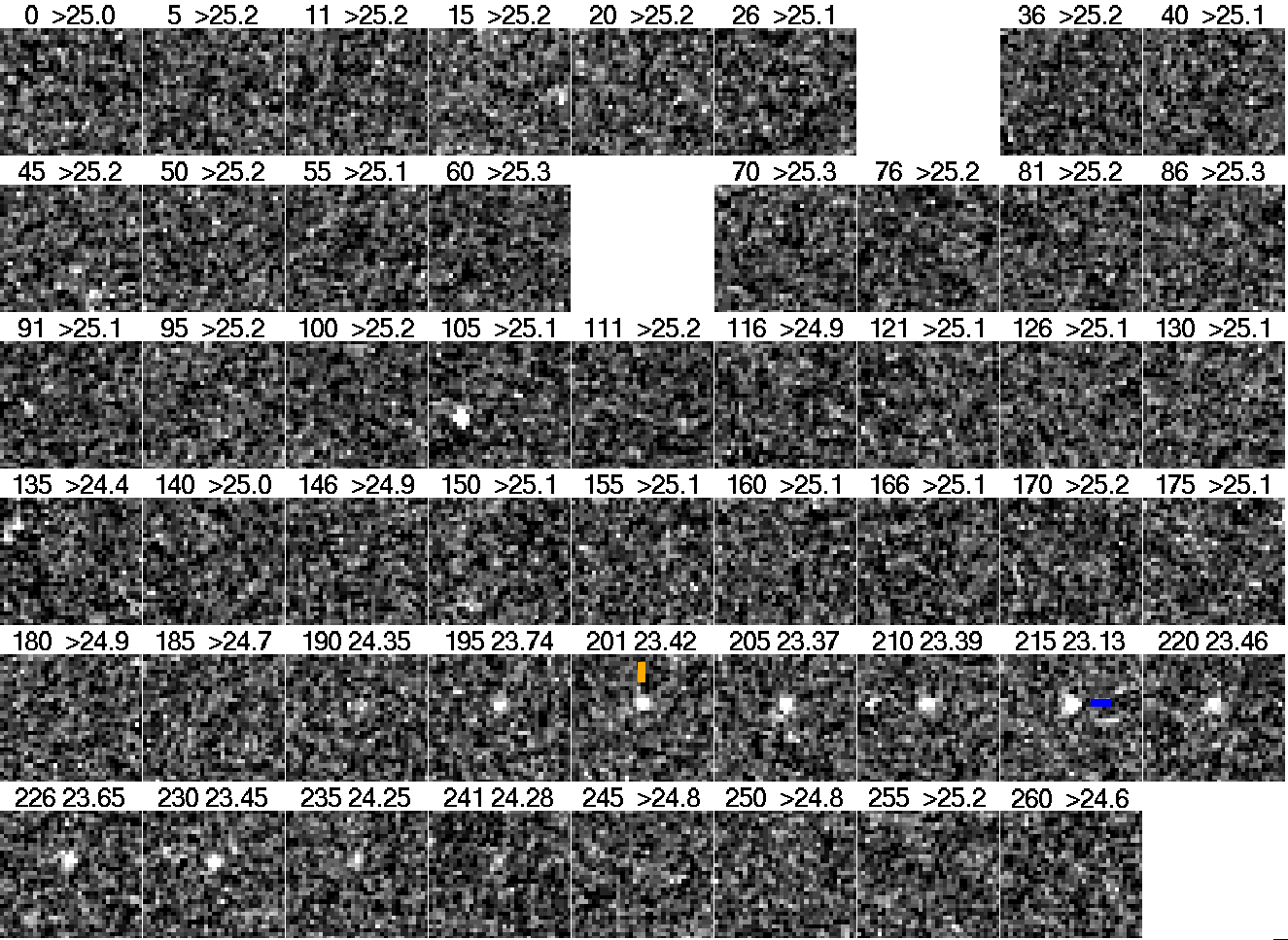}
    \end{tabular}
    \caption{Nova 28}
\end{sidewaysfigure}
\begin{sidewaysfigure}
    \begin{tabular}{cc}
     \multirow{2}{*}[2.6in]{\includegraphics[height=6in]{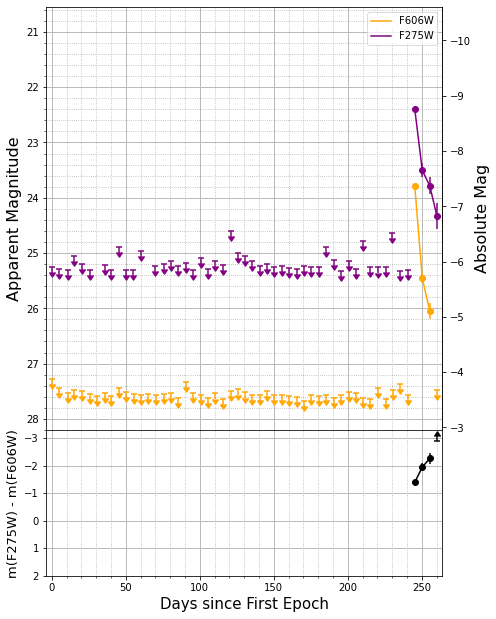}} & \includegraphics[height=3.6in]{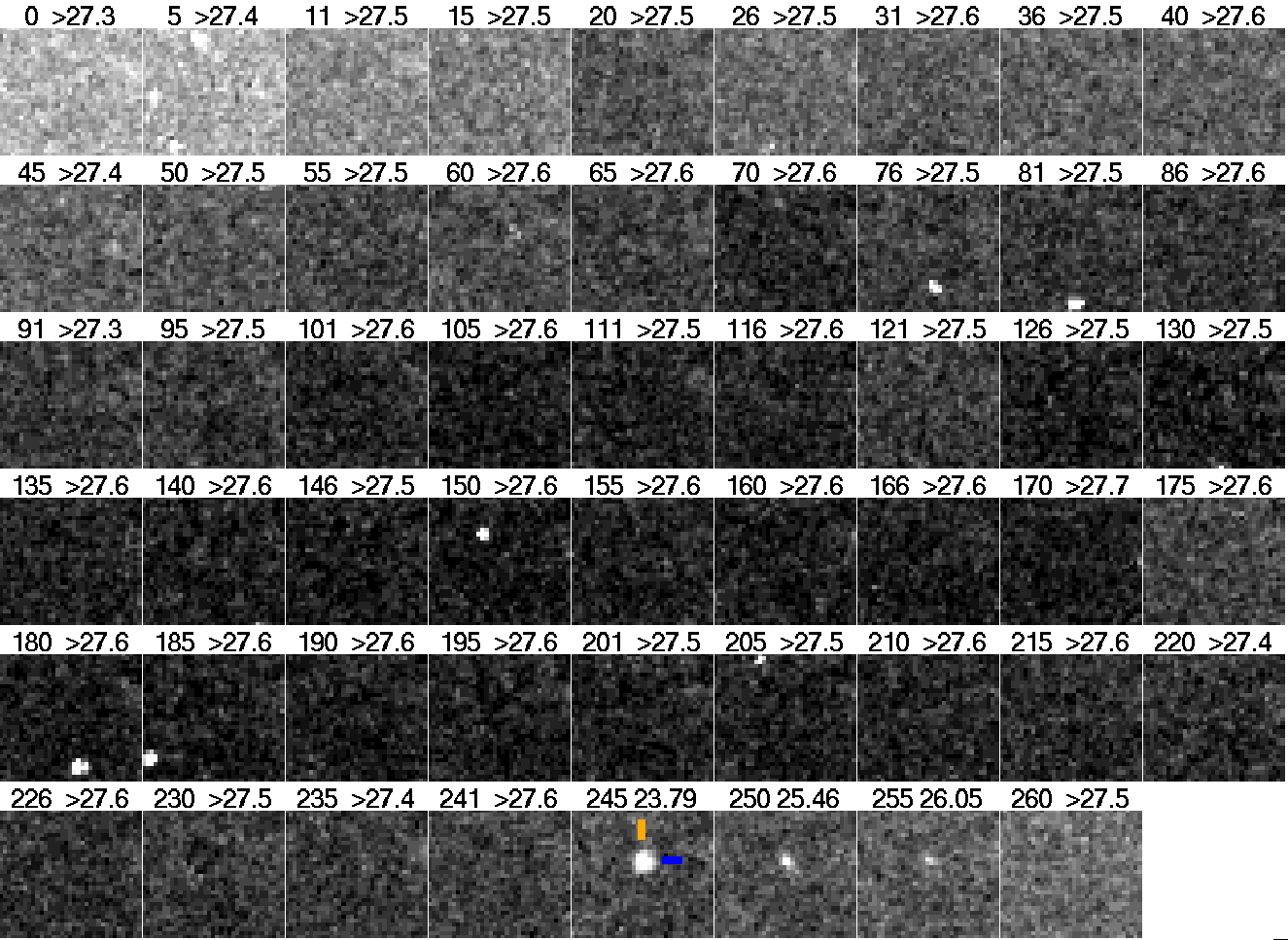} \\
     & \includegraphics[height=3.6in]{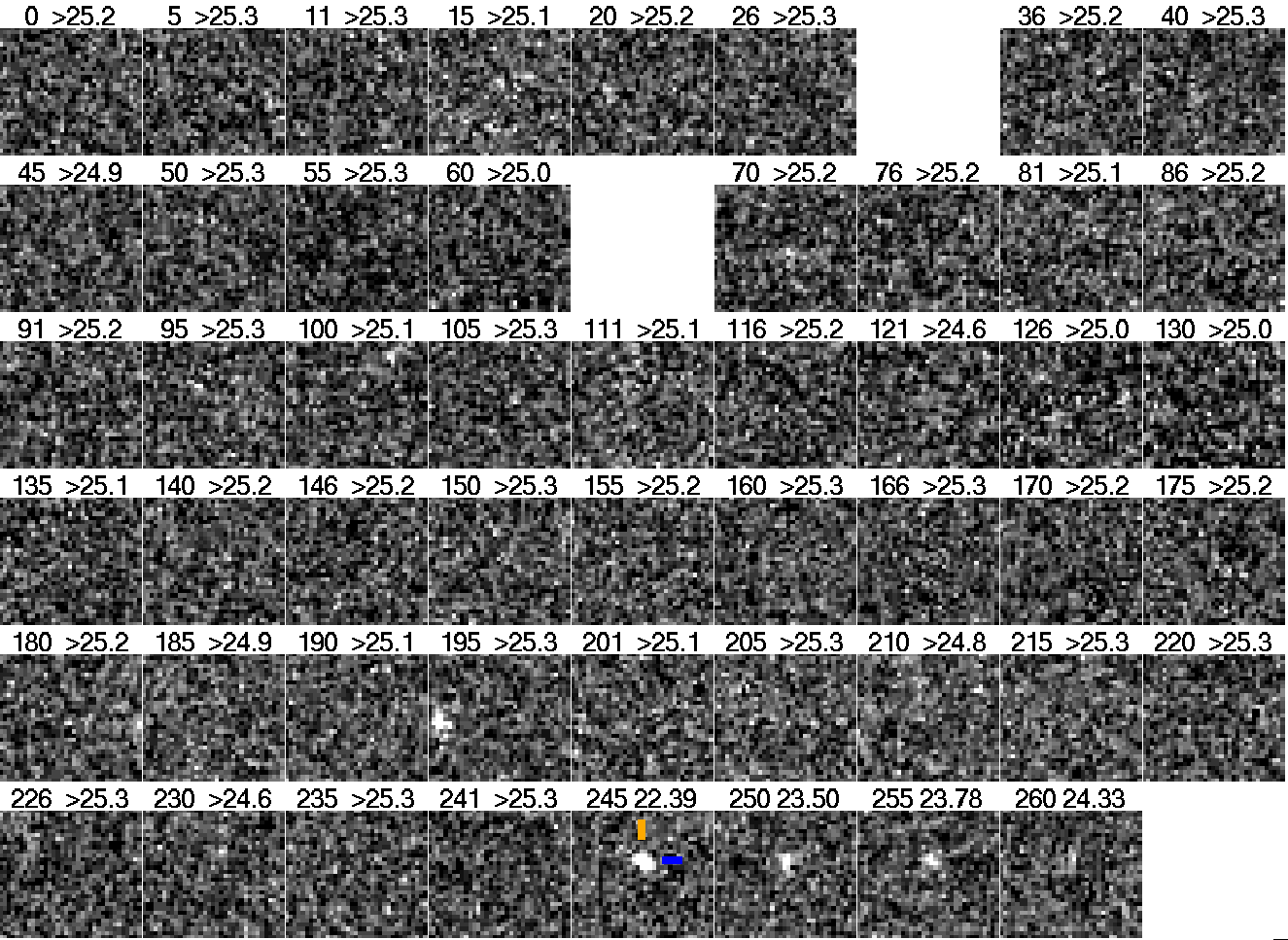}
    \end{tabular}
    \caption{Nova 29}
\end{sidewaysfigure}
\begin{sidewaysfigure}
    \begin{tabular}{cc}
     \multirow{2}{*}[2.6in]{\includegraphics[height=6in]{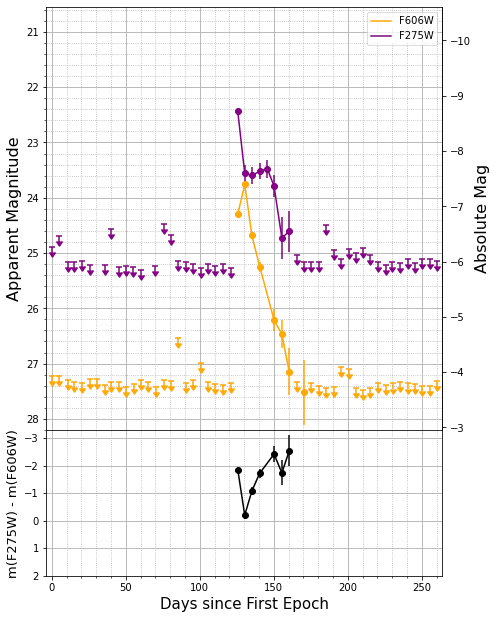}} & \includegraphics[height=3.6in]{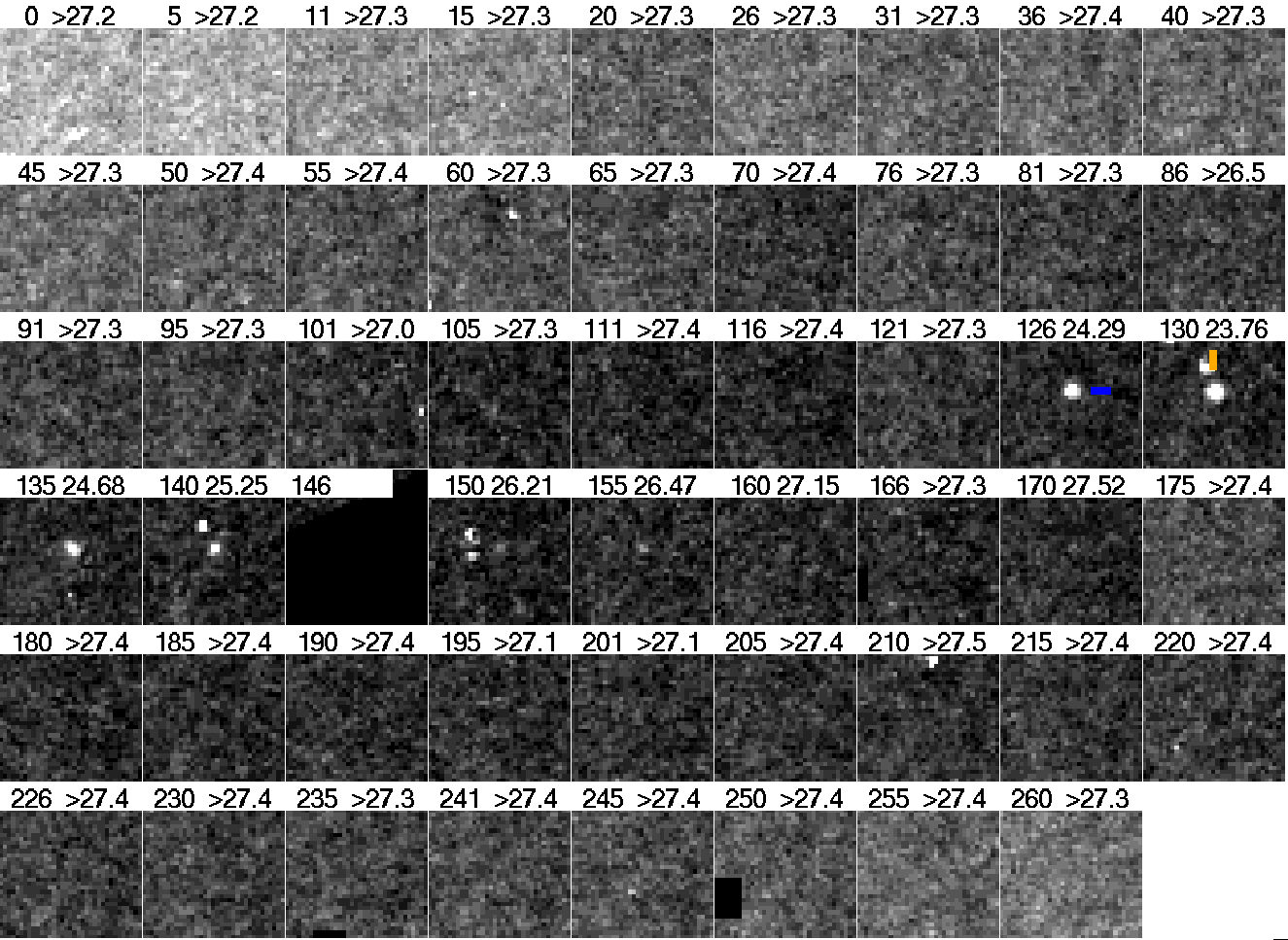} \\
     & \includegraphics[height=3.6in]{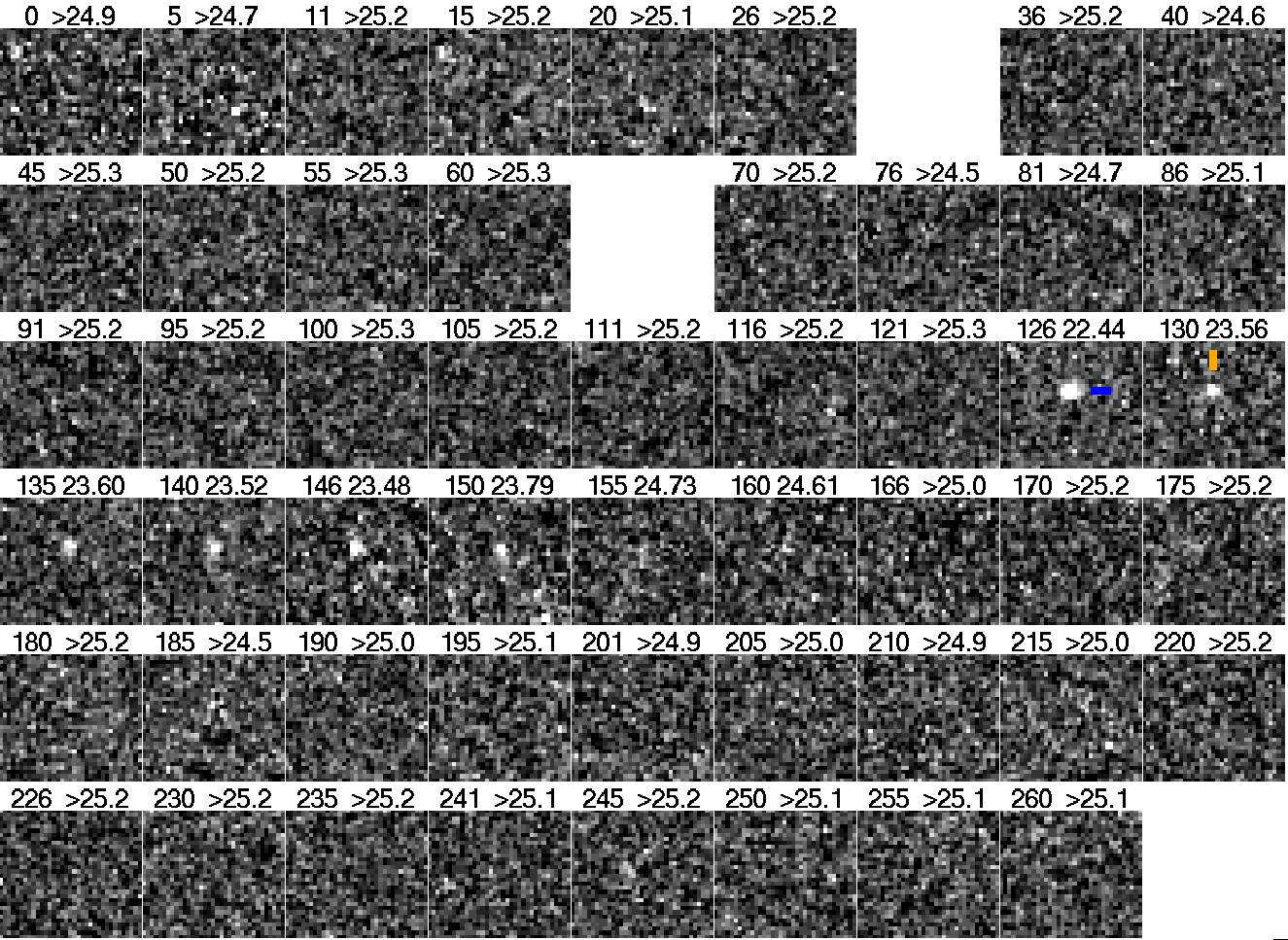}
    \end{tabular}
    \caption{Nova 30}
\end{sidewaysfigure}
\begin{sidewaysfigure}
    \begin{tabular}{cc}
     \multirow{2}{*}[2.6in]{\includegraphics[height=6in]{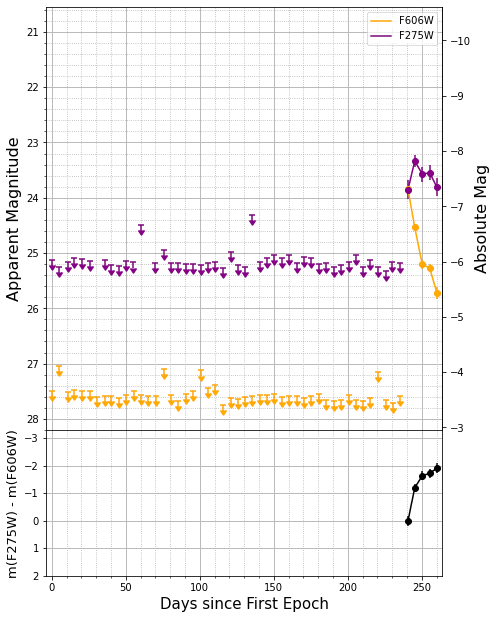}} & \includegraphics[height=3.6in]{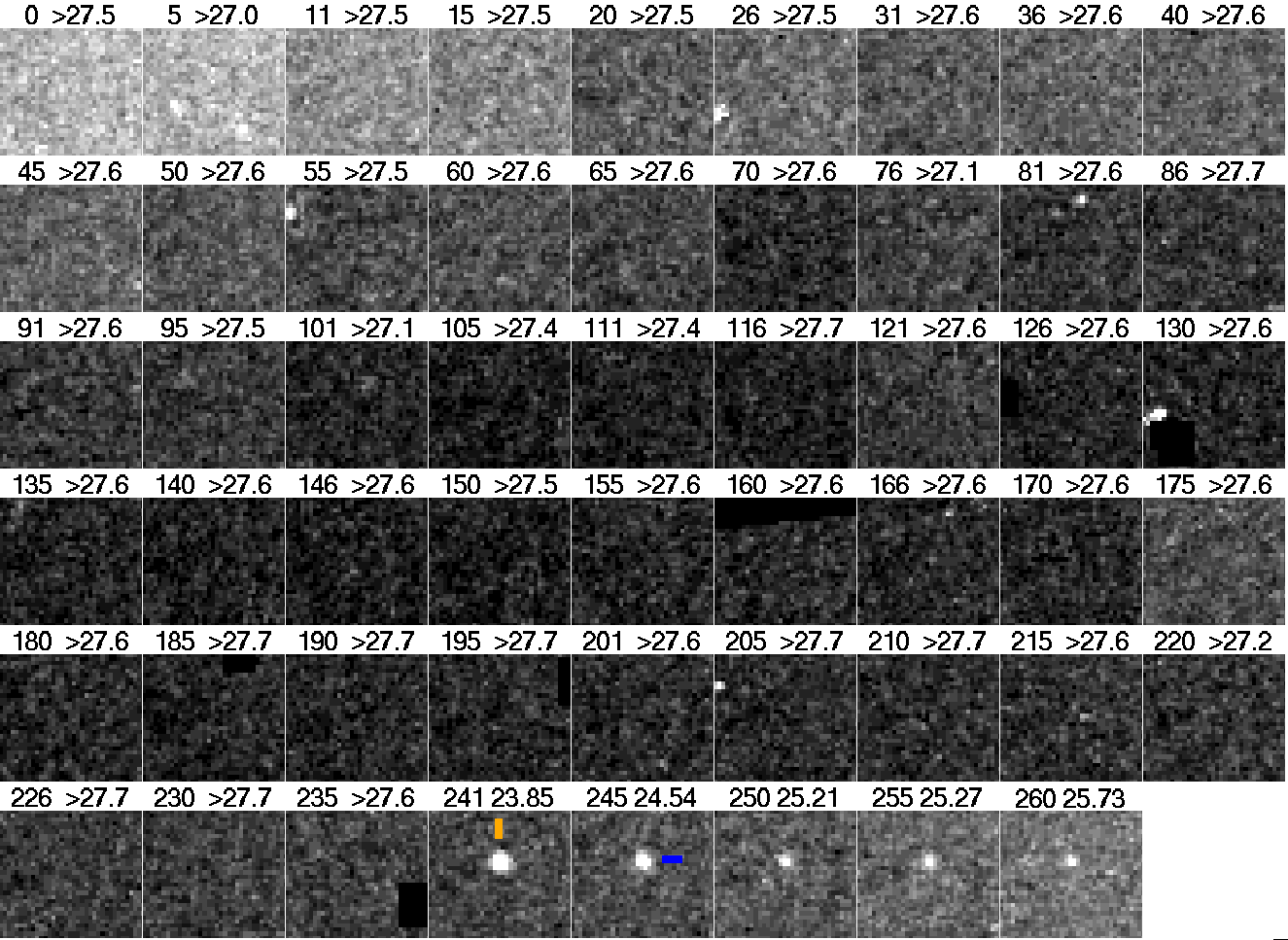} \\
     & \includegraphics[height=3.6in]{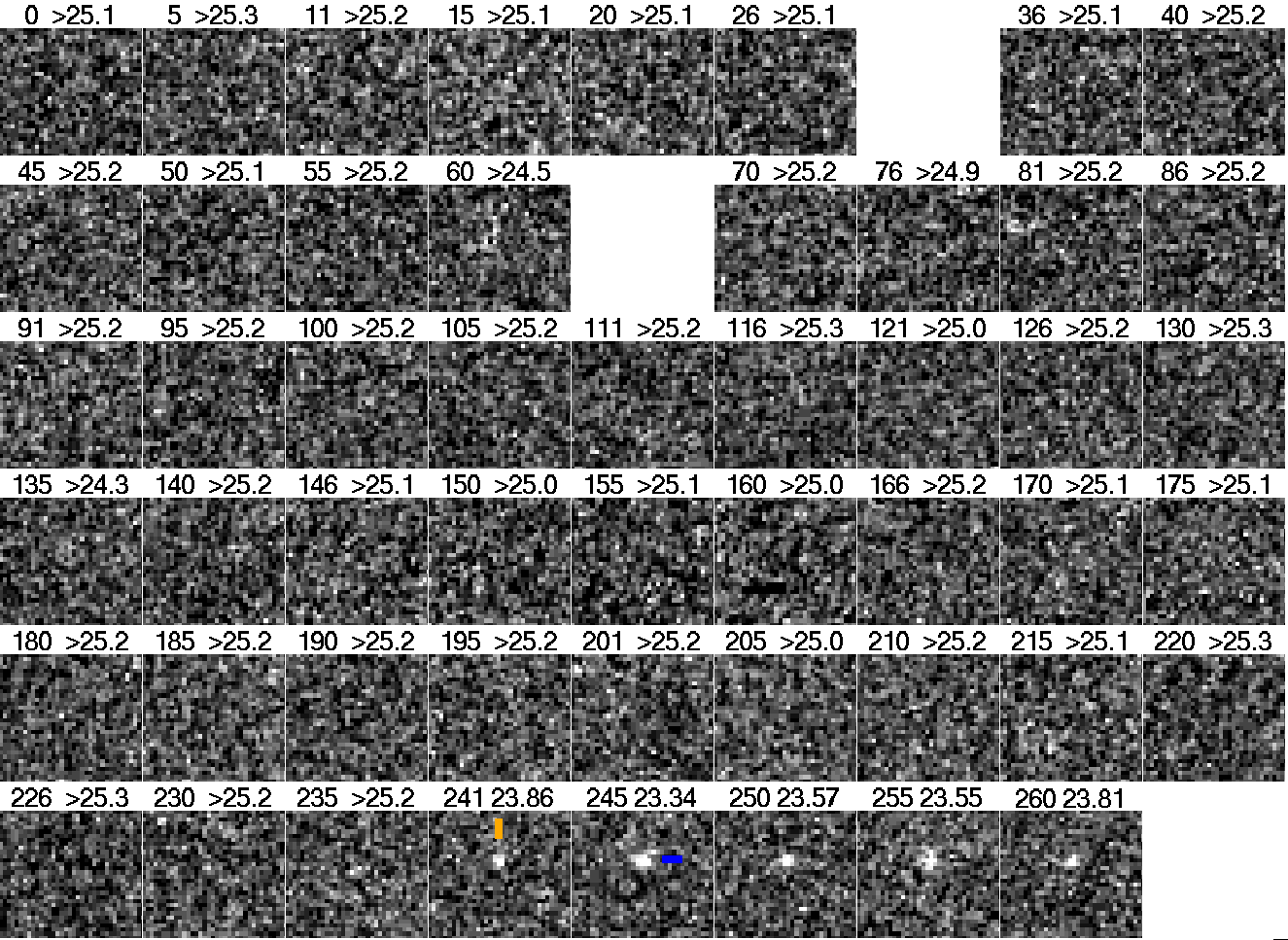}
    \end{tabular}
    \caption{Nova 31}
\end{sidewaysfigure}
\begin{sidewaysfigure}
    \begin{tabular}{cc}
     \multirow{2}{*}[2.6in]{\includegraphics[height=6in]{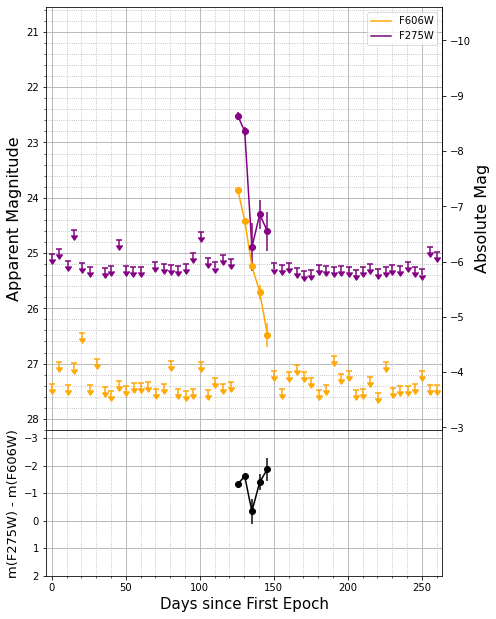}} & \includegraphics[height=3.6in]{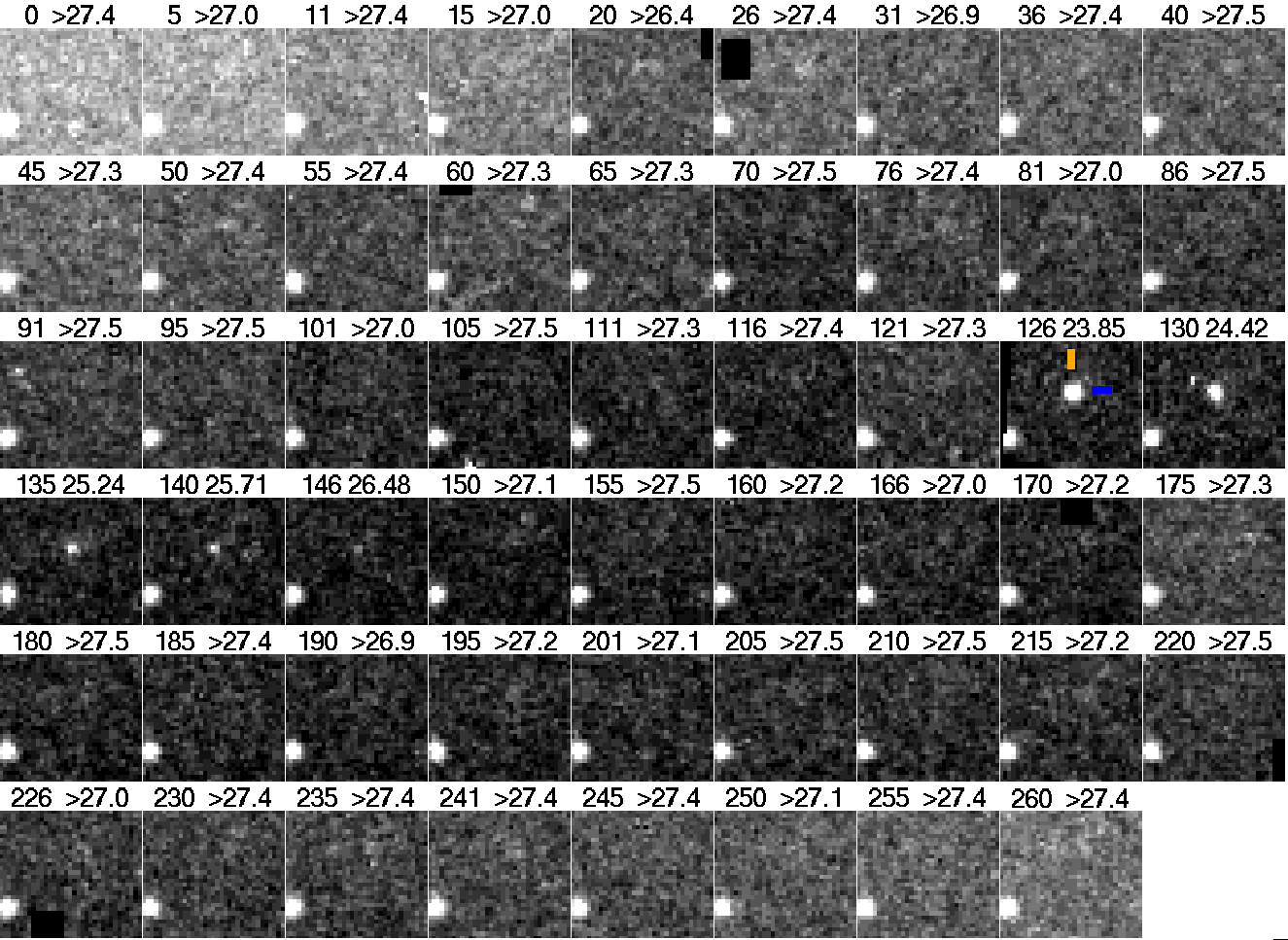} \\
     & \includegraphics[height=3.6in]{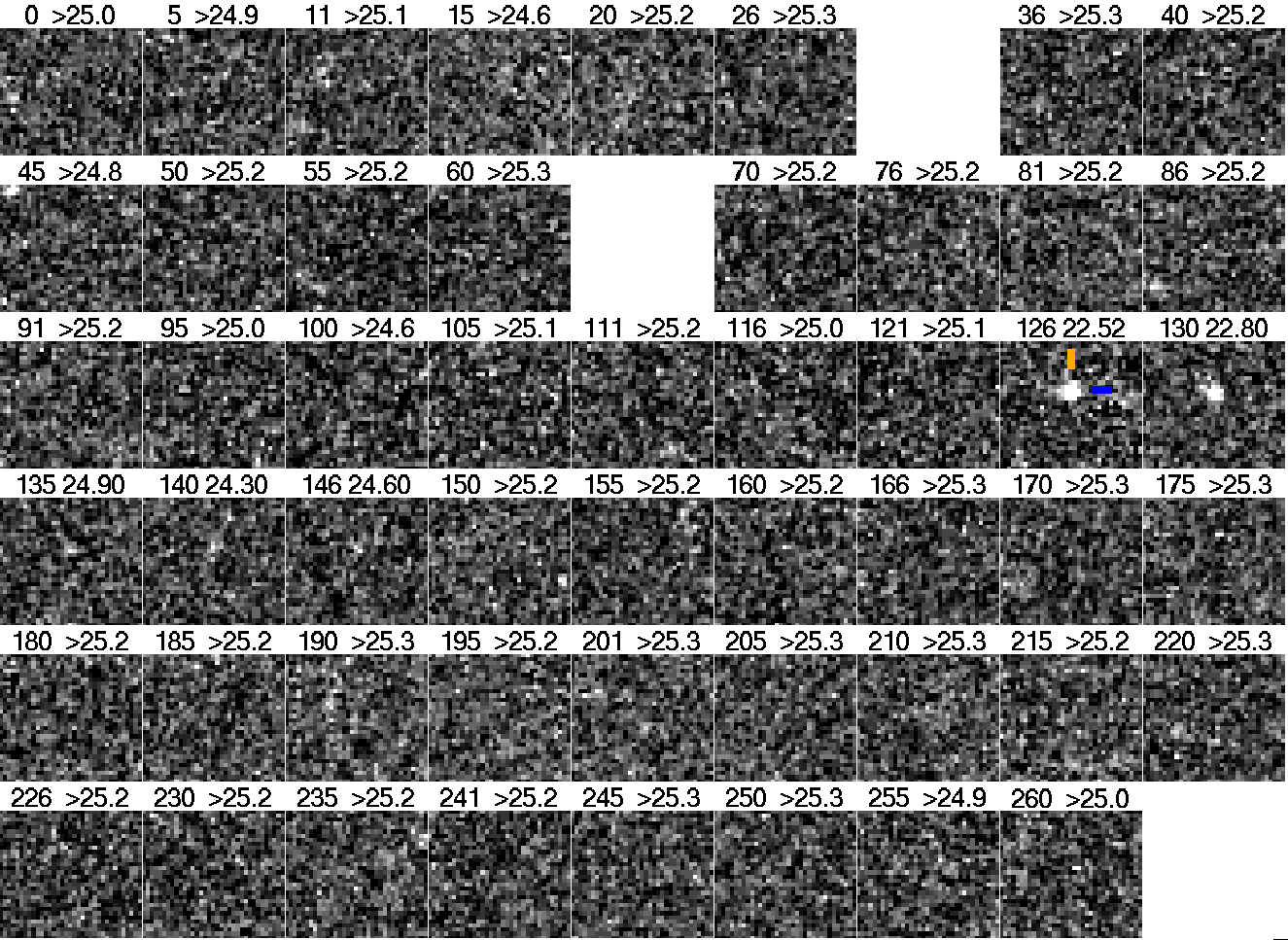}
    \end{tabular}
    \caption{Nova 32}
\end{sidewaysfigure}
\begin{sidewaysfigure}
    \begin{tabular}{cc}
     \multirow{2}{*}[2.6in]{\includegraphics[height=6in]{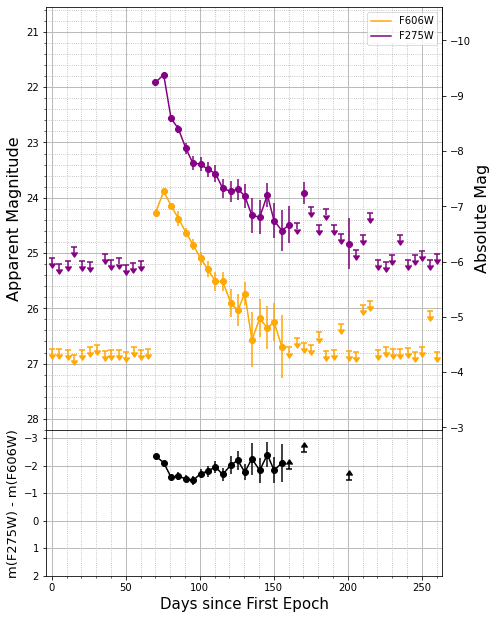}} & \includegraphics[height=3.6in]{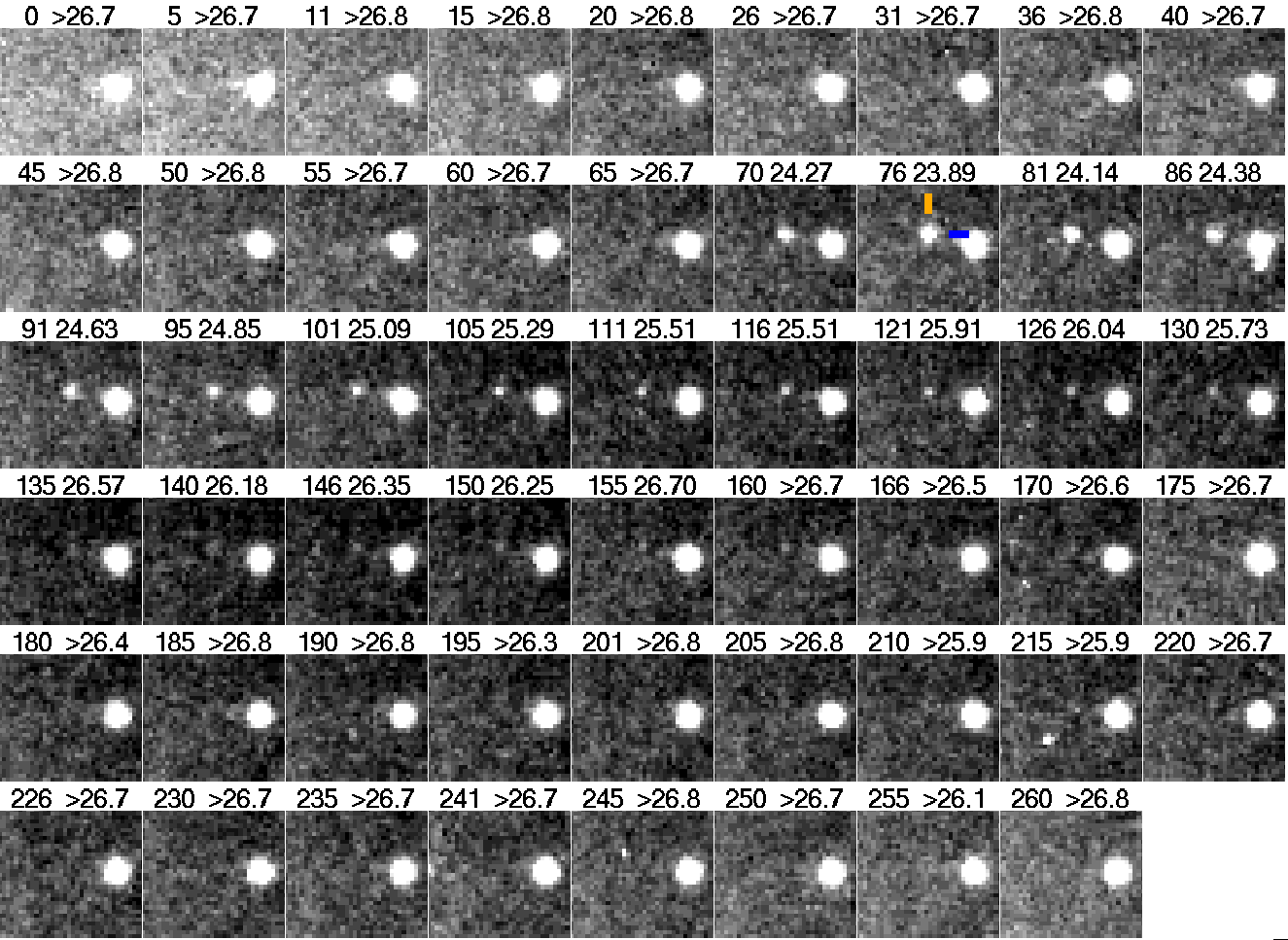} \\
     & \includegraphics[height=3.6in]{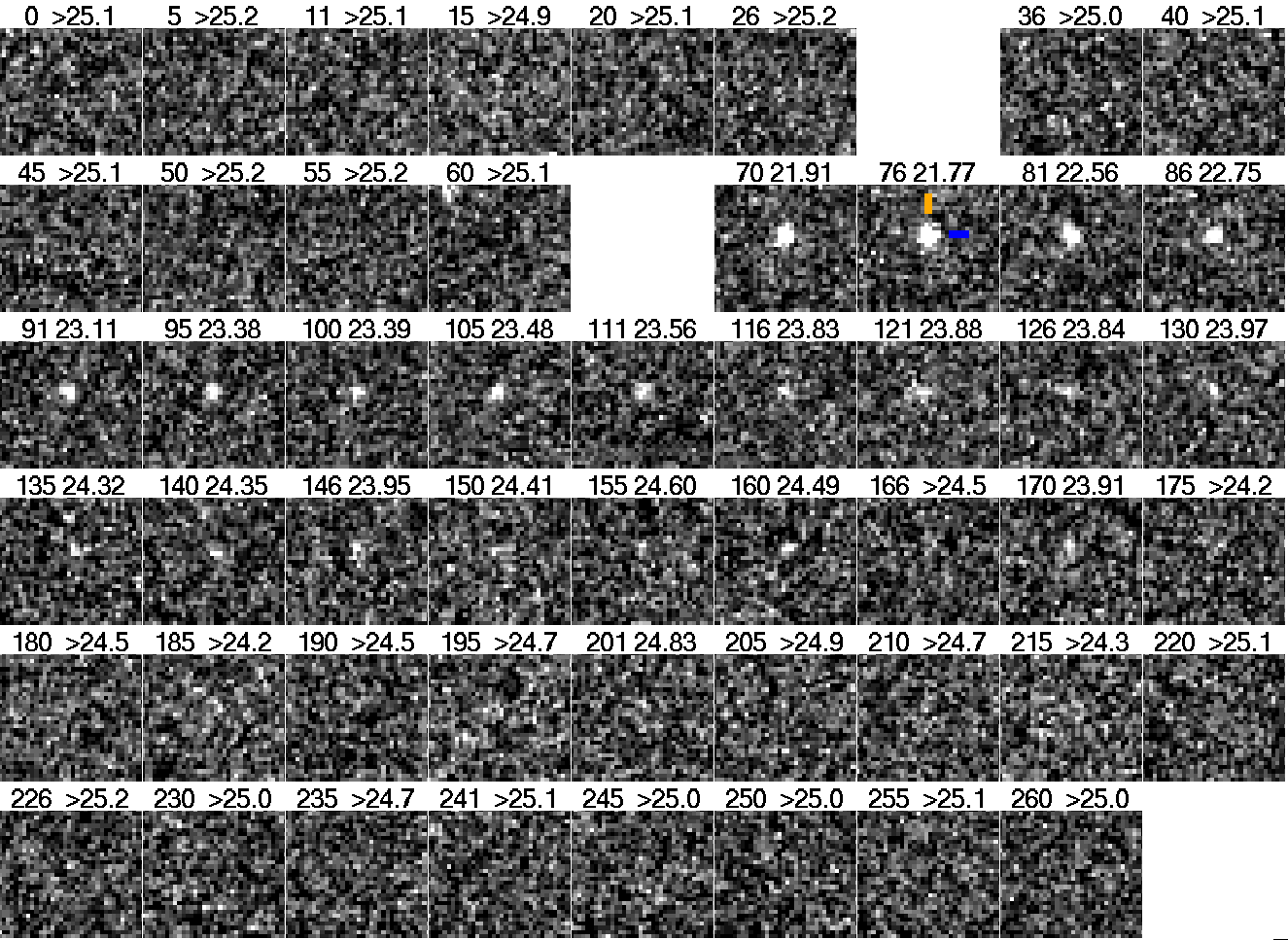}
    \end{tabular}
    \caption{Nova 33}
\end{sidewaysfigure}
\begin{sidewaysfigure}
    \begin{tabular}{cc}
     \multirow{2}{*}[2.6in]{\includegraphics[height=6in]{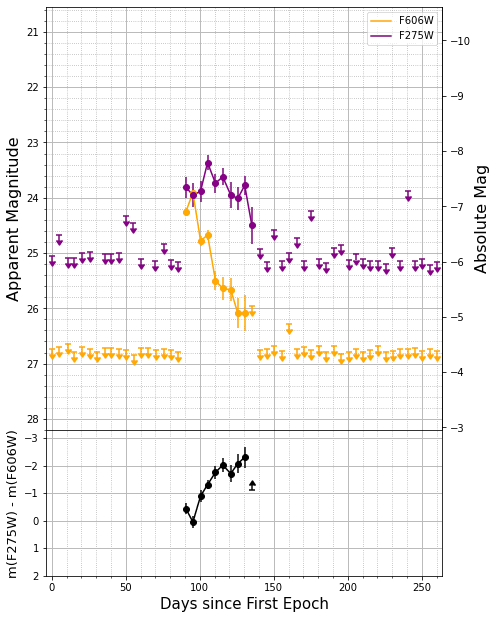}} & \includegraphics[height=3.6in]{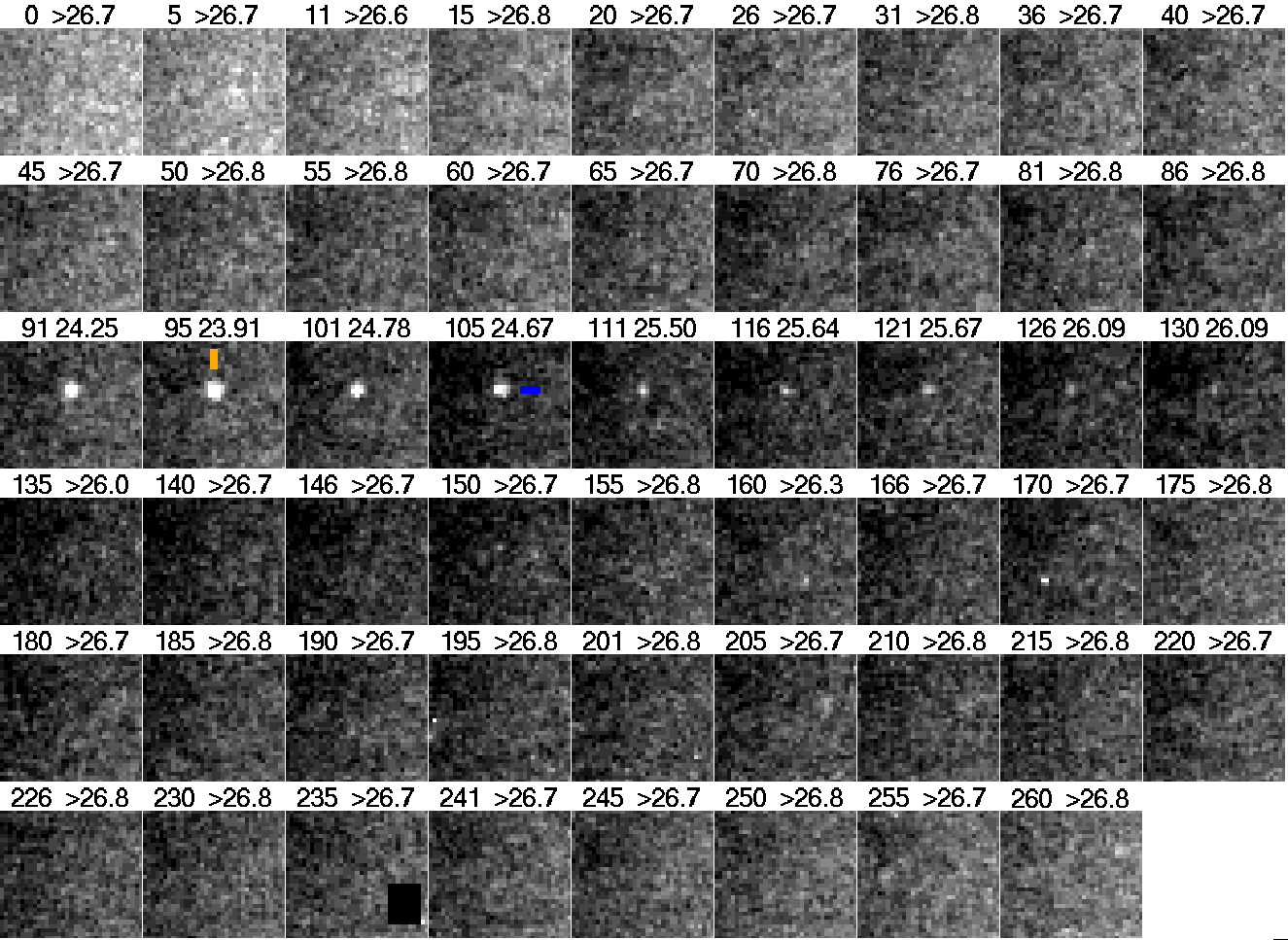} \\
     & \includegraphics[height=3.6in]{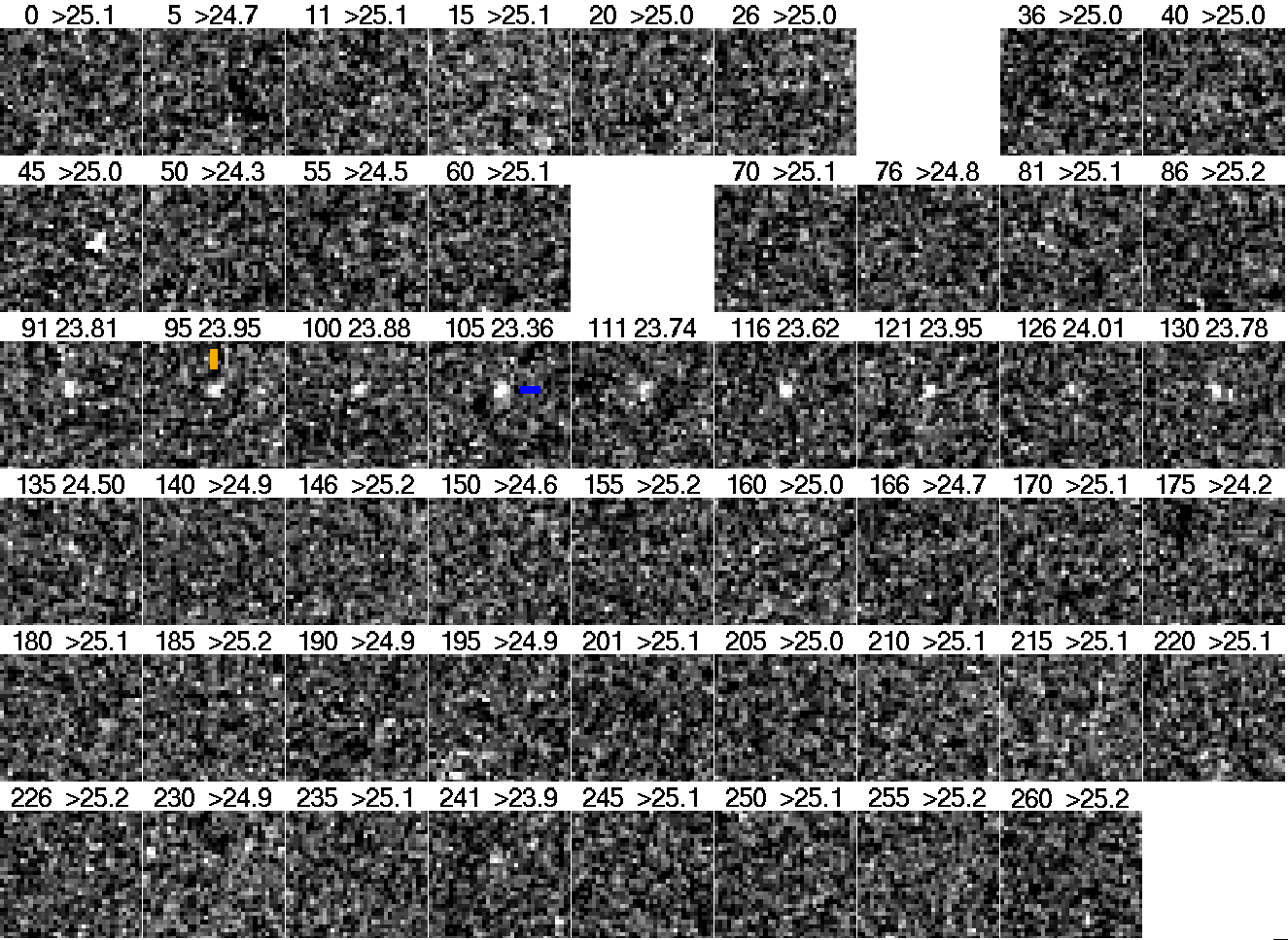}
    \end{tabular}
    \caption{Nova 34}
\end{sidewaysfigure}
\begin{sidewaysfigure}
    \begin{tabular}{cc}
     \multirow{2}{*}[2.6in]{\includegraphics[height=6in]{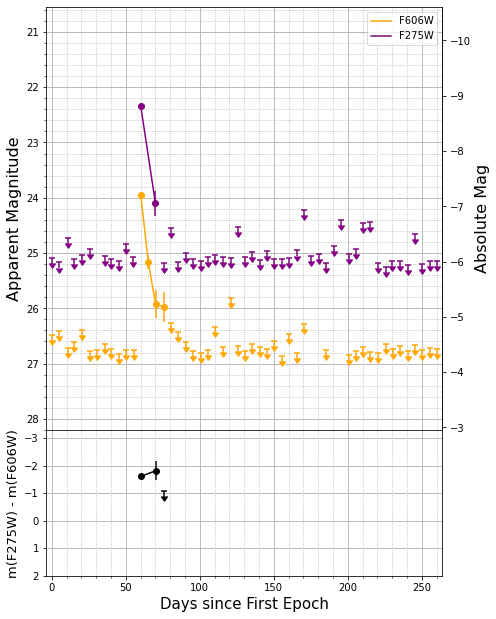}} & \includegraphics[height=3.6in]{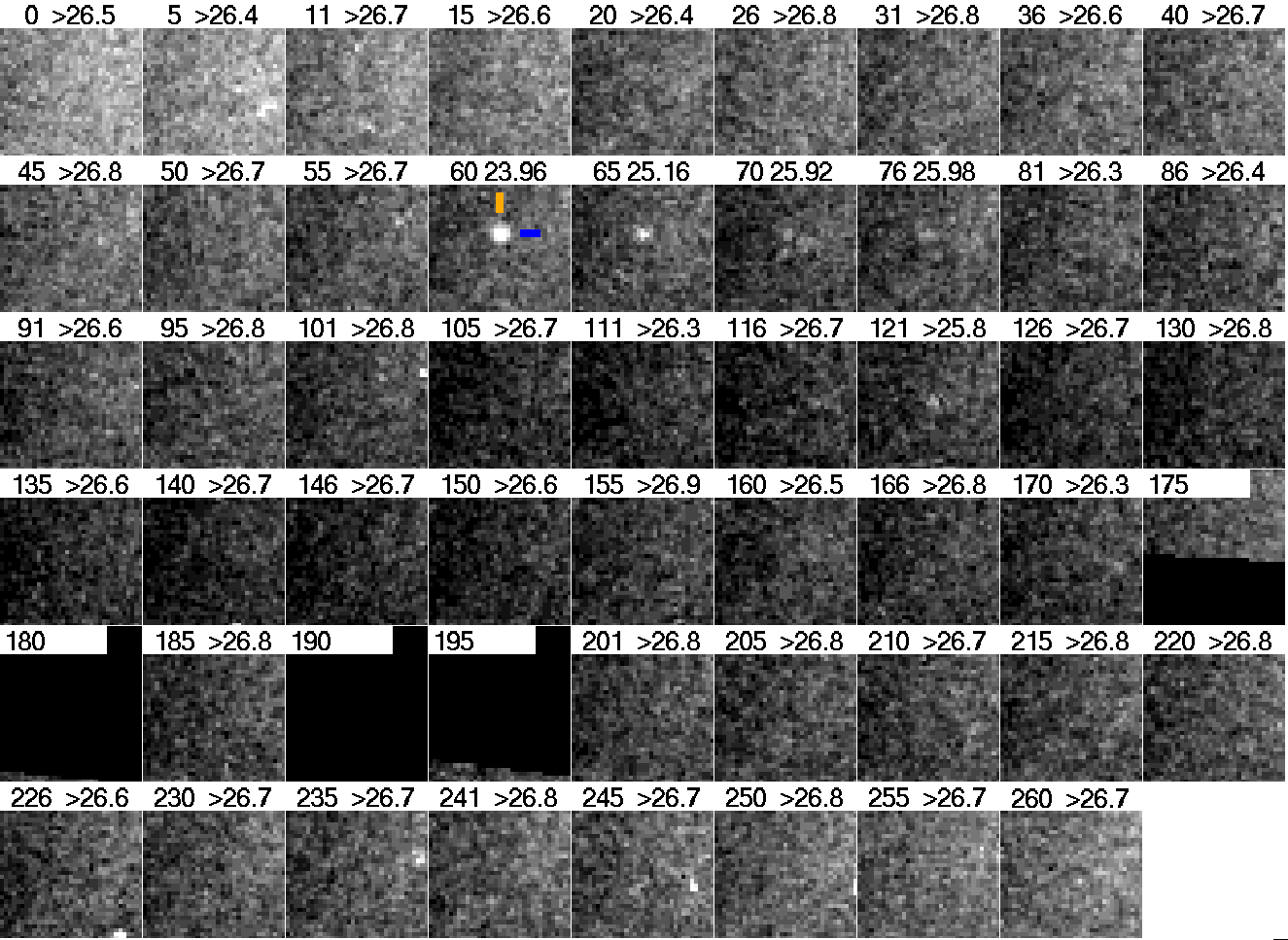} \\
     & \includegraphics[height=3.6in]{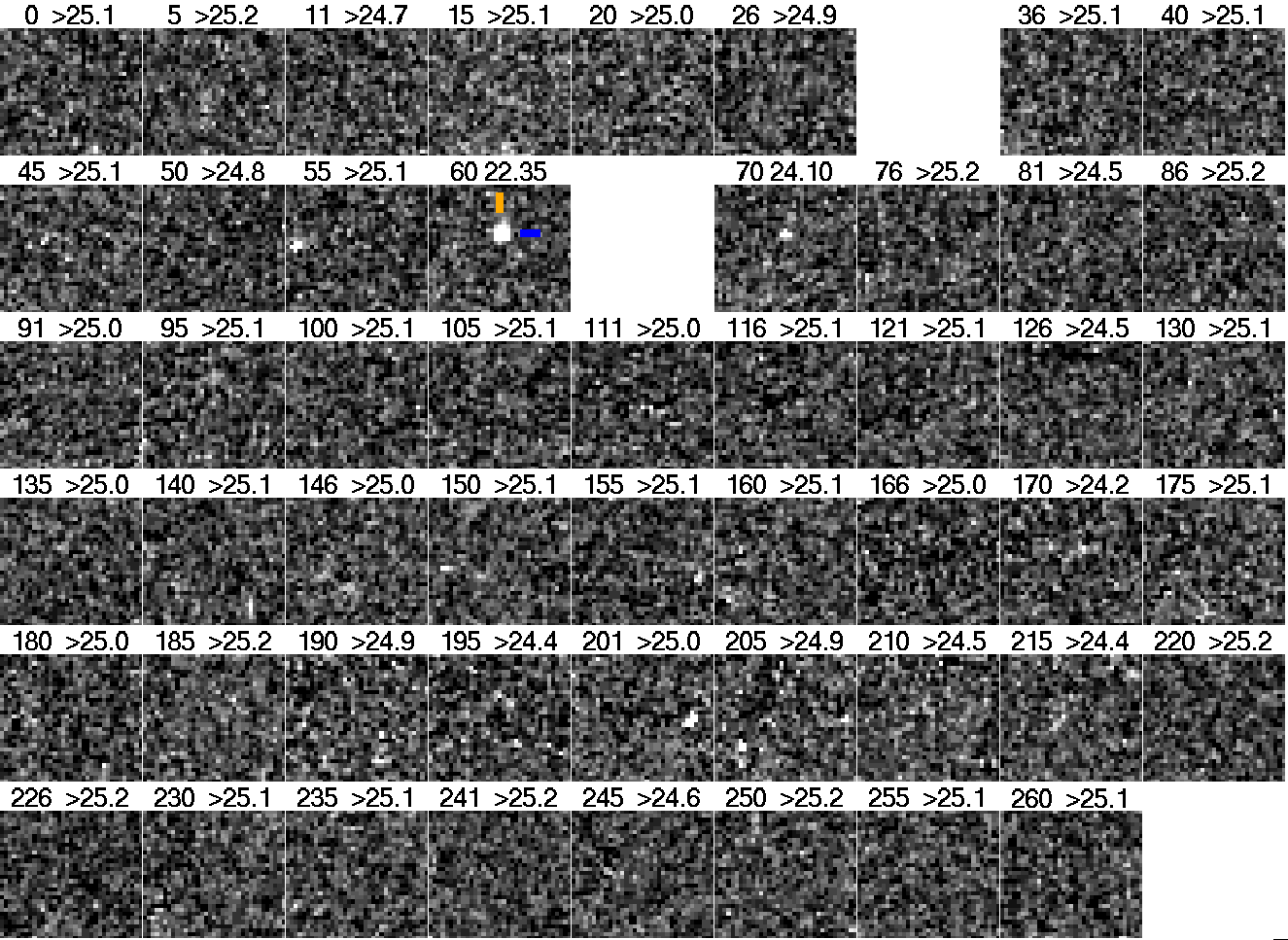}
    \end{tabular}
    \caption{Nova 35}
\end{sidewaysfigure}
\begin{sidewaysfigure}
    \begin{tabular}{cc}
     \multirow{2}{*}[2.6in]{\includegraphics[height=6in]{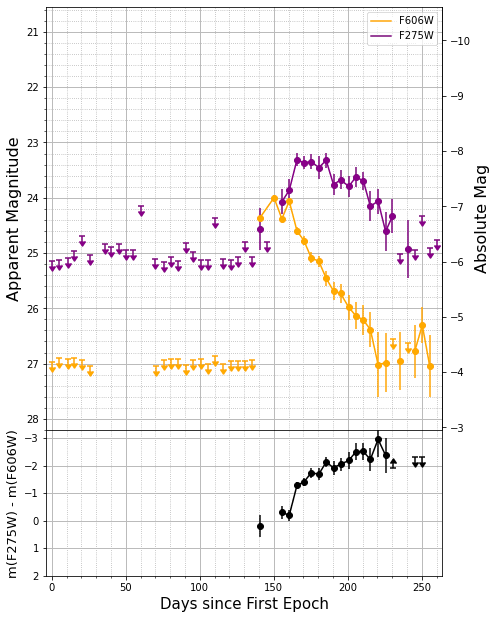}} & \includegraphics[height=3.6in]{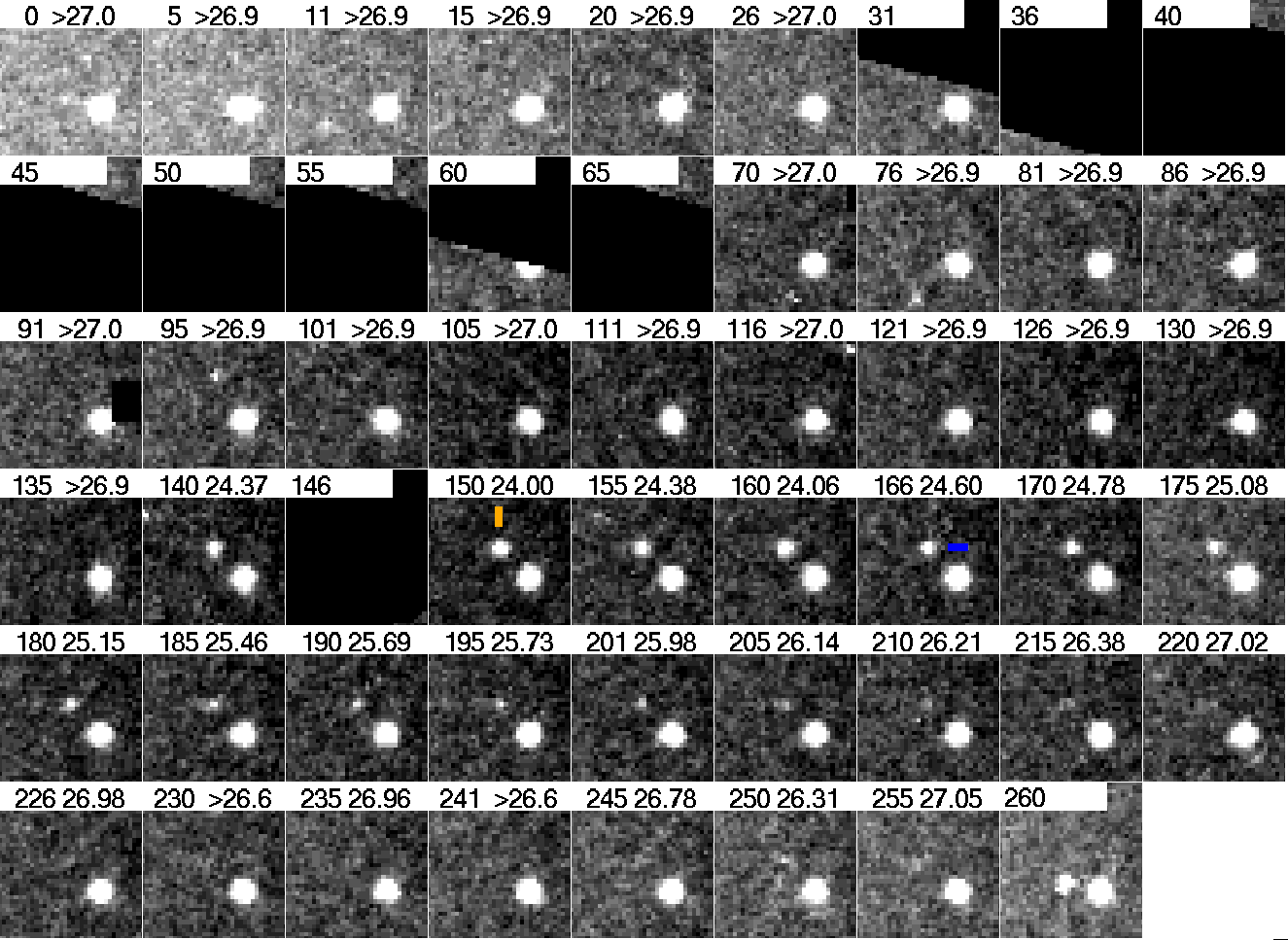} \\
     & \includegraphics[height=3.6in]{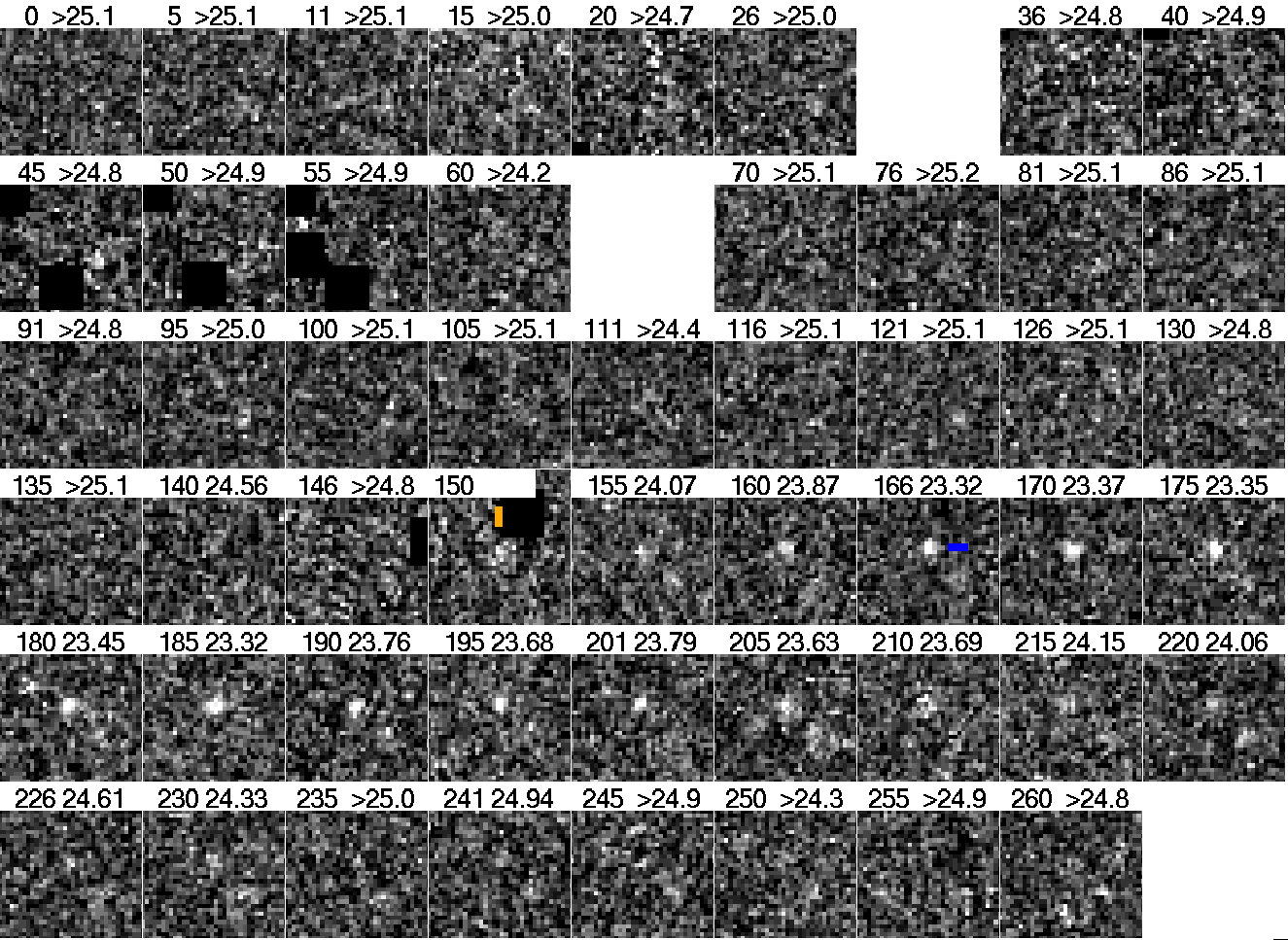}
    \end{tabular}
    \caption{Nova 36}
\end{sidewaysfigure}
\begin{sidewaysfigure}
    \begin{tabular}{cc}
     \multirow{2}{*}[2.6in]{\includegraphics[height=6in]{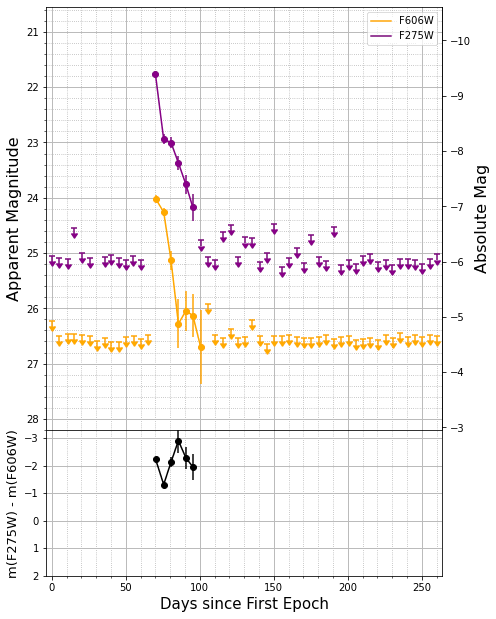}} & \includegraphics[height=3.6in]{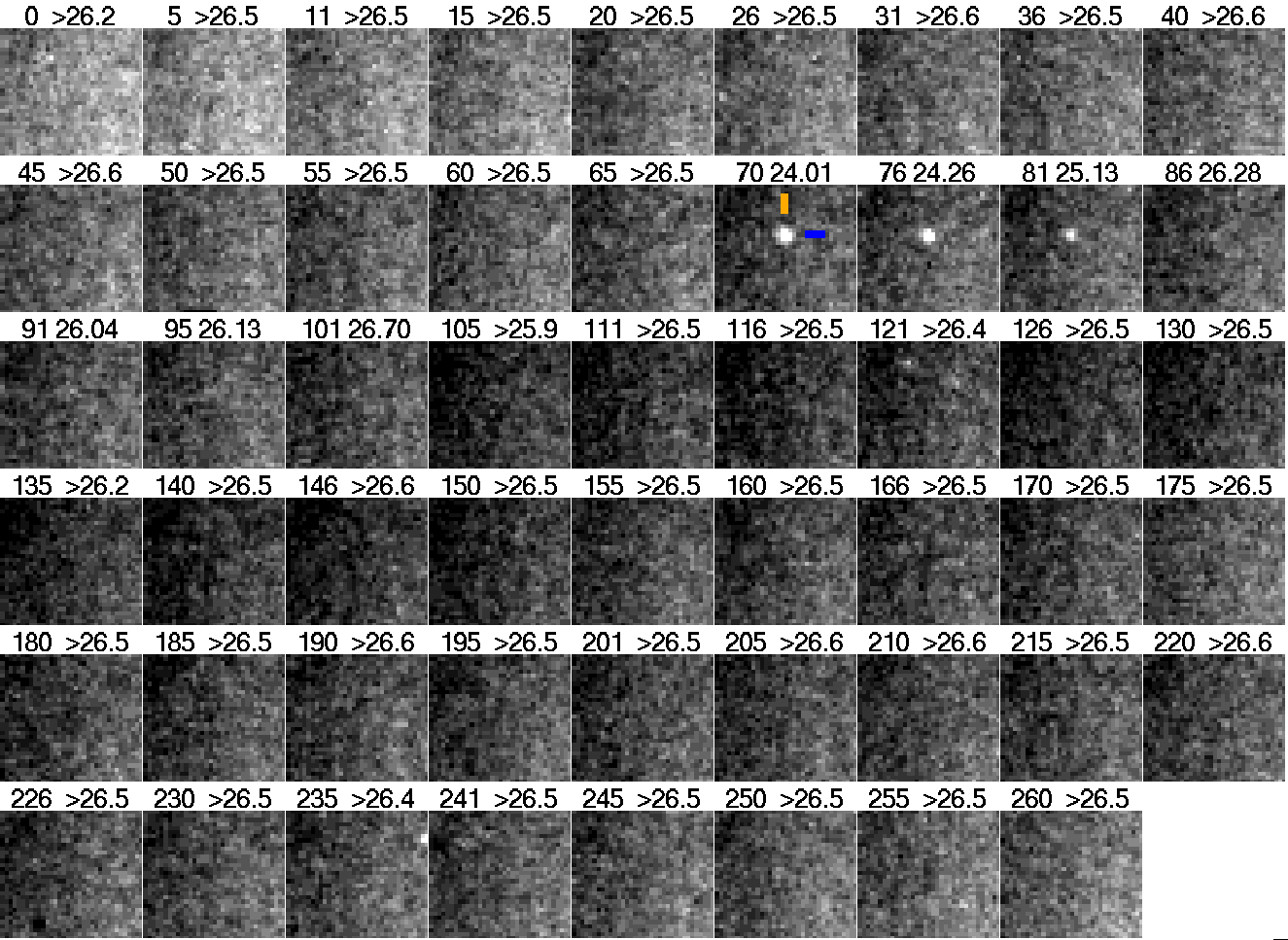} \\
     & \includegraphics[height=3.6in]{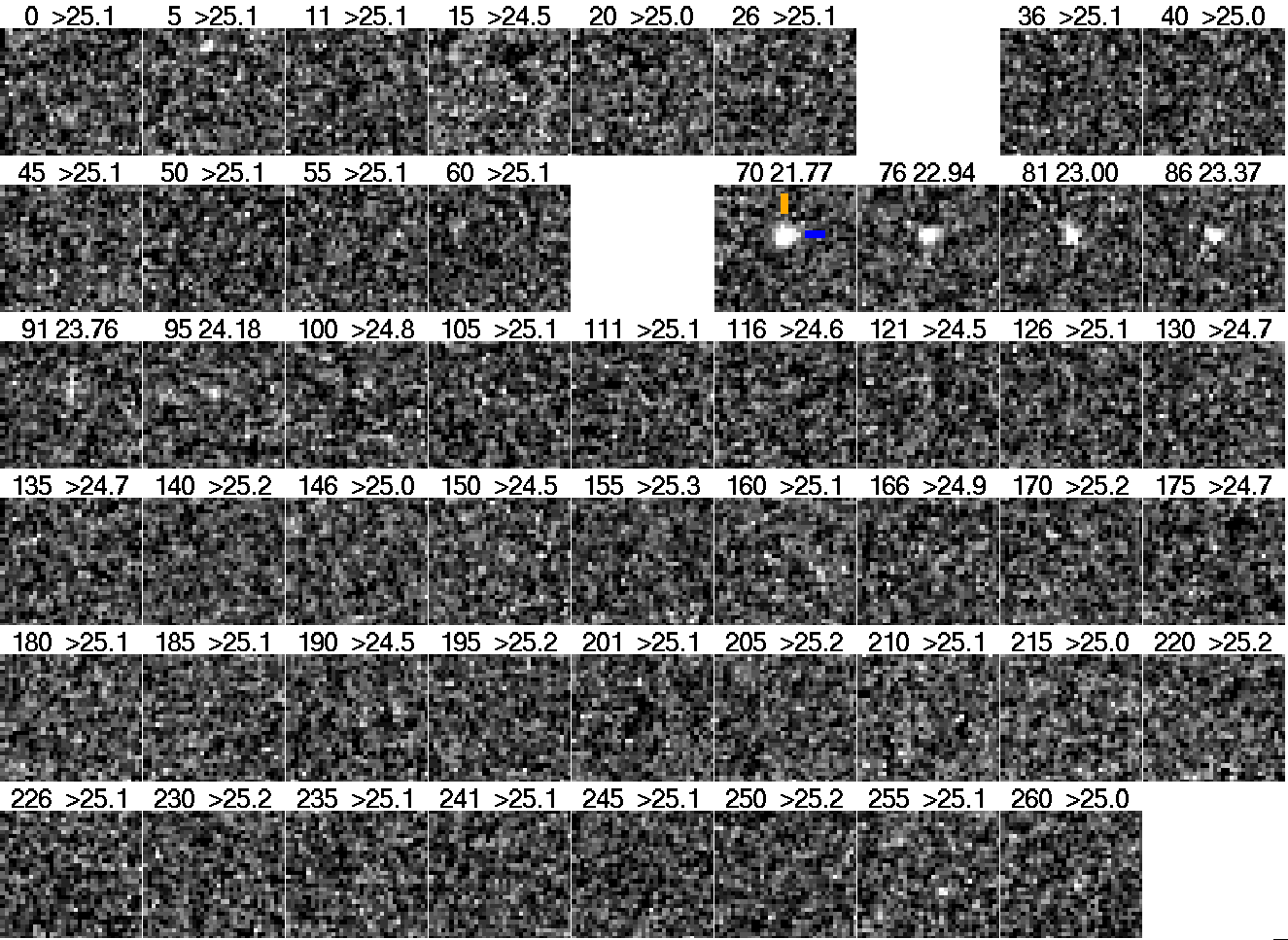}
    \end{tabular}
    \caption{Nova 37}
\end{sidewaysfigure}
\begin{sidewaysfigure}
    \begin{tabular}{cc}
     \multirow{2}{*}[2.6in]{\includegraphics[height=6in]{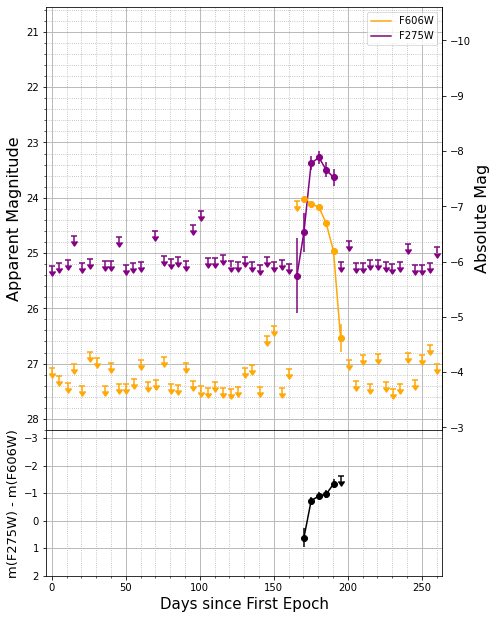}} & \includegraphics[height=3.6in]{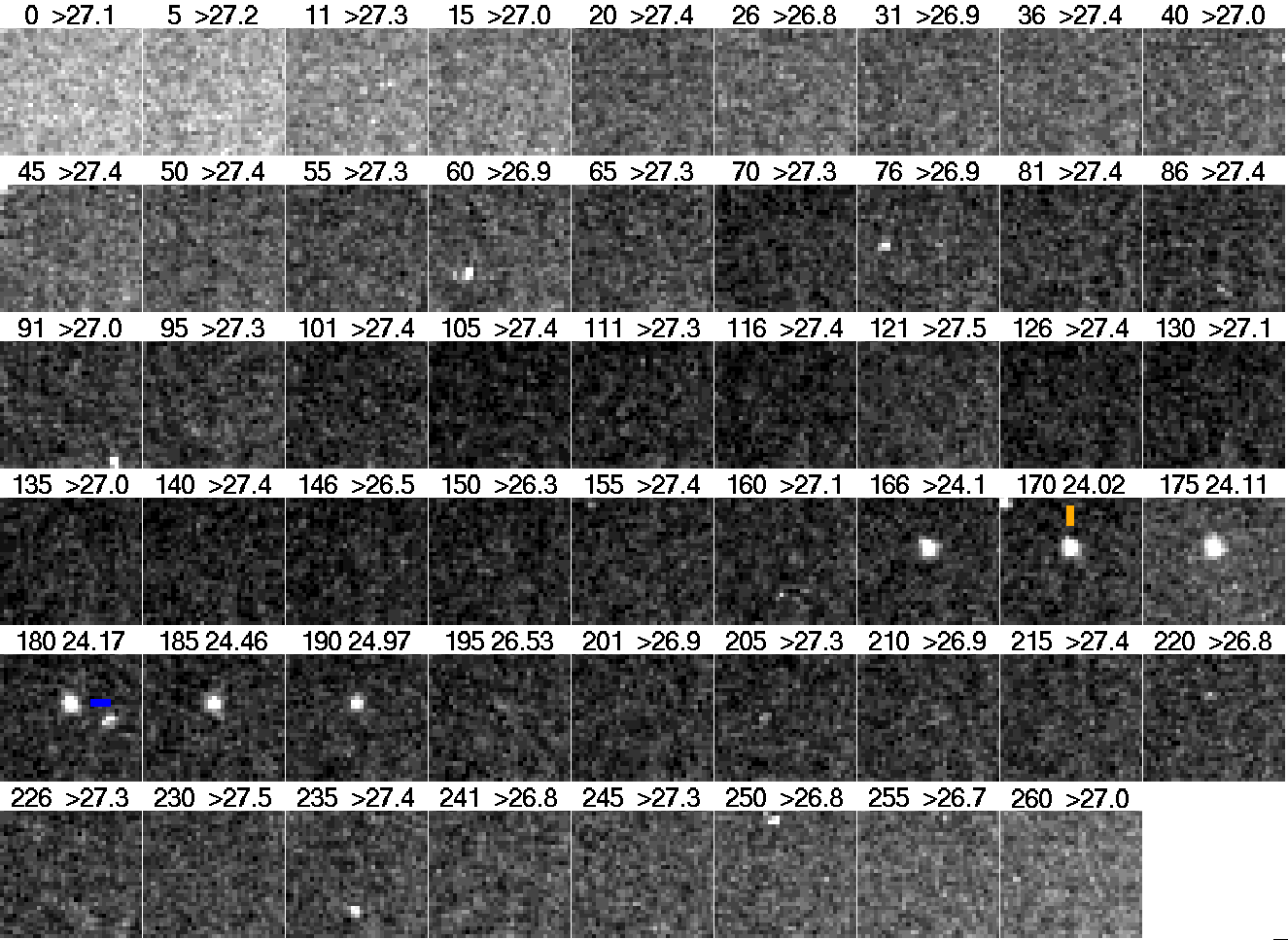} \\
     & \includegraphics[height=3.6in]{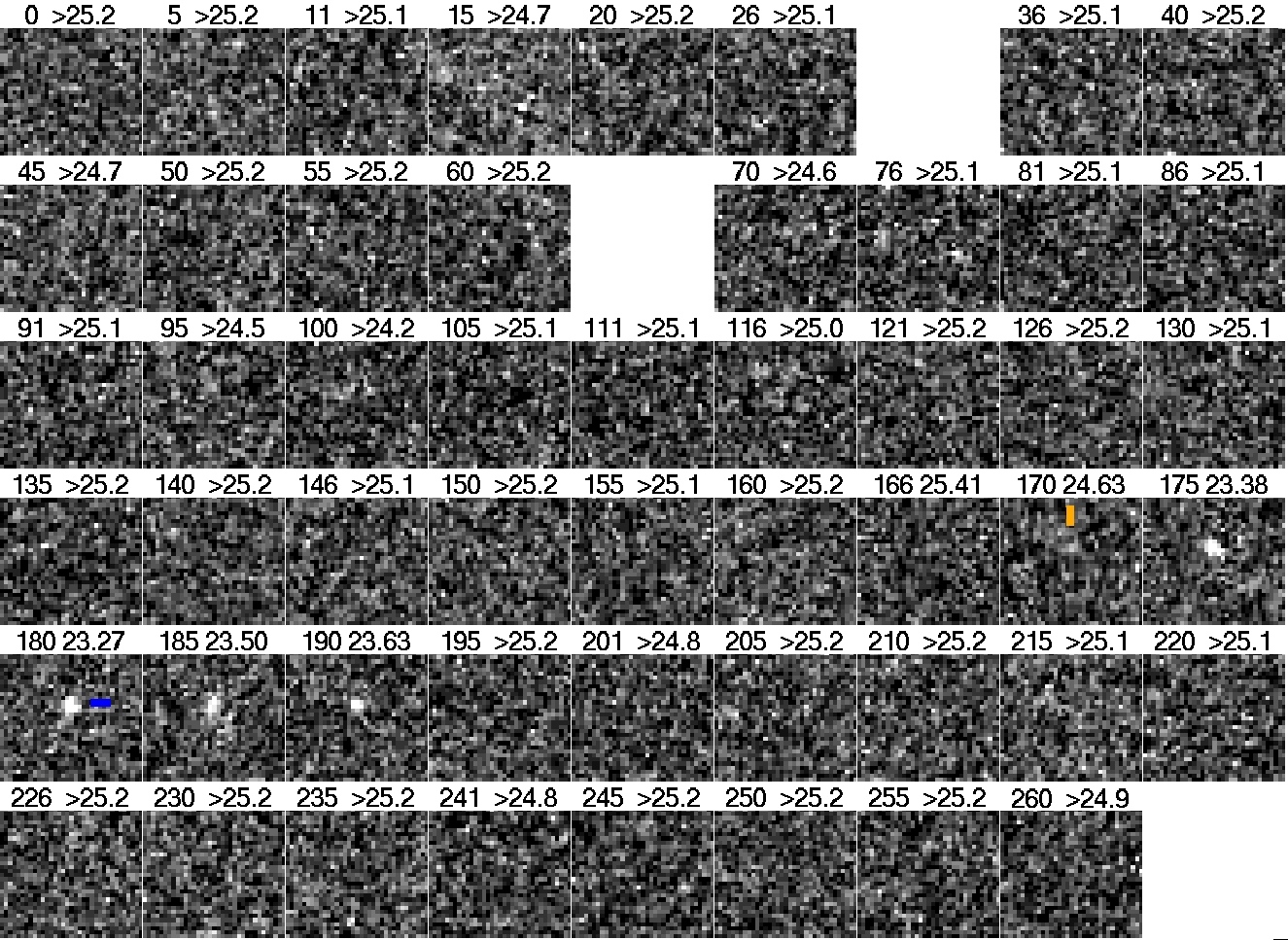}
    \end{tabular}
    \caption{Nova 38}
\end{sidewaysfigure}
\begin{sidewaysfigure}
    \begin{tabular}{cc}
     \multirow{2}{*}[2.6in]{\includegraphics[height=6in]{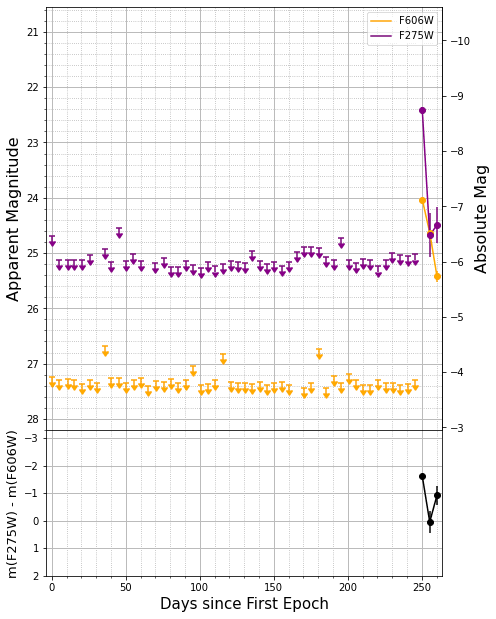}} & \includegraphics[height=3.6in]{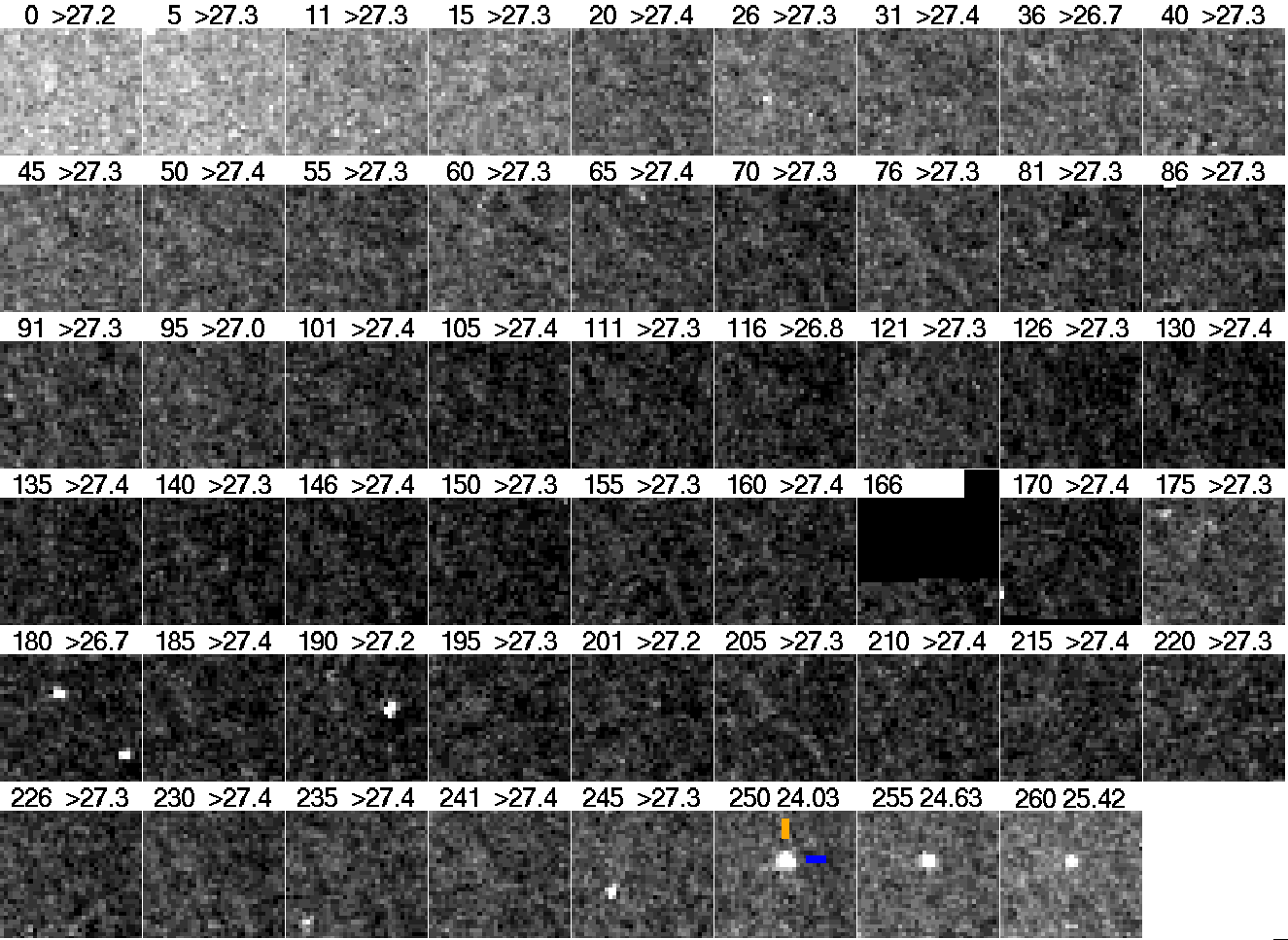} \\
     & \includegraphics[height=3.6in]{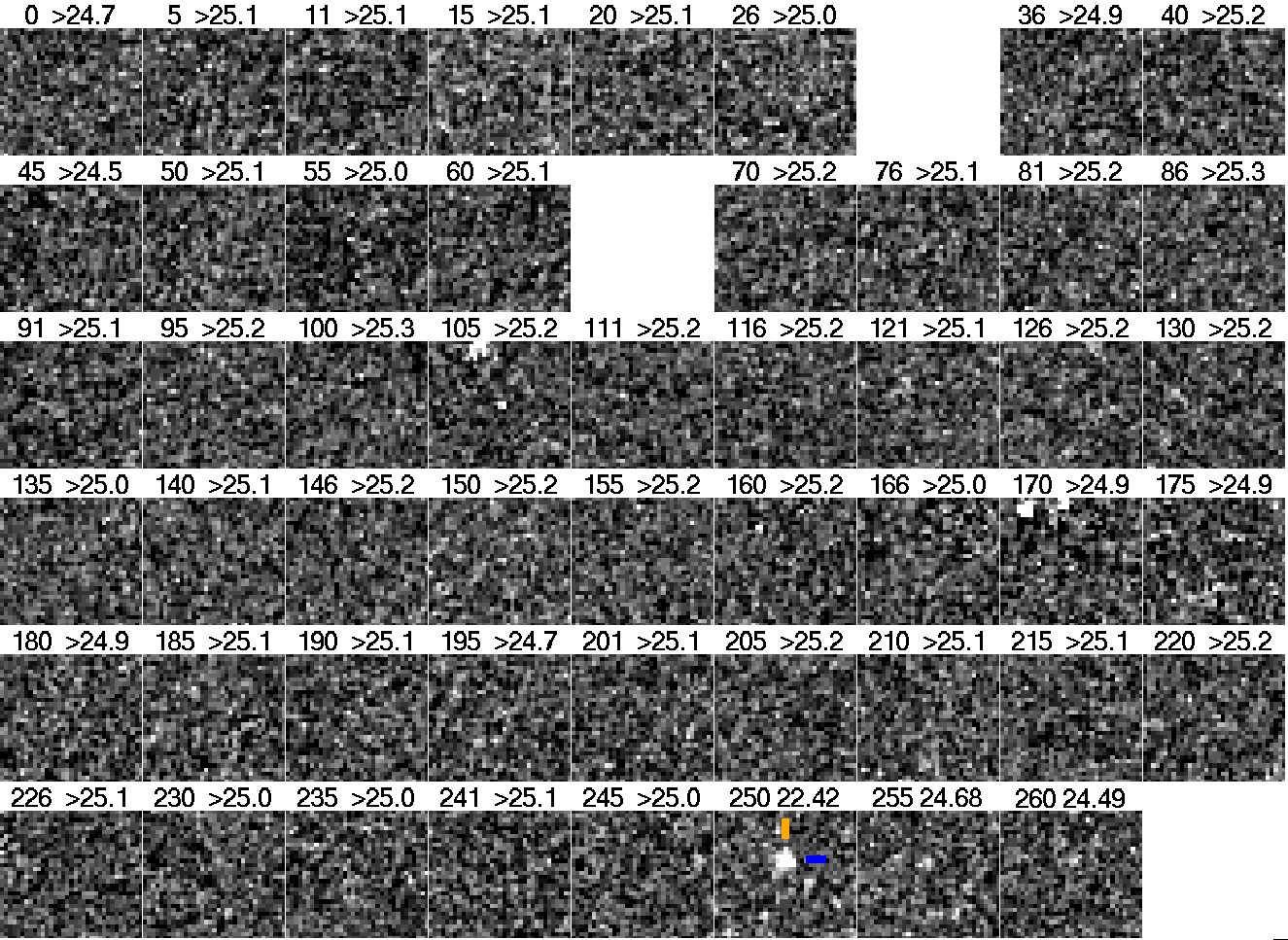}
    \end{tabular}
    \caption{Nova 39}
\end{sidewaysfigure}
\begin{sidewaysfigure}
    \begin{tabular}{cc}
     \multirow{2}{*}[2.6in]{\includegraphics[height=6in]{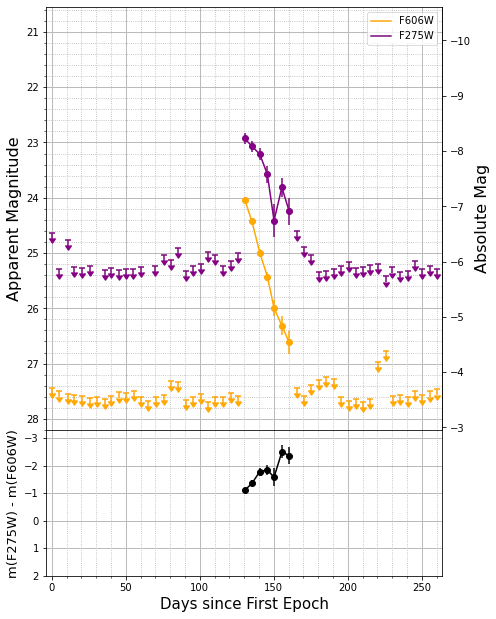}} & \includegraphics[height=3.6in]{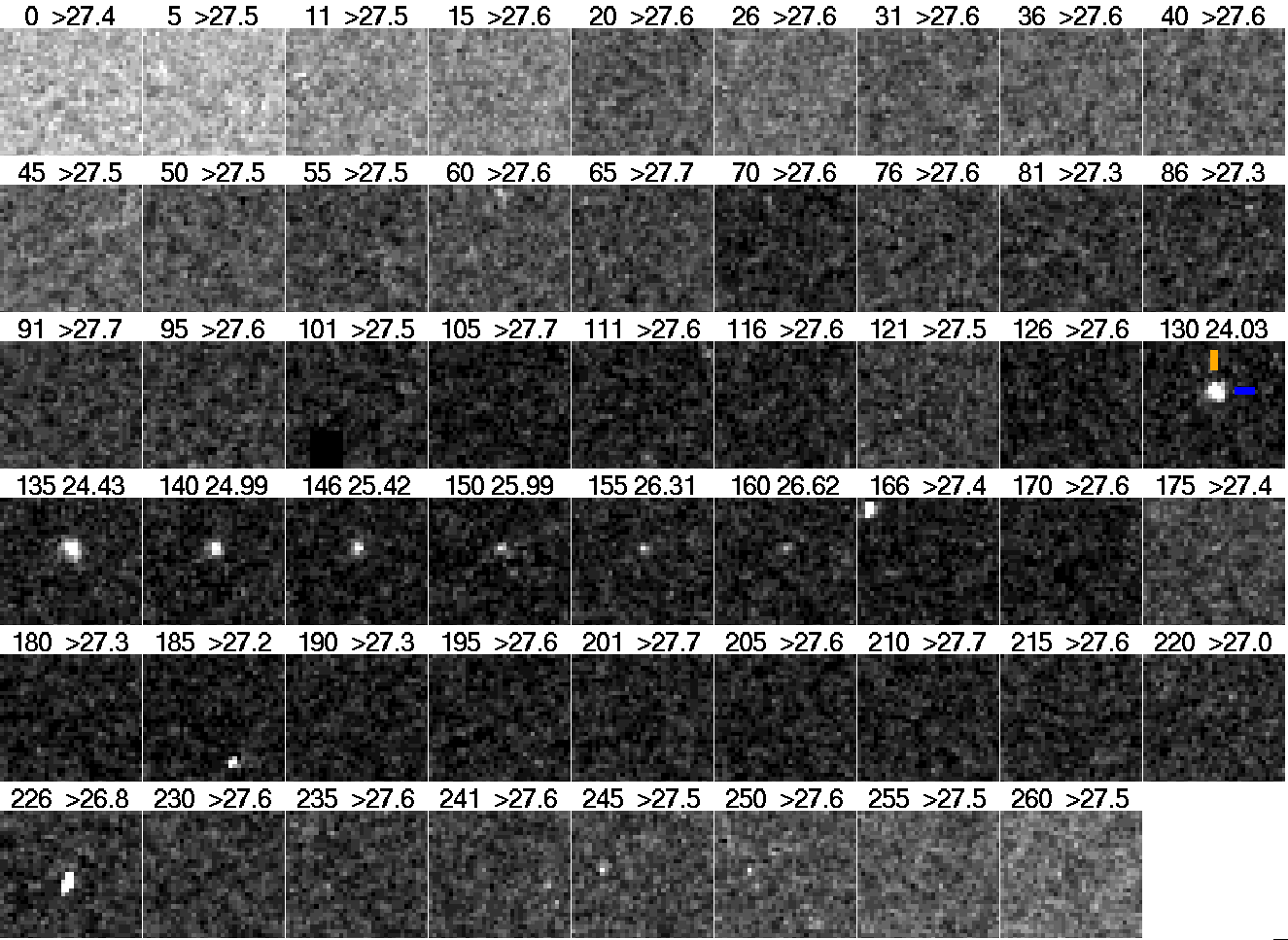} \\
     & \includegraphics[height=3.6in]{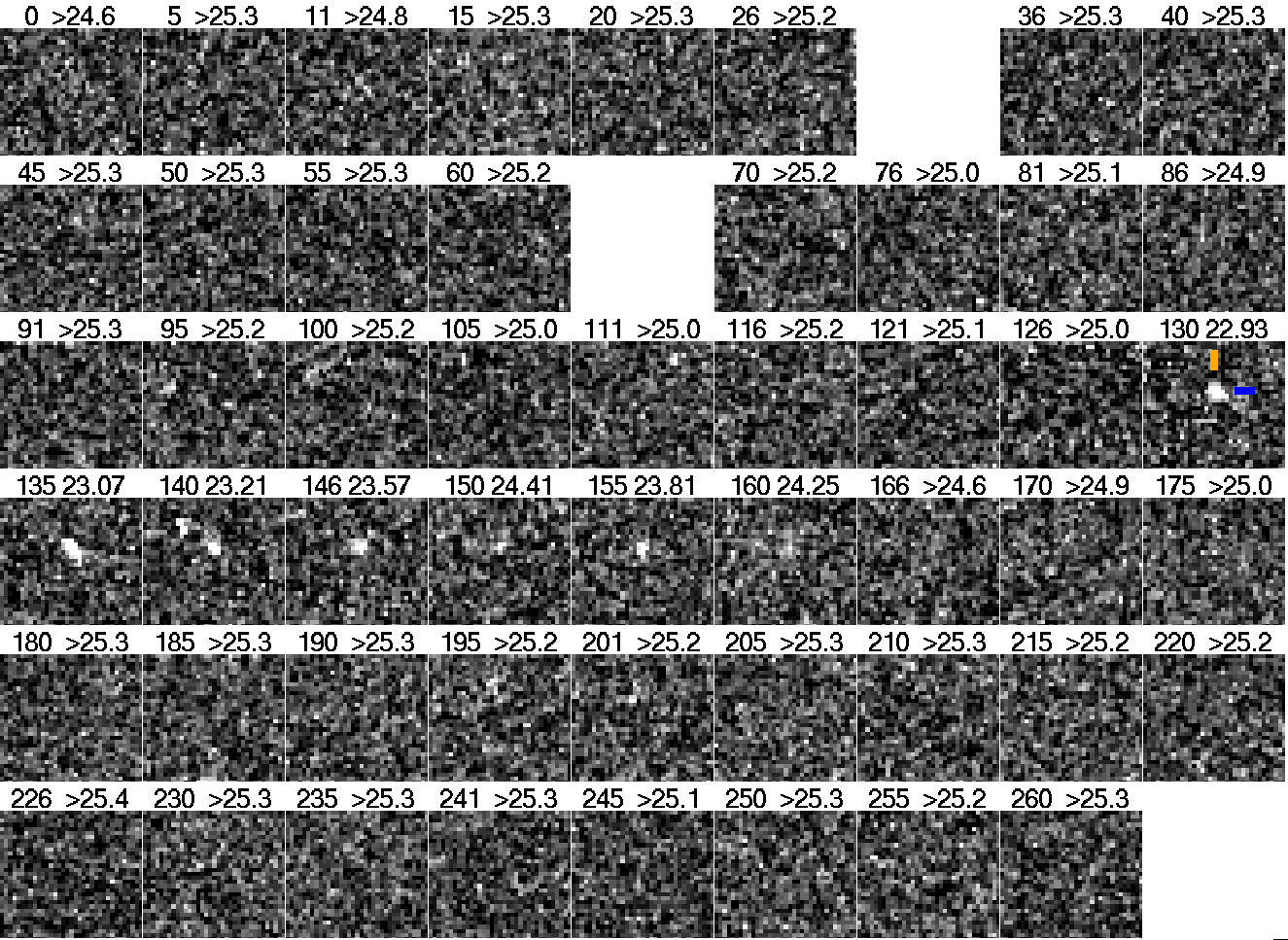}
    \end{tabular}
    \caption{Nova 40}
\end{sidewaysfigure}
\begin{sidewaysfigure}
    \begin{tabular}{cc}
     \multirow{2}{*}[2.6in]{\includegraphics[height=6in]{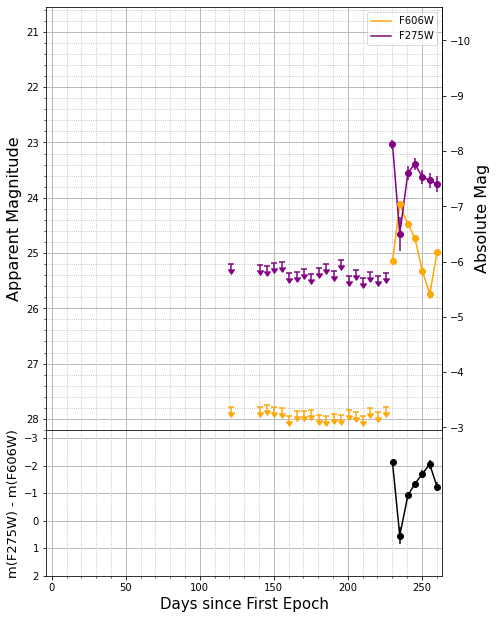}} & \includegraphics[height=3.6in]{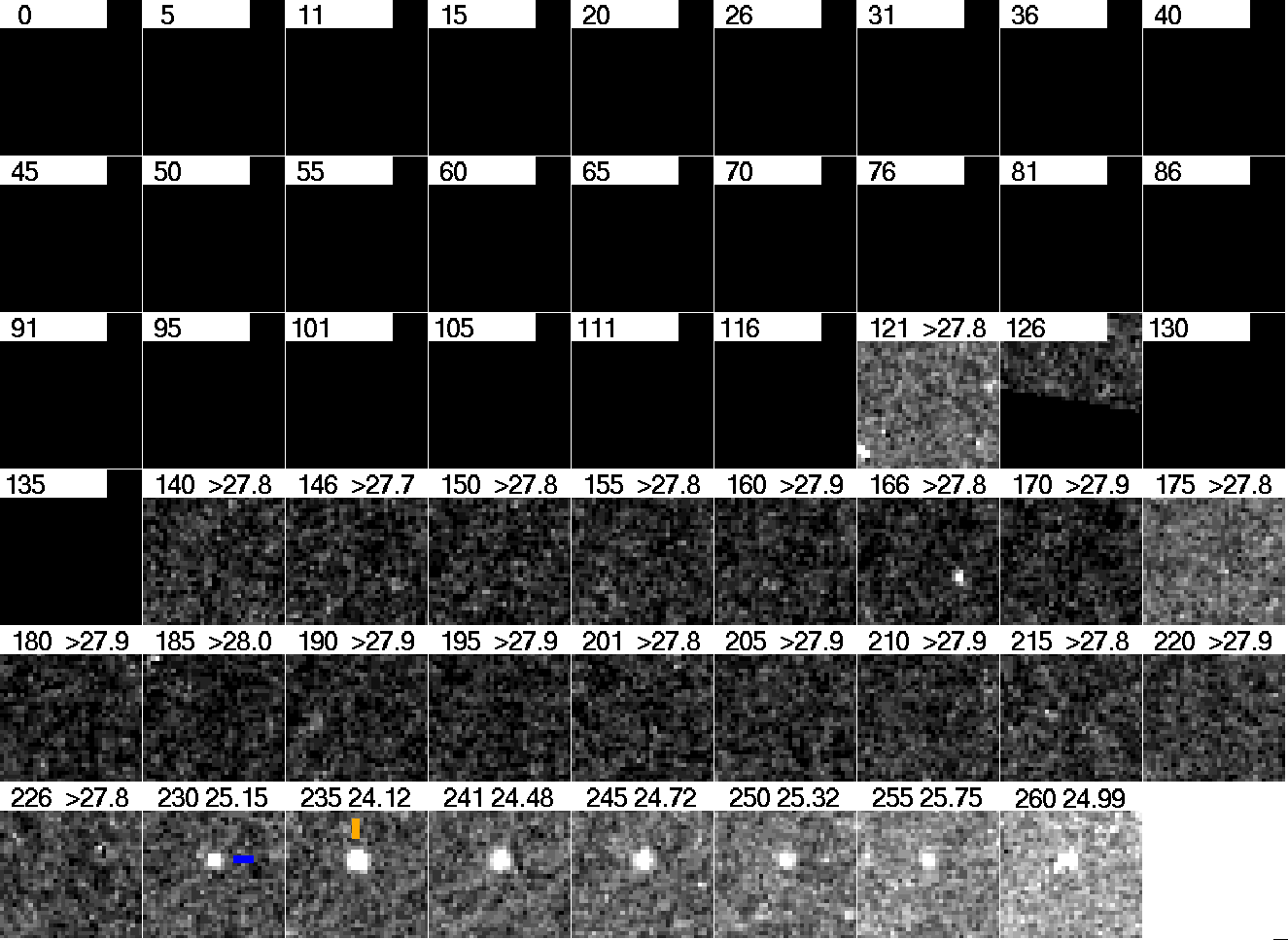} \\
     & \includegraphics[height=3.6in]{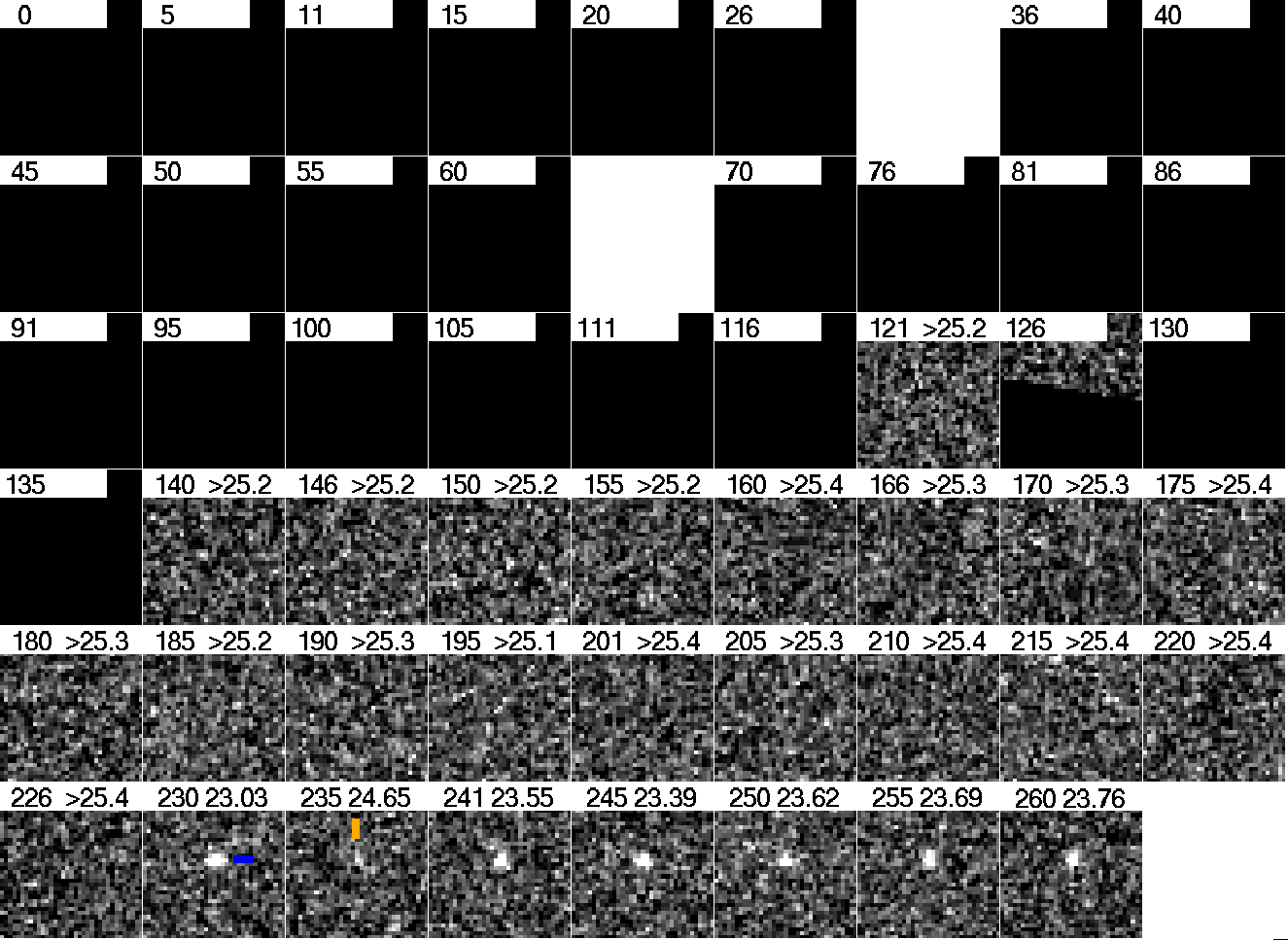}
    \end{tabular}
    \caption{Nova 41}
\end{sidewaysfigure}
\begin{sidewaysfigure}
    \begin{tabular}{cc}
     \multirow{2}{*}[2.6in]{\includegraphics[height=6in]{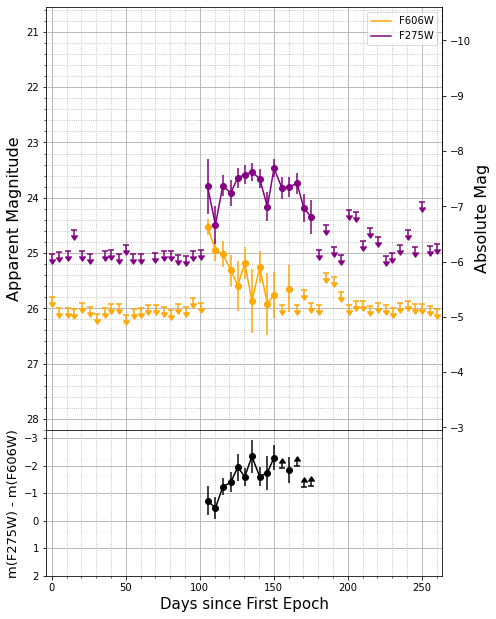}} & \includegraphics[height=3.6in]{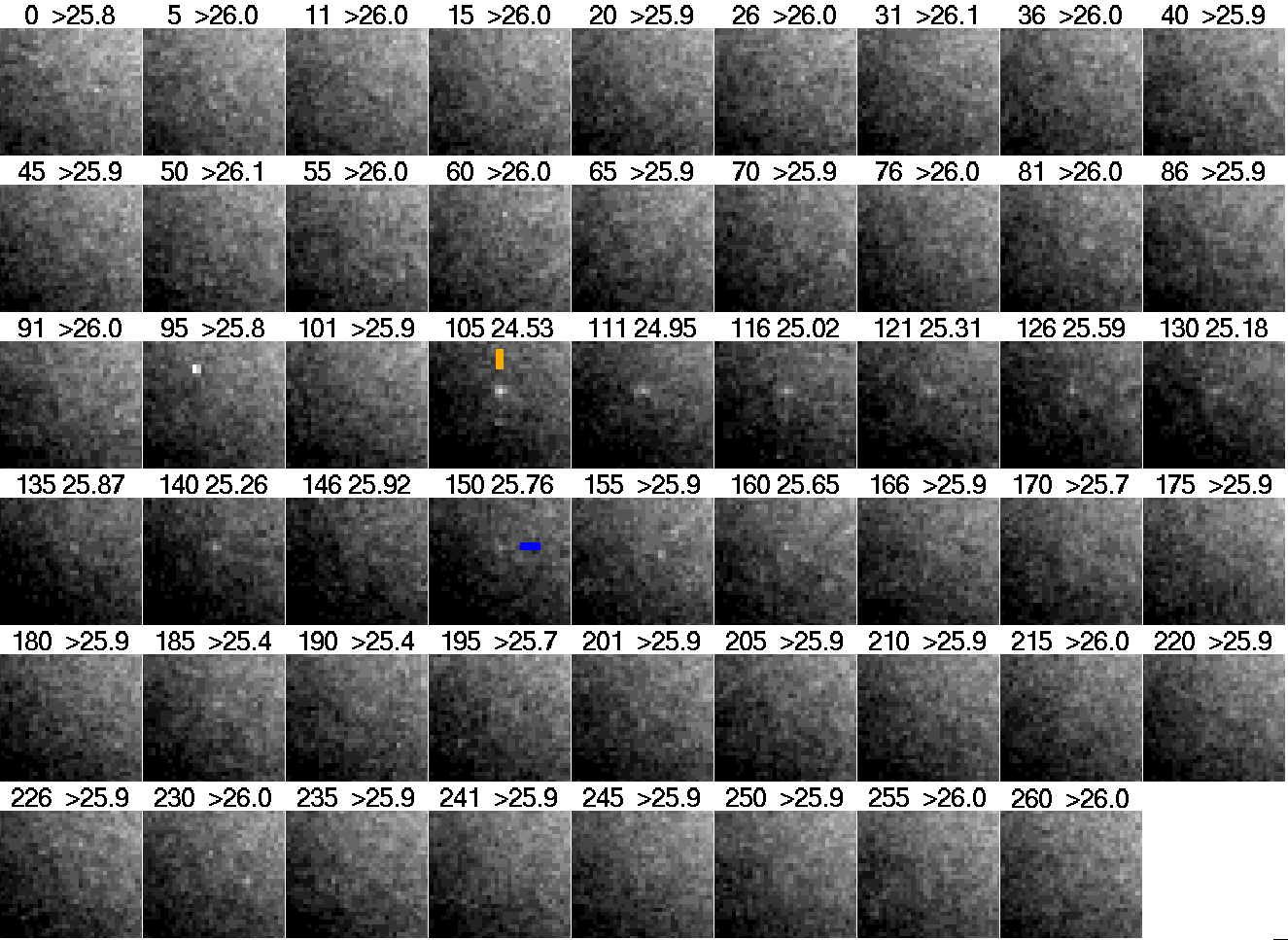} \\
     & \includegraphics[height=3.6in]{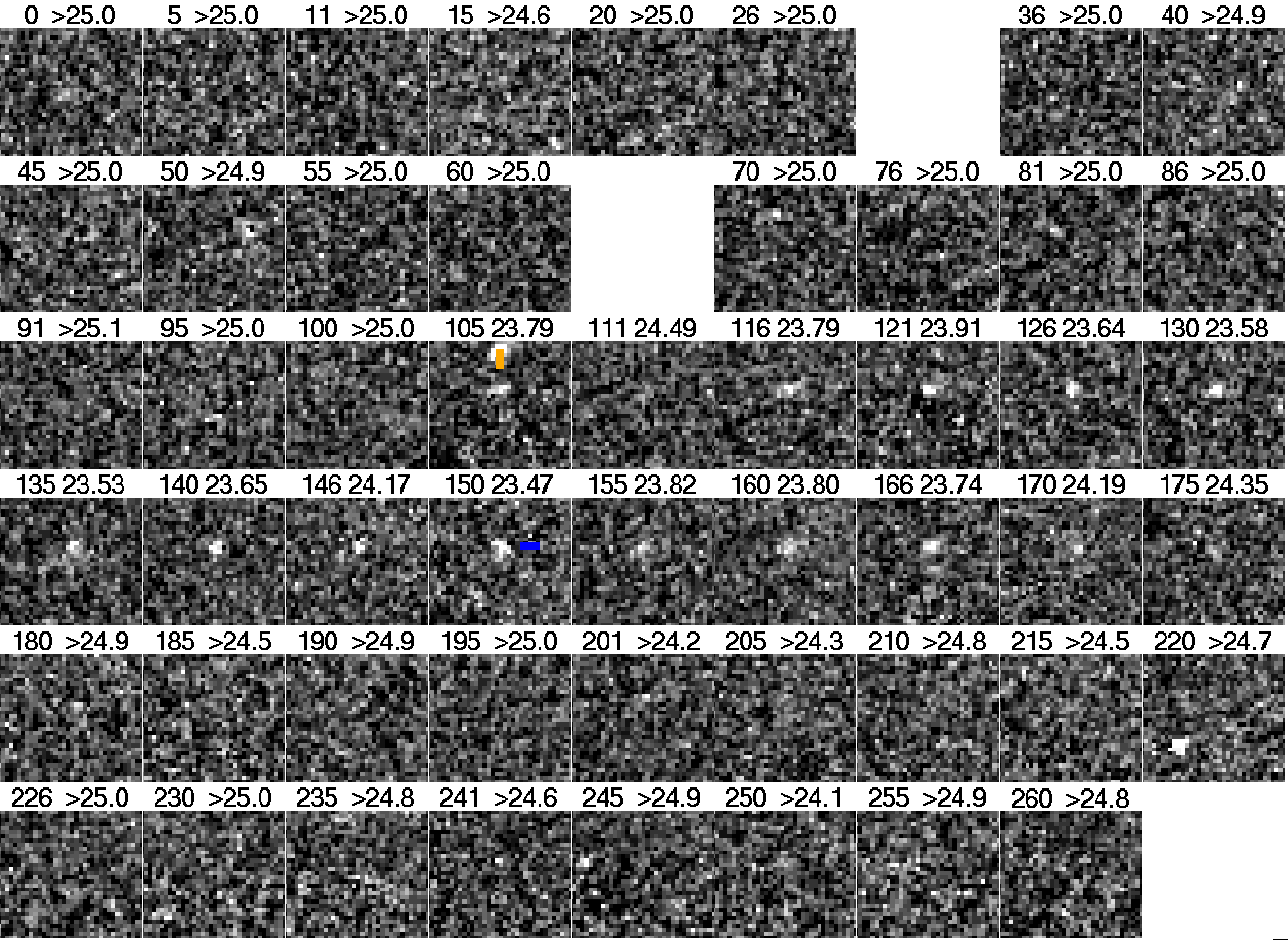}
    \end{tabular}
    \caption{Nova 42}
\end{sidewaysfigure}
\begin{sidewaysfigure}
    \begin{tabular}{cc}
     \multirow{2}{*}[2.6in]{\includegraphics[height=6in]{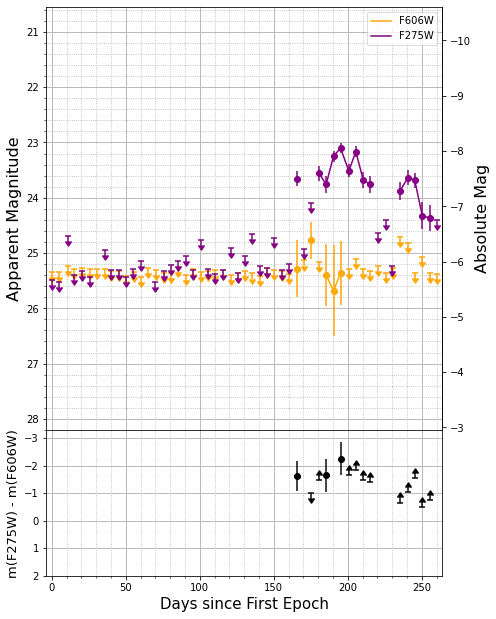}} & \includegraphics[height=3.6in]{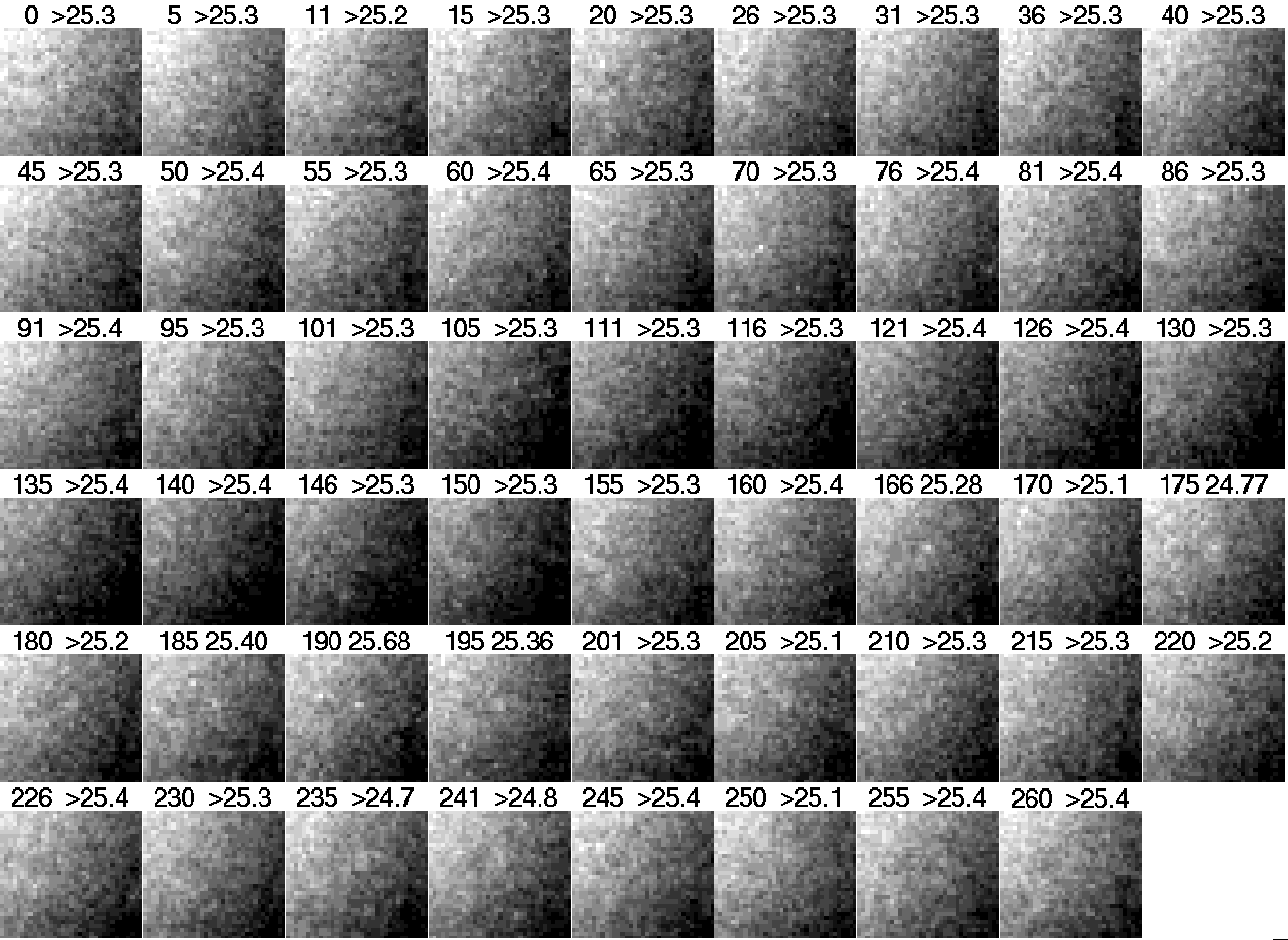} \\
     & \includegraphics[height=3.6in]{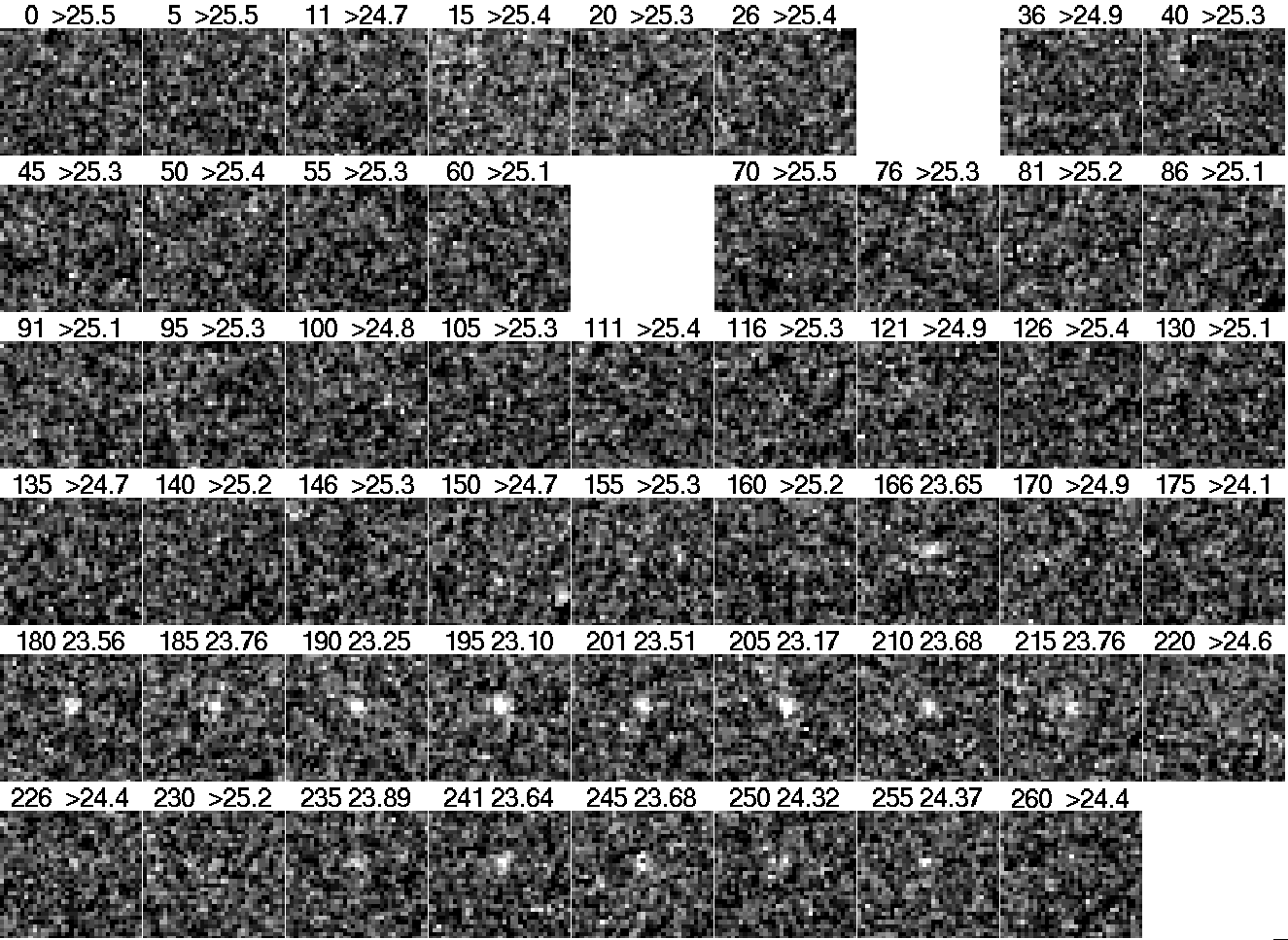}
    \end{tabular}
    \caption{Nova 43}
\end{sidewaysfigure}
\begin{sidewaysfigure}
    \begin{tabular}{cc}
     \multirow{2}{*}[2.6in]{\includegraphics[height=6in]{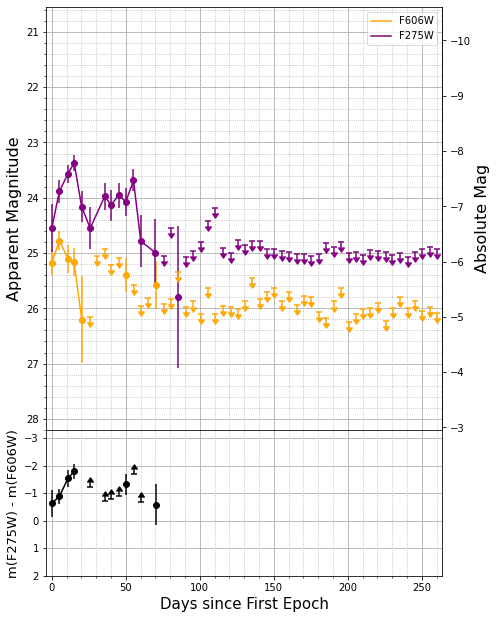}} & \includegraphics[height=3.6in]{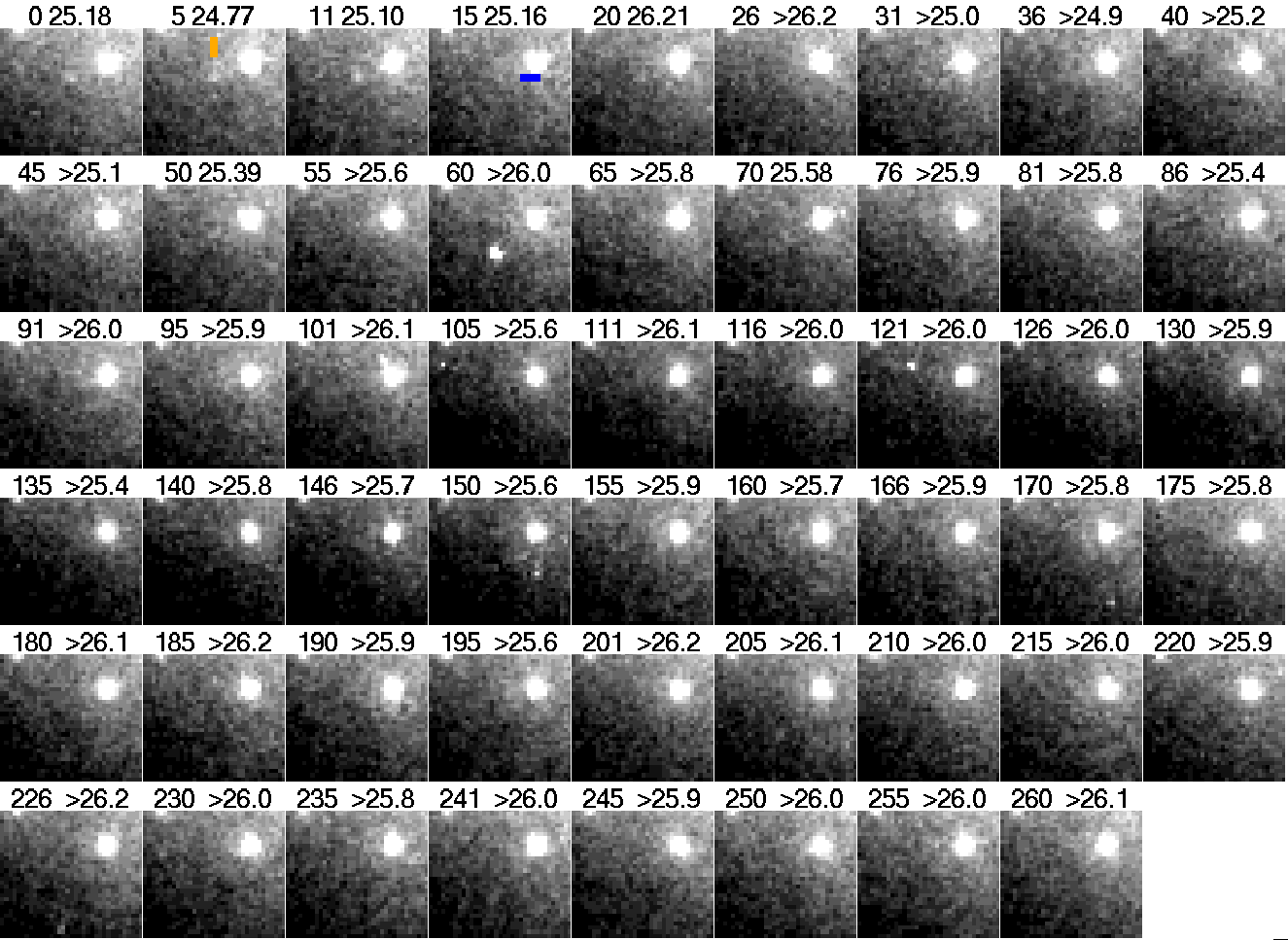} \\
     & \includegraphics[height=3.6in]{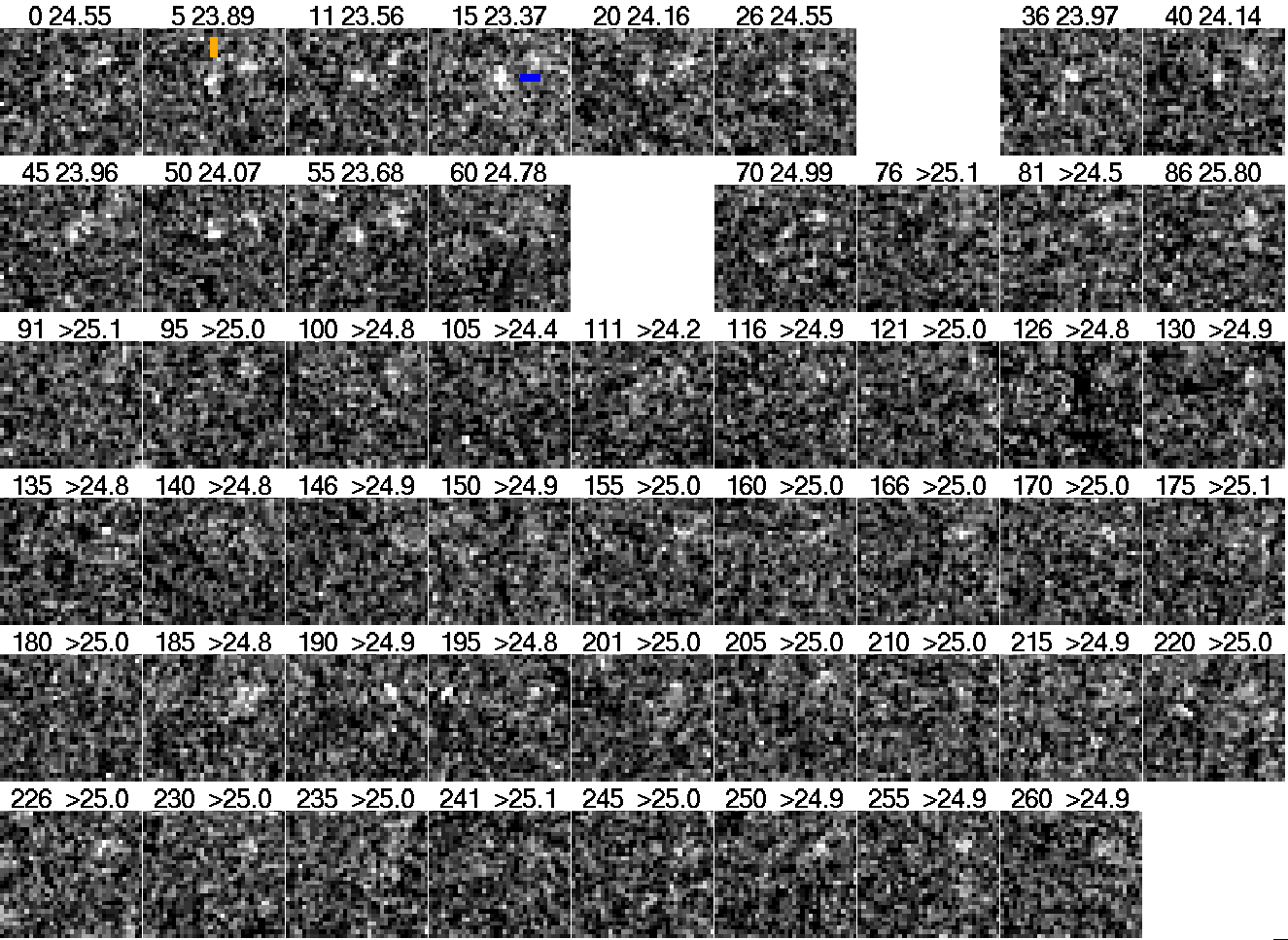}
    \end{tabular}
    \caption{Nova 44}
\end{sidewaysfigure}
\begin{sidewaysfigure}
    \begin{tabular}{cc}
     \multirow{2}{*}[2.6in]{\includegraphics[height=6in]{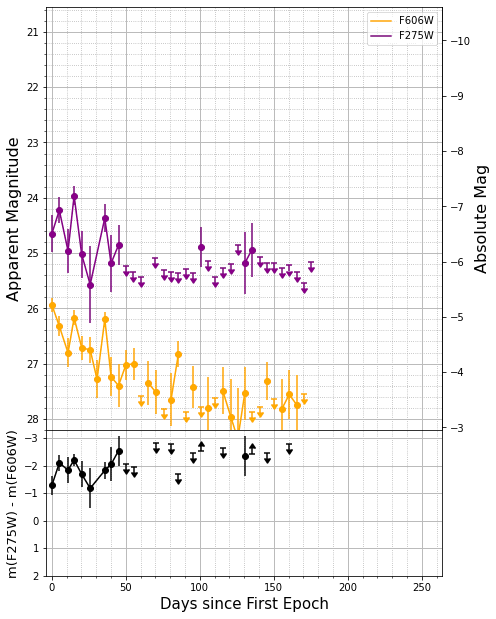}} & \includegraphics[height=3.6in]{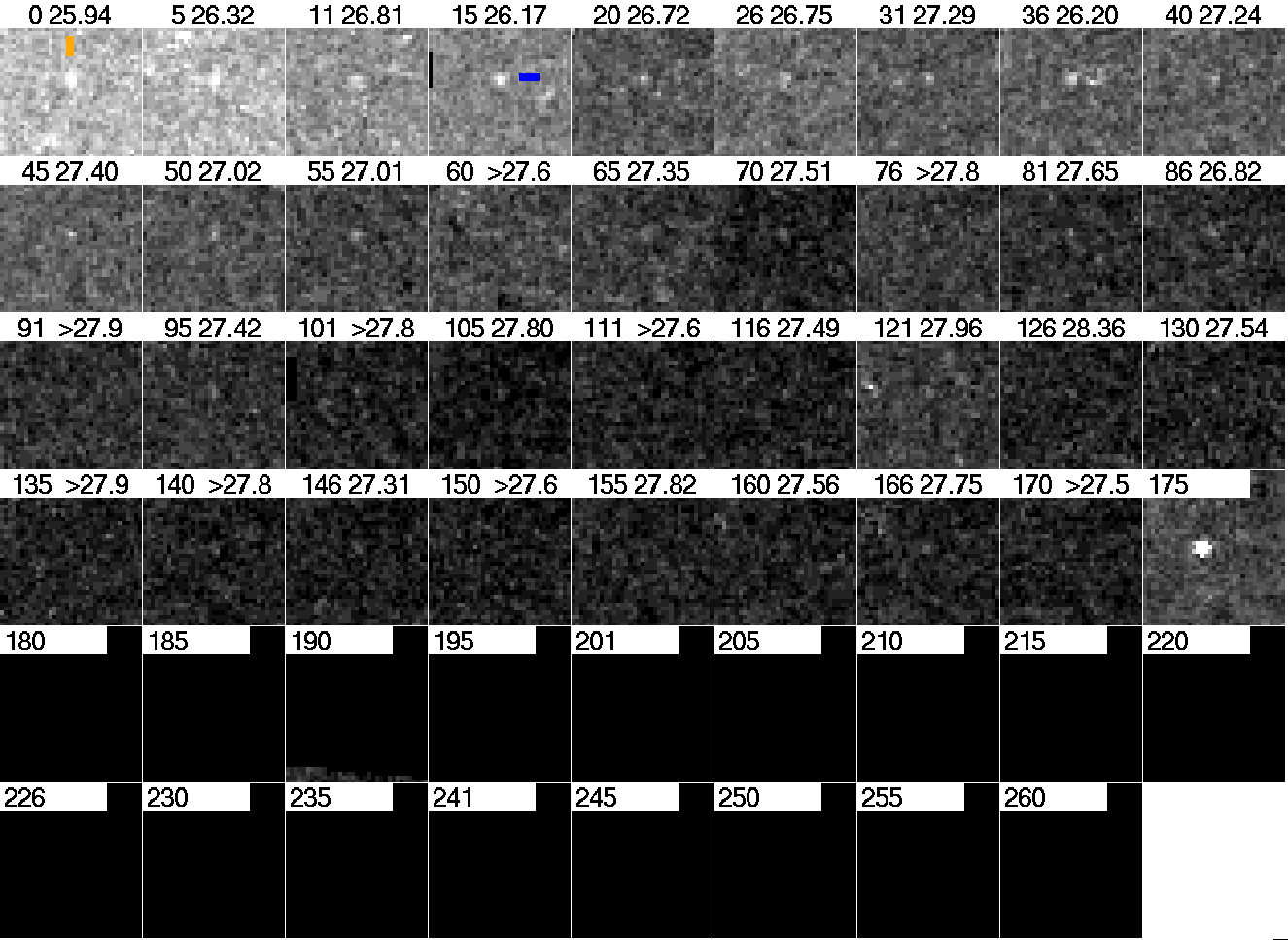} \\
     & \includegraphics[height=3.6in]{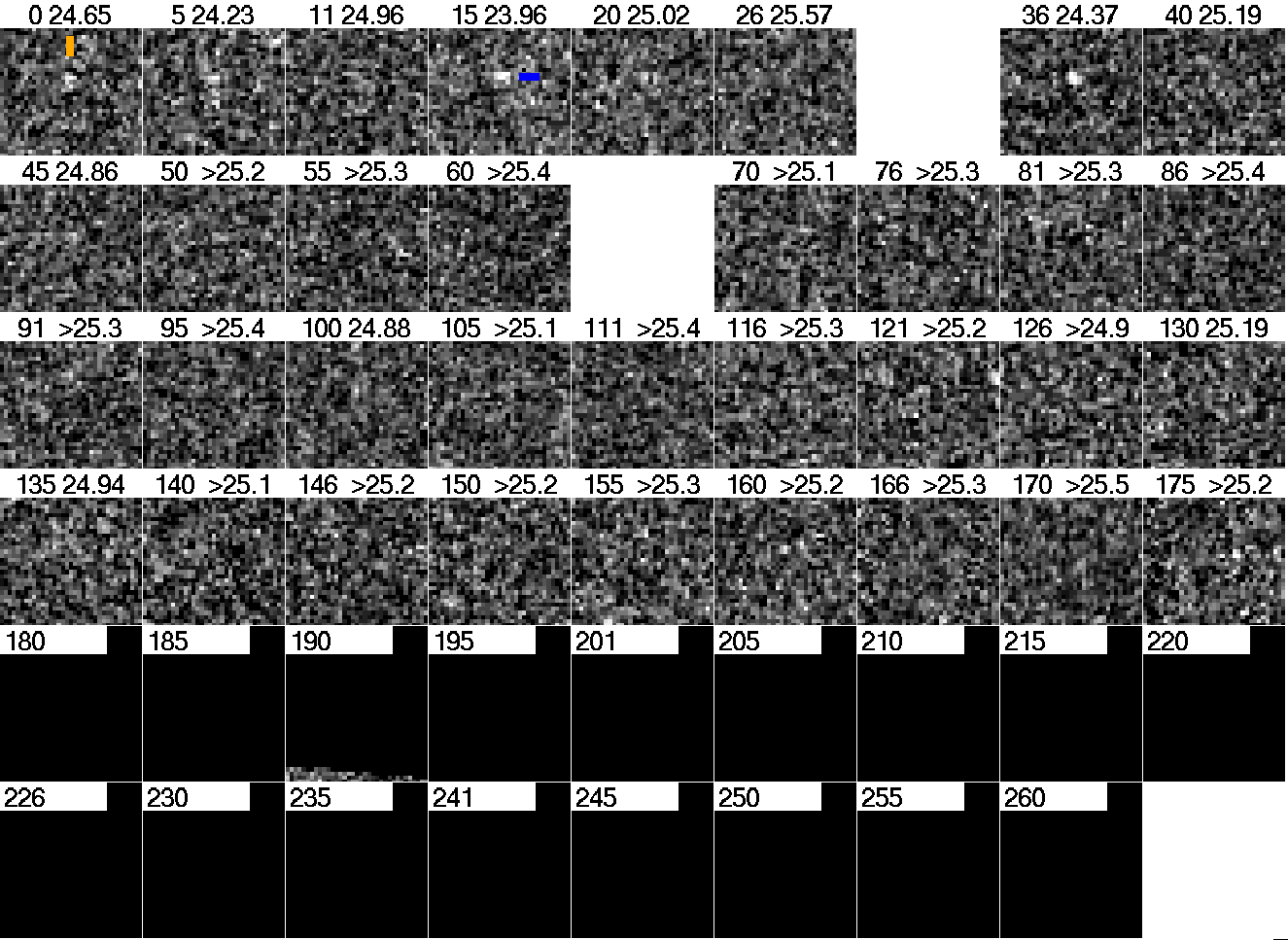}
    \end{tabular}
    \caption{Nova 45}
\end{sidewaysfigure}
\begin{sidewaysfigure}
    \begin{tabular}{cc}
     \multirow{2}{*}[2.6in]{\includegraphics[height=6in]{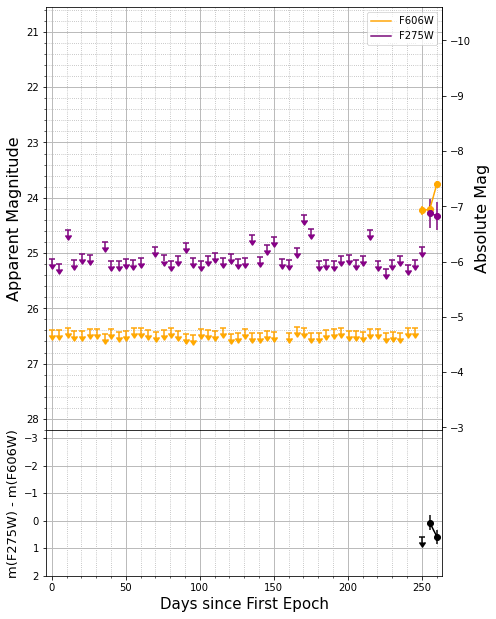}} & \includegraphics[height=3.6in]{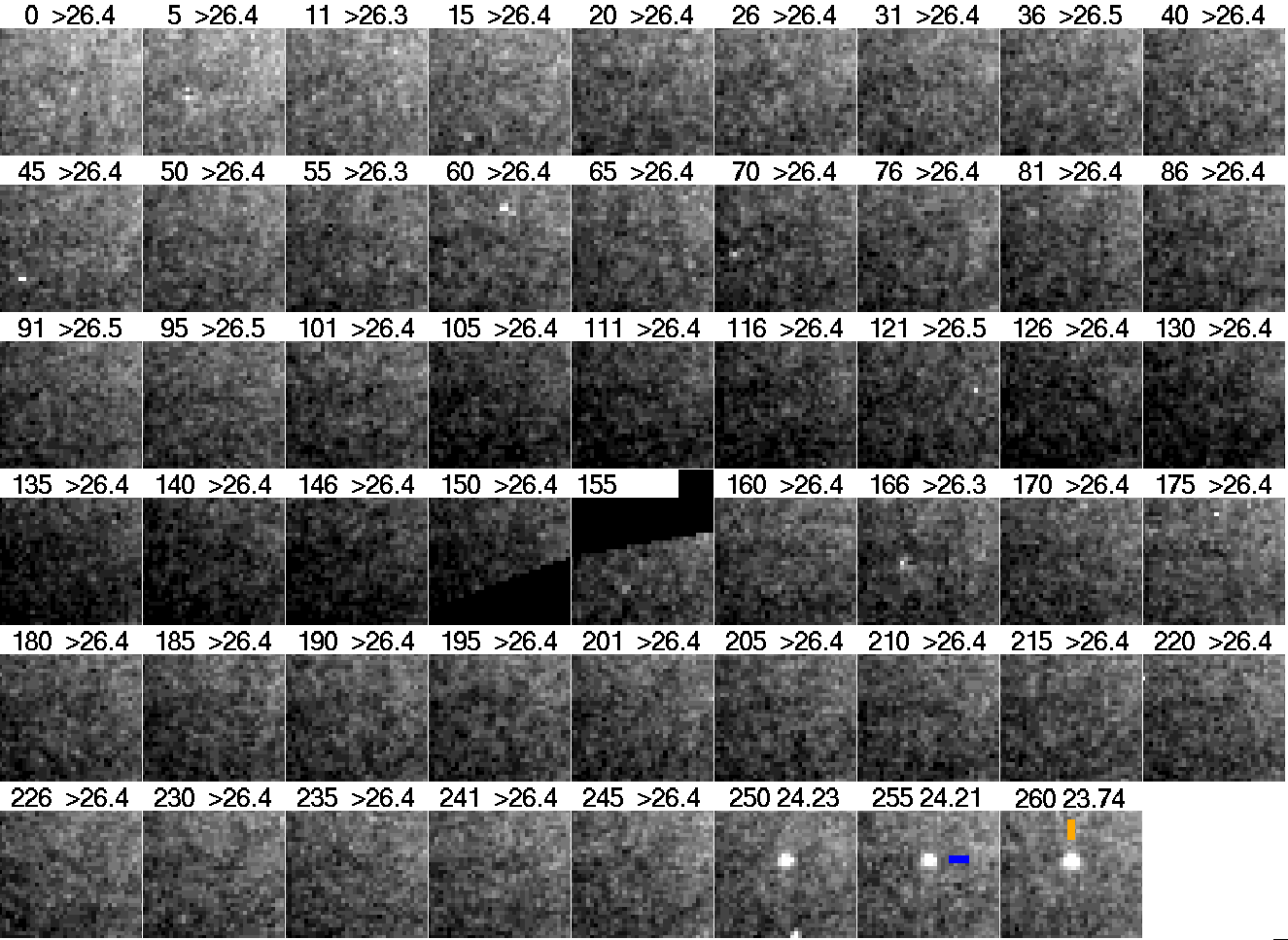} \\
     & \includegraphics[height=3.6in]{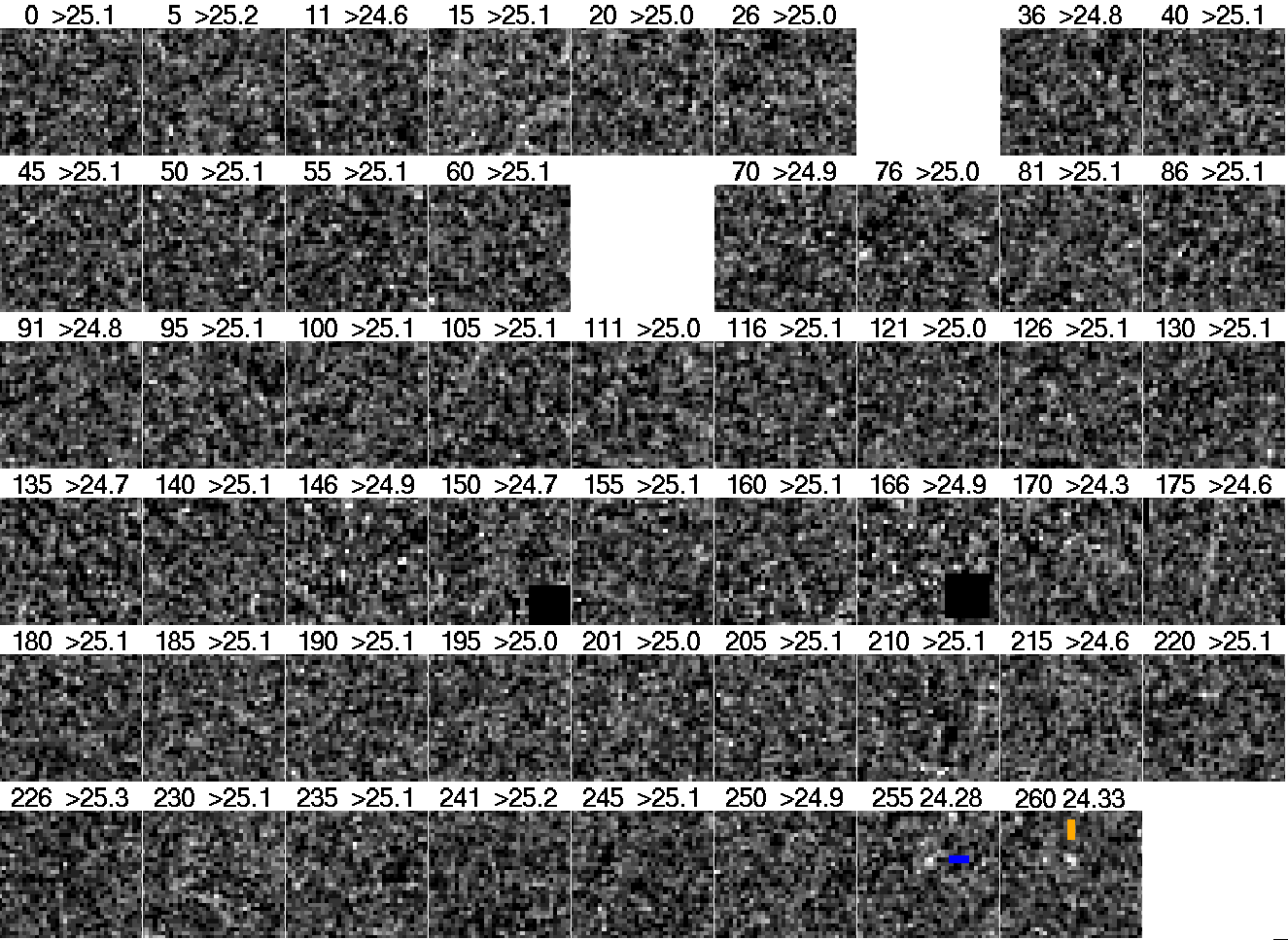}
    \end{tabular}
    \caption{Nova 46}
\end{sidewaysfigure}
\begin{sidewaysfigure}
    \begin{tabular}{cc}
     \multirow{2}{*}[2.6in]{\includegraphics[height=6in]{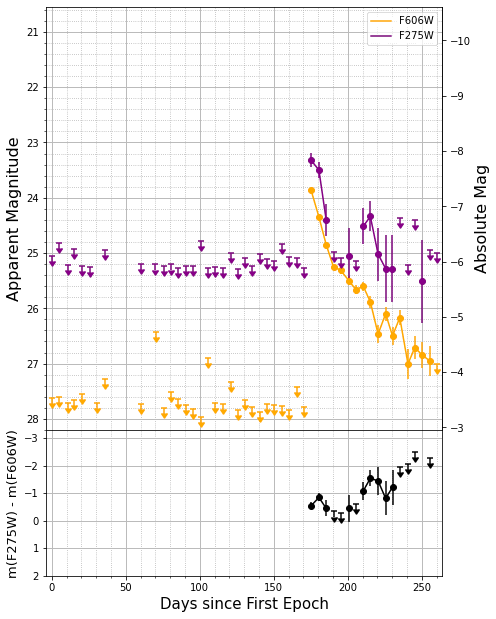}} & \includegraphics[height=3.6in]{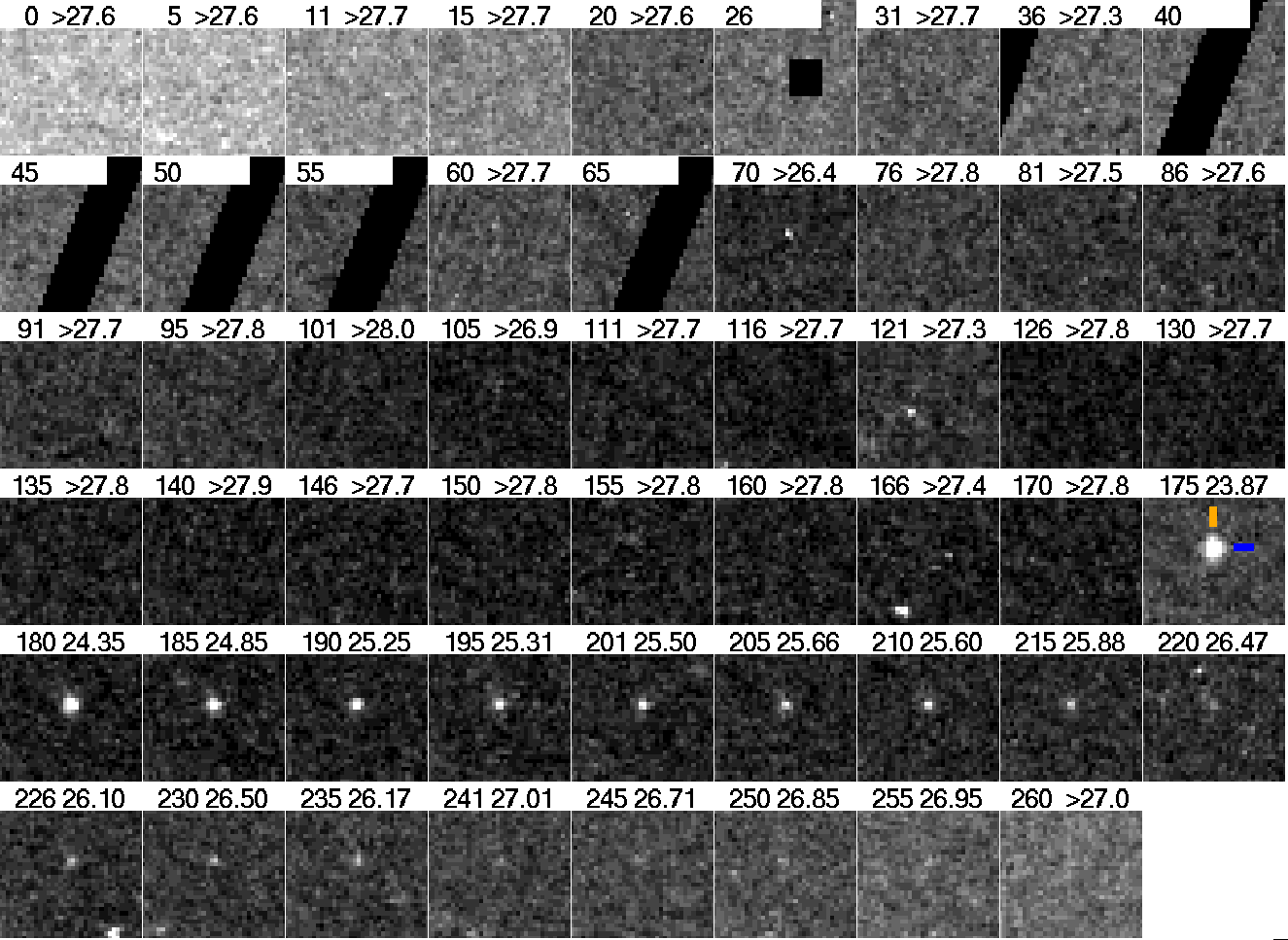} \\
     & \includegraphics[height=3.6in]{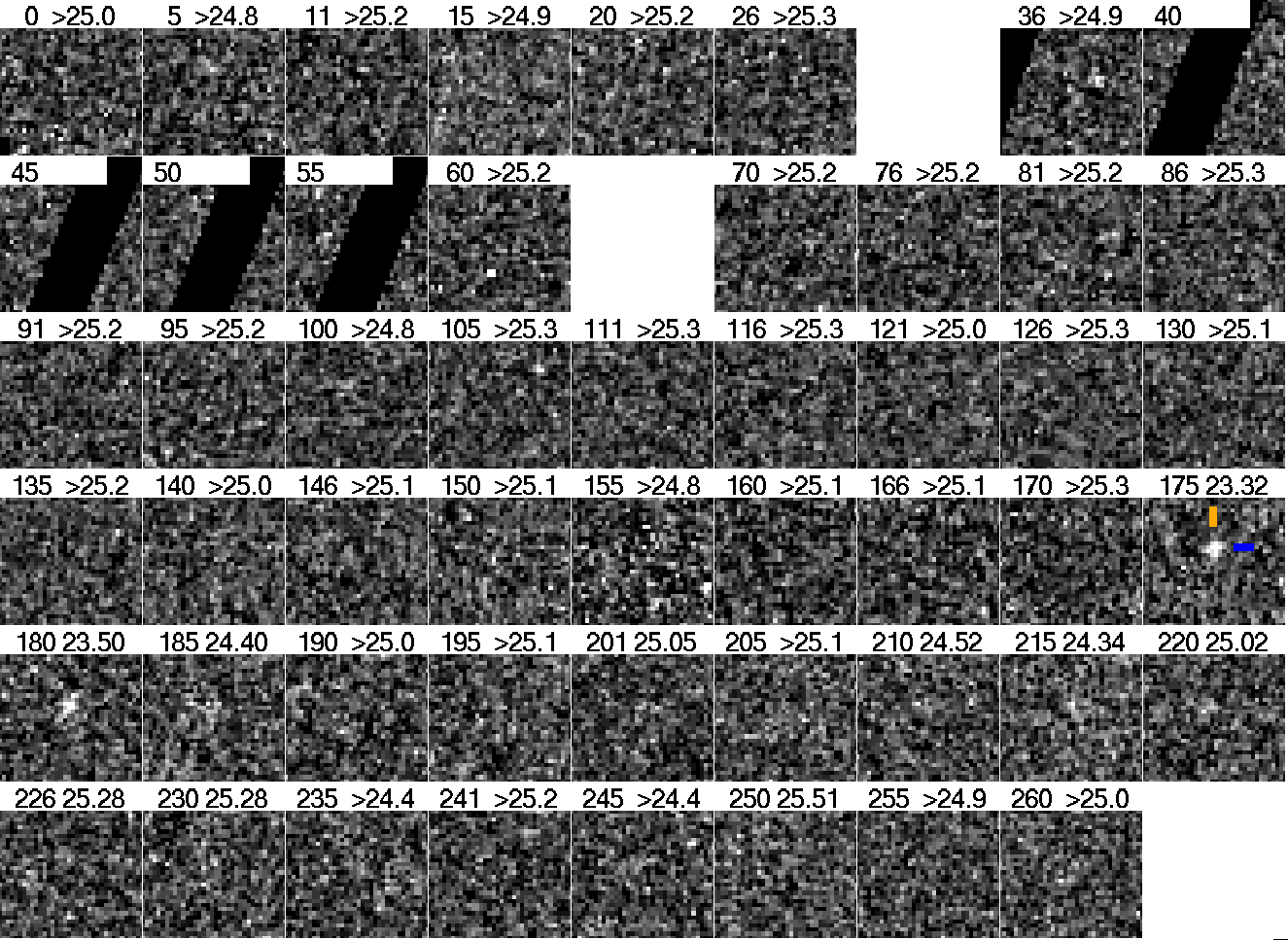}
    \end{tabular}
    \caption{Nova 47}
\end{sidewaysfigure}
\begin{sidewaysfigure}
    \begin{tabular}{cc}
     \multirow{2}{*}[2.6in]{\includegraphics[height=6in]{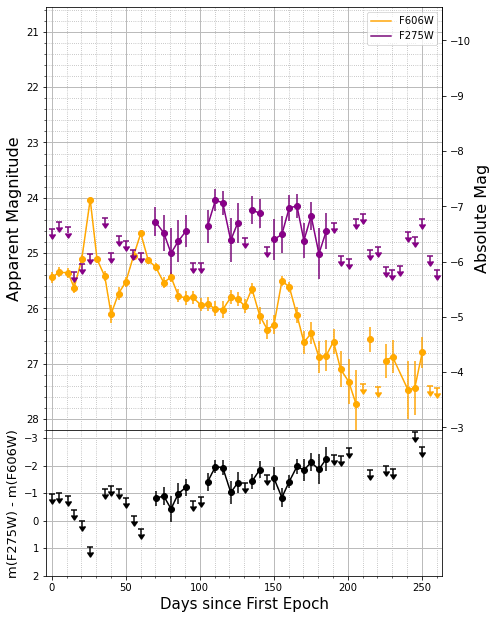}} & \includegraphics[height=3.6in]{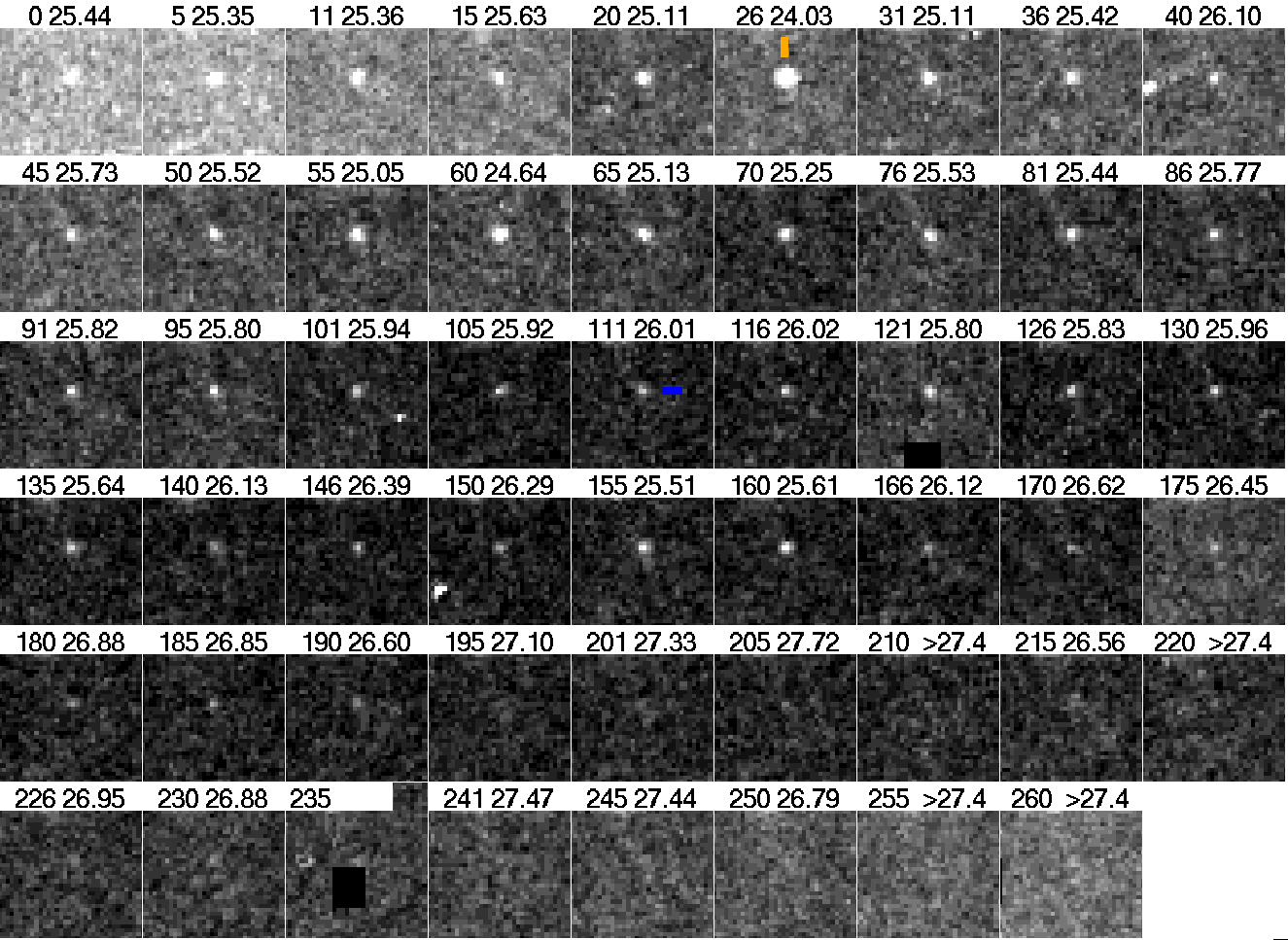} \\
     & \includegraphics[height=3.6in]{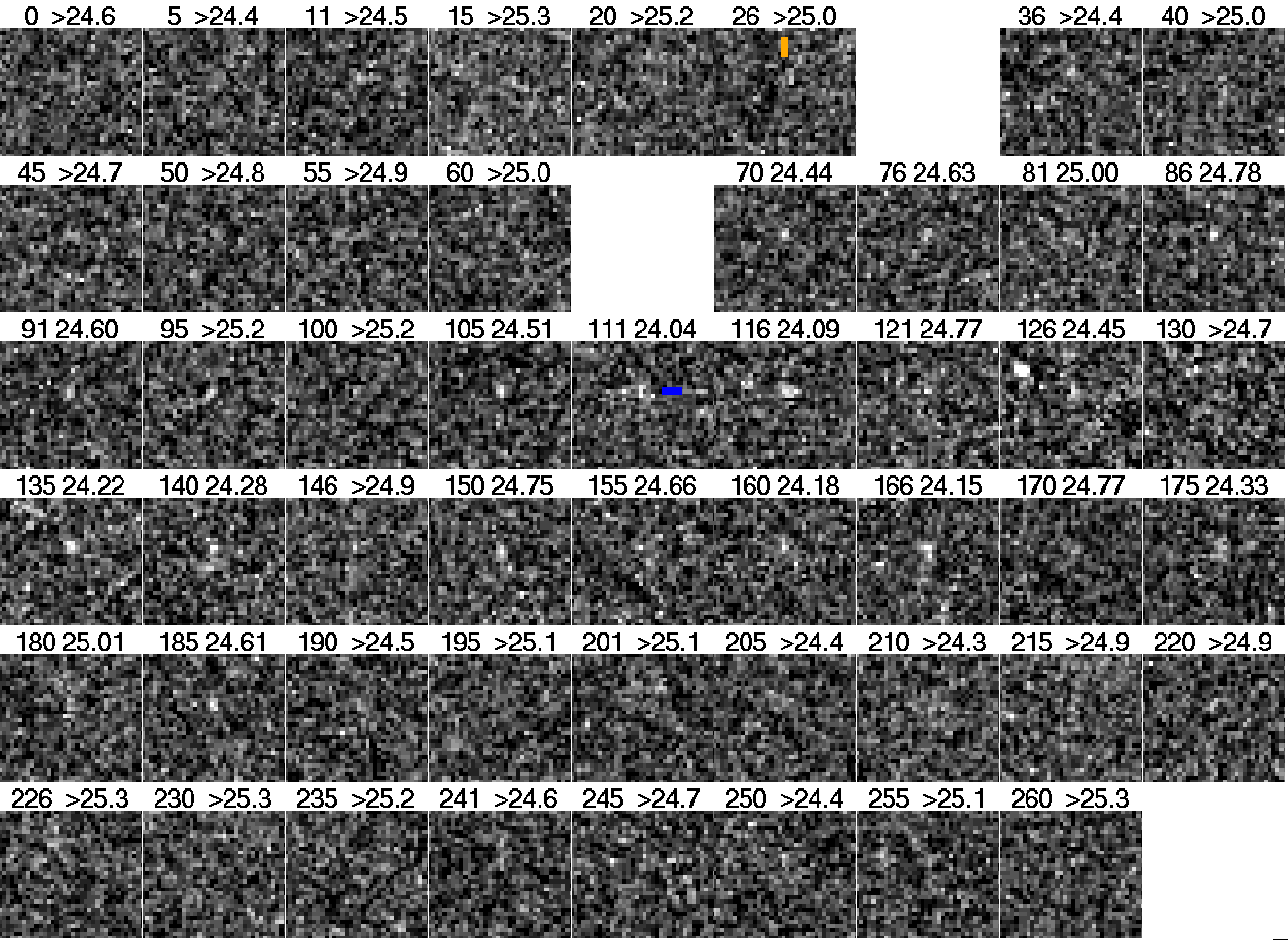}
    \end{tabular}
    \caption{Nova 48}
\end{sidewaysfigure}
\begin{sidewaysfigure}
    \begin{tabular}{cc}
     \multirow{2}{*}[2.6in]{\includegraphics[height=6in]{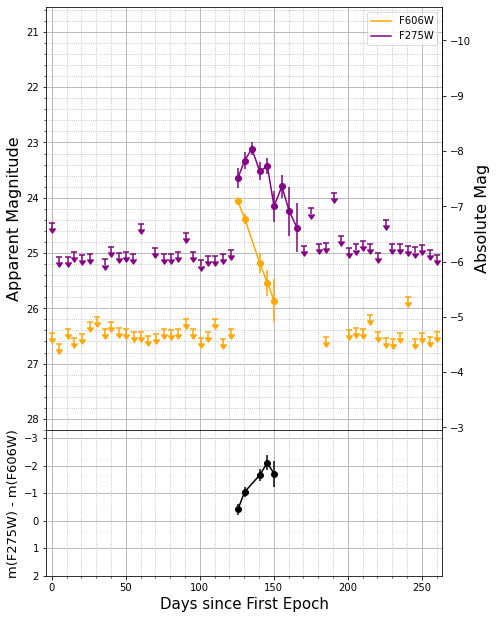}} & \includegraphics[height=3.6in]{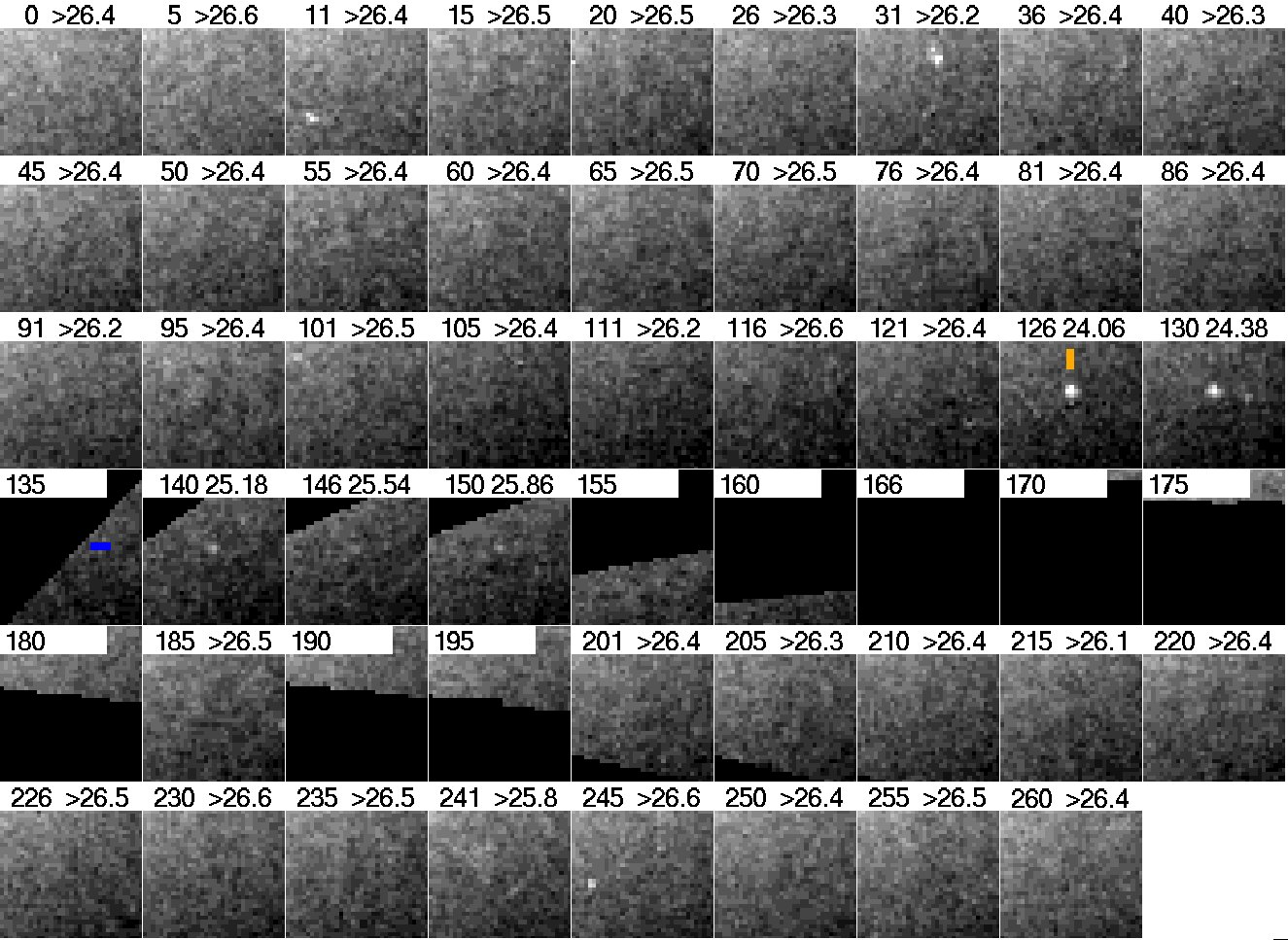} \\
     & \includegraphics[height=3.6in]{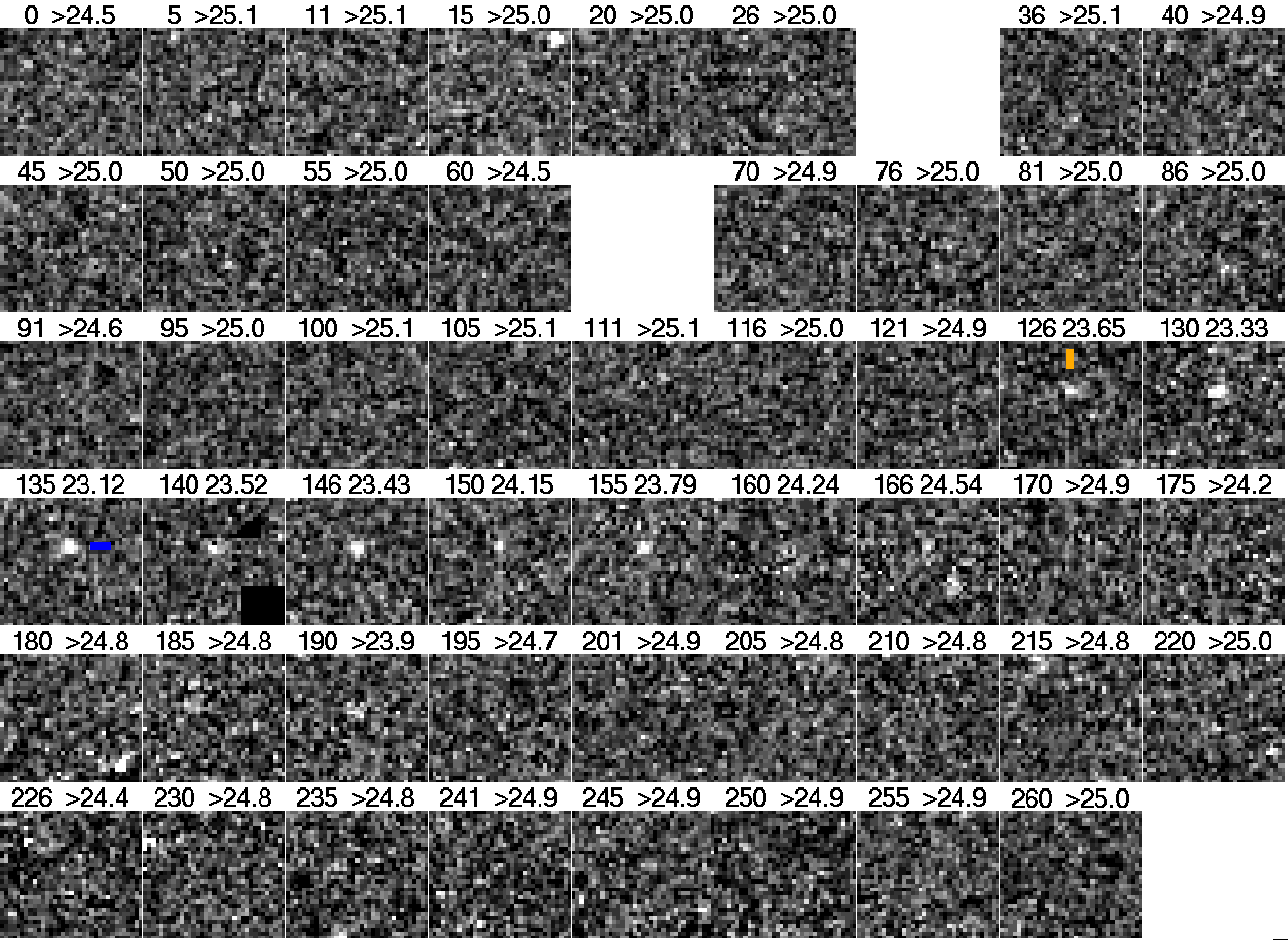}
    \end{tabular}
    \caption{Nova 49}
\end{sidewaysfigure}
\begin{sidewaysfigure}
    \begin{tabular}{cc}
     \multirow{2}{*}[2.6in]{\includegraphics[height=6in]{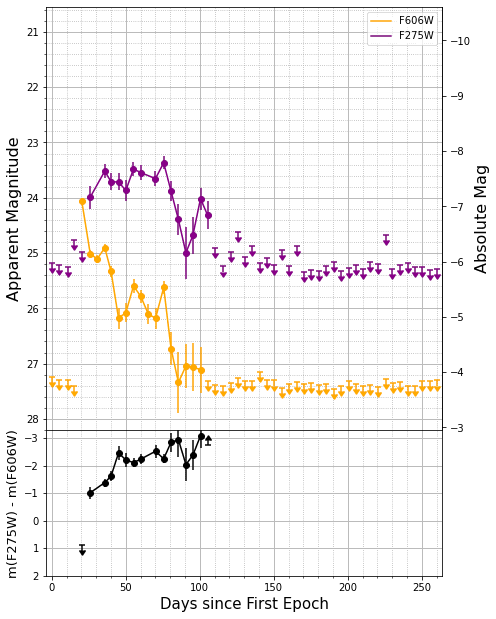}} & \includegraphics[height=3.6in]{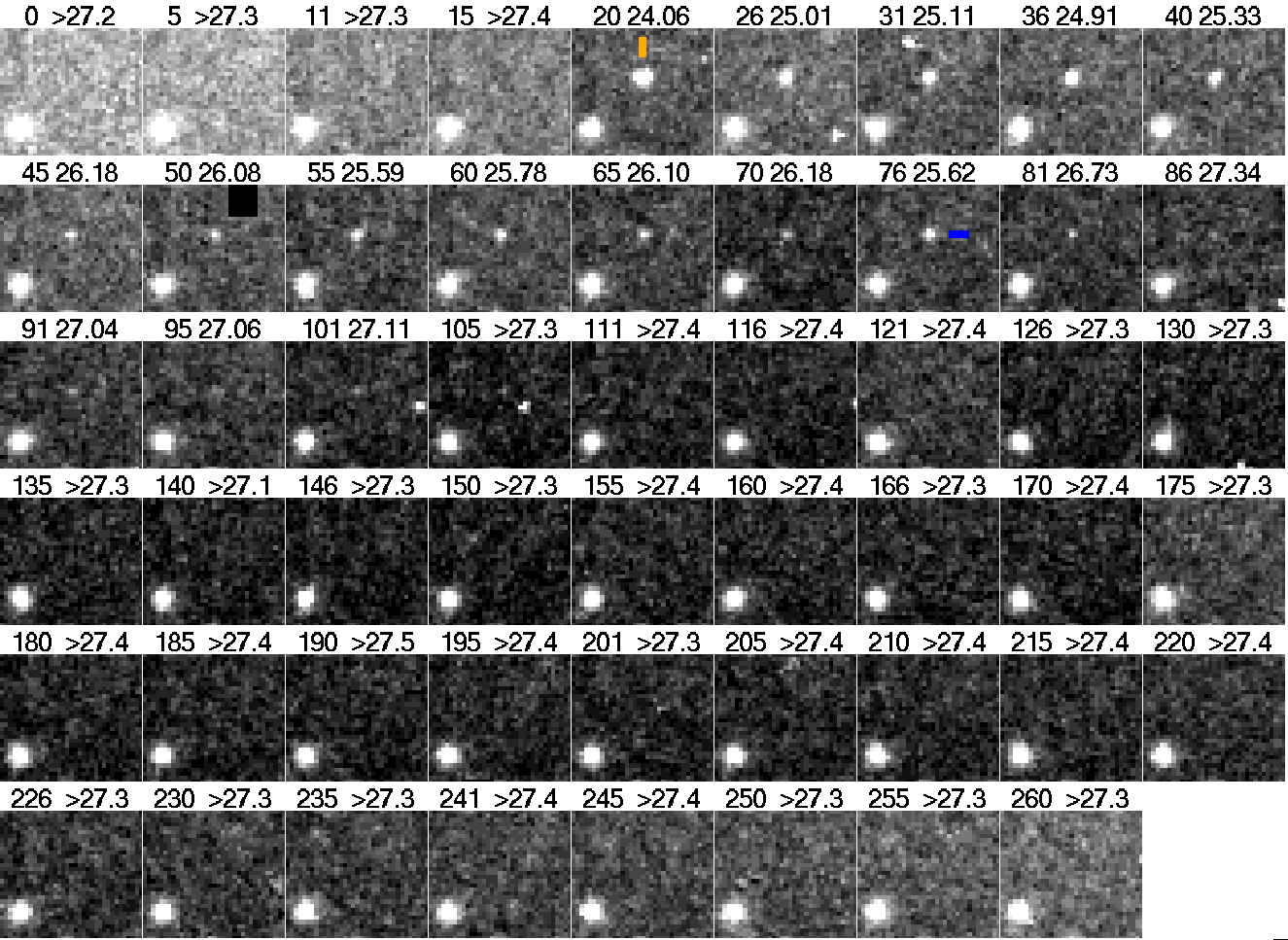} \\
     & \includegraphics[height=3.6in]{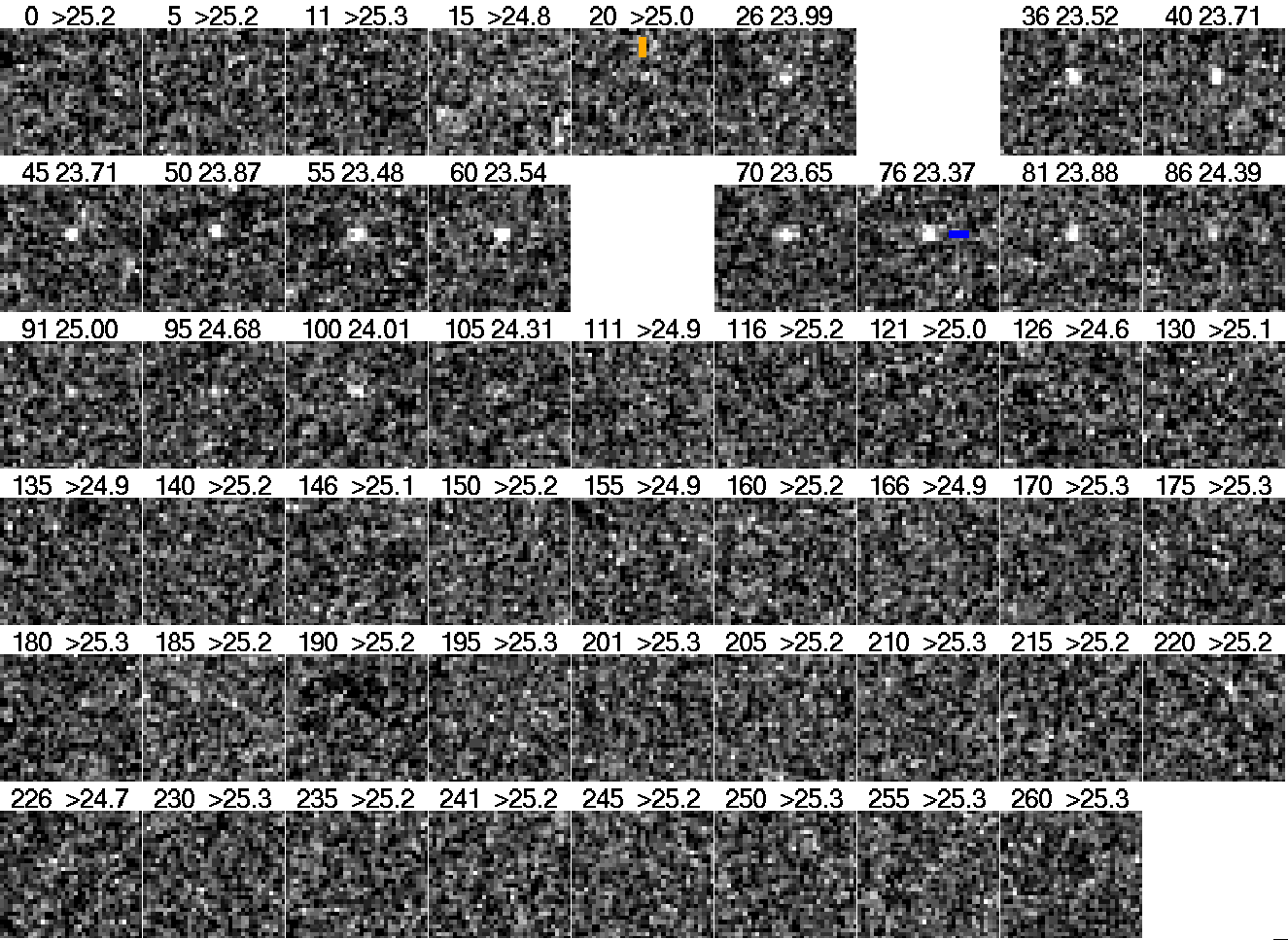}
    \end{tabular}
    \caption{Nova 50}
\end{sidewaysfigure}
\begin{sidewaysfigure}
    \begin{tabular}{cc}
     \multirow{2}{*}[2.6in]{\includegraphics[height=6in]{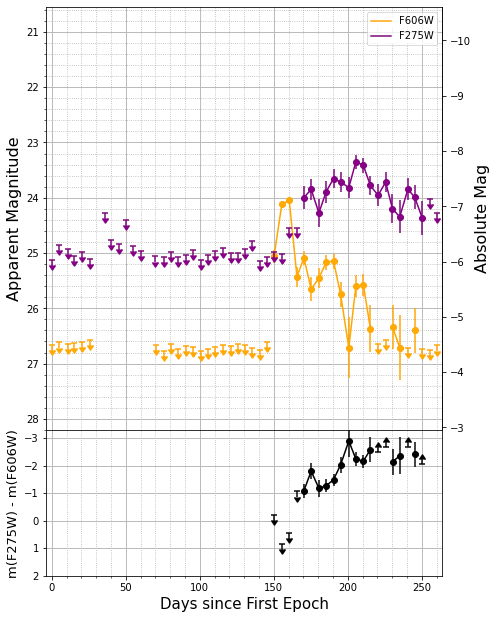}} & \includegraphics[height=3.6in]{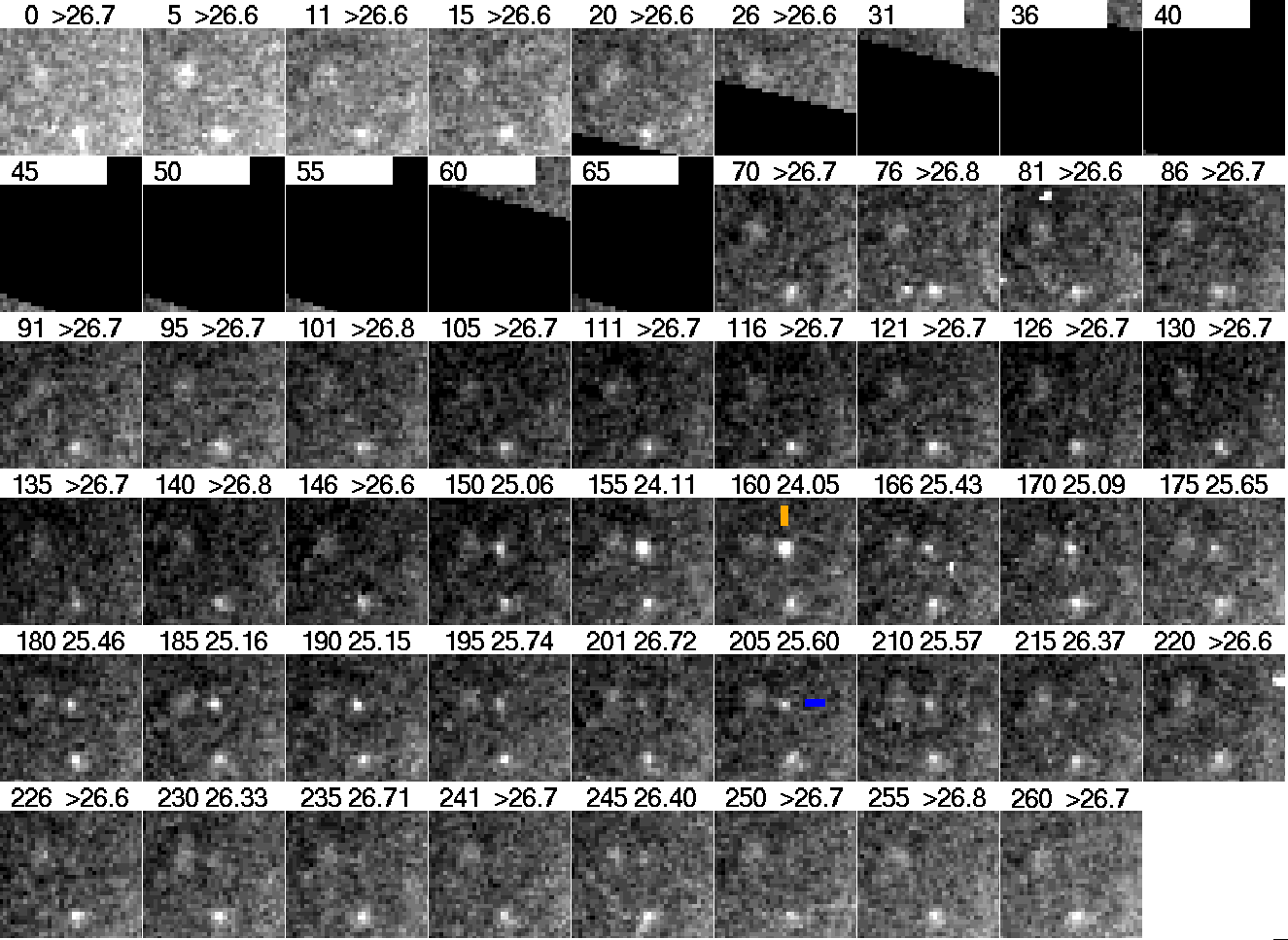} \\
     & \includegraphics[height=3.6in]{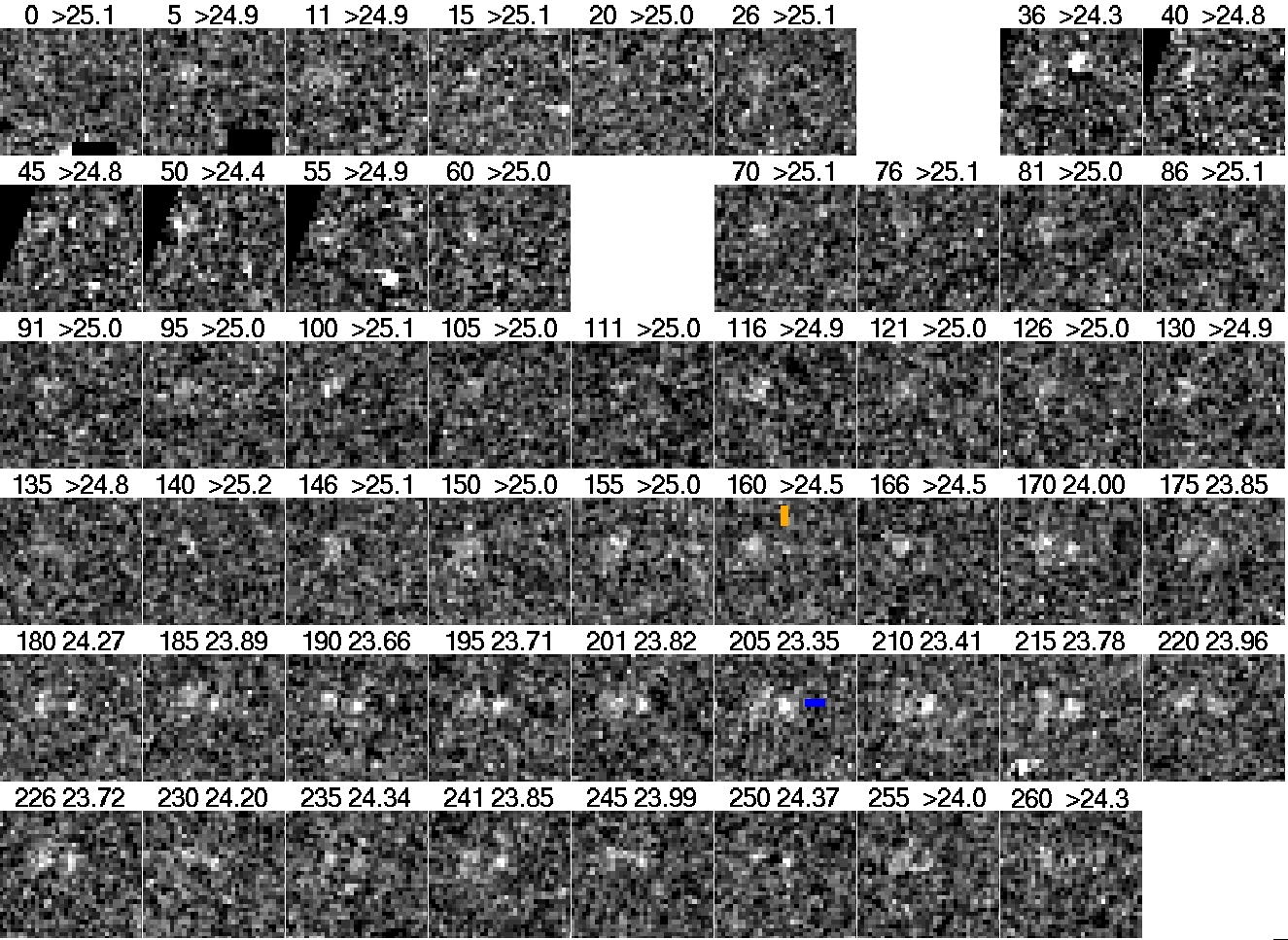}
    \end{tabular}
    \caption{Nova 51}
\end{sidewaysfigure}
\begin{sidewaysfigure}
    \begin{tabular}{cc}
     \multirow{2}{*}[2.6in]{\includegraphics[height=6in]{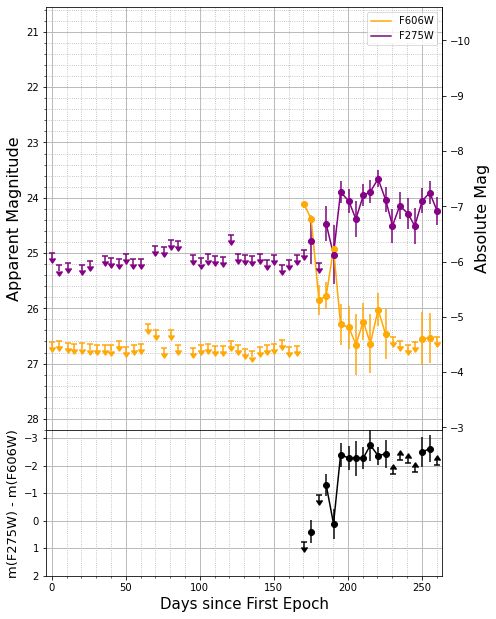}} & \includegraphics[height=3.6in]{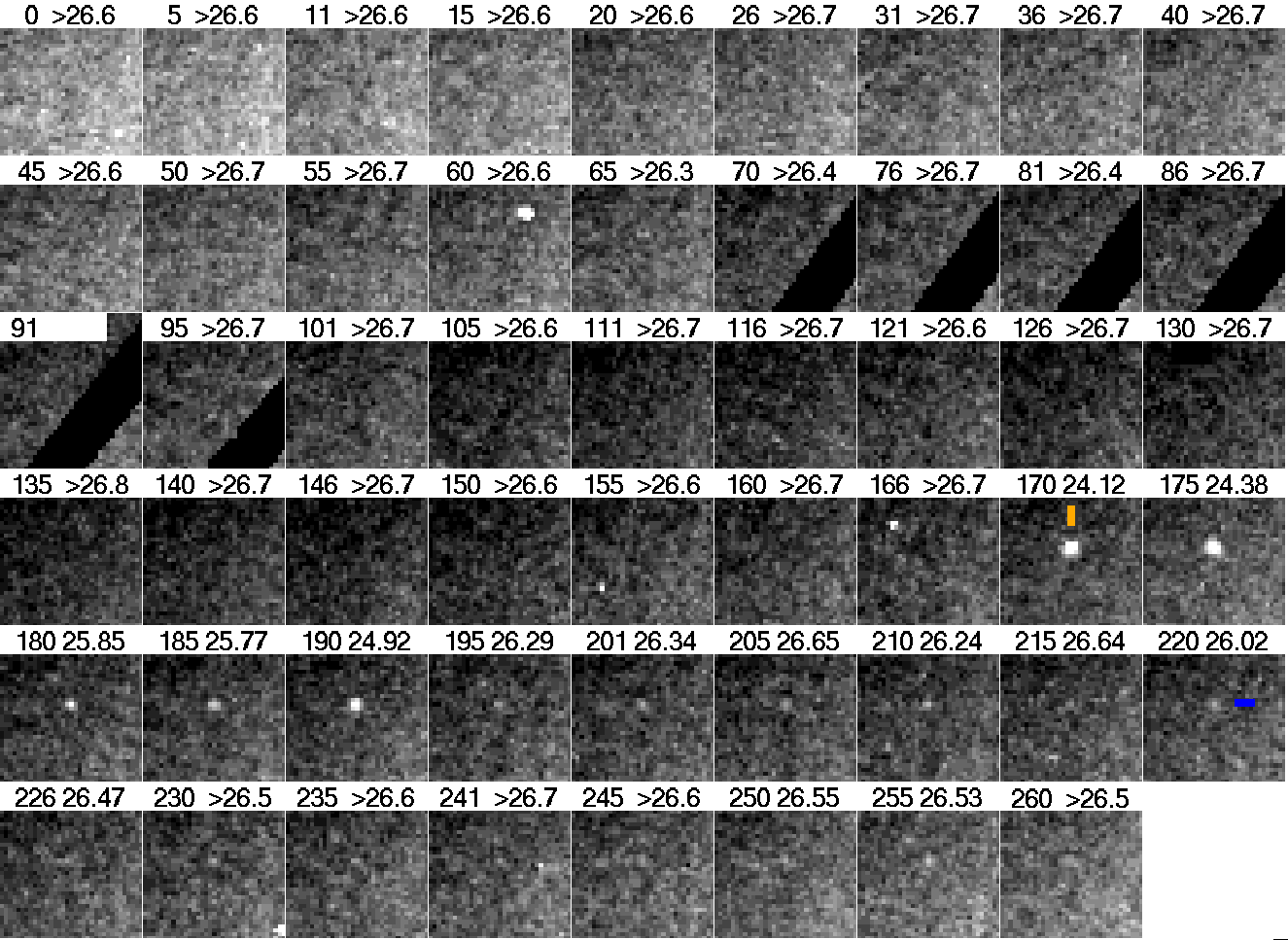} \\
     & \includegraphics[height=3.6in]{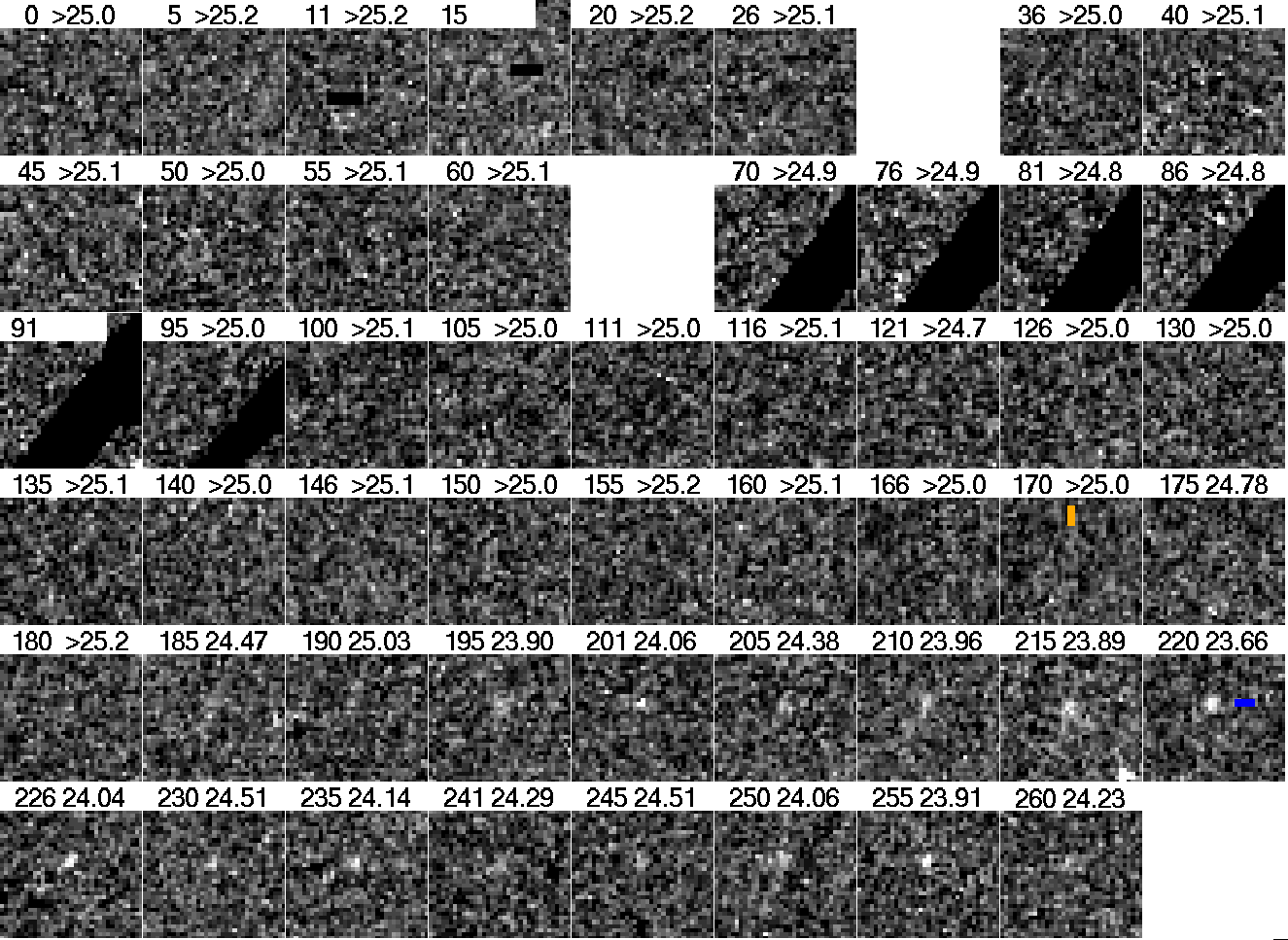}
    \end{tabular}
    \caption{Nova 52}
\end{sidewaysfigure}
\begin{sidewaysfigure}
    \begin{tabular}{cc}
     \multirow{2}{*}[2.6in]{\includegraphics[height=6in]{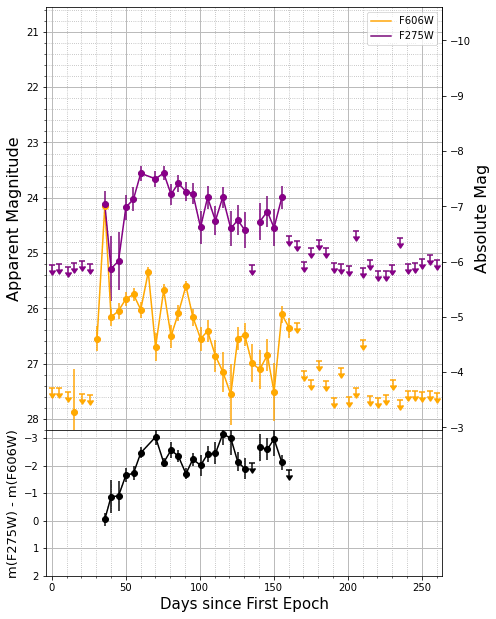}} & \includegraphics[height=3.6in]{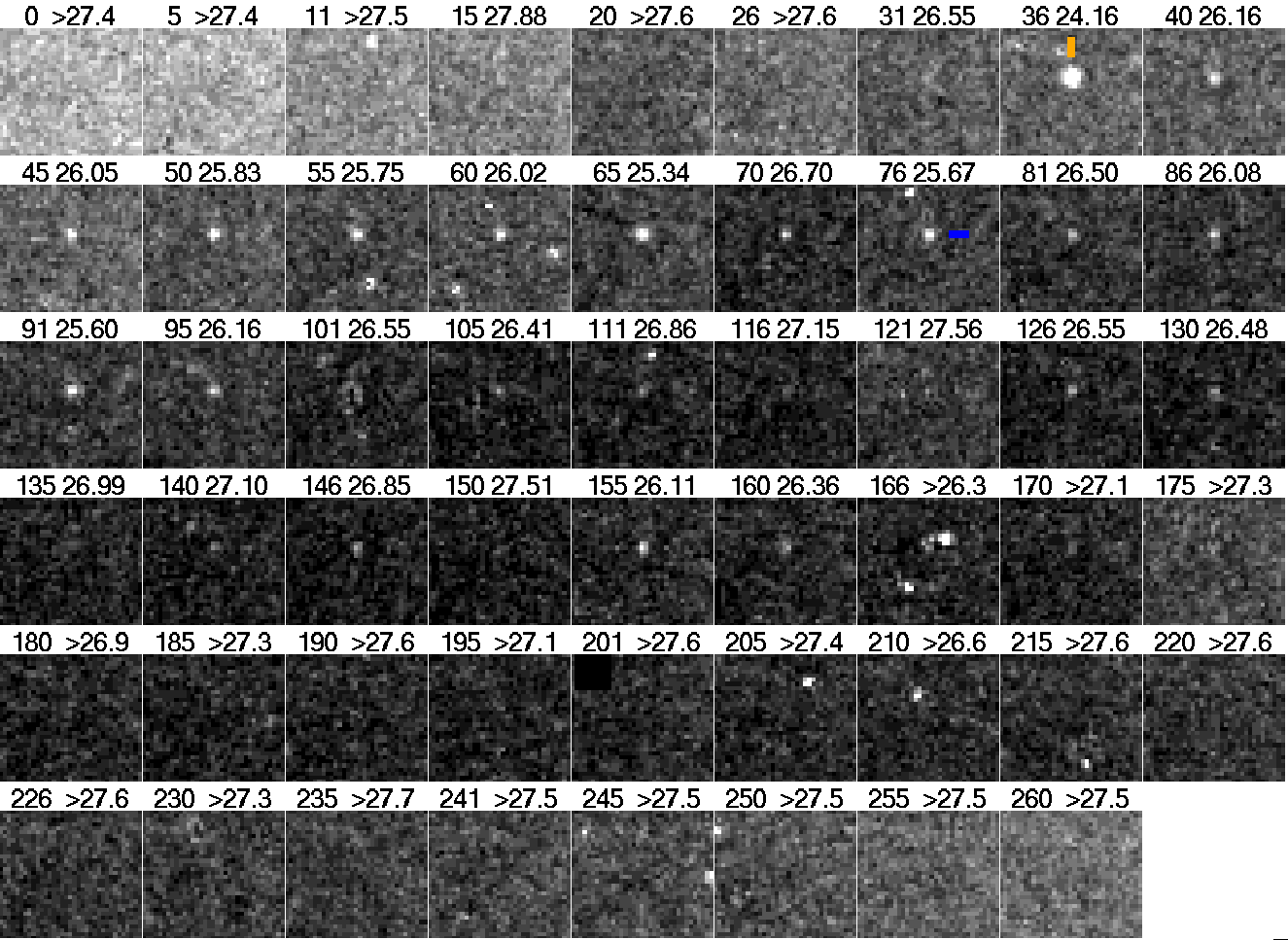} \\
     & \includegraphics[height=3.6in]{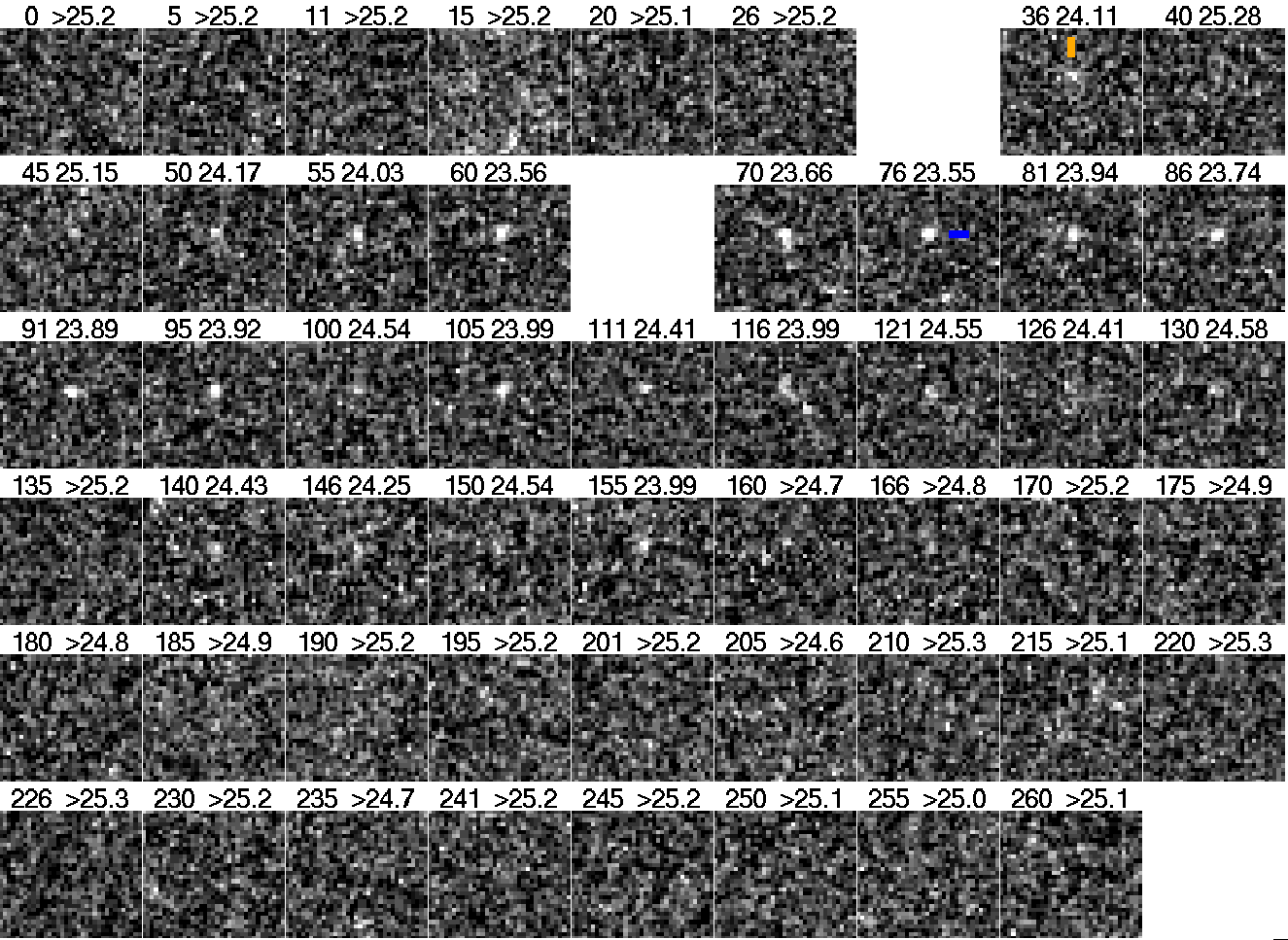}
    \end{tabular}
    \caption{Nova 53}
\end{sidewaysfigure}
\begin{sidewaysfigure}
    \begin{tabular}{cc}
     \multirow{2}{*}[2.6in]{\includegraphics[height=6in]{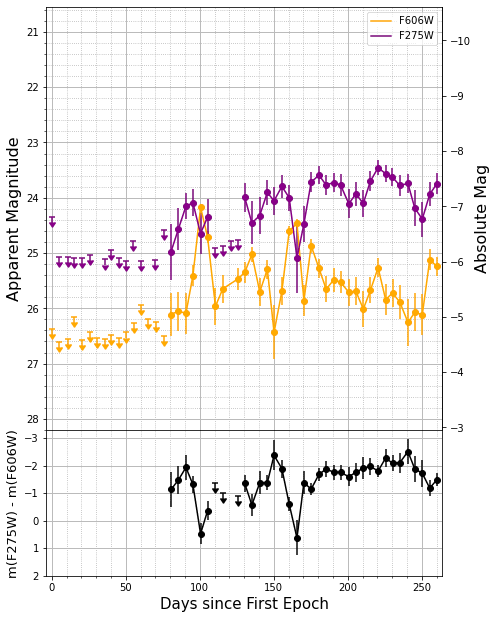}} & \includegraphics[height=3.6in]{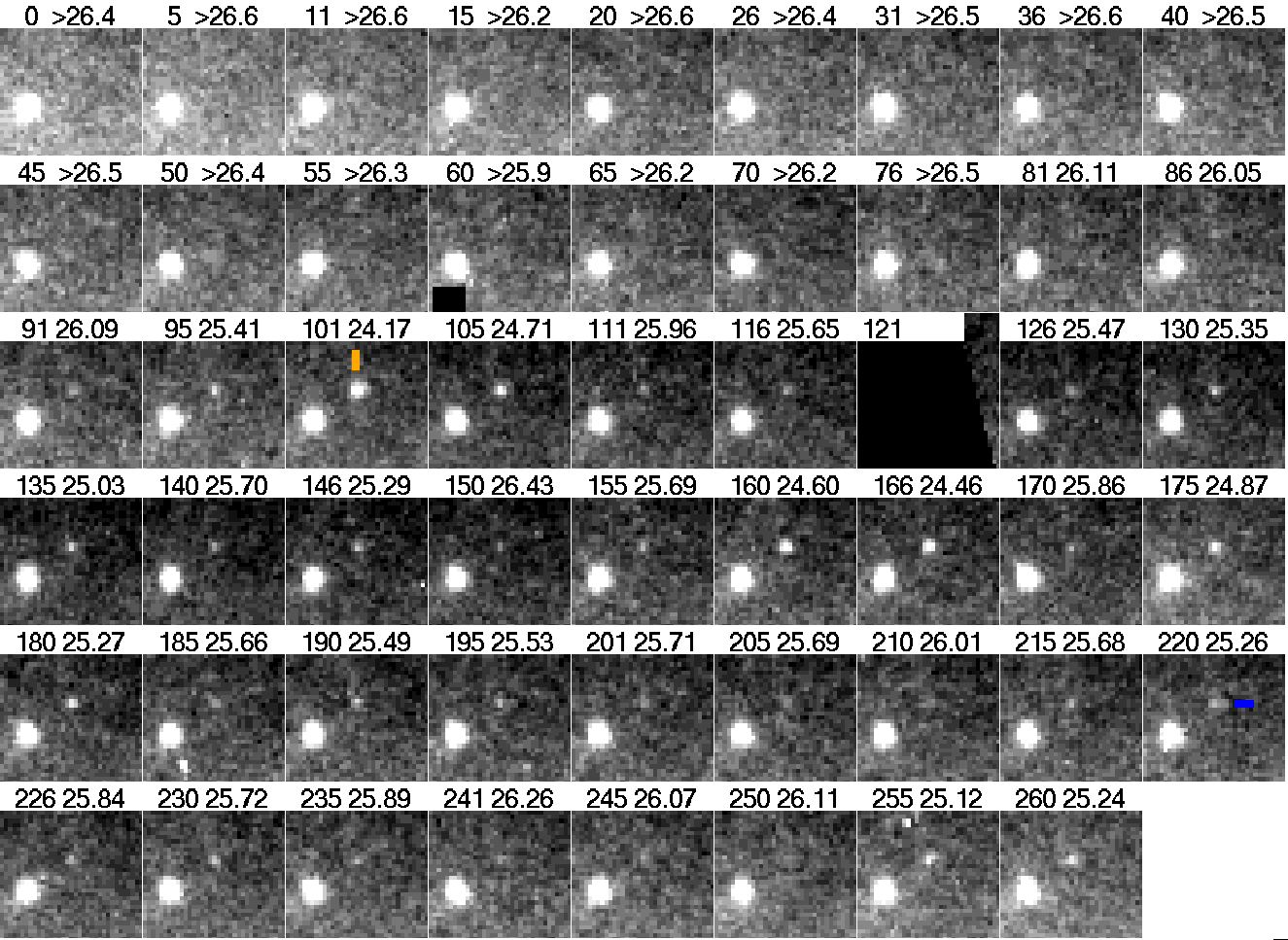} \\
     & \includegraphics[height=3.6in]{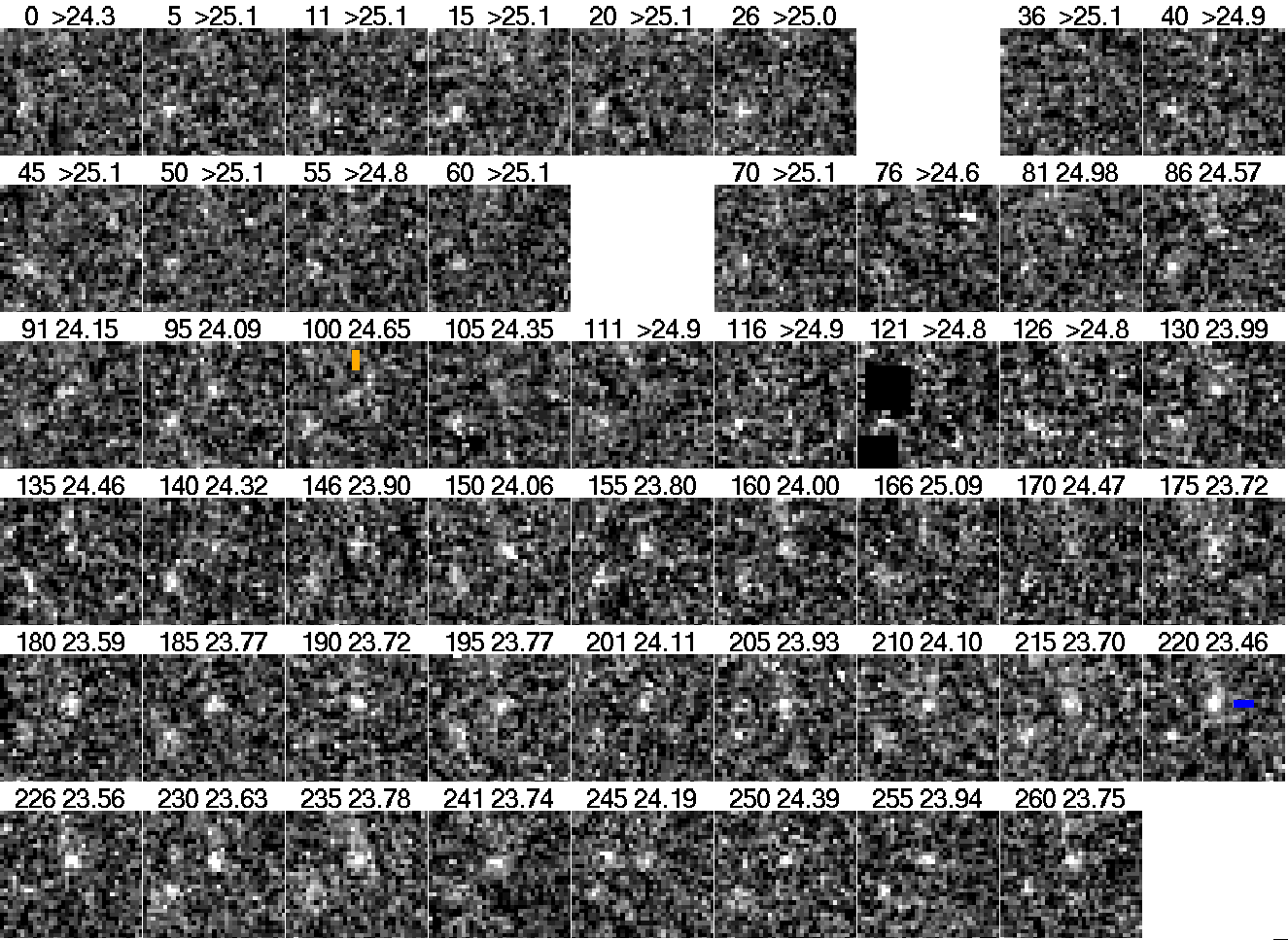}
    \end{tabular}
    \caption{Nova 54}
\end{sidewaysfigure}
\begin{sidewaysfigure}
    \begin{tabular}{cc}
     \multirow{2}{*}[2.6in]{\includegraphics[height=6in]{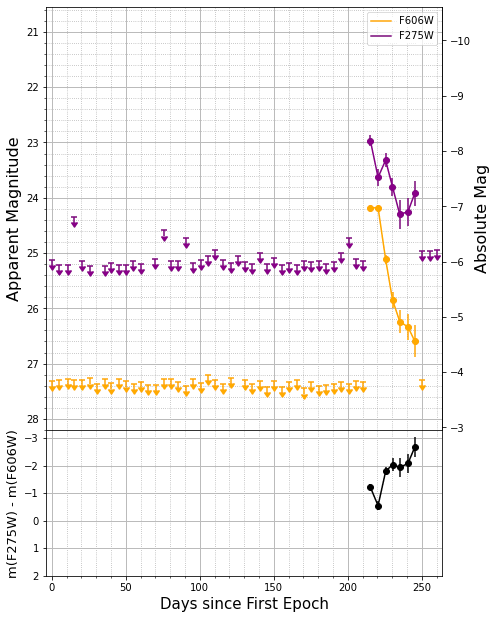}} & \includegraphics[height=3.6in]{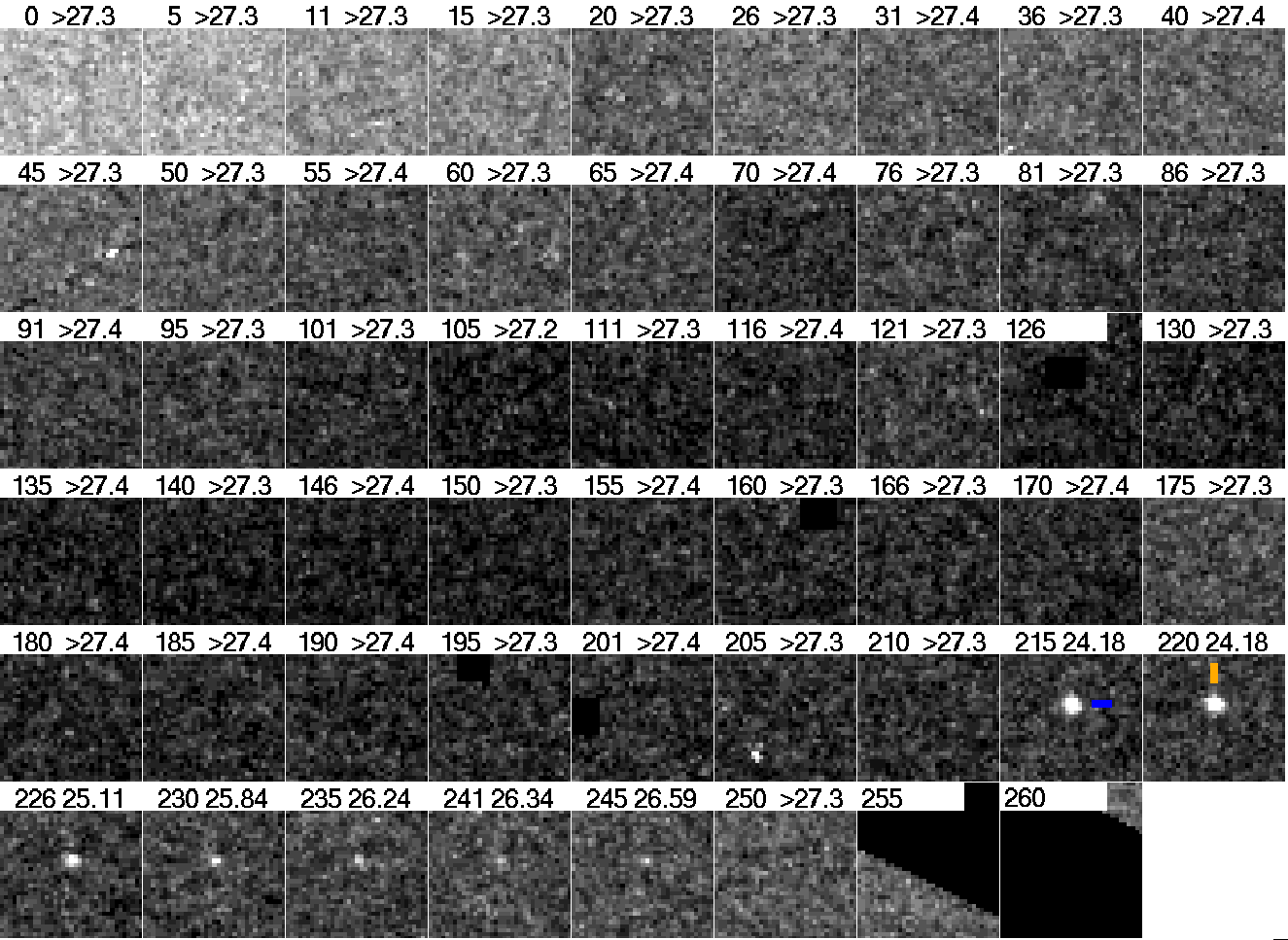} \\
     & \includegraphics[height=3.6in]{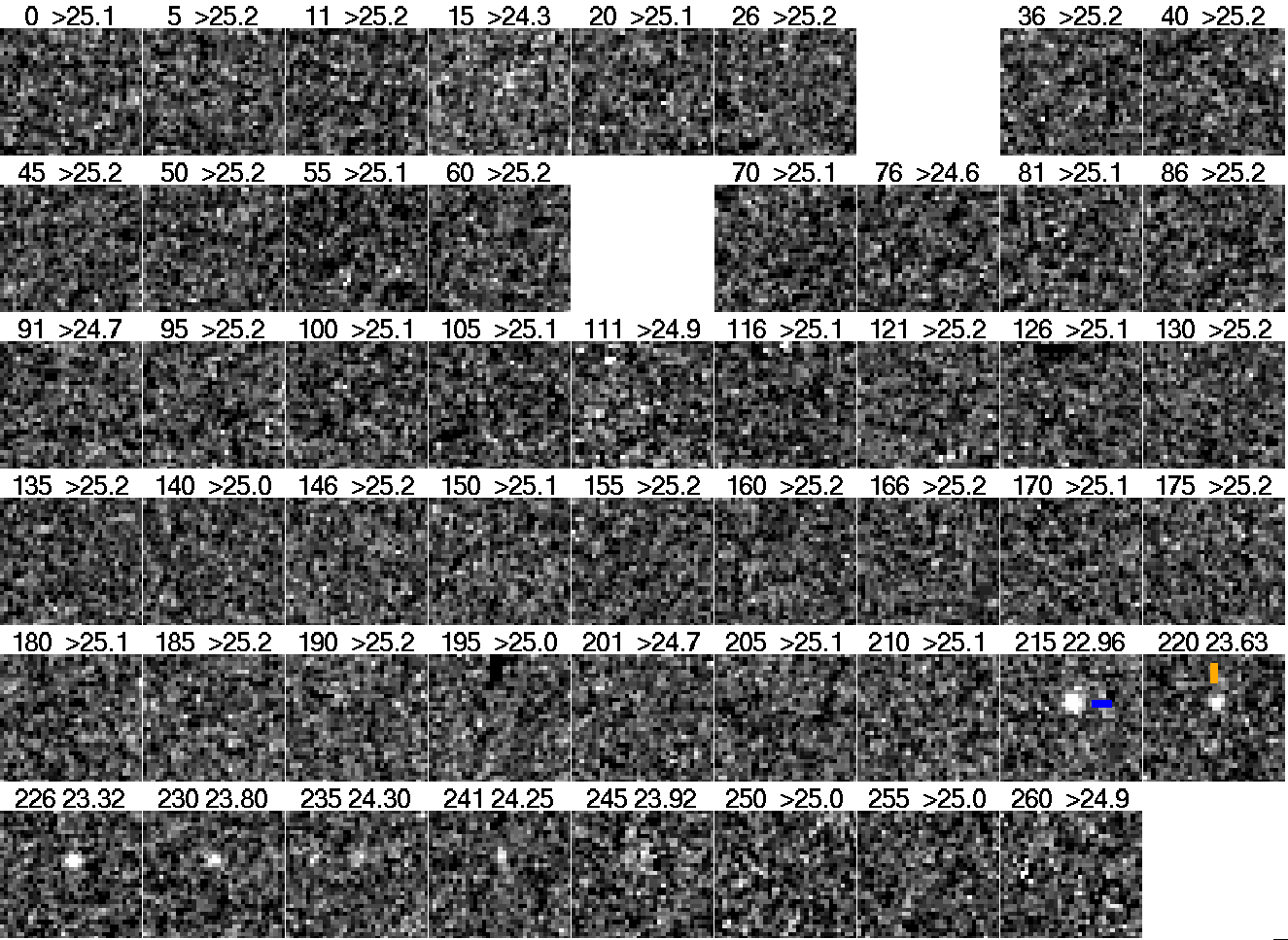}
    \end{tabular}
    \caption{Nova 55}
\end{sidewaysfigure}
\begin{sidewaysfigure}
    \begin{tabular}{cc}
     \multirow{2}{*}[2.6in]{\includegraphics[height=6in]{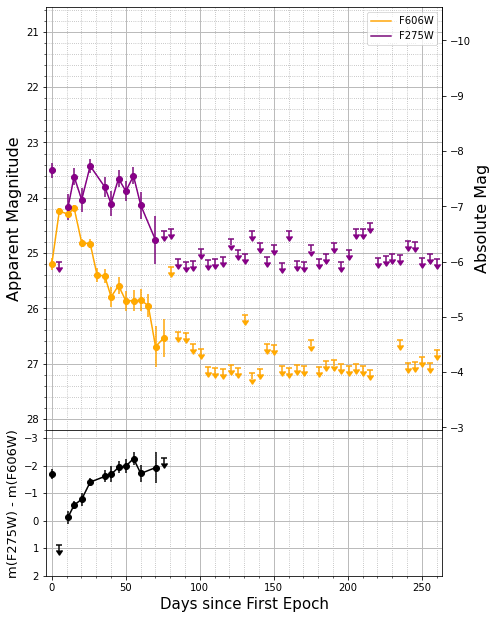}} & \includegraphics[height=3.6in]{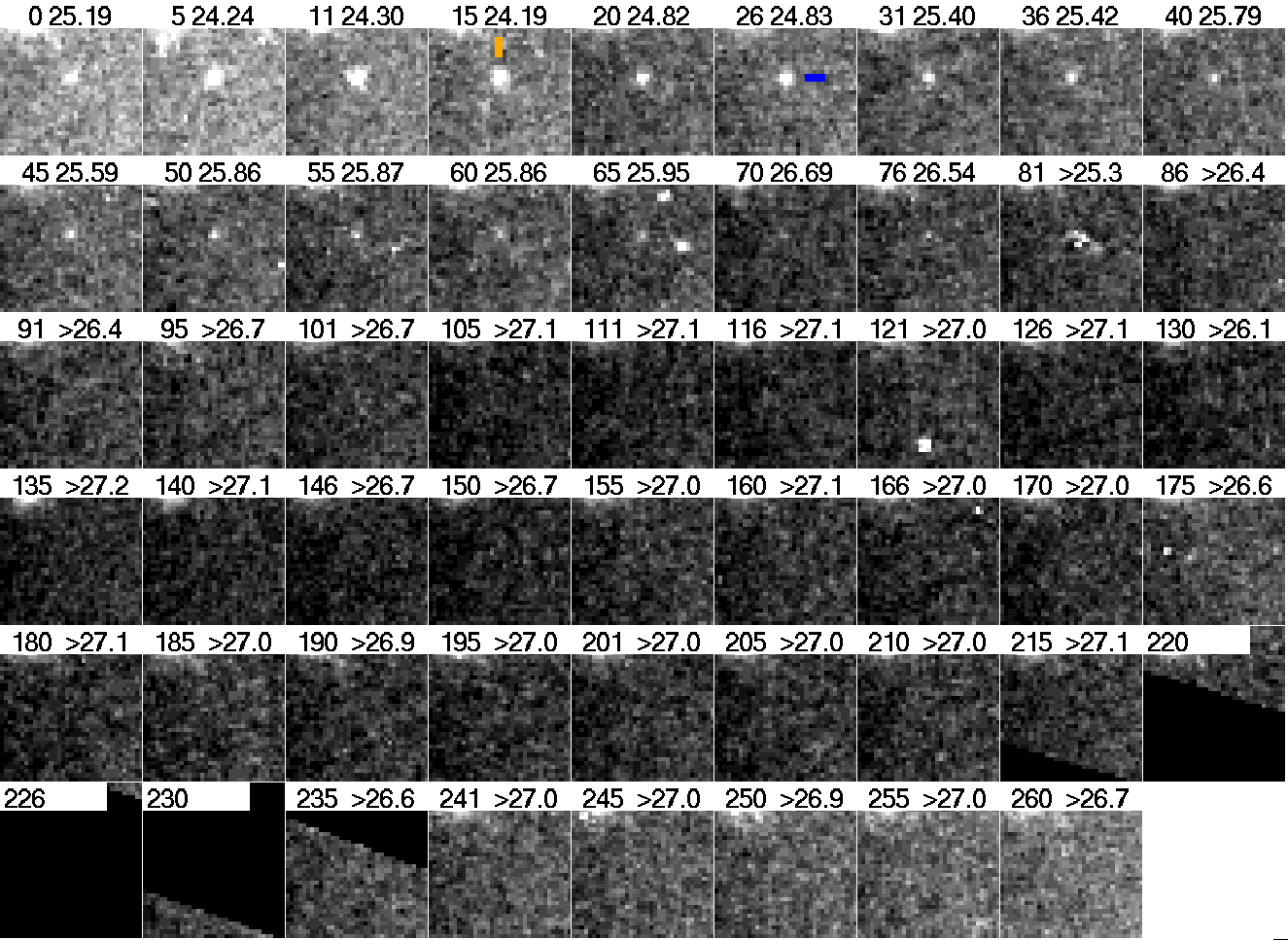} \\
     & \includegraphics[height=3.6in]{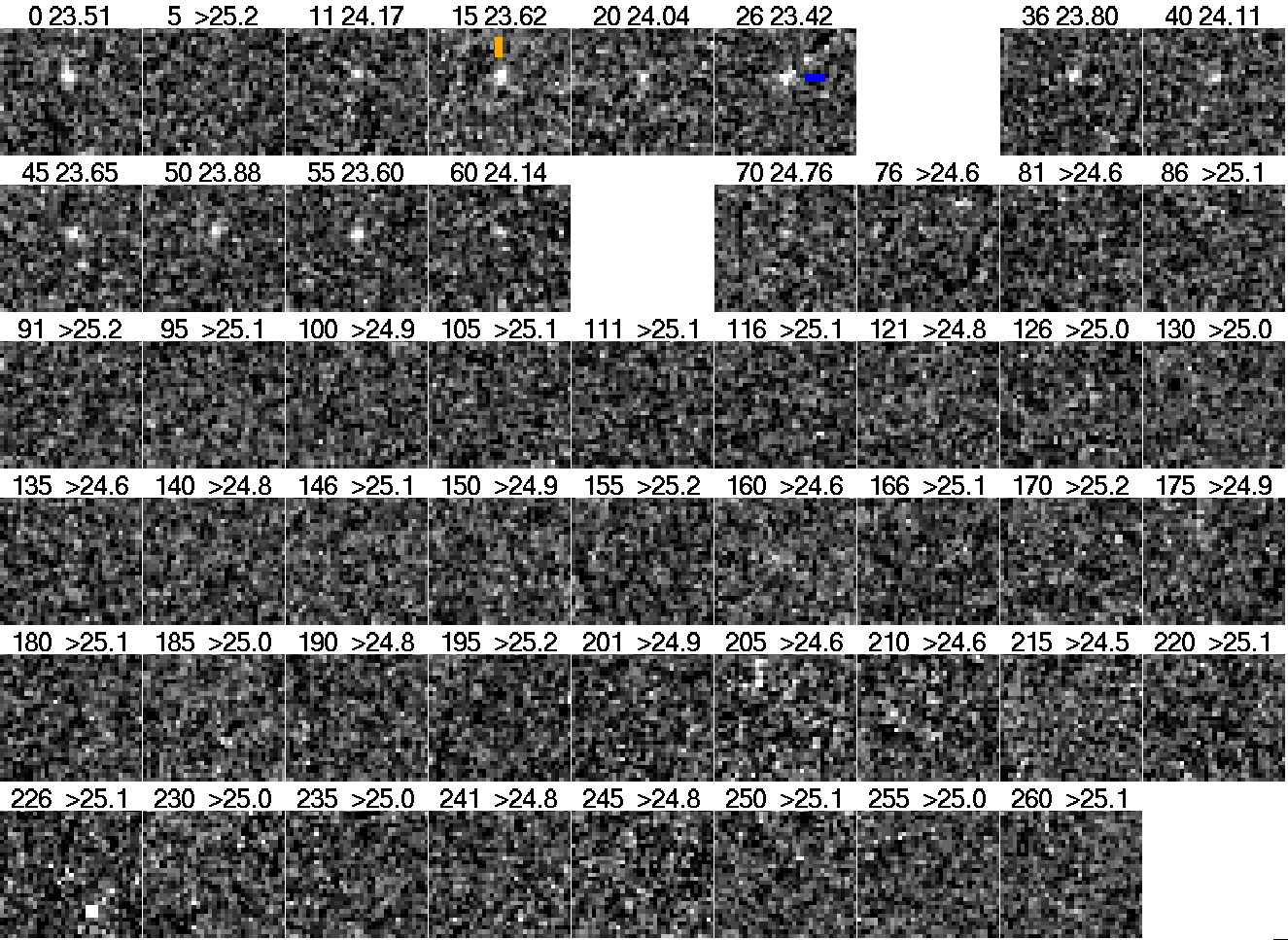}
    \end{tabular}
    \caption{Nova 56}
\end{sidewaysfigure}
\begin{sidewaysfigure}
    \begin{tabular}{cc}
     \multirow{2}{*}[2.6in]{\includegraphics[height=6in]{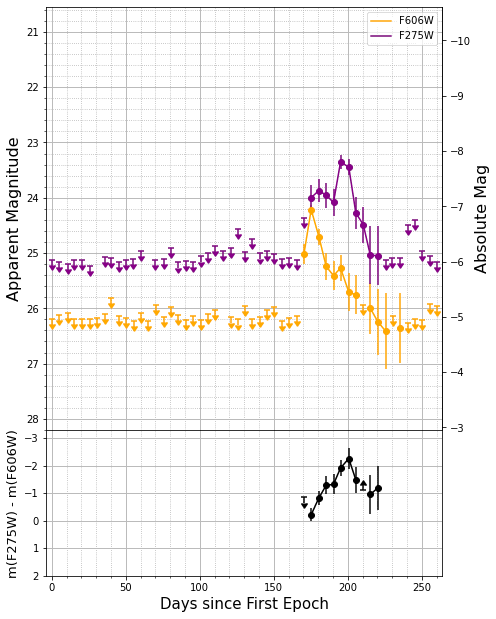}} & \includegraphics[height=3.6in]{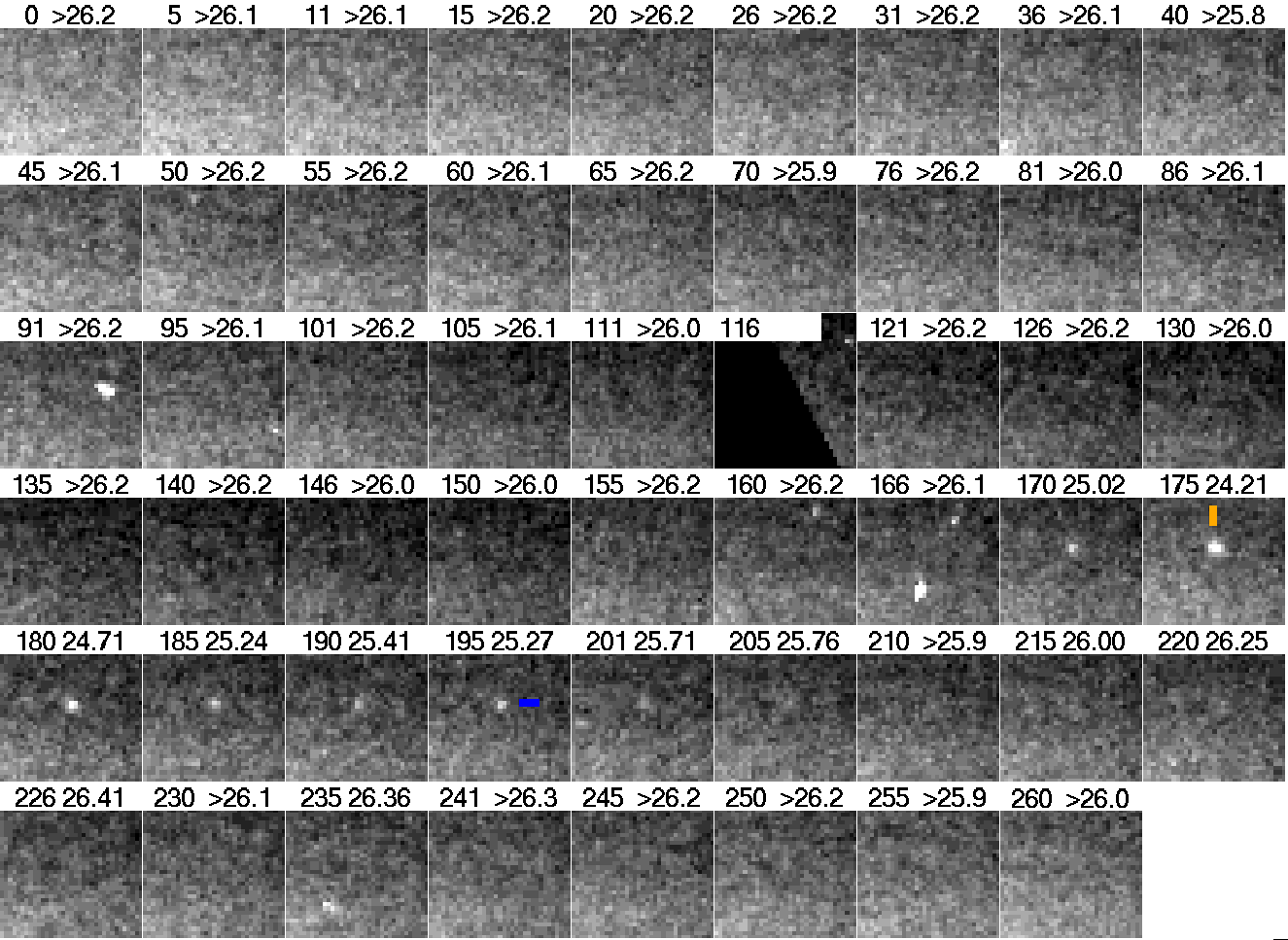} \\
     & \includegraphics[height=3.6in]{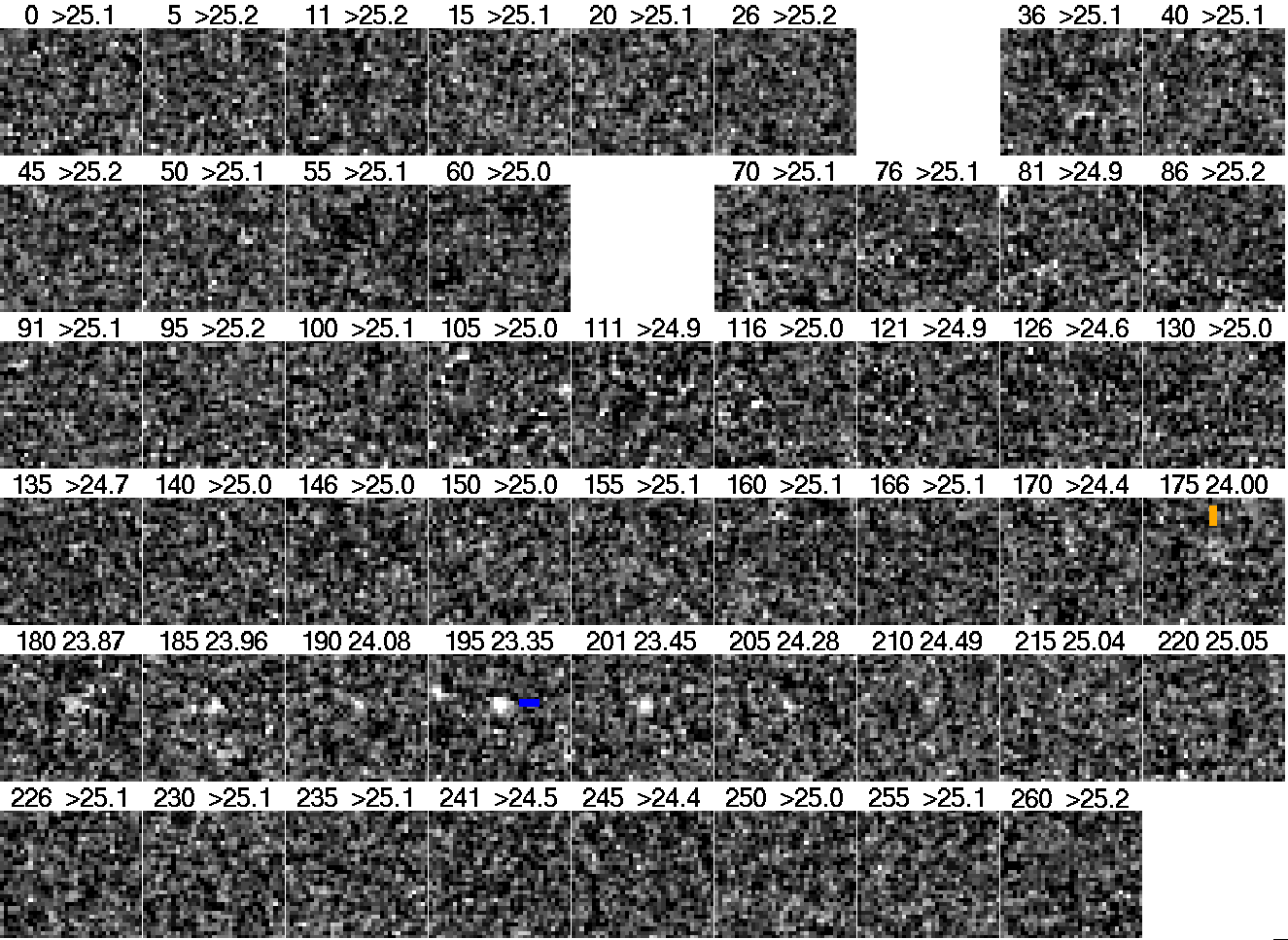}
    \end{tabular}
    \caption{Nova 57}
\end{sidewaysfigure}
\begin{sidewaysfigure}
    \begin{tabular}{cc}
     \multirow{2}{*}[2.6in]{\includegraphics[height=6in]{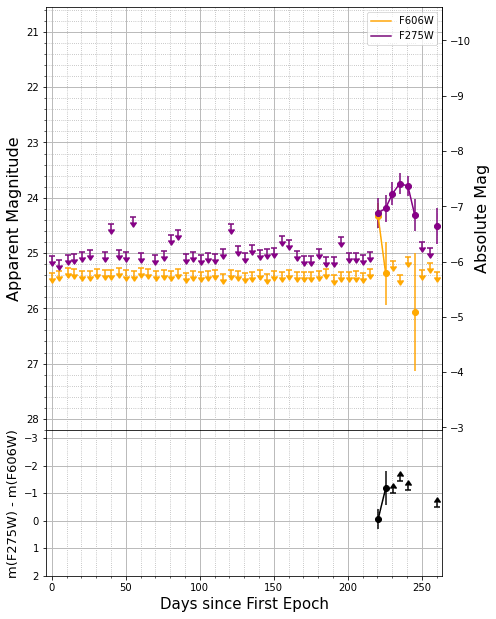}} & \includegraphics[height=3.6in]{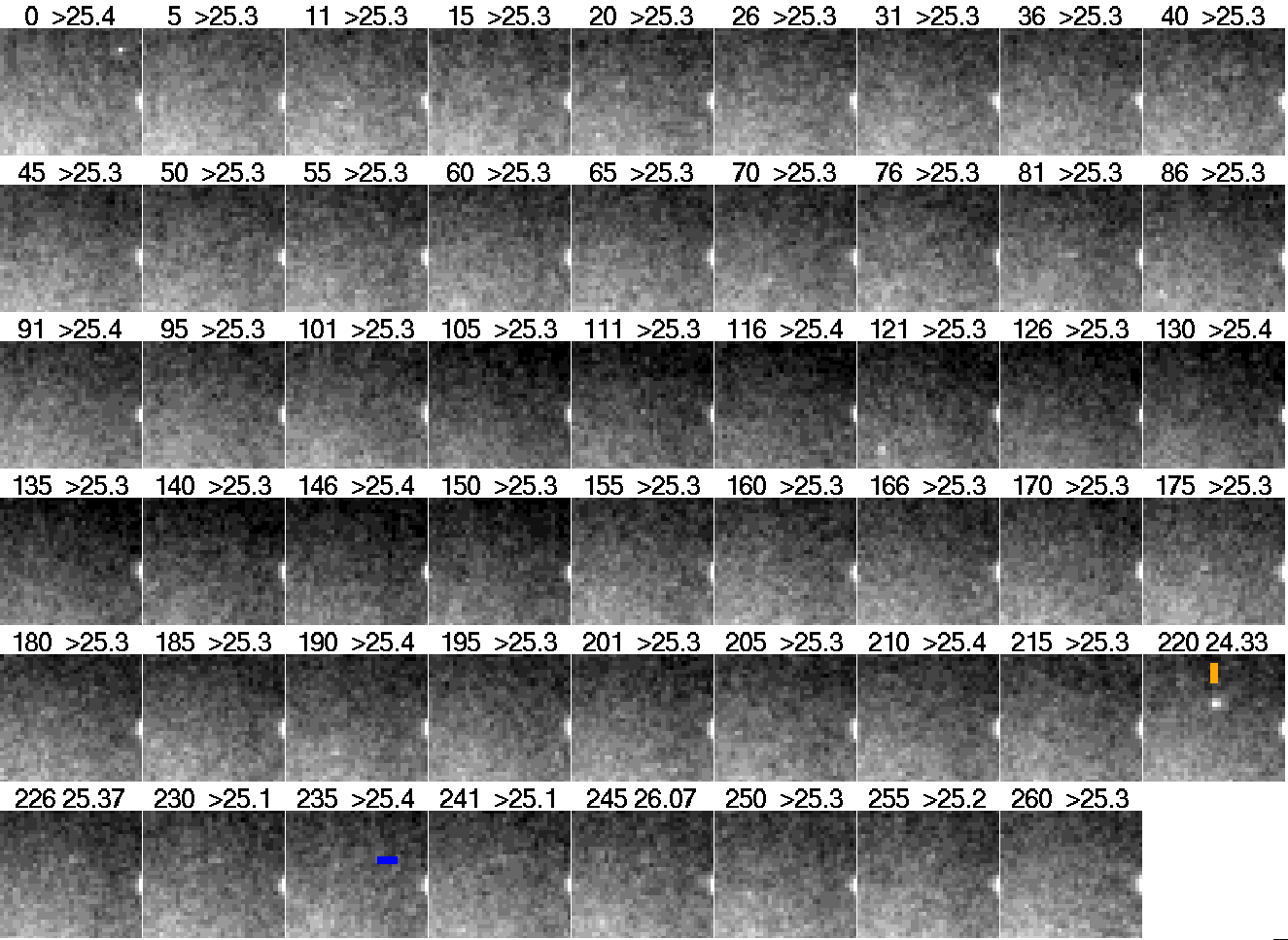} \\
     & \includegraphics[height=3.6in]{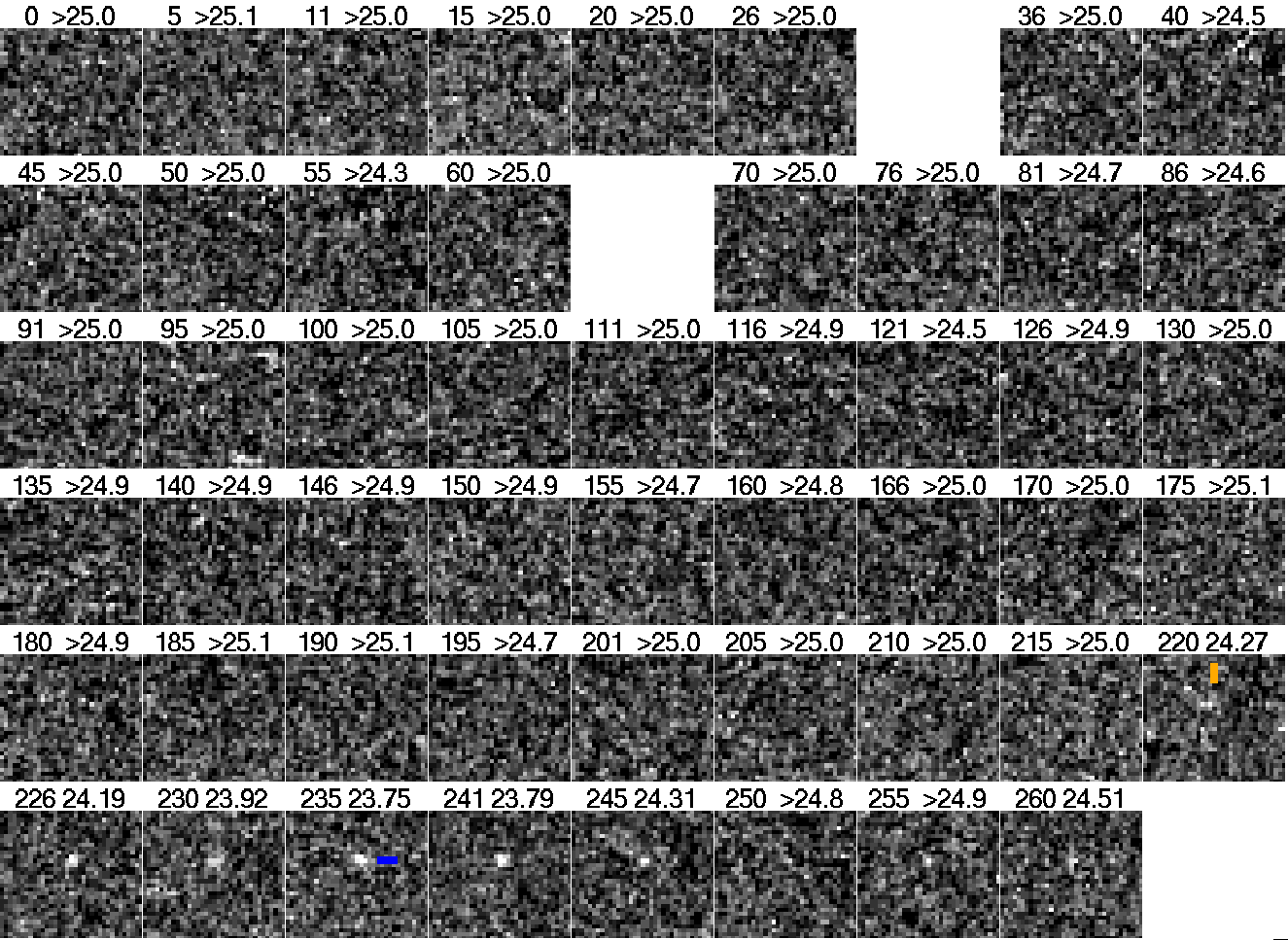}
    \end{tabular}
    \caption{Nova 58}
\end{sidewaysfigure}
\begin{sidewaysfigure}
    \begin{tabular}{cc}
     \multirow{2}{*}[2.6in]{\includegraphics[height=6in]{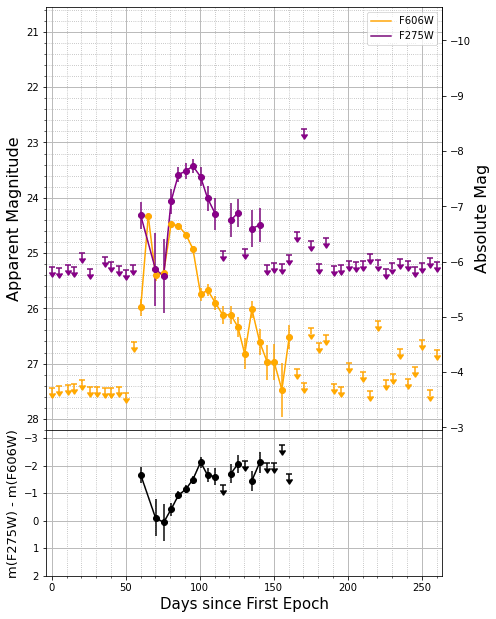}} & \includegraphics[height=3.6in]{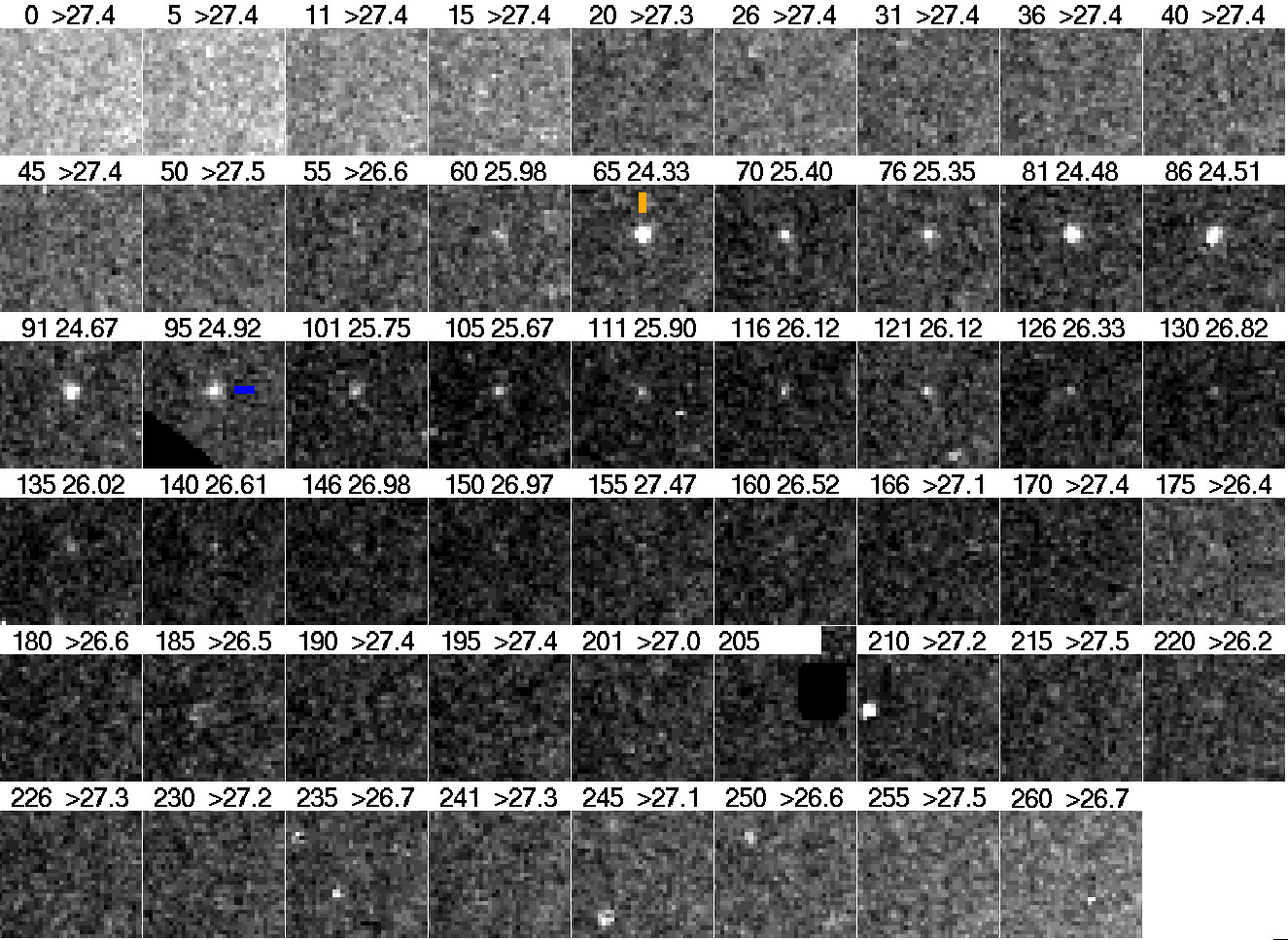} \\
     & \includegraphics[height=3.6in]{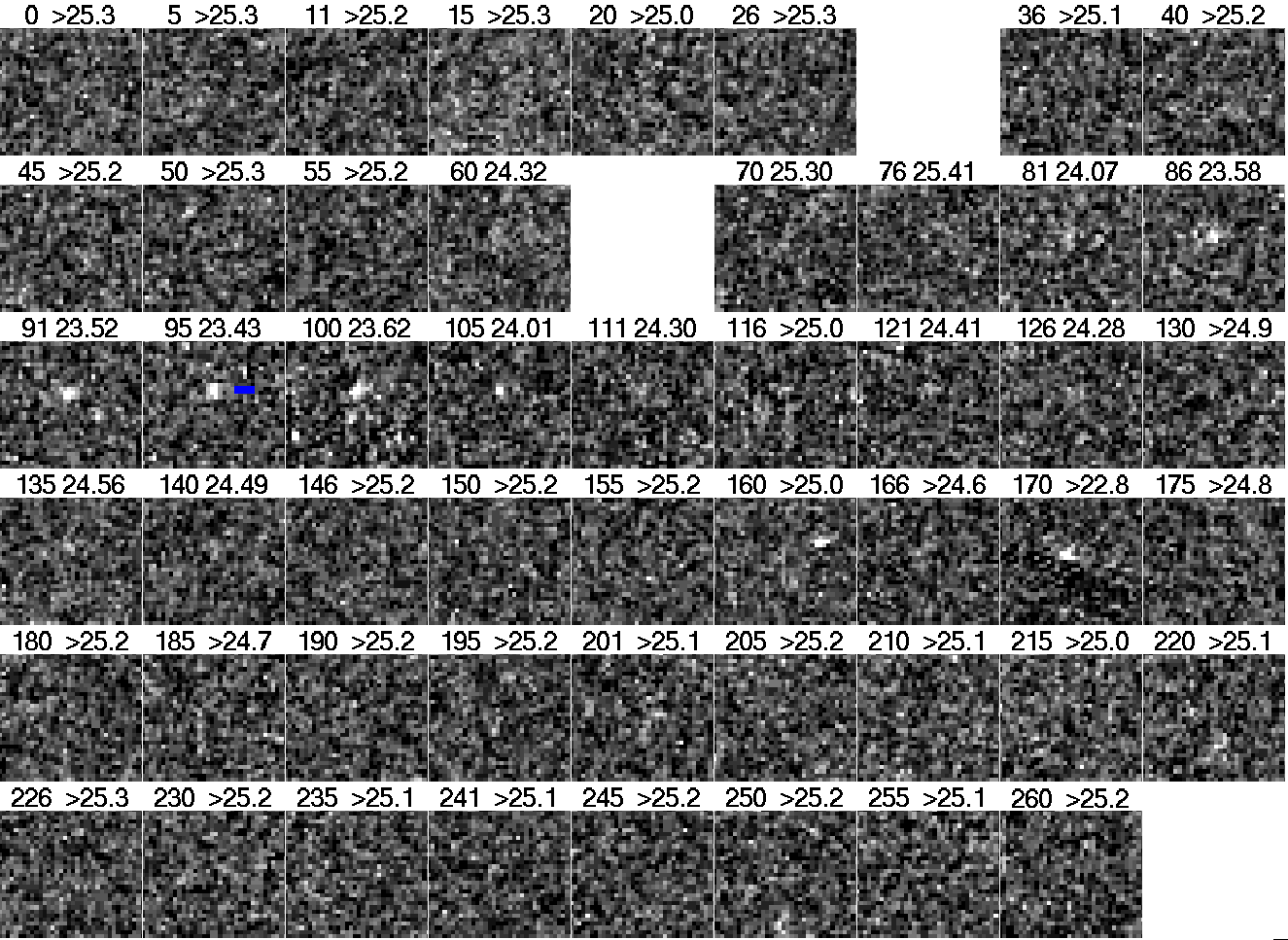}
    \end{tabular}
    \caption{Nova 59}
\end{sidewaysfigure}
\begin{sidewaysfigure}
    \begin{tabular}{cc}
     \multirow{2}{*}[2.6in]{\includegraphics[height=6in]{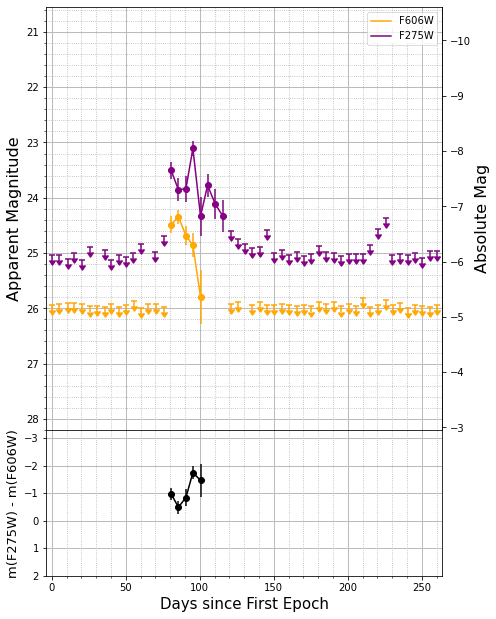}} & \includegraphics[height=3.6in]{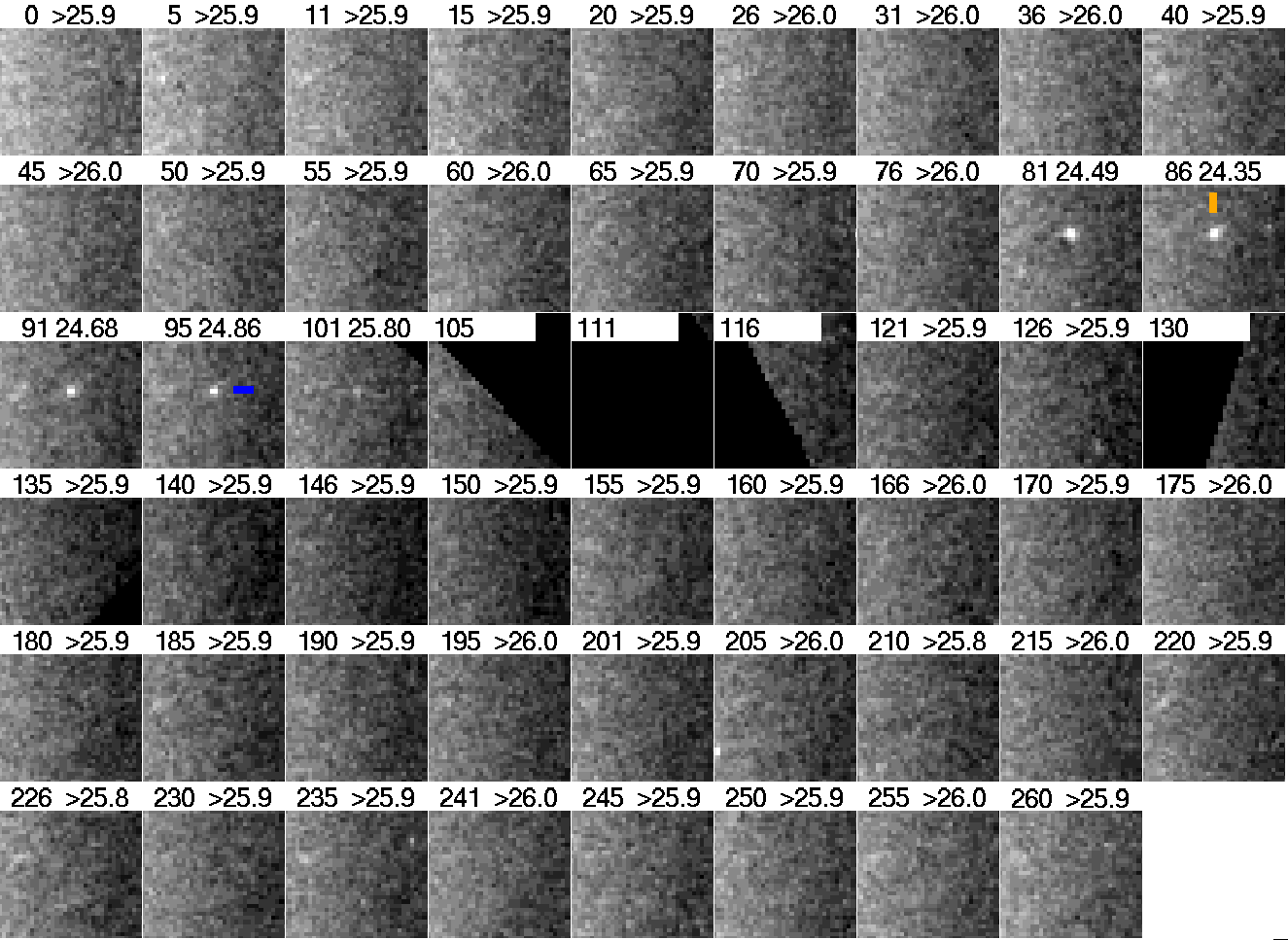} \\
     & \includegraphics[height=3.6in]{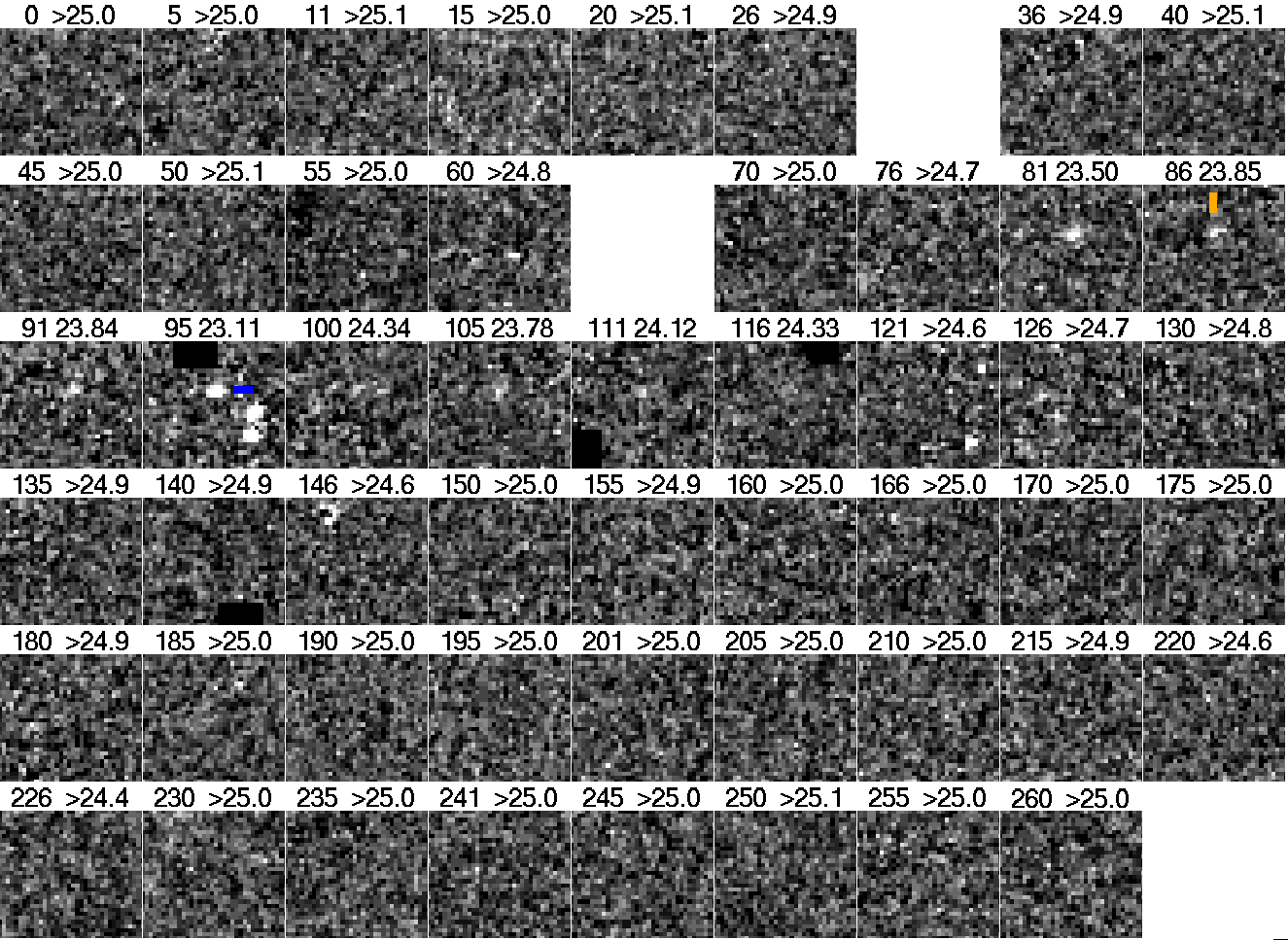}
    \end{tabular}
    \caption{Nova 60}
\end{sidewaysfigure}
\begin{sidewaysfigure}
    \begin{tabular}{cc}
     \multirow{2}{*}[2.6in]{\includegraphics[height=6in]{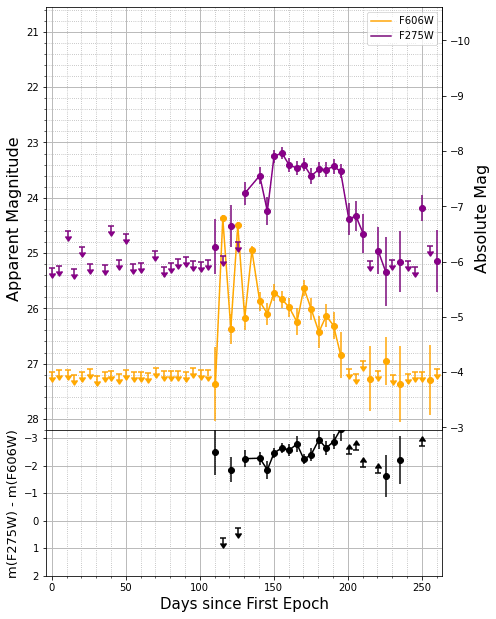}} & \includegraphics[height=3.6in]{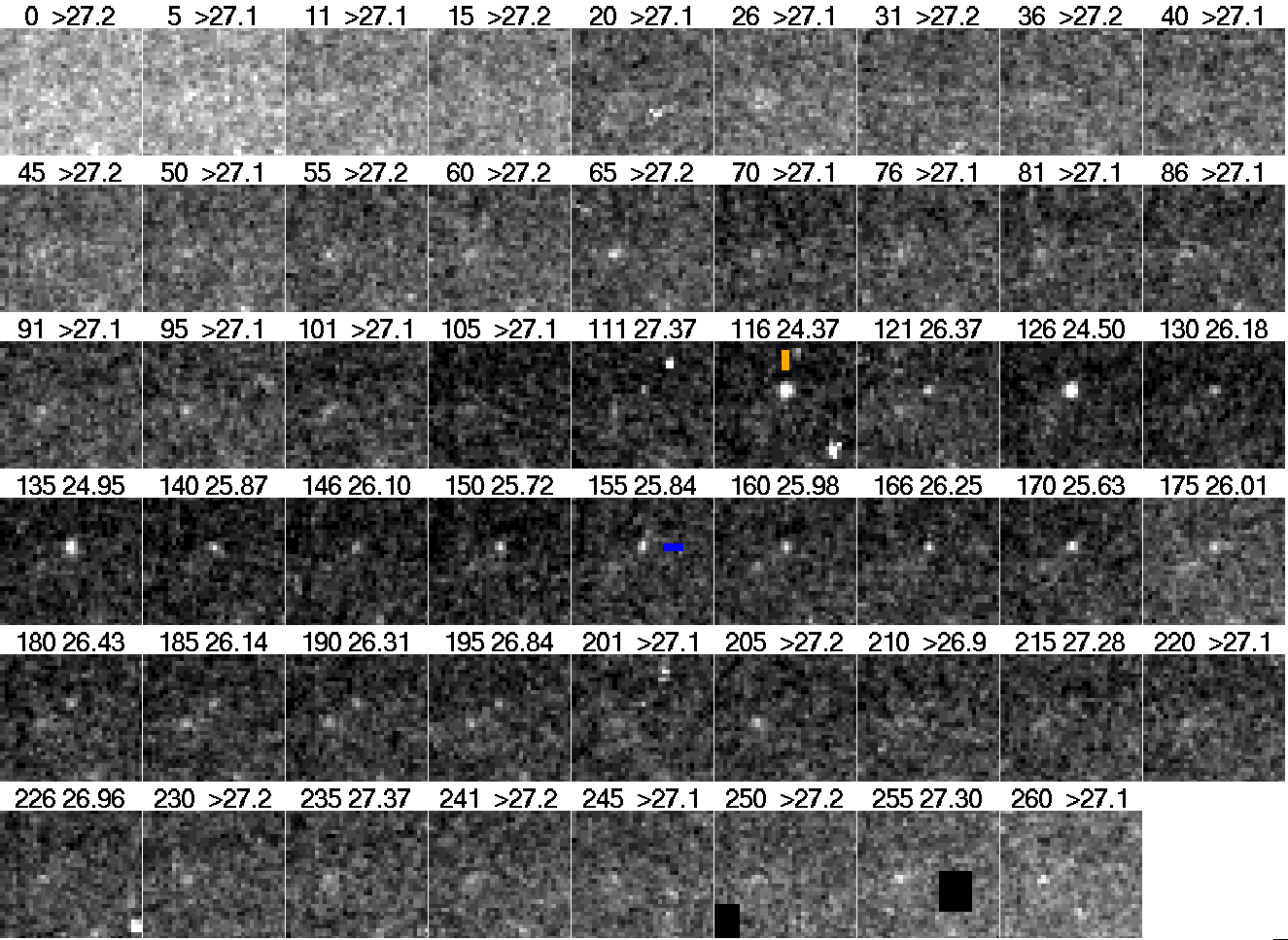} \\
     & \includegraphics[height=3.6in]{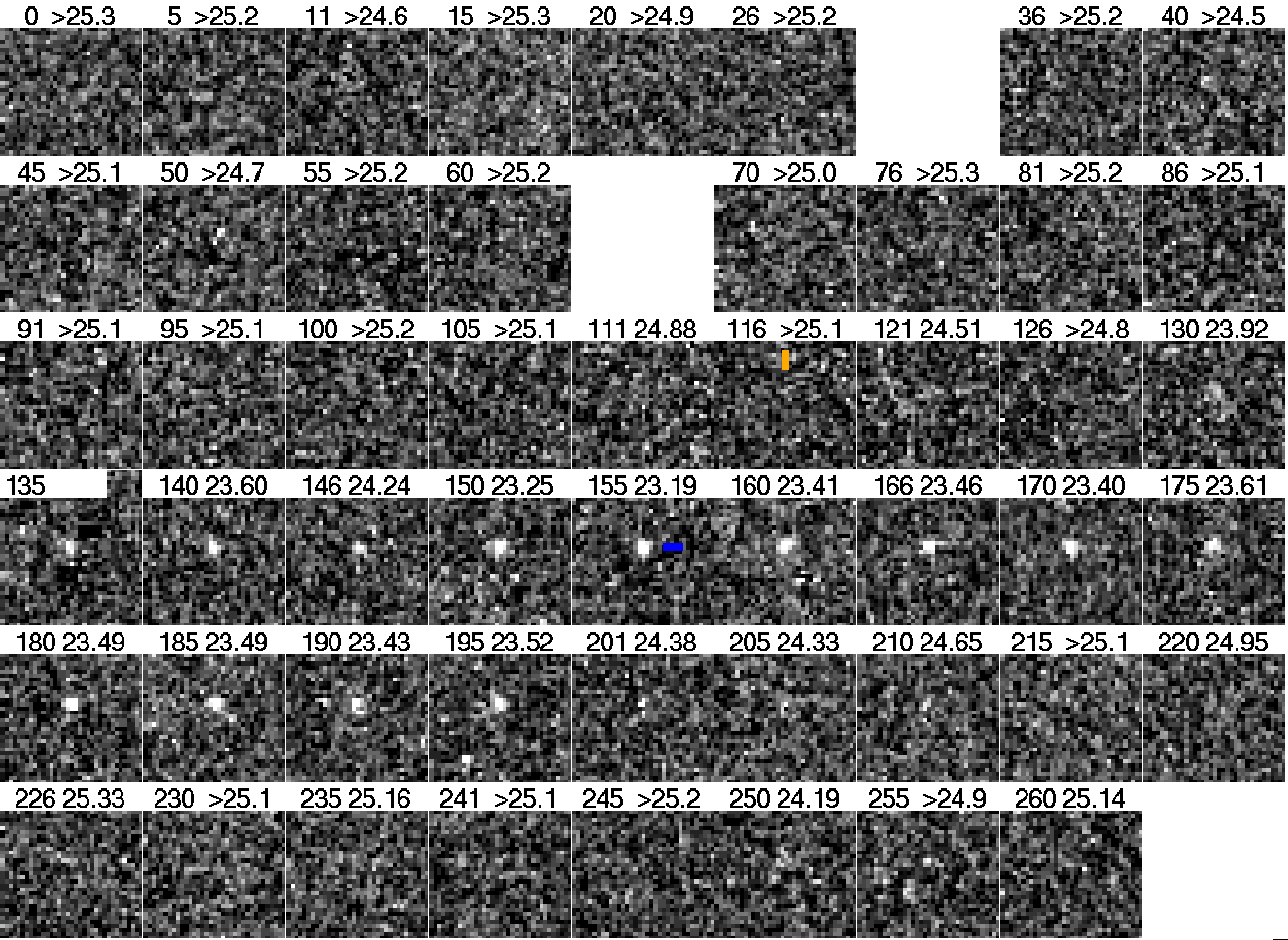}
    \end{tabular}
    \caption{Nova 61}
\end{sidewaysfigure}
\begin{sidewaysfigure}
    \begin{tabular}{cc}
     \multirow{2}{*}[2.6in]{\includegraphics[height=6in]{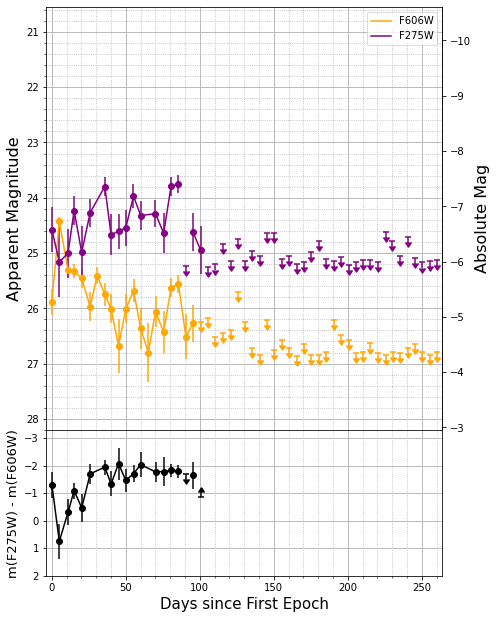}} & \includegraphics[height=3.6in]{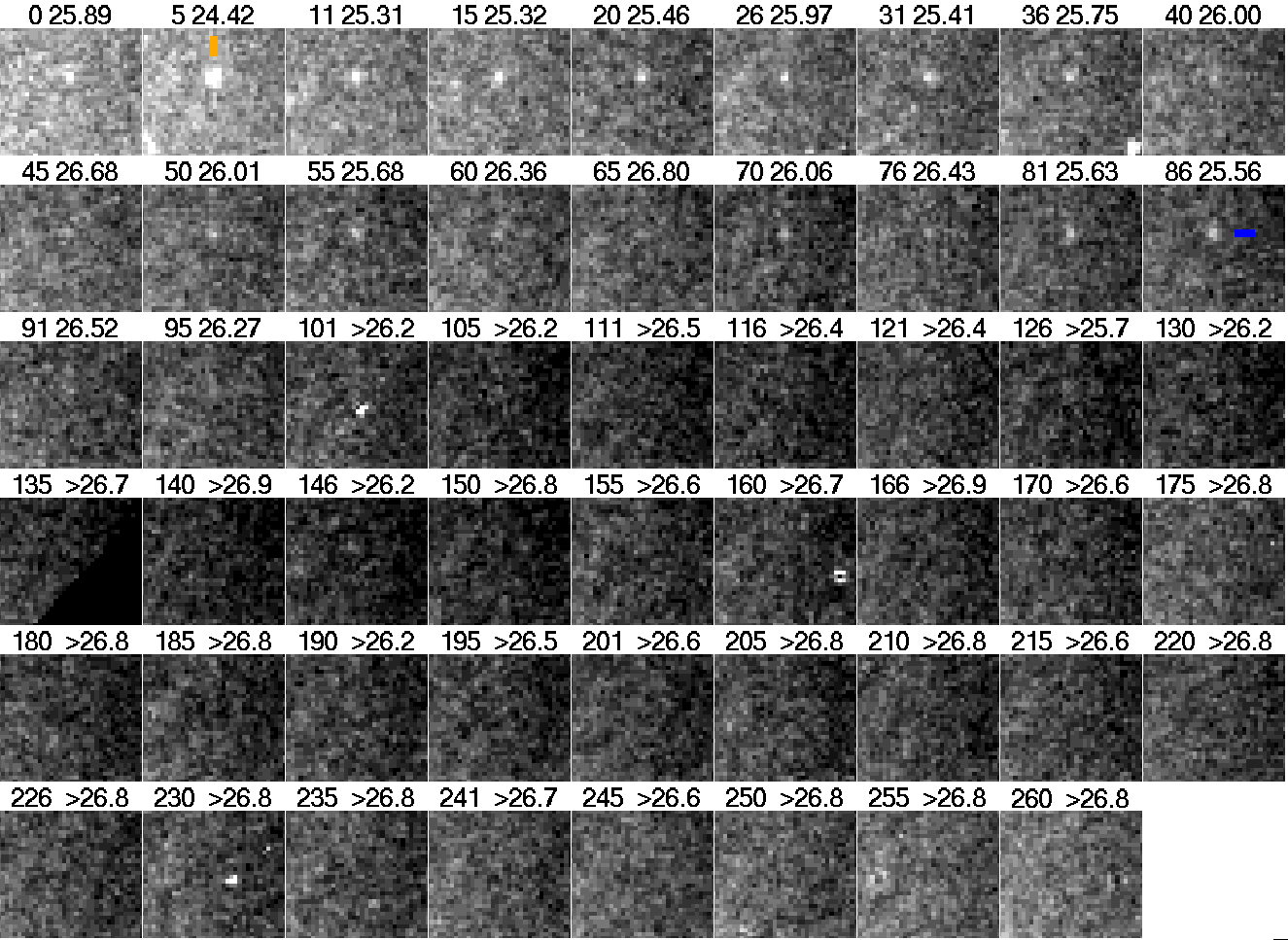} \\
     & \includegraphics[height=3.6in]{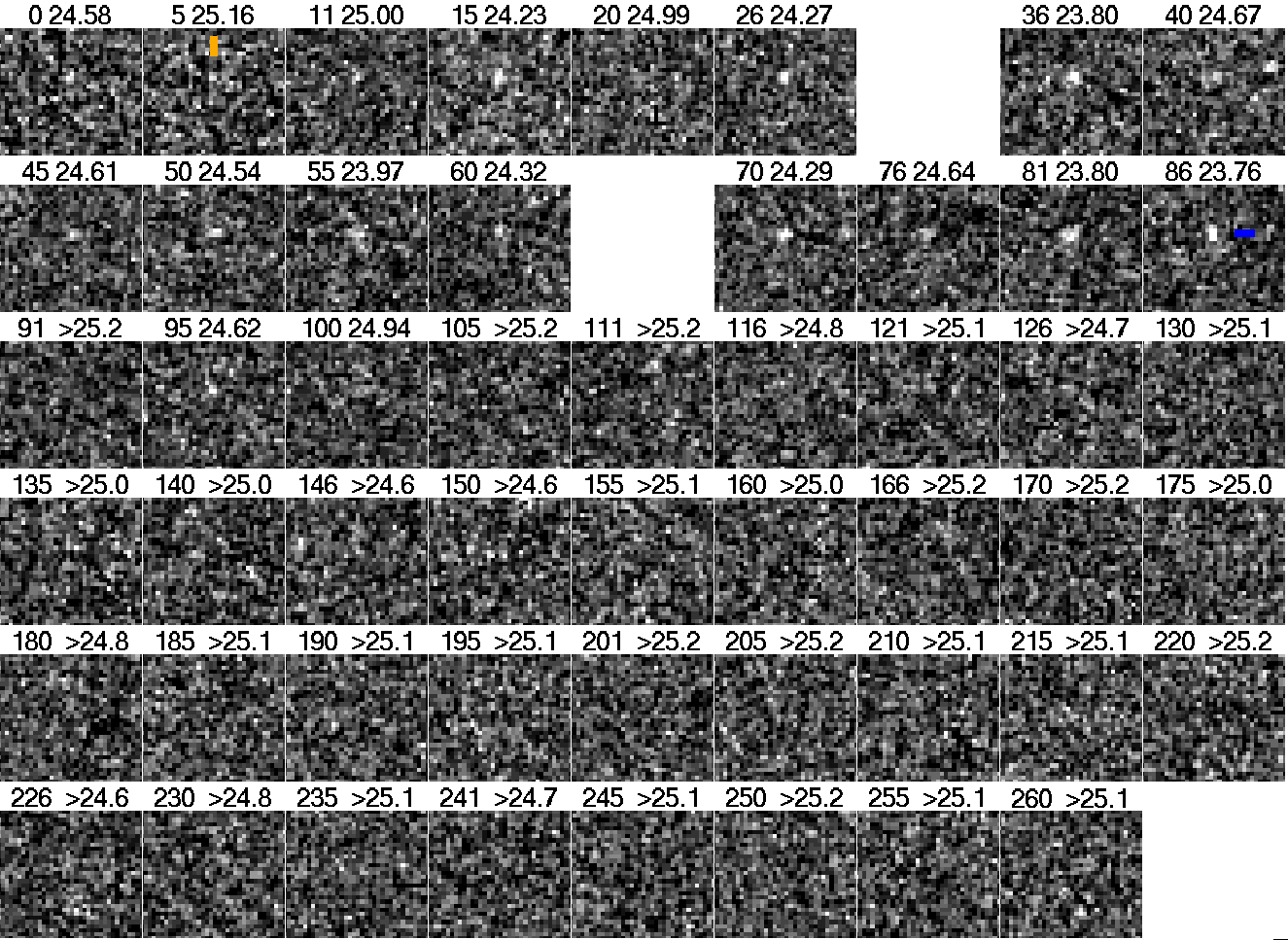}
    \end{tabular}
    \caption{Nova 62}
\end{sidewaysfigure}
\begin{sidewaysfigure}
    \begin{tabular}{cc}
     \multirow{2}{*}[2.6in]{\includegraphics[height=6in]{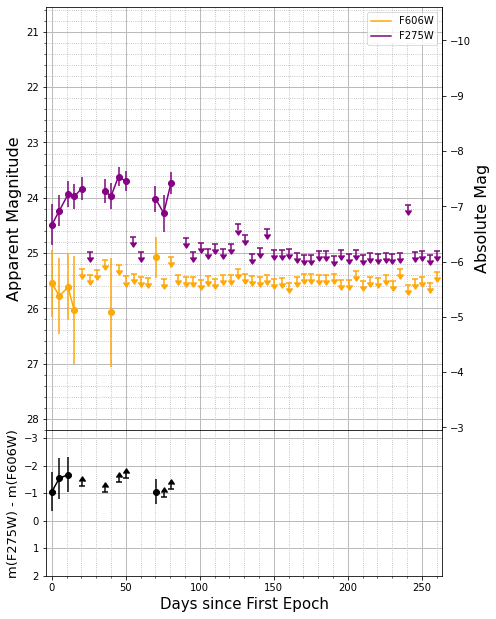}} & \includegraphics[height=3.6in]{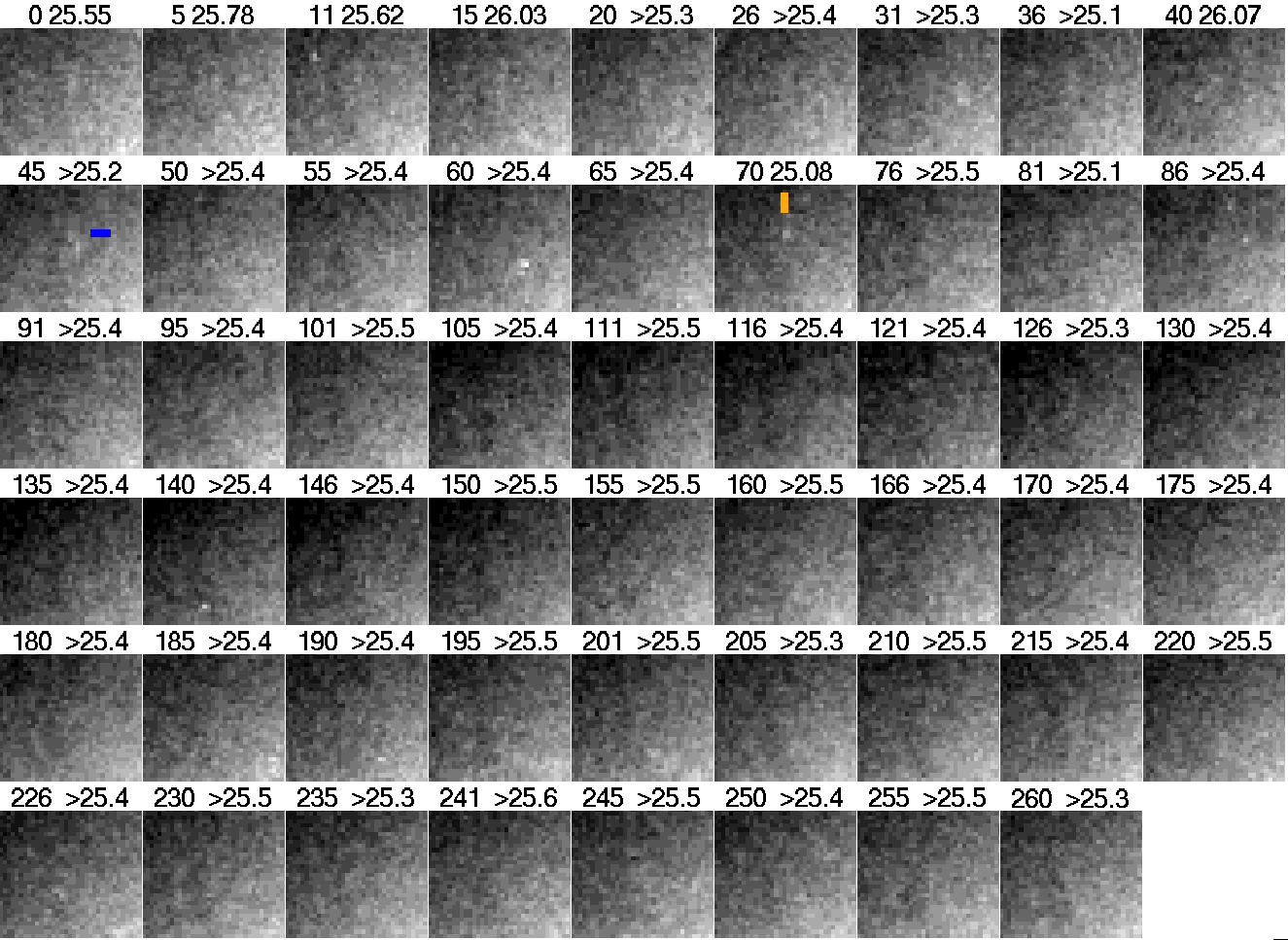} \\
     & \includegraphics[height=3.6in]{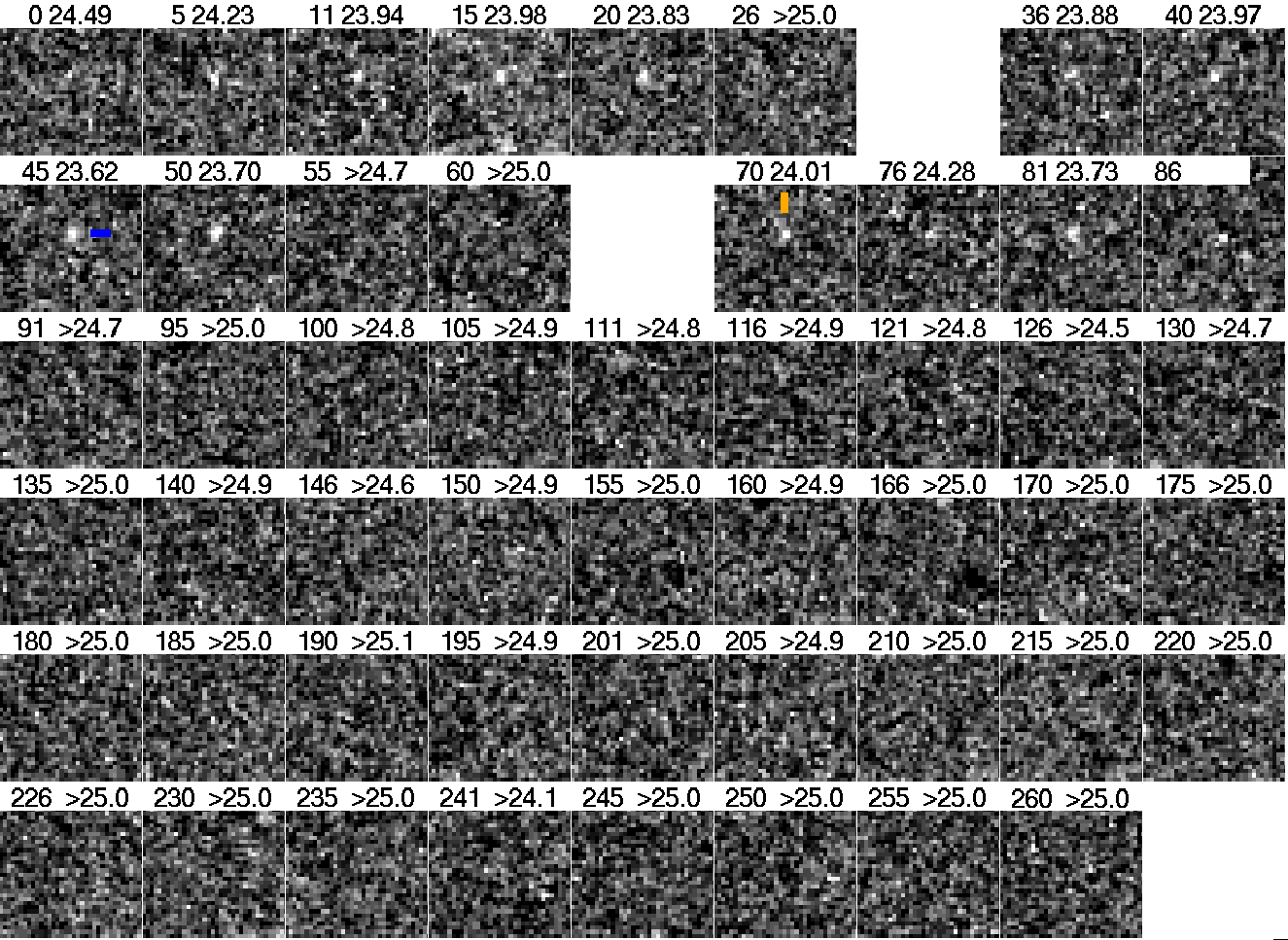}
    \end{tabular}
    \caption{Nova 63}
\end{sidewaysfigure}
\begin{sidewaysfigure}
    \begin{tabular}{cc}
     \multirow{2}{*}[2.6in]{\includegraphics[height=6in]{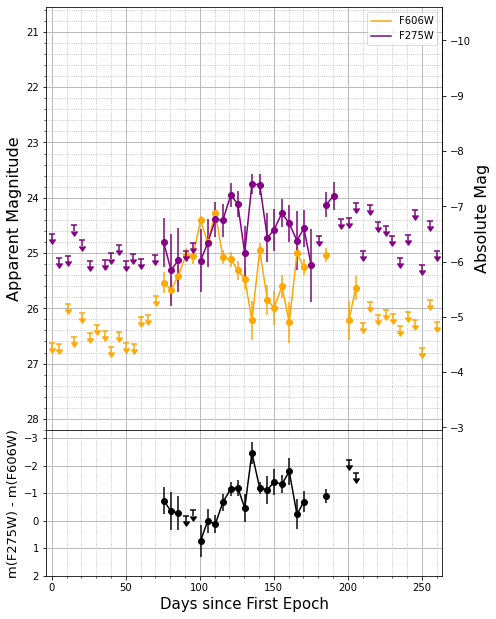}} & \includegraphics[height=3.6in]{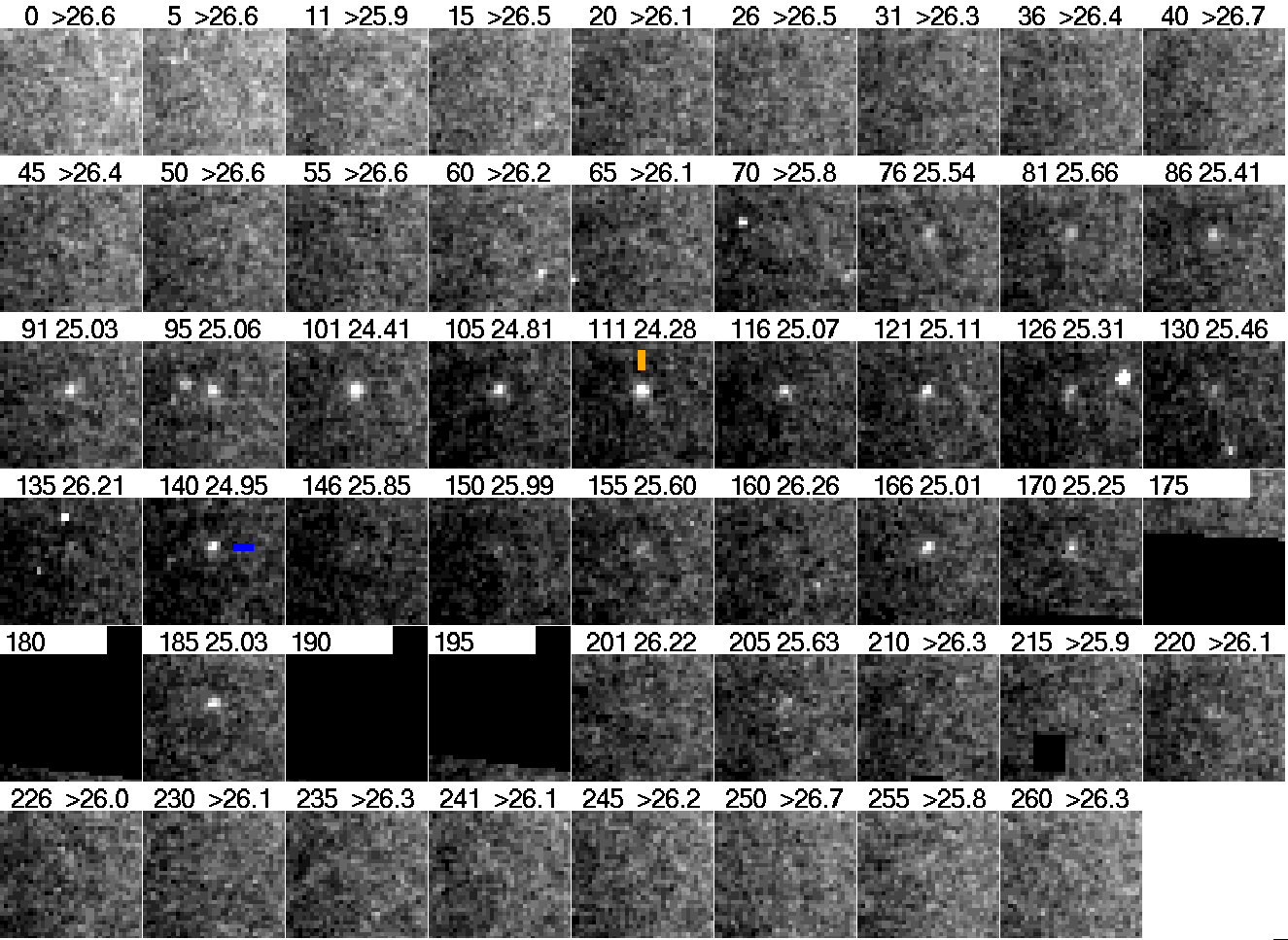} \\
     & \includegraphics[height=3.6in]{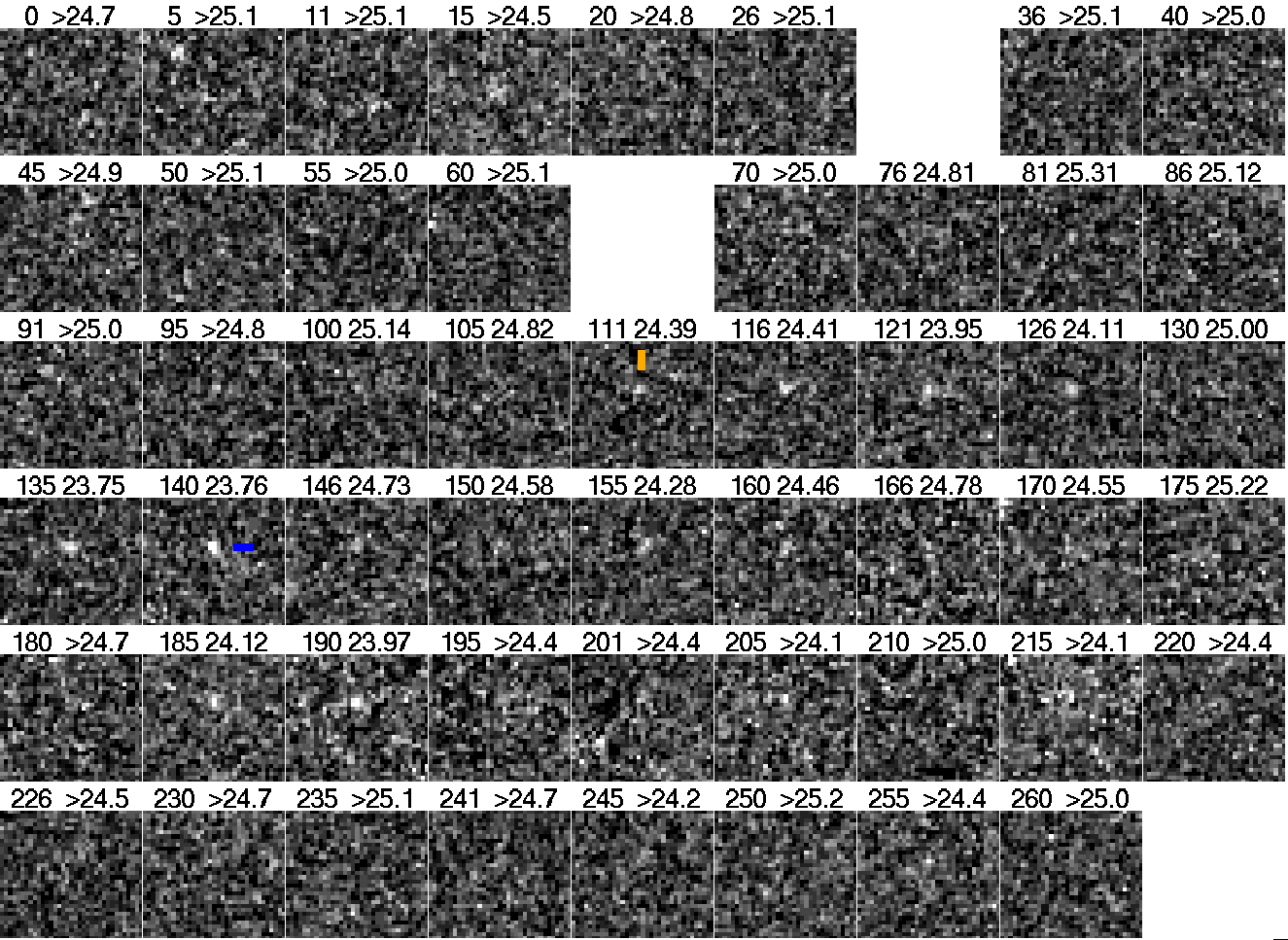}
    \end{tabular}
    \caption{Nova 64}
\end{sidewaysfigure}
\begin{sidewaysfigure}
    \begin{tabular}{cc}
     \multirow{2}{*}[2.6in]{\includegraphics[height=6in]{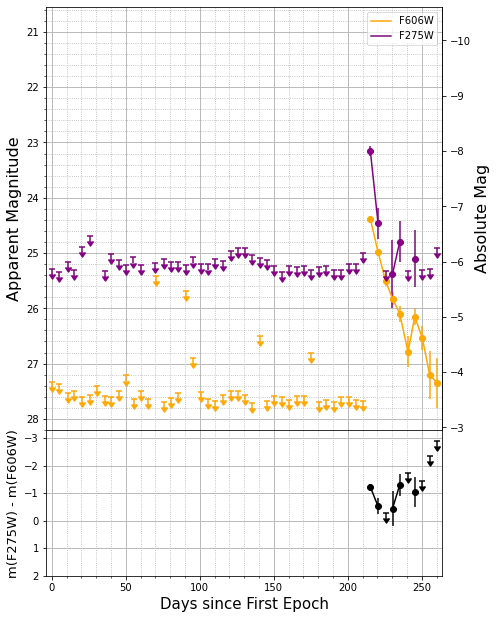}} & \includegraphics[height=3.6in]{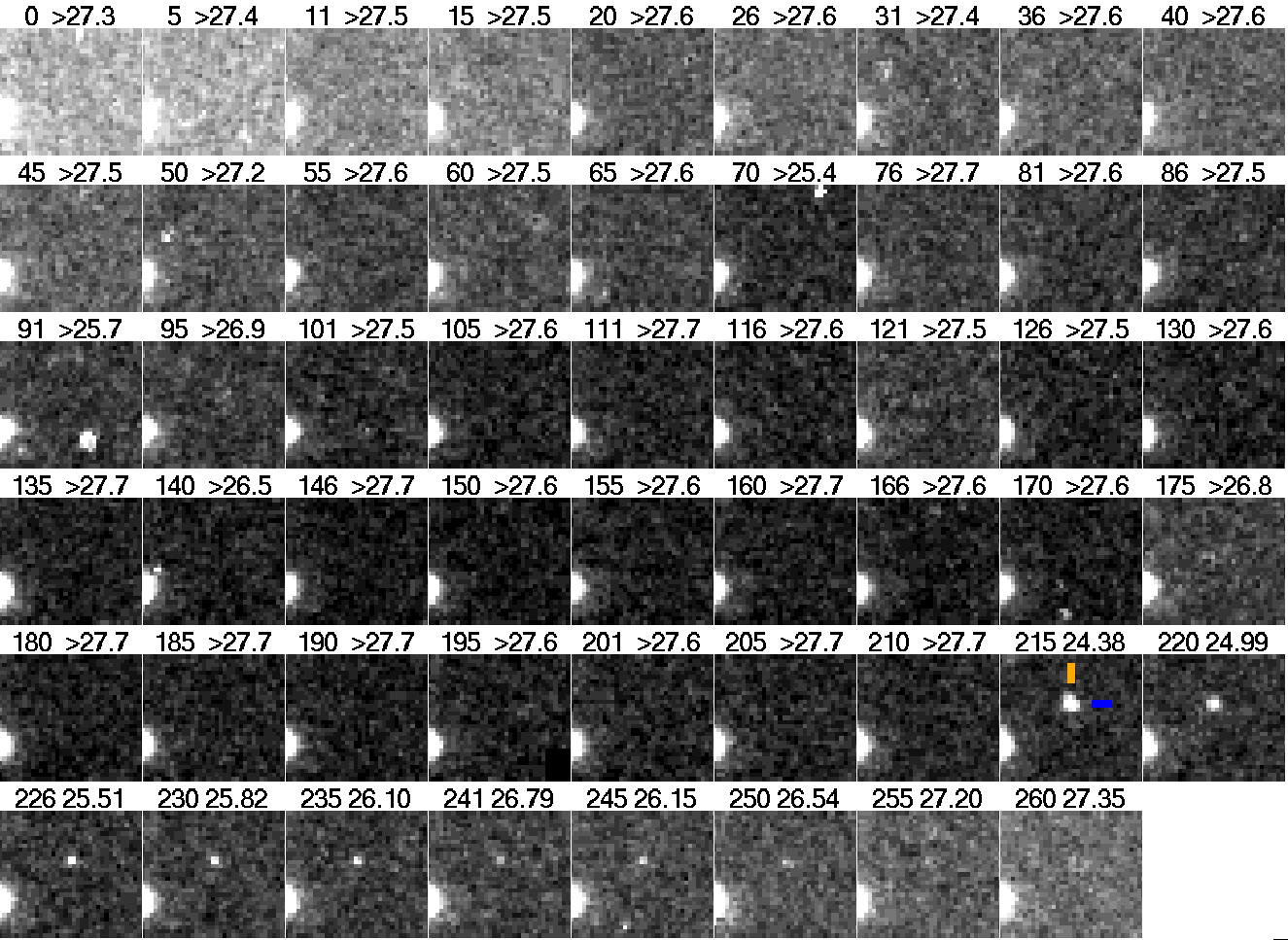} \\
     & \includegraphics[height=3.6in]{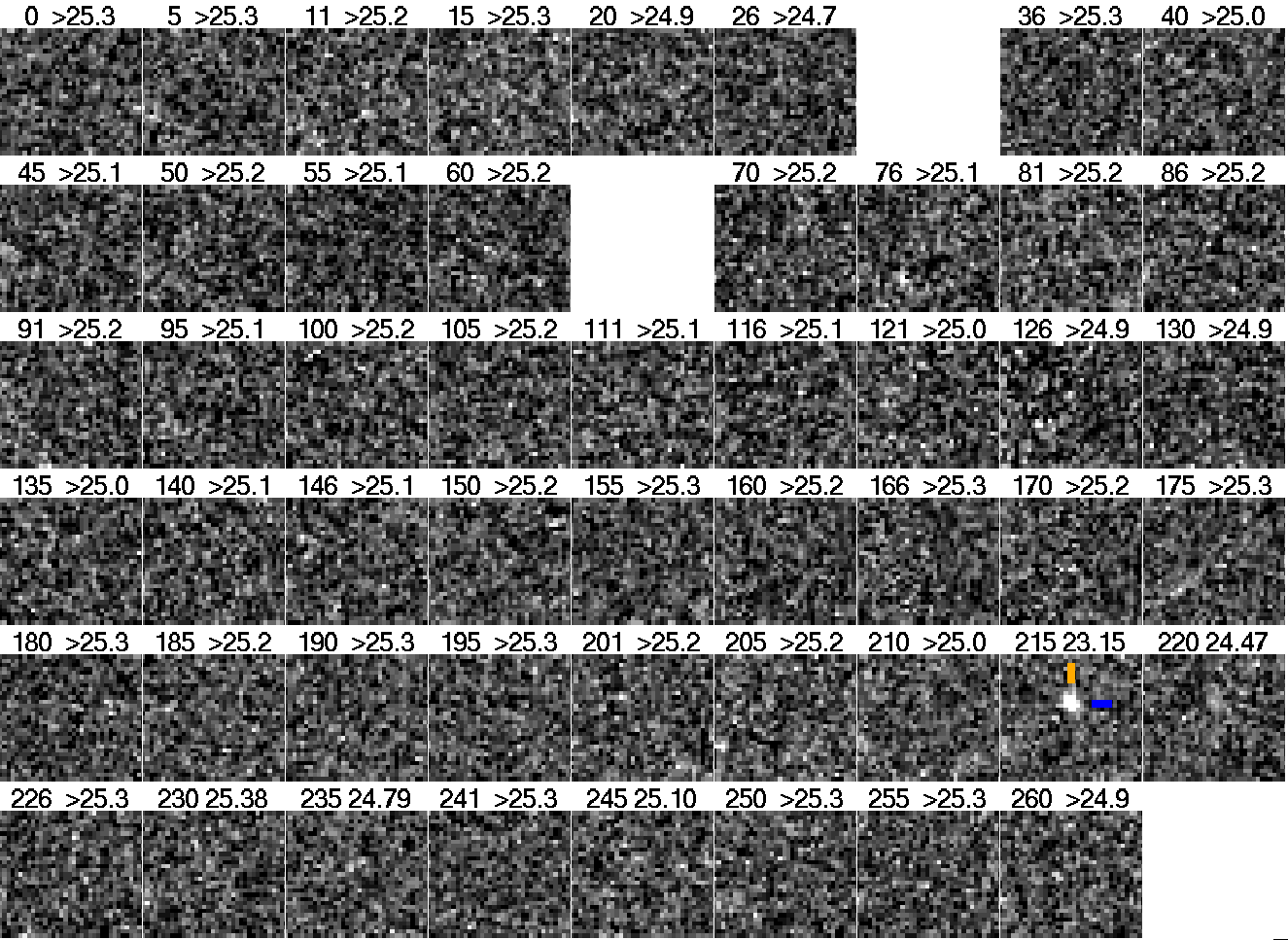}
    \end{tabular}
    \caption{Nova 65}
\end{sidewaysfigure}
\begin{sidewaysfigure}
    \begin{tabular}{cc}
     \multirow{2}{*}[2.6in]{\includegraphics[height=6in]{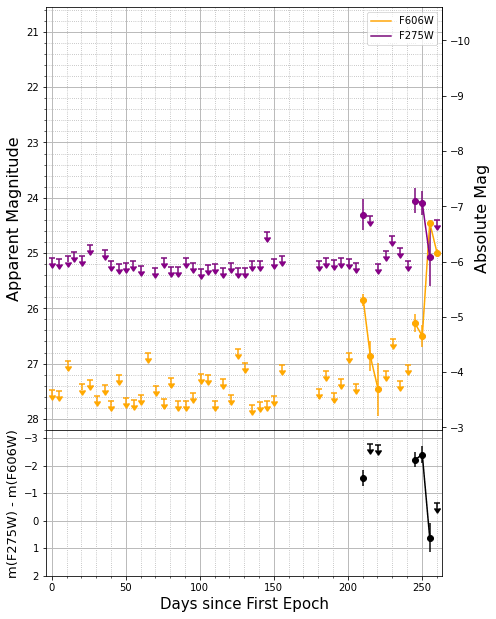}} & \includegraphics[height=3.6in]{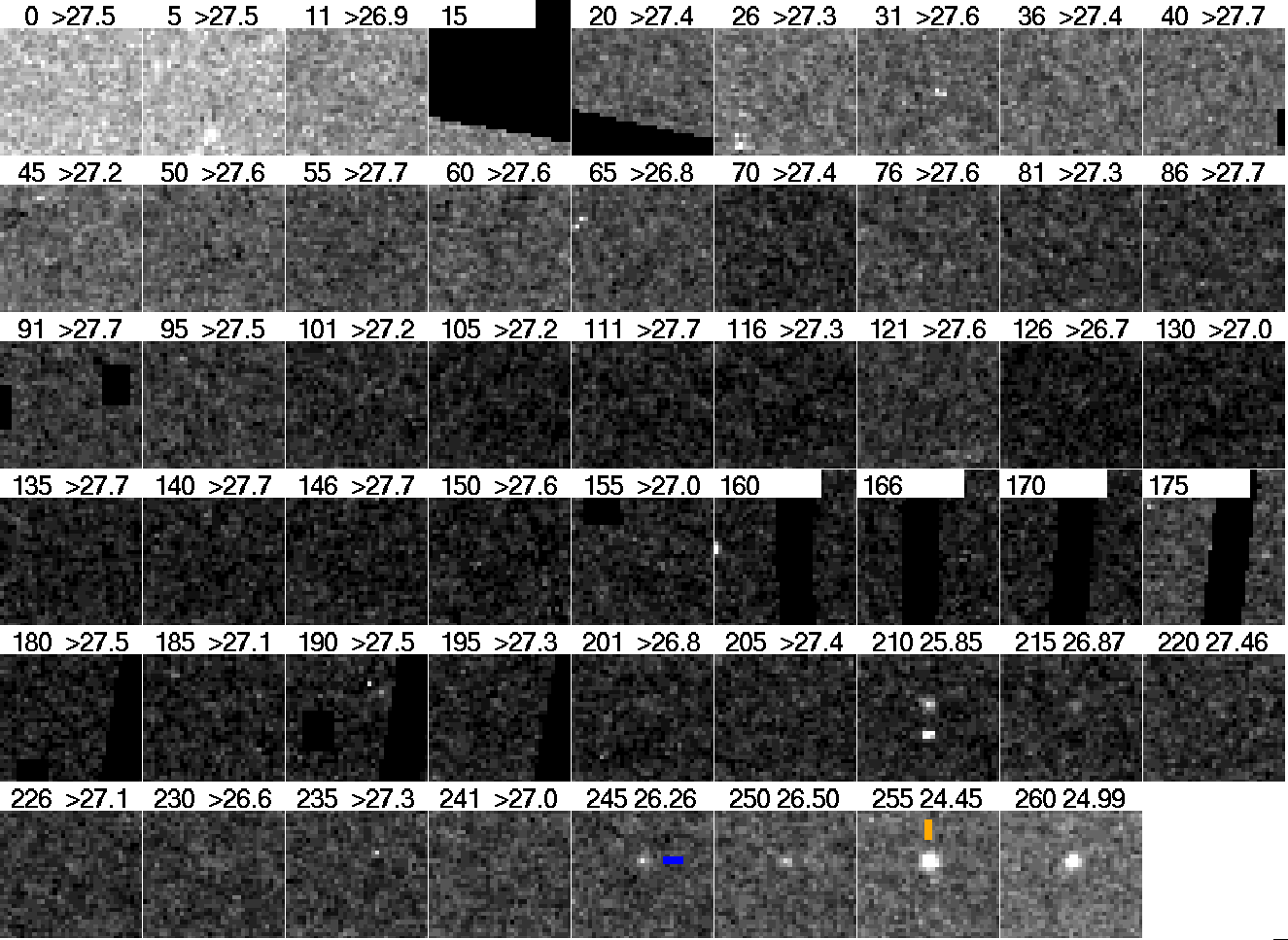} \\
     & \includegraphics[height=3.6in]{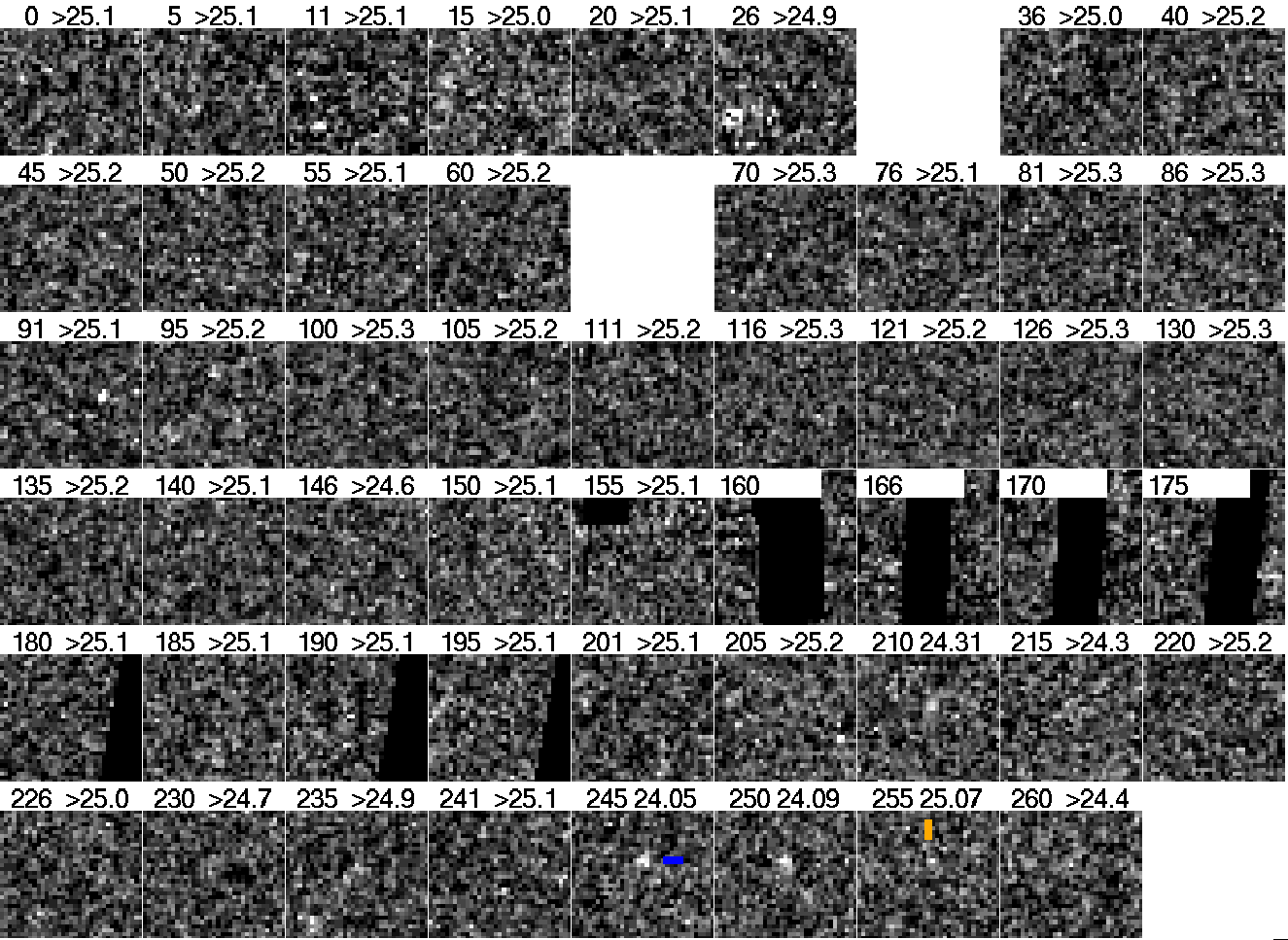}
    \end{tabular}
    \caption{Nova 66}
\end{sidewaysfigure}
\begin{sidewaysfigure}
    \begin{tabular}{cc}
     \multirow{2}{*}[2.6in]{\includegraphics[height=6in]{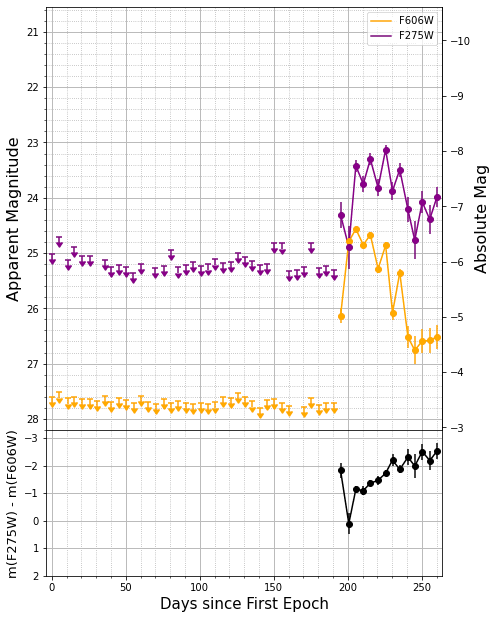}} & \includegraphics[height=3.6in]{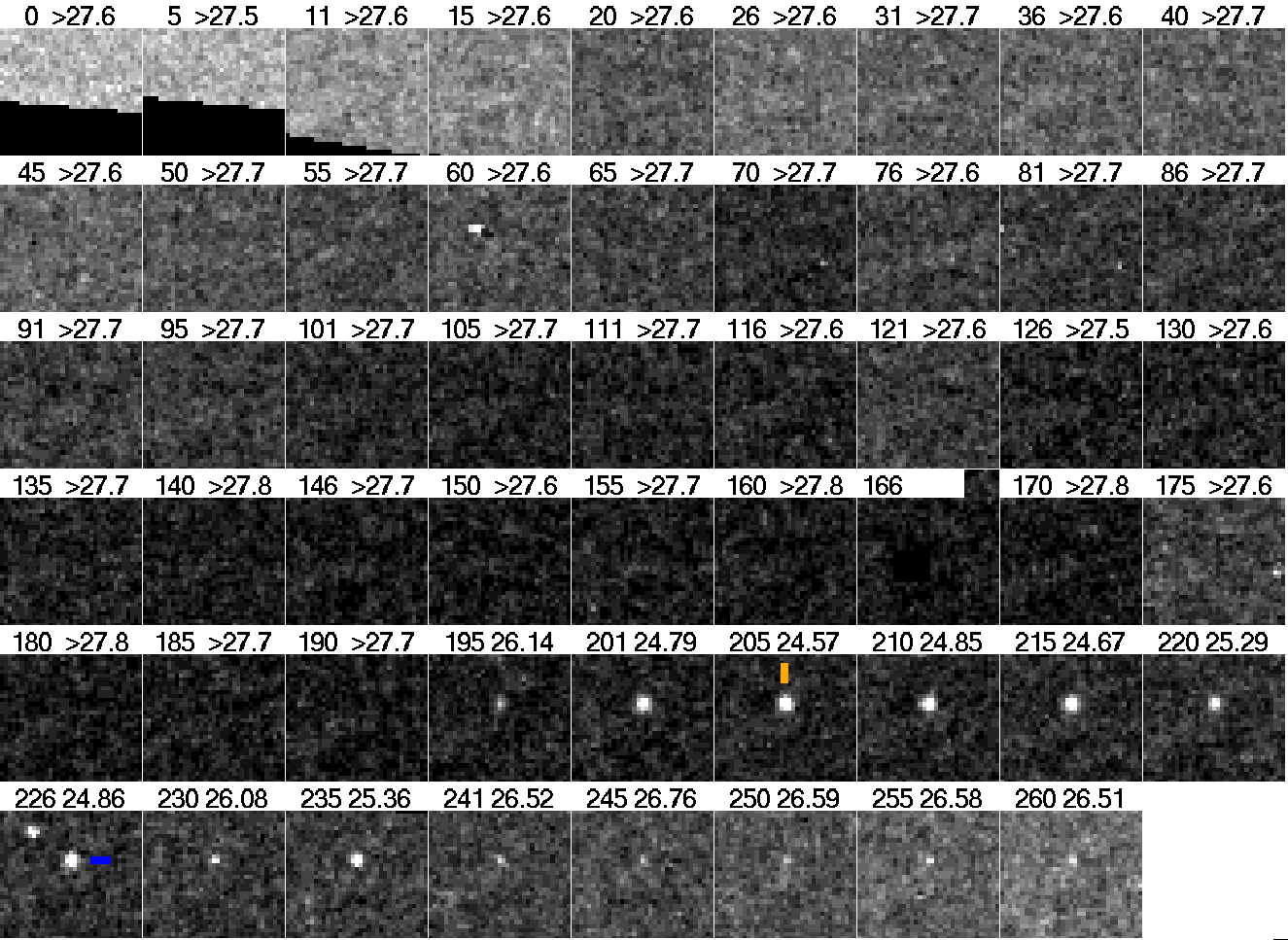} \\
     & \includegraphics[height=3.6in]{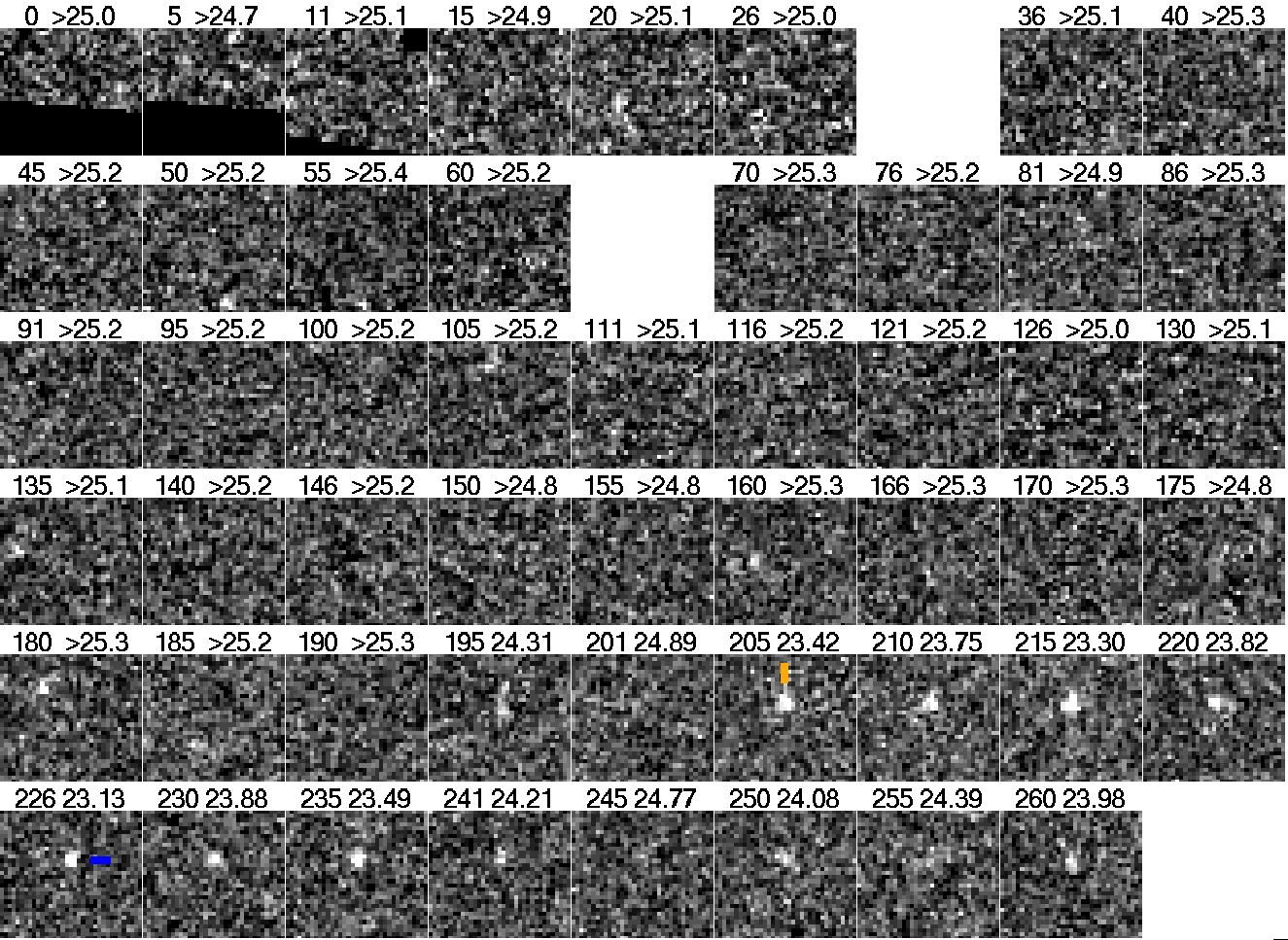}
    \end{tabular}
    \caption{Nova 67}
\end{sidewaysfigure}
\begin{sidewaysfigure}
    \begin{tabular}{cc}
     \multirow{2}{*}[2.6in]{\includegraphics[height=6in]{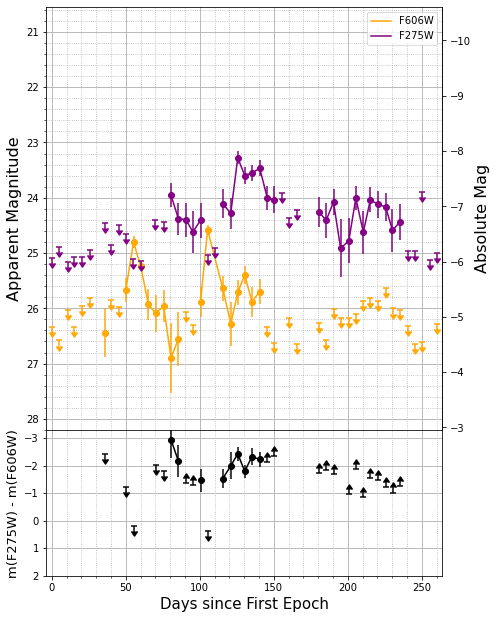}} & \includegraphics[height=3.6in]{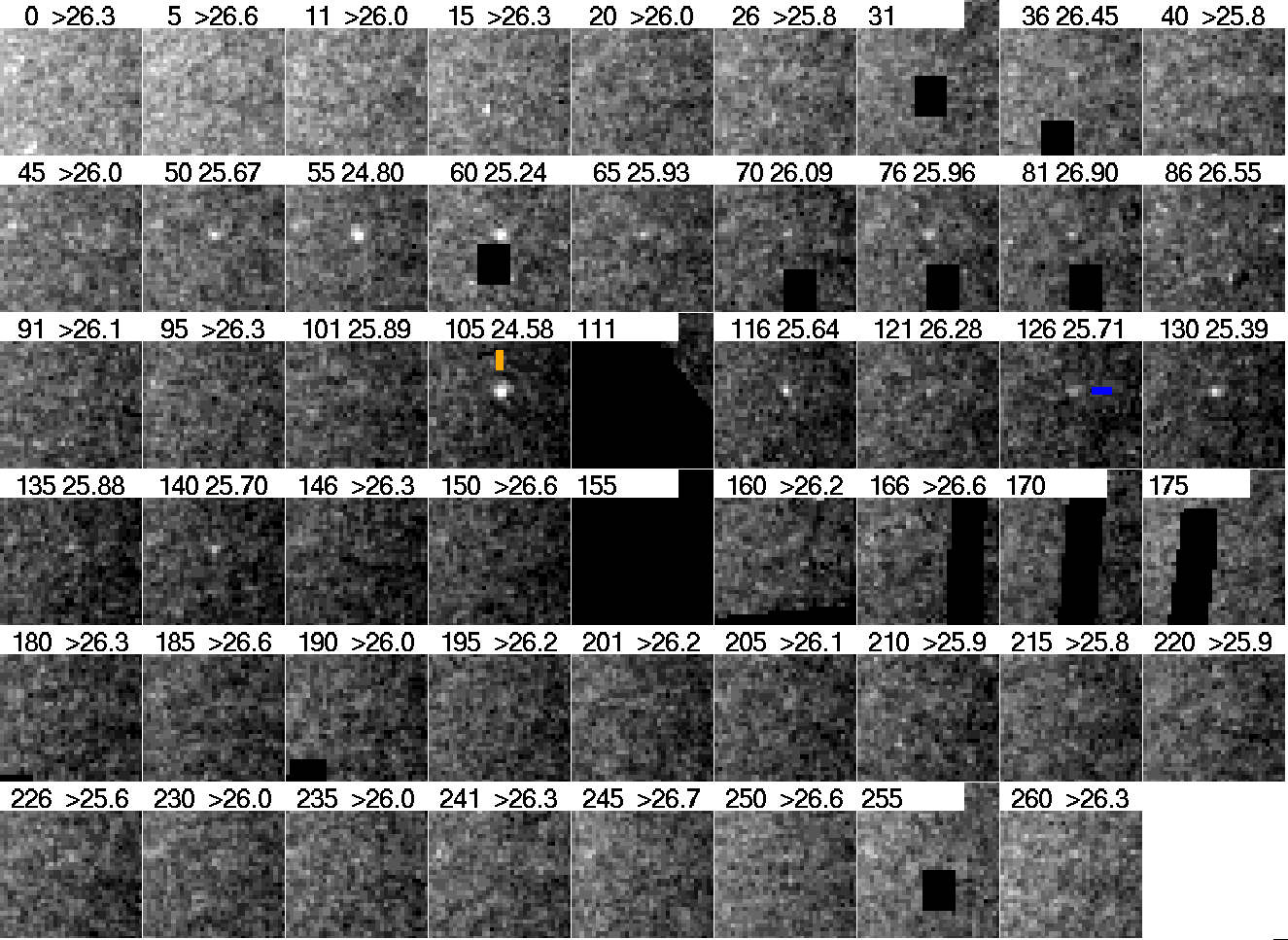} \\
     & \includegraphics[height=3.6in]{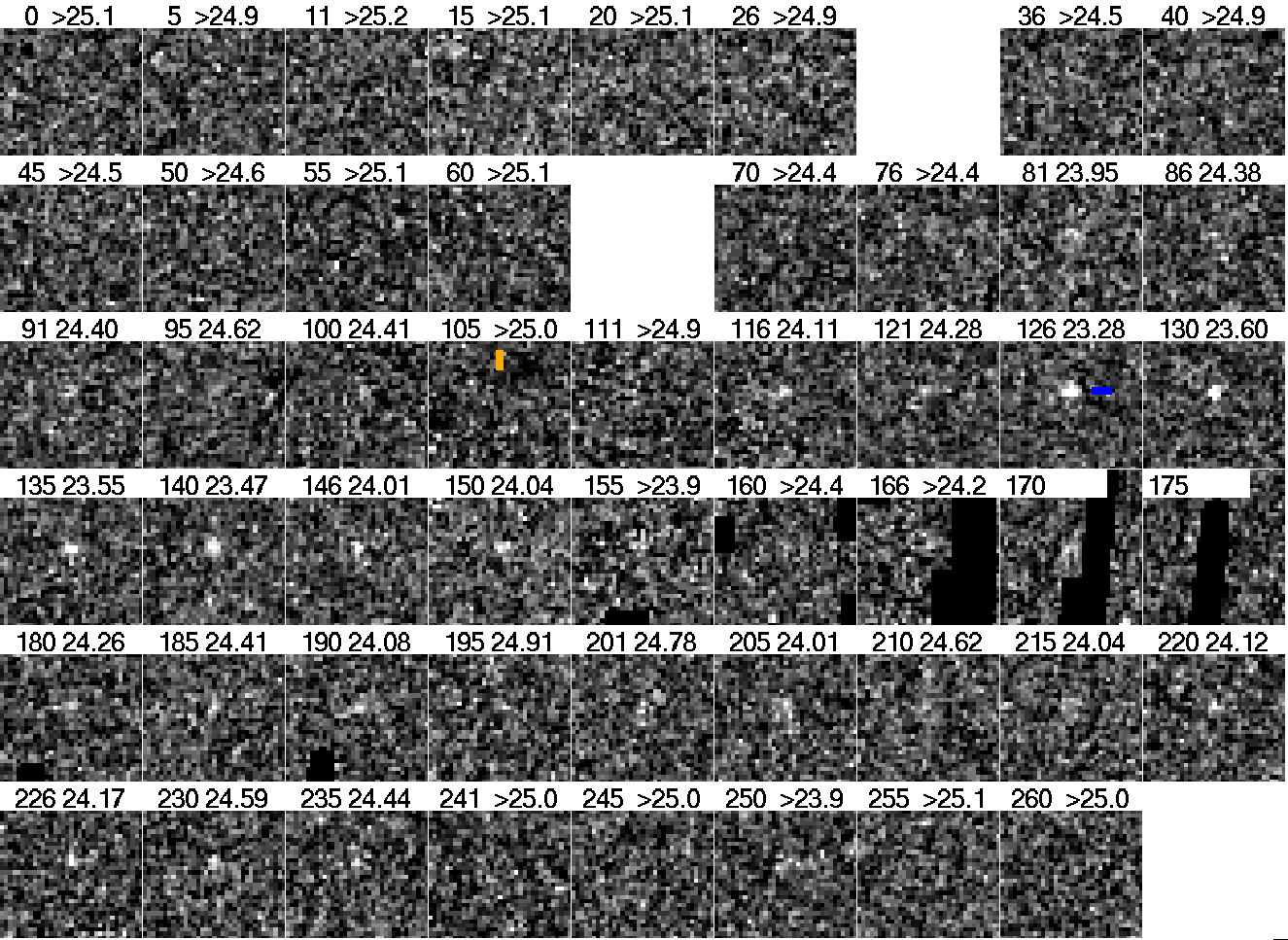}
    \end{tabular}
    \caption{Nova 68}
\end{sidewaysfigure}
\begin{sidewaysfigure}
    \begin{tabular}{cc}
     \multirow{2}{*}[2.6in]{\includegraphics[height=6in]{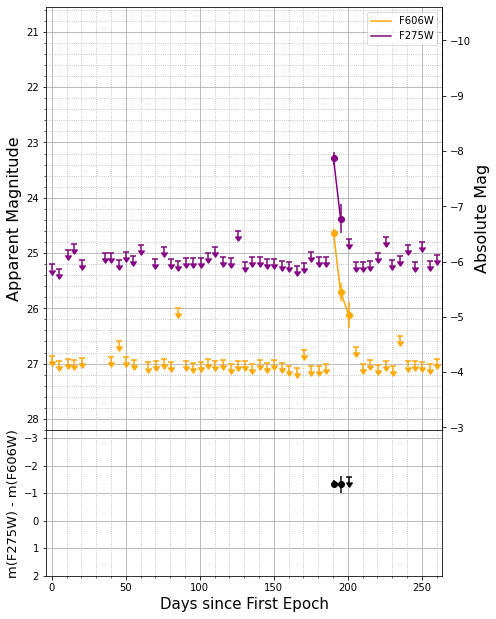}} & \includegraphics[height=3.6in]{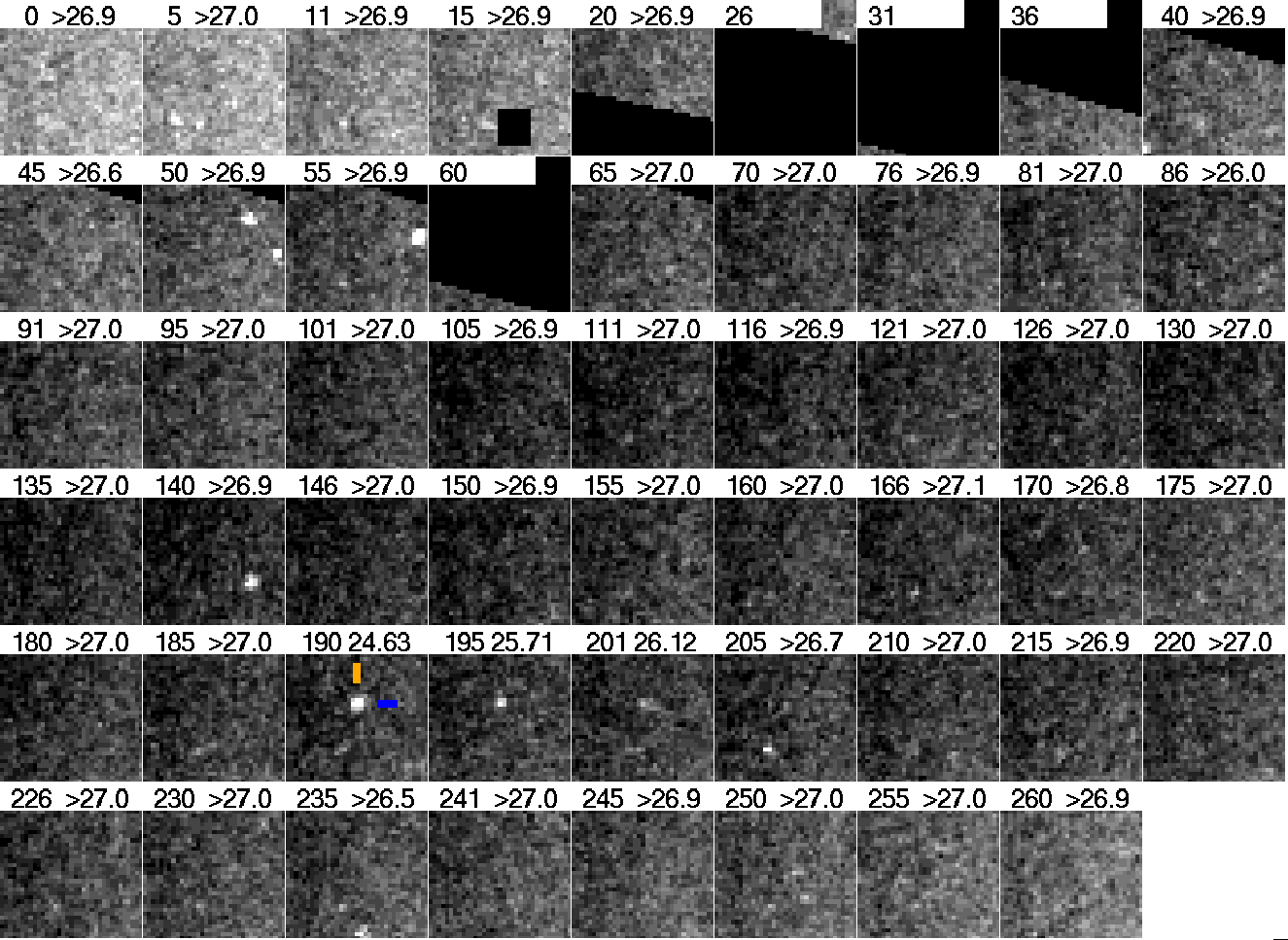} \\
     & \includegraphics[height=3.6in]{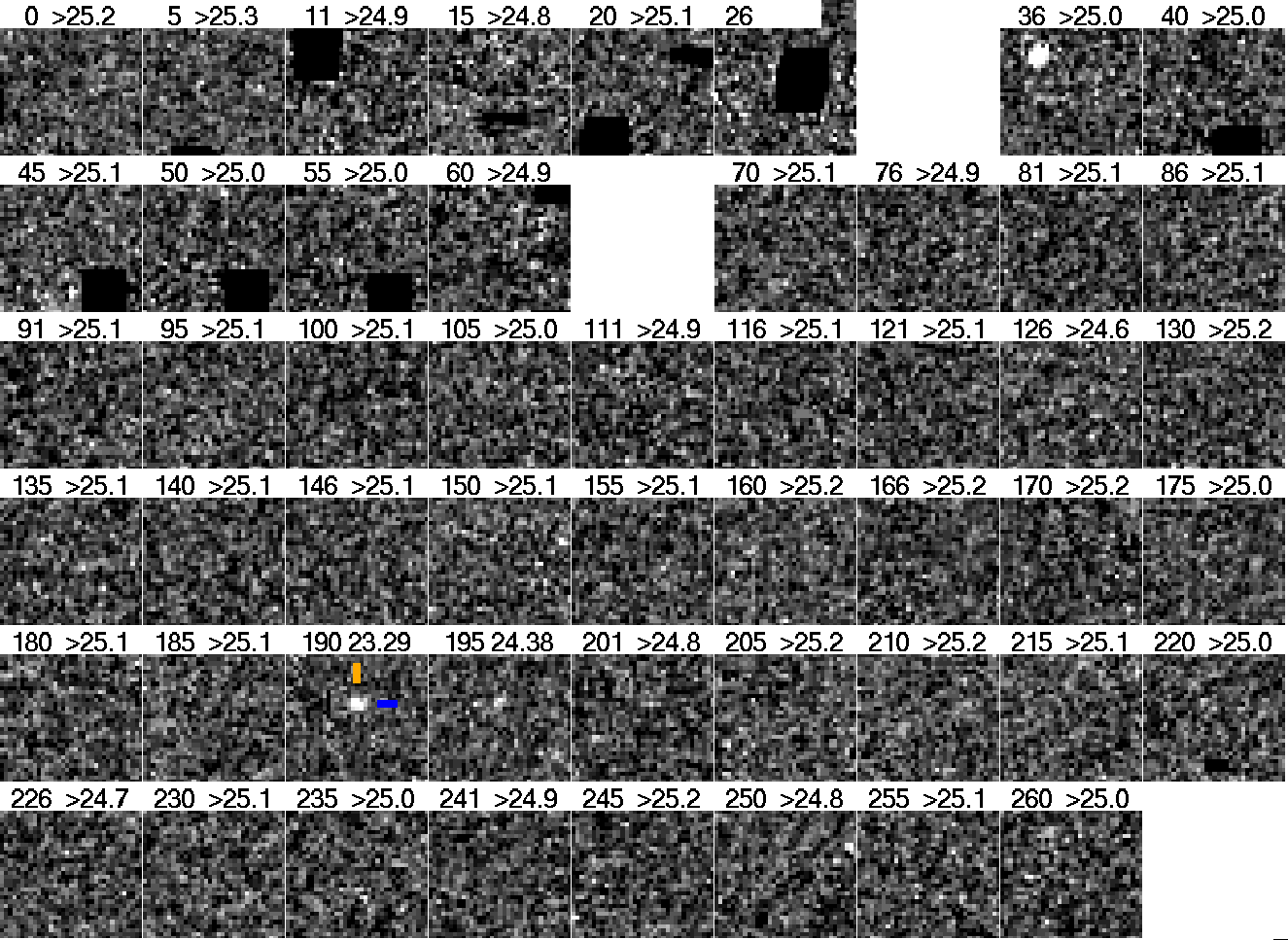}
    \end{tabular}
    \caption{Nova 69}
\end{sidewaysfigure}
\begin{sidewaysfigure}
    \begin{tabular}{cc}
     \multirow{2}{*}[2.6in]{\includegraphics[height=6in]{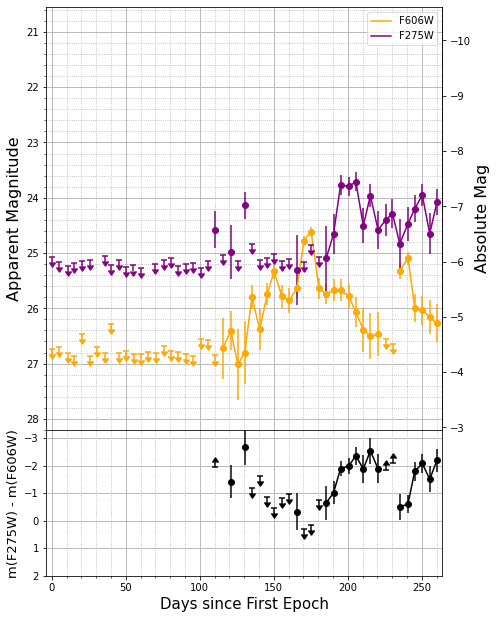}} & \includegraphics[height=3.6in]{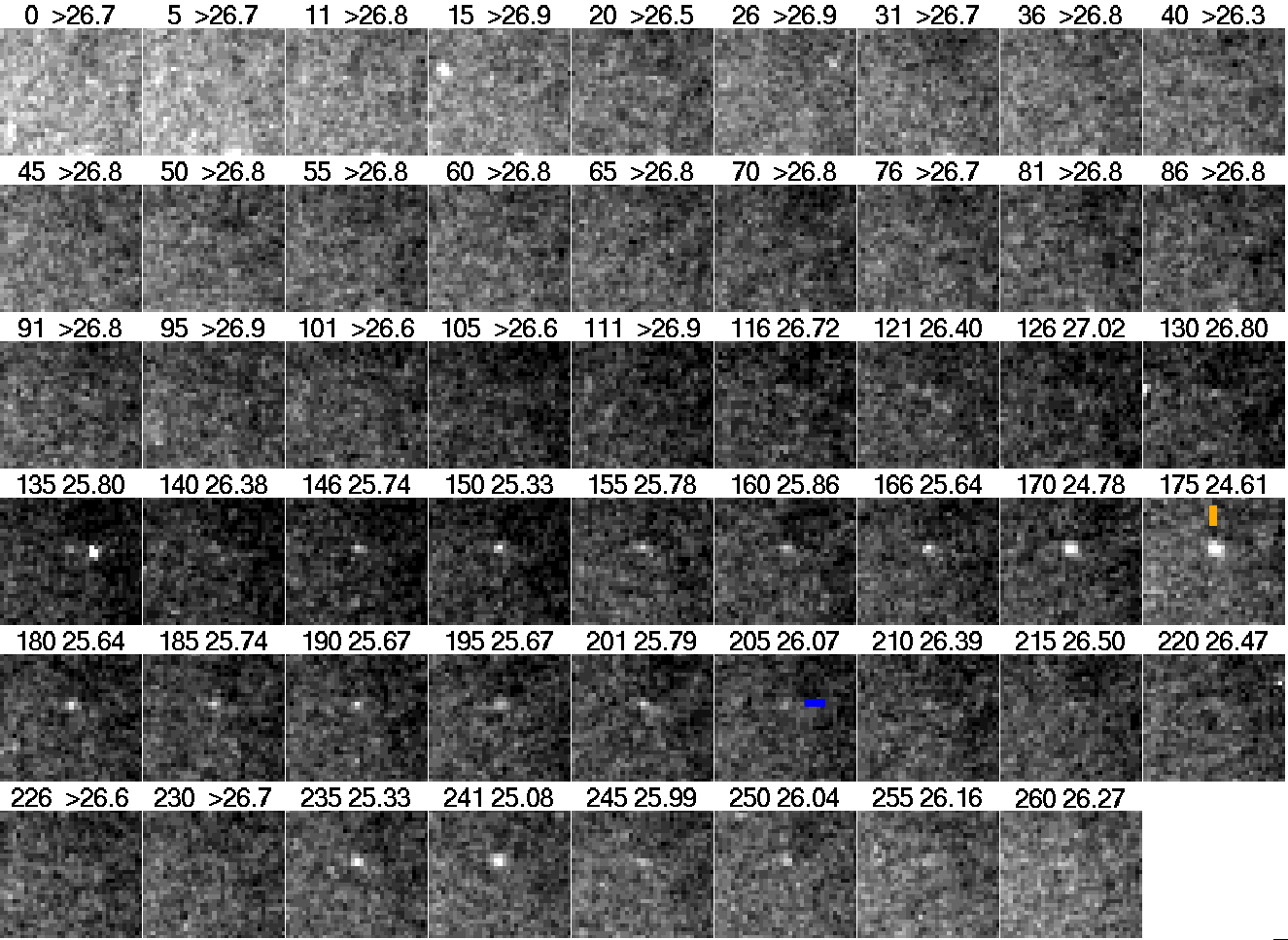} \\
     & \includegraphics[height=3.6in]{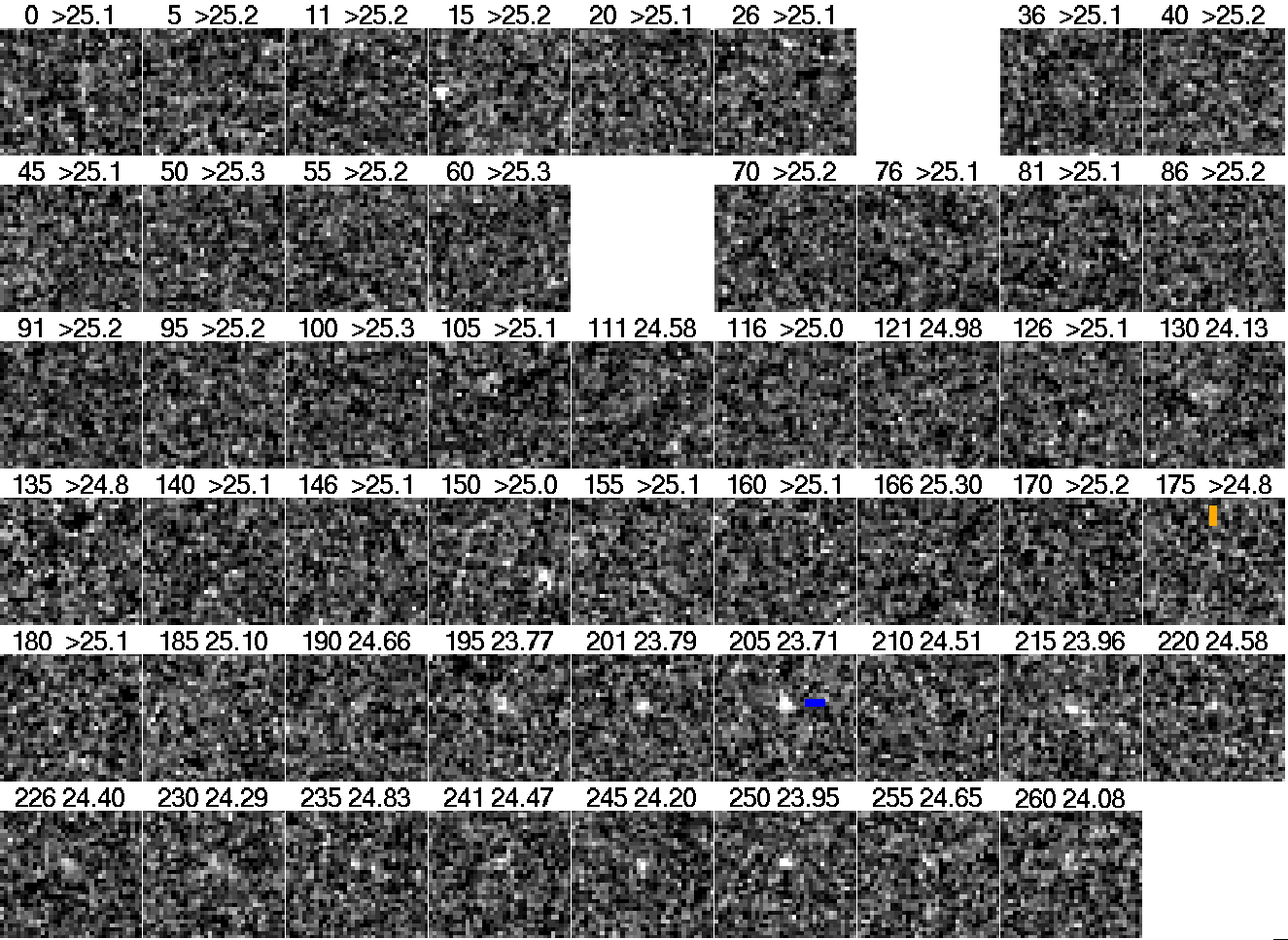}
    \end{tabular}
    \caption{Nova 70}
\end{sidewaysfigure}
\begin{sidewaysfigure}
    \begin{tabular}{cc}
     \multirow{2}{*}[2.6in]{\includegraphics[height=6in]{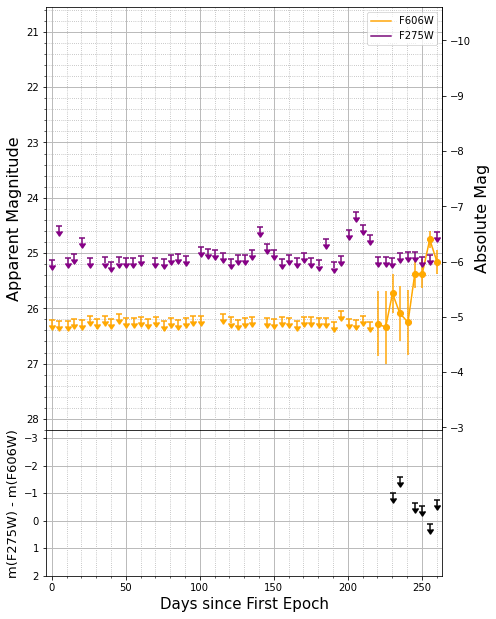}} & \includegraphics[height=3.6in]{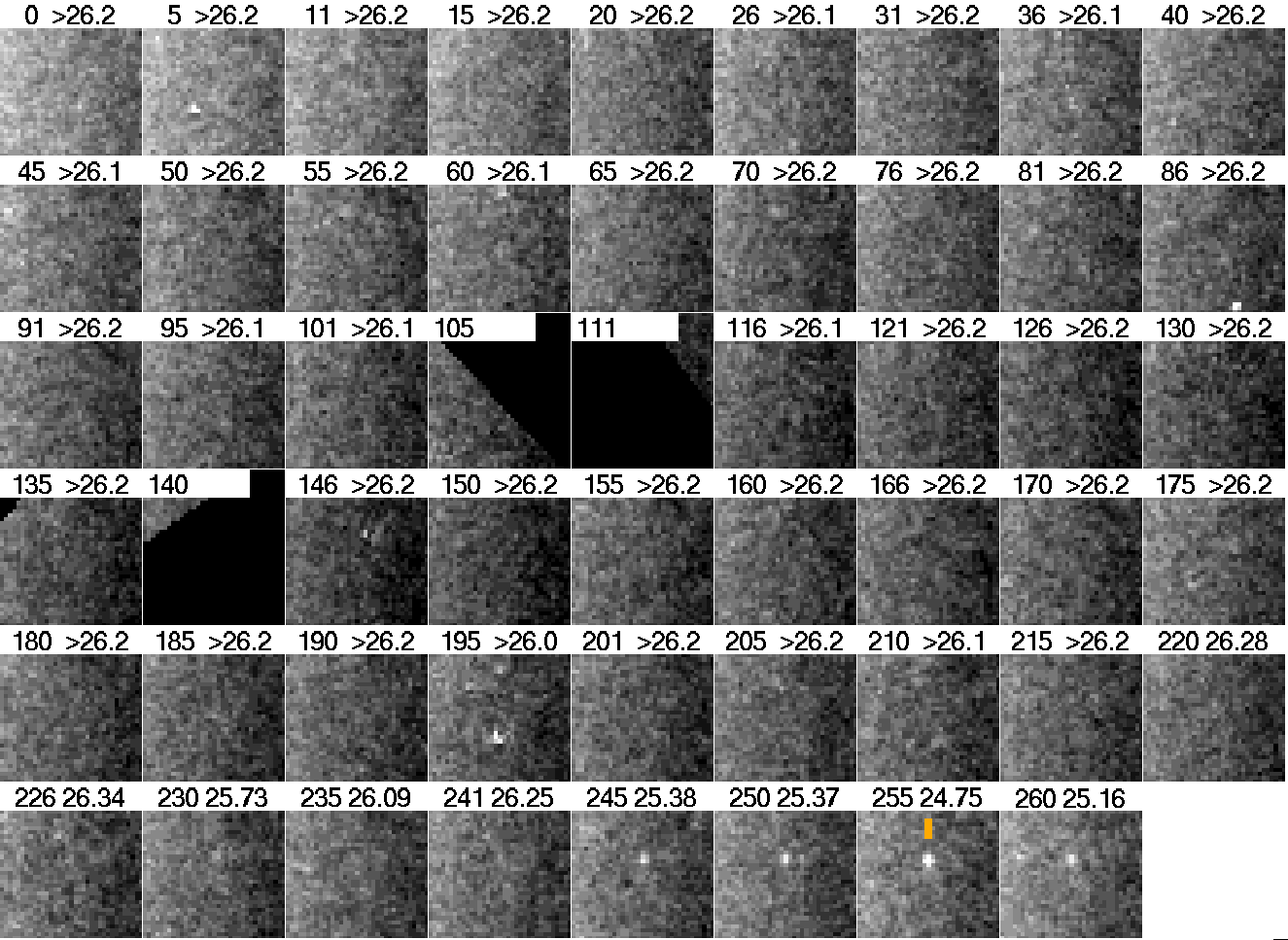} \\
     & \includegraphics[height=3.6in]{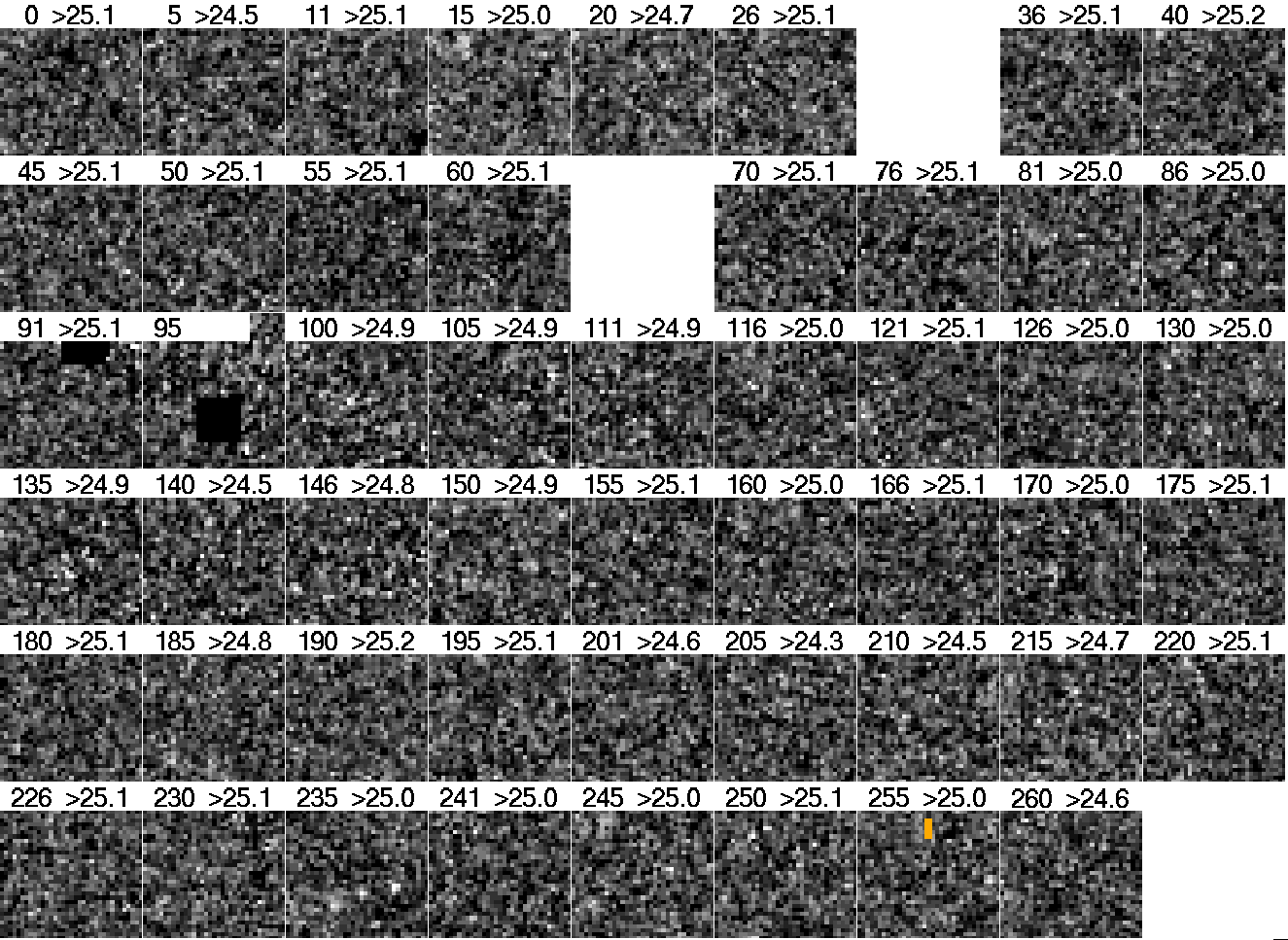}
    \end{tabular}
    \caption{Nova 71}
\end{sidewaysfigure}
\begin{sidewaysfigure}
    \begin{tabular}{cc}
     \multirow{2}{*}[2.6in]{\includegraphics[height=6in]{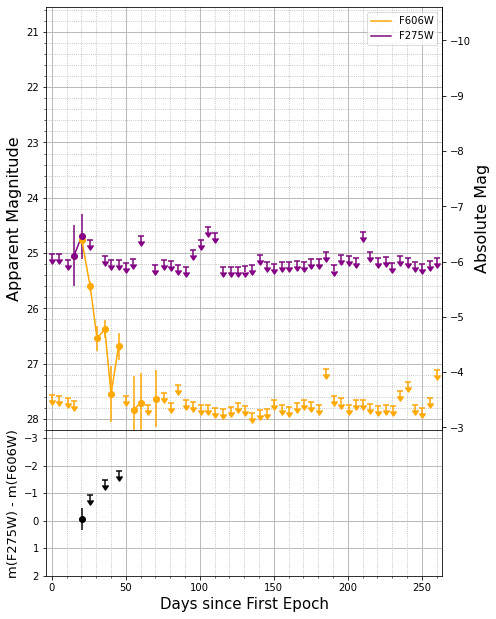}} & \includegraphics[height=3.6in]{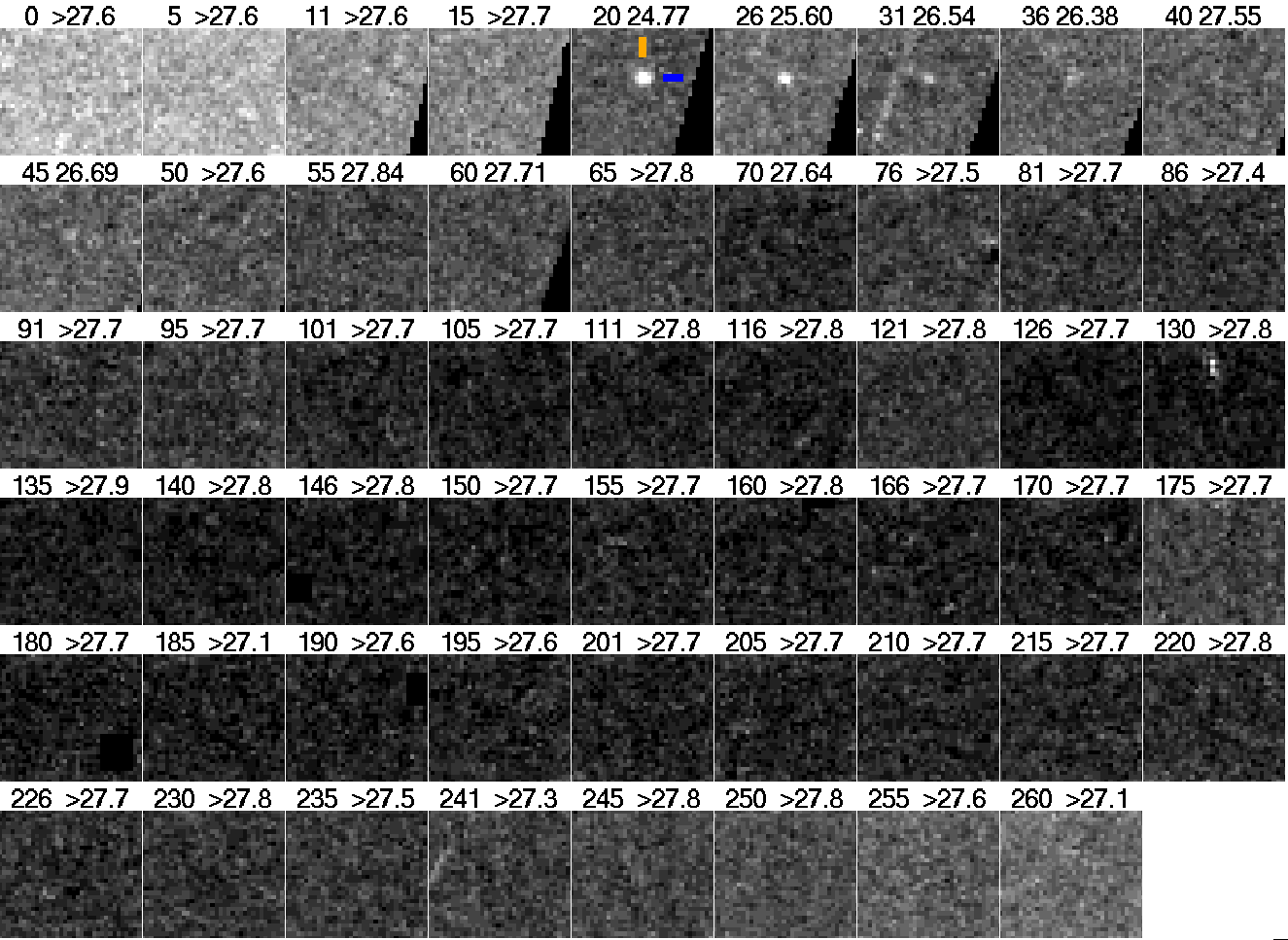} \\
     & \includegraphics[height=3.6in]{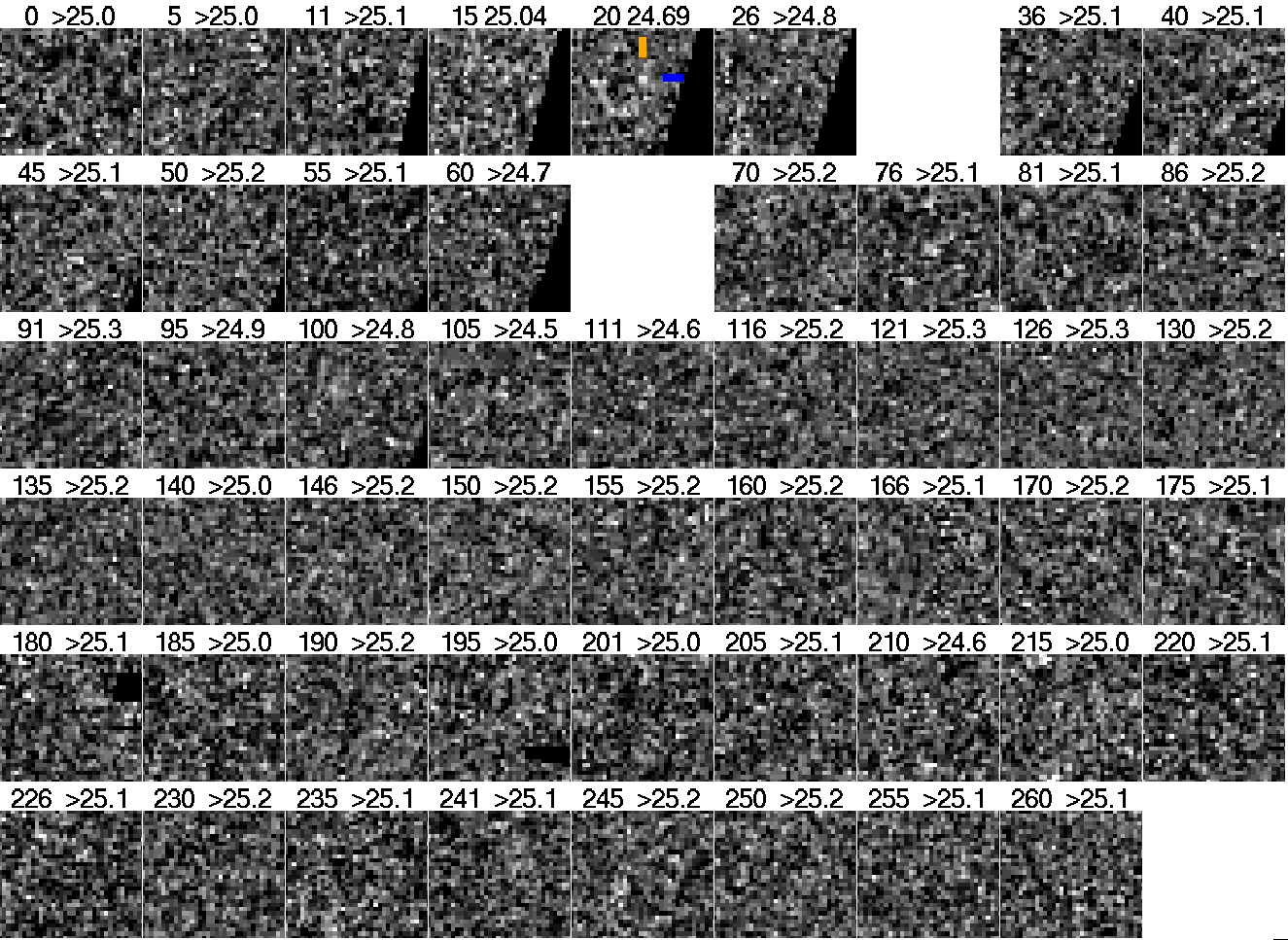}
    \end{tabular}
    \caption{Nova 72}
\end{sidewaysfigure}
\begin{sidewaysfigure}
    \begin{tabular}{cc}
     \multirow{2}{*}[2.6in]{\includegraphics[height=6in]{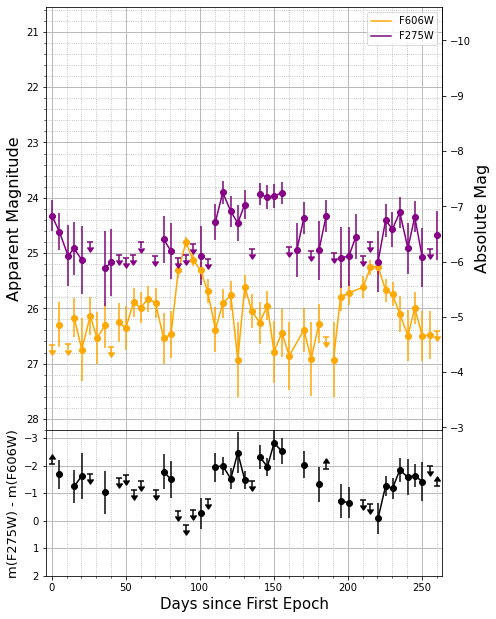}} & \includegraphics[height=3.6in]{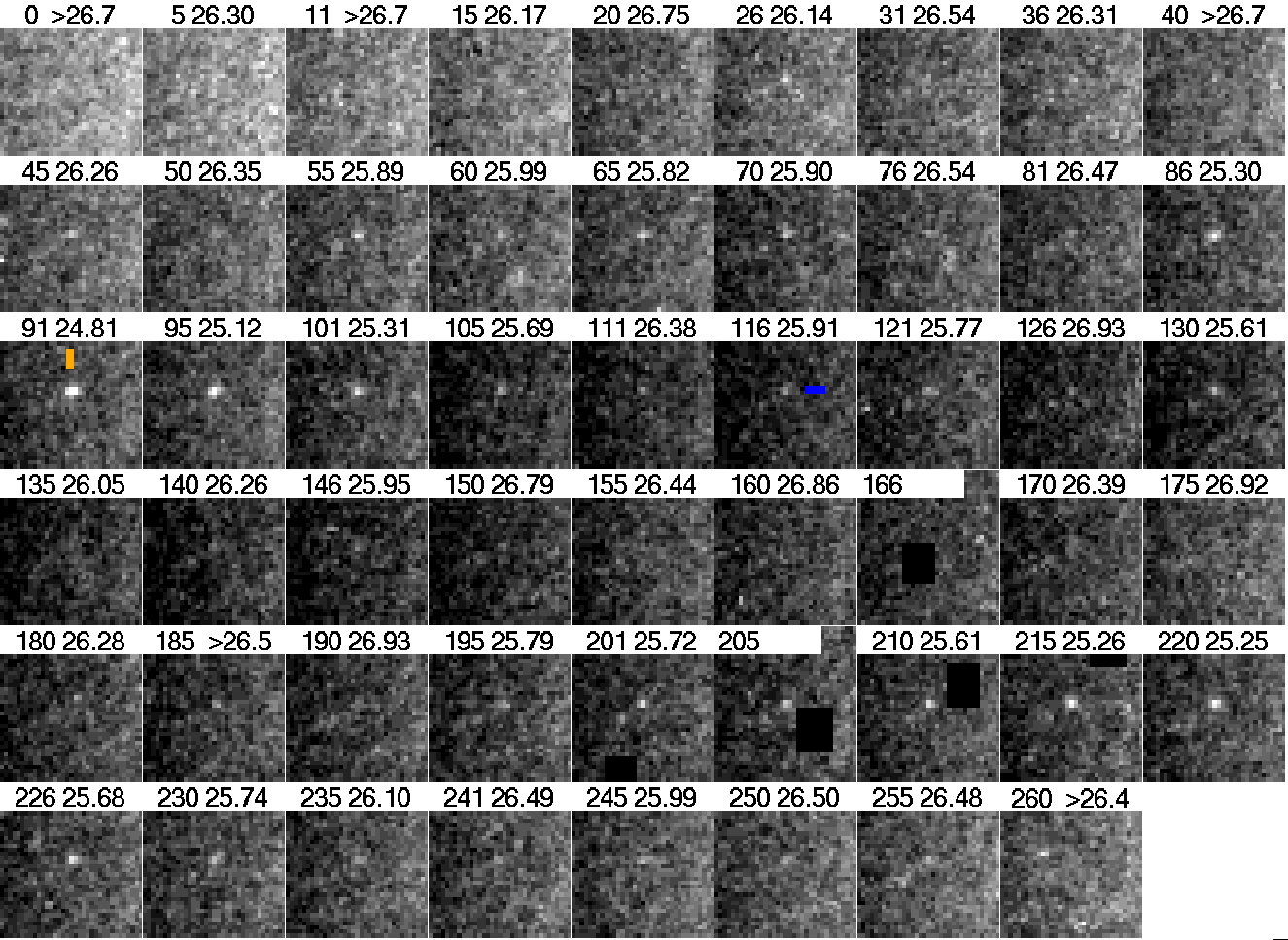} \\
     & \includegraphics[height=3.6in]{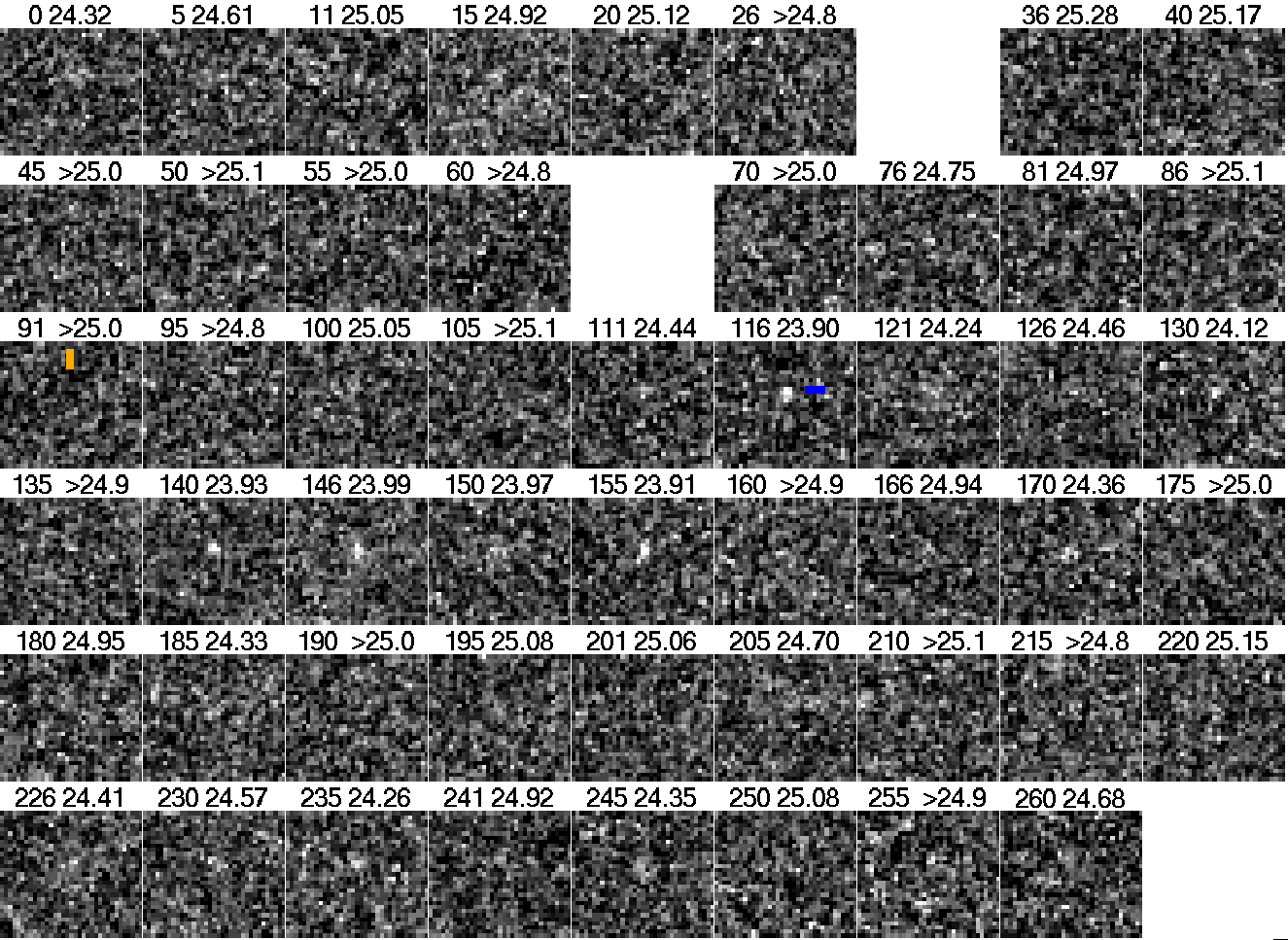}
    \end{tabular}
    \caption{Nova 73}
\end{sidewaysfigure}
\begin{sidewaysfigure}
    \begin{tabular}{cc}
     \multirow{2}{*}[2.6in]{\includegraphics[height=6in]{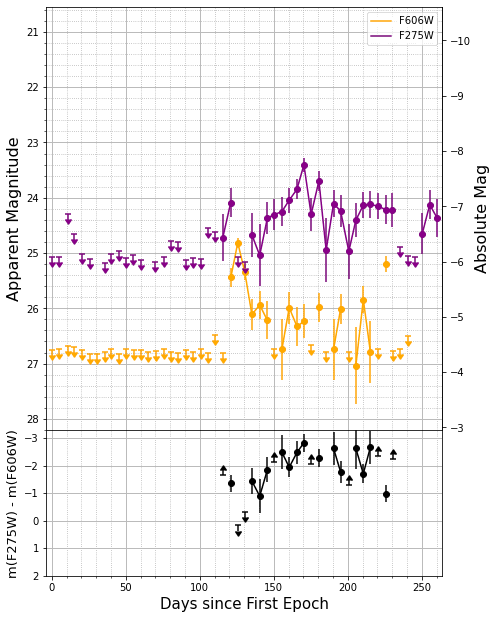}} & \includegraphics[height=3.6in]{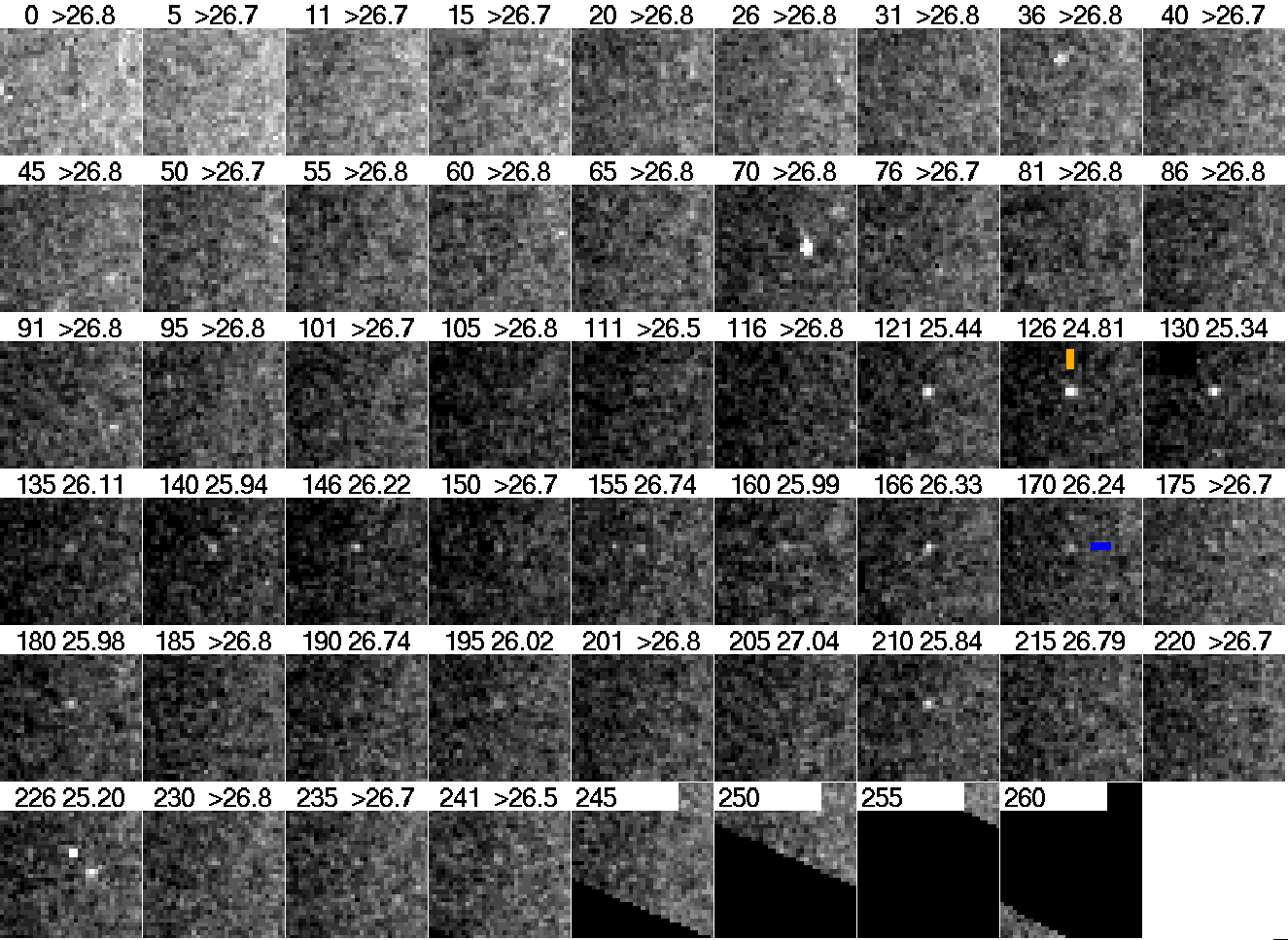} \\
     & \includegraphics[height=3.6in]{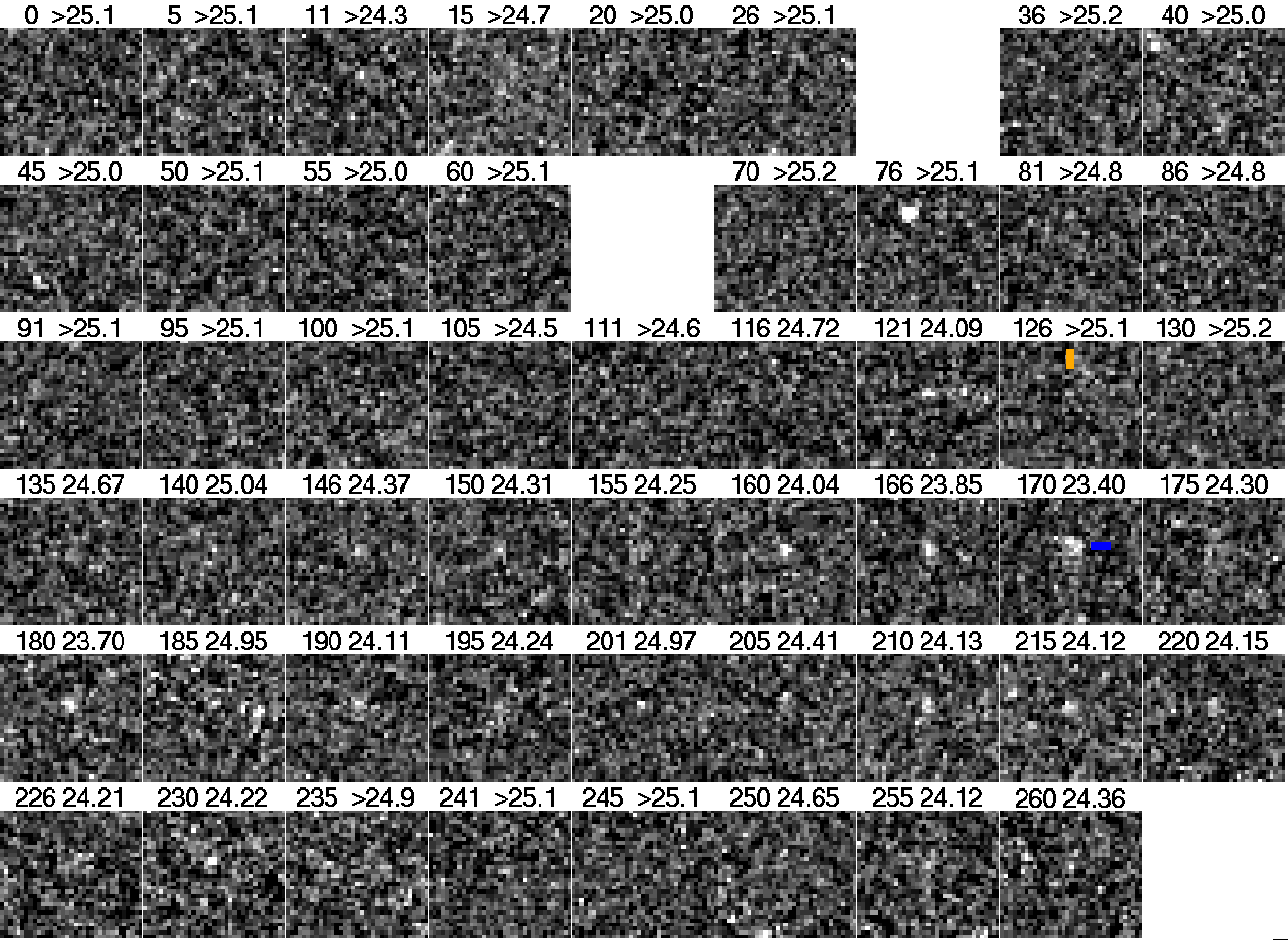}
    \end{tabular}
    \caption{Nova 74}
\end{sidewaysfigure}
\begin{sidewaysfigure}
    \begin{tabular}{cc}
     \multirow{2}{*}[2.6in]{\includegraphics[height=6in]{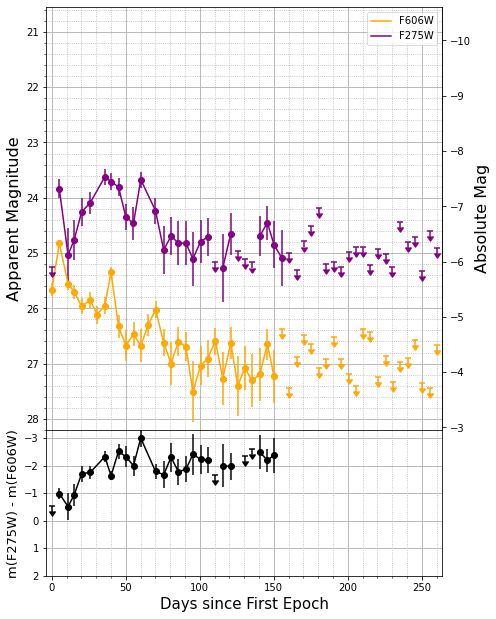}} & \includegraphics[height=3.6in]{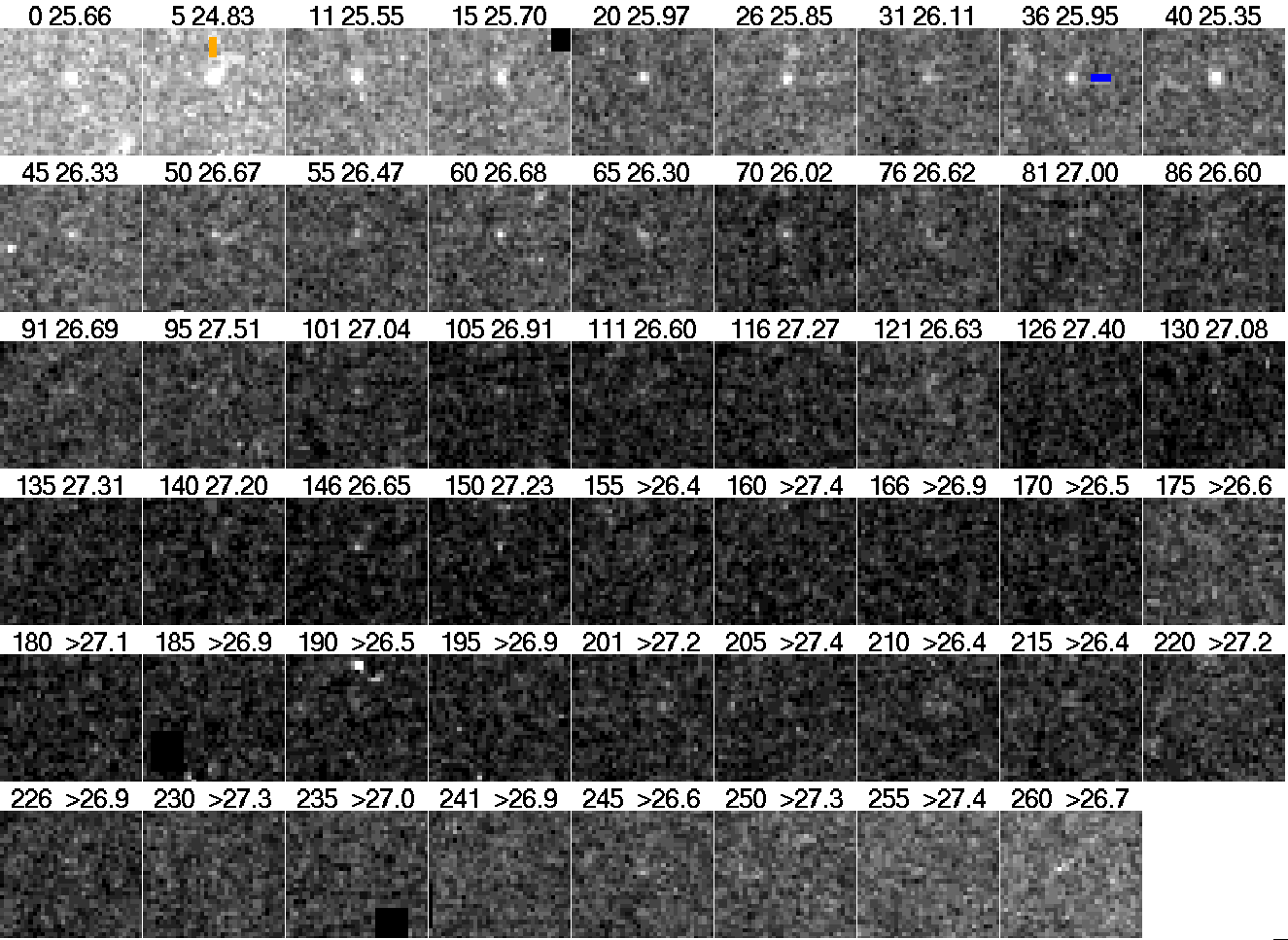} \\
     & \includegraphics[height=3.6in]{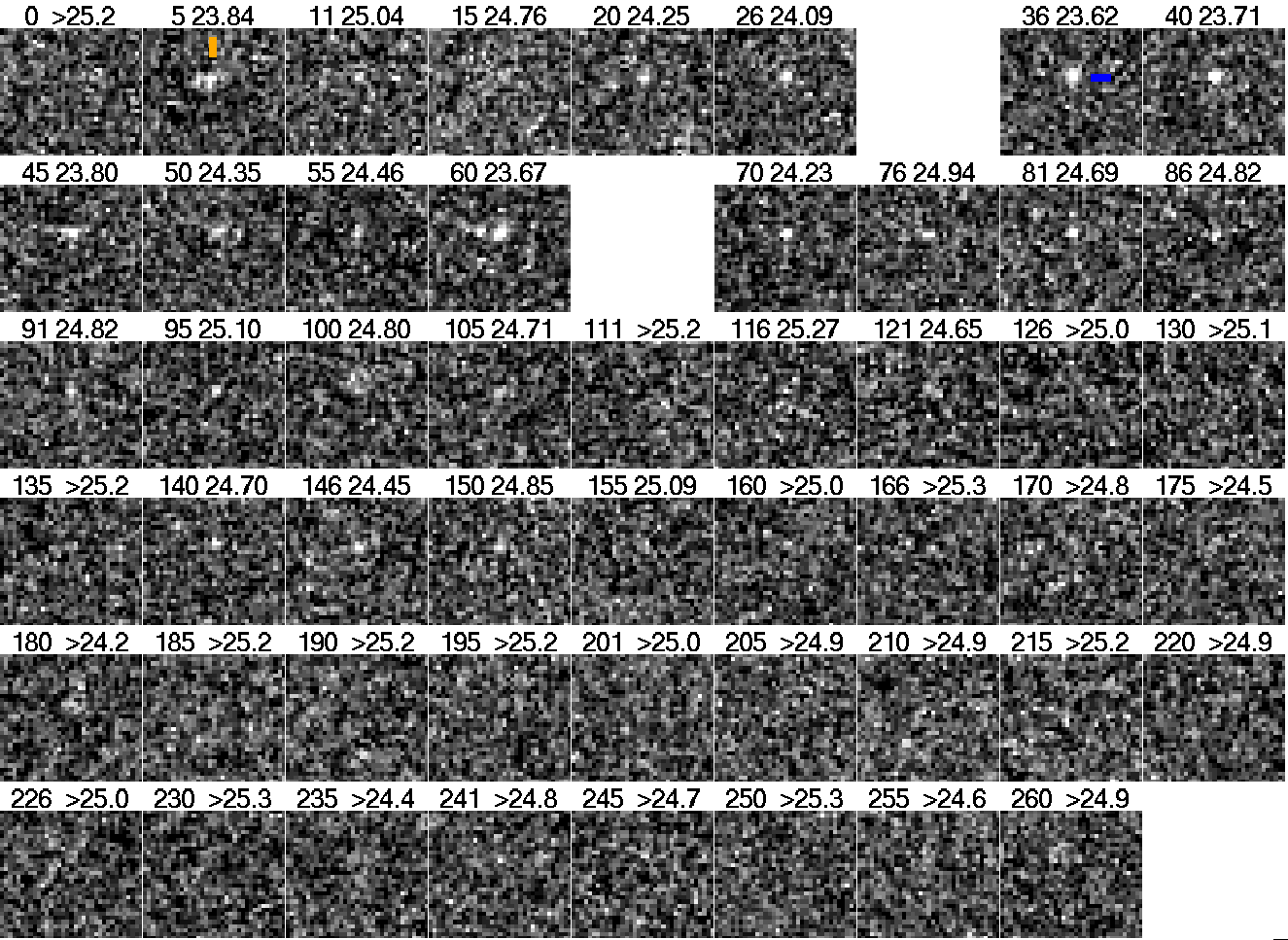}
    \end{tabular}
    \caption{Nova 75}
\end{sidewaysfigure}
\begin{sidewaysfigure}
    \begin{tabular}{cc}
     \multirow{2}{*}[2.6in]{\includegraphics[height=6in]{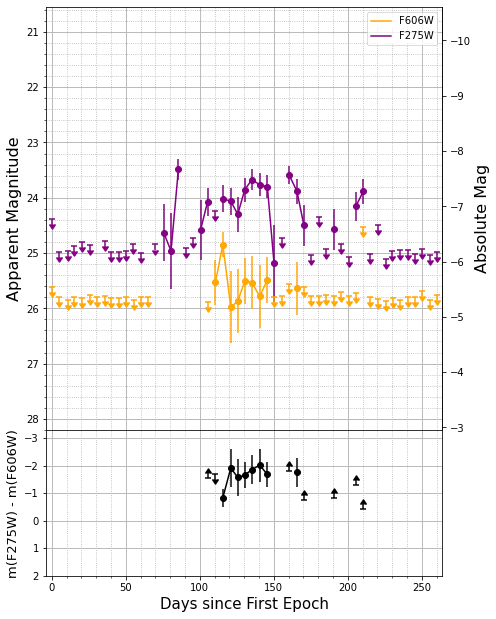}} & \includegraphics[height=3.6in]{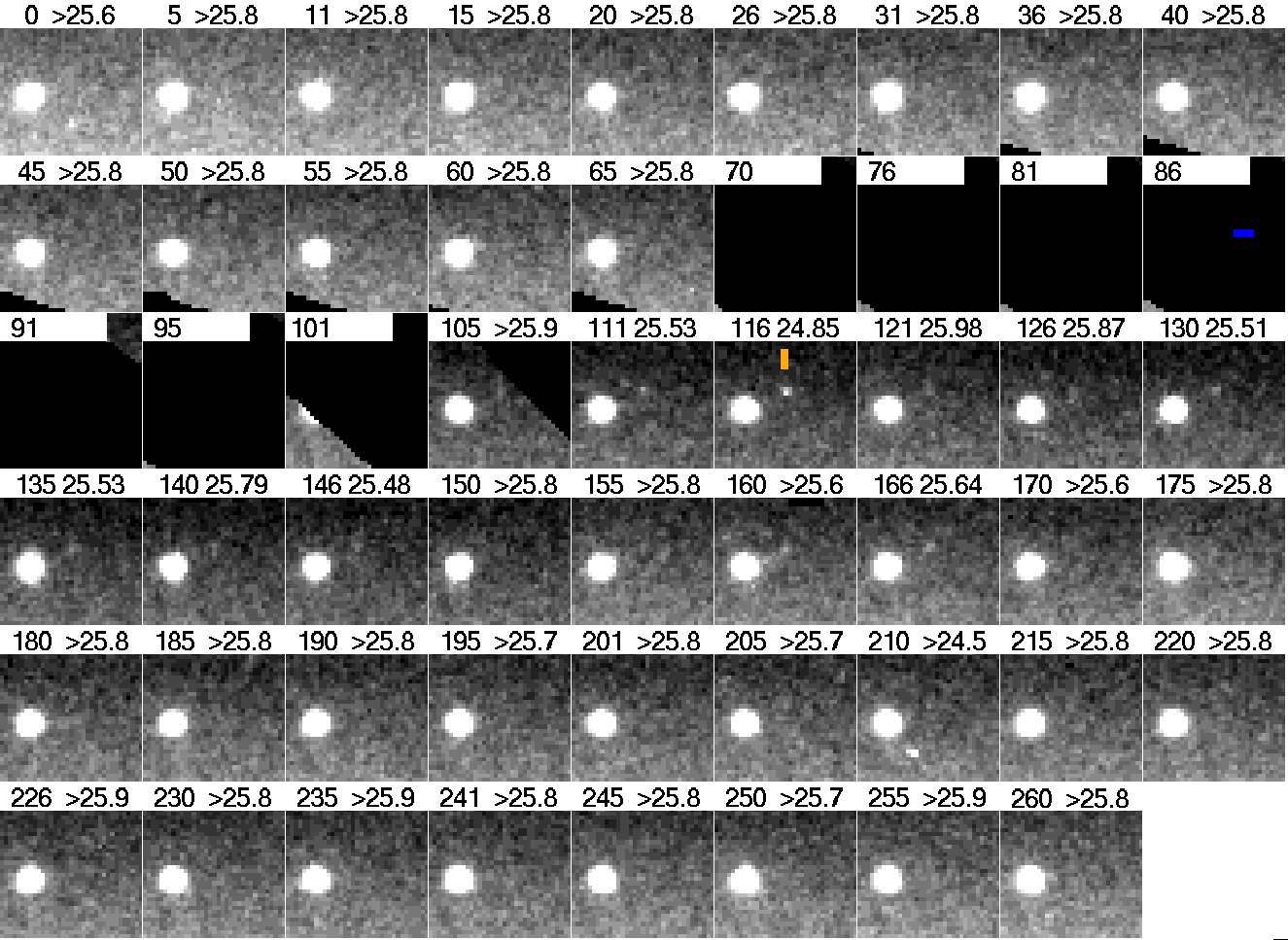} \\
     & \includegraphics[height=3.6in]{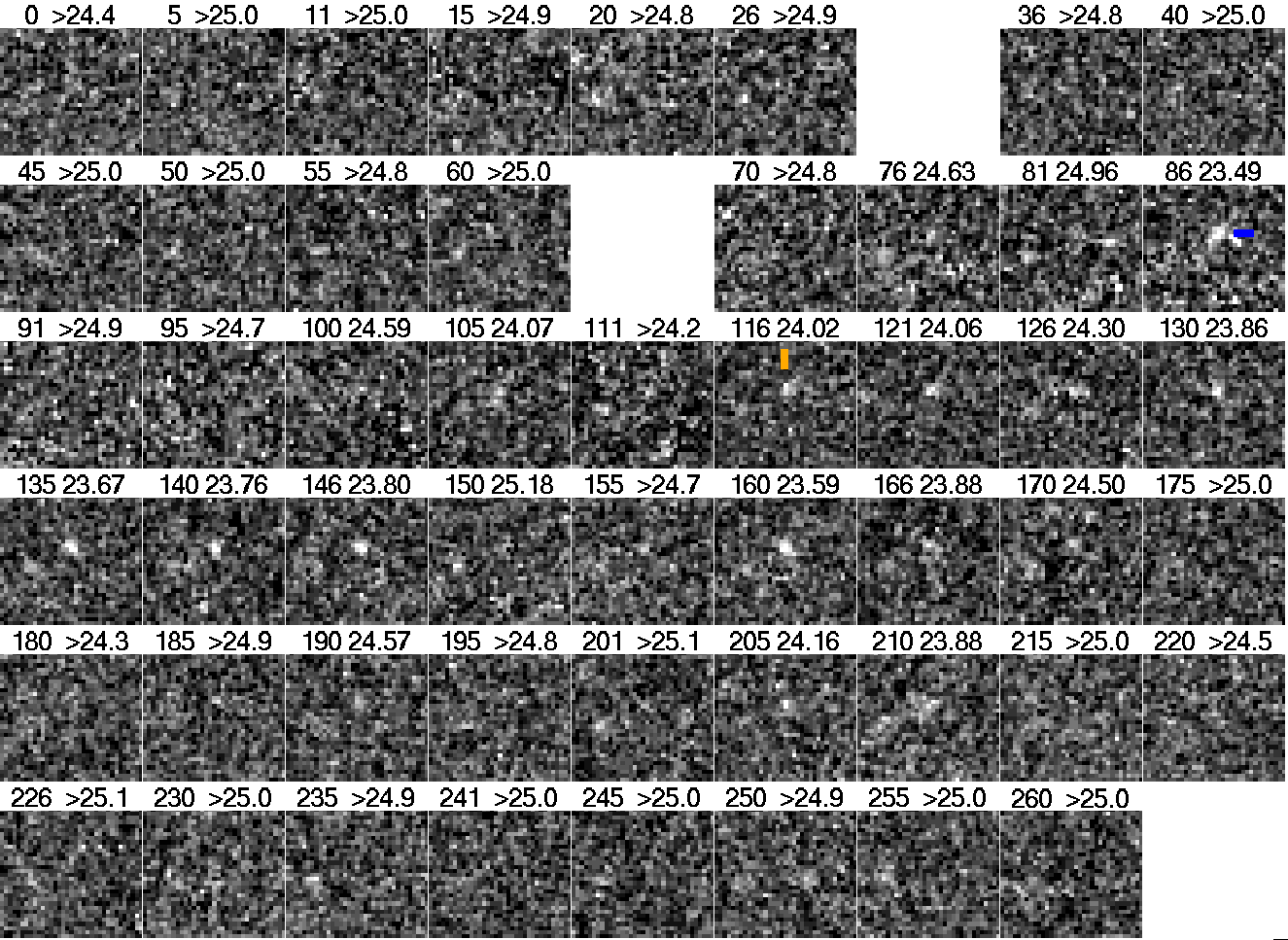}
    \end{tabular}
    \caption{Nova 76}
\end{sidewaysfigure}
\begin{sidewaysfigure}
    \begin{tabular}{cc}
     \multirow{2}{*}[2.6in]{\includegraphics[height=6in]{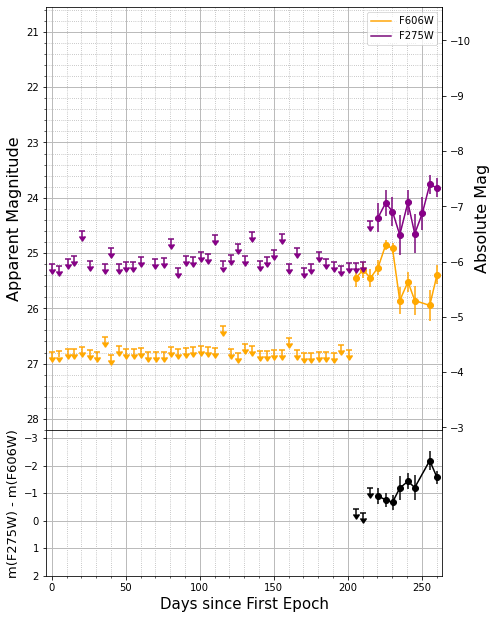}} & \includegraphics[height=3.6in]{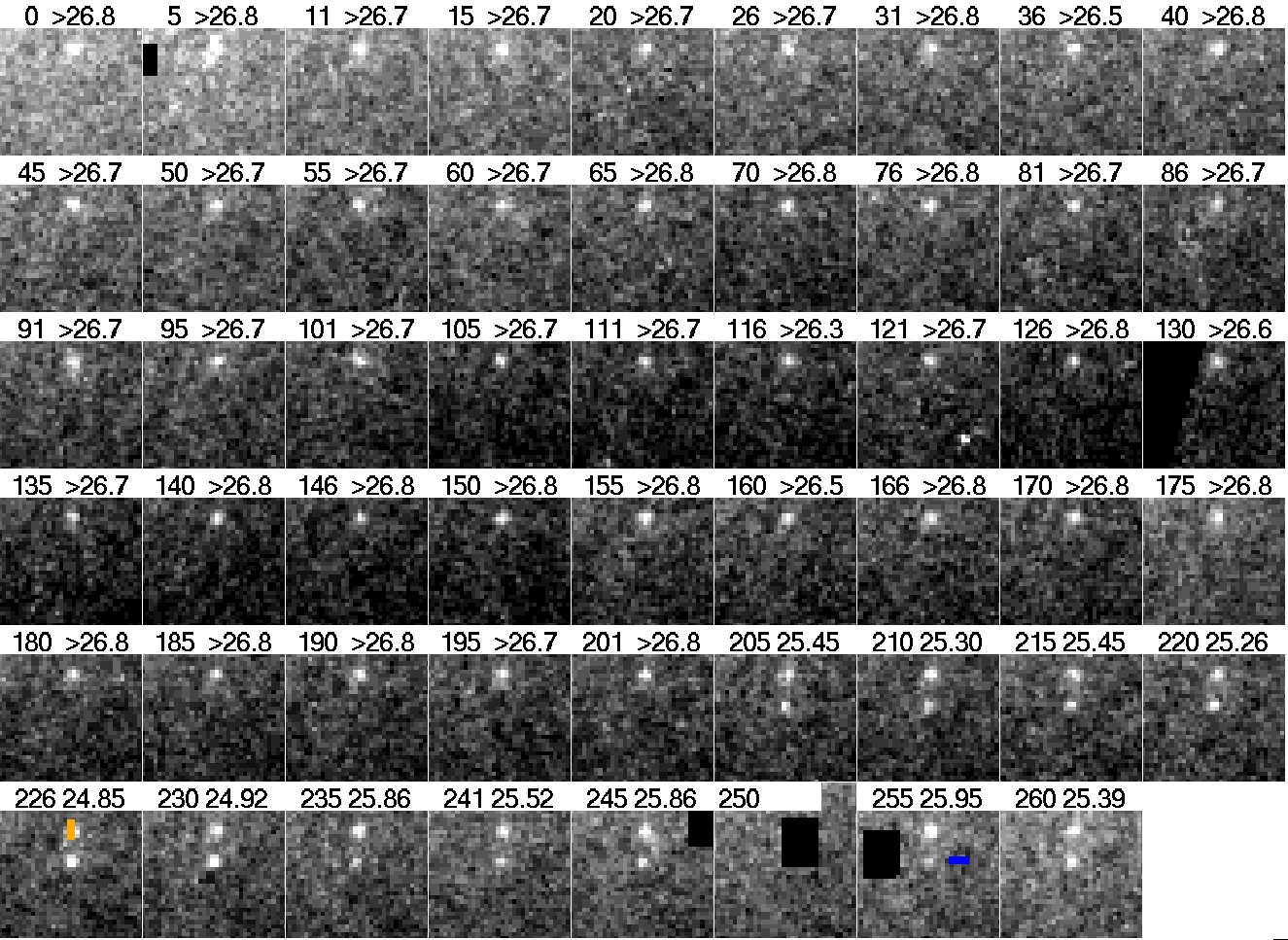} \\
     & \includegraphics[height=3.6in]{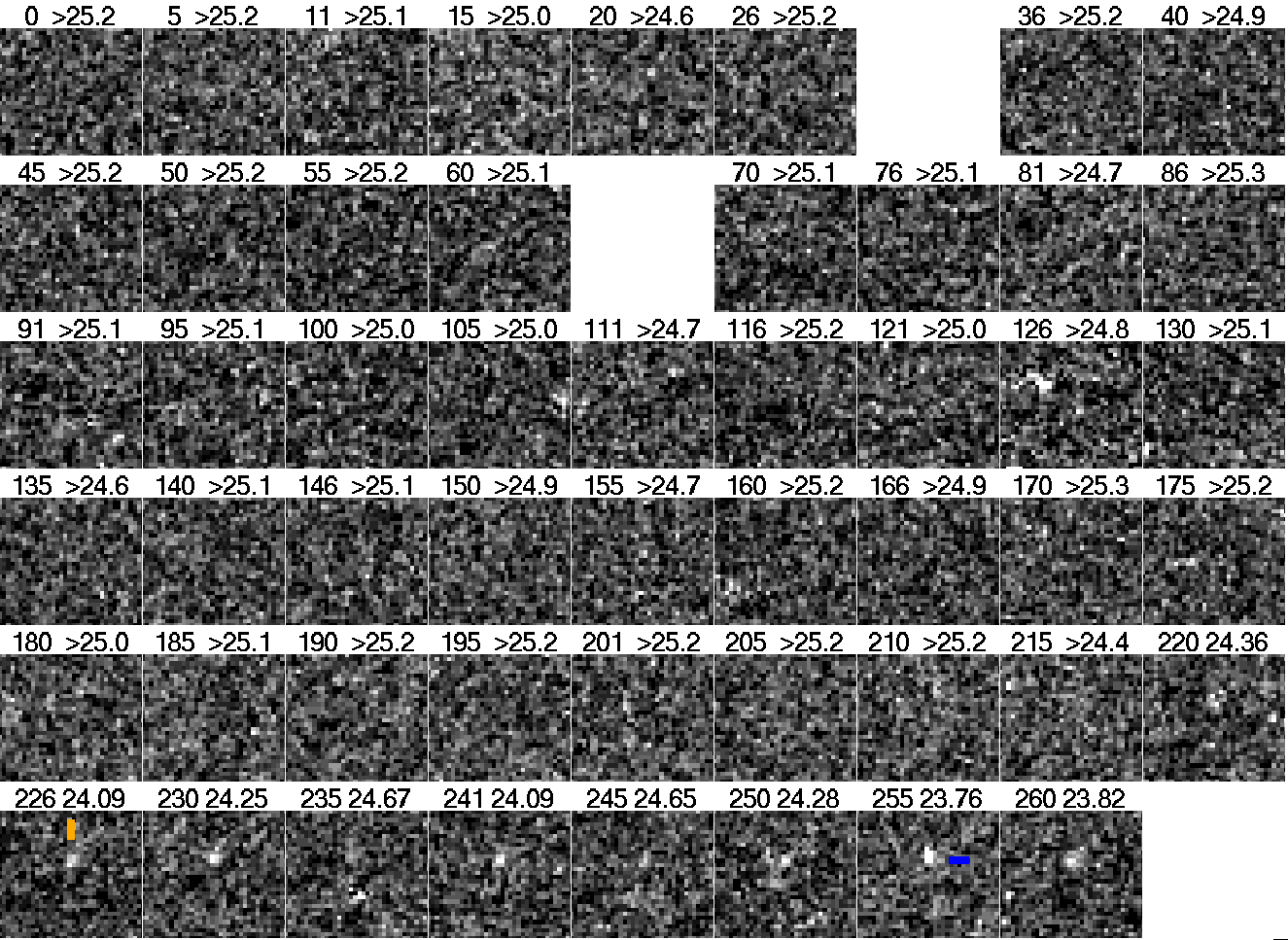}
    \end{tabular}
    \caption{Nova 77}
\end{sidewaysfigure}
\begin{sidewaysfigure}
    \begin{tabular}{cc}
     \multirow{2}{*}[2.6in]{\includegraphics[height=6in]{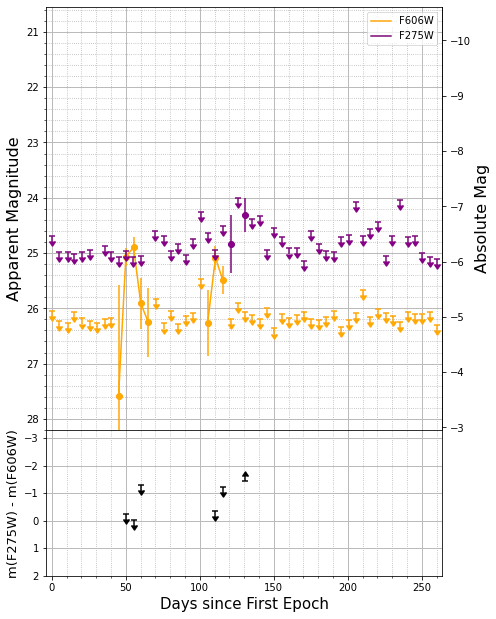}} & \includegraphics[height=3.6in]{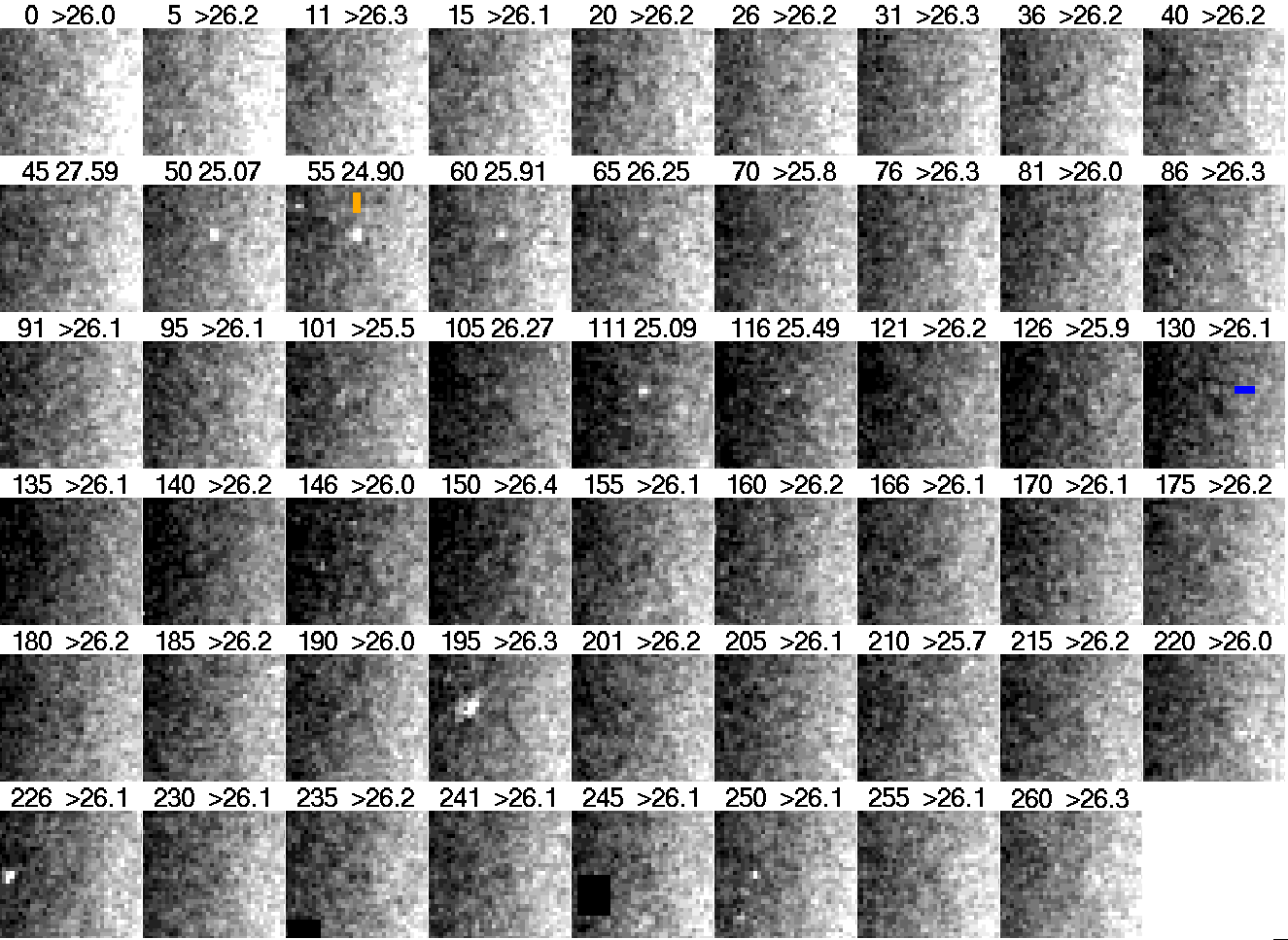} \\
     & \includegraphics[height=3.6in]{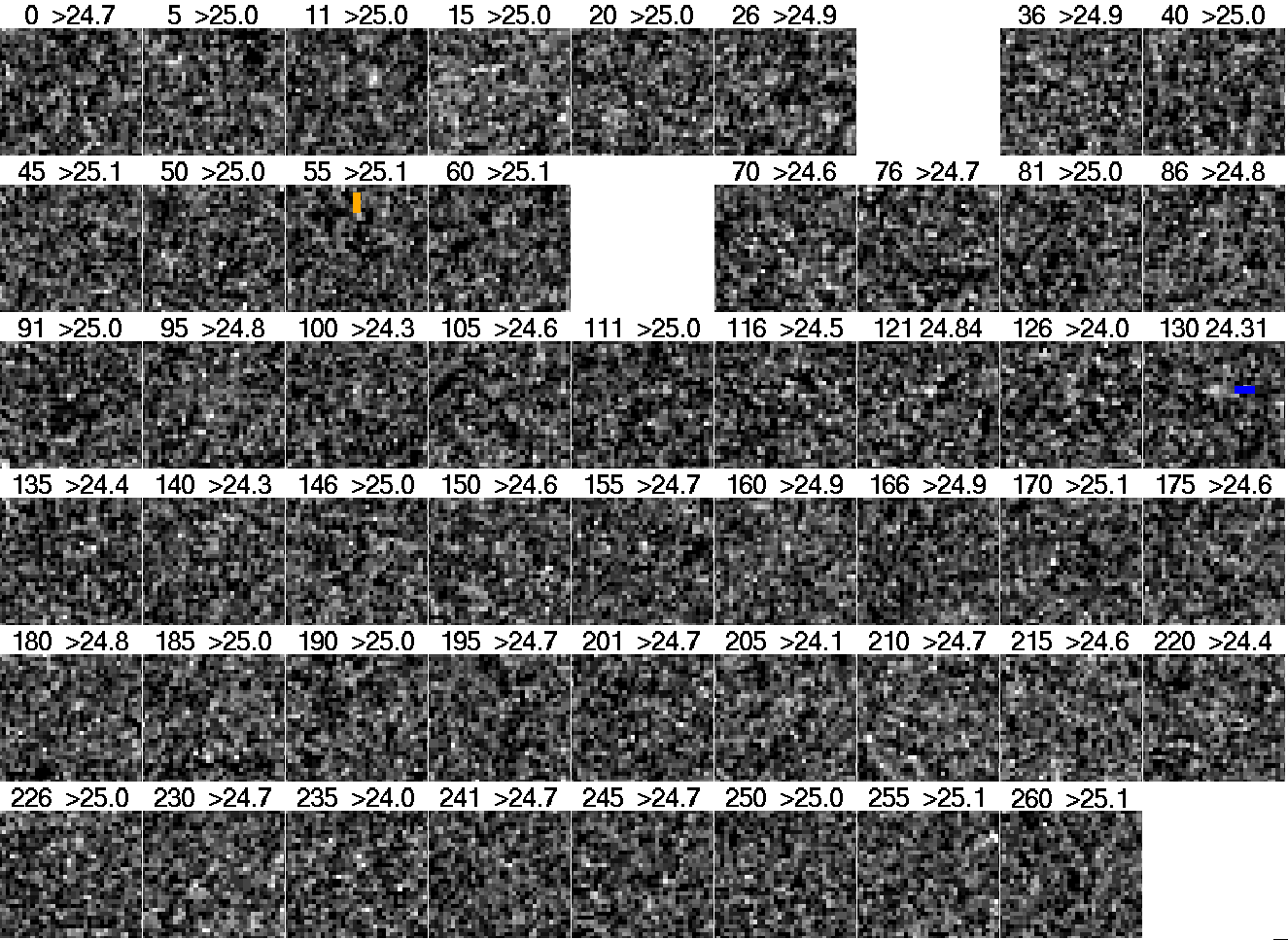}
    \end{tabular}
    \caption{Nova 78}
\end{sidewaysfigure}
\begin{sidewaysfigure}
    \begin{tabular}{cc}
     \multirow{2}{*}[2.6in]{\includegraphics[height=6in]{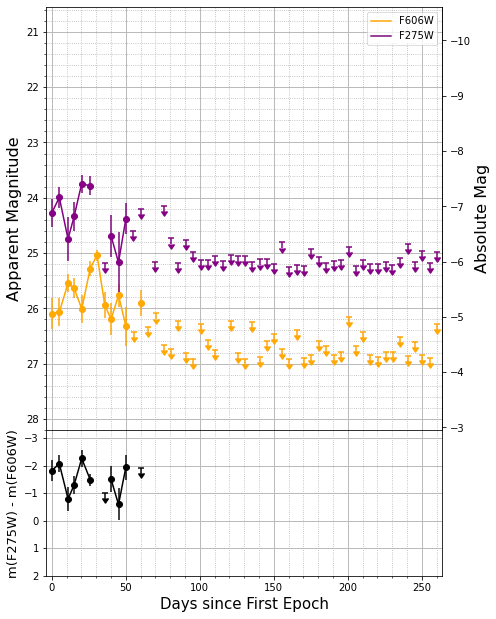}} & \includegraphics[height=3.6in]{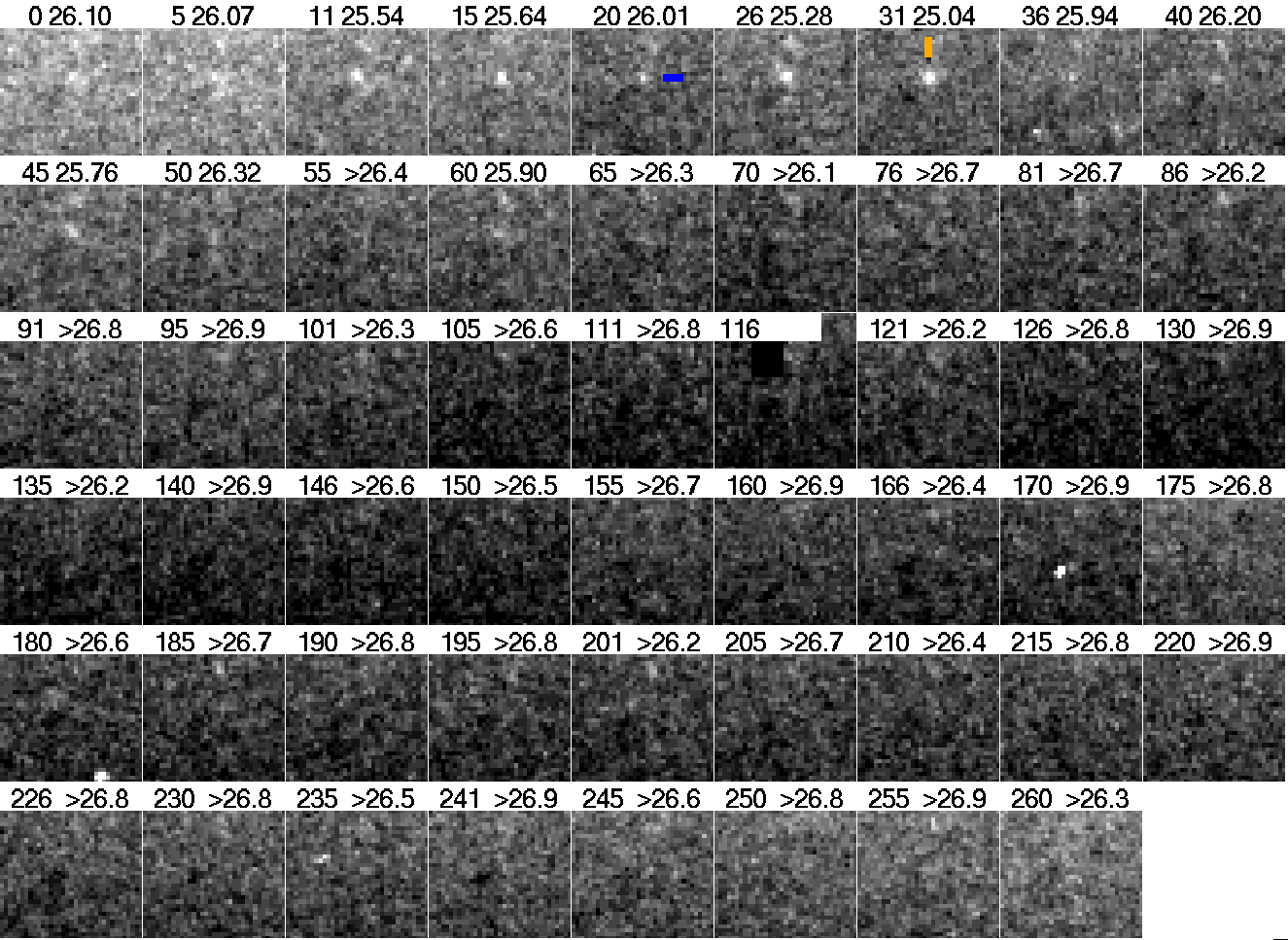} \\
     & \includegraphics[height=3.6in]{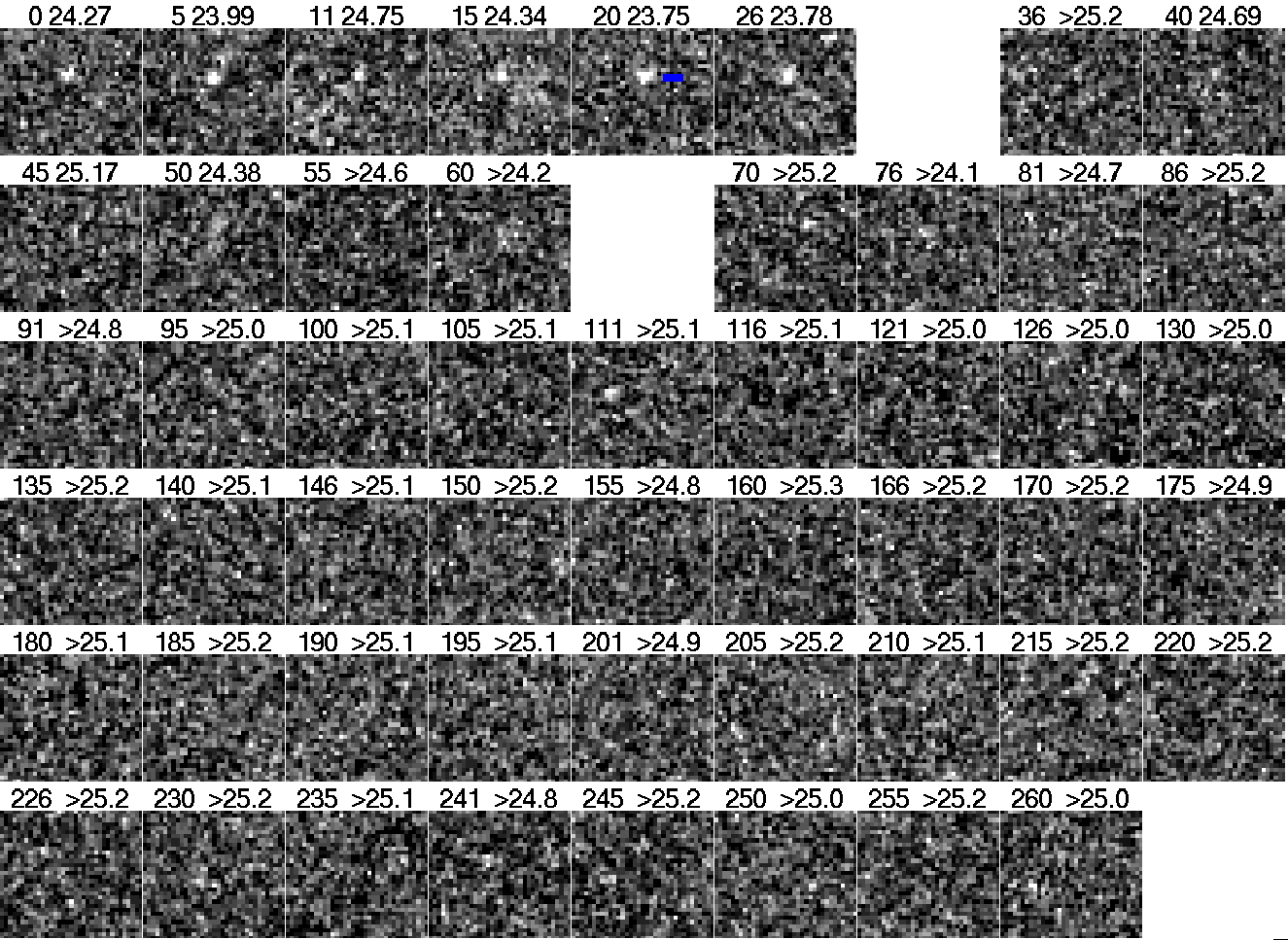}
    \end{tabular}
    \caption{Nova 79}
\end{sidewaysfigure}
\begin{sidewaysfigure}
    \begin{tabular}{cc}
     \multirow{2}{*}[2.6in]{\includegraphics[height=6in]{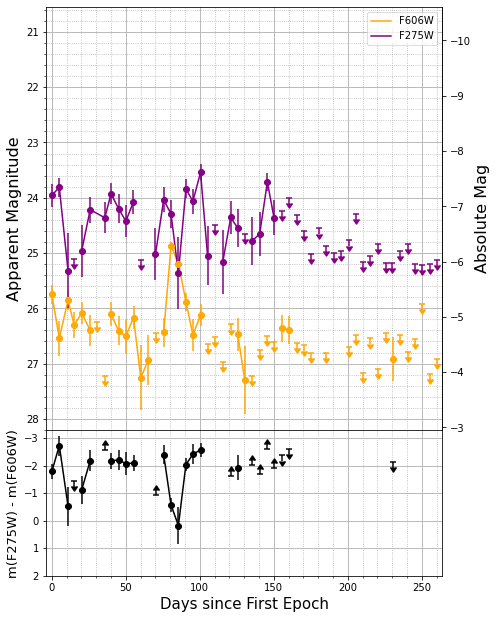}} & \includegraphics[height=3.6in]{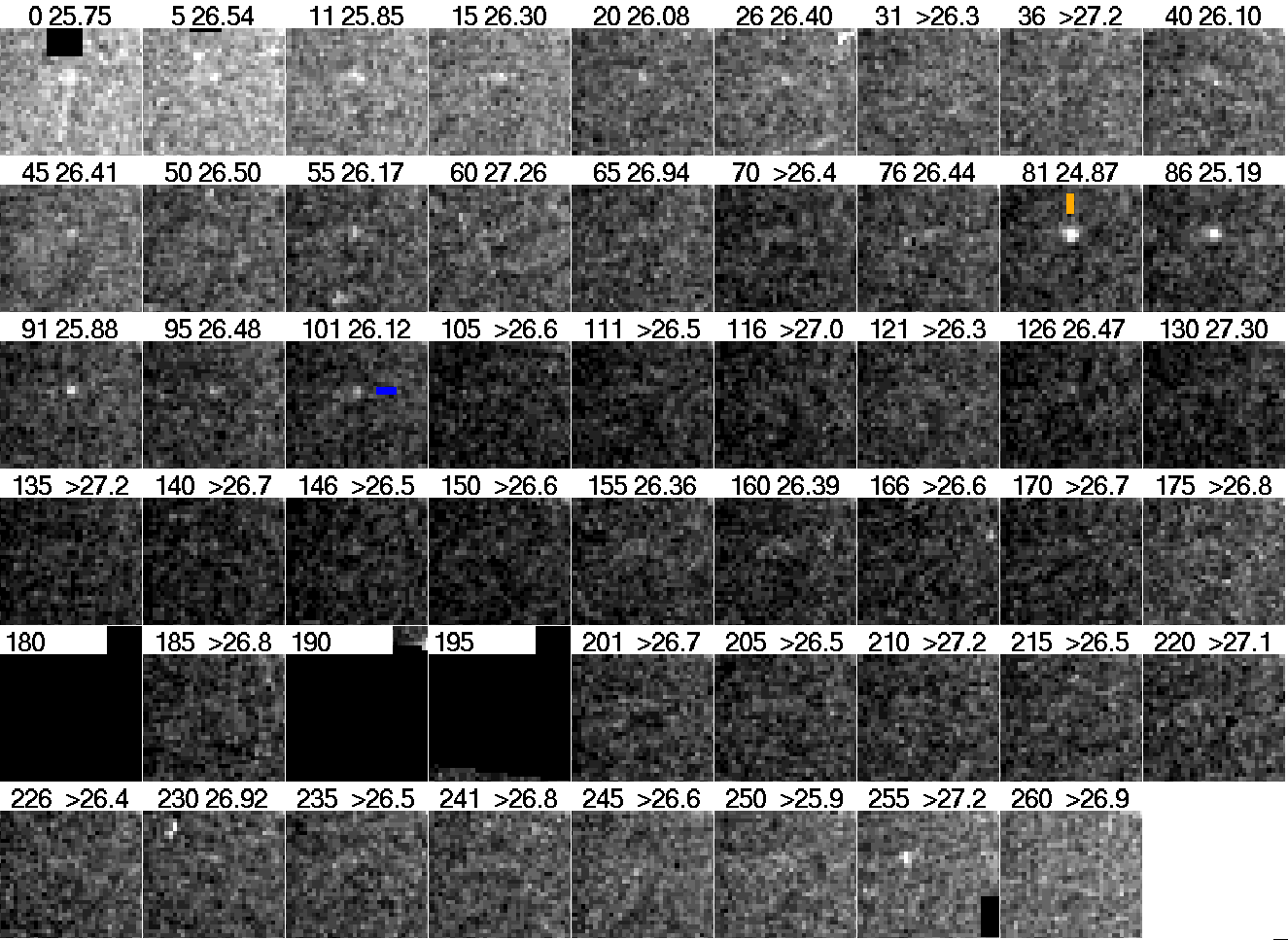} \\
     & \includegraphics[height=3.6in]{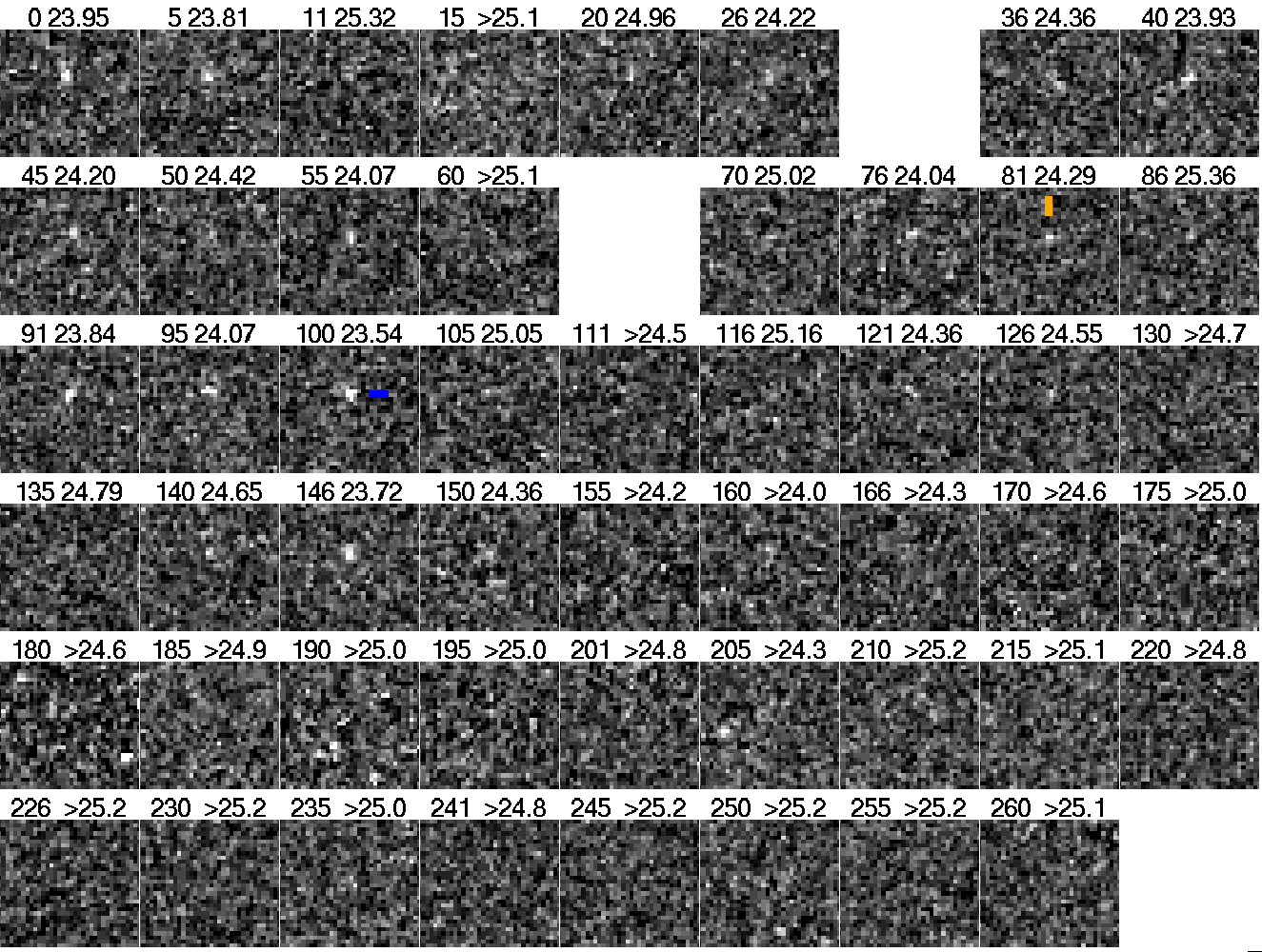}
    \end{tabular}
    \caption{Nova 80}
\end{sidewaysfigure}
\begin{sidewaysfigure}
    \begin{tabular}{cc}
     \multirow{2}{*}[2.6in]{\includegraphics[height=6in]{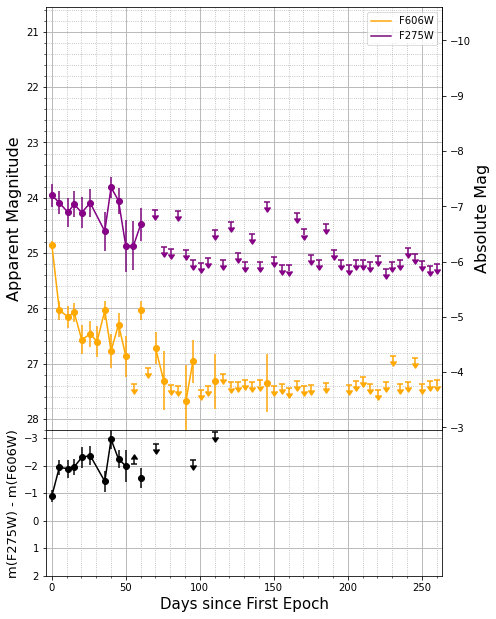}} & \includegraphics[height=3.6in]{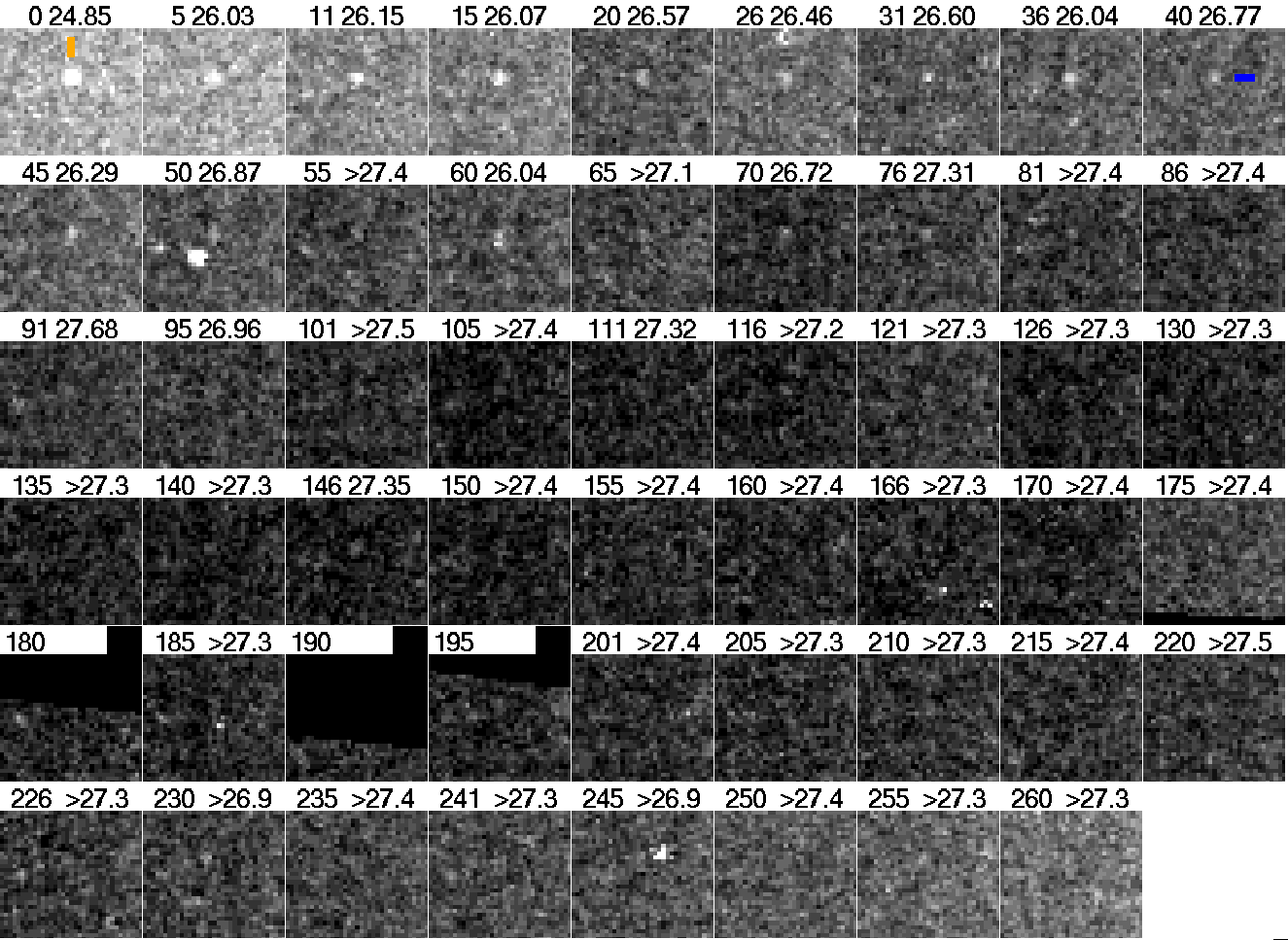} \\
     & \includegraphics[height=3.6in]{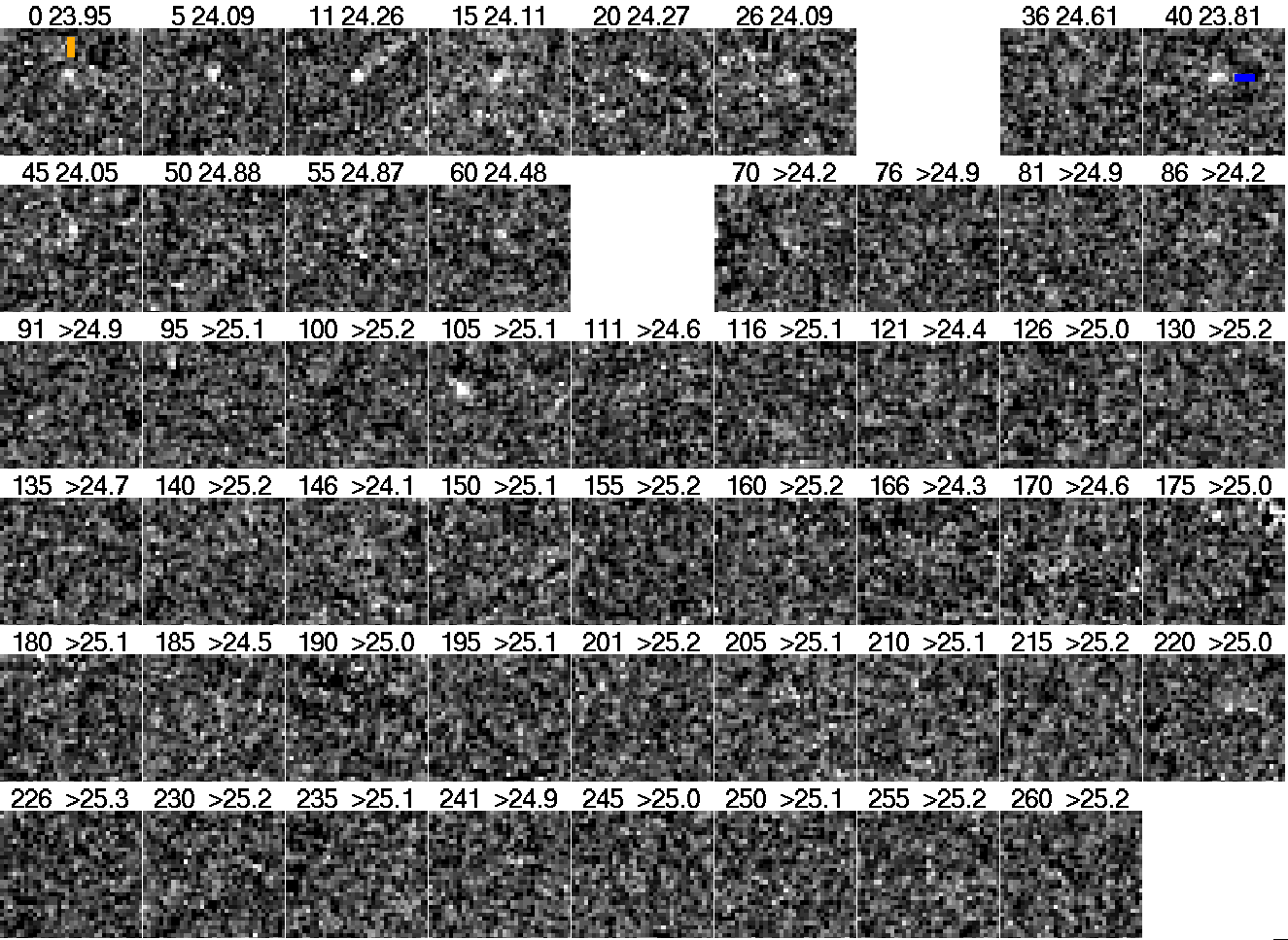}
    \end{tabular}
    \caption{Nova 81}
\end{sidewaysfigure}
\begin{sidewaysfigure}
    \begin{tabular}{cc}
     \multirow{2}{*}[2.6in]{\includegraphics[height=6in]{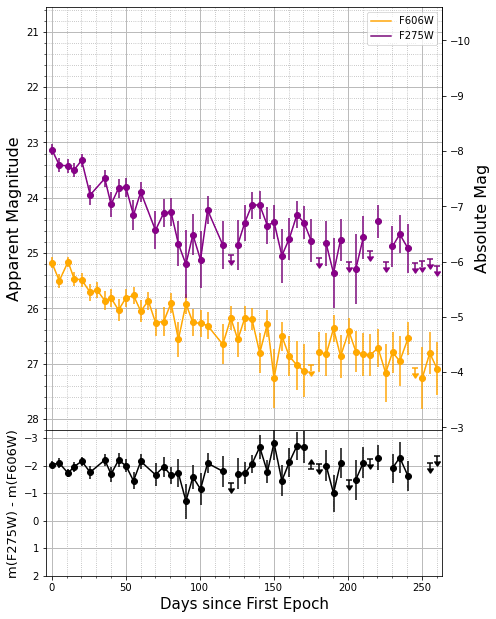}} & \includegraphics[height=3.6in]{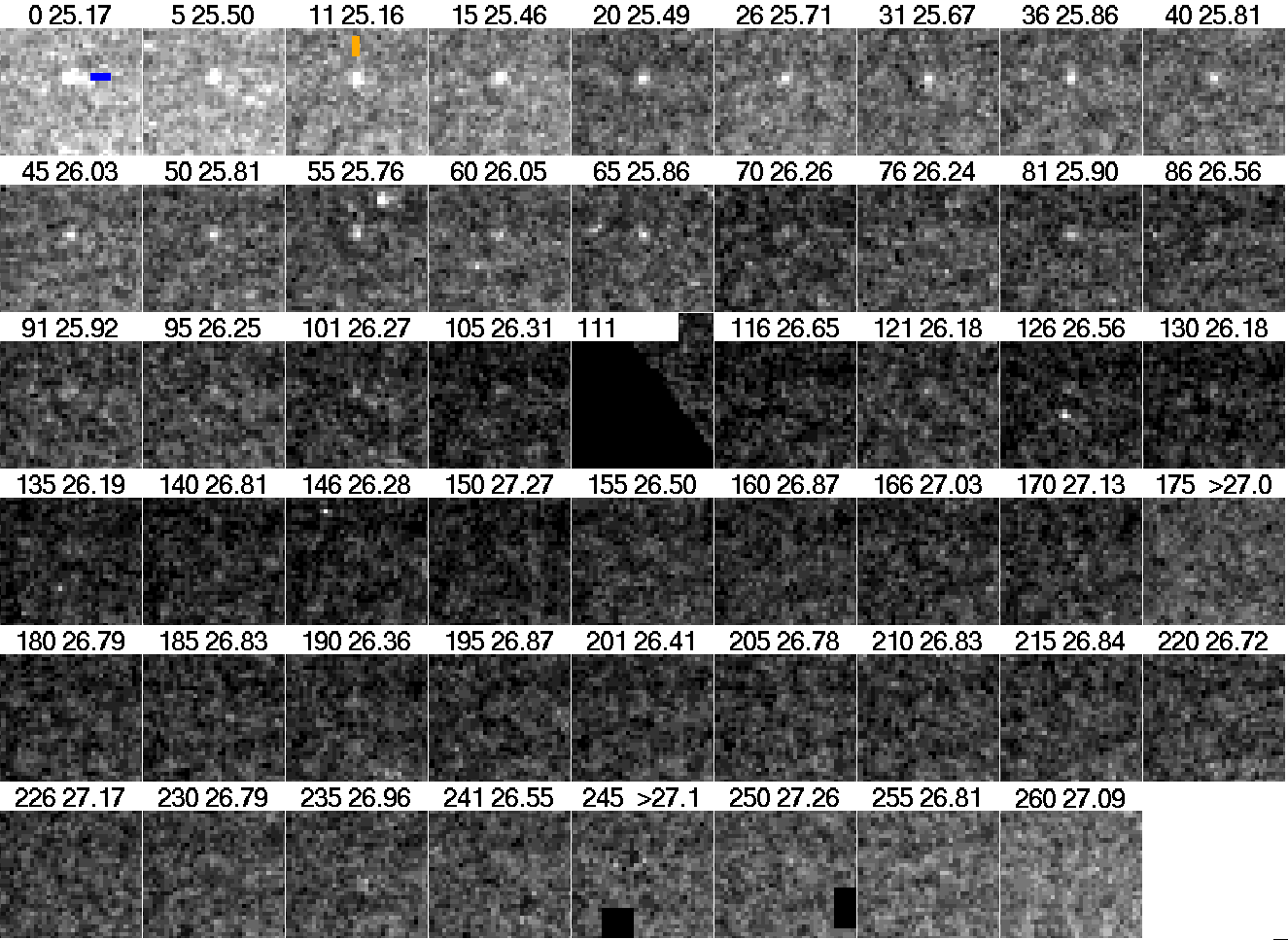} \\
     & \includegraphics[height=3.6in]{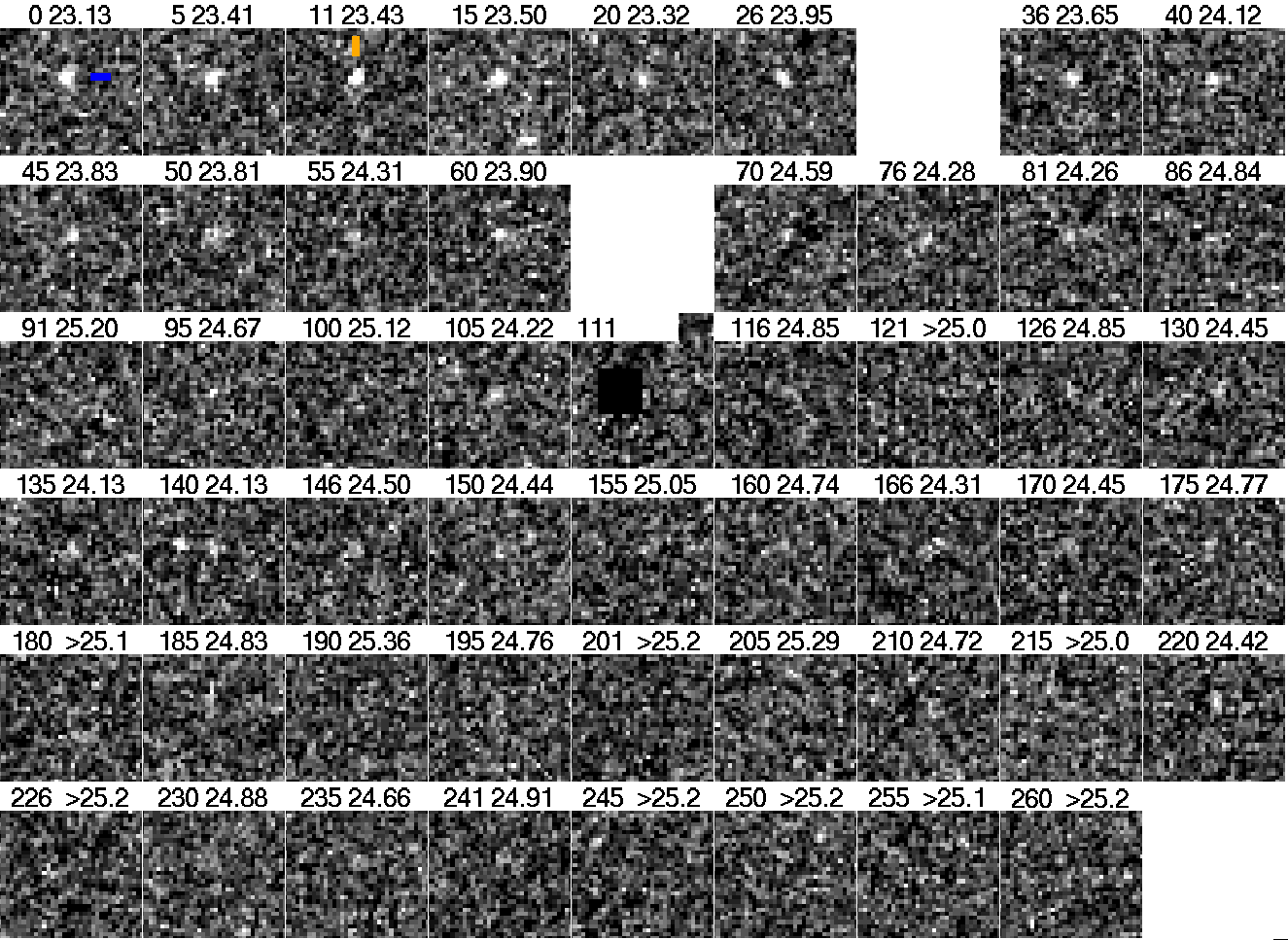}
    \end{tabular}
    \caption{Nova 82}
\end{sidewaysfigure}
\begin{sidewaysfigure}
    \begin{tabular}{cc}
     \multirow{2}{*}[2.6in]{\includegraphics[height=6in]{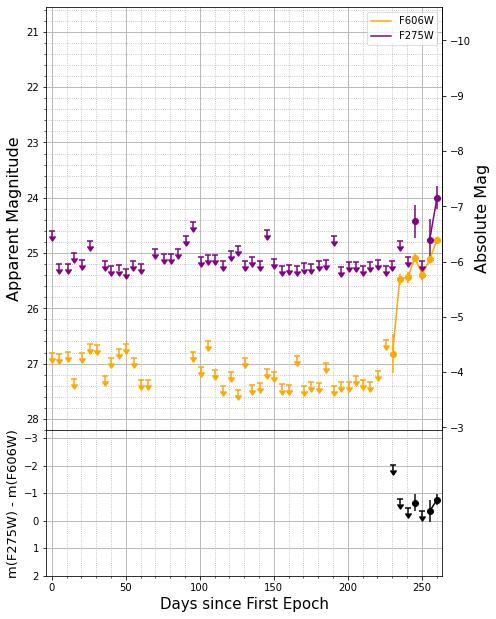}} & \includegraphics[height=3.6in]{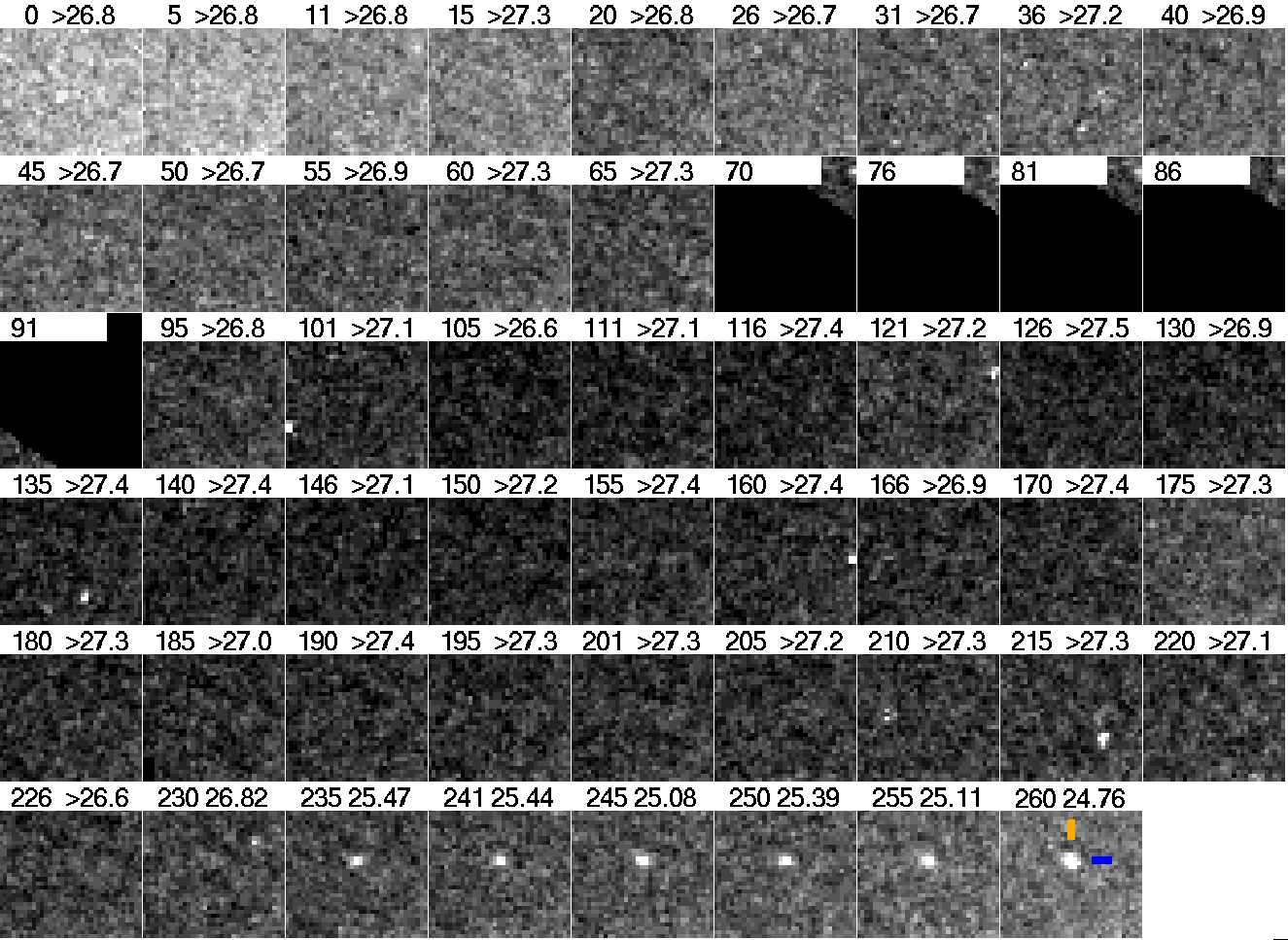} \\
     & \includegraphics[height=3.6in]{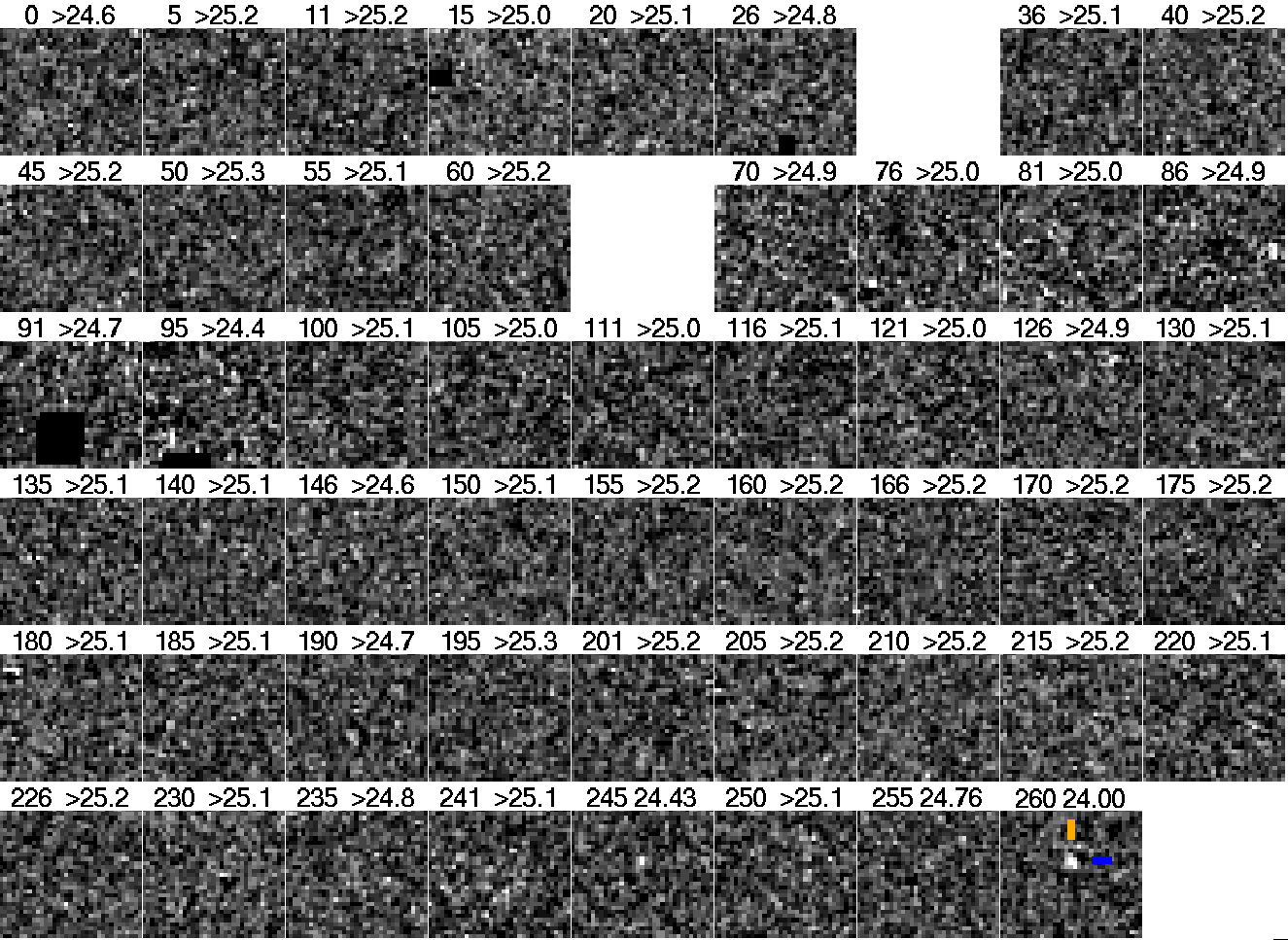}
    \end{tabular}
    \caption{Nova 83}
\end{sidewaysfigure}
\begin{sidewaysfigure}
    \begin{tabular}{cc}
     \multirow{2}{*}[2.6in]{\includegraphics[height=6in]{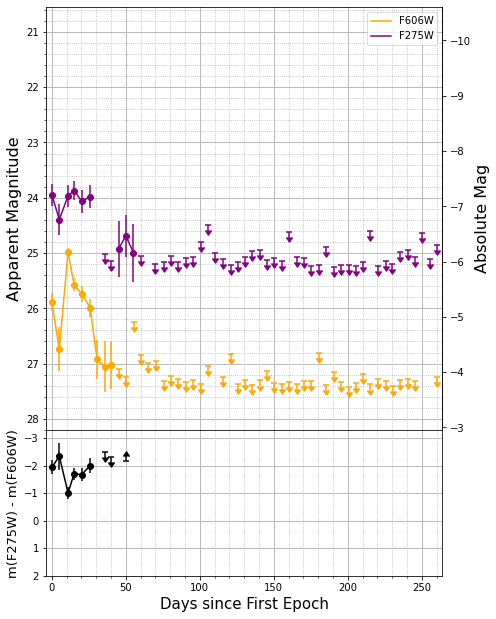}} & \includegraphics[height=3.6in]{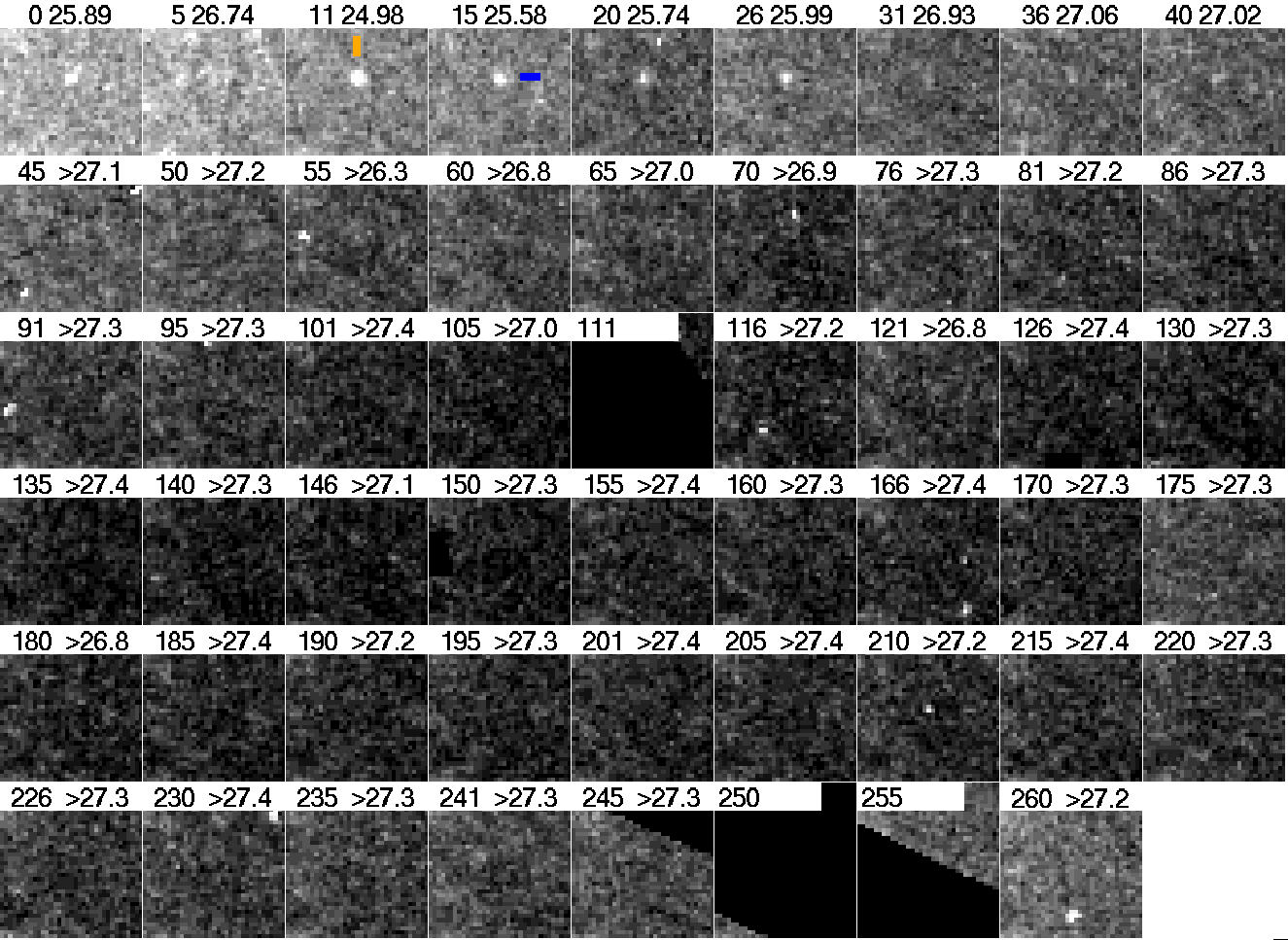} \\
     & \includegraphics[height=3.6in]{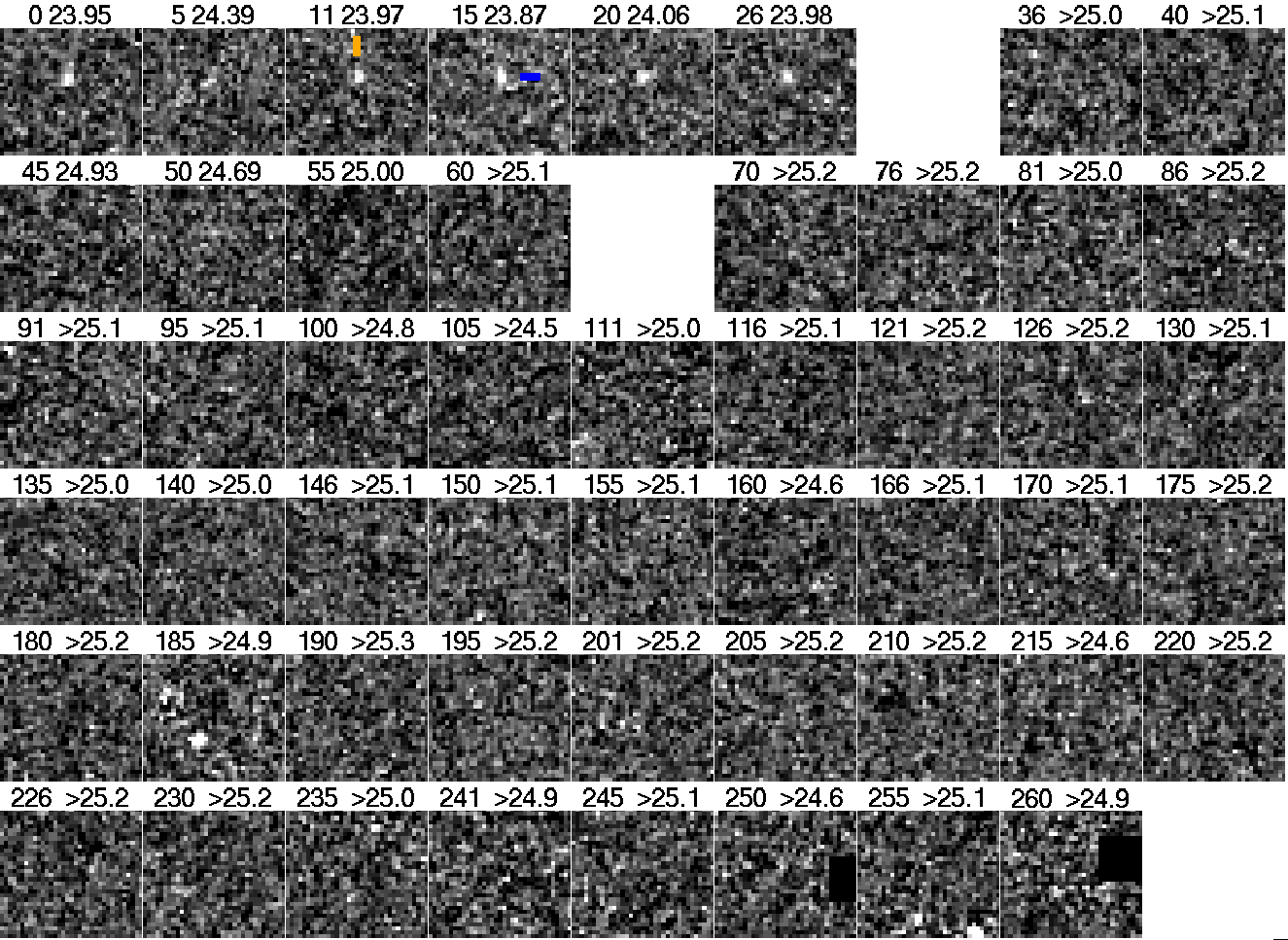}
    \end{tabular}
    \caption{Nova 84}
\end{sidewaysfigure}
\begin{sidewaysfigure}
    \begin{tabular}{cc}
     \multirow{2}{*}[2.6in]{\includegraphics[height=6in]{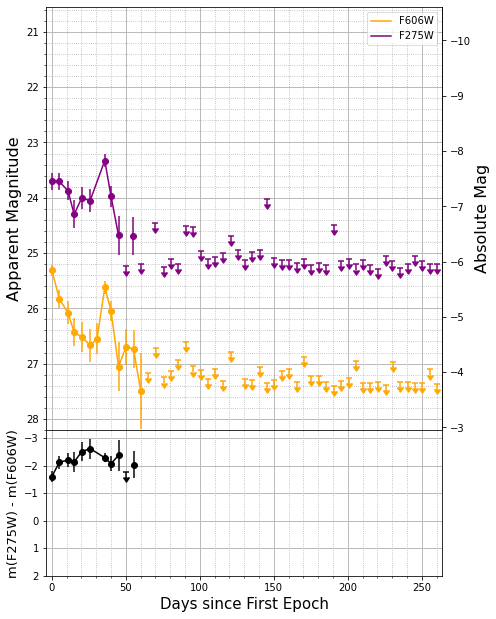}} & \includegraphics[height=3.6in]{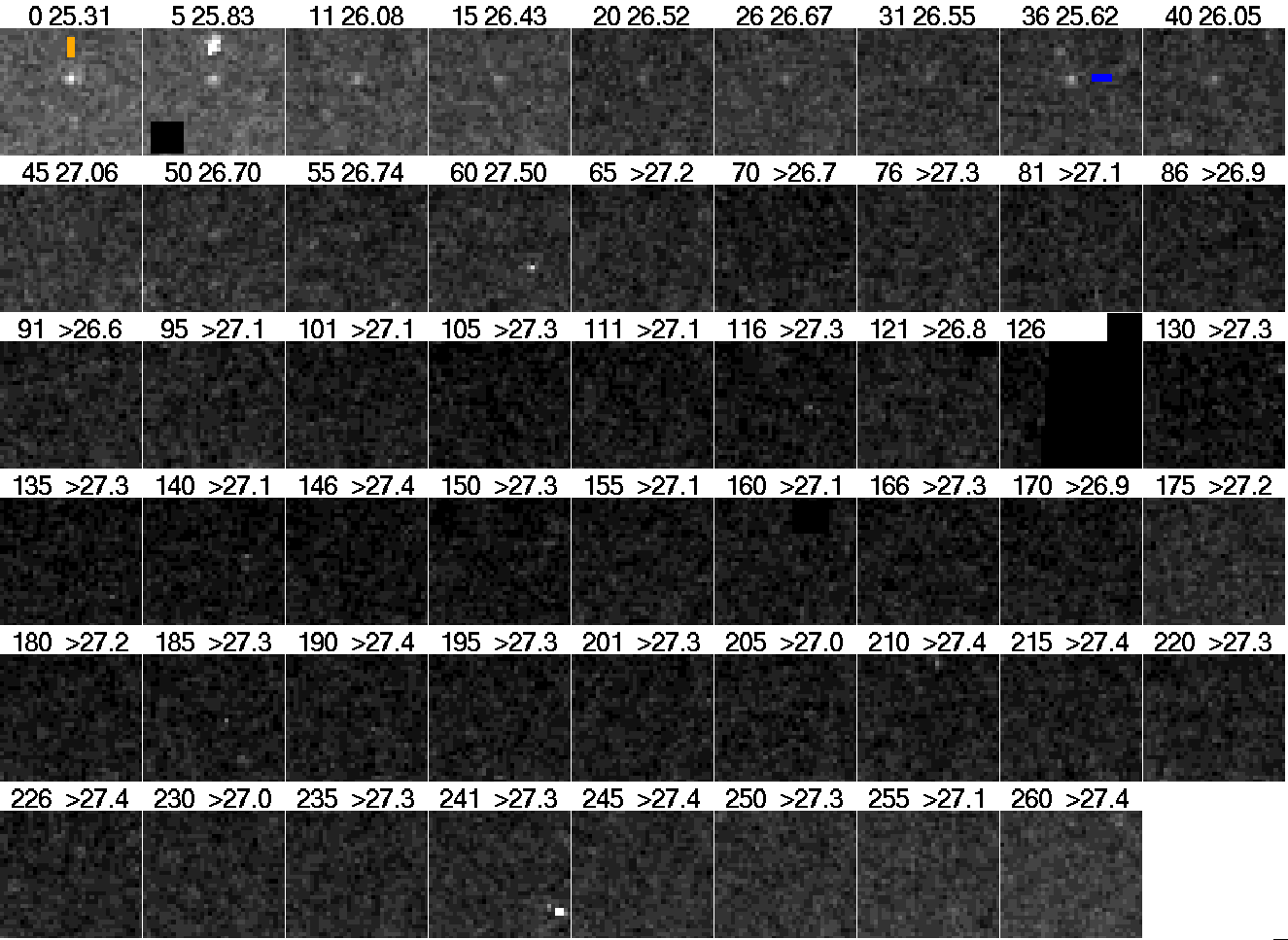} \\
     & \includegraphics[height=3.6in]{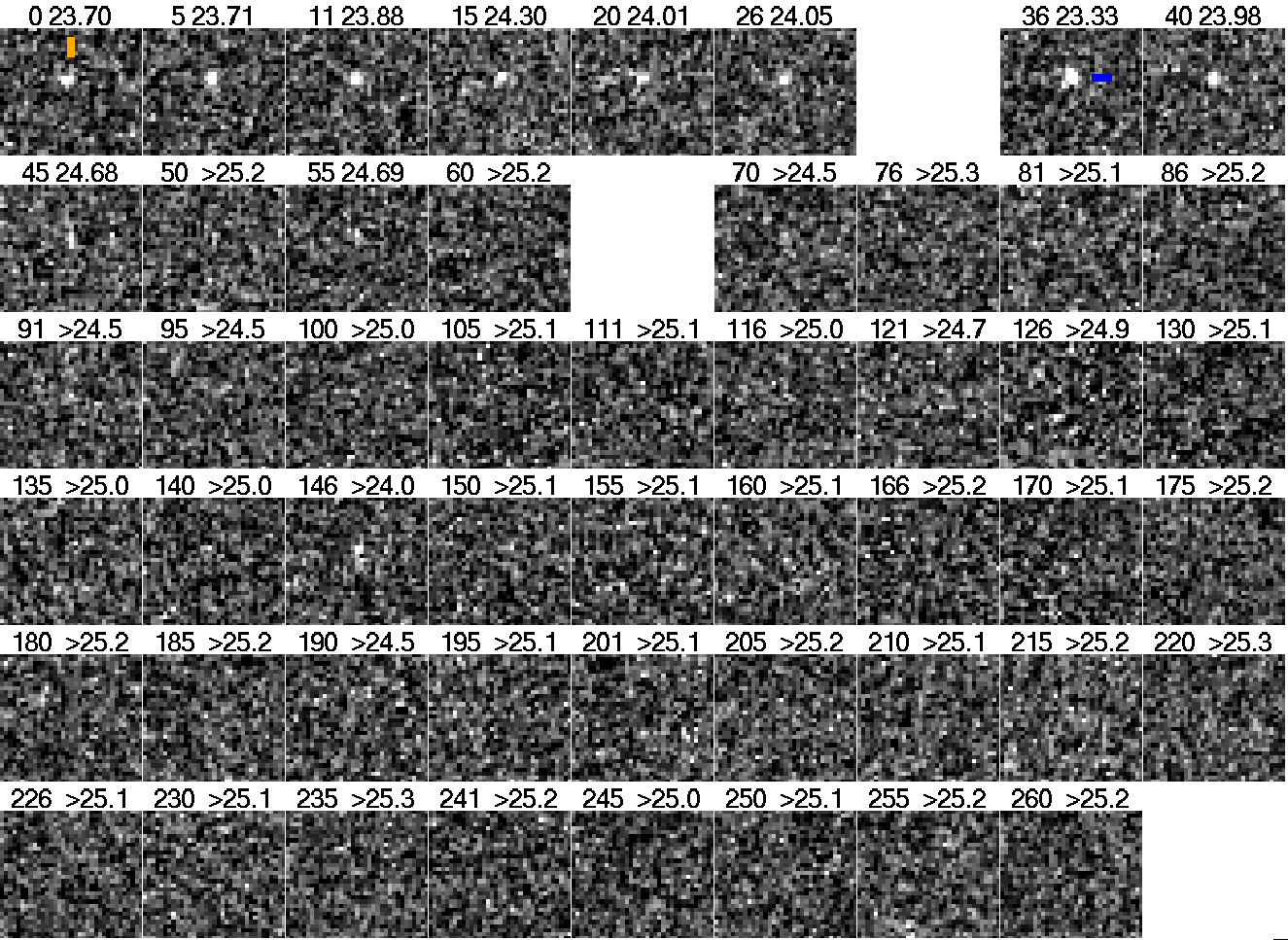}
    \end{tabular}
    \caption{Nova 85}
\end{sidewaysfigure}
\begin{sidewaysfigure}
    \begin{tabular}{cc}
     \multirow{2}{*}[2.6in]{\includegraphics[height=6in]{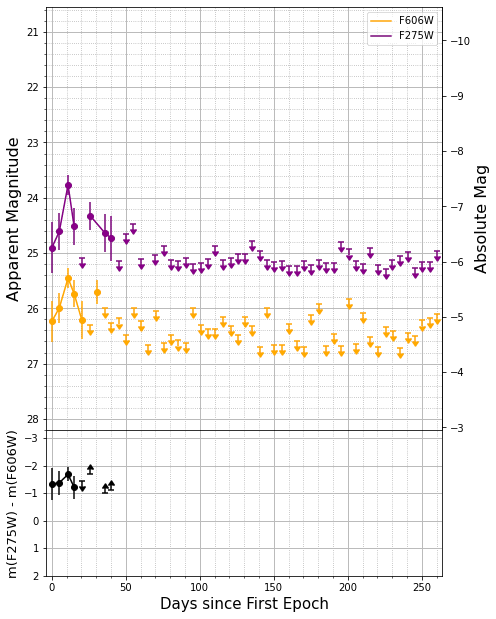}} & \includegraphics[height=3.6in]{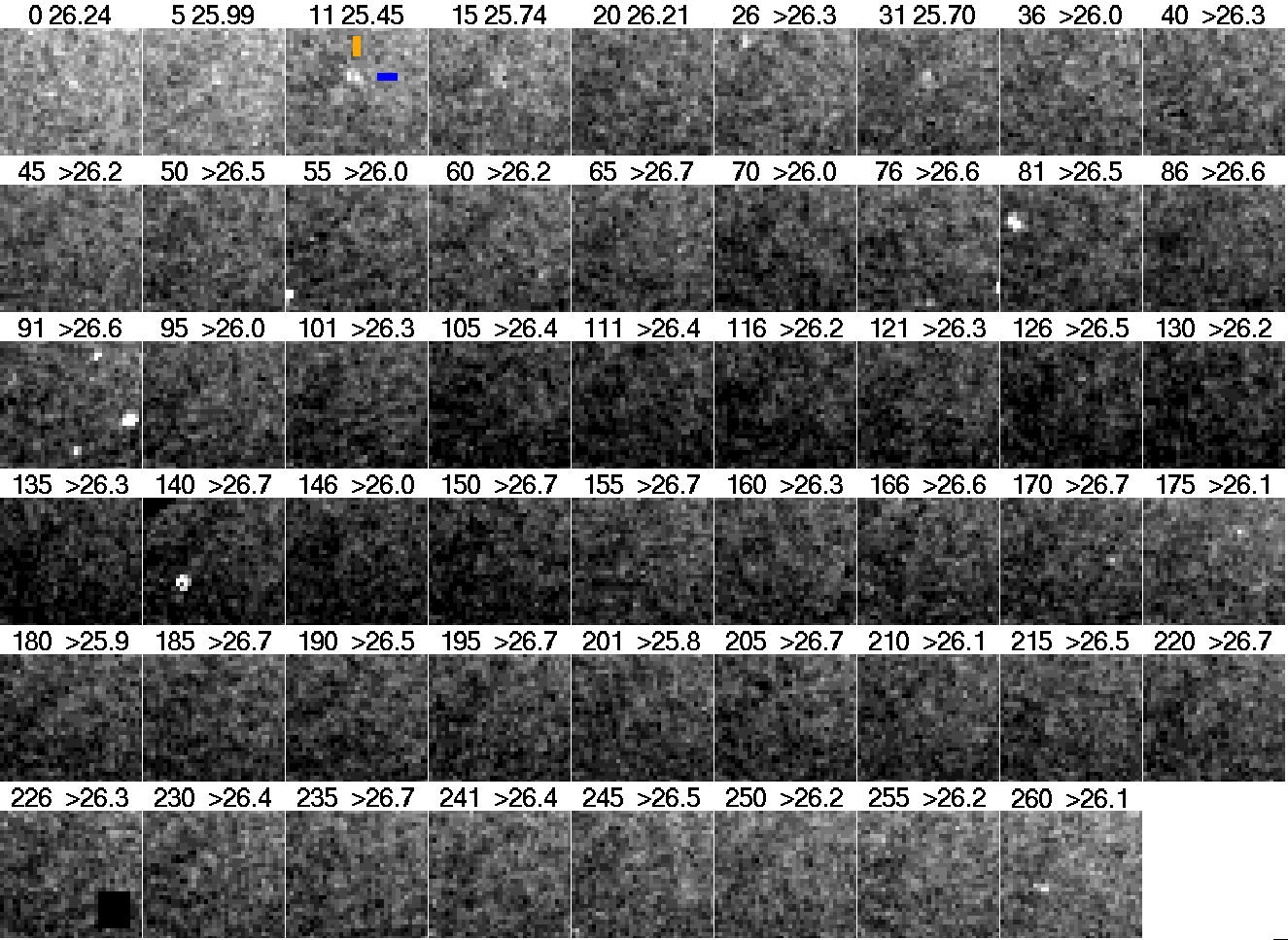} \\
     & \includegraphics[height=3.6in]{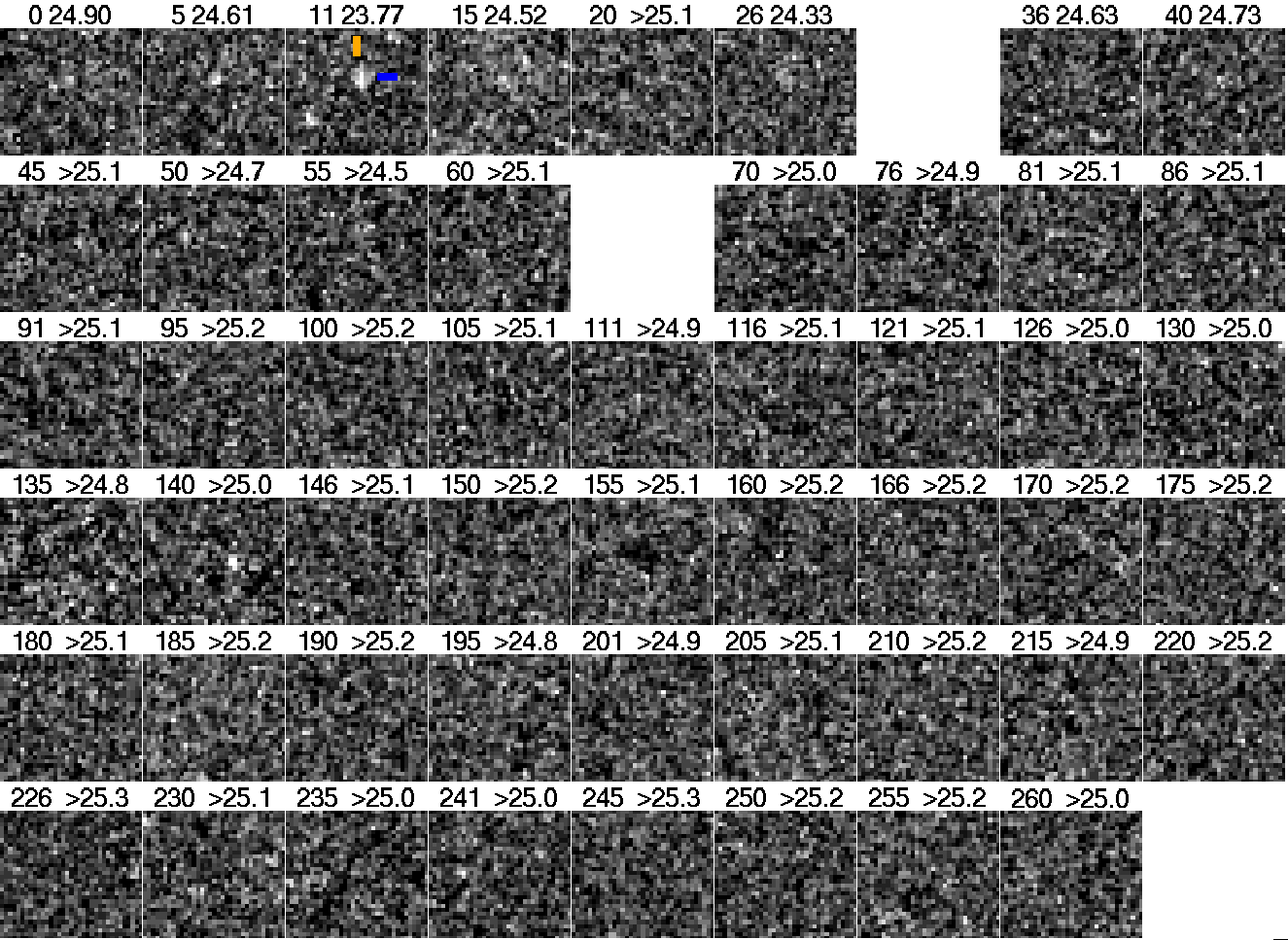}
    \end{tabular}
    \caption{Nova 86}
\end{sidewaysfigure}
\begin{sidewaysfigure}
    \begin{tabular}{cc}
     \multirow{2}{*}[2.6in]{\includegraphics[height=6in]{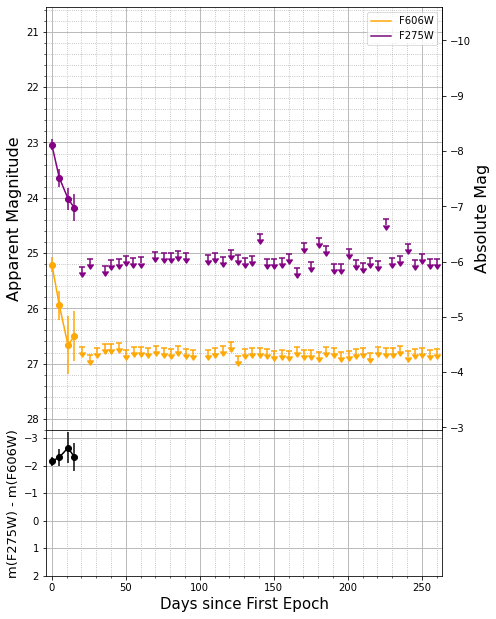}} & \includegraphics[height=3.6in]{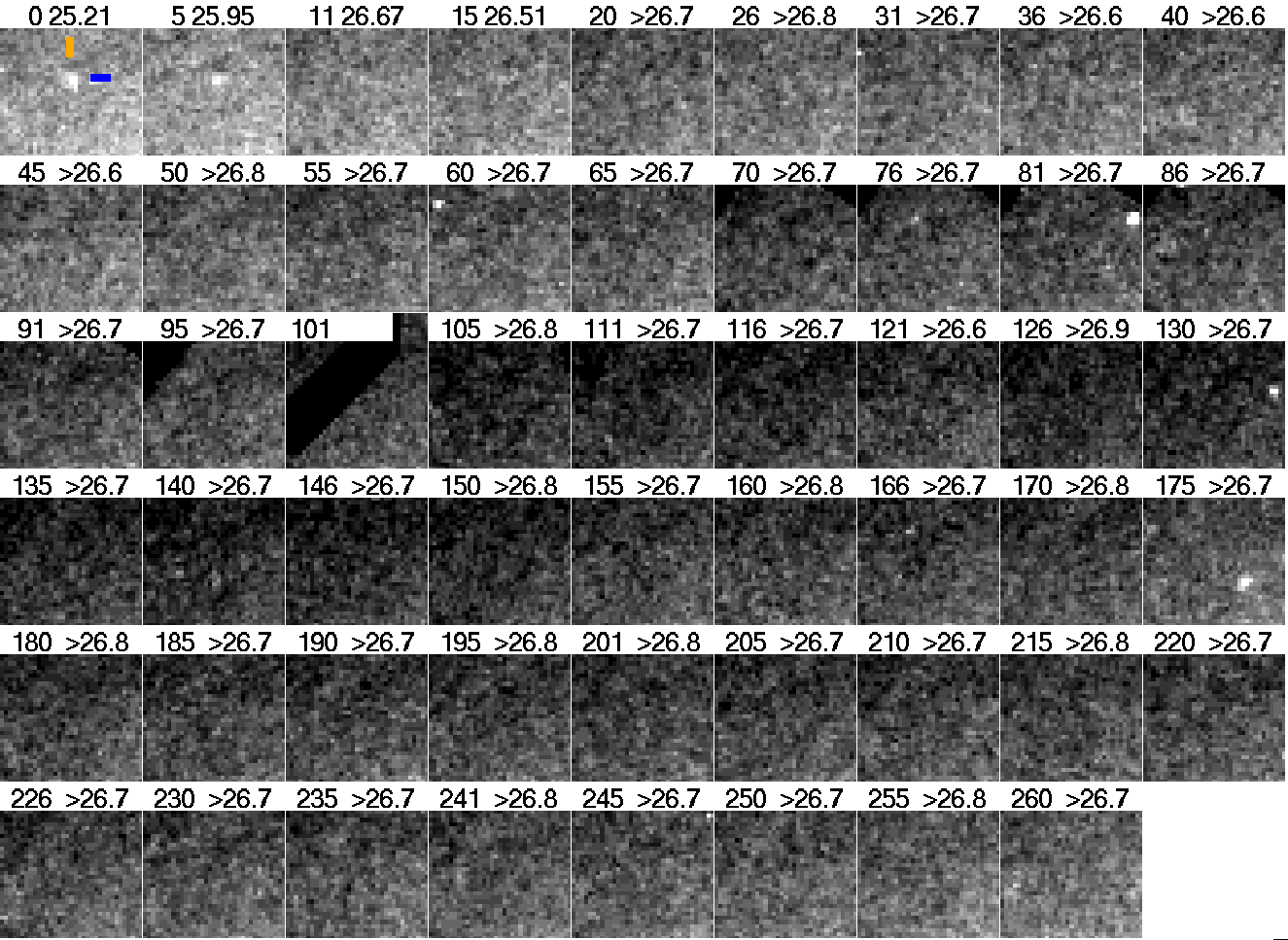} \\
     & \includegraphics[height=3.6in]{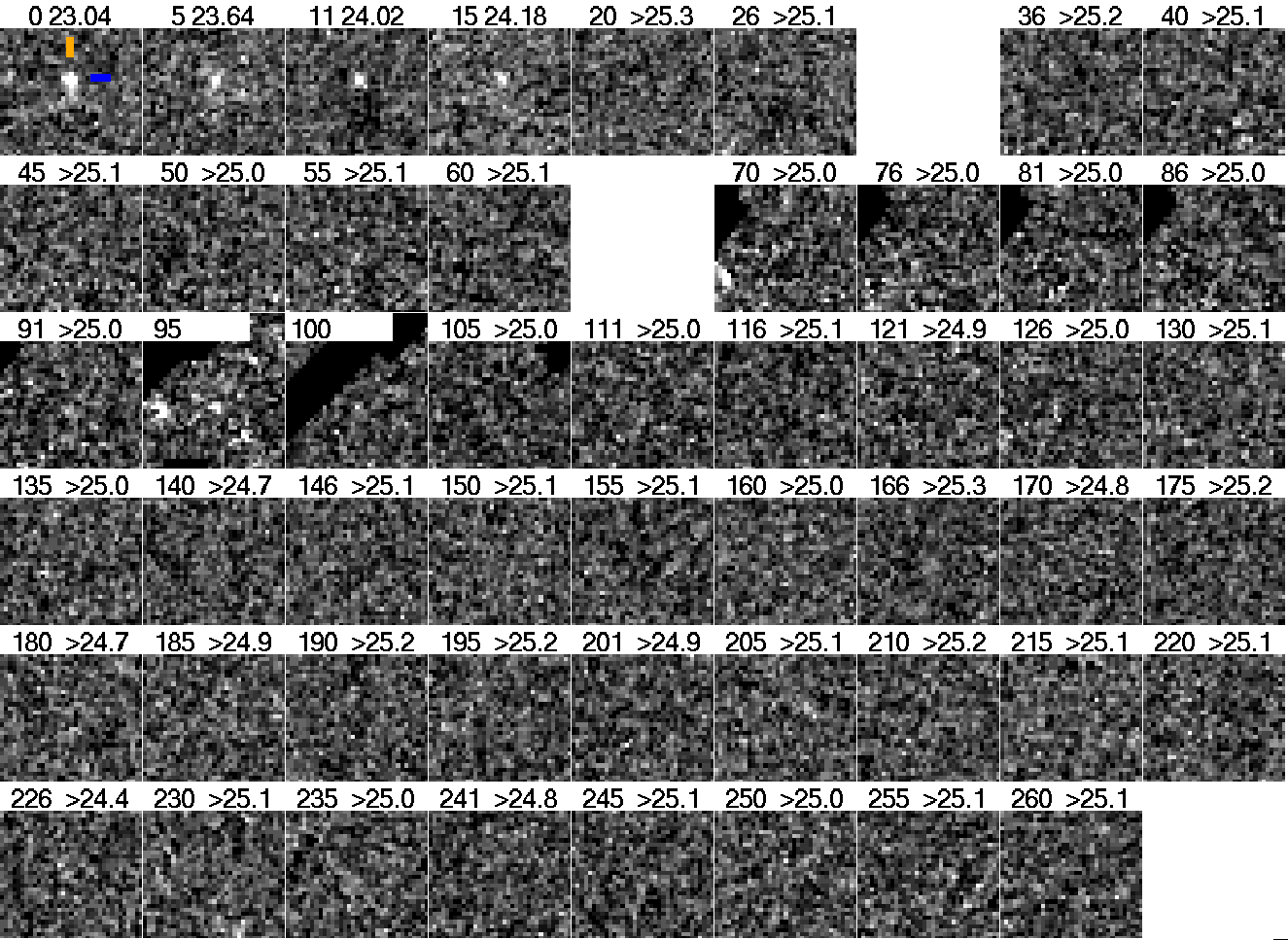}
    \end{tabular}
    \caption{Nova 87}
\end{sidewaysfigure}
\begin{sidewaysfigure}
    \begin{tabular}{cc}
     \multirow{2}{*}[2.6in]{\includegraphics[height=6in]{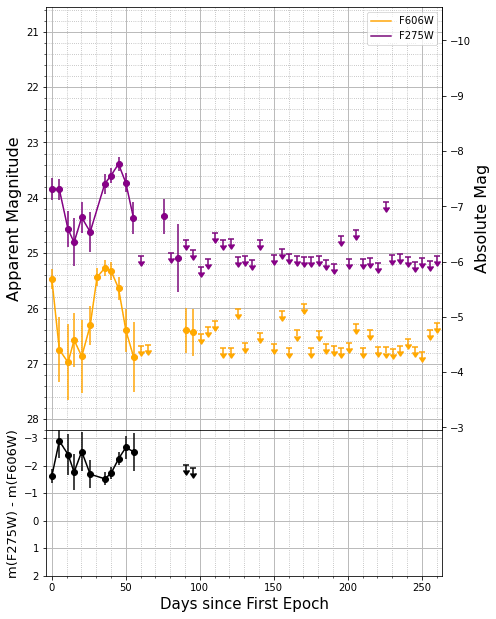}} & \includegraphics[height=3.6in]{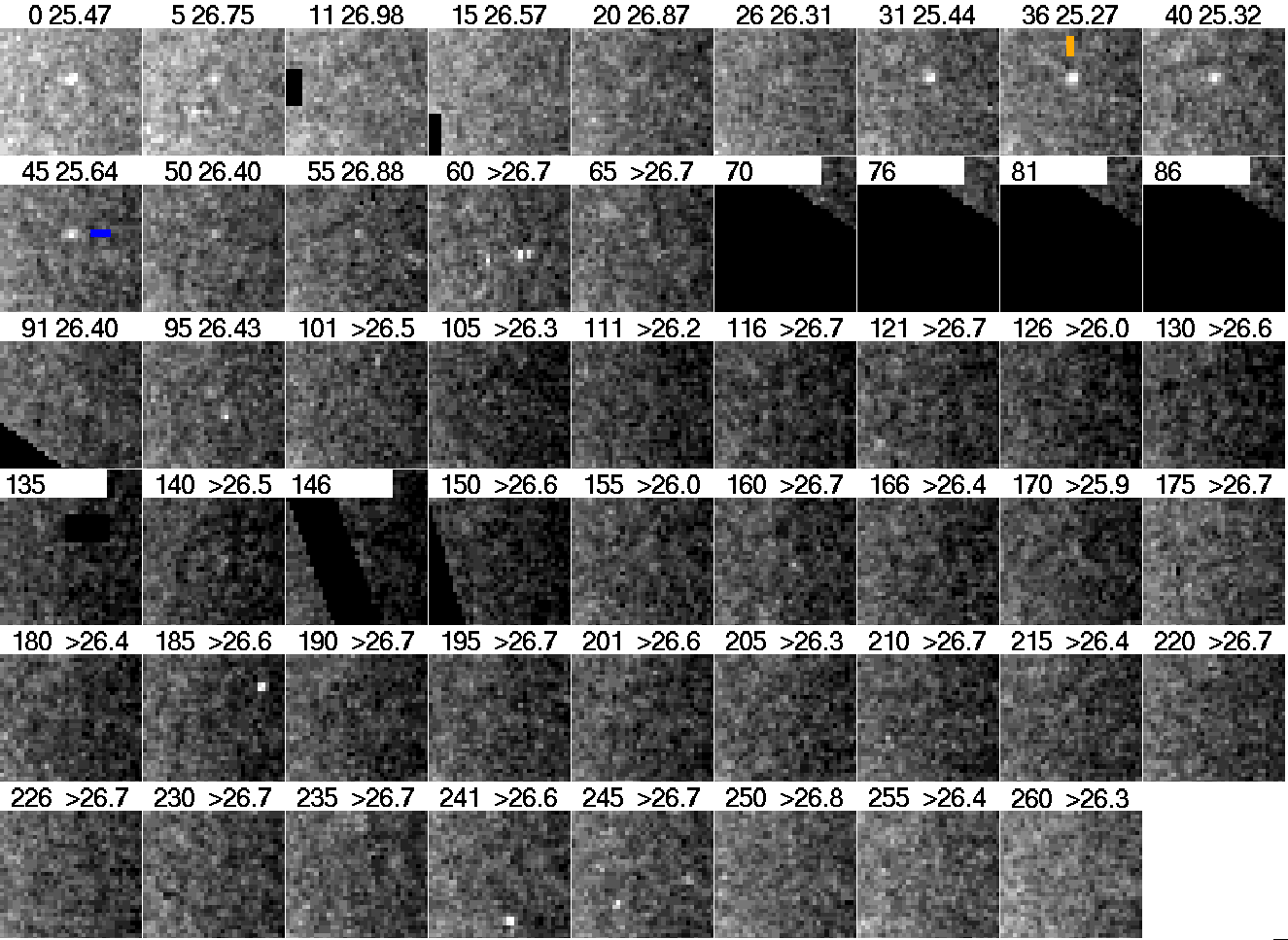} \\
     & \includegraphics[height=3.6in]{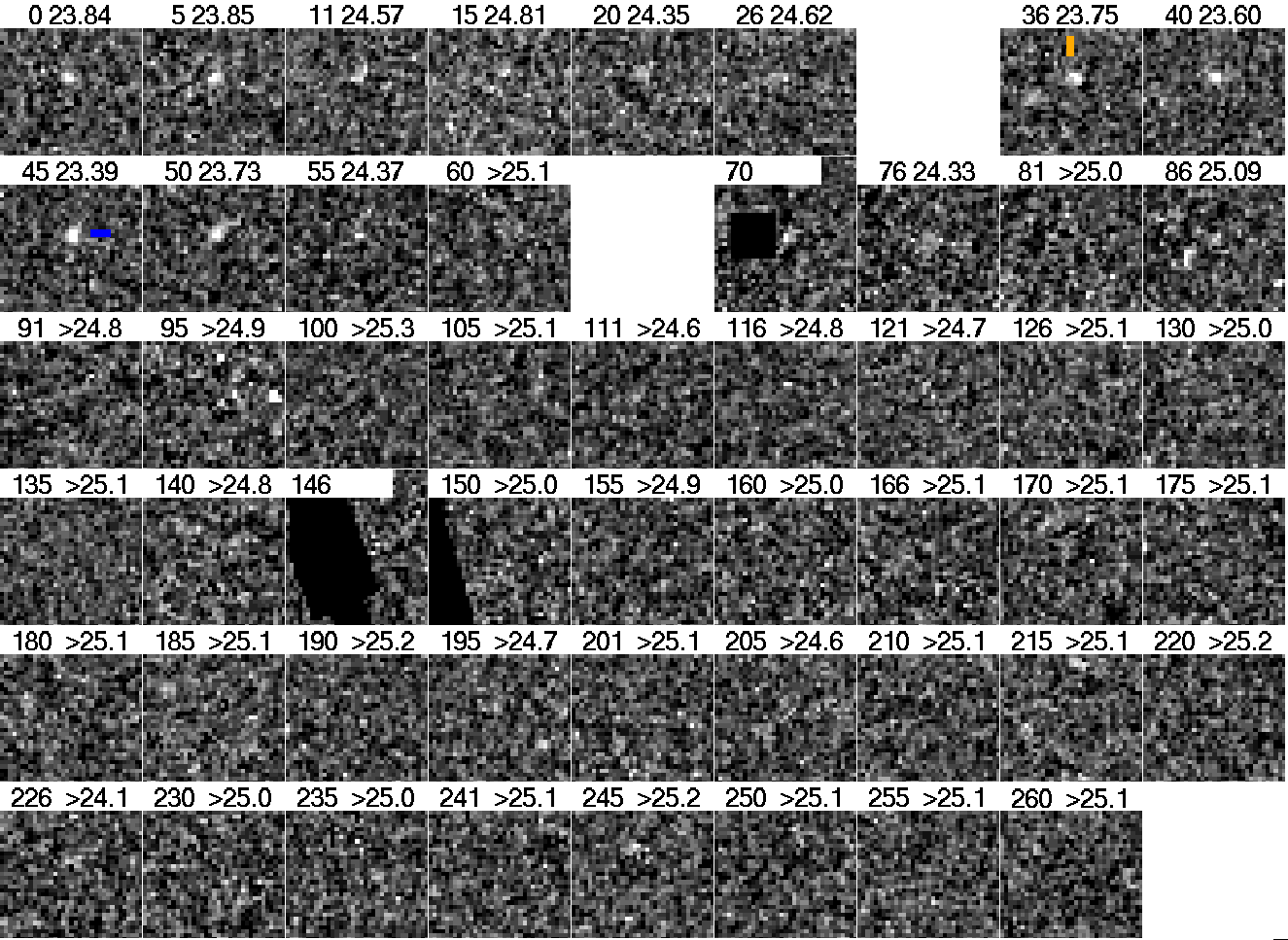}
    \end{tabular}
    \caption{Nova 88}
\end{sidewaysfigure}
\begin{sidewaysfigure}
    \begin{tabular}{cc}
     \multirow{2}{*}[2.6in]{\includegraphics[height=6in]{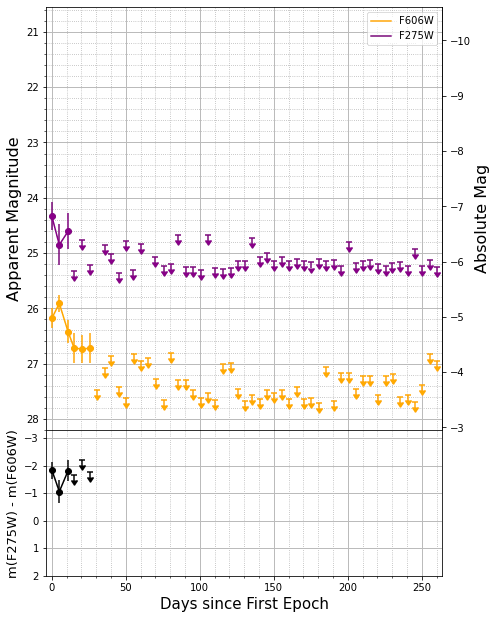}} & \includegraphics[height=3.6in]{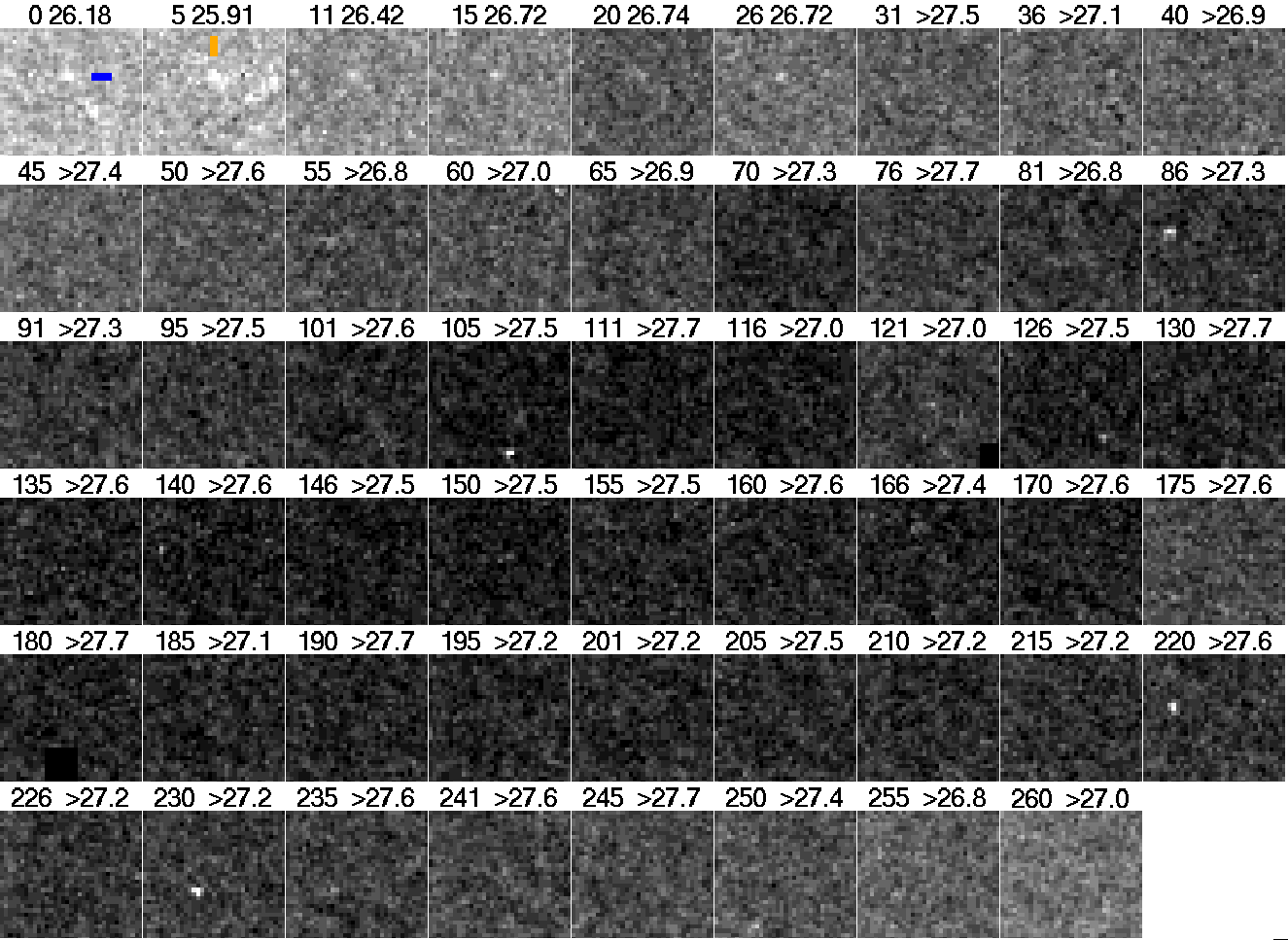} \\
     & \includegraphics[height=3.6in]{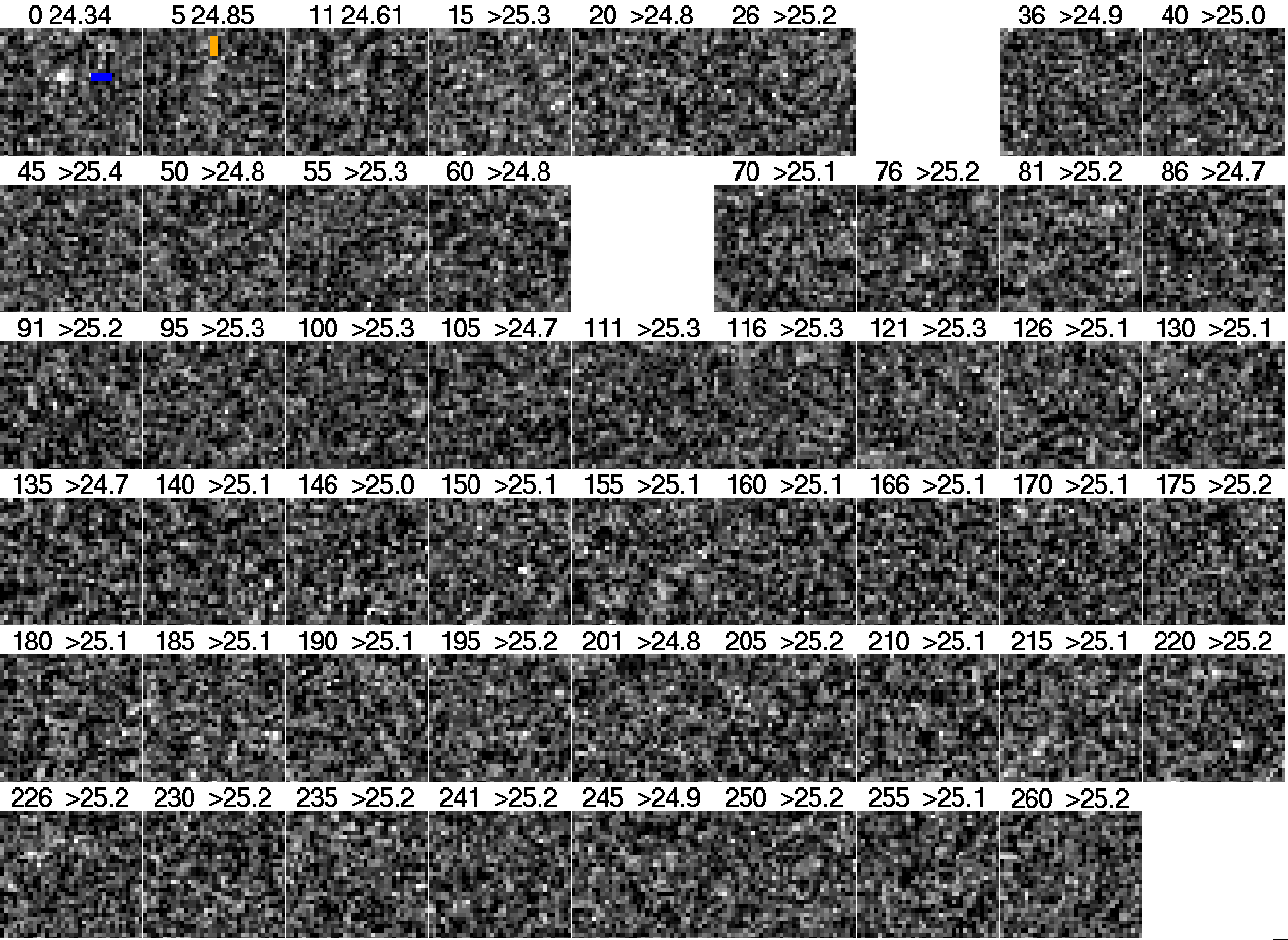}
    \end{tabular}
    \caption{Nova 89}
\end{sidewaysfigure}
\begin{sidewaysfigure}
    \begin{tabular}{cc}
     \multirow{2}{*}[2.6in]{\includegraphics[height=6in]{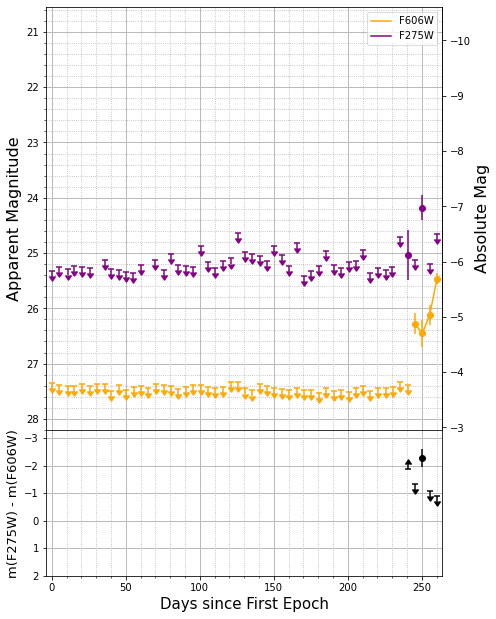}} & \includegraphics[height=3.6in]{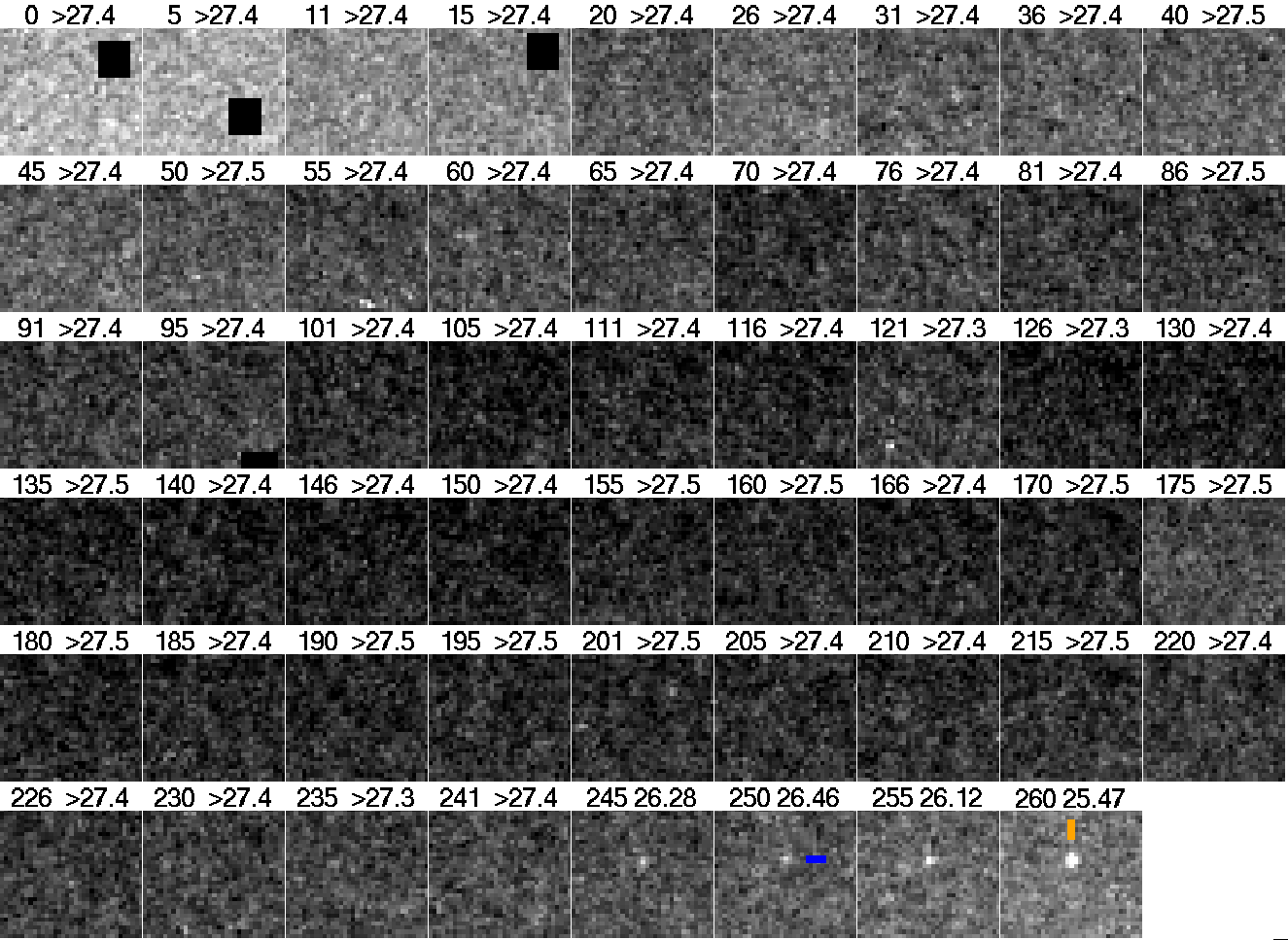} \\
     & \includegraphics[height=3.6in]{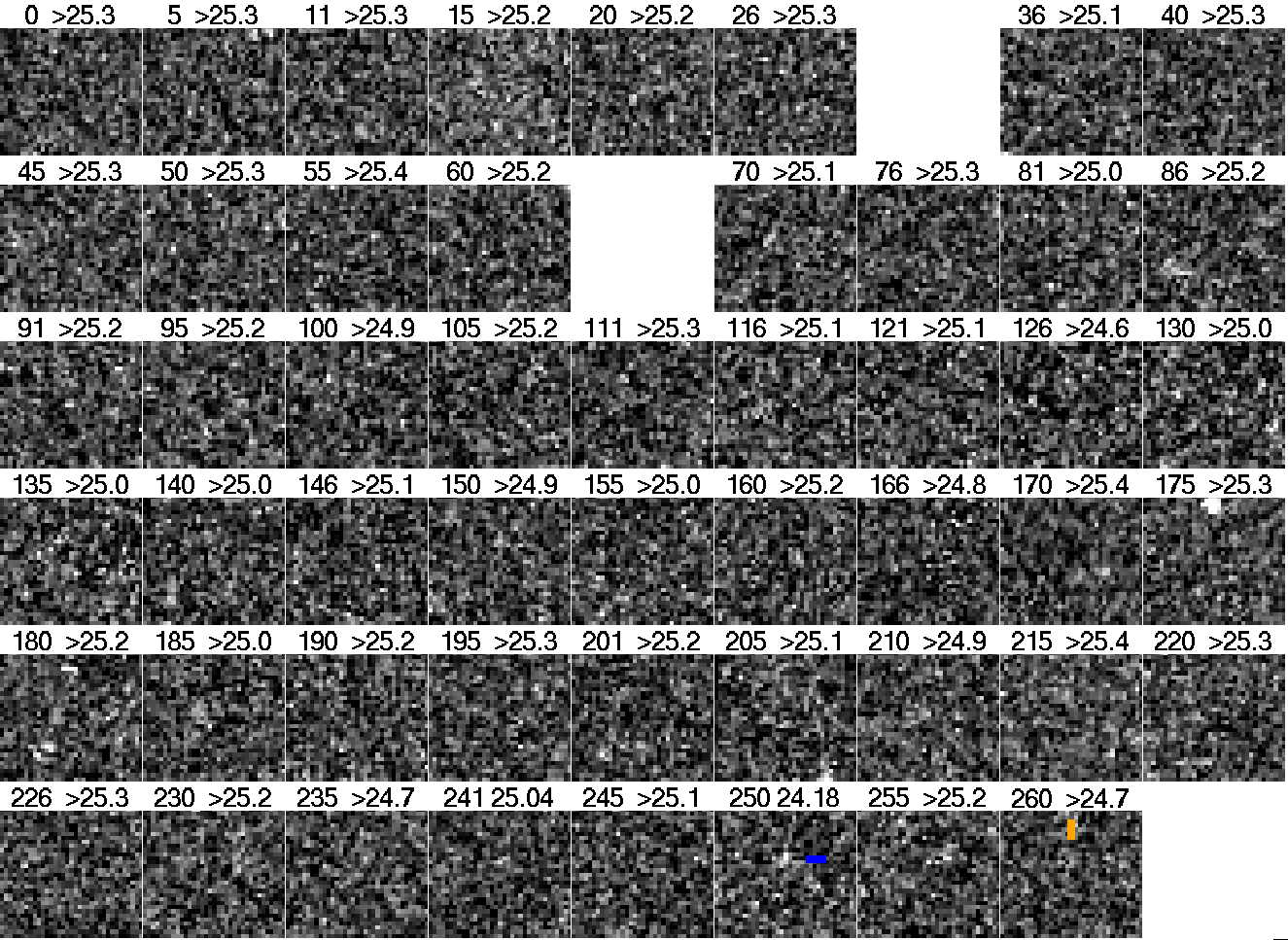}
    \end{tabular}
    \caption{Nova 90}
\end{sidewaysfigure}
\begin{sidewaysfigure}
    \begin{tabular}{cc}
     \multirow{2}{*}[2.6in]{\includegraphics[height=6in]{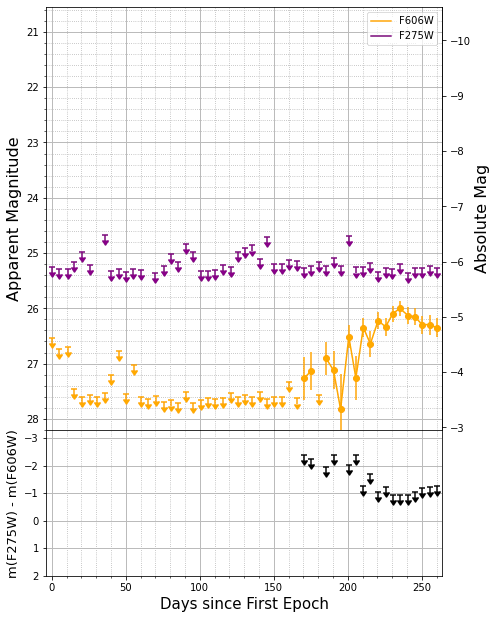}} & \includegraphics[height=3.6in]{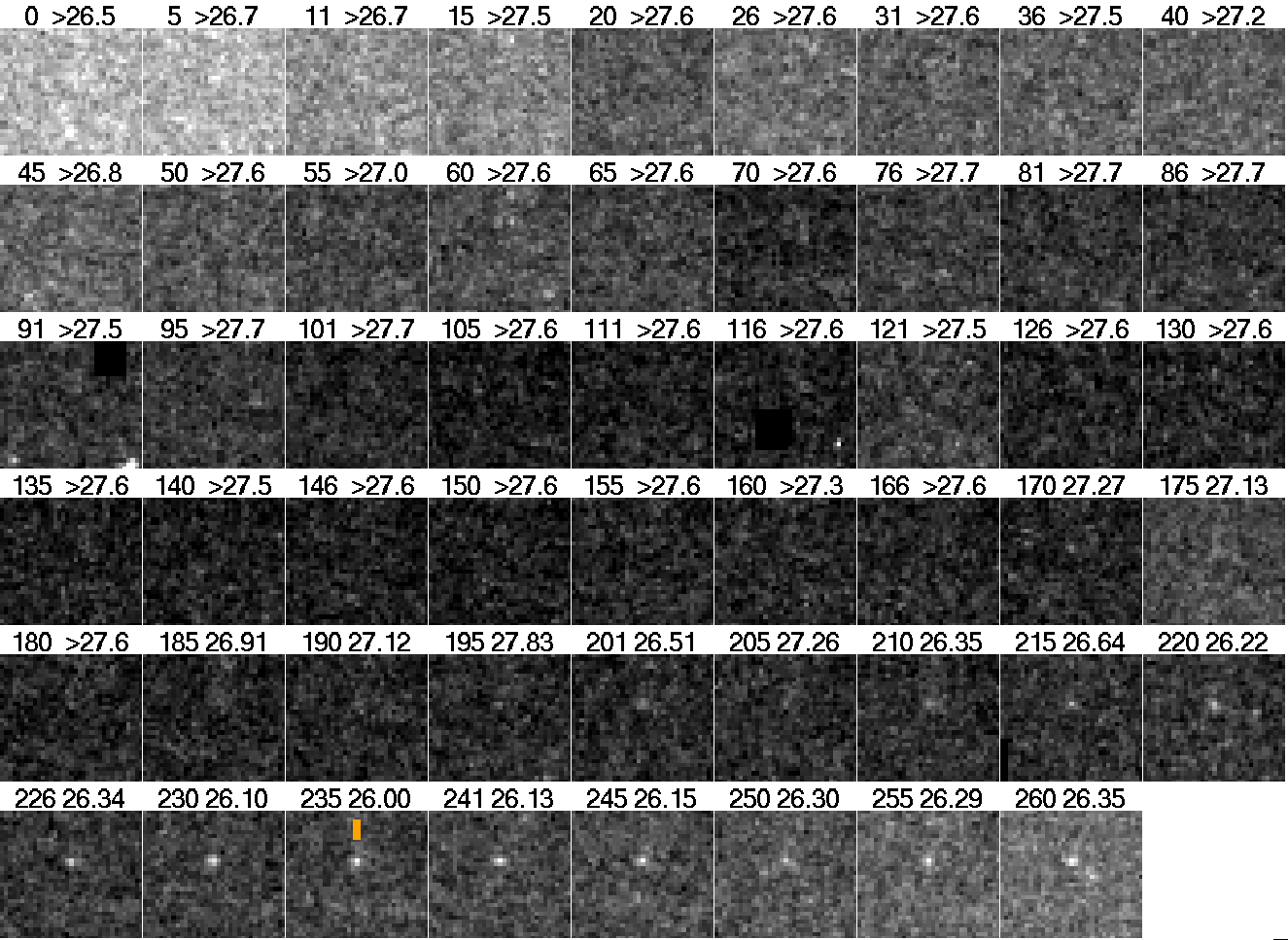} \\
     & \includegraphics[height=3.6in]{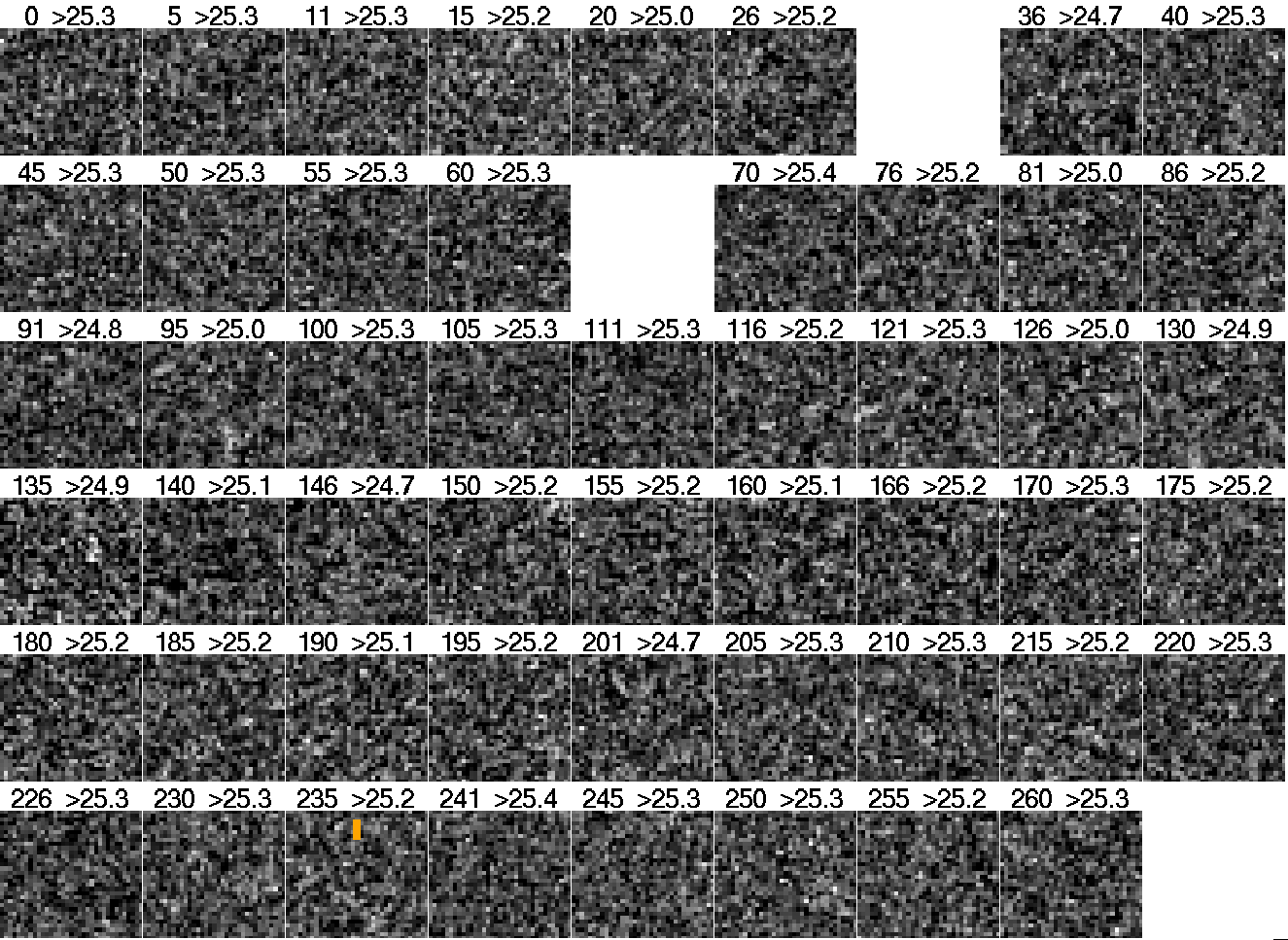}
    \end{tabular}
    \caption{Nova 91}
\end{sidewaysfigure}
\begin{sidewaysfigure}
    \begin{tabular}{cc}
     \multirow{2}{*}[2.6in]{\includegraphics[height=6in]{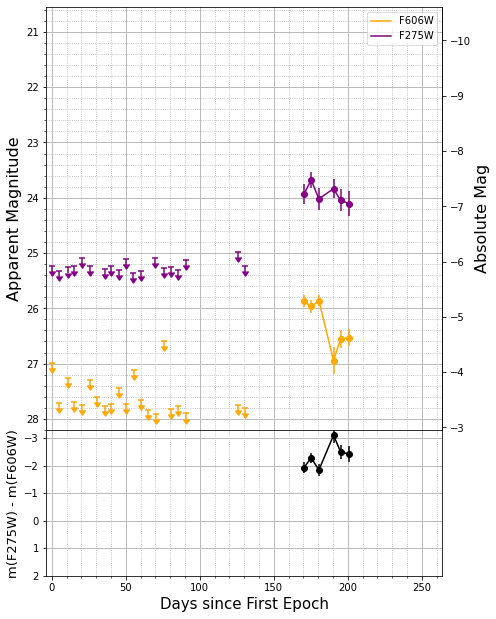}} & \includegraphics[height=3.6in]{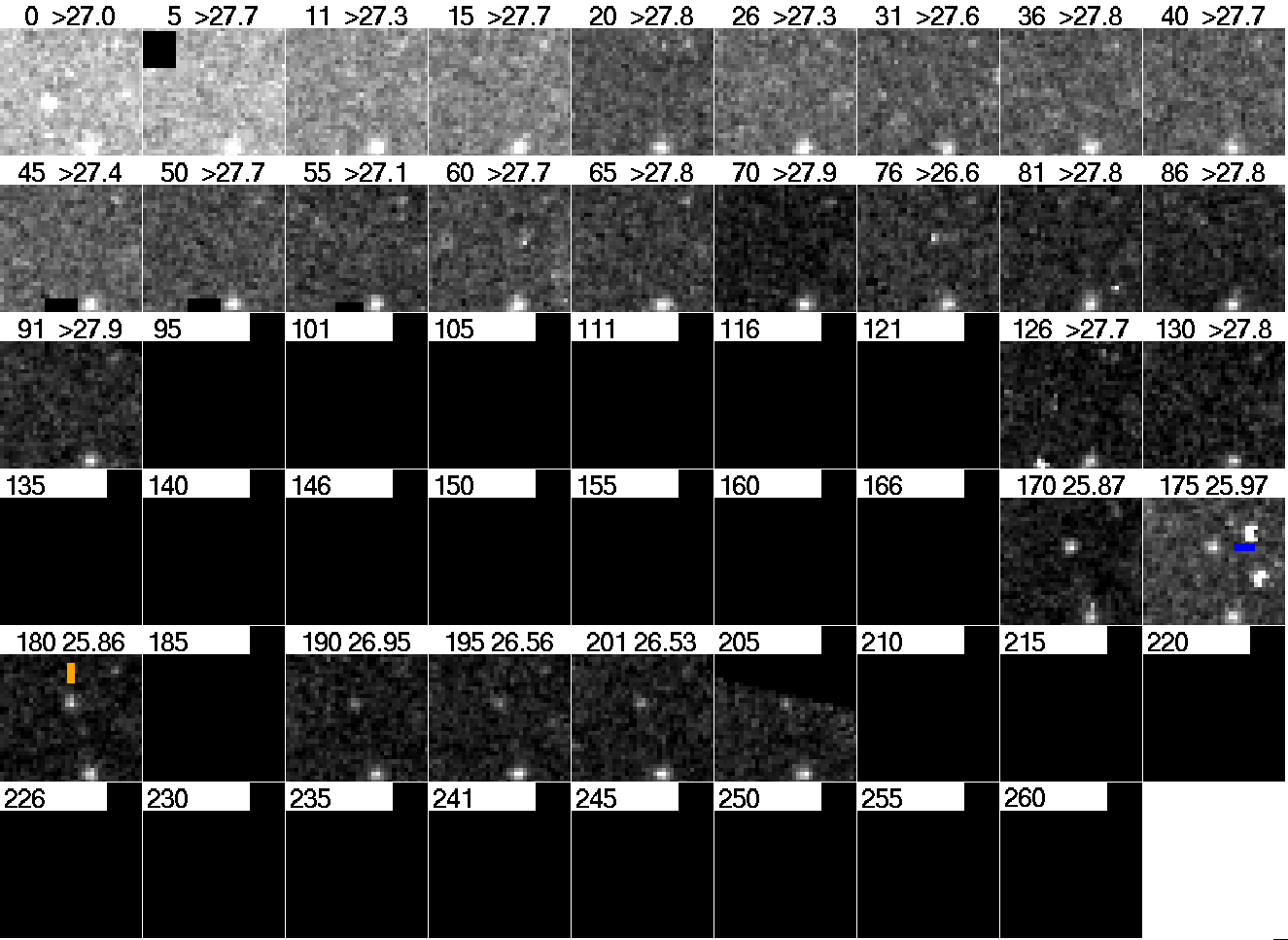} \\
     & \includegraphics[height=3.6in]{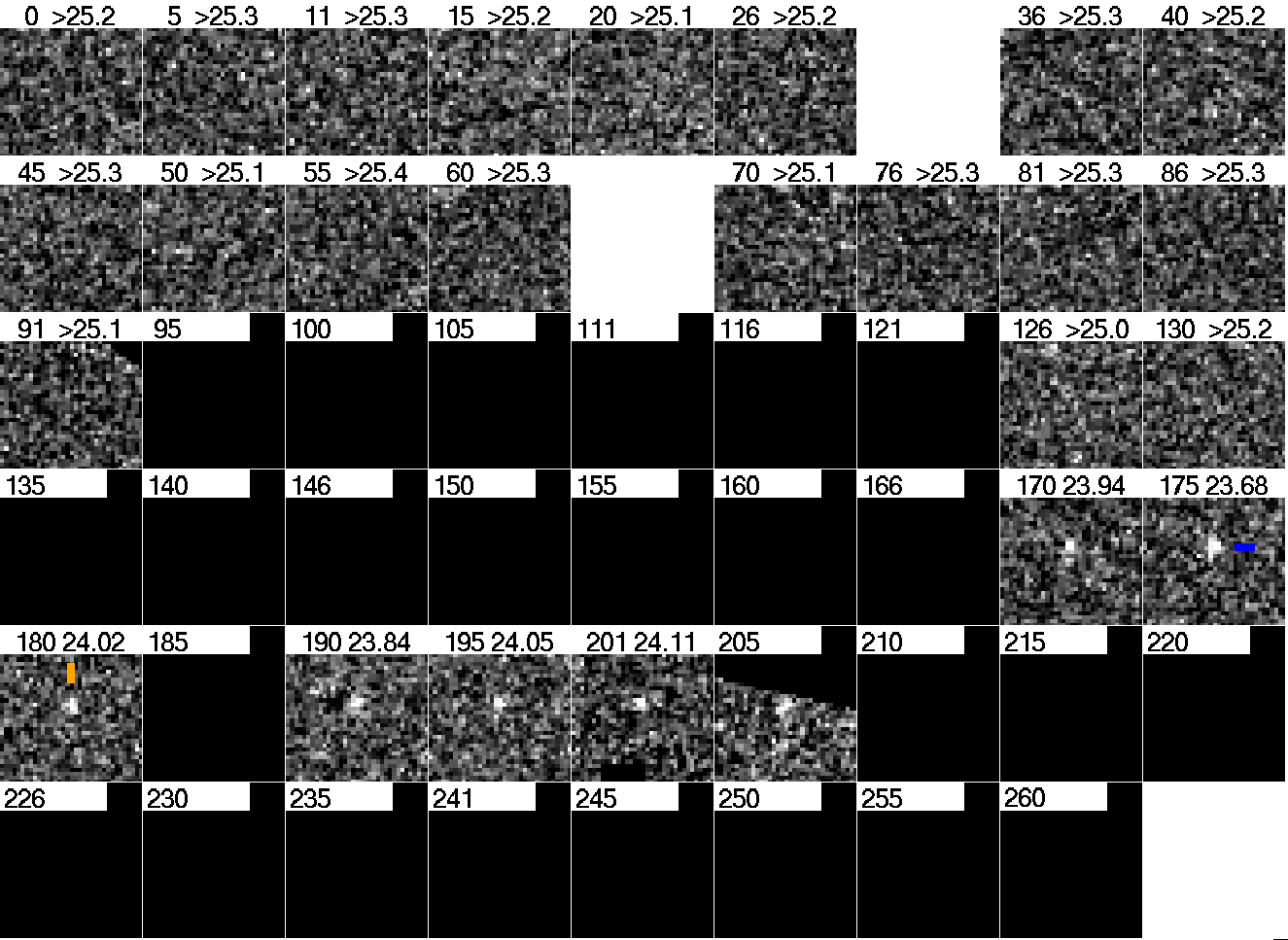}
    \end{tabular}
    \caption{Nova 92}
\end{sidewaysfigure}
\begin{sidewaysfigure}
    \begin{tabular}{cc}
     \multirow{2}{*}[2.6in]{\includegraphics[height=6in]{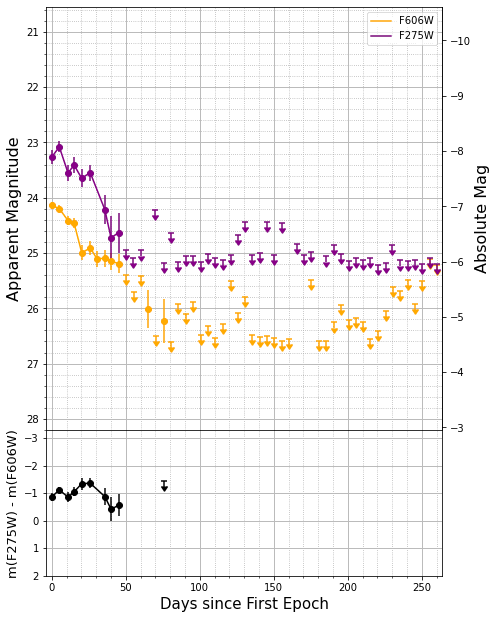}} & \includegraphics[height=3.6in]{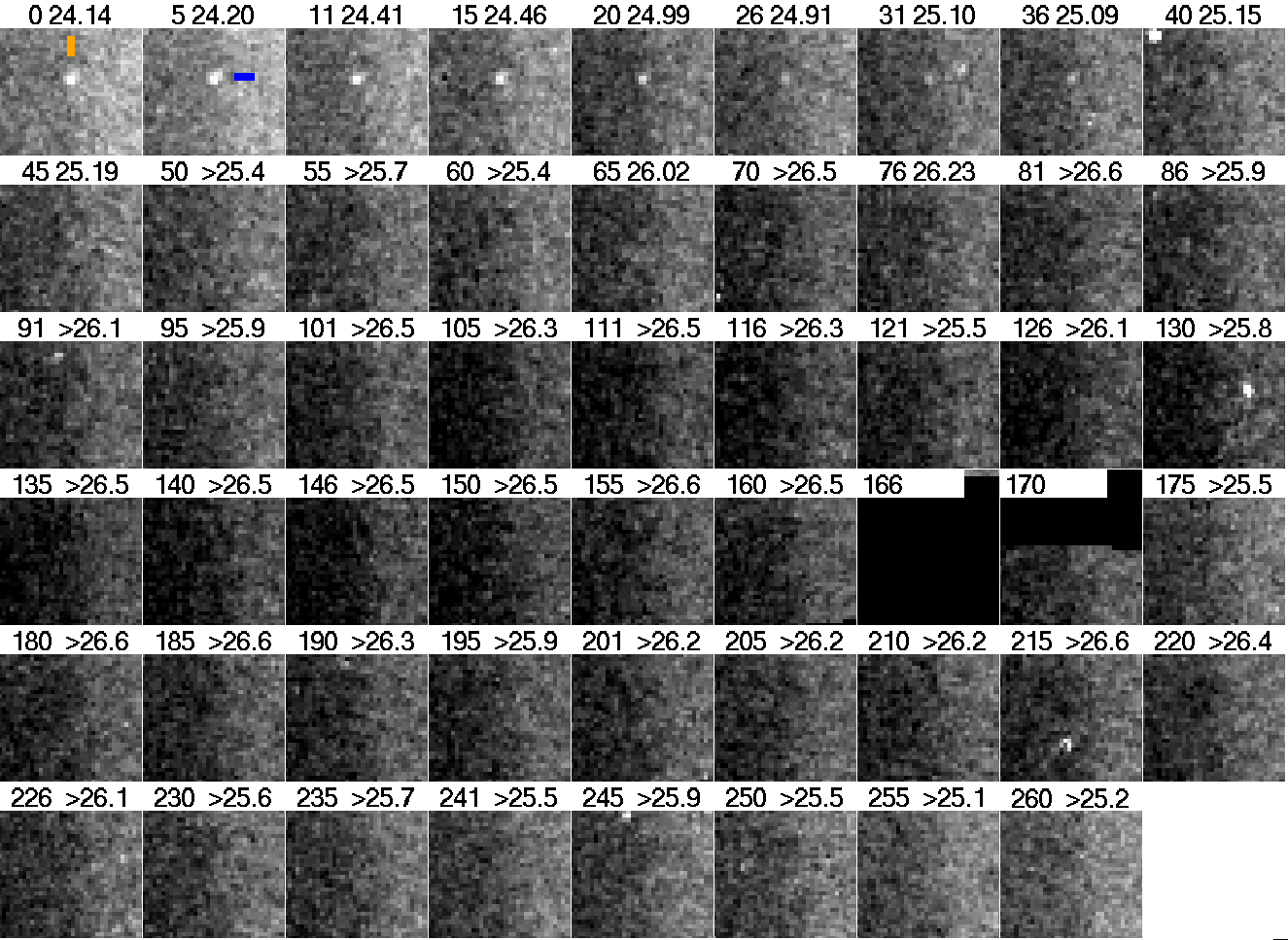} \\
     & \includegraphics[height=3.6in]{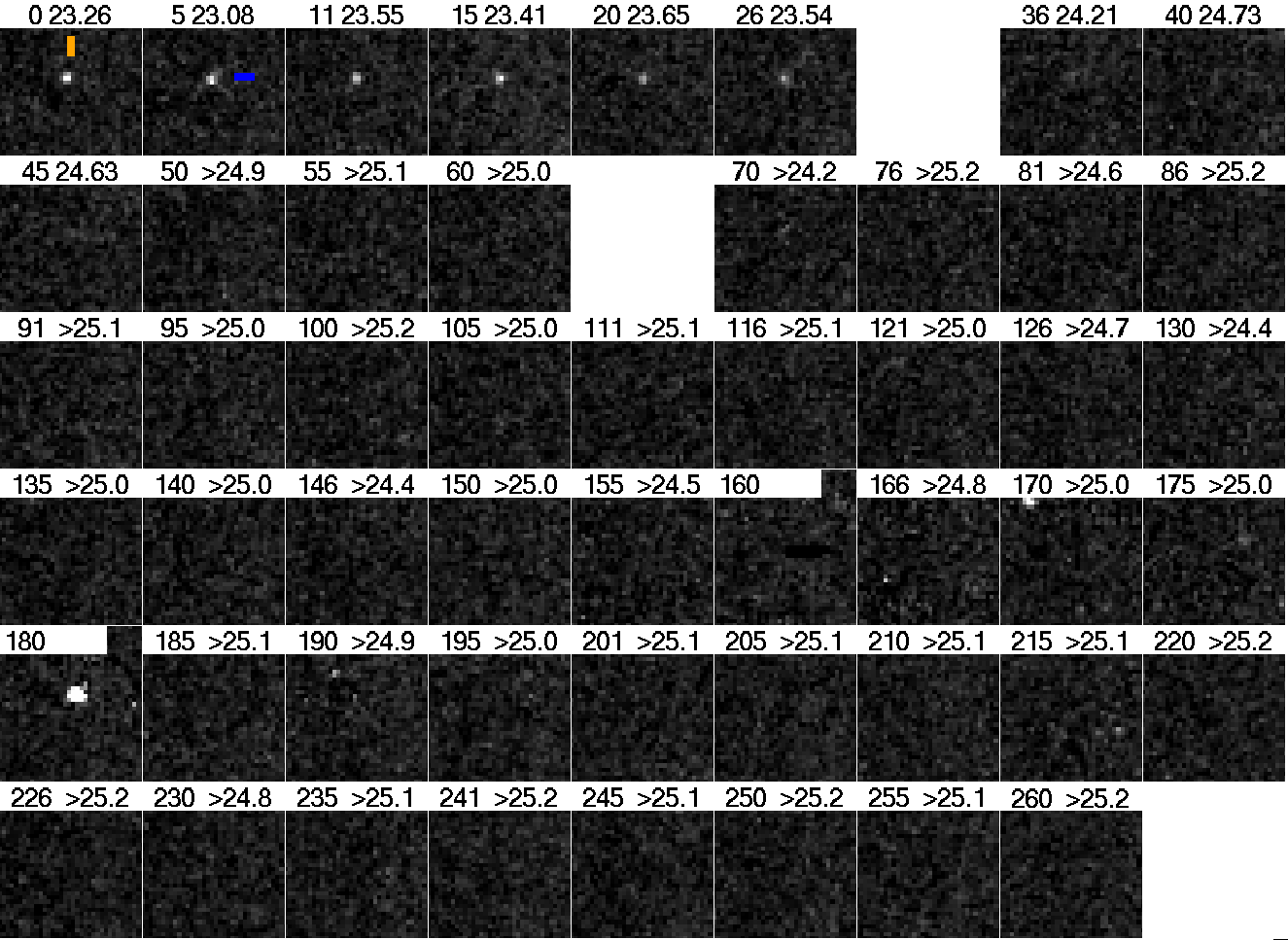}
    \end{tabular}
    \caption{Nova 93}
\end{sidewaysfigure}
\begin{sidewaysfigure}
    \begin{tabular}{cc}
     \multirow{2}{*}[2.6in]{\includegraphics[height=6in]{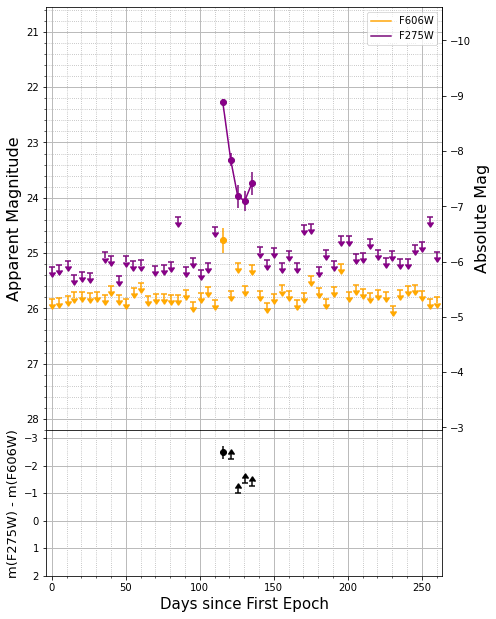}} & \includegraphics[height=3.6in]{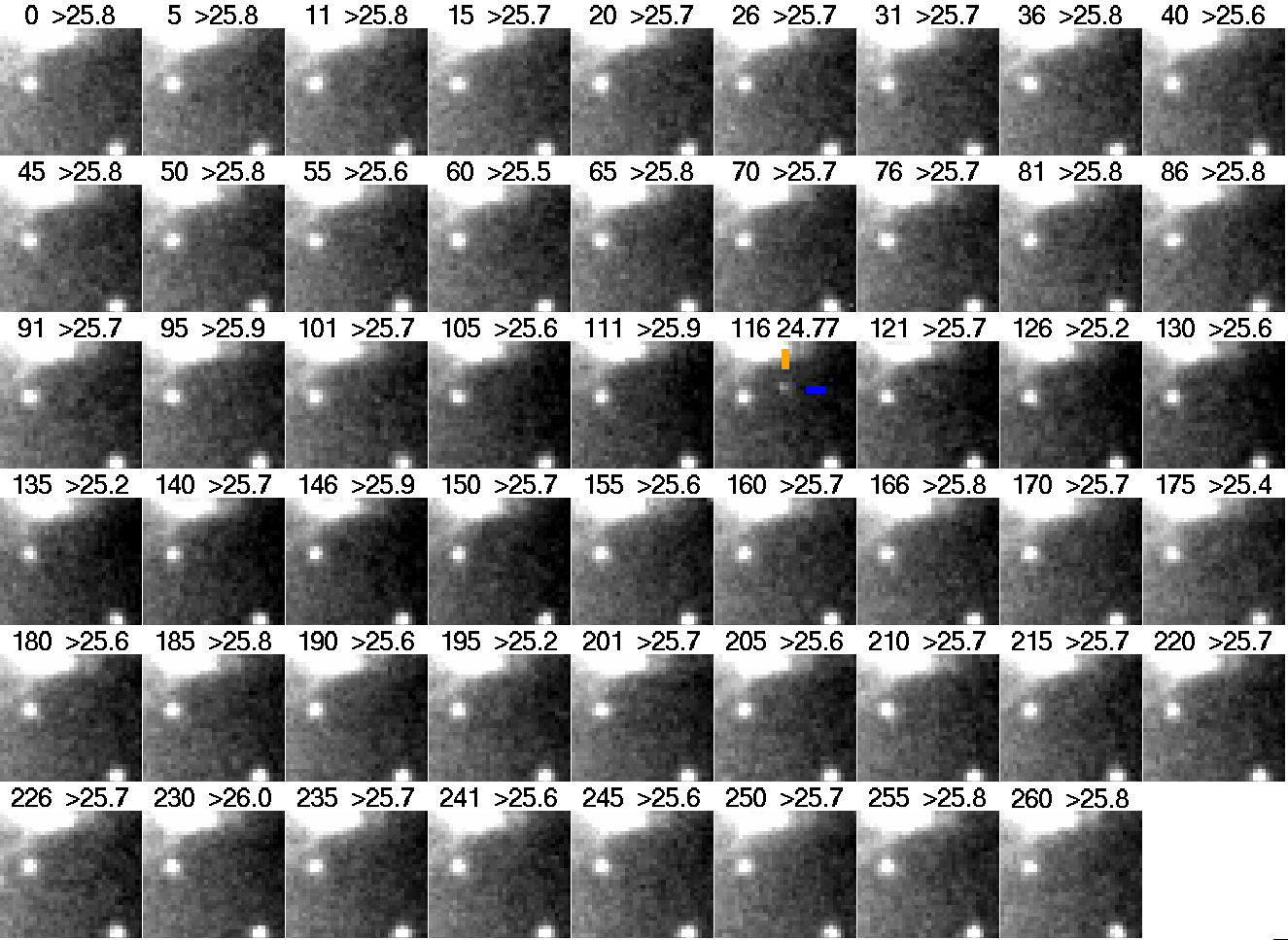} \\
     & \includegraphics[height=3.6in]{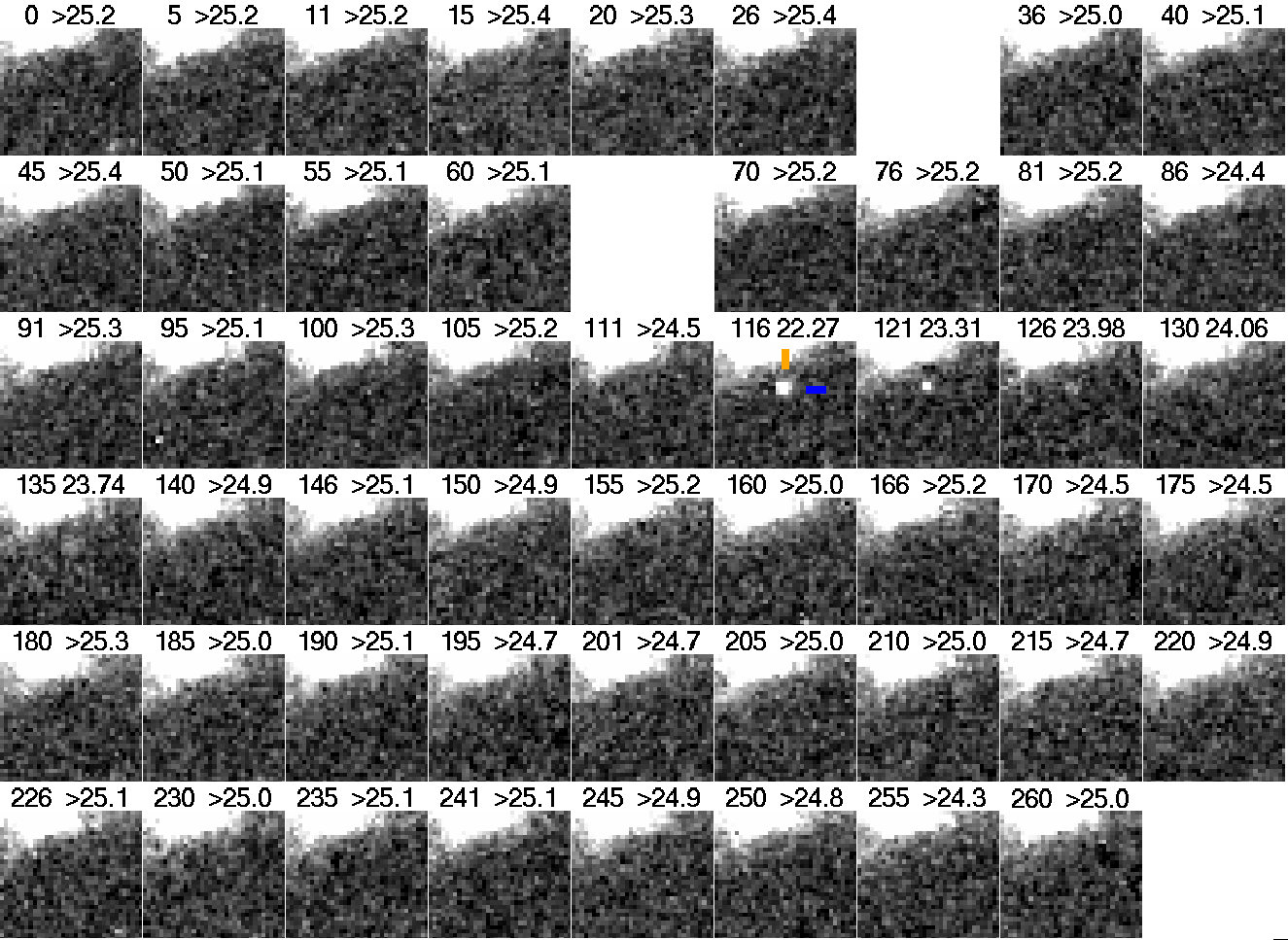}
    \end{tabular}
    \caption{Nova 94}
    \label{fig:nova-94}
\end{sidewaysfigure}

\end{document}